%% file: dl4mg-arxiv-v4.tex
\pdfoutput=1

\RequirePackage{amsmath}
\RequirePackage{mathptmx}

\documentclass[graybox,envcountchap]{svmono}

\usepackage{helvet}
\usepackage{courier}
\usepackage{type1cm}
\usepackage{makeidx}         
\usepackage{graphicx}        
\usepackage[bottom]{footmisc}	
\usepackage{harmony}
\usepackage{musixtex}

\usepackage{a4}
\usepackage{fullpage}

\begin{document}

\begin{center}
{\LARGE
Deep Learning Techniques for Music Generation
{\em -- A Survey}}\\[1cm]
{\large
Jean-Pierre Briot$^{*,}$\footnote{Also
	Visiting Professor at UNIRIO (Universidade Federal do Estado do Rio de Janeiro)
	and Permanent Visiting Professor at PUC-Rio (Pontif\'icia Universidade Cat\'olica do Rio de Janeiro),
	Rio de Janeiro, Brazil.},
Ga\"etan Hadjeres$^\dagger$
and Fran\c{c}ois-David Pachet$^\ddagger$}\\[1cm]
$^*$ Sorbonne Universit\'e, CNRS,
LIP6, F-75005 Paris, France\\
$^\dagger$ Sony Computer Science Laboratories, CSL-Paris, F-75005 Paris, France\\
$^\ddagger$ Spotify Creator Technology Research Lab, CTRL, F-75008 Paris, France\\[1cm]
\end{center}



This paper is a survey and an analysis of different ways of using deep learning\index{Deep!learning} (deep artificial neural networks)
to generate musical content\index{Musical!content}.
We propose a methodology based on five dimensions for our analysis:

\begin{itemize}

\item {\em Objective}\index{Objective}

\begin{itemize}

\item {\em What musical content is to be generated?}

Examples are: melody, polyphony, accompaniment or counterpoint.

\item {\em For what destination and for what use?}

To be performed by a human(s) (in the case of a musical score),
or by a machine (in the case of an audio file).

\end{itemize}

\item {\em Representation}\index{Representation}

\begin{itemize}

\item {\em What are the concepts to be manipulated?}

Examples are: waveform, spectrogram, note, chord, meter and beat.

\item {\em What format is to be used?}

Examples are:  MIDI, piano roll or text.

\item {\em How will the representation be encoded?}

Examples are: scalar, one-hot or many-hot.


\end{itemize}

\item {\em Architecture}\index{Architecture}

\begin{itemize}

\item {\em What type(s) of deep neural network is (are) to be used?}

Examples are: feedforward network, recurrent network, autoencoder or generative adversarial networks.

\end{itemize}

\item {\em Challenge}\index{Challenge}

\begin{itemize}

\item {\em What are the limitations\index{Limitation} and open challenges?}

Examples are: variability, interactivity and creativity.

\end{itemize}

\item {\em Strategy}\index{Strategy}

\begin{itemize}

\item {\em How do we model and control the process of generation?}

Examples are: single-step feedforward, iterative feedforward, sampling or input manipulation.

\end{itemize}

\end{itemize}

For each dimension, we conduct a comparative analysis of various models and techniques
and we propose some tentative multidimensional typology\index{Typology}.
This typology is {\em bottom-up}, based on the analysis of many existing deep-learning based systems for music generation selected from the relevant literature.
These systems are described
and are used to exemplify
the various choices of objective, representation, architecture, challenge and strategy.
The last section
includes some discussion and some prospects.

Supplementary material is provided at the following companion web site:

\begin{center}
www.briot.info/dlt4mg/
\end{center}

This paper is a simplified (weak DRM\footnote{In addition to including high quality color figures,
	the book
	includes:
	a table of contents,
	a list of tables,
	a list of figures,
	a table of acronyms,
	a glossary
	and an index.})
version of the following book \cite{briot:dlt4mg:book:2019}:
Jean-Pierre Briot, Ga{\"e}tan Hadjeres and Fran{\c{c}}ois-David Pachet,
Deep Learning Techniques for Music Generation,
Computational Synthesis and Creative Systems,
Springer,
2019.
Hardcover ISBN: 978-3-319-70162-2.
eBook ISBN: 978-3-319-70163-9.
Series ISSN: 2509-6575.

\include{introduction}

\include{method}

\include{objective}
\include{representation}

\include{architecture}
\include{challenge-strategy}
\include{analysis}

\include{discussion-conclusion}

\bibliographystyle{plain}
\bibliography{dl4mg}

\end{document}

%% file: introduction.tex
\chapter{Introduction}
\label{section:chapter:introduction}

\abstract*{Chapter~\ref{section:chapter:introduction} Introduction provides motivation for this book.
It includes a short summary of the history of computer music and music generation
(including previous uses of artificial neural networks as well as other models such as grammars, rules and Markov chains)
up to the recent rise of deep learning.
It also describes the organization and the public of the book as well as related work.}

\label{section:introduction}

\label{section:introduction:context}

Deep learning\index{Deep!learning} has recently become a fast growing domain
and is now used routinely for classification\index{Classification} and prediction\index{Prediction} tasks,
such as image recognition\index{Image!recognition}, voice recognition or translation\index{Translation}.
It became popular in 2012, when a deep learning architecture significantly outperformed standard techniques relying on handcrafted features
in an image classification competition, see more details in Section~\ref{section:architectures:history}.

We may explain this success and reemergence of artificial neural network\index{Artificial!neural network} techniques by the combination of:

\begin{itemize}

\item availability of {\em massive data};

\item
	availability of {\em efficient and affordable computing power}\footnote{Notably, thanks to graphics processing units\index{Graphics processing unit} (GPU\index{GPU}),
	initially designed for video games,
	which have now one of their biggest markets in data science and deep learning applications.};

\item {\em technical advances},
such as:

\begin{itemize}

\item {\em pre-training\index{Pre-training}},
which resolved initially inefficient training of neural networks with many layers \cite{hinton:fast:algorithm:2006}\footnote{Although nowadays
	it has being replaced by other techniques, such as batch normalization \cite{ioffe:batch:normalization:arxiv:2015} and deep residual learning \cite{he:resnet:deep:residual:arxiv:2015}.};

\item {\em convolutions\index{Convolution}},
which provide motif translation invariance \cite{le:cun:convolutional:handbook:1998};

\item LSTM\index{LSTM} (long short-term memory\index{Long!short-term memory}),
which resolved
initially inefficient training of recurrent neural networks \cite{hochreiter:lstm:1997}.

\end{itemize}


\end{itemize}

There is no consensual definition for deep learning\index{Deep!learning}.
It is a repertoire of machine learning\index{Machine learning} (ML\index{ML}) techniques,
based on artificial neural networks\index{Artificial!neural network}.
The key aspect and common ground is the term {\em deep}.
This means that there are multiple layers\index{Layer} processing multiple hierarchical\index{Hierarchical} levels of abstractions\index{Abstraction},
which are automatically extracted from data\footnote{That said,
	although deep learning will automatically extract significant features from the data,
	manual choices of input representation, e.g., spectrum vs raw wave signal for audio,
	may be very significant for the accuracy of the learning and for the quality of the generated content,
	see Section~\ref{section:representation:feature:extraction}.}.
Thus a deep architecture can manage and decompose complex representations\index{Representation} in terms of simpler representations.
The technical foundation is mostly artificial neural networks, as we will see in Chapter~\ref{section:chapter:architecture},
with many extensions,
such as:
convolutional networks\index{Convolutional!network|see{Convolutional neural network}},
recurrent networks\index{Recurrent!network|see{Recurrent neural network}},
autoencoders\index{Autoencoder},
and restricted Boltzmann machines\index{Restricted Boltzmann machine}.
For more information about the history and various facets of deep learning, see, e.g., a recent comprehensive book on the domain
\cite{goodfellow:deep:learning:book:2016}.

Driving applications of deep learning are traditional machine learning tasks\index{Task}\footnote{Tasks in machine learning are types of problems
	and may also be described in terms of how the machine learning system should process an example
	\cite[Section~5.1.1]{goodfellow:deep:learning:book:2016}.
	Examples 	are: classification, regression and anomaly detection\index{Anomaly detection}.}:
{\em classification}\index{Classification} (for instance, identification\index{Identification} of images)
and {\em prediction}\index{Prediction}\footnote{As a testimony of the initial DNA of neural networks:
	{\em linear regression}\index{Linear!regression}\index{Regression}
	and {\em logistic regression}\index{Logistic!regression}, see Section~\ref{section:architecture:linear:regression}.}
(for instance, of the weather)
and also more recent ones such as {\em translation\index{Translation}}.

But a growing area of application of deep learning techniques is the {\em generation of  content\index{Content}\index{Content!generation}}.
Content can be of various kinds:
images\index{Image}, text\index{Text} and music\index{Music}, the latter being the focus of our analysis.
The motivation is in using now widely available various corpora\index{Corpus} to automatically learn\index{Learning} musical {\em styles\index{Style}}
and to generate\index{Generation} {\em new} musical content based on this.

\section{Motivation}
\label{section:introduction:motivation}

\subsection{Computer-Based Music Systems}
\label{section:introduction:motivation:topic}

The first music generated by computer appeared in 1957.
It was a 17 seconds long melody named ``The Silver Scale'' by its author Newman Guttman
and was generated by a software for sound synthesis named Music I\index{Music I}, developed by Mathews at Bell Laboratories.
The same year,
``The Illiac Suite\index{Illiac Suite}'' was the first score composed by a computer \cite{lejaren:illiac:book:1959}.
It was named
after
the ILLIAC I computer at the University of Illinois at Urbana-Champaign (UIUC)
in the United States.
The human ``meta-composers'' were Lejaren A. Hiller and Leonard M. Isaacson, both musicians and scientists.
It was an early example of algorithmic composition\index{Algorithmic composition},
making use of stochastic\index{Stochastic} models (Markov chains\index{Markov!chain}) for generation 
as well as rules\index{Rule} to filter generated material according to desired properties.

In the domain of sound synthesis, a landmark was the release in 1983 by Yamaha of the DX 7\index{DX7} synthesizer\index{Synthesizer},
building on groundwork by Chowning on a model of synthesis\index{Synthesis}
based on frequency modulation\index{Frequency!modulation} (FM\index{FM}).
The same year, the MIDI\index{MIDI}\footnote{Musical instrument digital interface\index{Musical!instrument digital interface},
	to be introduced in Section~\ref{section:representation:midi}.}
interface was launched, as a way to interoperate various software and instruments (including the Yamaha DX 7 synthesizer\index{Synthesizer}).
Another landmark was the development by Puckette at IRCAM of the Max/MSP\index{Max/MSP} real-time interactive processing environment,
used for real-time synthesis and for interactive performances.

Regarding algorithmic composition,
in the early 1960s
Iannis Xenakis\index{Xenakis}
explored the idea of stochastic composition\footnote{One of the first documented case of {\em stochastic music\index{Stochastic!music}},
	long before computers,
	is the Musikalisches Wurfelspiel (Dice Music), attributed to Wolfgang Amadeus Mozart\index{Mozart}.
	It was designed for using dice to generate music
	by concatenating randomly selected predefined music segments
	composed in a given style (Austrian waltz in a given key).}
\cite{xenakis:formalized:music:book:1963},
in his composition named ``Atr\'ees'' in 1962.
The idea involved using computer fast computations to calculate various possibilities
from a set of probabilities designed by the composer
in order to generate samples of musical pieces to be selected.
In another approach following the initial direction of ``The Illiac Suite'',
grammars and rules were used to specify the style of a given corpus or more generally tonal music theory.
An example is the generation in the 1980s by Ebcio\u{g}lu's composition software named CHORAL\index{CHORAL}
of a four-part chorale in the style of Johann Sebastian Bach\index{Bach},
according to over 350 handcrafted rules \cite{ebcioglu:expert:system:bach:cmj:1988}.
In the late 1980s David Cope's system named Experiments in Musical Intelligence\index{Experiments in musical intelligence} (EMI\index{EMI})
extended that approach with the capacity to learn from a corpus of scores of a composer to create its own grammar and database of rules
\cite{cope:algorithmic:composer:book:2000}.

Since then, computer music\index{Computer music} has continued developing
for the general public, if we consider, for instance, the GarageBand\index{GarageBand}
music composition and production
application for Apple platforms (computers, tablets and cellphones),
as an offspring of the initial Cubase sequencer software, released by Steinberg in 1989.

For more details about the history and principles of computer music in general,
see, for example, the book by Roads \cite{roads:computer:music:tutorial:book:1996}.
For more details about the history and principles of algorithmic composition,
see, for example, \cite{maurer:brief:history:algorithmic:composition:1999}
and the books by Cope \cite{cope:algorithmic:composer:book:2000}
or Dean and McLean \cite{dean:oxford:handbook:2018}.

\subsection{Autonomy versus Assistance}
\label{section:introduction:motivation:assistance:versus}

When talking about computer-based music generation, there is actually some ambiguity about whether the objective is

\begin{itemize}

\item to design and construct {\em autonomous} music-making systems --
two recent examples being the deep-learning based Amper\texttrademark\,and Jukedeck systems/companies
aimed at the creation of original music for commercials and documentary; or

\item to design and construct computer-based environments to {\em assist} human musicians
(composers, arrangers, producers, etc.) --
two examples being the FlowComposer environment developed at Sony CSL-Paris \cite{papadopoulos:flow:composer:cp:2016}
(introduced in Section~\ref{section:systems:flow:composer})
and the OpenMusic environment developed at IRCAM \cite{assayag:openmusic:cmj:1999}.

\end{itemize}

The quest for autonomous music-making systems may be an interesting perspective
for exploring the process of composition\footnote{As Richard Feynman\index{Richard Feynman} coined it:
	``What I cannot create, I do not understand.''}
and it also serves as an evaluation method.
An example of a musical Turing test\footnote{Initially codified in 1950 by Alan Turing\index{Alan Turing}\index{Turing|see{Alan Turing}}
	and named by him the ``imitation game'' \cite{turing:test:mind:1950},
	the ``Turing test''\index{Turing!test} is a test of the ability for a machine to exhibit intelligent behavior equivalent to
	(and more precisely, indistinguishable from) the behavior of a human.
	In his imaginary experimental setting, Turing proposed the test to be a natural language conversation
	between a human (the evaluator) and a hidden actor (another human or a machine).
	If the evaluator cannot reliably tell the machine from the human, the machine is said to have passed the test.}
will be introduced in Section~\ref{section:experiment:deep:bach}.
It consists in presenting to various members of the public (from beginners to experts) chorales composed
by J. S. Bach\index{Bach}
or generated by a deep learning system and played by human musicians\footnote{This is to avoid
	the bias (synthetic flavor) of a computer rendered generated music.}.
As we will see in the following, deep learning techniques turn out to be very efficient at succeeding in such tests,
due to their capacity to learn musical style from a given corpus and to generate new music that fits into this style.
That said, we consider that such a test is more a means than an end.

A broader perspective is in assisting human musicians during the various steps
of music creation: composition, arranging, orchestration, production, etc.
Indeed,
to compose or to improvise\footnote{Improvisation
	is a form of real time composition.},
a musician rarely creates new music from scratch.
S/he reuses and adapts, consciously or unconsciously, features from various music that s/he already knows
or has heard,
while following some principles and guidelines, such as theories about harmony and scales.
A computer-based musician assistant may act during different stages of the composition,
to initiate, suggest, provoke and/or complement
the inspiration of the human composer.

That said, as we will see,
the majority of current deep-learning based systems for generating music
are still focused on autonomous generation, although more and more systems are addressing
the issue of human-level control and interaction.

\subsection{Symbolic versus Sub-Symbolic AI}
\label{section:introduction:motivation:symbolic:versus}

Artificial Intelligence\index{Artificial!intelligence} (AI\index{AI}) is often divided into two main streams\footnote{With some precaution,
	as this division is not that strict.}:

\begin{itemize}

\item symbolic\index{Symbolic} AI -- dealing with high-level symbolic representations (e.g., chords, harmony\ldots)
and processes (harmonization, analysis\ldots); and

\item sub-symbolic\index{Sub-symbolic} AI -- dealing with low-level representations (e.g., sound, timbre\ldots) and processes
(pitch recognition, classification\ldots).

\end{itemize}

Examples of symbolic models used for music are rule-based systems\index{Rule!-based system} or grammars\index{Grammar} to represent harmony.
Examples of sub-symbolic models used for music are machine learning algorithms for automatically learning musical styles from a corpus of musical pieces.
%
These models can then be used in a generative and interactive manner, to help musicians in creating new music,
by taking advantage of this added ``intelligent'' memory (associative, inductive and generative)
to suggest proposals, sketches, extrapolations, mappings, etc.
This is now feasible because of the growing availability of music in various forms,
e.g., sound, scores and MIDI files, which can be automatically processed by computers.

A recent example of an integrated music composition environment is FlowComposer\index{FlowComposer} \cite{papadopoulos:flow:composer:cp:2016},
which we will introduce in Section~\ref{section:systems:flow:composer}.
It offers various symbolic and sub-symbolic techniques,
e.g., Markov chains for modeling
style,
a constraint solving module for expressing constraints,
a rule-based module to produce harmonic analysis;
and an audio mapping module to produce rendering.
Another example of an integrated music composition environment is OpenMusic\index{OpenMusic} \cite{assayag:openmusic:cmj:1999}.

However, a deeper integration of sub-symbolic techniques, such as deep learning,
with symbolic techniques, such as constraints and reasoning,
is still an open issue\footnote{The general objective of integrating sub-symbolic and symbolic levels
	into a complete AI system
	is among the ``Holy Grails'' of AI.},
although some partial integrations in restricted contexts already exist
(see, for example, Markov constraints in \cite{pachet:markov:constraints:constraints:2011,barbieri:lyrics:style:2012}
and an example of use for FlowComposer in Section~\ref{section:systems:flow:composer}).

\subsection{Deep Learning}
\label{section:introduction:motivation:techniques}

The motivation for using deep learning (and more generally machine learning techniques)
to generate musical content is its {\em generality\index{Generality}}.
As opposed to handcrafted models\index{Handcrafted!model},
such as
grammar-based\index{Grammar}
\cite{steedman:generative:grammar:blues:mp:1984}
or rule-based\index{Rule!-based system} music generation systems
\cite{ebcioglu:expert:system:bach:cmj:1988},
a machine learning-based generation system
can be agnostic\index{Agnostic}, as it learns a model from an arbitrary corpus\index{Corpus} of music.
As a result, the same system may be used for various musical genres\index{Musical!genre}.

Therefore, as more large scale musical datasets
are made available,
a machine learning-based generation system will be able to automatically learn\index{Learning} a musical style\index{Style} from a corpus
and to generate\index{Generation} new musical content.
As stated by Fiebrink and Caramiaux \cite{fiebrink:ml:creative:tool:arxiv:2016},
some benefits are

\begin{itemize}

\item it can make creation feasible when the desired application is too complex to be described by analytical formulations or
manual brute force design, and

\item learning algorithms are often less brittle\index{Brittle} than manually designed rule sets
and learned rules are more likely to generalize\index{Generalization} accurately to new contexts in which inputs may change.


\end{itemize}

Moreover, as opposed to structured representations like rules\index{Rule} and grammars\index{Grammar},
deep learning is good at processing raw unstructured data, from which its hierarchy of layers will extract higher level representations adapted to the task.

\subsection{Present and Future}
\label{section:introduction:motivation:present:future}

As we will see, the research domain in deep learning-based music generation has turned hot recently,
building on initial work using artificial neural networks to generate music
(e.g., the pioneering experiments by Todd in 1989 \cite{todd:connectionist:composition:1989}
and the CONCERT system developed by Mozer in 1994 \cite{mozer:composition:prediction:1994}),
while creating an active stream of new ideas and challenges made possible thanks to the progress of deep learning.
Let us also note the growing interest by some private big actors of digital media
in the computer-aided generation of artistic content,
with the creation by Google in June 2016 of the Magenta research project \cite{google:magenta:project:web}
and the creation by Spotify in September 2017 of the Creator Technology Research Lab (CTRL) \cite{spotify:ctrl:2017}.
This is likely to contribute to blurring the line between music creation and music consumption
through the personalization of musical content \cite{amato:ai:media:arxiv:2019}.

\section{This Book}
\label{section:introduction:motivation:book}

The lack (to our knowledge) of a comprehensive survey and analysis of this active research domain motivated the writing of this book,
built in a {\em bottom-up\index{Bottom-up}} way from the analysis of numerous recent research works.
The objective is to provide a comprehensive description of the issues and techniques for using deep learning to generate music,
illustrated through the analysis of various architectures, systems and experiments presented in the literature.
We also propose a conceptual framework and typology aimed at a better understanding of the design decisions for current
as well as future systems.



\subsection{Other Books and Sources}
\label{section:introduction:related:work:books}

To our knowledge, there are only a few partial attempts at analyzing the use of deep learning for generating music.
In \cite{briot:dlt4mg:arxiv:2017}, a very preliminary version of this work,
Briot {\em et al.} proposed a first survey of various systems through a multicriteria analysis
(considering as dimensions the objective, representation, architecture and strategy).
We have extended and consolidated this study
by integrating as an additional dimension the challenge (after having analyzed it in \cite{briot:mgbdl:cd:ncaa:2018}).

In \cite{graves:generating:sequences:rnn:arxiv:2014}, Graves presented an analysis
focusing on recurrent neural networks and text generation.
In \cite{humphrey:feature:learning:music:2013},
Humphrey {\em et al.} presented another analysis,
sharing some issues about music representation (see Section~\ref{section:representation})
but dedicated to music information retrieval\index{Music!information retrieval} (MIR\index{MIR}) tasks,
such as chord recognition, genre recognition and mood estimation.
On MIR applications of deep learning, see also the recent tutorial paper by Choi {\em et al.} \cite{choi:tutorial:deep:learning:mir:arxiv:2017}.

One could also consult the proceedings of some recently created international workshops on the topic, such as

\begin{itemize}

\item the Workshop on Constructive Machine Learning (CML 2016),
held during the 30th Annual Conference on Neural Information Processing Systems (NIPS 2016)
\cite{cml:nips:2016};

\item the Workshop on Deep Learning for Music (DLM),
held during the International Joint Conference on Neural Networks (IJCNN 2017)
\cite{dl4m:ijcnn:2017},
followed by a special journal issue
\cite{herremans:deep:music:ncaa:2019}; and

\item on the deep challenge of {\em creativity\index{Creativity}},
the related Series of International Conferences on Computational Creativity (ICCC)
\cite{iacc:iccc:web}.

\end{itemize}

For a more general survey of computer-based techniques to generate music,
the reader can refer to
general books such as

\begin{itemize}

\item Roads' book about computer music\index{Computer music} \cite{roads:computer:music:tutorial:book:1996};

\item Cope's \cite{cope:algorithmic:composer:book:2000}, Dean and McLean's \cite{dean:oxford:handbook:2018}
and/or Nierhaus' books \cite{nierhaus:algorithmic:composition:book:2009}
about algorithmic composition\index{Algorithmic composition};

\item a recent survey about AI methods in algorithmic composition \cite{fernandez:ai:methods:algorithmic:composition:survey:jair:2013}; and

\item Cope's book about models of musical creativity \cite{cope:musical:creativity:book:2005}.

\end{itemize}

About machine learning in general,
some examples of textbooks are

\begin{itemize}

\item the textbook by Mitchell \cite{mitchell:ml:book:1997};

\item a nice introduction and summary by Domingos \cite{domingos:things:know:2012}; and

\item a recent, complete and comprehensive book about deep learning by Goodfellow {\em et al.} \cite{goodfellow:deep:learning:book:2016}.

\end{itemize}

\subsection{Other Models}
\label{section:introduction:other:models}

We have to remember that there are various other models and techniques for using computers to generate music,
such as rules, grammars, automata, Markov models and graphical models.
These models are either {\em manually} defined by experts or are automatically {\em learnt} from examples by using various machine learning techniques.
They will not be addressed in this book as we are concerned here with deep learning techniques.
However, in the following section
we make a quick comparison of deep learning and Markov models.

\subsection{Deep Learning versus Markov Models}
\label{section:introduction:versus:markov}

Deep learning models are not the only models able to learn musical style from examples.
Markov chain models are also widely used, see, for example, \cite{pachet:continuator:ieee:cga:2004}. 
A quick comparison
(inspired by the analysis of Mozer in \cite{mozer:composition:prediction:1994}\footnote{Note that he made his analysis in in 1994,
	long before the deep learning wave.})
of the pros (+) and cons (--) of deep neural network models and Markov chain models
is as follows:


\begin{description}[--]

\item[+] Markov models are conceptually simple.

\item[+] Markov models have a simple implementation and a simple learning algorithm,
as
the model is a transition probability table\footnote{Statistics
	are collected from the dataset of examples in order to compute the probabilities.}.

\item[--] Neural network models are conceptually simple but the optimized implementations of current deep network architectures
may be complex and need a lot of tuning.

\item[--] Order 1 Markov models (that is, considering only the previous state)
do not capture long-term temporal structures.

\item[--] Order {\em n} Markov models (considering {\em n} previous states) are possible but
require an explosive training set size\footnote{See the discussion
	in \cite[page~249]{mozer:composition:prediction:1994}.}
and can lead to plagiarism\index{Plagiarism}\footnote{By recopying too long sequences
	from the corpus.
	Some promising solution
	is to consider a variable order Markov model
	and to constrain the generation (through min order and max order constraints) on some sweet spot between junk and plagiarism
	\cite{papadopoulos:maxorder:universality:book:2016}.}.

\item[+] Neural networks can capture various types of relations, contexts and regularities.

\item[+] Deep networks can learn long-term and high-order dependencies.

\item[+] Markov models can learn from a few examples.

\item [--] Neural networks need a lot of examples in order to be able to learn well.

\item[--] Markov models do not generalize very well.

\item[+] Neural networks generalize better through the use of distributed representations \cite{hinton:boltzmann:machines:pdp:1986}.

\item[+] Markov models are operational models (automata) on which some control on the generation could be attached\footnote{Examples are
	Markov constraints \cite{pachet:markov:constraints:constraints:2011}
	and factor graphs \cite{pachet:variations:structured:ismir:2017}.}.

\item[--] Deep networks are generative models with a distributed representation and therefore with no direct control to be attached\footnote{This issue
	as well as some possible solutions will be discussed in Section~\ref{section:challenges:strategies:control:dimensions:strategies}.}.

\end{description}

As deep learning implementations are now mature and a large number of examples are available,
deep learning-based models are in high demand
for their characteristics.
That said, other models (such as Markov chains, graphical models, etc.) are still useful and used
and the choice of a model and its tuning
depends on the characteristics of the problem.

\subsection{Requisites and Roadmap}
\label{section:introduction:roadmap}

This book does not require prior knowledge about deep learning and neural networks nor music.


{\bf Chapter~\ref{section:chapter:introduction} Introduction}
(this chapter) introduces the purpose and rationale of the book.

{\bf Chapter~\ref{section:chapter:method} Method}
introduces the method of analysis (conceptual framework)
and the five dimensions at its basis
(objective, representation, architecture, challenge and strategy), dimensions that we discuss within the next four chapters.

{\bf Chapter~\ref{section:chapter:objective} Objective}
concerns the different types of musical content that we want to generate
(such as a melody or an accompaniment to an existing melody)\footnote{Our proposed typology
	of possible objectives
	will turn out to be useful for our analysis
	because, as we will see,
	different objectives can lead to different architectures and strategies.},
as well as their expected use (by a human and/or a machine).


{\bf Chapter~\ref{section:chapter:representation} Representation}
provides an analysis of the different types of representation and techniques for encoding
musical content (such as notes, durations or chords) for a deep learning architecture.
This chapter may be skipped by a reader already expert in computer music,
although some of the encoding strategies are specific to neural networks and deep learning architectures.

{\bf Chapter~\ref{section:chapter:architecture} Architecture}
summarizes the most common deep learning architectures
(such as feedforward, recurrent or autoencoder) used for the generation of music.
This includes a short reminder of the very basics of a simple neural network.
This chapter may be skipped by a reader already expert in artificial neural networks and deep learning architectures.

%
%

{\bf Chapter~\ref{section:chapter:challenges:strategies} Challenge and Strategy}
provides an analysis of the various challenges
that occur when applying deep learning techniques to music generation,
as well as various strategies for addressing them.
We will ground our study in the analysis of various systems and experiments surveyed from the literature.
This chapter is the core of the book.

{\bf Chapter~\ref{section:chapter:analysis} Analysis}
summarizes the survey and analysis conducted in
Chapter~\ref{section:chapter:challenges:strategies}
through some tables
as a way to identify the design decisions and their interrelations
for the different systems surveyed\footnote{And hopefully also
	for the future ones.
	If we draw the analogy (at some meta-level) with the expected ability for a model learnt from a corpus by a machine
	to be able to generalize to future examples
	(see Section~\ref{section:architecture:training:overfitting}),
	we hope that the conceptual framework presented in this book,
	(manually) inducted from a corpus of scientific and technical literature
	about deep-learning-based music generation systems,
	will also be able to help in the design and the understanding of future systems.}.


{\bf Chapter~\ref{section:chapter:discussion:conclusion} Discussion
and Conclusion}
revisits some of the open issues that were touched in during the
analysis of challenges and strategies presented in Chapter~\ref{section:chapter:challenges:strategies},
before concluding this book.



A table of contents, a table of acronyms, a list of references, a glossary and an index complete this book.

Supplementary material is provided at the following companion web site:

\begin{center}
www.briot.info/dlt4mg/
\end{center}

\subsection{Limits}
\label{section:introduction:limits}

This book does not intend to be a general introduction to deep learning
-- a recent and
broad spectrum
book on this topic is \cite{goodfellow:deep:learning:book:2016}.
We do not intend to get into all technical details of implementation, like engineering and tuning, as well as theory\footnote{For instance,
	we will not develop the probability theory and information theory frameworks for formalizing and interpreting
	the behavior of neural networks and deep learning.
	However, Section~\ref{section:architecture:neural:network:entropy} will introduce the intuition behind the notions of entropy and cross-entropy,
	used for measuring the progress made during learning.},
as we
wish to focus on the conceptual level,
whilst providing a sufficient degree of precision.
Also, although having a clear pedagogical objective,
we do not provide some end-to-end tutorial with all the steps and details on how to implement and tune
a complete deep learning-based music generation system. 

Last, as this book is about a very active domain and as our survey and analysis is based on existing systems,
our analysis is obviously not exhaustive.
We have tried to select the most representative proposals and experiments,
while new proposals are being presented at the time of our writing.
Therefore, we encourage readers and colleagues to provide any feedback and suggestions for improving this survey and analysis
which is a still ongoing project.





%% file: method.tex
\chapter{Method}
\label{section:chapter:method}

\abstract*{Chapter~\ref{section:chapter:method} Method describes the conceptual framework proposed in this book to analyze, classify and compare various
deep learning-based music generation systems.
The five dimensions considered are:
objective, representation, architecture, challenge and strategy.
The typologies associated to each dimension have been constructed in a bottom-up way,
from the analysis of numerous deep learning-based music generation systems from the literature.}

\label{section:method}

In our analysis, we consider five main {\em dimensions} to characterize different ways
of applying deep learning techniques to generate musical content.
This typology is aimed at helping the analysis of the various perspectives (and elements)
leading to the design of different deep learning-based music generation systems\footnote{In this book,
	{\em systems}\index{System} refers to the various proposals (architectures, systems and experiments)
	about deep learning-based music generation that we have surveyed from the literature.}.

\section{Dimensions}
\label{section:method:dimension}

The five dimensions that we consider are as follows.

\subsection{Objective}
\label{section:method:objective}

The {\em objective\index{Objective}}\footnote{We could have used the term {\em task}\index{Task} in place of {\em objective}.
	However, as task is a relatively well-defined and common term in the machine learning community
	(see Section~\ref{section:introduction:context} and \cite[Chapter 5]{goodfellow:deep:learning:book:2016}),
	we preferred an alternative term.}
consists in:

\begin{itemize}

\item The musical {\em nature} of the content to be generated.

Examples are
a melody,
a polyphony
or an accompaniment; and
\vspace{0.2cm}

\item The {\em destination} and {\em use} of the content generated.

Examples are
a musical score to be performed by some human musician(s)
or an audio file to be played.

\end{itemize}

\subsection{Representation}
\label{section:method:representation}

The {\em representation\index{Representation}}
is the nature and format of the information (data) used to {\em train} and to {\em generate} musical content.

Examples are
signal\index{Signal}, transformed signal (e.g., a spectrum, via a Fourier transform), piano roll, MIDI or text.


\subsection{Architecture}
\label{section:method:architecture}

The {\em architecture\index{Architecture}}
is the nature of the assemblage of processing {\em units}
(the artificial neurons) and their {\em connexions}.

Examples are
a feedforward architecture,
a recurrent architecture,
an autoencoder architecture
and generative adversarial networks.

\subsection{Challenge}
\label{section:method:challenges}

A {\em challenge\index{Challenge}}
is one of the qualities (requirements) that may be desired for music generation.

Examples are
content variability,
interactivity
and originality.


\subsection{Strategy}
\label{section:method:strategy}

The {\em strategy\index{Strategy}}
represents the way the architecture will process representations in order to {\em generate}\footnote{Note, that we consider here
	the strategy relating to the {\em generation phase} and not the strategy relating to the training phase,
	as they could be different.}
the objective while matching desired requirements.

Examples are
single-step feedforward,
iterative feedforward,
decoder feedforward,
sampling
and input manipulation.


\section{Discussion}
\label{section:method:discussion}

Note that these five dimensions are not orthogonal.
The choice of representation is partially determined by the objective
and it also constrains the input and output
(interfaces)
of the architecture.
A given type of architecture also usually leads to a default strategy of use,
while
new strategies may be designed in order to target specific challenges.



The exploration of these five different dimensions and of their interplay is 
actually at the core of our analysis\footnote{Let us remember
	that our proposed typology has been constructed in a {\em bottom-up\index{Bottom-up}} manner
	from the survey and analysis of numerous systems retrieved from the literature,
	most of them being very recent.}.
Each of the first three dimensions (objective, representation and architecture)
will be analyzed with its associated typology in a specific chapter,
with various illustrative examples and discussion.
The challenge\index{Challenge} and strategy\index{Strategy} dimensions will be jointly analyzed within the same chapter
(Chapter~\ref{section:chapter:challenges:strategies})
in order to jointly illustrate potential issues (challenges) and possible solutions (strategies).
As we will see,
the same strategy may relate to more than one challenge and vice versa.

Last, we do not expect our proposed conceptual framework (and its associated five dimensions and related typologies)
to be a final result,
but rather a first step towards a better understanding of design decisions and challenges for deep learning-based music generation.
In other words, it is likely to be further amended and refined, but we hope that it could help bootstrap
what we believe to be a necessary comprehensive study.



%% file: objective.tex
\chapter{Objective}
\label{section:chapter:objective}

\abstract*{Chapter~\ref{section:chapter:objective} Objective presents the first dimension of the conceptual framework proposed in this book to analyze,
classify and compare various deep learning-based music generation systems.
This first dimension is the objective, which is the nature of the musical content to be generated.
We consider some facets (type -- the main one --, destination, use, mode and style) with related taxonomies.
For example, main types are: melody, polyphony, multivoice and accompaniment.}

\label{section:objective}

The first dimension, the {\em objective\index{Objective}}, is the nature of the musical content to be generated.

\section{Facets}
\label{section:objective:facets}

We may consider five main {\em facets\index{Objective!facet}} of an objective:

\begin{itemize}

\item {\em Type\index{Objective!type}}

The musical nature of the generated content.

Examples are a melody, a polyphony or an accompaniment.
\vspace{0.2cm}

\item {\em Destination\index{Objective!destination}\index{Destination}}

The entity aimed at using (processing) the generated content.

Examples are
a human musician, a software or an audio system.
\vspace{0.2cm}

\item {\em Use\index{Objective!use}\index{Use}}

The way the destination entity will process the generated content.

Examples are
playing an audio file or performing a music score.
\vspace{0.2cm}

\item {\em Mode\index{Objective!mode}\index{Mode}}

The way the {\em generation} will be conducted,
i.e. with some human intervention ({\em interaction})
or without any intervention ({\em automation}).
\vspace{0.2cm}

\item {\em Style\index{Style}}

The musical style of the content to be generated.

Examples are
Johann Sebastian Bach chorales, Wolfgang Amadeus Mozart sonatas, Cole Porter songs or Wayne Shorter music.
The style will actually be set though the choice of the dataset of musical examples (corpus) used as the training examples.

\end{itemize}

\subsection{Type}
\label{section:objective:type}

Main examples of musical types are as follows:

\begin{itemize}

\item {\em Single-voice monophonic melody\index{Single!-voice monophonic melody}},
abbreviated as {\em Melody}
\index{Melody|see{Single-voice monophonic melody}}
\index{Monophonic!melody|see{Single-voice monophonic melody}}

It is a sequence\index{Sequence} of notes\index{Note}
for a single instrument\index{Instrument} or vocal\index{Vocal},
with {\em at most} one note at the same time.

An example is the music produced by a monophonic\index{Monophonic} instrument like a flute\index{Flute}\footnote{Although
	there are non-standard techniques to produce more than one note,
	the simplest one being to sing simultaneously as playing.
	There are also non-standard diphonic techniques for voice.}.
\vspace{0.2cm}

\item {\em Single-voice polyphony\index{Single!-voice polyphony}}
(also named {\em Single-track polyphony}),
abbreviated as {\em Polyphony\index{Polyphony|see{Single-voice polyphony}}}
\index{Single!-track polyphony|see{Single-voice polyphony}}

It is a sequence of notes
for a single instrument,
where more than one note can be played at the same time.

An example is the music produced by a polyphonic\index{Polyphonic} instrument such as a piano\index{Piano} or guitar\index{Guitar}.
\vspace{0.2cm}

\item {\em Multivoice polyphony\index{Multivoice!polyphony}}
(also named {\em Multitrack polyphony\index{Multitrack!polyphony|see{Multivoice polyphony}}}),
abbreviated as {\em Multivoice\index{Multivoice|see{Multivoice polyphony}}}
or {\em Multitrack\index{Multitrack|see{Multivoice polyphony}}}

It is a set of multiple {\em voices\index{Voice}/tracks\index{Track}},
which is intended for more than one voice or instrument\index{Instrument}.

Examples are:
a chorale\index{Chorale} with soprano, alto, tenor and bass voices
or a jazz\index{Jazz} trio with piano, bass and drums.
\vspace{0.2cm}

\item {\em Accompaniment\index{Accompaniment}} to a given melody

Such as

\begin{itemize}

\item {\em Counterpoint\index{Counterpoint}}, composed of one or more melodies (voices); or
\vspace{0.2cm}

\item {\em Chord\index{Chord} progression\index{Chord!progression}}, which provides some associated {\em harmony\index{Harmony}}.

\end{itemize}

\item {\em Association of a melody with a chord progression}

An example is what is named a {\em lead sheet}\index{Lead sheet}\footnote{Figure~\ref{figure:lead:sheet}
	in Chapter~\ref{section:chapter:representation} Representation will show an example of a lead sheet.}
and is common in jazz\index{Jazz}.
It may also include {\em lyrics}\index{Lyrics}\footnote{Note that lyrics could be generated too.
	Although this target is beyond the scope of this book,
	we will see later in Section~\ref{section:representation:text} that,
	in some systems, music is encoded as a text.
	Thus, a similar technique could be applied to lyric generation.}.

\end{itemize}


Note that the {\em type} facet is actually the most important facet, as it captures the musical nature of the objective for content generation.
In this book,
we will frequently identify an objective according to its {\em type},
e.g.,
a melody,
as a matter of simplification.
The next three facets -- {\em destination}, {\em use} and {\em mode} -- will turn out important when regarding the dimension of the {\em interaction}
of human user(s) with the process of content generation.

\subsection{Destination and Use}
\label{section:objective:destination:use}

Main examples of destination and use are as follows:

\begin{itemize}

\item {\em Audio system}

Which will {\em play} the generated content,
as in the case of the generation of an audio file.
\vspace{0.2cm}

\item {\em Sequencer software\index{Sequencer}}

Which will {\em process} the generated content,
as in the case of the generation of a MIDI\index{MIDI} file.
\vspace{0.2cm}

\item {\em Human(s)}

Who will perform and {\em interpret} the generated content,
as in the case of the generation of a music score.

\end{itemize}


\subsection{Mode}
\label{section:objective:mode}

There are two main modes of music generation:

\begin{itemize}

\item {\em Autonomous\index{Autonomous}} and {\em Automated\index{Automated}}

Without any human intervention; or
\vspace{0.2cm}

\item {\em Interactive\index{Interactive}} (to some degree)

With some control interface\index{Control!interface} for the human user(s)
to have some interactive control over the process of generation.

\end{itemize}

As deep learning for music generation is recent and basic neural network techniques are non-interactive,
the majority of systems that we have analyzed are not
yet very interactive\footnote{Some examples of
	interactive systems will be introduced in Section~\ref{section:challenges:strategies:interactivity}.}.
Therefore, an important goal appears to be the design of fully interactive support systems for musicians
(for composing, analyzing, harmonizing, arranging, producing, mixing, etc.),
as pioneered by the FlowComposer prototype \cite{papadopoulos:flow:composer:cp:2016}
to be introduced in Section~\ref{section:systems:flow:composer}.


\subsection{Style}
\label{section:objective:style}

As stated previously, the musical style of the content to be generated
will be governed by the choice of the dataset of musical examples that will be used as training examples.
As will be discussed further in Section~\ref{section:representation:dataset},
we will see that the choice of a dataset,
notably properties like {\em coherence\index{Coherence}}, {\em coverage\index{Coverage}} (versus {\em sparsity\index{Sparsity}})
and {\em scope\index{Scope}} (specialized versus large breadth),
is actually fundamental for good music generation.


%% file: representation.tex
\chapter{Representation}
\label{section:chapter:representation}

\abstract*{Chapter~\ref{section:chapter:representation} Representation presents the second dimension of the conceptual framework proposed in this book to analyze,
classify and compare various deep learning-based music generation systems.
The representation is about the way a musical content is specified and then encoded.
A big divide is between an audio or a symbolic representation, the latter type being more frequent
as it focuses on the compositional level.
Main concepts are: note, duration, rest, chord and meter.
Various formats for representation may be used, the most frequent ones being piano roll, MIDI and text.
Encoding is the final and lowest-level decision about how concepts or values will be represented and computed by the deep network.
This chapter may be skipped by a reader already expert in computer music,
although some of the encoding strategies are specific to neural networks and deep learning architectures.}

\label{section:representation}


The second dimension of our analysis, the {\em representation}, is about the way the musical content is represented.
The choice of representation and its encoding
is tightly connected to the configuration of the input\index{Input} and the output\index{Output} of the architecture,
i.e. the number of input and output variables\index{Variable} as well as their corresponding types.

We will see that, although a deep learning architecture can automatically extract\index{Extraction}
significant {\em features\index{Feature}} from the data,
the choice of representation
may be significant for the accuracy of the learning and for the quality of the generated content.

%
%
%
%
%
%
%

For example, in the case of an audio representation, 
we could use a spectrum\index{Spectrum} representation (computed by a Fourier transform\index{Fourier transform})
instead of a raw waveform\index{Waveform} representation.
In the case of a symbolic representation\index{Symbolic!representation},
we could consider (as in most systems) enharmony\index{Enharmony}, i.e. A$\sharp$ being equivalent to B$\flat$
and C$\flat$ being equivalent to B,
or instead preserve the distinction in order to keep the harmonic and/or voice leading meaning.
%
%

\section{Phases and Types of Data}
\label{section:representation:stages}

Before getting into the choices of representation for the various data to be processed by a deep learning architecture,
it is important to identify the two main phases
related to
the activity of
a deep learning architecture:
the {\em training phase\index{Training!phase}} and the {\em generation phase\index{Generation!phase}},
as well as the related four\footnote{There may be more types of data
	depending on the complexity of the architecture, which may include {\em intermediate} processing steps.}
main types of data to be considered:

\begin{itemize}

\item {\em Training phase}

\begin{itemize}

\item {\em Training data\index{Training!data}}

The set of examples used for training the deep learning system;
\vspace{0.2cm}

\item {\em Validation data\index{Validation!data}} (also\footnote{Actually,
		a difference could be made, as will be later explained in Section~\ref{section:architecture:training:overfitting}.}
	named {\em Test data\index{Test!data}})

The set of examples used for testing the deep learning system\footnote{The motivation will
	be introduced in Section~\ref{section:architecture:training:overfitting}.}.

\end{itemize}

\item {\em Generation phase}

\begin{itemize}

\item {\em Generation (input) data\index{Generation!data}}

The data that will be used as input for the generation,
e.g., a melody for which the system will generate an accompaniment, or a note that will be the first note of a generated melody;
\vspace{0.2cm}

\item {\em Generated (output) data\index{Generated data}}

The data produced by the generation, as specified by the objective\index{Objective}.

\end{itemize}

\end{itemize}

Depending on the objective\footnote{As stated in Section~\ref{section:objective:type},
	we identify an objective by its type as a matter of simplification.},
these four types of data may be equal or different\footnote{Actually,
	training data and validation data are of the same kind, being both input data of the same architecture.}.
For instance:

\begin{itemize}

\item in the case of the generation of
a melody
(for example, in Section~\ref{section:experiment:sturm:celtic:lstm}),
both the training data and the generated data are
melodies; whereas

\item in the case of the generation of a counterpoint accompaniment
(for example,
in Section~\ref{section:experiment:mini:bach}),
the generated data is a set of melodies.

\end{itemize}


\section{Audio versus Symbolic}
\label{section:representation:signal:vs:symbolic}

A big divide in terms of the choice of representation (both for input and output) is {\em audio\index{Audio}} versus {\em symbolic\index{Symbolic}}.
This also corresponds to the divide between {\em continuous\index{Continuous}} and {\em discrete\index{Discrete}} variables.
As we will see, their respective raw material is very different in nature,
as are the types of techniques for possible processing and transformation of the initial representation\footnote{The initial
	representation may be transformed, through, e.g.,
	data compression
	or extraction of higher-level representations,
	in order to improve learning and/or generation.}.
They in fact correspond to different scientific and technical communities,
namely {\em signal processing\index{Signal!processing}} and {\em knowledge representation\index{Knowledge representation}}.

However, the actual processing of these two main types of representation
by a deep learning architecture is basically the {\em same}\footnote{Indeed, at the level of processing by a deep network architecture,
	the initial distinction between audio and symbolic representation boils down,
	as only {\em numerical} values and operations are considered.}.
Therefore, actual audio and symbolic architectures for music generation may be pretty similar.
For example, the WaveNet\index{WaveNet} audio generation architecture (to be introduced in Section~\ref{section:systems:wavenet})
has been transposed into the MidiNet\index{MidiNet} symbolic music generation architecture (in Section~\ref{section:systems:midinet}).
This polymorphism\index{Polymorphism} (possibility of multiple representations leading to genericity)
is an additional advantage of the deep learning approach.

That said, we will focus in this book
on {\em symbolic} representations and on deep learning techniques for generation of {\em symbolic} music.
There are various reasons for this choice:

\begin{itemize}

\item the grand majority of the current deep learning systems for music generation are symbolic;

\item we believe that the essence of music (as opposed to sound\footnote{Without minimizing the importance of the orchestration
	and the production.})
is in the compositional process,
which is exposed via symbolic representations (like musical scores or lead sheets) and is subject to analysis (e.g., harmonic analysis);

\item covering the details and variety of techniques for processing and transforming audio representations
(e.g., spectrum\index{Sceptrum}, cepstrum\index{Cepstrum}, MFCC\index{MFCC}\footnote{Mel-frequency
	cepstral coefficients\index{Mel-frequency cepstral coefficients}.}, etc.)
would necessitate an additional book\footnote{An example entry point is
	the recent review by Wyse of audio representations for deep convolutional networks \cite{wyse:spectrogram:convolutional:dlm:2017}.};
and

\item as stated previously,
independently of considering audio or symbolic music generation,
the principles of deep learning architectures as well as the encoding techniques used
are
actually pretty similar.

\end{itemize}

Last, let us mention a recent deep learning-based architecture which combines audio and symbolic representations.
In this proposal from Manzelli {\em et al.} \cite{manzelli:combining:raw:symbolic:audio:networks:ismir:2018},
a symbolic representation is used as a conditioning input\footnote{Conditioning
	will be introduced in Section~\ref{section:challenges:strategies:control:conditioning}.}
in addition to the audio representation main input,
in order to better guide and structure the generation of (audio) music (see more details in Section~\ref{section:systems:wavenet}).

\section{Audio}
\label{section:representation:signal}

The first type of representation of musical content is audio\index{Audio} {\em signal\index{Signal}},
either in its raw form (waveform) or transformed.

\subsection{Waveform}
\label{section:representation:signal:wave:form}

\begin{figure}
\includegraphics[scale=0.5]{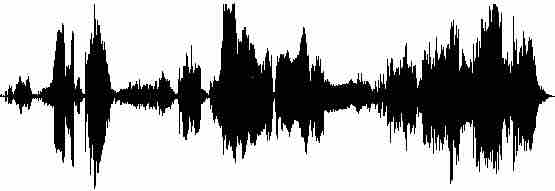}
\caption{Example of a waveform}
\label{figure:example:waveform}
\end{figure}

\begin{figure}
\includegraphics[scale=0.18]{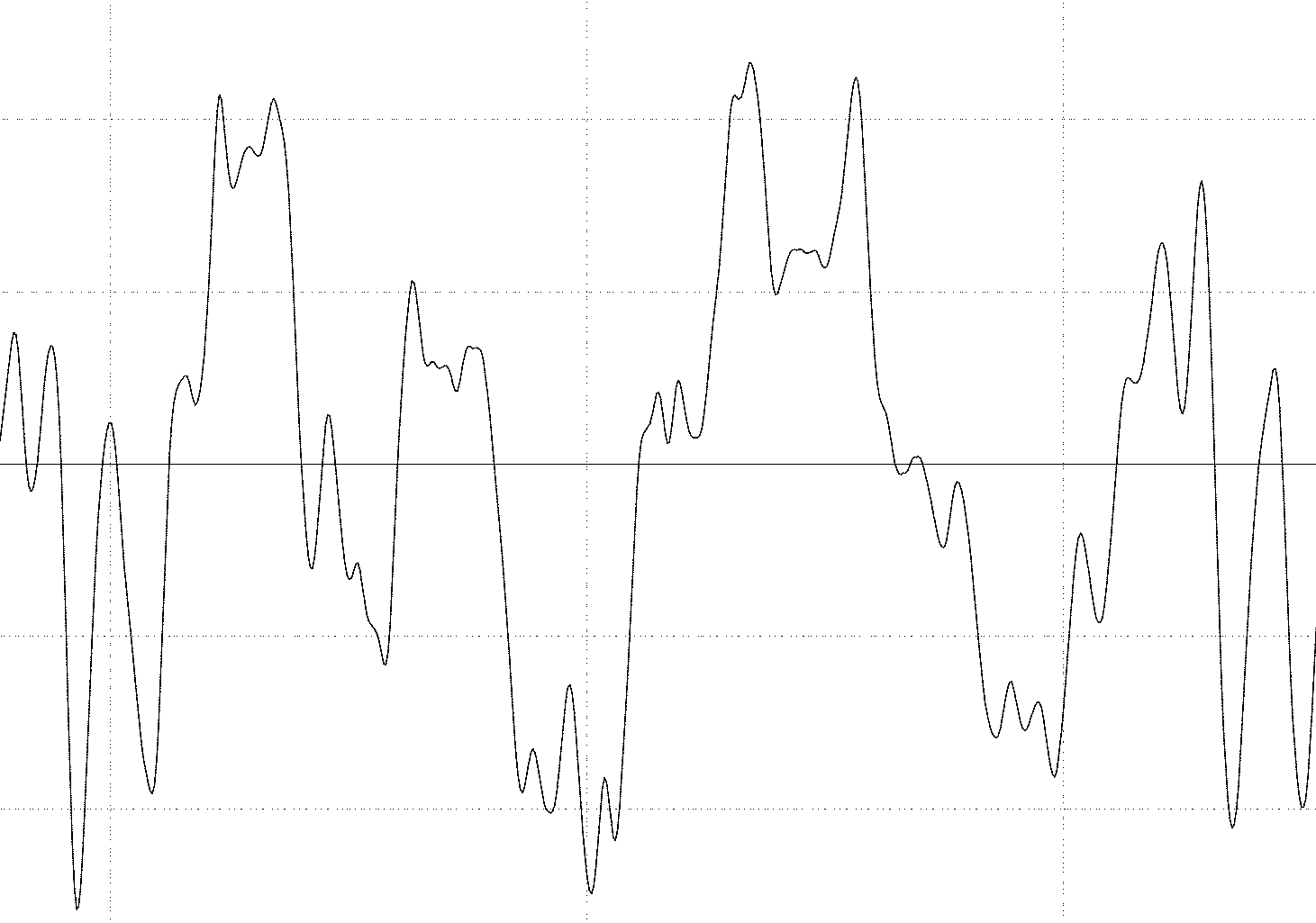}
\caption{Example of a waveform with a fine grain resolution.
Excerpt from a waveform visualization (sound of a guitar) by Michael Jancsy
reproduced from ``https://plot.ly/{\textasciitilde}michaeljancsy/205.embed''
with permission of the author}
\label{figure:example:waveform:focus}
\end{figure}

The most direct representation is the raw audio signal: the {\em waveform\index{Waveform}}.
The visualization of a waveform is shown in Figure~\ref{figure:example:waveform}
and another one with a finer grain resolution is shown in Figure~\ref{figure:example:waveform:focus}.
In both figures, the x axis represents time and the y axis represents the amplitude of the signal.

The advantage of using a waveform is in considering the raw material untransformed, with its full initial resolution.
Architectures that process the raw signal are sometimes named {\em end-to-end\index{End-to-end architecture}} architectures\footnote{The term
	{\em end-to-end} emphasizes that a system learns all features from raw unprocessed data -- without any pre-processing, transformation of representation,
	or extraction of features -- to produce the final output.}.
The disadvantage is in the computational load: low level raw signal is demanding in terms of both memory and processing.

\subsection{Transformed Representations}
\label{section:representation:signal:transformed}

Using transformed representations of the audio signal
usually leads to data compression and higher-level information,
but as noted previously, at the cost of losing some information and introducing some bias\index{Bias}.

\subsection{Spectrogram}
\label{section:representation:signal:spectrogram}

\begin{figure}
\includegraphics[width=\textwidth]{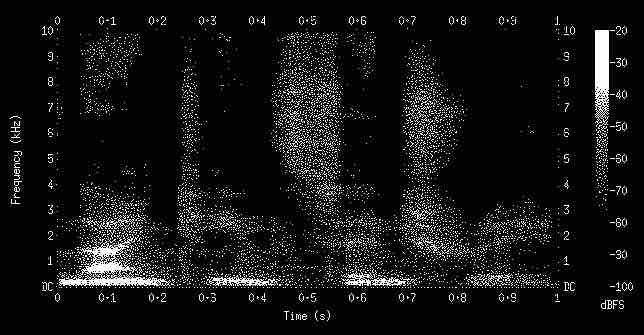}
\caption{Example of a spectrogram of the spoken words ``nineteenth century''.
Reproduced from Aquegg's original image at ``https://en.wikipedia.org/wiki/Spectrogram''}
\label{figure:example:spectrogram}
\end{figure}

A common transformed representation for audio is the {\em spectrum\index{Spectrum}}, obtained
via a {\em Fourier transform\index{Fourier transform}}\footnote{The objective of the Fourier transform
	(which could be continuous or discrete)
	is the decomposition\index{Decomposition} of an arbitrary signal\index{Signal} into its elementary components
	(sinusoidal\index{Sinusoidal} waveforms).
	As well as compressing the information, its role is fundamental for musical purposes
	as it reveals the {\em harmonic\index{Harmonics}} components of the signal.}.
%
%
Figure~\ref{figure:example:spectrogram} shows an example of a {\em spectrogram\index{Spectrogram}},
a visual representation of a spectrum,
where the x axis represents time (in seconds), the y axis represents the frequency (in kHz)
and the third axis in color represents the intensity of the sound
(in dBFS\index{dBFS}\footnote{Decibel relative to full scale\index{Decibel!relative to full scale},
	a unit of measurement for amplitude levels in digital systems.}).

\subsection{Chromagram}
\label{section:representation:signal:chromagram}

\begin{figure}
\includegraphics[width=\textwidth]{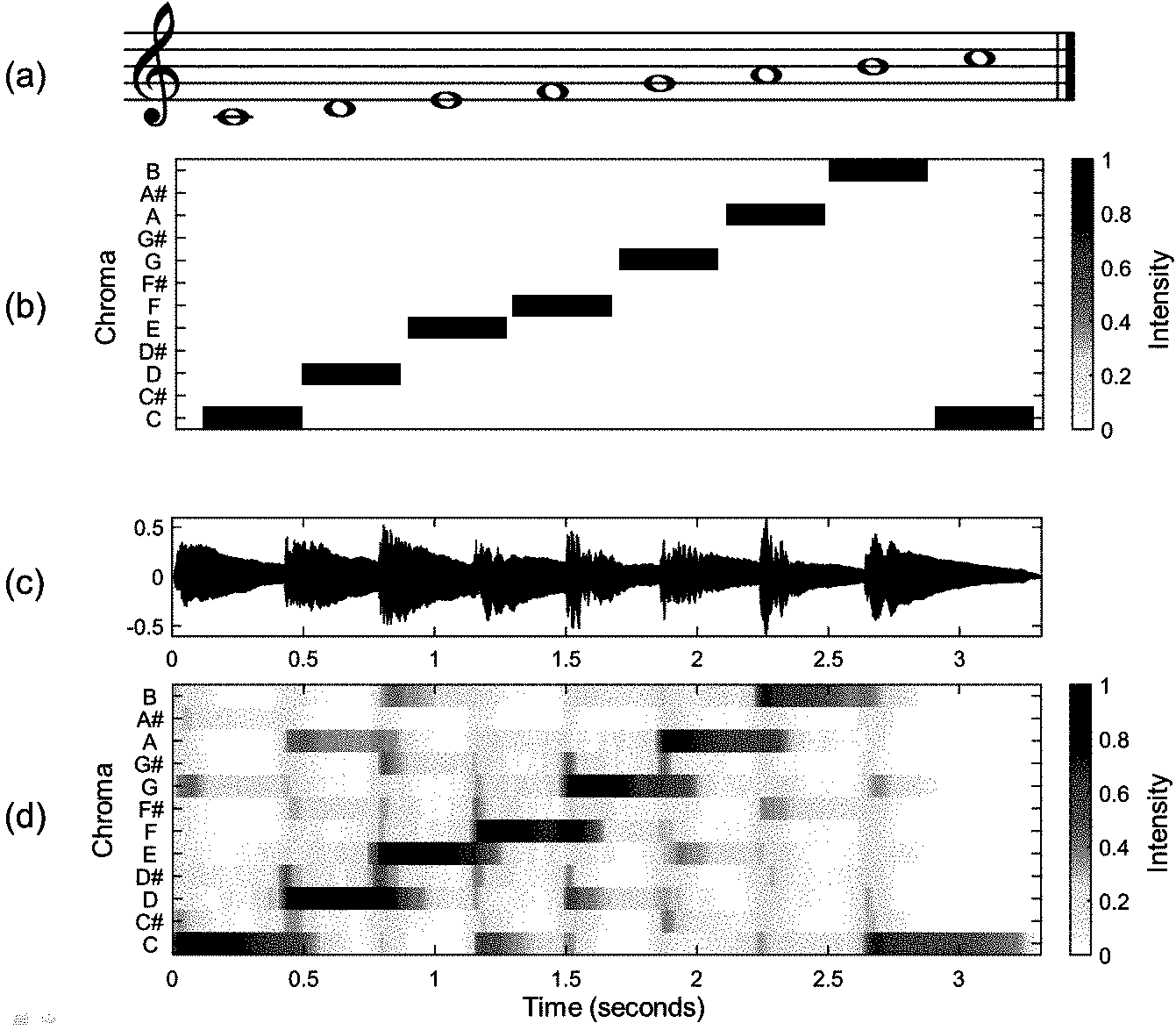}
\caption{Examples of chromagrams.
(a) Musical score of a C-major scale.
(b) Chromagram obtained from the score.
(c) Audio recording of the C-major scale played on a piano.
(d) Chromagram obtained from the audio recording.
Reproduced from Meinard Mueller's original image at ``https://en.wikipedia.org/wiki/Chroma\_feature'' under a CC BY-SA 3.0 licence}
\label{figure:example:chromagram}
\end{figure}

A variation of the spectrogram, discretized onto the tempered scale and
independent of the octave,
is a {\em chromagram\index{Chromagram}}.
It is
restricted to {\em pitch classes}\footnote{A {\em pitch class\index{Pitch!class}}
	(also named a {\em chroma\index{Chroma|see{Chromagram}}})
	represents the name of the corresponding note independently of the octave\index{Octave} position.
	Possible pitch classes are
	C, C$\sharp$ (or D$\flat$), D,~\ldots~A$\sharp$ (or B$\flat$) and B.}.
The chromagram of the C major scale played on a piano is illustrated in Figure~\ref{figure:example:chromagram}.
The x axis common to the four subfigures (a to d) represents time (in seconds).
The y axis of the score (a) represents the note,
the y axis of the chromagrams (b and d) represents the chroma (pitch class)
and the y axis of the signal (c) represents the amplitude.
For chromagrams (b and d), the third axis in color represents the intensity.

\section{Symbolic}
\label{section:representation:symbolic}

Symbolic representations are concerned with concepts like notes, duration and chords,
which will be introduced in the following sections.
%




\section{Main Concepts}
\label{section:representation:symbolic:concepts}

\subsection{Note}
\label{section:representation:note}

In a symbolic representation, a note\index{Note} is represented through the following main features,
and for each feature there are alternative ways of specifying its value:

\begin{itemize}

\item {\em Pitch\index{Pitch}} -- specified by

\begin{itemize}

\item {\em frequency\index{Frequency}}, in Hertz\index{Hertz} (Hz\index{Hz|see{Hertz}});

\item {\em vertical position} ({\em height}) on a score\index{Score}; or

\item
{\em pitch notation\index{Pitch!notation}}\footnote{Also named
	{\em international pitch notation} or {\em scientific pitch notation}.},
which combines a musical note name, e.g., A, A$\sharp$, B, etc. -- actually its pitch class --
and a number (usually notated in subscript)
identifying the pitch class octave\index{Octave} which belongs to the $[-1, 9]$ discrete interval.
An example is A$_4$, which corresponds to A440 -- with a frequency of 440 Hz -- and serves as a general pitch tuning standard.

\end{itemize}

\item {\em Duration} -- specified by

\begin{itemize}

\item {\em absolute} value, in milliseconds (ms); or

\item {\em relative} value, notated as a division or a multiple of a reference note duration,
i.e. the whole note \Ganz.
Examples are a quarter note\footnote{Named
	a {\em crotchet} in British English.}
\Vier~and
an eighth note\footnote{Named
	a {\em quaver} in British English.}
\Acht.

\end{itemize}

\item {\em Dynamics} -- specified by

\begin{itemize}

\item {\em absolute} and {\em quantitative} value, in decibels\index{Decibel} (dB); or

\item {\em qualitative} value, an annotation on a score about how to perform the note,
which belongs to the discrete set $\{$\ppp$,$ \pp$,$ \p$,$ \f$,$ \ff$,$ \fff$\}$, from pianissimo to fortissimo.

\end{itemize}

\end{itemize}

\subsection{Rest}
\label{section:representation:rest}

Rests are important in music as they represent intervals of silence
allowing a pause for breath\footnote{As much for appreciation of the music
	as for respiration by human performer(s)!}.
A {\em rest\index{Rest}} can be considered as a special case of a note, with only one feature,
its duration, and no pitch or dynamics.
The duration of a rest may be specified by

\begin{itemize}

\item {\em absolute} value, in milliseconds (ms); or

\item {\em relative} value, notated as a division or a multiple of a reference rest duration, the whole rest \GaPa~having
the same duration as a whole note \Ganz.
Examples are
a quarter rest~\ViPa~and
an eighth rest \AcPa,
corresponding respectively to
a quarter note \Vier~and an eighth note \Acht.
 
\end{itemize}

\subsection{Interval}
\label{section:representation:interval}

An interval\index{Interval} is a relative transition between two notes.
Examples are a major third (which includes 4 semitones), a minor third (3 semitones) and a (perfect) fifth (7~semitones).
Intervals are the basis of chords (to be introduced in the next section).
For instance, the two main chords in classical music are major (with a major third and a fifth)
and minor (with a minor third and a fifth).

In the pioneering experiments described in \cite{todd:connectionist:composition:1989},
Todd discusses an alternative way for representing the pitch of a note.
The idea is not to represent it in an {\em absolute} way as in Section~\ref{section:representation:note},
but in a {\em relative} way by specifying the relative transition (measured in semitones), i.e. the interval, between two successive notes.
For example, the melody C$_4$, E$_4$, G$_4$ would be represented as C$_4$, +4, +3.

In \cite{todd:connectionist:composition:1989}, Todd points out as two advantages
the fact that there is no fixed bounding of the pitch range
and the fact that it is independent of a given key (tonality).
However, he also points out that this second advantage may also be a major drawback,
because in case of an error in the generation of an interval (resulting in a change of key),
the wrong tonality (because of a wrong index) will be maintained in the rest of the melody generated.
Another limitation is that this strategy applies only to the specification of a monophonic melody
and cannot directly represent a single-voice polyphony, unless separating the parallel intervals into different voices.
Because of these pros and cons, an interval-based representation is actually rarely used in deep learning-based music generation systems.

\subsection{Chord}
\label{section:representation:chord}

A representation of a {\em chord\index{Chord}}, which is a set of at least 3 notes (a triad\index{Triad})\footnote{Modern music
	extends the original major and minor triads into a huge set of
	richer possibilities (diminished, augmented, dominant 7th, suspended, 9th, 13th, etc.)
	by adding and/or altering intervals/components.},
could be

\begin{figure}
\includegraphics[scale=0.6]{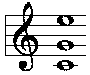}
\caption{C major chord with an open position/voicing: 1-5-3 (root, 5th and 3rd)}
\label{figure:example:chord:voicing}
\end{figure}

\begin{itemize}

\item {\em implicit} and {\em extensional}, enumerating the exact notes composing it.
This permits the specification of the precise octave as well as the position (voicing) for each note,
see
an example in
Figure~\ref{figure:example:chord:voicing}; or

\item {\em explicit} and {\em intensional}, by using a chord symbol combining

\begin{itemize}

\item the pitch class\index{Pitch!class} of its root note, e.g., C, and

\item the {\em type}, e.g., major\index{Major}, minor\index{Minor}, dominant seventh, or diminished\footnote{There are some abbreviated notations\index{Notation convention},
	frequent in jazz\index{Jazz} and popular music,
	for example C minor = Cmin = Cm = C-; C major seventh = CM7 = Cmaj7 = C$\Delta$,
	etc.}.

\end{itemize}

\end{itemize}


We will see that the extensional approach
(explicitly listing all component notes)
is more common for deep learning-based music generation systems,
but there are some examples of systems representing chords explicitly with the intensional approach,
as for instance the MidiNet\index{MidiNet} system to be introduced in Section~\ref{section:systems:midinet}.



\subsection{Rhythm}
\label{section:representation:rhythm}

{\em Rhythm\index{Rhythm}} is fundamental to music.
It conveys the pulsation as well as the stress on specific beats, indispensable for dance!
Rhythm introduces pulsation, cycles and thus structure in what would otherwise remain a flat linear sequence of notes.

\subsubsection{Beat and Meter}
\label{section:representation:rhythm:beat}

A {\em beat\index{Beat}} is the unit of pulsation in music.
Beats are grouped into measures\index{Measure}, separated by {\em bars\index{Bar}}\footnote{Although (and because) a bar is actually the {\em graphical entity}
	-- the line segment ``{\tt |}'' -- separating measures,
	the term bar is also often used, specially in the United States, in place of measure.
	In this book we will stick to the term {\em measure}.}.
The number of beats in a measure as well as the duration between two successive beats constitute the rhythmic signature of a measure
and consequently of a piece of music\footnote{For more elaborate music,
	the meter may change within different portions of the music.}.
This {\em time signature\index{Time!signature}} is also often named {\em meter\index{Meter}}.
It is expressed as the fraction $numberOfBeats/BeatDuration$, where

\begin{itemize}

\item $numberOfBeats$ is the number of beats within a measure; and

\item $beatDuration$ is the duration between two beats.
As with the relative duration of a note
(see Section~\ref{section:representation:note}) or of a rest,
it is expressed as a division of the duration of a whole note {\large \Ganz}.

\end{itemize}

More frequent meters are 2/4, 3/4 and 4/4.
For instance, 3/4 means 3 beats per measure, each one with the duration of a quarter note \Vier. It is the rhythmic signature of a Waltz\index{Waltz}.
%
%
The stress (or accentuation) on some beats or their subdivisions
may form the actual style of a rhythm for music as well as for a dance,
e.g., ternary\index{Ternary}
jazz
versus binary\index{Binary} rock.

\subsubsection{Levels of Rhythm Information}
\label{section:representation:rhythm:level}

We may consider
three different levels
in terms of the amount and granularity of information about rhythm to be included in a musical representation for a deep learning architecture:

\begin{itemize}

\item {\em None} -- only notes and their durations are represented, without any explicit representation of measures.
This is the case for most systems.

\item {\em Measures} -- measures are explicitly represented.
An example is
the system
described in Section~\ref{section:experiment:sturm:celtic:lstm}\footnote{It is interesting to note
	that, as pointed out by Sturm {\em et al.} in \cite{sturm:celtic:melody:csmc:2016},
	the generated music format also contains bars separating measures
	and that there is no guarantee that the number of notes in a measure will always fit to a measure.
	However, errors rarely occur, indicating that this representation
	is sufficient for the architecture to learn to count,
	see \cite{gers:rnn:time:count:ijcnn:2000} and Section~\ref{section:experiment:sturm:celtic:lstm}.}.

\item {\em Beats} -- information about meter, beats, etc. is included.
An example is the C-RBM system described in Section~\ref{section:experiment:c:rbm},
which allows us to impose a specific meter and beat stress for the music to be generated.


\end{itemize}

\section{Multivoice/Multitrack}
\label{section:multi:voice:track}

A {\em multivoice\index{Multivoice}} representation, also named {\em multitrack\index{Multitrack}},
considers independent various voices,
each being a different vocal range (e.g., soprano, alto\ldots) or a different instrument (e.g., piano, bass, drums\ldots).
Multivoice music is usually modeled as parallel tracks,
each one with a distinct sequence of notes\footnote{With possibly simultaneous notes
	for a given voice, see Section~\ref{section:objective:type}.},
sharing the same meter but possibly with different strong (stressed) beats\footnote{Dance music
	is good at this, by having some syncopated bass and/or guitar not aligned on the strong drum beats,
	in order to create some bouncing pulse.}.

Note that in some cases, although there are simultaneous notes, the representation will be a single-voice polyphony,
as introduced in Section~\ref{section:objective:type}.
Common examples are polyphonic instruments like a piano or a guitar.
Another example is a drum or percussion kit, where each of the various components,
e.g., snare, hi-hat, ride cymbal, kick, etc., will usually be considered as a distinct note for the same voice.

The different ways to encode single-voice polyphony and multivoice polyphony will be further discussed in Section~\ref{section:representation:input:encoding:multi:hot}.
	
\section{Format}
\label{section:representation:format}

The format is the 
language (i.e. grammar and syntax) 
in which a piece of music is expressed (specified)
in order to be interpreted by a computer\footnote{The standard format
	for humans is a musical score.}.

\subsection{MIDI}
\label{section:representation:midi}

Musical Instrument Digital Interface\index{Musical!instrument digital interface} (MIDI\index{MIDI})
is a technical standard that describes a protocol,
a digital interface and connectors for interoperability between various electronic musical instruments, softwares and devices
\cite{midi:web}.
%
%
%
%
%
MIDI carries event messages\index{Event message} that specify
real-time note performance data as well as control data.
%
%
We only consider here the two most important messages for our concerns:

\begin{itemize}

\item {\em Note on} -- to indicate that a note is
played.
It contains

\begin{itemize}


\item a {\em channel number},
which indicates the instrument or track\index{Track},
specified by an integer within the set $\{0, 1,\ldots~, 15\}$;

\item a MIDI {\em note number\index{MIDI!note number}},
which indicates the note {\em pitch\index{Pitch}},
specified by an integer within the set $\{0, 1,\ldots~, 127\}$; and

\item a {\em velocity\index{Velocity}\index{MIDI!note velocity}},
which indicates how loud the note is played\footnote{For a keyboard,
	it means the speed of pressing down the key and therefore corresponds to the volume\index{Volume}.},
specified by an integer within the set $\{0, 1,\ldots~, 127\}$.

\end{itemize}

An example is ``Note on, 0, 60, 50'' which means ``On channel 1, start playing a middle C with velocity 50'';
\vspace{0.2cm}

\item {\em Note off} -- to indicate that a note ends.
In this situation,
the velocity indicates how fast the note is released.
An example is ``Note off, 0, 60, 20'' which means ``On channel 1, stop playing a middle C with velocity 20''.

\end{itemize}

Each note event is actually embedded into a track chunk, a data structure containing a delta-time value which specifies the timing information
and the event itself.
A {\em delta-time value} represents the time position
of the event and could represent

\begin{itemize}

\item a {\em relative} {\em metrical} time -- the number of {\em ticks\index{Tick}} from the beginning.
A reference, named the {\em division} and defined in the file header, specifies the number of ticks per quarter note \Vier; or

\item an {\em absolute} time -- useful for real performances,
not
detailed here,
see \cite{midi:web}.


\end{itemize}

\begin{figure}
\includegraphics[scale=0.35]{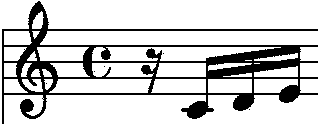}
\caption{Score corresponding to the MIDI excerpt}
\label{figure:extract:midi:file:score}
\end{figure}

An example of an excerpt from a MIDI file (turned into readable ascii)
and its corresponding score are shown in Figures~\ref{figure:extract:midi:file} and~\ref{figure:extract:midi:file:score}.
The division has been set to 384,
i.e. 384 ticks per quarter note \Vier~(which corresponds to 96 ticks for a sixteenth note \Sech).




\begin{figure}
\begin{verbatim}
2,  96, Note_on,  0, 60, 90
2, 192, Note_off, 0, 60,  0
2, 192, Note_on,  0, 62, 90
2, 288, Note_off, 0, 62,  0
2, 288, Note_on,  0, 64, 90
2, 384, Note_off, 0, 64,  0
\end{verbatim}
\caption{Excerpt from a MIDI file}
\label{figure:extract:midi:file}
\end{figure}

%

In \cite{huang:deep:learning:music:arxiv:2016}, Huang and Hu claim that one drawback of encoding MIDI messages directly is that it does not
effectively preserve the notion of multiple notes being played at once through the use of multiple tracks\index{Multitrack}.
In their experiment, they concatenate tracks end-to-end and thus posit that it will be difficult for such a model to
learn that multiple notes in the same position across different tracks can really be played at the same time.
Piano roll, to be introduced in
next section,
does not have this limitation but at the cost of another limitation.

\subsection{Piano Roll}
\label{section:representation:piano:roll}

\begin{figure}
\includegraphics[width=\textwidth]{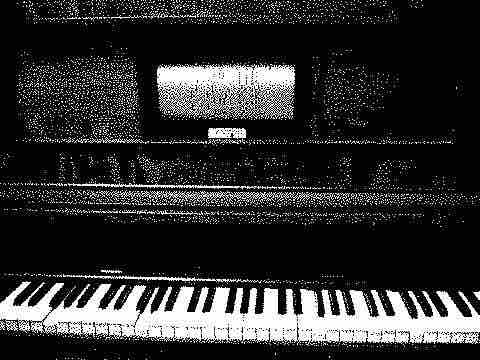}
\caption{Automated piano and piano roll.
Reproduced from Yaledmot's post ``https://www.youtube.com/watch?v=QrcwR7eijyc'' with permission of YouTube}
\label{figure:piano:roll}
\end{figure}

The {\em piano roll\index{Piano!roll}} representation of a melody (monophonic or polyphonic) is inspired from automated pianos
(see Figure~\ref{figure:piano:roll}).
This was a continuous roll of paper with perforations (holes) punched into it.
Each perforation represents a piece of {\em note control information}, to trigger a given note.
The {\em length} of the perforation corresponds to the duration\index{Duration} of a note.
In the other dimension, the {\em localization} of a perforation corresponds to its pitch\index{Pitch}.

\begin{figure}
\includegraphics[width=\textwidth]{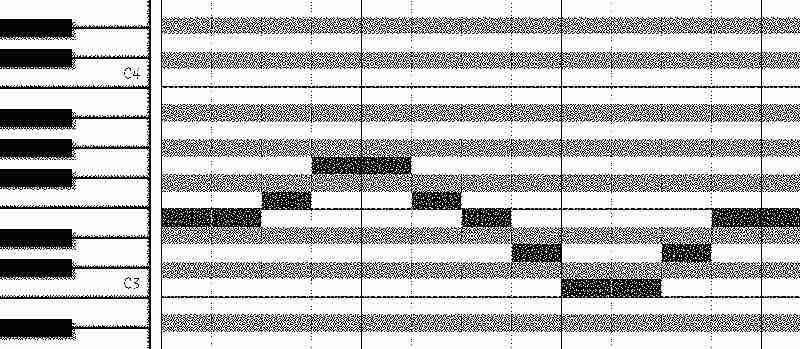}
\caption{Example of symbolic piano roll.
Reproduced from \cite{hao:staff}
with permission of Hao Staff Music Publishing (Hong Kong) Co Ltd.}
\label{figure:example:symbolic:piano:roll}
\end{figure}

An example of a modern piano roll representation (for digital music systems) is shown in Figure~\ref{figure:example:symbolic:piano:roll}.
The x axis represents time and the y axis the pitch.


There are several music environments using piano roll as a basic visual representation,
in place of or in {\em complement} to a score,
as it is more intuitive than the traditional score notation\footnote{Another notation specific to guitar or string instruments is a {\em tablature\index{Tablature}},
	in which the six lines represent the chords of a guitar (four lines for a bass)
	and the note is specified by the number of the fret used to obtain it.}.
An example is
Hao Staff piano roll sheet music \cite{hao:staff},
shown in Figure~\ref{figure:example:symbolic:piano:roll}
with the time axis being horizontal rightward and notes represented as green cells.
Another example is tabs\index{Tab}, where the melody is represented in a piano roll-like format \cite{theorytab:web:2017},
in complement to chords and lyrics.
Tabs are used as an input by the MidiNet\index{MidiNet} system, to be introduced in Section~\ref{section:systems:midinet}.

The piano roll is one of the most commonly used representations, although it has some limitations.
An important one, compared to MIDI representation, is that there is no note off information.
As a result, there is no way to distinguish between a long note and a repeated short note\footnote{Actually,
	in the original mechanical paper piano roll, the distinction is made: two holes are different from a longer single hole.
	The end of the hole is the encoding of the end of the note.}.
In Section~\ref{section:representation:note:ending},
we will look at different ways to address this limitation.
For a more detailed comparison between MIDI and piano roll,
see
\cite{huang:deep:learning:music:arxiv:2016}
and
\cite{2016arXiv160601368W}.


\subsection{Text}
\label{section:representation:text}

\subsubsection{Melody}
\label{section:representation:text:melody}

\begin{figure}
\includegraphics[width=\textwidth]{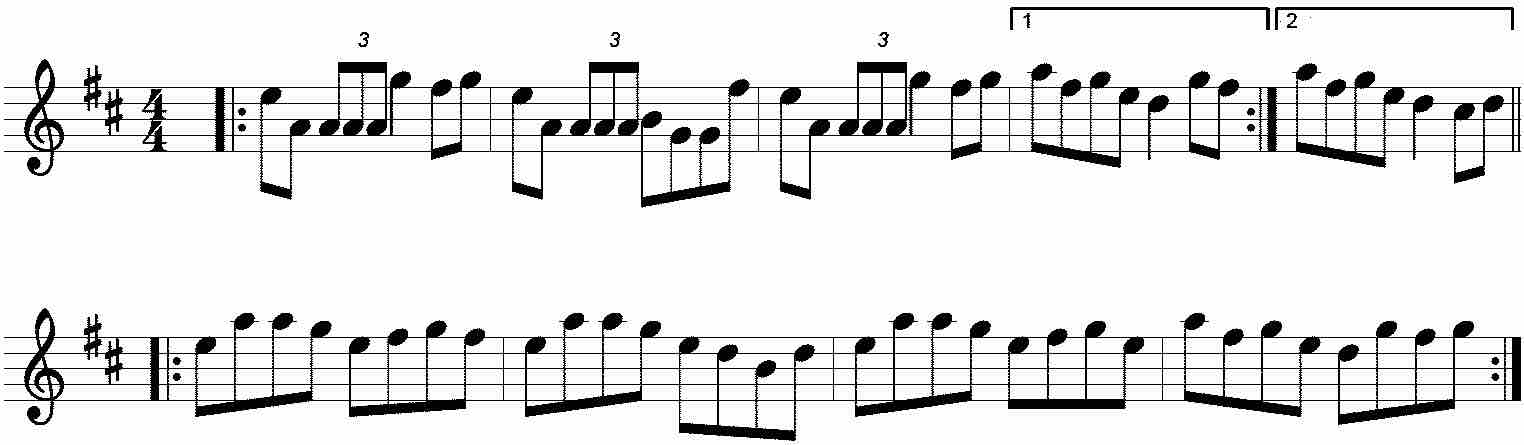}
\caption{Score of ``A Cup of Tea'' (Traditional).
Reproduced from The Session \cite{web:the:session} with permission of the manager}
\label{figure:score:cup:tea}
\end{figure}

A melody can be encoded in a textual representation\index{Textual representation} and processed as a {\em text\index{Text}}.
A significant example is the ABC notation\index{ABC notation} \cite{web:abc:notation}, a {\em de facto} standard for folk and traditional music\footnote{Note that
	the ABC notation has been
	designed {\em independently} of computer music and machine learning concerns.}.
Figures~\ref{figure:score:cup:tea}~and~\ref{figure:abc:cup:tea}
show the original score and its associated ABC notation
for a tune named ``A Cup of Tea'',
from the repository and discussion platform The Session \cite{web:the:session}.

\begin{figure}
\begin{verbatim}
X: 1
T: A Cup Of Tea
R: reel
M: 4/4
L: 1/8
K: Amix
|:eA (3AAA g2 fg|eA (3AAA BGGf|eA (3AAA g2 fg|1afge d2 gf:
|2afge d2 cd|| |:eaag efgf|eaag edBd|eaag efge|afge dgfg:|
\end{verbatim}
\caption{ABC notation of ``A Cup of Tea''.
Reproduced from The Session \cite{web:the:session} with permission of the manager}
\label{figure:abc:cup:tea}
\end{figure}

The first six lines are the header and represent {\em metadata\index{Metadata}}:
T is the title of the music,
M is the meter\index{Meter},
L is the default note length,
K is the key, etc.
The header is followed by the main text representing the melody.
Some basic principles of the encoding rules of the ABC notation are as follows\footnote{Please
	refer to \cite{web:abc:notation} for more details.}:

\begin{itemize}

\item the pitch class\index{Pitch!class}
	of a note is encoded as the letter corresponding to its English notation\index{Notation convention}, e.g., {\tt A} for A or La;

\item its pitch\index{Pitch} is encoded as following: {\tt A} corresponds to A$_4$,
{\tt a} to an A one octave up
and {\tt a'} to an A two octaves up;

\item the duration\index{Duration} of a note is encoded as following:
if the default length is marked as {\tt 1/8}
(i.e. an eighth note \Acht, the case for the ``A Cup of Tea'' example),
{\tt a} corresponds to an eighth note \Acht, {\tt a/2} to a sixteenth note \Sech\,
and {\tt a2} to a quarter note \Vier\footnote{Note that rests may be expressed in the ABC notation
	through the {\tt z} letter.
	Their durations are expressed as for notes, e.g., {\tt z2} is a double length rest.};
and

\item measures\index{Measure}
are separated by ``{\tt |}'' (bars).

\end{itemize}

Note that the ABC notation can only represent monophonic melodies.

In order to be processed by a deep learning architecture,
the ABC text is usually transformed from a character vocabulary\index{Vocabulary} text into a {\em token} vocabulary text
in order to properly consider concepts which could be noted on more than one character,
e.g., {\tt g2}.
Sturm {\em et al.}'s experiment, described in Section~\ref{section:experiment:sturm:celtic:lstm},
uses a token-based notation named the folk-rnn\index{Folk!-rnn} notation \cite{sturm:celtic:melody:csmc:2016}.
A tune is enclosed within a ``{\tt <s>}'' begin mark and an ``{\tt <{\textbackslash}s>}'' end mark.
Last, all example melodies are transposed to the same C root base,
resulting in the notation of the tune ``A Cup of Tea'' shown in Figure~\ref{figure:folk:rnn:cup:tea}.

\begin{figure}
\begin{verbatim}
<s> M:4/4 K:Cmix |: g c (3 c c c b 2 a b | g c (3 c c c d B B a
| g c (3 c c c b 2 a b |1 c' a b g f 2 b a :| |2 c' a b g f 2 e
f |: g c' c' b g a b a | g c' c' b g f d f | g c' c' b g a b g
| c' a b g f b a b :| <\s>
\end{verbatim}
\caption{Folk-rnn notation of ``A Cup of Tea''.
Reproduced from  \cite{sturm:celtic:melody:csmc:2016} with permission of the authors}
\label{figure:folk:rnn:cup:tea}
\end{figure}




\subsubsection{Chord and Polyphony}
\label{section:representation:text:chord}

When represented extensionally, chords are usually encoded with simultaneous notes as a vector.
An interesting alternative extensional representation of chords,
named Chord2Vec\index{Chord2Vec}\footnote{Chord2Vec is inspired by the Word2Vec\index{Word2Vec} model
	for natural language processing\index{Natural language processing}
	\cite{mikolov:word2vec:arxiv:1:2013}.},
has recently been proposed
in \cite{madjiheurem:chord2vec:cml:nips:2016}\footnote{For information,
	there is another similar model,
	also named Chord2Vec, proposed in \cite{huang:chord:ripple:iui:2016}.}.
%
%
Rather than thinking of chords\index{Chord} (vertically) as vectors, it represents chords (horizontally) as sequences of constituent notes\index{Note}.
More precisely,

\begin{itemize}

\item a chord is represented as an arbitrary length-ordered sequence of notes; and

\item chords are separated by a special symbol, as with sentence markers in natural language processing\index{Natural language processing}.

\end{itemize}

When using this representation for predicting neighboring chords, a specific compound architecture is used,
named RNN Encoder-Decoder\index{RNN Encoder-Decoder}
which will be described
in Section~\ref{section:experiment:rnn:encoder:decoder}.

Note that a somewhat similar model is also used for polyphonic music generation by the BachBot\index{BachBot} system \cite{liang:bach:bot:masters:2016}
which will be introduced in Section~\ref{section:experiment:bach:bot}.
In this model, for each time step, the various notes\index{Note}
(ordered in a descending pitch)
are represented as a sequence and a special delimiter symbol
``{\tt |||}'' indicates the next time frame\index{Time!frame}.


\subsection{Markup Language}
\label{section:representation:markup}

Let us mention the case of general text-based structured representations based on markup languages
(famous examples are HTML and XML).
Some markup languages have been designed for music applications,
like for instance the open standard MusicXML \cite{good:music:xml:book:2001}.
The motivation is to provide a common format to facilitate the sharing, exchange and storage of scores
by musical software systems (such as score editors and sequencers).
MusicXML\index{MusicXML}, as well as similar languages, is not intended for direct use by humans because of its verbosity,
which is the down side of its richness and effectiveness as an interchange language.
Furthermore, it is not very appropriate as a direct representation for machine learning tasks for the same reasons,
as its verbosity and richness would create too much overhead as well as bias.

\subsection{Lead Sheet}
\label{section:representation:lead:sheet}

\begin{figure}
\includegraphics[scale=0.19]{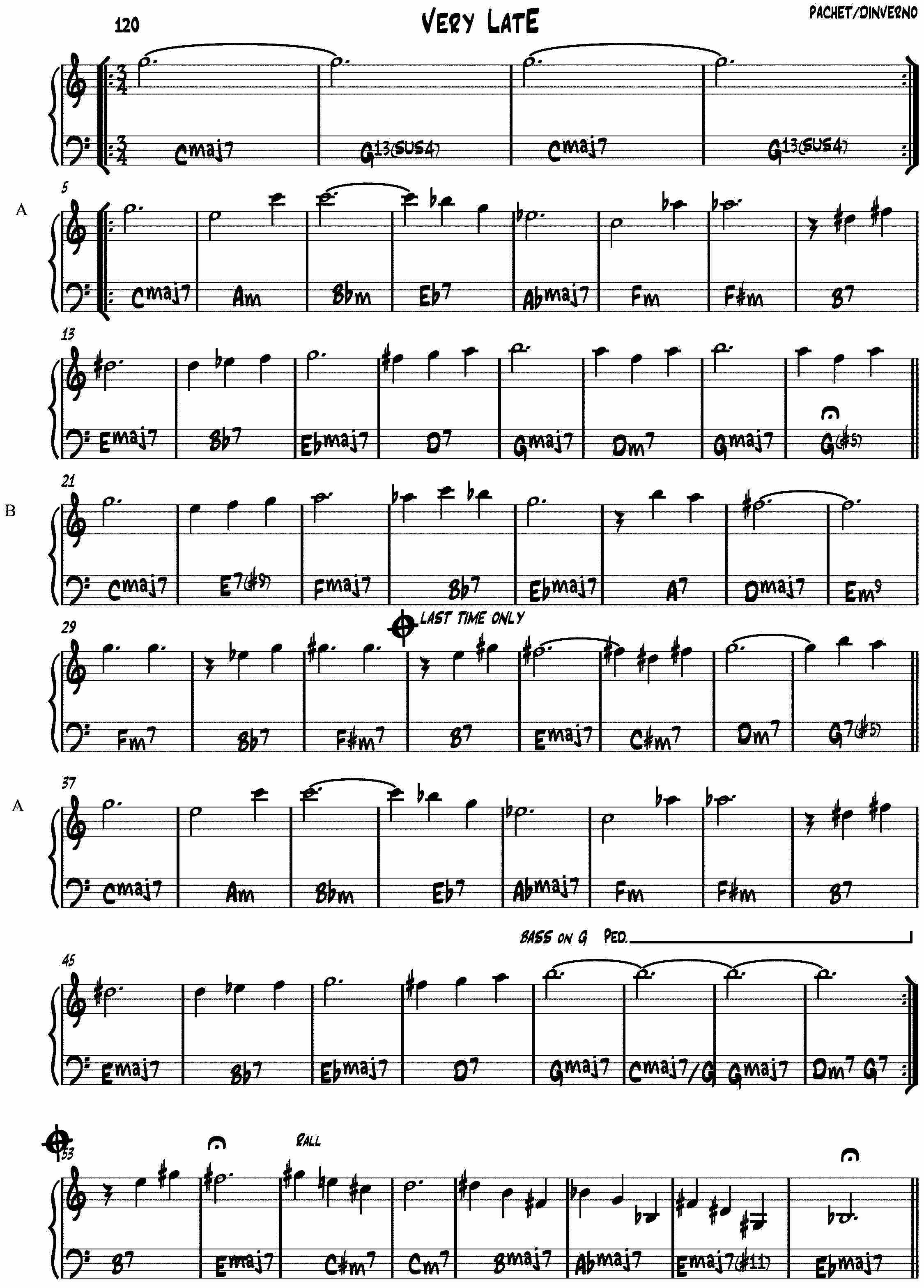}
\caption{Lead sheet of ``Very Late'' (Pachet and d'Inverno).
Reproduced with permission of the composers}
\label{figure:lead:sheet}
\end{figure}

Lead sheets are an important representation format for popular music (jazz, pop, etc.).
A {\em lead sheet\index{Lead sheet}} conveys in upto a few pages the score of a melody\index{Melody}
and its corresponding chord progression\index{Chord!progression} via an intensional notation\index{Notation convention}\footnote{See
	Section~\ref{section:representation:chord}.}.
Lyrics\index{Lyrics} may also be added.
Some important information for the performer, such as the composer, author, style and tempo\index{Tempo},
is often also present.
An example of lead sheet in shown in Figure~\ref{figure:lead:sheet}.

Paradoxically, few systems and experiments use this rich and concise representation, and most of the time they focus on the notes.
Note that Eck and Schmidhuber's Blues generation system,
to be introduced in Section~\ref{section:experiment:eck:blues:lstm},
outputs a combination of melody and chord progression, although not as an {\em explicit} lead sheet.
A notable contribution is the systematic encoding of lead sheets done in the Flow Machines\index{Flow Machines} project \cite{csl:flow:machines:web:2012},
resulting in the Lead Sheet Data Base (LSDB\index{LSDB}) repository \cite{lsdb:ismir:2013}, which includes more than 12,000 lead sheets.

Note that there are some alternative notations, notably
tabs\index{Tab} \cite{theorytab:web:2017},
where the melody is represented in a piano roll-like format (see Section~\ref{section:representation:piano:roll})
and complemented with the corresponding chords.
An example of use of tabs is the MidiNet\index{MidiNet} system to be analyzed in Section~\ref{section:systems:midinet}.





\section{Temporal Scope and Granularity}
\label{section:representation:global:local}

The representation of time\index{Time} is fundamental for
musical processes.

\subsection{Temporal Scope}
\label{section:representation:temporal:scope}

\begin{figure}
\includegraphics[width=\textwidth]{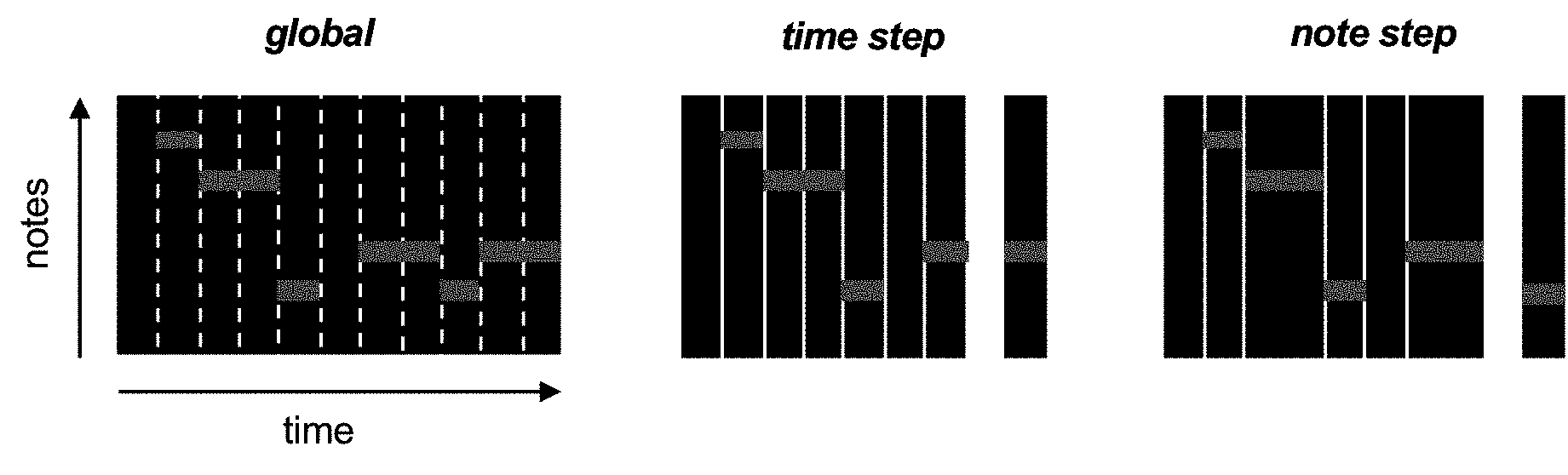}
\caption{Temporal scope for a piano roll-like representation}
\label{figure:temporal:granularity}
\end{figure}

An initial design decision concerns the {\em temporal scope\index{Temporal!scope}} of the representation
used for the generation data\index{Generation!data} and for the generated data\index{Generated data},
that is the way the representation will be interpreted by the architecture with respect to time,
as illustrated in Figure~\ref{figure:temporal:granularity}:

\begin{itemize}

\item {\em Global} --
in this first case, the temporal scope of the representation is the {\em whole} musical piece.
The deep network architecture
(typically a feedforward\index{Feedforward} or an autoencoder\index{Autoencoder} architecture,
see Sections~\ref{section:architecture:feedforward} and~\ref{section:architecture:autoencoder})
will process the input and produce the output
within a {\em global single step\index{Step}}\footnote{In Chapter~\ref{section:chapter:challenges:strategies},
	we will name
	it the {\em single-step feedforward strategy\index{Single!-step feedforward strategy}},
	see Section~\ref{section:strategy:single:step:feedforward}.}.
Examples are the MiniBach\index{MiniBach} and DeepHear\index{DeepHear} systems
introduced in Sections~\ref{section:experiment:mini:bach} and~\ref{section:experiment:deep:hear:melody},
respectively.

\item {\em Time step\index{Time!step}} (or {\em time slice\index{Time!slice}}) -- in this second case, the most frequent one,
the temporal scope of the representation
is a {\em local} {\em time slice\index{Slice|see{Time slice}}} of the musical piece,
corresponding to a specific temporal moment (time step).
The granularity of the processing by the deep network architecture
(typically a recurrent network\index{Recurrent!neural network}) is a {\em time step}
and generation is iterative\footnote{In Chapter~\ref{section:chapter:challenges:strategies},
	we will name
	it the {\em iterative feedforward strategy\index{Iterative feedforward strategy}},
	see Section~\ref{section:strategy:iterative:feedforward}.}.
Note that the time step\index{Time!step} is usually set to the {\em shortest note duration}
(see more details in Section~\ref{section:representation:quantization}),
but it may be larger, e.g., set to a measure in the system as discussed in \cite{todd:connectionist:composition:1989}.

\item {\em Note step\index{Note!step}} -- this third case
was proposed by Mozer in \cite {mozer:composition:prediction:1994}
in his CONCERT system \cite{mozer:composition:prediction:1994},
see Section~\ref{section:experiment:concert}.
In this approach there is {\em no fixed time step}.
The granularity of processing by the deep network architecture is a {\em note}.
This strategy uses a distributed encoding of duration that allows to process a note of any duration in a single network processing step.
Note that, by considering as a single processing step a note rather than a time step,
the number of processing steps to be bridged by the network is greatly reduced.
The approach proposed later on by Walder in \cite{2016arXiv160601368W} is similar.

\end{itemize}

Note that a global temporal scope representation actually also considers time steps
(separated by dash lines in Figure~\ref{figure:temporal:granularity}).
However, although time steps are present at the representation level,
they will not be interpreted as distinct processing steps by the neural network architecture.
Basically,
the encoding of the successive time slices will be concatenated into a global representation
considered as a whole by the network, as shown in Figure~\ref{figure:architecture:mini:bach:detailed}
of an example to be introduced in Section~\ref{section:experiment:mini:bach}.

Also note that in the case of a global temporal scope
the musical content generated has a {\em fixed length} (the number of time steps),
whereas in the case of a time step or a note step temporal scope
the musical content generated has an {\em arbitrary length},
because generation is iterative as we will see in
Section~\ref{section:strategy:iterative:feedforward}.

\subsection{Temporal Granularity}
\label{section:representation:quantization}
\label{section:representation:temporal:granularity}

In the case of a global or a time step temporal scope,
the granularity of the time step\index{Time!step},
corresponding to the granularity of the time {\em discretization\index{Discretization}},
must be defined.
%
%
%
%
There are two main strategies:

\begin{itemize}

\item The most common strategy is to set the time step to a {\em relative duration},
the smallest duration\index{Note!duration} of a note in the corpus
(training examples/dataset\index{Dataset}),
e.g., a sixteenth note \Sech.
To be more precise, as stated by Todd in \cite{todd:connectionist:composition:1989},
the time step should be the {\em greatest common factor} of the durations of all the notes to be learned.
This ensures that the duration of every note will be properly represented with a whole number of time steps.
One immediate consequence of this ``leveling down'' is the number of processing steps necessary,
independent of the duration of actual notes.

\item Another strategy is to set the time step to a fixed {\em absolute duration}, e.g., 10 milliseconds.
This strategy permits us to capture expressiveness in the timing of each note during a human performance,
as we will see in Section~\ref{section:representation:expressiveness}.

\end{itemize}

Note that in the case of a note step temporal scope,
there is no uniform discretization of time (no fixed time step) and no need for.

\section{Metadata}
\label{section:representation:meta:data}

In some systems, additional information from the score may also be explicitly represented and used
as {\em metadata\index{Metadata}},
such as

\begin{itemize}

\item note tie\index{Note!tie}\footnote{A tied note on a music score specifies
	how a note duration extends across a single measure.
	In our case, the issue is how to specify that the duration extends across a single {\em time step}.
	Therefore,
	we consider it as metadata information, as it is specific to the representation and its processing by a neural network architecture.},


\item fermata\index{Fermata},

\item harmonics\index{Harmonics},

\item key,

\item meter, and

\item the instrument associated to a voice.


\end{itemize}

This extra information may lead to more accurate learning and generation.


%
%


\subsection{Note Hold/Ending}
\label{section:representation:note:ending}

An important issue
is how to represent if a note is held\index{Hold}, i.e. tied\index{Tied} to the previous note.
This is actually equivalent to the issue of how to represent the ending of a note\index{Note!ending}.

In the MIDI\index{MIDI} representation format, the end of a note is explicitly stated (via a ``Note off'' event\footnote{Note that,
	in MIDI,
	a ``Note on'' message with a null (0) velocity\index{Velocity} is interpreted as a ``Note off'' message.}).
In the piano roll\index{Piano!roll}
format discussed in Section~\ref{section:representation:piano:roll},
there is no explicit representation of the ending of a note
and, as a result, one cannot distinguish between two repeated quarter notes \Vier\,\Vier\, and a half note \Halb.

The main possible techniques are

\begin{itemize}

\item to introduce a {\em hold/replay} representation, as a dual representation of the sequence of notes.
	This solution is used, for example, by Mao {\em et al.} in their DeepJ system \cite{mao:deepj:arxiv:2018}
	(to be analyzed in Section~\ref{section:systems:deepj}),
	by introducing a replay matrix similar to the piano roll-type matrix of notes;

\item to divide the size of the time step\index{Time!step}\footnote{See
		Section~\ref{section:representation:quantization}
		for details of how the value of the time step is defined.}
	by two
	and always mark a {\em note ending\index{Note!ending}} with a special tag, e.g., 0.
	This solution is used, for example, by Eck and Schmidh\"uber in \cite{eck:composition:lstm:2002},
	and will be analyzed in Section~\ref{section:experiment:eck:blues:lstm};

\item to divide the size of the time step as before but instead
	mark a {\em new note beginning\index{Note!beginning}}.
	This solution is used by Todd in \cite{todd:connectionist:composition:1989}; or
	
\item to use a special {\em hold} symbol ``\_\_'' in place of a note to specify when the previous note is held.
	This solution was proposed by Hadjeres {\em et al.} in
	their DeepBach\index{DeepBach} system \cite{hadjeres:deep:bach:arxiv:2017}
	to be analyzed in Section~\ref{section:experiment:deep:bach}.

\end{itemize}

\begin{figure}
\includegraphics[width=\textwidth]{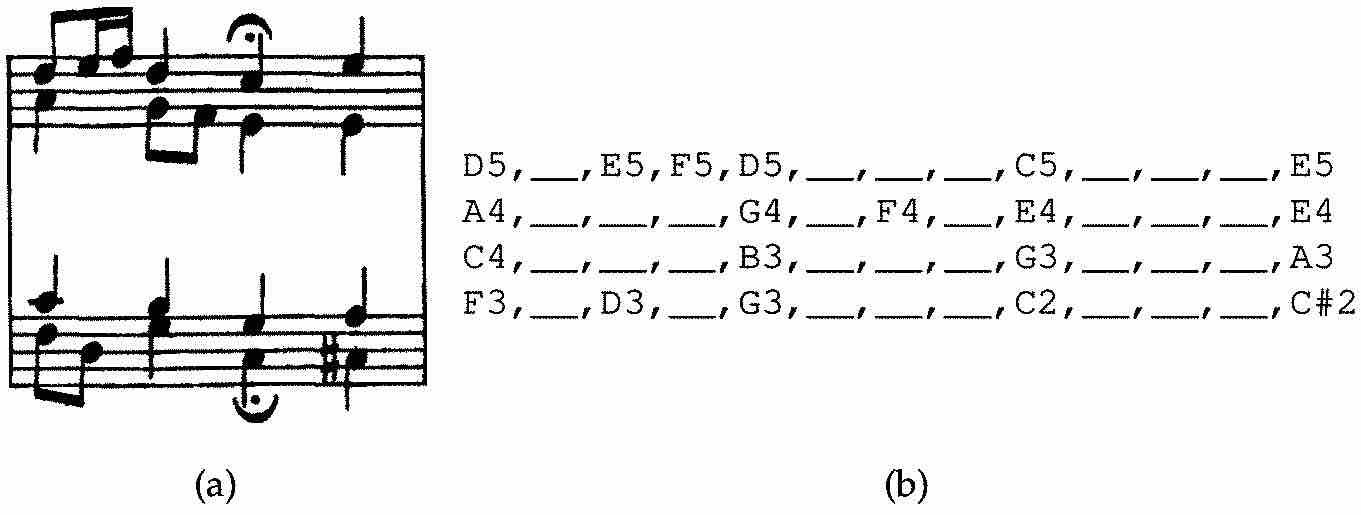}
\caption{a) Extract from a J. S. Bach chorale and b) its representation using the hold symbol ``\_\_''.
Reproduced from \cite{hadjeres:deep:bach:arxiv:2017} with permission of the authors}
\label{figure:deepbach:note:hold}
\end{figure}

This last solution considers the hold symbol as a note,
see an example in Figure~\ref{figure:deepbach:note:hold}.
The advantages of the hold symbol technique are

\begin{itemize}

\item it is simple and uniform as the hold symbol is considered as a note; and

\item there is no need to divide the value of the time step by two
and mark a note ending or beginning.

\end{itemize}

The authors of DeepBach
also emphasize that the good results they obtain using Gibbs sampling rely exclusively on their choice to integrate
the hold\index{Hold} symbol into the list of notes
(see \cite{hadjeres:deep:bach:arxiv:2017} and Section~\ref{section:experiment:deep:bach}).
An important limitation is that the hold symbol only applies to the case of a monophonic melody,
that is it cannot directly express held notes in an unambiguous way in the case of a single-voice polyphony.
In this case, the single-voice polyphony must be reformulated into a multivoice representation with each voice being a monophonic melody;
then a hold symbol is added separately for each voice.
Note that in the case of the replay matrix, the additional information (matrix row) is for each possible note and not for each voice.

We will discuss in Section~\ref{section:representation:input:encoding:hold:rest}
how to encode a hold symbol.
	
%
%
%
%
%
%


%

%
%



\subsection{Note Denotation (versus Enharmony)}
\label{section:note:encoding}

Most systems consider {\em enharmony\index{Enharmony}},
i.e. in the tempered system A$\sharp$ is {\em enharmonically equivalent} to
(i.e. has the same pitch as)
B$\flat$,
although harmonically and in the composer's intention they are different.
An exception is the DeepBach\index{DeepBach} system, described in Section~\ref{section:experiment:deep:bach},
which encodes notes using their real names and not their MIDI\index{MIDI} note numbers.
The authors of DeepBach state that this additional information leads to a more accurate model
and better results \cite{hadjeres:deep:bach:arxiv:2017}.

\subsection{Feature Extraction}
\label{section:representation:feature:extraction}

Although deep learning is good at processing raw unstructured data,
from which its hierarchy of layers will extract higher-level representations adapted to the task
(see Section~\ref{section:introduction:motivation:techniques}),
some systems include a preliminary step of automatic {\em feature extraction\index{Feature!extraction}},
in order to represent the data in a more compact, characteristic and discriminative form.
One motivation could be to gain efficiency and accuracy for the training and for the generation.
Moreover, this {\em feature-based representation\index{Feature!-based representation}} is also useful for indexing data,
in order to control generation through compact labeling
(see, for example, the DeepHear system in Section~\ref{section:experiment:deep:hear}),
or for indexing musical units to be queried and concatenated
(see Section~\ref{section:experiment:brentan:unit:selection}).

\label{section:representation:bag:of:words}

The set of {\em features\index{Feature}} can be defined {\em manually} ({\em handcrafted}) or {\em automatically}
(e.g. by an autoencoder\index{Autoencoder}, see Section~\ref{section:architecture:autoencoder}).
In the case of handcrafted features\index{Handcrafted!feature}, the bag-of-words\index{Bag-of-words} (BOW\index{BOW})
model is a common strategy for natural language text processing\index{Natural language processing},
which may also be applied to other types of data, including musical data,
as we will see in Section~\ref{section:experiment:brentan:unit:selection}.
It consists in transforming the original text (or arbitrary representation)
into a ``bag of words''
(the vocabulary composed of all occurring words, or more generally speaking, all possible tokens);
then various measures can be used to characterize the text\index{Text}.
The most common is {\em term frequency},
i.e. the number of times a term appears in the text\footnote{Note that this bag-of-words representation
	is
	a {\em lossy representation\index{Lossy representation}}
	(i.e. without effective means to perfectly reconstruct the original data representation).}.

\label{section:representation:embedding}

Sophisticated methods have been designed for neural network architectures to automatically compute a vector representation
which preserves, as much as possible, the relations between the items.
Vector representations of texts are named {\em word embeddings}\index{Word embedding}\footnote{The term {\em embedding}
	comes from the analogy with {\em mathematical embedding}, which is an injective and structure-preserving mapping.
	Initially used for natural language processing\index{Natural language processing},
	it is now often used in deep learning as a general term for {\em encoding\index{Encoding}}
	a given representation into a vector representation.
	Note that the term embedding, which is an abstract model representation,
	is often also used (we think, abusively) to define a specific instance of an embedding
	(which may be better named, for example, a {\em label\index{Label}},
	see \cite{sun:deep:hear} and  Section~\ref{section:experiment:deep:hear:melody}).}.
A recent reference model for natural language processing\index{Natural language processing} (NLP\index{NLP})
is the Word2Vec\index{Word2Vec} model \cite{mikolov:word2vec:arxiv:1:2013}.
It has recently been transposed to the Chord2Vec model for the vector encoding of chords\index{Chord}, as described in \cite{madjiheurem:chord2vec:cml:nips:2016}
(see Section~\ref{section:representation:chord}).


\section{Expressiveness}
\label{section:representation:expressiveness}

\subsection{Timing}
\label{section:representation:expressiveness:quantization}

If training examples are processed from conventional scores or MIDI-format libraries,
there is a good chance that the music is perfectly {\em quantized\index{Quantization}}
-- i.e., note onsets\index{Note!onset}\footnote{An onset refers to
	the beginning of a musical note (or sound).}
are exactly aligned onto the tempo --
resulting in a mechanical sound without {\em expressiveness\index{Expressiveness}}.
One approach is to consider symbolic records -- in most cases recorded directly in MIDI --
from real human {\em performances\index{Performance}},
with the musician interpreting the tempo\index{Tempo}.
An example of a system for this purpose is Performance RNN\index{Performance RNN} \cite{simon:performance:rnn:web:2017},
which will be analyzed in Section~\ref{section:experiment:performance:rnn}.
It follows the {\em absolute time duration} quantization strategy, presented in Section~\ref{section:representation:quantization}.

\subsection{Dynamics}
\label{section:representation:expressiveness:dynamics}

Another common limitation is that many MIDI-format libraries do not include {\em dynamics\index{Dynamics}}
(the volume\index{Volume} of the sound produced by an instrument),
which stays fixed throughout the whole piece.
One option is to take into consideration (if present on the score) the annotations made by the composer about the dynamics,
from pianissimo {\ppp} to fortissimo {\fff}, see Section~\ref{section:representation:note}.
As for tempo expressiveness, addressed in Section~\ref{section:representation:expressiveness:quantization},
another option is to use real human performances,
recorded with explicit dynamics variation -- the velocity\index{Velocity} field in MIDI.

\subsection{Audio}
\label{section:representation:expressiveness:audio}

Note that in the case of an audio\index{Audio} representation,
expressiveness
as well as tempo and dynamics are
entangled\index{Entangled} within the whole representation.
Although it is easy to control the global dynamics (global volume),
it is less easy to separately control the dynamics of a single instrument or voice\footnote{More generally speaking,
	audio source separation, often coined as the {\em cocktail party effect\index{Cocktail party effect}},
	has been known for a long time to be a very difficult problem,
	see the original article in \cite{cherry:cocktail:problem:1953}.
	Interestingly, this problem has been solved in 2015 by deep learning architectures
	\cite{deep:learning:solves:cocktail:party:2015},
	opening up ways for disentangling instruments or voices and their relative dynamics as well as tempo
	(by using audio time stretching\index{Time!stretching} techniques).}.


\section{Encoding}
\label{section:representation:input:encoding}

Once the format of a representation\index{Representation} has been chosen,
the issue still remains of how to {\em encode} this representation.
The {\em encoding\index{Encoding}} of a representation (of a musical content)
consists in the {\em mapping} of the representation
(composed of a set of {\em variables}, e.g., pitch or dynamics)
into a set of {\em inputs} (also named {\em input nodes\index{Input!node}} or {\em input variables\index{Input!variable}})
for the neural network architecture\footnote{See Section~\ref{section:architecture:basic:architectures}
	for more details about the input nodes of a neural network architecture.}.

\subsection{Strategies}
\label{section:representation:input:encoding:strategy}
\label{section:representation:input:encoding:one:hot}

At first, let us consider the three possible types for a variable:

\begin{itemize}

\item {\em Continuous\index{Continuous}} variables --
an example is the pitch\index{Pitch} of a note defined by its frequency in Hertz,
that is a real value within the $]0, +\infty[$ interval\footnote{The notation $]0, +\infty[$ is for an open interval excluding its endpoints.
	   An alternative notation is $(0, +\infty)$.}.

The straightforward way is to directly encode the variable\footnote{In practice,
	the different variables are also usually scaled and normalized,
	in order to have similar domains of values ($[0, 1]$ or $[-1, +1]$) for all input variables,
	in order to ease learning convergence.}
as a {\em scalar\index{Scalar}} whose domain is real values.
We call this strategy {\em value encoding\index{Value encoding}}.

\item {\em Discrete\index{Discrete} integer\index{Integer}} variables --
an example is the pitch of a note defined by its MIDI note number\index{MIDI!note number},
that is an integer value within the $\{0, 1,\ldots~, 127\}$ discrete set\footnote{See
	our summary of MIDI specification in Section~\ref{section:representation:midi}.}.

The straightforward way is to encode the variable as a real value {\em scalar},
by casting the integer into a real.
This is another case of {\em value encoding\index{Value encoding}}.

\item {\em Boolean\index{Boolean} (binary\index{Binary}}) variables --
an example is the specification of a note ending (see Section~\ref{section:representation:note:ending}).

The straightforward way is to encode the variable as a real value {\em scalar},
with two possible values: 1 (for true) and 0 (for false).

\item {\em Categorical\index{Categorical}} variables\footnote{In statistics,
	a {\em categorical variable} is a variable that can take one of a limited
	-- and usually fixed -- number of possible values.
	In computer science it is usually referred as an {\em enumerated type}.}
--
an example is a component of a drum kit;
an element within a set of possible values: $\{$snare, high-hat, kick, middle-tom, ride-cymbal, etc.$\}$.

The usual strategy is to encode a categorical variable as a {\em vector} having as its length the number of possible elements,
in other words the cardinality of the set of possible values.
Then, in order to represent a given element, the corresponding element of the encoding vector is set to 1 and all other elements to 0.
Therefore, this encoding strategy is usually called {\em one-hot encoding\index{One-hot encoding}}\footnote{The name comes from digital circuits,
	{\em one-hot} referring to a group of bits among which the only legal (possible) combinations of values are those
	with a single {\em high} (hot!) (1) bit, all the others being {\em low} (0).}.
This frequently used strategy is also often employed for encoding discrete integer variables, such as MIDI note numbers.

\end{itemize}

\subsection{From One-Hot to Many-Hot and to Multi-One-Hot}
\label{section:representation:input:encoding:multi:hot}

Note that a one-hot encoding of a note corresponds to a time slice of a piano roll\index{Piano!roll} representation
(see Figure~\ref{figure:example:symbolic:piano:roll}),
with as many lines as there are possible pitches.
Note also that while a one-hot encoding
of a piano roll representation of a {\em monophonic\index{Monophonic}} melody
(with one note at a time)
is straightforward,
a one-hot encoding of a {\em polyphony}
(with simultaneous notes,
as for a guitar playing a chord)
is not.
One could then consider

\begin{itemize}

\item {\em many-hot encoding\index{Many-hot encoding}} --
where all elements of the vector corresponding to the notes or to the active components are set to 1;


\item {\em multi-one-hot encoding\index{Multi!-one-hot encoding}} --
where different voices\index{Multivoice} or tracks\index{Multitrack} are considered
(for multivoice representation, see Section~\ref{section:multi:voice:track})
and a one-hot encoding\index{One-hot encoding} is used for each different voice\index{Voice}/track\index{Track}; or

\item {\em multi-many-hot encoding\index{Multi!-many-hot encoding}} --
which is a multivoice representation with simultaneous notes for at least one or all of the voices.

\end{itemize}



\subsection{Summary}
\label{section:representation:input:encoding:summary}

\begin{figure}
\includegraphics[width=\textwidth]{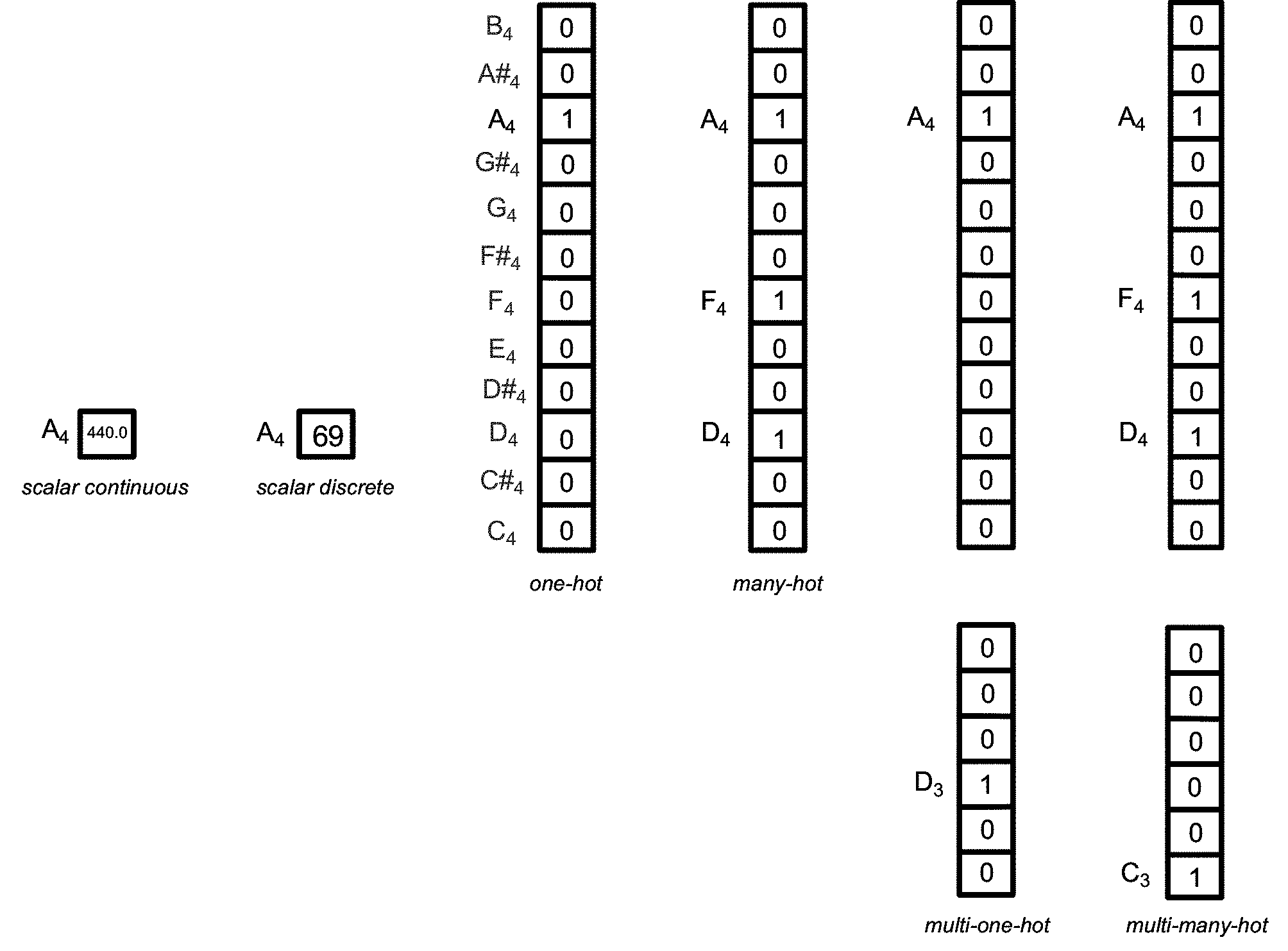}
\caption{Various types of encoding}
\label{figure:encoding:value:one:hot}
\end{figure}

The various approaches for encoding are illustrated in Figure~\ref{figure:encoding:value:one:hot},
showing from left to right

\begin{itemize}

\item a scalar\index{Scalar} continuous value encoding of A$_4$ (A440), the real number specifying its frequency in Hertz;

\item a scalar discrete integer value encoding\footnote{Note that,
	because the processing level of an artificial neural network only considers real values,
	an integer value will be casted into a real value.
	Thus, the case of a scalar integer value encoding boils down to the previous case of a scalar continuous value encoding.}
 of A$_4$, the integer number specifying its MIDI note number;

\item a one-hot encoding of A$_4$;

\item a many-hot encoding of a D minor chord (D$_4$, F$_4$, A$_4$);

\item a multi-one-hot encoding of a first voice with A$_4$ and a second voice with D$_3$; and

\item a multi-many-hot encoding of a first voice with a D minor chord (D$_4$, F$_4$, A$_4$)
and a second voice with C$_3$ (corresponding to a minor seventh on bass).

\end{itemize}

\subsection{Binning}
\label{section:representation:input:encoding:binning}

In some cases, a continuous variable is transformed into a discrete domain.
A common technique, named {\em binning\index{Binning}}, or also {\em bucketing\index{Bucketing}},
consists of

\begin{itemize}

\item dividing the original domain of values into smaller intervals\footnote{This can be automated
	through a learning process,
	e.g., by automatic construction of a decision tree\index{Decision tree}.},
named {\em bins\index{Bin}}; and

\item replacing each bin (and the values within it) by a {\em value representative}, often the central value.

\end{itemize}

Note that this binning technique may also be used to reduce the cardinality of the discrete domain of a variable.
An example is the Performance RNN\index{Performance RNN} system described in Section~\ref{section:experiment:performance:rnn},
for which the initial MIDI set of 127 values for note dynamics is reduced into 32 bins.

%



\subsection{Pros and Cons}
\label{section:representation:input:encoding:pros}

In general,
value encoding\index{Value encoding} is rarely used except for audio,
whereas one-hot encoding\index{One-hot encoding} is the most common strategy for symbolic representation\footnote{Let us remind
	(as pointed out in Section~\ref{section:representation:signal:vs:symbolic})
	that,
	at the level of the encoding of a representation
	and its processing by a deep network,
	the distinction between audio and symbolic representation boils down to nothing,
	as only numerical values and operations are considered.
	In fact the general principles of a deep learning architecture are independent of that distinction
	and this is one of the vectors of the generality of the approach.
	See also in \cite{manzelli:combining:raw:symbolic:audio:networks:ismir:2018}
	the example of an architecture
	(to be introduced in Section~\ref{section:systems:wavenet})
	which combines
	audio and symbolic representations.}.
	
A counterexample is the case of the DeepJ\index{DeepJ} symbolic generation system described in Section~\ref{section:systems:deepj},
which is, in part, inspired by the WaveNet\index{WaveNet} audio generation system.
DeepJ's authors state that:
``We keep track of the dynamics of every note in an N x T dynamics matrix that,
for each time step, stores values of each note's dynamics scaled between 0 and 1,
where 1 denotes the loudest possible volume.
In our preliminary work,
we also tried an alternate representation of dynamics as a categorical value with 128 bins\index{Bin} as suggested by Wavenet \cite{oord:wavenet:arxiv:2016}.
Instead of predicting a scalar value, our model would learn a multinomial distribution of note dynamics.
We would then randomly sample dynamics during generation from this multinomial distribution.
Contrary to Wavenet's results, our experiments concluded that the scalar representation yielded results that were more harmonious.''
\cite{mao:deepj:arxiv:2018}.
	
The advantage of value encoding\index{Value encoding} is its compact representation,
at the cost of sensibility because of numerical operations (approximations).
The advantage of one-hot encoding\index{One-hot encoding} is its robustness\index{Robustness} (discrete versus analog),
at the cost of a high cardinality and therefore a potentially large number of inputs.

It is also important to understand that the choice of one-hot encoding at the {\em output} of the network architecture
is often (albeit not always) associated to a {\em softmax\index{Softmax}} function\footnote{Introduced
	in Section~\ref{section:architecture:neural:network:output:activation:function}.}
in order to compute the probabilities of each possible value,
for instance the probability of a note being an A, or an A$\sharp$, a B, a C, etc.
This actually corresponds to a {\em classification task\index{Classification!task}} between the possible values of the categorical variable.
This will be further analyzed in Section~\ref{section:architecture:neural:network:output:activation:function}.

\subsection{Chords}
\label{section:representation:input:encoding:chord}

Two methods of encoding chords, corresponding to the two main alternative representations discussed in Section~\ref{section:representation:chord}, are

\begin{itemize}

\item {\em implicit} and {\em extensional} -- enumerating the exact notes composing the chord.
The natural encoding strategy is many-hot\index{Many-hot encoding}.
An example is the RBM-based polyphonic music generation system described in Section~\ref{section:experiment:rbm}; and

\item {\em explicit} and {\em intensional} -- using a chord symbol combining a pitch class and a type (e.g., D minor).
The natural encoding strategy is multi-one-hot\index{Multi!-one-hot encoding},
with an initial one-hot encoding\index{One-hot encoding}
of the pitch class
and a second one-hot encoding of the class type (major, minor, dominant seventh, etc.).
An example is the MidiNet system\footnote{In MidiNet, the possible chord types are actually reduced to only
	major and minor.
	Thus, a boolean variable can be used in place of one-hot encoding.} 
described in Section~\ref{section:systems:midinet}.

\end{itemize}

\subsection{Special Hold and Rest Symbols}
\label{section:representation:input:encoding:hold:rest}

We have to consider the case of special symbols for hold
(``hold previous note'', see Section~\ref{section:representation:note:ending}) and rest (``no note'', see Section~\ref{section:representation:rest})
and how they relate to the encoding of actual notes.

First, note that there are some rare cases where the rest is actually {\em implicit}:

\begin{itemize} 

\item in MIDI format -- when there is no ``active'' ``Note on'',
that is when they all have been ``closed'' by a corresponding ``Note off''; and

\item in one-hot encoding -- when all elements of the vector encoding the possible notes are equal to 0
(i.e. a ``zero-hot'' encoding, meaning that none of the possible notes is currently selected).
This is for instance the case in the experiments by Todd (to be described in Section~\ref{section:experiment:todd:time:windowed})\footnote{This
	may appear at first as an economical encoding of a rest,
	but at the cost of some ambiguity when interpreting probabilities (for each possible note)
	produced by the softmax\index{Softmax} output of the network architecture.
	A vector with low probabilities for each note may be interpreted as a rest or as an equiprobability between notes.
	See the threshold trick proposed in Section~\ref{section:experiment:todd:time:windowed} in order to discriminate between the two
	possible interpretations.}.

\end{itemize}


Now, let us consider how to encode hold and rest depending on how a note pitch is encoded:

\begin{itemize}

\item {\em value encoding\index{Value encoding}} --
In this case, one needs to add two extra boolean variables (and their corresponding input nodes) {\em hold} and {\em rest}.
This must be done for each possible independent voice in the case of a polyphony; or
	
\item {\em one-hot encoding\index{One-hot encoding}} --
In that case (the most frequent and manageable strategy),
one just needs to extend the vocabulary of the one-hot encoding with two additional possible values: {\em hold} and {\em rest}.
They will be considered at the same level, and of the same nature,
as possible notes (e.g., A$_3$ or C$_4$) for the input as well as for the output.

\end{itemize}

\subsection{Drums and Percussion}
\label{section:representation:rhythm:drums}

Some systems explicitly consider drums\index{Drums} and/or percussion\index{Percussion}.
A drum or percussion kit
is usually modeled 
%
as a single-track polyphony\index{Single!-track polyphony}
by considering distinct simultaneous ``notes'',
each ``note'' corresponding to a drum or percussion component
(e.g., snare, kick, bass tom, hi-hat, ride cymbal, etc.),
that is as a many-hot encoding\index{Many-hot encoding}.
%
%

An example of a system dedicated to rhythm generation is described in Section~\ref{section:experiment:makris:rhythm}.
It follows the single-track polyphony\index{Single!-track polyphony} approach.
In this system, each of the five components is represented through a binary value,
specifying whether or not there is a related event for current time step.
Drum events are represented as a binary word\footnote{In this system, encoding is made in text\index{Text},
	similar
	to the format described in Section~\ref{section:representation:text}
	and more precisely following the approach proposed in \cite{choi:text:lstm:arxiv:2016}.}
of length 5,
where each binary value corresponds to one of the five drum components;
for instance, $10010$ represents simultaneous playing
of the kick (bass drum) and the high-hat,
following a many-hot encoding\index{Many-hot encoding}.

Note that this system
also includes
-- as an additional voice\index{Voice}/track\index{Track} --
a condensed representation of the bass line\index{Bass!line} part
and some information
representing the meter,
see more details in Section~\ref{section:experiment:makris:rhythm}.
The authors \cite{makris:rhythm:composition:2017}
argue that this extra explicit information ensures that the network architecture is aware of the beat structure at any given point.

Another example is the MusicVAE\index{MusicVAE} system
(see Section~\ref{section:system:music:vae}),
where nine different drum/percussion components are considered, which gives $2^9$
possible combinations,
i.e. $2^9 = 512$ different tokens.

\section{Dataset}
\label{section:representation:dataset}

The choice
of a dataset is fundamental for good music generation.
At first, a dataset should be of sufficient size (i.e. contain a sufficient number of examples)
to guarantee accurate learning\footnote{Neural networks and deep learning architectures
	need lots of examples to function properly.
	However, one recent research area is about learning from scarce data.}. 
As noted by Hadjeres in \cite{hadjeres:thesis:2018}:
``I believe that this tradeoff between the size\index{Size} of a dataset\index{Dataset} and its coherence\index{Coherence}
is one of the major issues when building deep generative models.
If the dataset is very heterogeneous\index{Heterogeneous},
a good generative model should be able to distinguish the different subcategories and manage to generalize\index{Generalization} well.
On the contrary, if there are only slight differences between subcategories,
it is important to know if the ``averaged model'' can produce musically-interesting results.''

\subsection{Transposition and Alignment}
\label{section:representation:transposition}

A common technique in machine learning is to generate {\em synthetic data\index{Synthetic data}} as a way to artificially augment
the size of the dataset\index{Dataset} (the number of training examples)\footnote{This is
	named {\em dataset augmentation\index{Dataset!augmentation}}.},
in order to improve accuracy and generalization of the learnt model (see Section~\ref{section:architecture:training:regularization}). 
In the musical domain, a natural and easy way is {\em transposition\index{Transposition}},
i.e. to transpose all examples in all keys\index{Key}.
In addition to artificially augmenting the dataset,
this provides a key (tonality) invariance of all examples and thus makes the examples more generic.
Moreover, this also reduces sparsity\index{Sparsity} in the training data.
This transposition technique is, for instance,
used in the C-RBM system \cite{lattner:structure:polyphonic:generation:jcms:2018}
described in Section~\ref{section:experiment:c:rbm}.

An alternative approach is to transpose (align) all examples into a {\em single common key}.
This has been advocated for the RNN-RBM system \cite{boulanger:temporal:dependencies:icml:2012}
to facilitate learning, see Section~\ref{section:experiment:rnn:rbm}.

\subsection{Datasets and Libraries}
\label{section:representation:dataset:libraries}

A practical issue is the availability of datasets\index{Dataset} for training systems
and also for evaluating and comparing systems and approaches.
There are some reference datasets in the image\index{Image} domain
(e.g., the MNIST\index{MNIST}\footnote{MNIST
	stands for Modified National Institute of Standards and Technology.}
dataset about handwritten digits \cite{lecun:mnist:web:1998}),
but none yet
in the music domain.
However, various datasets or libraries\footnote{The difference between a dataset and a library
	is that a dataset is almost ready for use to train a neural network architecture,
	as all examples are encoded within a single file and in the same format,
	although some extra data processing may be needed in order to adapt the format to the encoding of the representation for the architecture or vice-versa;
	whereas a library is usually composed of a set of files, one for each example.}
have been made public, with some examples listed below:

\begin{itemize}


\item the Classical piano MIDI database\index{Classical piano MIDI database}
\cite{piano-midi.de:web};

\item the JSB Chorales dataset\index{JSB Chorales dataset}\footnote{Note that
	this dataset uses a quarter note quantization\index{Time!quantization},
	whereas a smaller quantization at the level of a sixteenth note should be used
	in order to capture the smallest note duration (eighth note),
	see Section~\ref{section:representation:note:ending}.}
\cite{allan2005harmonising};

\item the LSDB\index{LSDB} (Lead Sheet Data Base\index{Lead sheet data base}) repository \cite{lsdb:ismir:2013},
with more than 12,000 lead sheets\index{Lead sheet} (including from all jazz\index{Jazz} and bossa nova song books),
developed within the Flow Machines\index{Flow Machines} project \cite{csl:flow:machines:web:2012};

\item the MuseData library\index{MuseData library}, an electronic library of classical\index{Classical} music with more than 800 pieces,
from CCARH in Stanford University
\cite{ccarh:musedata:web};

\item the MusicNet dataset\index{MusicNet dataset}
\cite{thickstun:musicnet:arxiv:2016},
a collection of 330 freely-licensed classical music recordings
together with over 1 million annotated labels (indicating timing and instrumental information);

\item the Nottingham database\index{Nottingham database}, a collection of 1,200 folk\index{Folk} tunes in the ABC notation \cite{foxley:nottingham:database:web},
each tune consisting of a simple melody on top of chords,
in other words an ABC equivalent of a lead sheet;

\item the Session \cite{web:the:session}, a repository and discussion platform for Celtic\index{Celtic} music
in the ABC notation\index{ABC notation} containing more than 15,000 songs;

\item the Symbolic Music dataset\index{Symbolic Music dataset} by Walder \cite{2016arXiv160602542W}, a huge set of cleaned and preprocessed MIDI files;

\item the TheoryTab database\index{TheoryTab database} \cite{theorytab:web:2017},
a set of songs represented in a tab\index{Tab} format, a combination of a piano roll melody,
chords and lyrics,
in other words a piano roll equivalent of a lead sheet;

\item the Yamaha e-Piano Competition dataset\index{Yamaha e-Piano Competition dataset},
in which participants MIDI performance\index{Performance} records are made available \cite{yamaha:e-piano:competition:web}.

\end{itemize}

%% file: architecture.tex
\chapter{Architecture}
\label{section:chapter:architecture}

\abstract*{Chapter~\ref{section:chapter:architecture} Architecture presents the third dimension of the conceptual framework proposed in this book to analyze,
classify and compare various deep learning-based music generation systems.
The architecture represents the set of computational units (neurons), grouped in layers, and their weighted connexions
which process and generate a musical representation.
At first, we summarize the history and evolution of artificial neural network architectures.
An introduction to artificial neural networks is presented, starting from linear regression up to a typical neural network layer,
considered as a basic building block.
From it, various architectures are derived and introduced:
feedforward (MLP), recurrent (RNN), autoencoder, generative adversarial networks and others.
Various ways of composing architectures are also examined.
This chapter may be skipped by a reader already expert in neural networks and deep learning architectures.} 

\label{section:architectures}

\label{section:architectures:history}

\begin{figure}
\includegraphics[scale=0.8]{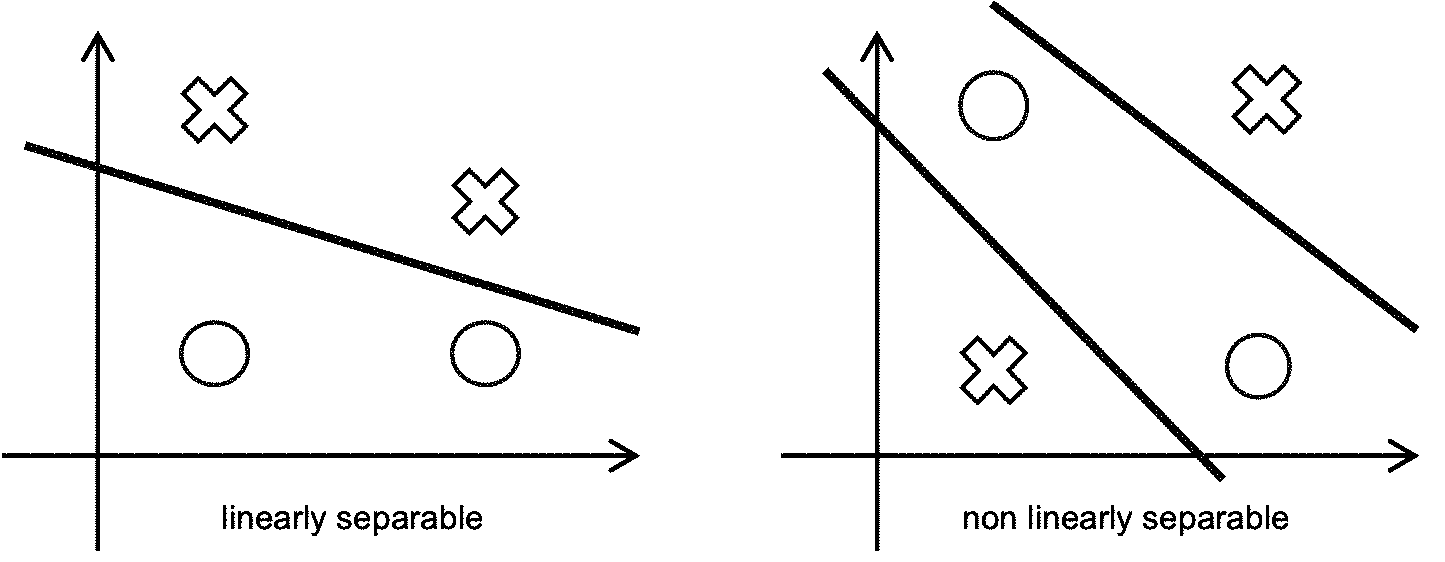}
\caption{Example and counterexample of linear separability}
\label{figure:non:linear:example}
\end{figure}

Deep networks\index{Deep!network|see{Deep neural network}} are a natural evolution of neural networks\index{Neural!network},
themselves being an evolution of the Perceptron\index{Perceptron},
proposed by Rosenblatt in 1957 \cite{rosenblatt:perceptron:1957}.
Historically speaking\footnote{See, for example, \cite[Section~1.2]{goodfellow:deep:learning:book:2016} for a more detailed analysis
	of key trends in the history of deep learning.},
the Perceptron was criticized by Minsky and Papert in 1969 \cite{book:perceptrons}
for its inability to classify {\em nonlinearly separable domains}\footnote{A simple example and a counterexample
	of linear separability
	(of a set of four points within a 2-dimensional space and belonging to green cross or red circle classes)
	are shown in Figure~\ref{figure:non:linear:example}.
	The elements of the two classes are linearly separable if there is at least one straight line separating them.
	Note that the discrete version of the counterexample corresponds to the case of the exclusive or (XOR) logical operator,
	which was used as an argument by Minsky and Papert in \cite{book:perceptrons}.}.
Their criticism also served in favoring an alternative
approach of Artificial Intelligence\index{Artificial!intelligence},
based on symbolic\index{Symbolic} representations and reasoning.

Neural networks reappeared in the 1980s, thanks to the idea of {\em hidden layers\index{Hidden!layer}}
joint with nonlinear units,
to resolve the initial linear separability limitation,
and to the {\em backpropagation\index{Backpropagation}} algorithm,
to train such multilayer neural networks \cite{rumelhart:backpropagation:1986}.

In the 1990s, neural networks suffered declining interest\footnote{Meanwhile,
	convolutional networks started to gain interest, notably though handwritten digit recognition applications \cite{lecun:document:recognition:1998}.
	As Goodfellow {\em et al.} in \cite[Section~9.11]{goodfellow:deep:learning:book:2016} put it:
	``In many ways, they carried the torch for the rest of deep learning
	and paved the way to the acceptance of neural networks in general.''}
because of the difficulty in training efficiently neural networks with many layers\footnote{Another related limitation,
	although specific to the case of recurrent networks\index{Recurrent!network},
	was the difficulty in training them efficiently on very long sequences.
	This was resolved in 1997 by Hochreiter and Schmidhuber
	with the {\em Long short-term memory\index{Long!short-term memory}} (LSTM\index{LSTM}) architecture
	\cite{hochreiter:lstm:1997}, presented in Section~\ref{section:architecture:lstm}.}
and due to the competition from {\em support vector machines\index{Support vector machine}} (SVM\index{SVM})
\cite{vapnik:book:svm:statistical:learning:theory:1995},
which were efficiently designed to maximize the {\em separation margin} and had a solid formal background.

An important advance was the invention of the {\em pre-training\index{Pre-training}} technique\footnote{Pre-training\index{Pre-training}
	consists in prior training in {\em cascade}
	(one layer at a time, also named {\em greedy layer-wise unsupervised training}\index{Greedy layer-wise unsupervised training}) of each hidden layer
	\cite{hinton:fast:algorithm:2006} \cite[page~528]{goodfellow:deep:learning:book:2016}.
	It turned out to be a significant improvement for the accurate training of neural networks with several layers
	\cite{erhan:pre:training:jmlr:2010}.
	That said, pre-training is now rarely used and has been replaced by other more recent techniques,
	such as {\em batch normalization} and {\em deep residual learning}.
	But its underlying techniques are useful for addressing some new concerns like {\em transfer learning}\index{Transfer learning},
	which deals with the issue of {\em reusability} (of what has been learnt,
	see Section~\ref{section:discussion:transfer}).}
by Hinton {\em et al.} in 2006
\cite{hinton:fast:algorithm:2006}, which resolved this limitation.
In 2012, an image recognition\index{Image!recognition} competition (the ImageNet Large Scale Visual Recognition Challenge \cite{ILSVRC15}) 
was won by a deep neural network algorithm named AlexNet\index{AlexNet}\footnote{AlexNet was designed by the SuperVision team headed by Hinton and composed of
	Alex Krizhevsky, Ilya Sutskever and Geoffrey E. Hinton \cite{Krizhevsky:2012:ICD:2999134.2999257}.
	AlexNet is a deep convolutional neural network\index{Convolutional!neural network}
	with 60 million parameters and 650,000 neurons,
	consisting of five convolutional layers, some followed by max-pooling layers,
	and three globally-connected layers.},
with a stunning margin\footnote{On the first task, AlexNet won the competition
	with a 15\% error rate whereas other teams did not achieve better than a 26\% error rate.}
over the other algorithms which were using handcrafted features.
This striking victory was the event which ended the prevalent opinion that neural networks with many hidden layers
	could not be efficiently trained\footnote{Interestingly,
	the title of Hinton {\em et al.}'s article about pre-training \cite{hinton:fast:algorithm:2006}
	is about ``deep belief nets\index{Deep!belief net}''
	and does not mention the term ``neural nets\index{Neural!net|see{Neural network}}''
	because, as Hinton
	remembers it in \cite{kurenkov:history:deep:learning:2015}:
	``At that time, there was a strong belief that deep neural networks were no good and could {\em never} be trained
	and that ICML (International Conference on Machine Learning) should {\em not} accept papers about neural networks.''}.


\section{Introduction to Neural Networks}
\label{section:architecture:linear:regression}
\label{section:architecture:basic:building:block}
\label{section:architecture:layer}

The purpose of this section is to review, or to introduce,
the basic principles of {\em artificial neural networks\index{Artificial!neural network}}.
Our objective is to define the key {\em concepts} and {\em terminology}
that we will use when analyzing various music generation systems.
Then, we will introduce the concepts and basic principles of various derived architectures,
like autoencoders, recurrent networks, RBMs, etc.,
which are used in musical applications.
We will not describe extensively the techniques of neural networks and deep learning,
for example covered in the recent book \cite{goodfellow:deep:learning:book:2016}.

\subsection{Linear Regression}
\label{section:statistics:linear:regression}

Although bio-inspired (biological neurons),
the foundation of neural networks and deep learning is {\em linear regression\index{Linear!regression}}.
In statistics, linear regression is an approach for modeling the (assumed linear) relationship between a scalar variable $\text{y} \in {\rm I\!R}$
and one\footnote{The case of one explanatory variable
	is called {\em simple linear regression\index{Simple linear regression}},
	otherwise it is named {\em multiple linear regression\index{Multiple linear regression}}.}
or more than one {\em explanatory variable(s)\index{Explanatory variable}} x$_1$ ... x$_n$,
with x$_i \in {\rm I\!R}$, jointly noted as vector x.
A simple example is to predict the value of a house, depending on some factors (e.g., size, height, location\ldots).

Equation~\ref{equation:linear:regression} gives the general model of a (multiple) linear regression, where

\begin{equation}
h(\text{x}) = b + \theta_1 \text{x}_1 + ... + \theta_n \text{x}_n = b + \sum\limits_{i=1}^{n} \theta_i \text{x}_i
\label{equation:linear:regression}
\end{equation}

\begin{itemize}

\item $h$ is the {\em model\index{Model}}, also named {\em hypothesis\index{Hypothesis}},
as this is the hypothetical best model to be discovered, i.e. learnt;

\item $b$ is the {\em bias\index{Bias}}\footnote{It could also be
	notated as $\theta_0$, see Section~\ref{section:architecture:linear:regression:architectural:view}.},
representing the {\em offset}; and

\item $\theta_1$ ... $\theta_n$ are the {\em parameters\index{Parameter}} of the model,
the {\em weights\index{Weight}},
corresponding to the explanatory variables x$_1$ ... x$_n$.

\end{itemize}


\subsection{Notations}
\label{section:architecture:notations}

We will use the following simple notation\index{Notation convention} conventions

\begin{itemize}

\item a {\em constant} is in roman (straight) font, e.g., integer 1 and note C$_4$.

\item a {\em variable\index{Variable}} of a model is in roman
font, e.g., input variable x and output variable y (possibly vectors).

\item a {\em parameter\index{Parameter}} of a model is in italics, e.g., bias $b$, weight parameter $\theta_1$,
model function $h$,
number of explanatory variables $n$ and index $i$ of a variable x$_i$.


\item a {\em probability\index{Probability}} as well as a {\em probability distribution\index{Probability!distribution}}
are in italics and upper case, e.g.,
probability $P(\text{note} = \text{A}_4)$ that the value of variable note is A$_4$ and
probability distribution $P(\text{note})$ of variable note over all possible notes (outcomes).



\end{itemize}

\subsection{Model Training}
\label{section:architecture:training}

The purpose of training\index{Training} a linear regression model
is to find the values for each weight $\theta_i$
and the bias $b$
that best fit the actual training data/exam\-ples, i.e. various pairs of values $(\text{x}, \text{y})$.
In other words, we want to find the parameters and bias values
such that for all values of x, $h(\text{x})$ is {\em as close as possible}\footnote{Actually,
	for the neural networks
	that are more complex (nonlinear models) than linear regression
	and that will be introduced in Section~\ref{section:architecture:feedforward},
	the best fit to the training data is not necessarily the best hypothesis
	because it may have a low {\em generalization\index{Generalization}},
	i.e. a low ability to predict {\em yet unseen data}.
	This issue, named {\em overfitting\index{Overfitting}}, will be introduced in Section~\ref{section:architecture:training:overfitting}.}
to y, according to some measure named the {\em cost\index{Cost}}.
This measure represents the {\em distance\index{Distance}}
between $h(\text{x})$ (the prediction, also notated as \^y) and y (the actual ground value), for {\em all} examples.

The cost,
also named the {\em loss\index{Loss}},
is usually\footnote{Or also $J(\theta)$, $\mathcal{L}_\theta$
	or $\mathcal{L}(\theta)$.}
notated $J_\theta(h)$
and could be measured, for example, by a mean squared error\index{Mean squared error}
(MSE),
which measures the average squared difference,
as shown in Equation~\ref{equation:linear:regression:cost:mean:squared:error},
where $m$ is the number of examples and $(\text{x}^{(i)}, \text{y}^{(i)})$ is the $i$th example pair.

\begin{equation}
J_\theta(h) = 1/m \sum_{i=1}^{m}{(\text{y}^{(i)} - h(\text{x}^{(i)}))^2} = 1/m \sum_{i=1}^{m}{(\text{y}^{(i)} - \text{\^y}^{(i)})^2}
\label{equation:linear:regression:cost:mean:squared:error}
\end{equation}

\begin{figure}
\includegraphics[scale=1]{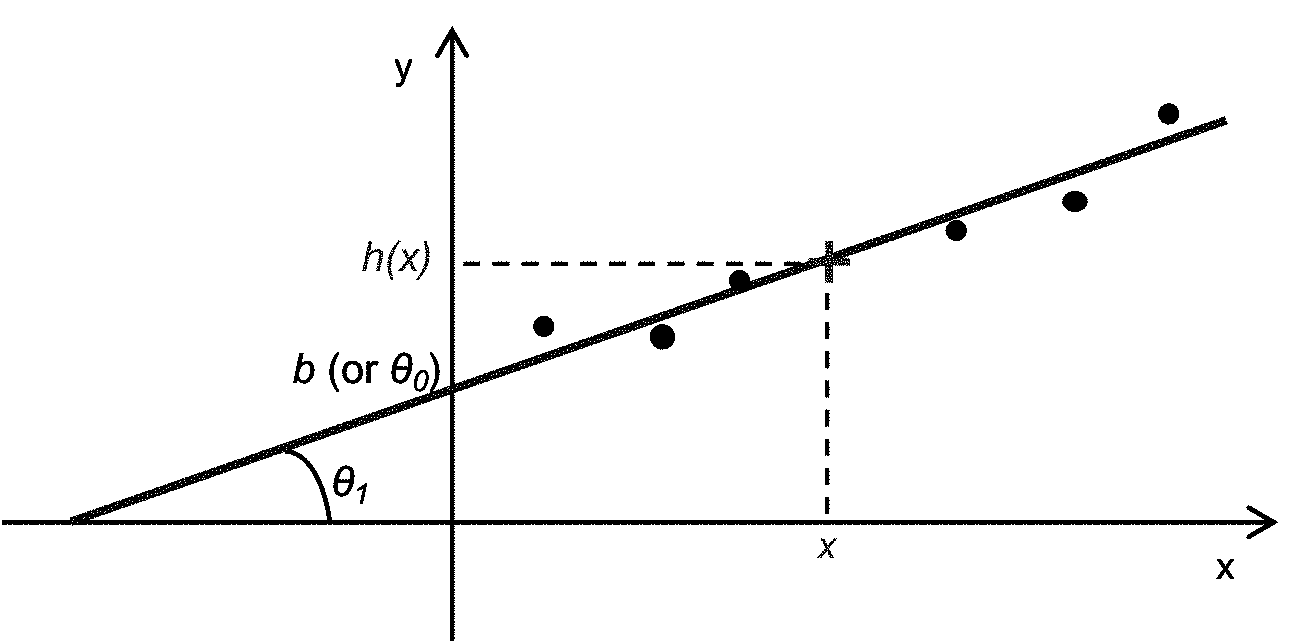}
\caption{Example of simple linear regression}
\label{figure:simple:linear:regression}
\end{figure}

An example is shown in Figure~\ref{figure:simple:linear:regression}
for the case of simple linear regression, i.e. with only one explanatory variable x.
Training data are shown as blue solid dots.
Once the model has been trained, values of the parameters are adjusted,
illustrated by the blue solid bold line
which mostly fits the examples.
Then, the model can be used for {\em prediction\index{Prediction}},
e.g., to provide a good estimate\index{Estimation} \^y of the actual value of y
for a given
value of x
by computing $h(\text{x})$ (shown in green).

\subsection{Gradient Descent Training Algorithm}
\label{section:training:algorithm}

\begin{figure}
\includegraphics[scale=0.4]{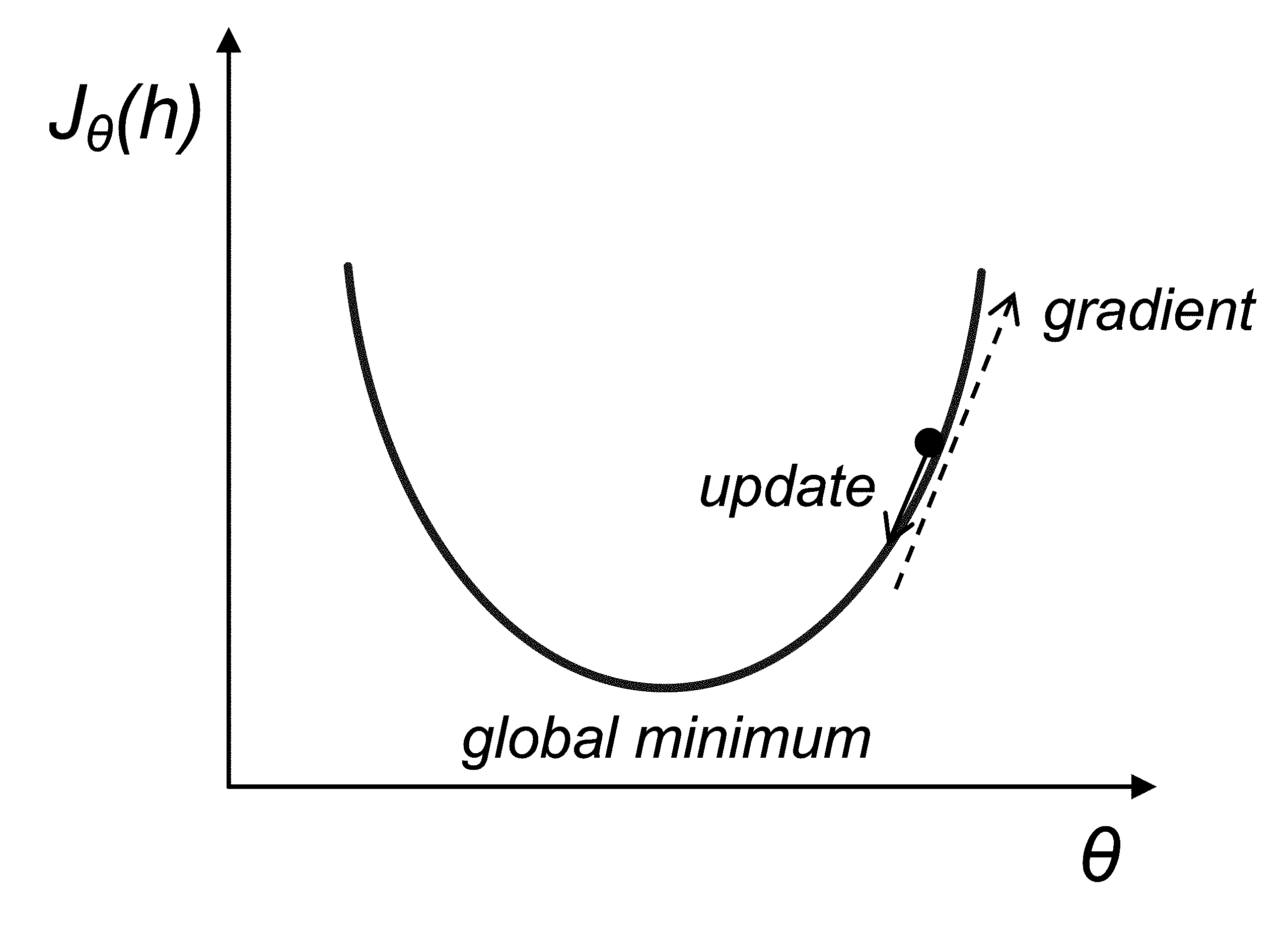}
\caption{Gradient descent}
\label{figure:linear:regression:gradient:descent}
\end{figure}

The basic algorithm for training a linear regression model, using the simple {\em gradient descent\index{Gradient!descent}} method,
is actually pretty simple\footnote{See,
	e.g., \cite{ng:course:notes:linear:regression:2016} for more details.}:

\begin{itemize}

\item initialize each parameter
$\theta_{i}$ and the bias $b$ to a random\index{Random}
or some heuristic\index{Heuristic} value\footnote{Pre-training\index{Pre-training}
	led to a significant advance,
	as it improved the initialization\index{Initialization} of the parameters\index{Parameter!initialization} by using actual training data,
	via sequential training of the successive layers\index{Layer} \cite{erhan:pre:training:jmlr:2010}.};

\item compute the values of the model $h$ for all examples\footnote{Computing the cost for all examples
	is the best method but also computationally costly.
	There are numerous heuristic alternatives to minimize the computational cost,
	e.g., {\em stochastic gradient descent\index{Stochastic!gradient descent}} (SGD\index{SGD}),
	where one example is randomly chosen,
	and
	{\em minibatch gradient descent\index{Minibatch!gradient descent}},
	where a subset of examples is randomly chosen.
	See, for example, \cite[Sections~5.9 and~8.1.3]{goodfellow:deep:learning:book:2016} for more details.};

\item compute the {\em cost\index{Cost}} $J_\theta(h)$,
e.g., by Equation~\ref{equation:linear:regression:cost:mean:squared:error};

\item compute the {\em gradients\index{Gradient}} $\frac{\partial J_\theta(h)}{\partial \theta_i}$
which are the {\em partial derivatives}
of the cost function $J_\theta(h)$ with respect to each $\theta_i$,
as well as to the bias $b$;


\item {\em update simultaneously}\footnote{A simultaneous update is necessary
	for the algorithm to behave correctly.}
all parameters $\theta_{i}$ and the bias according to the update rule\footnote{The update rule may also be notated as
	$\theta := \theta - \alpha {\nabla_\theta}J_\theta(h)$,
	where ${\nabla_\theta}J_\theta(h)$ is the vector of gradients $\frac{\partial J_\theta(h)}{\partial \theta_i}$.}
shown in Equation~\ref{equation:linear:regression:update:rule},
with $\alpha$ being the {\em learning rate\index{Learning!rate}}.

\begin{equation}
\theta_{i} := \theta_{i} - \alpha \frac{\partial J_\theta(h)}{\partial \theta_i}
\label{equation:linear:regression:update:rule}
\end{equation}

This represents an update in the opposite direction of the gradients in order to decrease the cost $J_\theta(h)$,
as illustrated in Figure~\ref{figure:linear:regression:gradient:descent}; and

\item {\em iterate} until the error reaches {\em a} {\em minimum}\footnote{If the cost function is {\em convex\index{Convex}}
	(the case for linear regression),
	there is only one {\em global minimum\index{Global!minimum}},
	and thus there is a guarantee of finding the {\em optimal} model.},
or after a certain number of iterations.

\end{itemize}

\subsection{From Model to Architecture}
\label{section:architecture:linear:regression:architectural:view}

\begin{figure}
\includegraphics[scale=0.8]{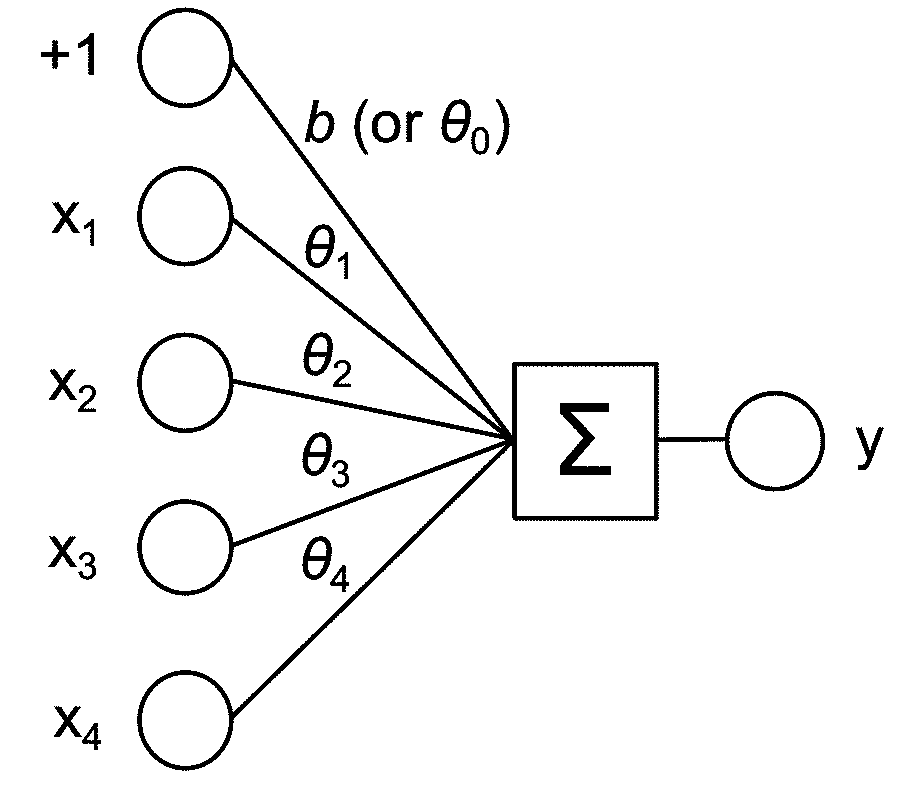}
\caption{Architectural model of linear regression}
\label{figure:linear:regression:architecture}
\end{figure}

Let us now introduce in Figure~\ref{figure:linear:regression:architecture} a graphical representation
of a linear regression model, as a precursor of a neural network.
The {\em architecture\index{Architecture}} represented
is actually the computational representation of the
model\index{Model}\footnote{We mostly use the term {\em architecture}
	as, in this book, we are concerned with the way to implement and compute a given model
	and also with the relation between an architecture and a representation\index{Representation}.}.

The weighted sum is represented as a {\em computational
unit\index{Unit}}\footnote{We use the term {\em node\index{Node}}
	for any component of a neural network,
	whether it is just an {\em interface} (e.g., an input node\index{Input!node})
	or a {\em computational unit} (e.g., a weighted sum or a function).
	We use the term {\em unit\index{Unit}} only in the case of a computational node.
	The term {\em neuron\index{Neuron}} is also often used in place of unit,
	as a way to emphasize the inspiration\index{Biological!inspiration} from biological neural networks\index{Biological!neural network}.},
drawn as a squared box with a $\sum$, taking its inputs from the x$_i$ nodes, drawn as circles.

In the example shown, there are four explanatory variables: x$_1$, x$_2$, x$_3$ and x$_4$.
Note that there is some convention of considering the bias\index{Bias} as a special case of weight
(thus alternatively notated as $\theta_0$)
and having a corresponding input node named the {\em bias node\index{Bias!node}},
which is {\em implicit}\footnote{However,
	as will be explained later in Section~\ref{section:architecture:neural:network},
	bias nodes rarely appear in illustrations of non-toy neural networks.}
and has a constant value notated as $+1$.
This actually corresponds to considering an implicit additional explanatory variable x$_0$ with constant value +1,
as shown in Equation~\ref{equation:linear:regression:theta:0},
alternative formulation of linear regression initially defined in Equation~\ref{equation:linear:regression}.

\begin{equation}
h(\text{x}) = \theta_0 + \theta_1 \text{x}_1 + ... + \theta_n \text{x}_n = \sum\limits_{i=0}^{n} \theta_i \text{x}_i
\label{equation:linear:regression:theta:0}
\end{equation}

\subsection{From Model to Linear Algebra Representation}
\label{section:architecture:linear:regression:linear:algebra}

The initial linear regression equation (in Equation~\ref{equation:linear:regression})
may also be made more compact thanks to a linear algebra\index{Linear!algebra} notation
leading to Equation~\ref{equation:linear:regression:linear:algebra} where

\begin{equation}
h(\text{x}) = b + \theta \text{x}
\label{equation:linear:regression:linear:algebra}
\end{equation}

\begin{itemize}


\item $b$ and $h(\text{x})$ are scalars;

\item $\theta$ is a row vector\index{Row vector}\footnote{That is
	a matrix\index{Matrix} which has a single row,
	i.e. a matrix of dimension $1{\times}n$.}
consisting of a single row of $n$ elements:
$\begin{bmatrix} \theta_1 &\theta_2 &\dots &\theta_n \end{bmatrix}$;


\item x is a column vector\index{Column vector}\footnote{That is
	a matrix which has a single column,
	i.e. a matrix of dimension $n{\times}1$.}
consisting of a single column of $n$ elements: $\begin{bmatrix} \text{x}_1\\ \text{x}_2\\ \vdots\\ \text{x}_n \end{bmatrix}$.

\end{itemize}

\subsection{From Simple to Multivariate Model}
\label{section:architecture:from:simple:multiple:linear}

\begin{figure}
\includegraphics[scale=0.8]{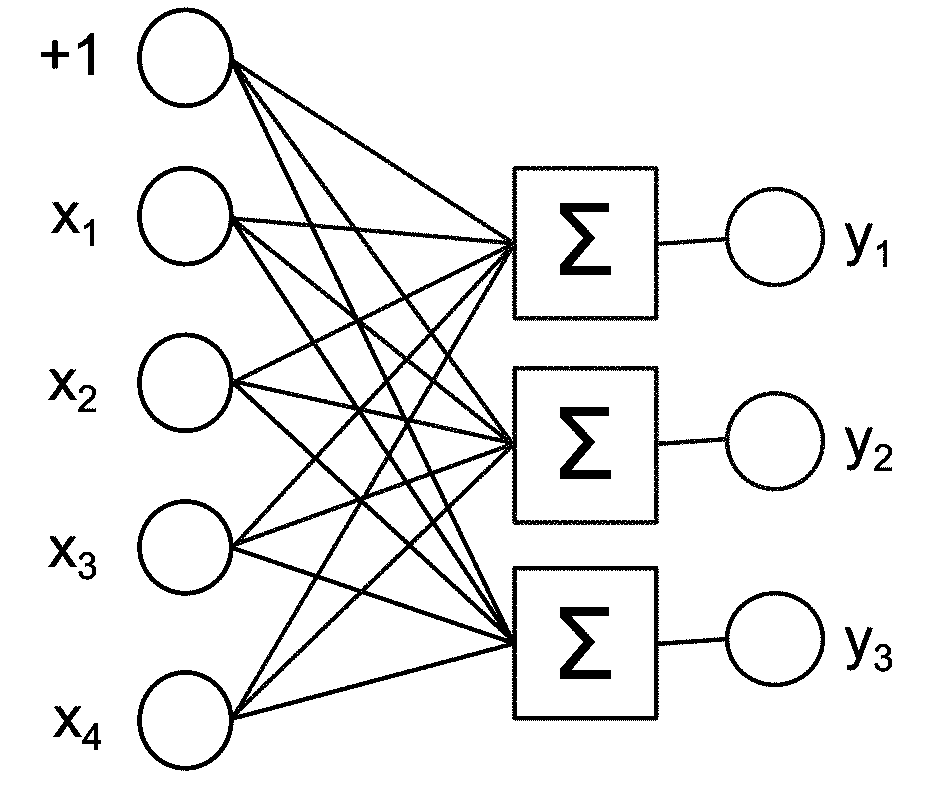}
\caption{Architectural model of multivariate linear regression}
\label{figure:multivariate:linear:regression:architecture}
\end{figure}

Linear regression can be generalized to {\em multivariate linear regression\index{Multivariate linear regression}},
the case when there are multiple variables y$_1$ ... y$_p$ to be predicted,
as illustrated in Figure~\ref{figure:multivariate:linear:regression:architecture}
with three predicted variables: y$_1$, y$_2$ and y$_3$,
each subnetwork represented in a different color.

\label{section:architecture:building:block:linear:algebra}

The corresponding linear algebra equation is
Equation~\ref{equation:linear:regression:linear:algebra:multivariate},
where

\begin{equation}
h(\text{x}) = b + W \text{x}
\label{equation:linear:regression:linear:algebra:multivariate}
\end{equation}


\begin{itemize}

\item the $b$ bias vector is a column vector of dimension $p{\times}1$,
with $b_j$ representing the weight of the connexion between the bias input node and the $j$th sum operation
corresponding to the $j$th output node;

\item the $W$ weight matrix is a matrix of dimension $p{\times}n$,
that is with $p$ rows and $n$ columns, with
$W_{i,j}$ representing the weight of the connexion between the $j$th input node
and the $i$th sum operation corresponding to the $i$th output node;

\item $n$ is the number of input nodes (without considering the bias node); and

\item $p$ is the number of output nodes.

\end{itemize}

\begin{figure}
\includegraphics[scale=0.8]{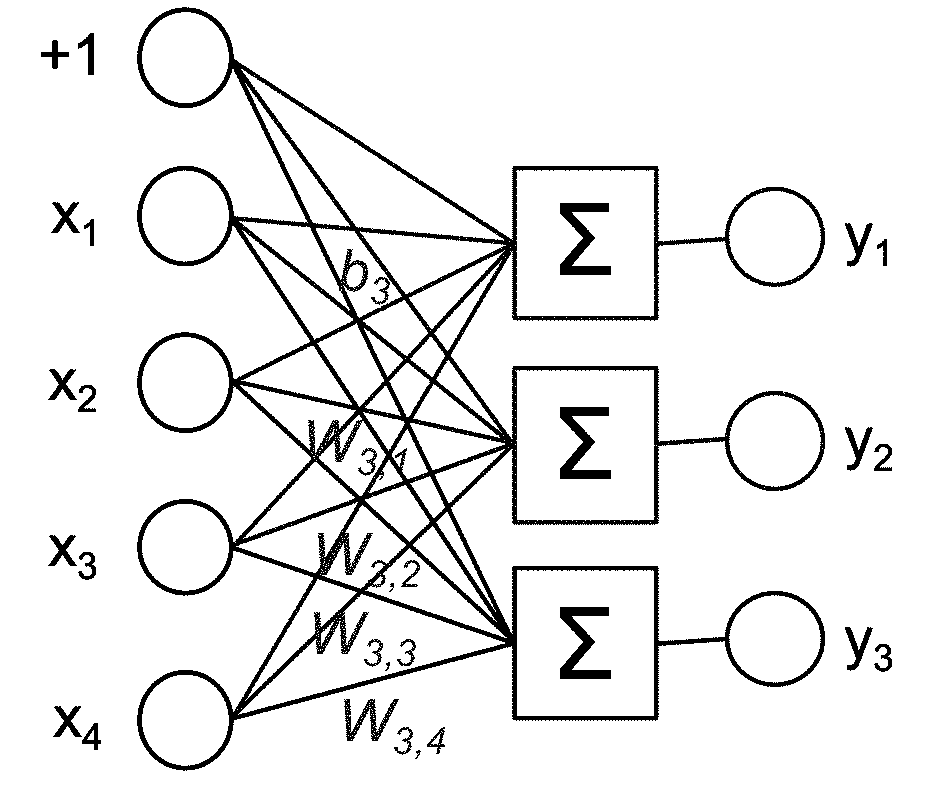}
\caption{Architectural model of multivariate linear regression showing the bias and the weights corresponding to the connexions to the third output}
\label{figure:multivariate:linear:regression:with:af:architecture:weights}
\end{figure}

For the architecture shown in
Figure~\ref{figure:multivariate:linear:regression:architecture},
$n = 4$ (the number of input nodes and of columns of $W$)
and $p = 3$ (the number of output nodes and of rows of $W$).
The corresponding $b$ bias vector and $W$ weight matrix are shown
in Equations~\ref{equation:basic:block:feedforward:example:bias} and~\ref{equation:basic:block:feedforward:example:matrix}\footnote{Indeed,
	$b$ and $W$ are generalizations of $b$ and $\theta$
	for the case of univariate linear regression
	(as shown
	in Section~\ref{section:architecture:linear:regression:linear:algebra})
	to the case of multivariate and thus to multiple rows, each row corresponding to an output node.}
and\footnote{By showing only the connexions to one of the output node,
	in order to keep readability.}
in Figure~\ref{figure:multivariate:linear:regression:with:af:architecture:weights}.

\begin{equation}
b =
\left[
\begin{array}{c}
b_1\\
b_2\\
b_3
\end{array}
\right]
\label{equation:basic:block:feedforward:example:bias}
\end{equation}

\begin{equation}
W =
\left[
\begin{array}{cccc}
W_{1,1}		&W_{1,2}		&W_{1,3}		&W_{1,4}\\
W_{2,1}		&W_{2,2}		&W_{2,3}		&W_{2,4}\\
W_{3,1}		&W_{3,2}		&W_{3,3}		&W_{3, 4}
\end{array}
\right]
\label{equation:basic:block:feedforward:example:matrix}
\end{equation}

\subsection{Activation Function}
\label{section:architecture:activation:function}

\begin{figure}
\includegraphics[scale=0.4]{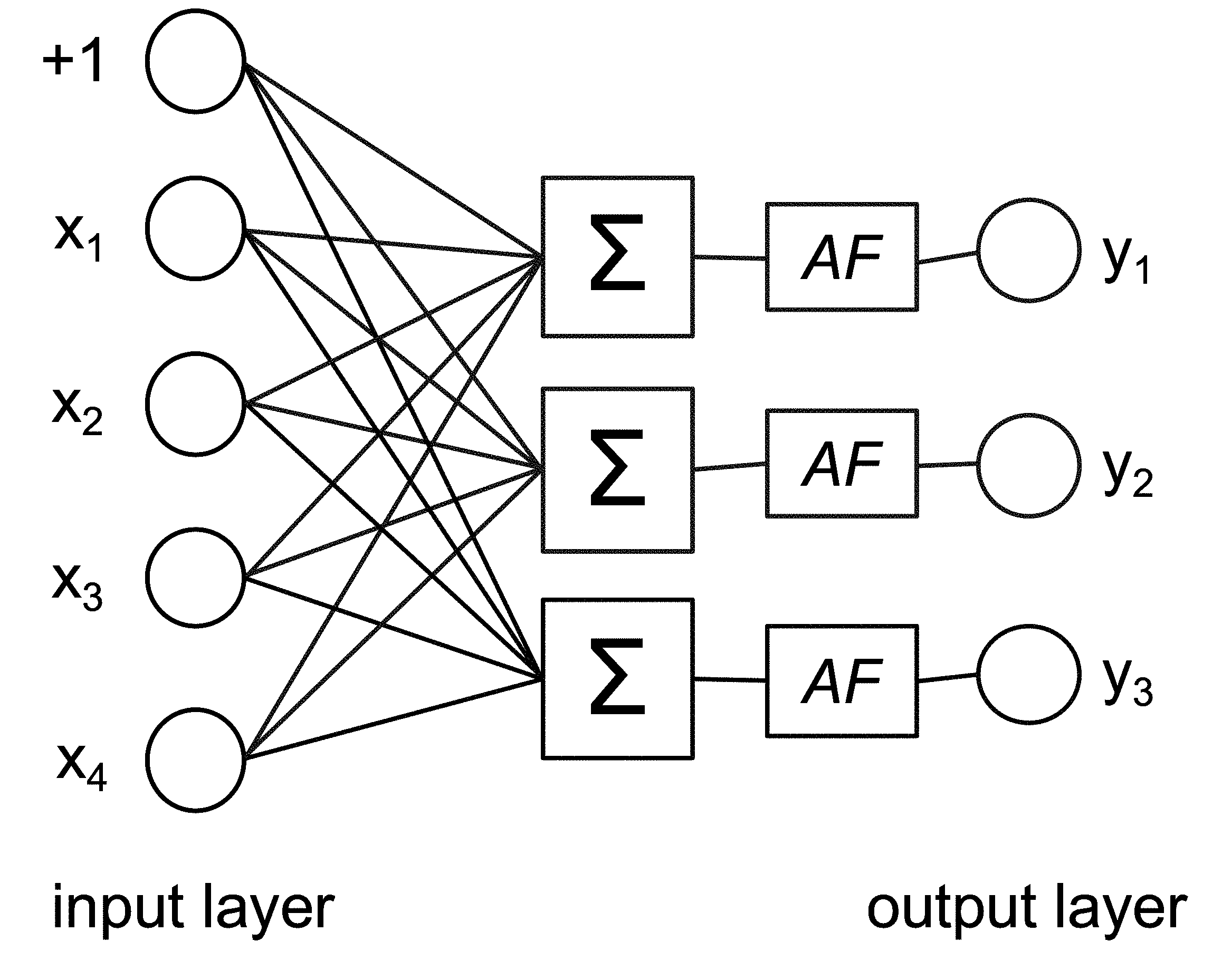}
\caption{Architectural model of multivariate linear regression with activation function}
\label{figure:multivariate:linear:regression:with:af:architecture}
\end{figure}

Let us now also apply an {\em activation function\index{Activation!function}} ({\em AF\index{AF}})
to each weighted sum\index{Weighted sum} unit\index{Unit},
as shown in Figure~\ref{figure:multivariate:linear:regression:with:af:architecture}.

This activation function allows us to introduce arbitrary {\em nonlinear functions\index{Nonlinear!function}}.

\begin{itemize}

\item From an {\em engineering\index{Engineering}} perspective,
a nonlinear function is necessary to overcome the linear separability limitation of the single layer Perceptron
(see Section~\ref{section:architectures:history}).

\item From a {\em biological inspiration\index{Biological!inspiration}} perspective,
a nonlinear function can capture the {\em threshold\index{Threshold}} effect for the activation of a neuron
through its incoming signals (via its dendrites\index{Dendrite}),
determining whether it fires
along its output (axone\index{Axone}).

\item From a {\em statistical} perspective,
when the activation function is the sigmoid function,
a model corresponds to {\em logistic regression\index{Logistic!regression}},
which models the probability of a certain class or event 
and thus performs binary classification\footnote{For each output node/variable.
	See more details in
	Section~\ref{section:architecture:neural:network:output:activation:function}.}.

\end{itemize}

\begin{figure}
\includegraphics[scale=0.20]{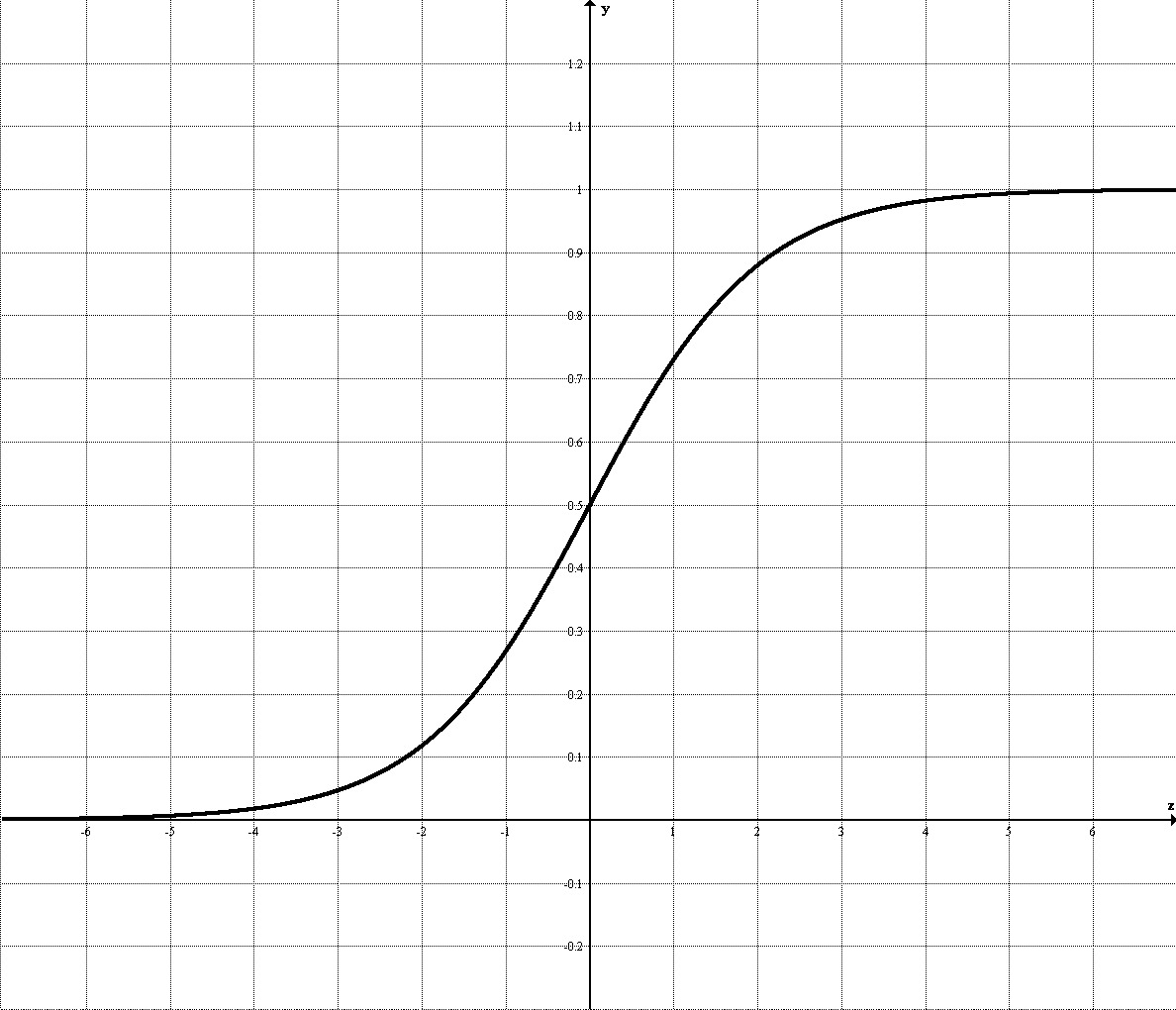}
\caption{Sigmoid function}
\label{figure:sigmoid:curve}
\end{figure}

Historically speaking,
the sigmoid\index{Sigmoid} function
(which is used for {\em logistic regression})
is the most common.
The sigmoid function (usually written $\sigma$) is defined in Equation~\ref{equation:sigmoid:formula}
and is shown in Figure~\ref{figure:sigmoid:curve}.
It will be further analyzed in Section~\ref{section:architecture:neural:network:output:activation:function}.

\begin{figure}
\includegraphics[scale=0.22]{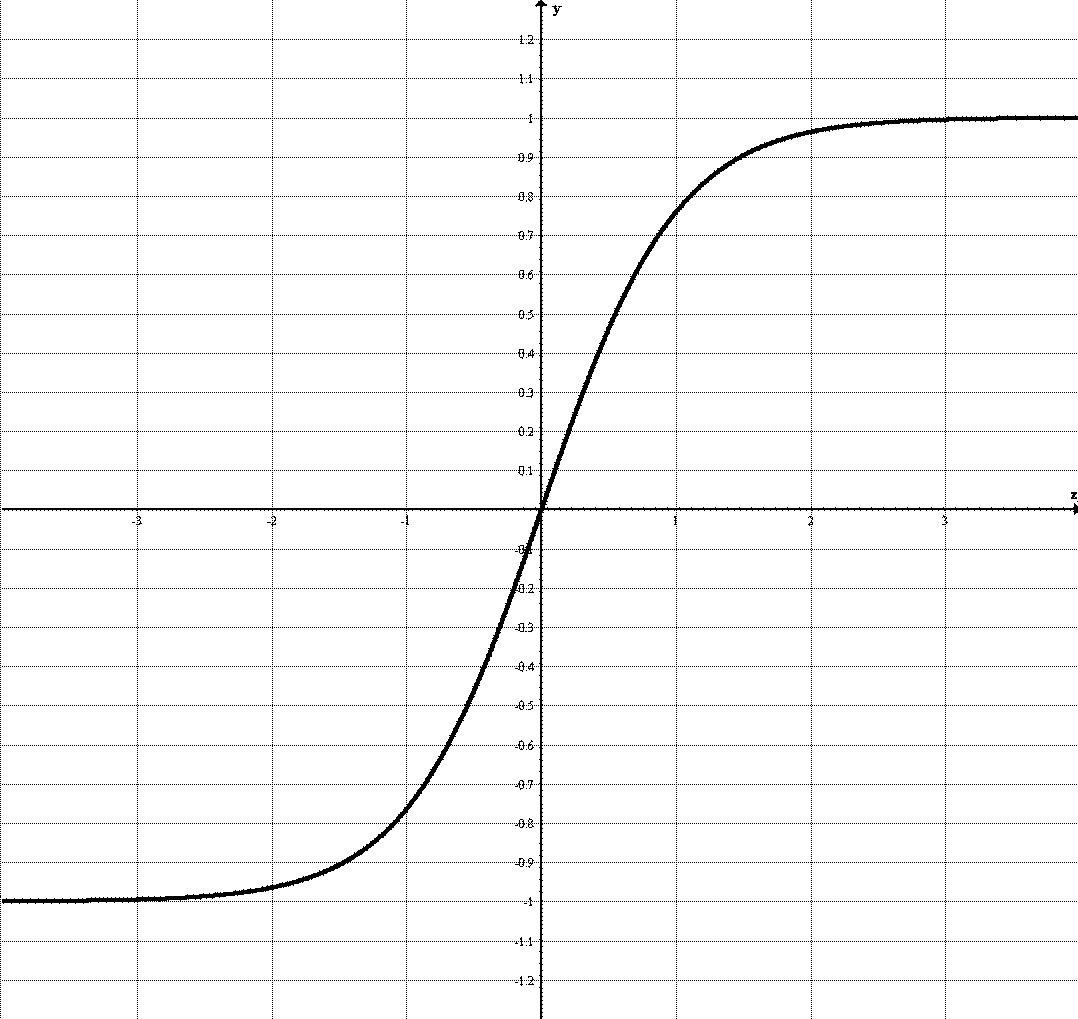}
\caption{Tanh function}
\label{figure:tanh:curve}
\end{figure}

An alternative is the hyperbolic tangent\index{Hyperbolic tangent}, often noted tanh\index{Tanh},
similar to sigmoid but having $[-1, +1]$ as its domain interval ($[0, 1]$ for sigmoid).
Tanh is defined in Equation~\ref{equation:tanh:formula}
and shown in Figure~\ref{figure:tanh:curve}.

\begin{figure}
\includegraphics[scale=0.19]{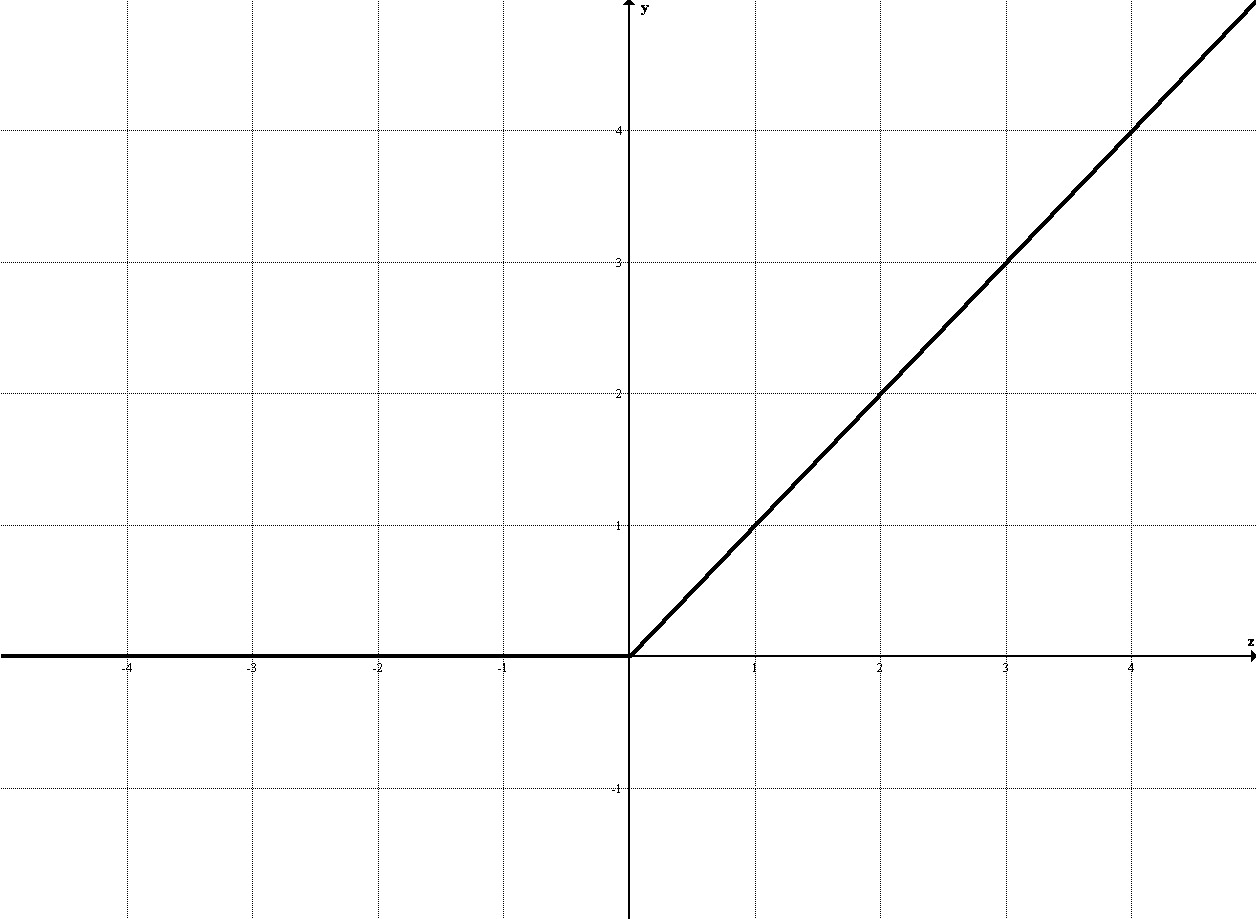}
\caption{ReLU function}
\label{figure:relu:curve}
\end{figure}

But ReLU is now
widely used
for its simplicity and effectiveness.
ReLU\index{ReLU}, which stands for {\em rectified linear unit\index{Rectified linear unit}},
is defined in Equation~\ref{equation:sigmoid:formula}
and is shown in Figure~\ref{figure:relu:curve}.
Note that, as some notation convention\index{Notation convention}
we use z as the name of the variable of an activation function,
as x is usually reserved for input variables.



\begin{equation}
\text{sigmoid}(\text{z}) = \sigma(\text{z}) = \frac{1}{1 + e^{-\text{z}}}
\label{equation:sigmoid:formula}
\end{equation}

\begin{equation}
\text{tanh}(\text{z}) = \frac{e^\text{z} - e^{-\text{z}}}{e^\text{z} + e^{-\text{z}}}
\label{equation:tanh:formula}
\end{equation}

\begin{equation}
\text{ReLU}(\text{z}) = \text{max}(0, \text{z})
\label{equation:relu:formula}
\end{equation}

%

\section{Basic Building Block}
\label{section:architecture:building:block}

The architectural representation
(of multivariate linear regression with activation function)
shown in Figure~\ref{figure:multivariate:linear:regression:with:af:architecture}
is an instance (with 4 input nodes and 3 output nodes)
of a {\em basic building block\index{Basic building block}} of neural networks\index{Neural!network} and deep learning\index{Deep!learning} architectures.
Although simple, this basic building block is actually a working neural network.

It has two layers\footnote{Although, as we will see in Section~\ref{section:architecture:feedforward:depth},
	it will be considered as a single-layer neural network architecture.
	As it has no hidden layer,
	it still suffers from the linear separability limitation of the Perceptron.}:
	
\begin{itemize}

\item The {\em input layer\index{Input!layer}}, on the left of the figure, is composed of the {\em input nodes\index{Input!node}} x$_i$
and the {\em bias\index{Bias} node} which is an {\em implicit} and specific input node
with a constant value of $1$,
therefore usually denoted as $+1$.

\item The {\em output layer\index{Output!layer}}, on the right of the figure, is composed of the {\em output nodes\index{Output!node}}
y$_j$.

\end{itemize}

\label{section:architecture:building:block:training}

Training a basic building block is essentially the same as training a linear regression model,
which has been described in Section~\ref{section:architecture:training}.

\subsection{Feedforward Computation}
\label{section:architecture:building:block:feedforward:computation}

After it has been trained, we can use this
basic building block neural network
for prediction\index{Prediction}.
Therefore, we simply {\em feedforward\index{Feedforward}} the network,
i.e. provide input data to the network ({\em feed in}) and compute the output values.
This corresponds to
Equation~\ref{equation:basic:block:feedforward:vectorized:one:example:equation}.

\begin{equation}
\text{\^y} = h(\text{x}) = AF(b + W \text{x})
\label{equation:basic:block:feedforward:vectorized:one:example:equation}
\end{equation}



The feedforward computation of the prediction
(for the architecture shown in
Figure~\ref{figure:multivariate:linear:regression:architecture})
is illustrated in Equation~\ref{equation:basic:block:feedforward:vectorized:one:example:matrix},
where $h_j(\text{x})$ (i.e. \^y$_j$) is the prediction of the $j$th variable $y_j$.

\begin{equation}
\begin{split}
\text{\^y} = h(\text{x}) =
h(\left[
\begin{array}{c}
\text{x}_1\\
\text{x}_2\\
\text{x}_3\\
\text{x}_4
\end{array}
\right])
= AF(b + W \text{x})\\
=
AF(\left[
\begin{array}{c}
b_1\\
b_2\\
b_3
\end{array}
\right]
+
\left[
\begin{array}{cccc}
W_{1,1}		&W_{1,2}		&W_{1,3}		&W_{1,4}\\
W_{2,1}		&W_{2,2}		&W_{2,3}		&W_{2,4}\\
W_{3,1}		&W_{3,2}		&W_{3,3}		&W_{3,4}
\end{array}
\right]
\times
\left[
\begin{array}{c}
\text{x}_1\\
\text{x}_2\\
\text{x}_3\\
\text{x}_4
\end{array}
\right]
)\\
=
AF(\left[
\begin{array}{c}
b_1\\
b_2\\
b_3
\end{array}
\right]
+
\left[
\begin{array}{c}
W_{1,1}\,\text{x}_1 + W_{1,2}\,\text{x}_2 + W_{1,3}\,\text{x}_3 + W_{1,4}\,\text{x}_4\\
W_{2,1}\,\text{x}_1 + W_{2,2}\,\text{x}_2 + W_{2,3}\,\text{x}_3 + W_{2,4}\,\text{x}_4\\
W_{3,1}\,\text{x}_1 + W_{3,2}\,\text{x}_2 + W_{3,3}\,\text{x}_3 + W_{3,4}\,\text{x}_4
\end{array}
\right]
)\\
=
AF(\left[
\begin{array}{c}
b_1 + W_{1,1}\,\text{x}_1 + W_{1,2}\,\text{x}_2 + W_{1,3}\,\text{x}_3 + W_{1,4}\,\text{x}_4\\
b_2 + W_{2,1}\,\text{x}_1 + W_{2,2}\,\text{x}_2 + W_{2,3}\,\text{x}_3 + W_{2,4}\,\text{x}_4\\
b_3 + W_{3,1}\,\text{x}_1 + W_{3,2}\,\text{x}_2 + W_{3,3}\,\text{x}_3 + W_{3,4}\,\text{x}_4
\end{array}
\right]
)\\
=
\left[
\begin{array}{ccc}
h_1(\text{x})\\
h_2(\text{x})\\
h_3(\text{x})
\end{array}
\right]
=
\left[
\begin{array}{ccc}
\text{\^y}_1\\
\text{\^y}_2\\
\text{\^y}_3
\end{array}
\right]
\end{split}
\label{equation:basic:block:feedforward:vectorized:one:example:matrix}
\end{equation}

\subsection{Computing Multiple Input Data Simultaneously}
\label{section:architecture:building:block:feedforward:computation:multiple:input:data:simultaneous}

Feedforwarding simultaneously a set of examples is easily expressed as a matrix\index{Matrix}
by
matrix
multiplication,
by substituting the single vector example x
in Equation~\ref{equation:basic:block:feedforward:vectorized:one:example:equation}
with a matrix of examples (usually notated as X),
leading to Equation~\ref{equation:basic:block:feedforward:vectorized:examples:equation}.

Successive columns of the matrix of examples X correspond to the different examples.
We use a superscript notation\index{Notation convention} X$^{(k)}$ to denote the $k$th example,
the $k$th column of the X matrix,
to avoid confusion with the subscript notation x$_i$ which is used to denote the $i$th input variable.
Therefore, X$^{(k)}_i$ denotes the $i$th input value of the $k$th example.
The feedforward computation of a set of examples is illustrated in
Equation~\ref{equation:basic:block:feedforward:vectorized:examples:matrix},
with
predictions
$h(\text{X}^{(k)})$
being successive columns of the resulting output matrix.

\begin{equation}
h(\text{X}) = AF(b + W \text{X})
\label{equation:basic:block:feedforward:vectorized:examples:equation}
\end{equation}


\begin{equation}
\begin{split}
h(\text{X})
=
h(
\left[
\begin{array}{cccc}
\text{X}^{(1)}_1		&\text{X}^{(2)}_1	&\hdots		&\text{X}^{(m)}_1\\
\text{X}^{(1)}_2		&\text{X}^{(2)}_2	&\hdots		&\text{X}^{(m)}_2\\
\text{X}^{(1)}_3		&\text{X}^{(2)}_3	&\hdots		&\text{X}^{(m)}_3\\
\text{X}^{(1)}_4		&\text{X}^{(2)}_4	&\hdots		&\text{X}^{(m)}_4
\end{array}
\right])
= AF(b + W \text{X})\\
=
AF(\left[
\begin{array}{c}
b_1\\
b_2\\
b_3
\end{array}
\right]
+
\left[
\begin{array}{cccc}
W_{1,1}		&W_{1,2}		&W_{1,3}		&W_{1,4}\\
W_{2,1}		&W_{2,2}		&W_{2,3}		&W_{2,4}\\
W_{3,1}		&W_{3,2}		&W_{3,3}		&W_{3,4}
\end{array}
\right]
\times
\left[
\begin{array}{cccc}
\text{X}^{(1)}_1		&\text{X}^{(2)}_1	&\hdots		&\text{X}^{(m)}_1\\
\text{X}^{(1)}_2		&\text{X}^{(2)}_2	&\hdots		&\text{X}^{(m)}_2\\
\text{X}^{(1)}_3		&\text{X}^{(2)}_3	&\hdots		&\text{X}^{(m)}_3\\
\text{X}^{(1)}_4		&\text{X}^{(2)}_4	&\hdots		&\text{X}^{(m)}_4
\end{array}
\right]
)\\
=
\left[
\begin{array}{cccc}
h_1(\text{X}^{(1)})	&h_1(\text{X}^{(2)})	&\hdots	&h_1(\text{X}^{(m)})\\
h_2(\text{X}^{(1)})	&h_2(\text{X}^{(2)})	&\hdots	&h_2(\text{X}^{(m)})\\
h_3(\text{X}^{(1)})	&h_3(\text{X}^{(2)})	&\hdots	&h_3(\text{X}^{(m)})
\end{array}
\right]
=
\left[
\begin{array}{cccc}
h(\text{X}^{(1)})		&h(\text{X}^{(2)})	&\hdots	&h(\text{X}^{(m)})
\end{array}
\right]
\end{split}
\label{equation:basic:block:feedforward:vectorized:examples:matrix}
\end{equation}

Note that the main computation taking place\footnote{Apart
	from the computation of the $AF$ activation function\index{Activation!function}.
	In the case of ReLU this is fast.}
is a product of matrices\index{Matrix}.
This can be computed very efficiently,
by using linear algebra\index{Linear!algebra} vectorized\index{Vectorized} implementation libraries
and furthermore with specialized hardware like graphics processing units\index{Graphics processing unit} (GPUs\index{GPU}).

%
%
%
%
%

\section{Machine Learning}
\label{section:architecture:machine:learning}

\subsection{Definition}
\label{section:architecture:machine:learning:definition}



Let us now reflect a bit on the meaning of training a model,
whether it is a linear regression model (Section~\ref{section:statistics:linear:regression})
or the basic building block architecture presented in Section~\ref{section:architecture:building:block}.
Therefore, let us consider what machine learning actually means.
Our starting point is the following concise and general definition of machine learning
provided by Mitchell in \cite{mitchell:ml:book:1997}:
``A computer program is said to learn from experience $E$
with respect to some class of tasks $T$ and performance measure $P$, 
if its performance at tasks in $T$, as measured by $P$, improves with experience $E$.''

At first, note that the word {\em performance} actually covers different meanings,
specially regarding the computer music context of the book:

\begin{enumerate}

\item the {\em execution} of (the action to perform) an action, notably an artistic act
such as a musician playing a piece of music;

\item a {\em measure} (criterium of evaluation) of that action, notably for a computer system its {\em efficiency} in performing a task, in terms of time and memory\footnote{With the corresponding
	analysis measurements, time complexity and space complexity,
	for the corresponding algorithms.}
measurements; or

\item
a measure of the {\em accuracy} in performing a task,
i.e. the ability to predict or classify with minimal errors.

\end{enumerate}

In the remainder of the book,
in order to try to minimize ambiguity,
we will use the terms as following:

\begin{itemize}

\item {\em performance} as an act by a musician,

\item {\em efficiency} as a measure of computational ability, and

\item {\em accuracy} as a measure of the quality of a prediction or a classification\footnote{In fact,
	accuracy may not be a pertinent metric for a classification task
	with {\em skewed} classes, i.e. with one class being vastly more represented in the data than other(s),
	e.g., in the case of the detection of a rare disease.
	Therefore a confusion matrix and additional metrics like {\em precision} and {\em recall},
	and possible combinations like F-score, are used
	(see, e.g., \cite[Section~11.1]{goodfellow:deep:learning:book:2016} for details).
	We will not address them in the book, because we are primarily concerned with content generation
	and not in pattern recognition (classification).}.

\end{itemize}

Thus, we could rephrase the definition as:
``A computer program is said to learn from experience $E$
with respect to some class of tasks $T$ and accuracy measure $A$, 
if its accuracy at tasks in $T$, as measured by $A$, improves with experience $E$.''

\subsection{Categories}
\label{section:architecture:supervised:learning}
\label{section:architecture:machine:learning:definition:categories}

\label{section:architecture:unsupervised:learning}

We may now consider the three main categories of machine learning
with regard to the nature of the experience\index{Experience} conveyed by the examples:
	
\begin{itemize}

\item {\em supervised learning\index{Supervised learning}} -- the dataset is fixed
and a correct (expected) answer\footnote{It is usually named a {\em label\index{Label}}
	in the case of a {\em classification\index{Classification}} task
	and a {\em target\index{Target}} in the case of a {\em prediction/regression\index{Prediction}\index{Regression}} task.}
is associated to each example,
the general objective being to {\em predict\index{Prediction} answers} for new examples.
Examples of tasks are
regression (prediction),
classification and
translation;

\item {\em unsupervised learning\index{Unsupervised learning}} -- the dataset is fixed and the general objective is in {\em extracting\index{Extraction} information}.
Examples of tasks are
feature extraction,
data compression (both performed by {\em autoencoders\index{Autoencoder}}, to be introduced in Section~\ref{section:architecture:autoencoder}),
probability distribution learning (performed by {\em RBMs\index{RBM}}, to be introduced in Section~\ref{section:architecture:rbm}),
series modeling (performed by {\em recurrent} networks\index{Recurrent!network}, to be introduced in Section~\ref{section:architecture:recurrent:network}),
clustering and
anomaly detection; and


\item {\em reinforcement learning\index{Reinforcement!learning}}\footnote{To be introduced
	in Section~\ref{section:architecture:reinforcement:learning}.}
-- the experience is {\em incremental} through successive actions of an {\em agent} within an {\em environment},
with some feedback (the {\em reward\index{Reward}}) providing information about the {\em value} of the action,
the general objective being to learn a near optimal {\em policy\index{Policy}} (strategy), i.e.
a suite of actions maximizing its cumulated rewards (its {\em gain\index{Gain}}).
Examples of tasks are
game playing and robot navigation.

\end{itemize}

\subsection{Components}
\label{section:learning:components}

In his introduction to machine learning \cite{domingos:things:know:2012},
Domingos describes machine learning algorithms through three components:

\begin{itemize}

\item {\em representation\index{Representation}} --
the way to represent the model -- in our case, a {\em neural network},
as it has been introduced
and will be further developed in
the following sections;

\item {\em evaluation\index{Evaluation}} --
the way to evaluate and compare models -- via a {\em cost function},
that will be analyzed in Section~\ref{section:architecture:neural:network:cost:function}; and
 
\item {\em optimization\index{Optimization}} --
the way to identify (search among models for) a best model.

\end{itemize}

\subsection{Optimization}
\label{section:architecture:training:optimization}

Searching for values (of the parameters of a model) that minimize the cost function is indeed
an {\em optimization\index{Optimization}} problem.
One of the most simple optimization algorithms is gradient descent\index{Gradient!descent},
as it has been introduced in Section~\ref{section:training:algorithm}.

There are various more sophisticated algorithms,
such as stochastic gradient descent\index{Stochastic!gradient descent}
(SGD\index{SGD}),
Nesterov accelerated gradient (NAG),
Adagrad,
BFGS, etc.
(see, for example, \cite[Chapter~9]{goodfellow:deep:learning:book:2016} for more details).

\section{Architectures}
\label{section:architecture:basic:architectures}

From this basic building block, we will describe
in the following sections
the main {\em types} of {\em deep learning architectures}
used for music generation (as well as for other purposes):

\begin{itemize}

\item feedforward,

\item autoencoder,

\item restricted Boltzmann machine (RBM), and

\item recurrent
(RNN).

\end{itemize}

We will also introduce {\em architectural patterns\index{Architectural!pattern}\index{Pattern|see{Architectural pattern}}}
(see Section~\ref{section:architecture:compound:composition:types})
which could be applied to them:

\begin{itemize}

\item convolutional,

\item conditioning, and

\item adversarial.

\end{itemize}

\section{Multilayer Neural Network {\em aka} Feedforward Neural Network}
\label{section:architecture:neural:network}
\label{section:architecture:feedforward}

A {\em multilayer neural network}\index{Multilayer!neural network|see{Neural network}},
also named a {\em feedforward neural network\index{Feedforward!neural network}},
is an assemblage of successive layers\index{Layer} of basic building blocks\index{Basic building block}:

\begin{itemize}

\item the {\em first layer}, composed of input nodes, is called the {\em input layer\index{Input!layer}};

\item the {\em last layer}, composed of output nodes, is called the {\em output layer\index{Output!layer}}; and

\item any layer {\em between} the input layer and the output layer is named a {\em hidden layer\index{Hidden!layer}}.

\end{itemize}

\begin{figure}
\includegraphics[width=\textwidth]{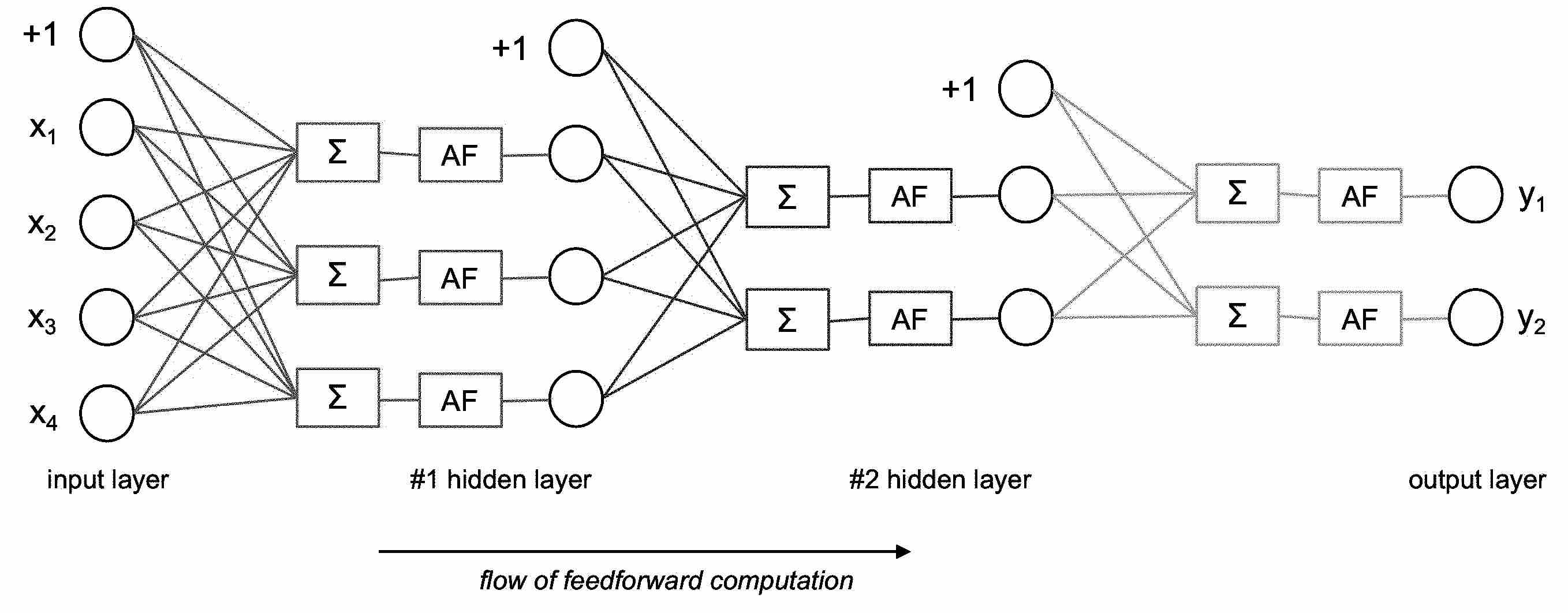}
\caption{Example of a feedforward neural network (detailed)}
\label{figure:multilayer:neural:network:architecture}
\end{figure}

An example of a multilayer neural network with two hidden layers is illustrated in Figure~\ref{figure:multilayer:neural:network:architecture}.

The combination of a hidden layer and a nonlinear activation function makes the neural network
a {\em universal approximator\index{Universal approximator}},
able to overcome the {\em linear separability\index{Linear!separability}} limitation\footnote{The universal approximation theorem
	\cite{hornik:approximation:theorem:nn:1991}
	states that a feedforward network
	with a single hidden layer containing a finite number of neurons
	can approximate a wide variety of interesting functions when given appropriate parameters (weights).
	However, there is no guarantee that the neural network will be able to learn them!}.

\subsection{Abstract Representation}
\label{section:architecture:feedforward:abstract}

\begin{figure}
\includegraphics[scale=0.21]{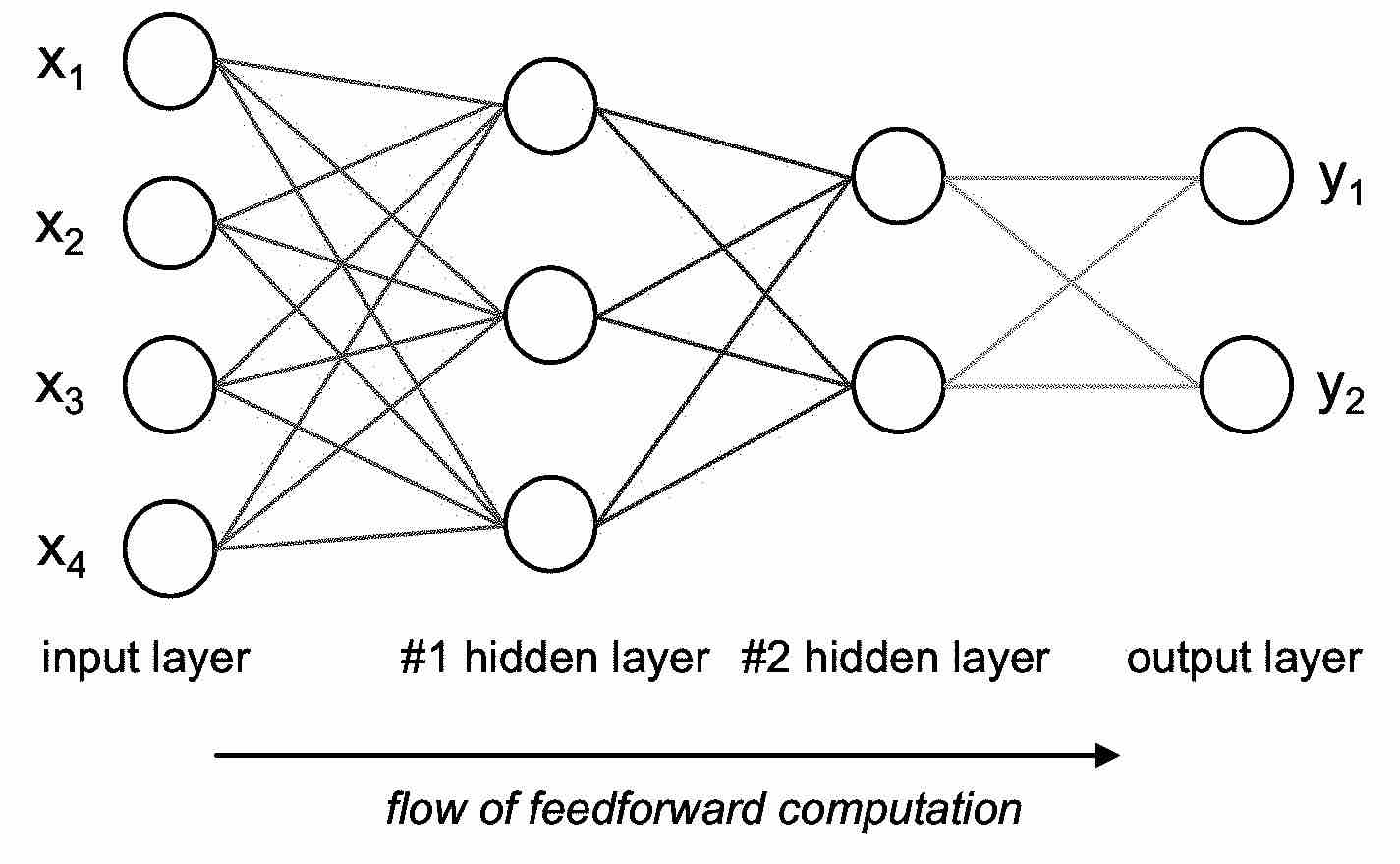}
\caption{Example of feedforward neural network (simplified)}
\label{figure:multilayer:neural:network:architecture:wo:bias}
\end{figure}

Note that, in the case of practical (non-toy) illustrations of neural network architectures,
in order to simplify the figures, bias nodes are very rarely illustrated.
With a similar objective, the sum units and the activation function units are also almost always omitted,
resulting in a more abstract view such as that shown in Figure~\ref{figure:multilayer:neural:network:architecture:wo:bias}.

\begin{figure}
\includegraphics[scale=0.85]{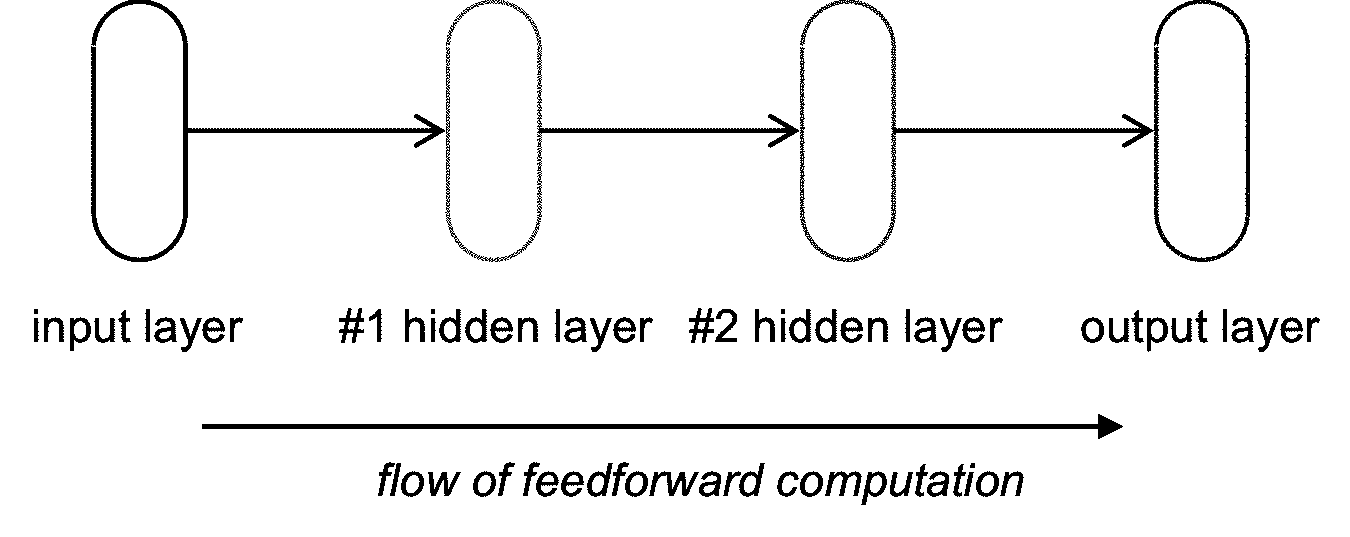}
\caption{Example of a feedforward neural network (abstract)}
\label{figure:abstract:abstract:multilayer:neural:network:architecture}
\end{figure}

We can further abstract each layer by representing it as an oblong form (by hiding its nodes)\footnote{It is sometimes
	pictured as a rectangle, see Figure~\ref{figure:architecture:googlenet:resnet},
	or even as a circle, notably in the case of recurrent networks,
	see Figure~\ref{figure:recurrent:network:folded:abstract:alternative}.}
as shown in Figure~\ref{figure:abstract:abstract:multilayer:neural:network:architecture}.

\subsection{Depth}
\label{section:architecture:feedforward:depth}

The architecture illustrated in Figure~\ref{figure:abstract:abstract:multilayer:neural:network:architecture}
is called a 3-layer\index{Layer} neural network architecture,
also indicating that the {\em depth\index{Depth}} of the architecture is three.
Note that the number of layers (depth)
is indeed three and {\em not} four,
irrespective of the fact that summing up the input layer, the output layer and the two hidden layers gives four and not three.
This is because, by convention, only layers with weights (and units) are considered when counting the number of layers
in a multilayer neural network;
therefore, the input layer is not counted.
Indeed, the input layer only acts as an input interface, without any weight or computation.

In this book, we will use a superscript (power) notation\index{Notation convention}\footnote{The set
	of compact notations
	for expressing the dimension of an architecture or a representation will be introduced in Section~\ref{section:challenge:strategy:architecture:notation:depth}.}
to denote the number of layers of a neural network architecture.
For instance, the architecture illustrated in Figure~\ref{figure:abstract:abstract:multilayer:neural:network:architecture}
could be denoted as Feedforward$^3$.

The depth of the first neural network architectures was small.
The original Perceptron\index{Perceptron} \cite{rosenblatt:perceptron:1957},
the ancestor of neural networks,
has only an input layer and an output layer
without any hidden layer,
i.e. it is a single-layer neural network.
In the 1980s, conventional neural networks were mostly 2-layer or 3-layer architectures.

\begin{figure}
\includegraphics[scale=1.01]{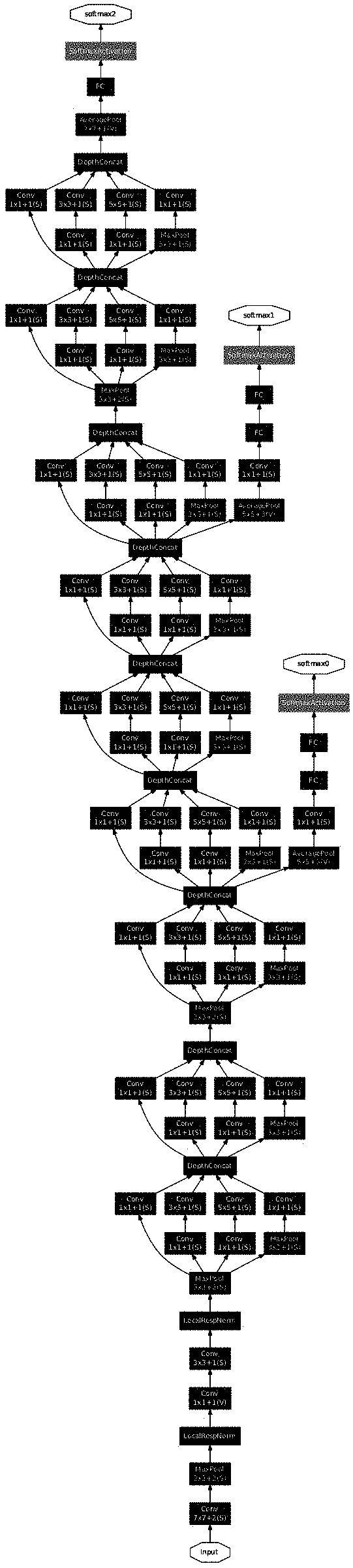}
\hspace{2cm}
\includegraphics[scale=1.11]{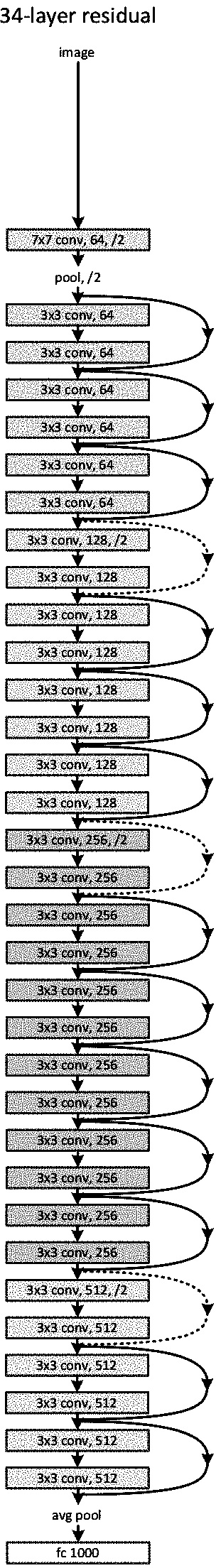}
\caption{(left) GoogLeNet 27-layer deep network architecture.
Reproduced from \cite{szegedy:googlenet:arxiv:2014} with permission of the authors.
(right) ResNet 34-layer deep network architecture.
Reproduced from \cite{he:resnet:deep:residual:arxiv:2015} with permission of the authors}
\label{figure:architecture:googlenet:resnet}
\end{figure}

For modern deep networks, the depth can indeed be very large,
deserving the name of {\em deep\index{Deep}} (or even {\em very deep}) networks.
Two recent examples, both illustrated in Figure~\ref{figure:architecture:googlenet:resnet}, are

\begin{itemize}

\item the 27-layer GoogLeNet\index{GoogLeNet} architecture \cite{szegedy:googlenet:arxiv:2014}; and

\item the 34-layer (up to 152-layer!)
ResNet\index{ResNet} architecture\footnote{It introduces the technique of {\em residual learning},
	reinjecting the input between levels and estimating the residual function $h(\text{x}) - \text{x}$,
	a technique aimed at very deep networks,
	see \cite{he:resnet:deep:residual:arxiv:2015} for more details.}
\cite{he:resnet:deep:residual:arxiv:2015}.

\end{itemize}

Note that depth {\em does} matter.
A recent theorem \cite{eldan:power:depth:arxiv:2016}
states that there is a simple radial function\footnote{A {\em radial function\index{Radial function}} is a function whose value at each point depends
	only on the distance between that point and the origin.
	More precisely, it is radial {\em if and only if} it is invariant under all rotations while leaving the origin fixed.}
on ${\rm I\!R}^d$, expressible by a 3-layer neural network, which cannot be approximated by any 2-layer network
to more than a constant accuracy
unless its width is exponential in the dimension $d$.
Intuitively, this means that reducing the depth (removing a layer)
means exponentially augmenting the width (the number of units) of the layer left.
On this issue, the interested reader may also wish to review the analyses
in \cite{urban:do:deep:arxiv:2014} and \cite{urban:do:deep:2:arxiv:2016}.

Note that for both networks pictured in Figure~\ref{figure:architecture:googlenet:resnet},
the flow of computation is vertical, upward for Goog\-Le\-Net and downward for ResNet.
These are different usages than the convention for the flow of computation that we have introduced and used so far,
which is horizontal, from left to right.
Unfortunately,
there is no consensus in the literature about the notation\index{Notation convention} for the flow of computation.
Note that in the specific case of recurrent networks, to be introduced in Section~\ref{section:architecture:recurrent:network},
the consensus notation is vertical, upward.

\subsection{Output Activation Function}
\label{section:architecture:neural:network:output:activation:function}

We have seen in Section~\ref{section:architecture:building:block} that,
in modern neural networks,
the activation function ($AF$) chosen for introducing nonlinearity at the output of each hidden layer is often the ReLU function.
But the output layer of a neural network has a special status.
Basically, there are three main possible types of activation function for the output layer,
named in the following, the {\em output activation function\index{Output!activation function}}\footnote{A shorthand
	for output layer activation function\index{Output!layer activation function|see{Output activation function}}.}:

\begin{itemize}

\item identity -- the case
	for a prediction\index{Prediction} (regression\index{Regression}) task.
	It has continuous (real) output values.
	Therefore, we do not need and we do not {\em want} a nonlinear\index{Nonlinear} transformation at the last layer;

\item sigmoid\index{Sigmoid} -- the case of a binary classification\index{Classification} task,
	as in logistic regression\index{Logistic!regression}\footnote{For details about logistic regression,
		see, for example,
		\cite[page~137]{goodfellow:deep:learning:book:2016}
		or \cite[Section~4.4]{hastie:elements:statistical:learning:book:2009}.
		For this reason,
		the sigmoid function is also called the {\em logistic function\index{Logistic!function|see{Sigmoid function}}}.}.
	The sigmoid function (usually written $\sigma$) has been defined in Equation~\ref{equation:sigmoid:formula}
	and shown in Figure~\ref{figure:sigmoid:curve}.
	Note its specific shape, which provides a ``separation'' effect, used for binary decision between two options represented by values $0$ and $1$; and

\item softmax\index{Softmax}
	-- the most common approach for a classification task
	with more than two classes but with only one label to be selected\footnote{A very common example
	is the estimation by a neural network architecture of the next note,
	modeled as a classification task of a single note label within the set of possible notes.}
	(and where a one-hot\index{One-hot encoding} encoding is generally used, see Section~\ref{section:representation:input:encoding}).

\end{itemize}



The softmax\index{Softmax} function actually represents a {\em probability distribution\index{Probability!distribution}}
over a discrete output variable with $n$ possible values
(in other words, the probability of the occurrence of each possible value $v$, knowing the input variable x, i.e. $P(\text{y} = v | \text{x})$).
Therefore, softmax ensures that the sum of the probabilities for each possible value is equal to $1$.
The softmax function is defined
in Equation~\ref{equation:softmax:formula}
and an example of its use is shown in Equation~\ref{figure:softmax:example}.
Note that the $\sigma$ notation\index{Notation convention} is used for the softmax function, as for the sigmoid function,
because softmax is actually the generalization of sigmoid\index{Sigmoid} to the case of multiple values,
being a variadic function, that is one which accepts a variable number of arguments.

\begin{equation}
\sigma(\text{z})_i = \frac{e^{\text{z}_i}}{\sum_{i=1}^n e^{\text{z}_i}}
\label{equation:softmax:formula}
\end{equation}


\begin{equation}
\sigma \left[
\begin{array}{ccc}
1.2\\
0.9\\
0.4
\end{array}
\right]
=
\left[
\begin{array}{ccc}
0.46\\
0.34\\
0.20
\end{array}
\right]
\label{figure:softmax:example}
\end{equation}

For a classification or prediction task, we can simply select the value with the highest probability\index{Probability}
(i.e. via the {\em argmax\index{Argmax}} function,
the indice of the one-hot vector with the highest value).
But the distribution\index{Distribution|see{Probability distribution}} produced by the softmax function
can also be used as the basis for {\em sampling\index{Sampling}},
in order to add nondeterminism and thus content variability to the generation
(this will be detailed in Section~\ref{section:challenges:strategies:variability}).


\subsection{Cost Function}
\label{section:architecture:neural:network:cost:function}

The choice of a cost (loss) function\index{Cost!function} is actually correlated to the choice of the output activation function\index{Output!activation function}
and to the choice of the encoding\index{Encoding} of the target y (the true value).
Table~\ref{table:cost:activation:functions}\footnote{Inspired
	by Ronaghan's concise pedagogical presentation in \cite{ronaghan:which:loss:activation:function:web:2018}.}
summarizes the main cases.

%
%
%
%
%

\begin{table}
\begin{tabular}{|l|l|l|l|l|}
\hline
{\em Task}			&{\em Type of the output (}\^y{\em )}	&{\em Encoding of}	&{\em Output activation}	&{\em Cost (loss)}\\
				&						&{\em the target (}y{\em )}	&{\em function}		&\\
\hline
\hline
Regression		&Real					&${\rm I\!R}$		&Identity (Linear)	&Mean squared error\index{Mean squared error}\\
\hline
Classification		&Binary					&\{0, 1\}			&Sigmoid			&Binary cross-entropy\\
\hline
Classification		&Multiclass single label		&One-hot			&Softmax			&Categorical cross-entropy\\
\hline
Classification		&Multiclass multilabel		&Many-hot		&Sigmoid			&Binary cross-entropy\\
\hline
				&			 			&				& 				&\\
Multiple			&Multi					&Multi			&Sigmoid			&Binary cross-entropy\\
														\cline{4-5}
Classification		&Multiclass single label		&One-hot			&Multi			&Multi\\
				&						&				&Softmax			&Categorical cross-entropy\\
\hline
\end{tabular}	
\caption{Relation between output activation function and cost (loss) function}
\label{table:cost:activation:functions}
\end{table}


A cross-entropy\index{Cross-entropy} function measures the difference
between two probability distributions, in our case (of a classification task)
between the target (true value) distribution (y)
and the predicted distribution (\^y).
Note that there are two types of cross-entropy cost\index{Cross-entropy!cost} functions:

\begin{itemize}

\item binary cross-entropy\index{Binary!cross-entropy}, when the classification is binary (Boolean),
	and

\item categorical cross-entropy\index{Categorical!cross-entropy}, when the classification is multiclass with a single label
	to be selected.

\end{itemize}

In the case of a classification with multiple labels,
binary cross-entropy must be chosen joint with sigmoid
(because in such cases we want to compare the distributions independently, class per class\footnote{In case of multiple labels,
	the probability of each class is independent from the other class probabilities
	-- the sum is greater than 1.})
and the costs for each class are summed up.

In the case of multiple simultaneous classifications (multi multiclass single label),
each classification is now independent from the other classifications,
thus we have two approaches:
apply sigmoid and binary cross-entropy for each element and sum up the costs,
or
apply softmax and categorical cross-entropy {\em independently} for each classification and sum up the costs.

\subsection{Interpretation}
\label{section:architecture:neural:network:interpretation}

\begin{figure}
\includegraphics[width=\textwidth]{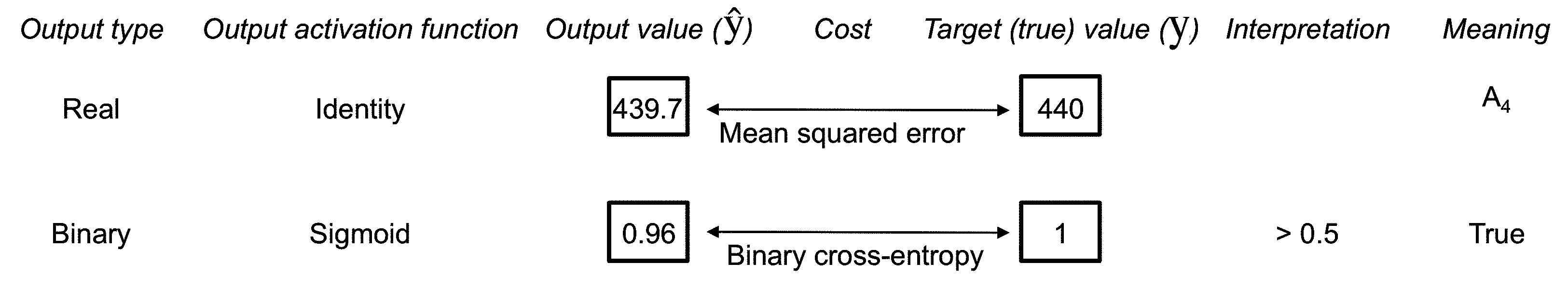}
\caption{Cost functions and interpretation for real and binary values}
\label{figure:architecture:interpretation:numerical:binary}
\end{figure}

Let us take some examples to illustrate these subtle but important differences,
starting with the cases of real and binary values in Figure~\ref{figure:architecture:interpretation:numerical:binary}.
They also include the basic interpretation of the result\footnote{The interpretation is actually part of the {\em strategy}
	of the generation of music content.
	It will be explored in Chapter~\ref{section:chapter:challenges:strategies}.
	For instance, sampling from the probability distribution may be used in order to ensure content generation variability,
	as will be explained in Section~\ref{section:challenges:strategies:variability}.}.

\begin{figure}
\includegraphics[width=\textwidth]{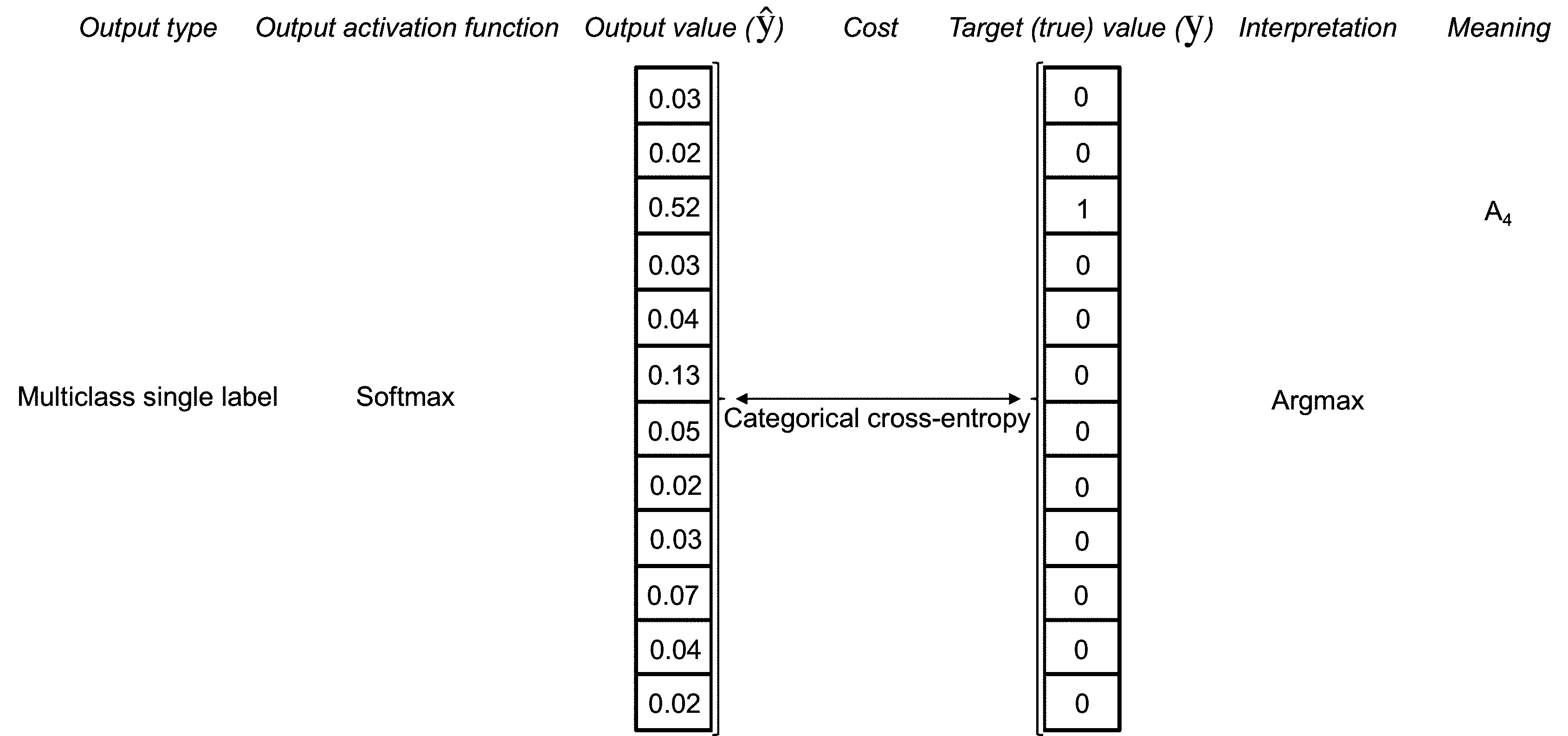}
\caption{Cost function and interpretation for a multiclass single label}
\label{figure:architecture:interpretation:multiclass:single:label}
\end{figure}

\begin{figure}
\includegraphics[width=\textwidth]{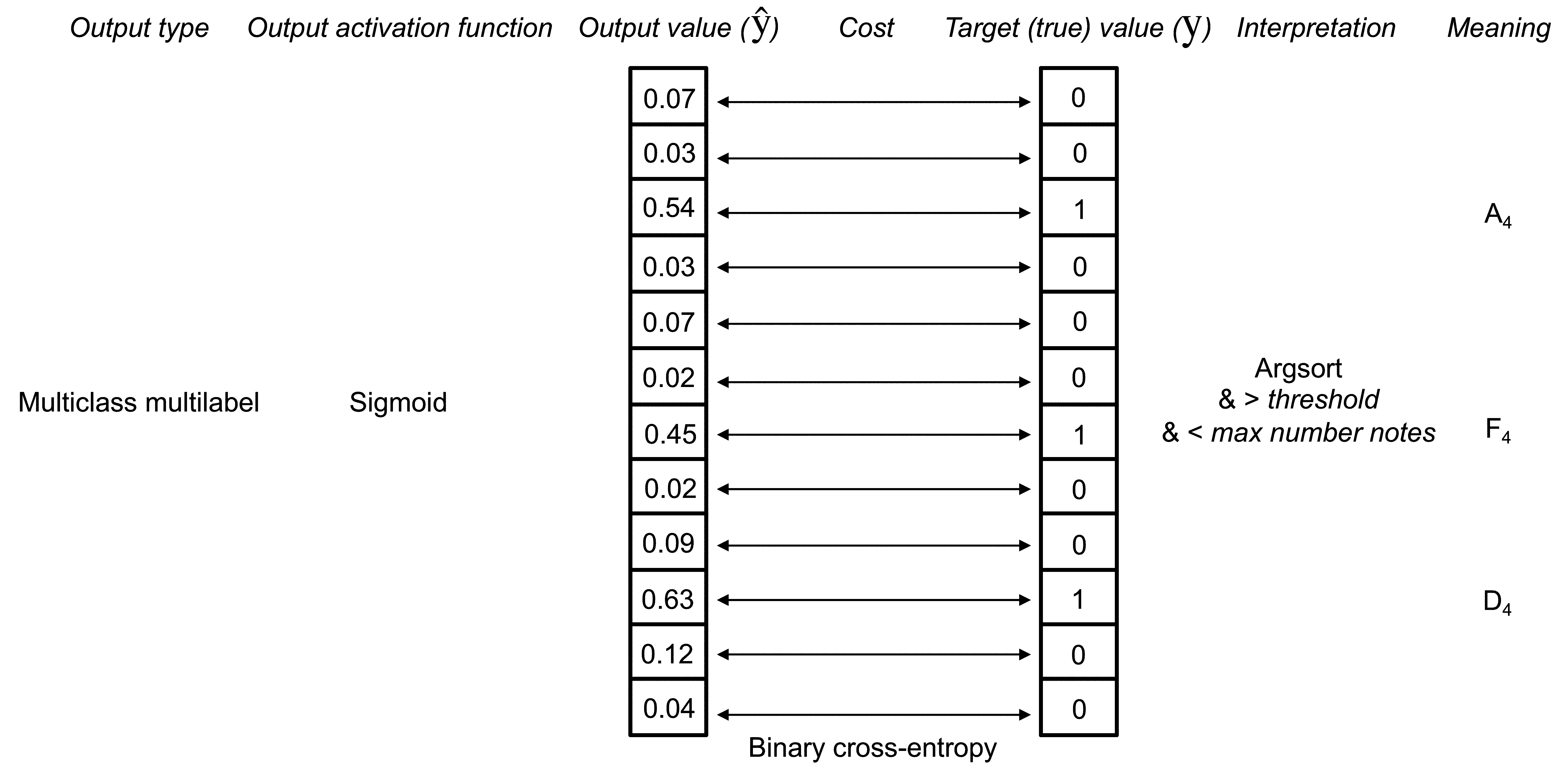}
\caption{Cost function and interpretation for a multiclass multilabel}
\label{figure:architecture:interpretation:multiclass:multilabel}
\end{figure}

\begin{figure}
\includegraphics[width=\textwidth]{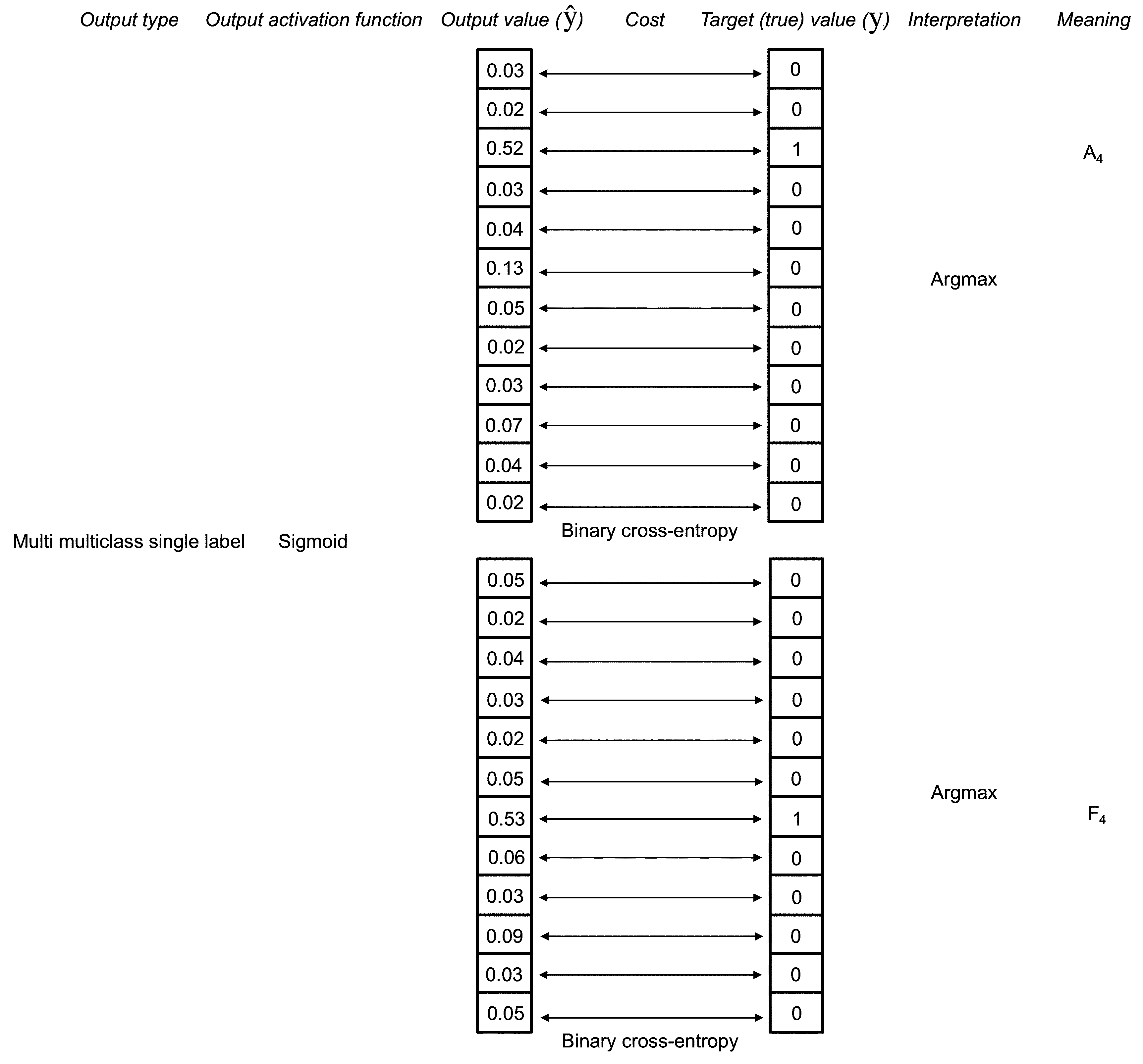}
\caption{Cost function and interpretation for a multi multiclass single label}
\label{figure:architecture:interpretation:multi:multiclass}
\end{figure}

\begin{itemize}

\item An example of use of the {\em multiclass single label\index{Multiclass!single label}} type is a classification among a set of possible notes for a monophonic melody,
therefore with only one single possible note choice (single label),
as shown in Figure~\ref{figure:architecture:interpretation:multiclass:single:label}.
See, for example, the Blues$_C$ system in Section~\ref{section:experiment:eck:blues:lstm:first:experiment}.

\item An example of use of the {\em multiclass multilabel\index{Multiclass!multilabel}} type is a classification among a set of possible notes for a single-voice polyphonic melody,
therefore with several possible note choices (several labels),
as shown in Figure~\ref{figure:architecture:interpretation:multiclass:multilabel}.
See, for example, the Bi-Axial LSTM system in Section~\ref{section:experiment:biaxial}.

\item An example of use of the {\em multi multiclass single label\index{Multi!multiclass single label}} type is a multiple classification among a set of possible notes for multivoice monophonic melodies,
therefore with only one single possible note choice for each voice,
as shown in Figure~\ref{figure:architecture:interpretation:multi:multiclass}.
See, for example, the Blues$_{MC}$ system in Section~\ref{section:experiment:eck:blues:lstm:second:experiment}.

\item Another example of use of the {\em multi multiclass single label} type is a multiple classification among a set of possible notes
for a set of time steps (in a piano roll representation) for a monophonic melody,
therefore with only one single possible note choice for each time step.
See, for example, the DeepHear$_M$ system in Section~\ref{section:experiment:deep:hear:melody}.

\item An example of use of a {\em multi$^2$ multiclass single label} type is a 2-level multiple classification among a set of possible notes
for a set of time steps for a multivoice set of monophonic melodies.
See, for example, the MiniBach system in Section~\ref{section:experiment:mini:bach}.

\end{itemize}

The three main interpretations used\footnote{In various
	systems to be analyzed in Chapter~\ref{section:chapter:challenges:strategies}.}
are

\begin{itemize}

\item argmax\index{Argmax} (the index of the output vector with the largest value),
	in the case of a one-hot multiclass single label
	(in order to select the most likely note),

\item {\em sampling\index{Sampling}} from the probability represented by the output vector,
	in the case of a one-hot multiclass single label
	(in order to select a note sorted along its likelihood),
	and

\item argsort\index{Argsort}\footnote{argsort
		is a numpy library Python function.}
	(the indexes of the output vector sorted according to their diminishing values),
	in the case of a many-hot multiclass multi label,
	filtered by some thresholds 
	(in order to select the most likely notes above a probability threshold and under
	a maximum number of simultaneous notes).

\end{itemize}

\subsection{Entropy and Cross-Entropy}
\label{section:architecture:neural:network:entropy}

Mean squared error has been defined in Equation~\ref{equation:linear:regression:cost:mean:squared:error}
in Section~\ref{section:architecture:training}.
Without getting into details about information theory, we now introduce the notion and the formulation of cross-entropy\footnote{With some
	inspiration
	from Preiswerk's introduction in \cite{preiswerk:entropy:ml:web:2018}.}.

The intuition behind information theory is that the information content about an event with a likely (expected) outcome is low,
while the information content about an event with an unlikely (unexpected, i.e. a surprise) outcome is high.

Let us take the example of a neural network architecture used to estimate the next note of a melody.
Suppose that the outcome is $\text{note} = \text{B}$ and that it has a probability $P(\text{note} = \text{B})$.
We can then introduce the {\em self-information\index{Self!-information}} (notated $I$) of that event
in Equation~\ref{equation:self:information}.

\begin{equation}
I(\text{note} = \text{B}) = \text{log}(1/P(\text{note} = \text{B})) = - \text{log}\,P(\text{note} = \text{B})
\label{equation:self:information}
\end{equation}

\begin{figure}
\includegraphics[scale=0.25]{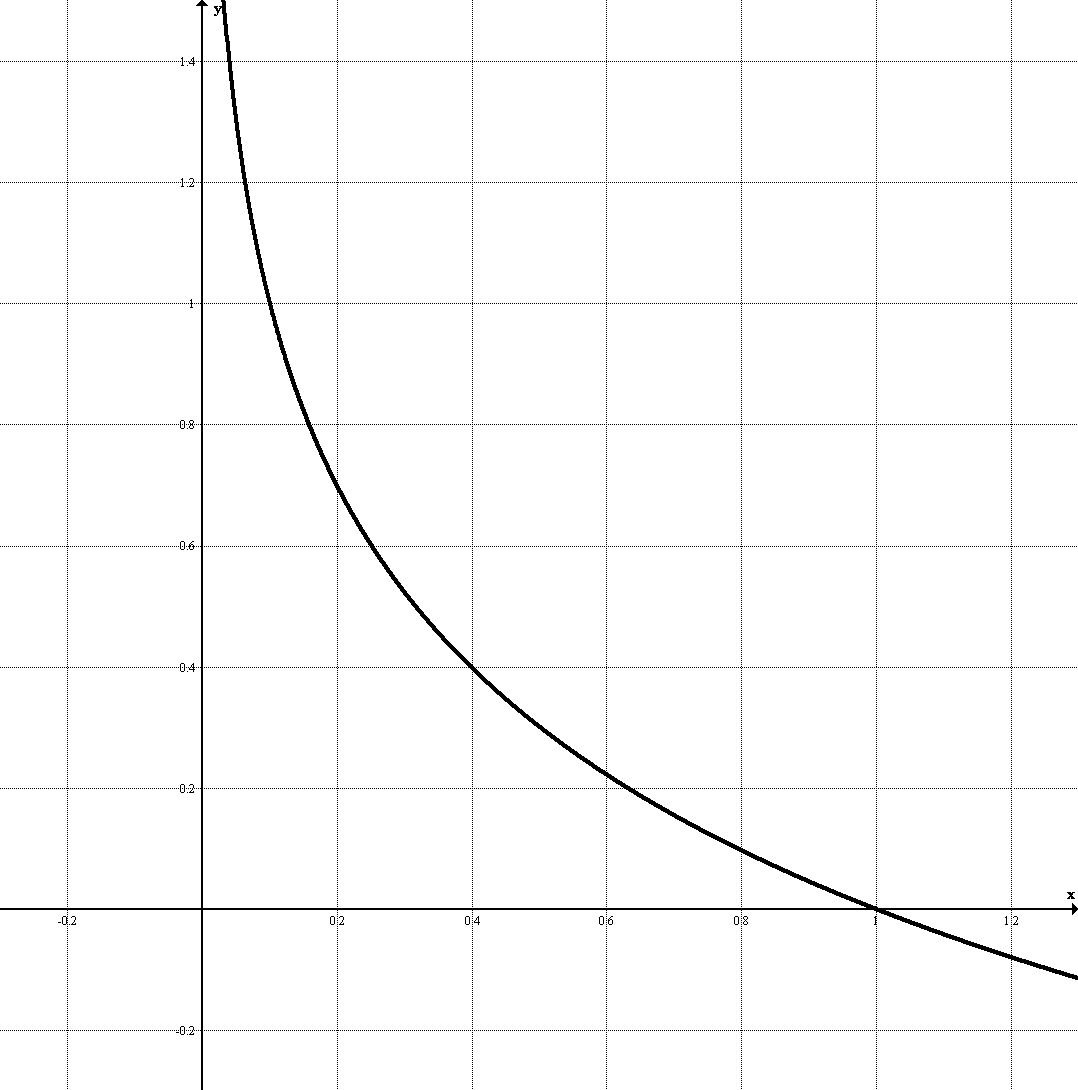}
\caption{-log function}
\label{figure:-log:curve}
\end{figure}

Remember that a probability is by definition within $[0, 1]$ interval.
If we look at -log function in Figure~\ref{figure:-log:curve},
we could see that its value is high for a low probability value (unlikely outcome) and its value is null
for a probability value equal to 1 (certain outcome), which corresponds to the objective introduced above.
Note that the use of a logarithm
also makes self-information additive for independent events,
i.e. $I(P_1\,P_2) = I(P_1) + I(P_2)$.

Then, let us consider all possible outcomes $\text{note} = \text{Note}_i$,
each outcome having $P(\text{note} = \text{Note}_i)$ as its associated probability,
and $P(\text{note})$ being the probability distribution for all possible outcomes.
The intuition is to define the {\em entropy\index{Entropy}} (notated $H$)
of the probability distribution for all possible outcomes
as the sum of the self-information for each possible outcome,
weighted by
the probability of the outcome.
This leads to Equation~\ref{equation:entropy}.

\begin{equation}
\begin{split}
H(P) = \sum_{i=0}^{n}{P(\text{note} = \text{Note}_i)~I(\text{note} = \text{Note}_i)}\\
= - \sum_{i=0}^{n}{P(\text{note} = \text{Note}_i)\,\text{log}\,P(\text{note} = \text{Note}_i)}
\end{split}
\label{equation:entropy}
\end{equation}

Note that we can further rewrite the definition
by using the notion of expectation\index{Expectation}\footnote{An expectation,
	or expected value, of some function $f(\text{x})$
	with respect to a probability distribution\index{Probability!distribution} $P(\text{x})$,
	usually notated as $\mathbb{E}_{\text{x} \sim P}[f(\text{x})]$,
	is the average (mean) value that $f$ takes on when x is drawn from $P$,
	i.e. $\mathbb{E}_{\text{x} \sim P}\,[f(\text{x})] = \sum_{\text{x}}^{}{P(\text{x}) f(\text{x})}$
	(we are here considering the case of discrete variables,
	which is the case for classification within a set of possible notes).},
which leads to Equation~\ref{equation:entropy:expectation}.

\begin{equation}
H(P) = \mathbb{E}_{\text{note} \sim P}\,[I(\text{note})]
= - \mathbb{E}_{\text{note} \sim P}\,[\text{log}\,P(\text{note})]
\label{equation:entropy:expectation}
\end{equation}

Now, let us introduce in Equation~\ref{equation:kl:divergence}
the {\em Kullback-Leibler divergence\index{Kullback-Leibler divergence}}
(often abbreviated as {\em KL-divergence\index{KL-divergence}},
and notated $D_\text{KL}$),
as some measure\footnote{Note that it is not a true distance measure
	as it not symmetric.}
of how different are two separate probability distributions $P$ and $Q$ over a same variable (note).

\begin{equation}
\begin{split}
D_\text{KL}(P || Q) = \mathbb{E}_{\text{note} \sim P}\,[\text{log}\,\frac{P(\text{note})}{Q(\text{note})}]\\
= \mathbb{E}_{\text{note} \sim P}\,[\text{log}\,P(\text{note}) - \text{log}\,Q(\text{note})]\\
= \mathbb{E}_{\text{note} \sim P}\,[\text{log}\,P(\text{note})] - \mathbb{E}_{\text{note} \sim P}\,[\text{log}\,Q(\text{note})]
\end{split}
\label{equation:kl:divergence}
\end{equation}

$D_\text{KL}$ may be rewritten
as
in Equation~\ref{equation:kl:divergence:cross:entropy:2}\footnote{By using
	$H(P)$ definition in Equation~\ref{equation:entropy:expectation}.},
where $H(P, Q)$,
named the {\em categorical cross-entropy\index{Categorical!cross-entropy}},
is defined in Equation~\ref{equation:cross:entropy}.


\begin{equation}
D_\text{KL}(P || Q) = - H(P) + H(P, Q)
\label{equation:kl:divergence:cross:entropy:2}
\end{equation}

\begin{equation}
H(P, Q) = - \mathbb{E}_{\text{note} \sim P}\,[\text{log}\,Q(\text{note})]
\label{equation:cross:entropy}
\end{equation}

Note that categorical cross-entropy is similar to KL-divergence\footnote{And, just like KL-divergence,
	it is not symmetric.},
while lacking the $H(P)$ term.
But minimizing $D_\text{KL}(P || Q)$ or minimizing $H(P, Q)$,
with respect to $Q$,
are equivalent,
because the omitted term $H(P)$
is a constant with respect to $Q$.

%

Now, remember\footnote{See
	Section~\ref{section:architecture:neural:network:cost:function}.}
that the objective of the neural network is
to predict the \^y probability distribution,
which is an estimation of the y true ground
probability distribution,
by minimizing the difference between them.
This leads to Equations~\ref{equation:kl:divergence:y} and~\ref{equation:cross:entropy:y}.


\begin{equation}
D_\text{KL}(\text{y} || \text{\^y})
= \mathbb{E}_{\text{y}}\,[\text{log}\,\text{y} - \text{log}\,\text{\^y}]
= \sum_{i=0}^{n}{\text{y}_i\,(\text{log}\,\text{y}_i - \text{log}\,\text{\^y}_i})
\label{equation:kl:divergence:y}
\end{equation}


\begin{equation}
H(\text{y}, \text{\^y})
= -  \mathbb{E}_{\text{y}}\,[\text{log}\,\text{\^y}]
= - \sum_{i=0}^{n}{\text{y}_i\,\text{log}\,\text{\^y}_i}
\label{equation:cross:entropy:y}
\end{equation}

As mentioned above, minimizing $D_\text{KL}(\text{y} || \text{\^y})$ or minimizing $H(\text{y}, \text{\^y})$,
with respect to \^y,
are equivalent,
because the omitted term
$H(\text{y})$
is a constant with respect to \^y.

Last, deriving the {\em binary cross-entropy\index{Binary!cross-entropy}} (that we notate $H_\text{B}$) is easy,
as there are only two possible outcomes, which leads to Equation~\ref{equation:binary:cross:entropy:rewriting:1}.


\begin{equation}
H_\text{B}(\text{y}, \text{\^y}) = - (\text{y}_0\,\text{log}\,\text{\^y}_0 + \,\text{y}_1\,\text{log}\,\text{\^y}_1)
\label{equation:binary:cross:entropy:rewriting:1}
\end{equation}


Because
$\text{y}_1 = 1 - \text{y}_0$
and
$\text{\^y}_1 = 1 - \text{\^y}_0$
(as the sum of the probabilities of the two possible outcomes is 1),
this ends up into
Equation~\ref{equation:binary:cross:entropy:rewriting:scalar}.

\begin{equation}
H_\text{B}(\text{y}, \text{\^y}) = - (\text{y}\,\text{log}\,\text{\^y} + (1 - \text{y})\,\text{log}\,(1 - \text{\^y}))
\label{equation:binary:cross:entropy:rewriting:scalar}
\end{equation}

More details and principles for the cost functions\footnote{The underlying principle
	of {\em maximum likelihood estimation\index{Maximum likelihood estimation}\index{Likelihood}}, not explained here.}
can be found, for example, in \cite[Section~6.2.1]{goodfellow:deep:learning:book:2016} and \cite[Section~5.5]{goodfellow:deep:learning:book:2016},
respectively.
In addition,
the information theory foundation of cross-entropy as the number of bits needed for encoding information
is introduced, for example, in \cite{dipietro:introduction:cross:entropy:web:2016}.

\subsection{Feedforward Propagation}
\label{architecture:network:feedforward}

Feedforward propagation\index{Feedforward!propagation} in a multilayer neural network consists in injecting input data\footnote{The x part of an example,
	for the generation phase as well as for the training phase.}
into the input layer
and propagating the computation through its successive layers\index{Layer}
until the output is produced.
This can be implemented very efficiently
because it consists in a pipelined computation of successive vectorized matrix\index{Matrix} products (intercalated with $AF$ activation function\index{Activation!function} calls).

\begin{figure}
\includegraphics[scale=0.85]{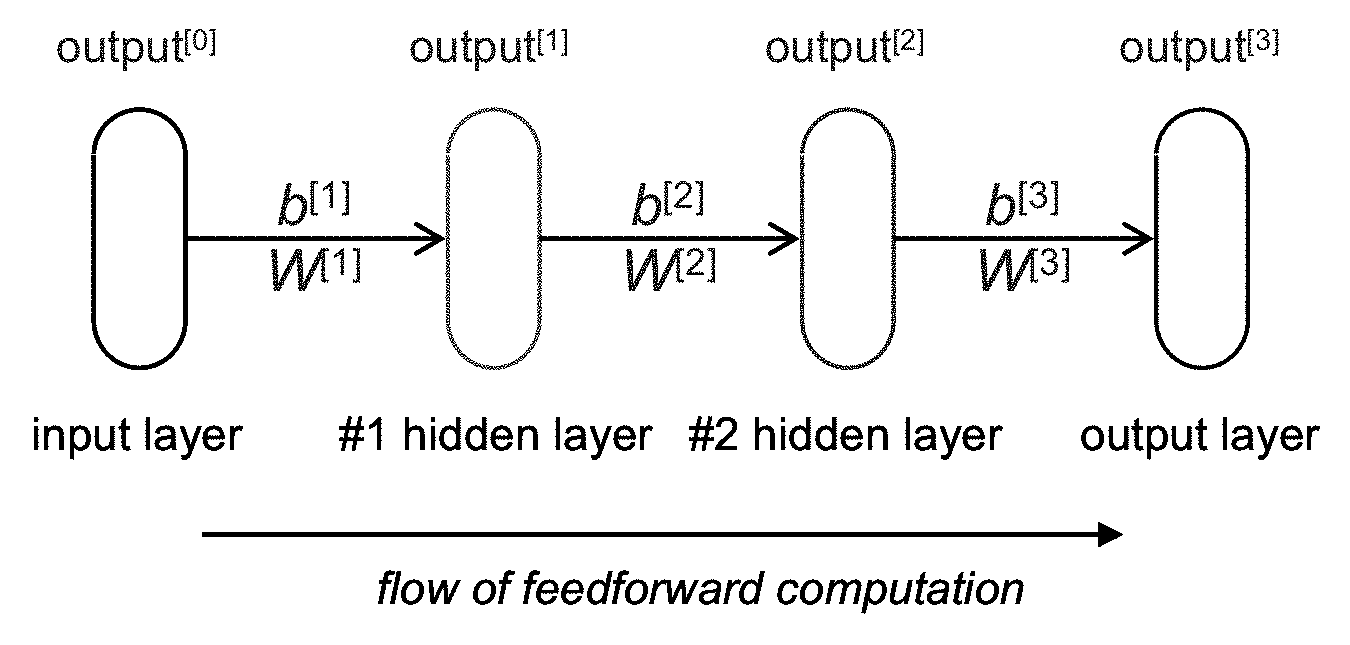}
\caption{Example of a feedforward neural network (abstract) pipelined computation}
\label{figure:abstract:abstract:multilayer:neural:network:architecture:k:layer}
\end{figure}

Each computation from layer $k-1$ to layer $k$ is processed as in Equation~\ref{equation:multilayer:feedforward:vectorized:formula},
which is a generalization of Equation~\ref{equation:basic:block:feedforward:vectorized:one:example:equation}\footnote{Feedforward computation
	for one layer has been introduced in Section~\ref{section:architecture:building:block:feedforward:computation}.},
where $b^{[k]}$ and $W^{[k]}$\footnote{We use
	a superscript notation\index{Notation convention} with brackets $^{[k]}$ to denote the $k$th layer,
	to avoid confusion with the superscript notation with parentheses $^{(k)}$ to denote the $k$th example
	and the subscript notation $_i$ to denote the $i$th input variable.}
are respectively the bias and the weight matrix
between layer $k-1$ and layer $k$,
and where $\text{output}^{[0]}$ is the input layer,
as shown in Figure~\ref{figure:abstract:abstract:multilayer:neural:network:architecture:k:layer}.

\begin{equation}
\text{output}^{[k]} = AF(b^{[k]} + W^{[k]} \text{output}^{[k-1]})
\label{equation:multilayer:feedforward:vectorized:formula}
\end{equation}

%

Multilayer neural networks are therefore often also named {\em feedforward neural networks\index{Feedforward!neural network}}
or {\em multilayer Perceptron\index{Multilayer!Perceptron}} (MLP\index{MLP})\footnote{The original Perceptron\index{Perceptron}
	was a neural network with no hidden layer, and thus equivalent to our basic building block,
	with only one output node and with the step function as the activation function.}.

Note that neural networks are {\em deterministic}.
This means that
the same input will deterministically {\em always} produce the {\em same} output.
This is a useful guarantee for prediction and classification purposes
but may be a limitation for generating new content.
However, this may be compensated by {\em sampling\index{Sampling}} from the resultant probability distribution
(see Sections~\ref{section:architecture:neural:network:output:activation:function}
and~\ref{section:challenges:strategies:variability}).


\subsection{Training}
\label{architecture:network:training}

For the training\index{Training} phase\footnote{Let us remember
	that this a case of supervised learning\index{Supervised learning}
	(see Section~\ref{section:architecture:building:block:training}).},
computing the derivatives becomes a bit more complex than for the basic building block (with no hidden layer)
presented in Section~\ref{section:training:algorithm}.
{\em Backpropagation\index{Backpropagation}} is the standard method
of estimating\index{Estimation} the derivatives (gradients\index{Gradient}) for a multilayer neural network.
It is based on the {\em chain rule\index{Chain rule}} principle \cite{rumelhart:backpropagation:1986},
in order to estimate\index{Estimation} the contribution of each weight\index{Weight} to the final prediction error\index{Error},
that is the cost\index{Cost}.
See, for example, \cite[Chapter~6]{goodfellow:deep:learning:book:2016} for more details.

Note that, in the most common case, the cost function of a multilayer neural network is {\em not convex}\index{Convex},
meaning that there may be {\em multiple local minima}\index{Local!minimum}.
Gradient descent, as well as other more sophisticated heuristic optimization methods, does not guarantee the global optimum will be reached.
But in practice a clever configuration of the model
(notably, its {\em hyperparameters\index{Hyperparameter}}, see Section~\ref{section:architecture:hyperparameters})
and well-tuned optimization heuristics, such as stochastic gradient descent\index{Stochastic!gradient descent} (SGD\index{SGD}),
	will lead to accurate solutions\footnote{On this issue, see \cite{choromanska:loss:surfaces:arxiv:2015},
	which shows that
	1) local minima are located in a well-defined band,
	2) SGD converges to that band,
	3) reaching the global minimum becomes harder as the network size increases and
	4) in practice this is irrelevant as the global minimum often leads to overfitting
	(see next section).}.

\subsection{Overfitting}
\label{section:architecture:training:overfitting}

A fundamental issue for neural networks (and more generally speaking for machine learning algorithms)
is their {\em generalization\index{Generalization}} ability,
that is their capacity to perform well on {\em yet unseen data}.
In other words, we do not want a neural network to just perform well on the training data\footnote{Otherwise, the best and simpler algorithm
	would be a memory-based\index{Memory!-based} algorithm,
	which simply {\em memorizes\index{Memory}} all (x, y) pairs.
	It has the best fit to the training data but it does not have any generalization ability.}
but also on future data\footnote{Future data is not yet known but that does not mean that it is {\em any kind} of (random) data,
	otherwise a machine learning algorithm would not be able to learn and generalize well.
	There is indeed a fundamental assumption of regularity of the data corresponding to a task
	(e.g., images of human faces, jazz chord progressions, etc.) that neural networks will exploit.}.
This is actually a fundamental dilemma, the two opposing risks being

\begin{itemize}

\item {\em underfitting} -- when the {\em training error\index{Training!error}} (error measure on the {\em training data}) is large; and

\item {\em overfitting} -- when the {\em generalization error\index{Generalization!error}} (expected error on {\em yet unseen data}) is large.

\end{itemize}

\begin{figure}
\includegraphics[width=\textwidth]{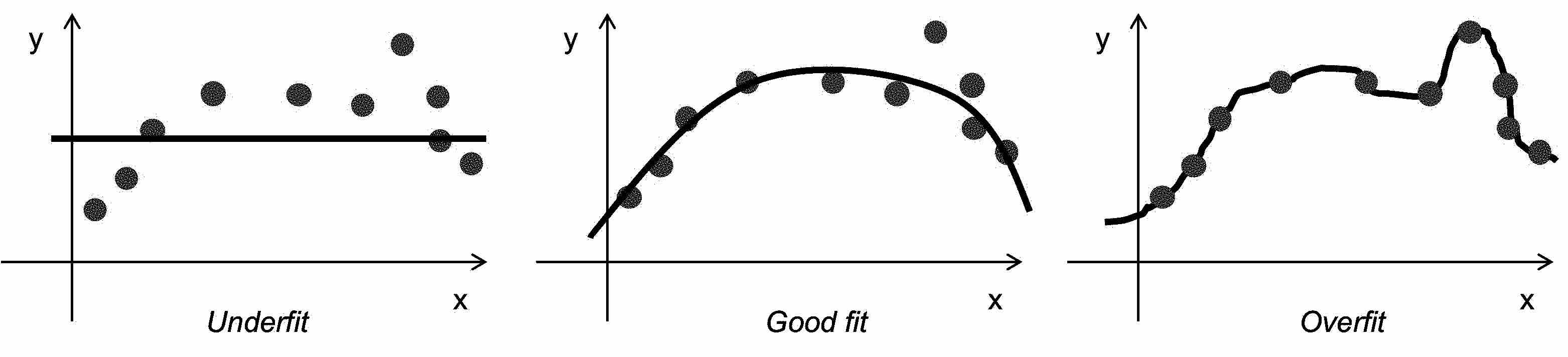}
\caption{Underfit, good fit and overfit models}
\label{figure:architecture:overfit}
\end{figure}

A simple illustrative example of underfit, good fit and overfit models for the same training data (the green solid dots) is shown in Figure~\ref{figure:architecture:overfit}.

In order to be able to estimate the potential for generalization,
the dataset is actually divided into two portions, with a ratio of approximately 70/30:

\begin{itemize}

\item the {\em training set\index{Training!set}} -- which will be used for training the neural network; and

\item the {\em validation set\index{Validation!set}}, also named {\em test set\index{Test!set}}\footnote{Actually,
	a difference could (should) be made, as explained by Hastie {\em et al.} in \cite[page~222]{hastie:elements:statistical:learning:book:2009}:
	``It is important to note that there are in fact two separate goals that we
	might have in mind:
%
%
	
	Model selection:
	estimating the performance of different models in order
	to choose the best one.
%
	
	Model assessment:
	having chosen a final model, estimating its prediction error (generalization error) on new data.
	
	
	If we are in a data-rich situation, the best approach for both problems is
	to randomly divide the dataset into three parts: a training set, a validation
	set, and a test set. The training set is used to fit the models; the validation
	set is used to estimate prediction error for model selection; the test set is
	used for assessment of the generalization error of the final chosen model.''
	However, as a matter of simplification, we will not consider that difference in the book.}
-- which will be used to estimate the capacity of the model for generalization.

\end{itemize}

	
\subsection{Regularization}
\label{section:architecture:training:regularization}

There are various techniques to control overfitting\index{Overfitting}, i.e., to improve generalization.
They are usually named {\em regularization\index{Regularization}} and some examples of well-known techniques are

\begin{itemize}

\item {\em weight decay\index{Weight!decay}} (also known as L$^2$\index{L$^2$}),
by penalizing over-preponderant weights;

\item {\em dropout\index{Dropout}}, by introducing random disconnections;

\item {\em early stopping\index{Early stopping}},
by storing a copy of the model parameters every time the error on the validation set
reduces,
then terminating after an absence of progress during a pre-specified number of iterations,
and returning these parameters; and

\item {\em dataset augmentation\index{Dataset!augmentation}}, by data synthesis\index{Data synthesis}
(e.g., by mirroring, translation and rotation for images;
by transposition for music,
see Section~\ref{section:representation:transposition}),
in order to augment the number of training examples.

\end{itemize}

We will not further detail regularization techniques, see, for example, \cite[Section~7]{goodfellow:deep:learning:book:2016}.

\subsection{Hyperparameters}
\label{section:architecture:hyperparameters}

In addition to the {\em parameters\index{Parameter}} of the model,
which are the weights\index{Weight} of the connexions\index{Connexion} between nodes,
a model also includes {\em hyperparameters\index{Hyperparameter}}, which are parameters at an
{\em architectural meta-level\index{Architectural!meta-level}\index{Meta-level}},
concerning both {\em structure}\index{Structure} and {\em control}\index{Control}.

Examples of {\em structural} hyperparameters, mainly concerned with the architecture, are

\begin{itemize}

\item number of layers,

\item number of nodes, and

\item nonlinear activation function\index{Activation!function}.

\end{itemize}

Examples of {\em control} hyperparameters, mainly concerned with the learning process, are

\begin{itemize}

\item optimization procedure\index{Optimization},

\item learning rate\index{Learning!rate}, and

\item regularization strategy\index{Regularization} and associated parameters.

\end{itemize}

Choosing proper values for (tuning) the various hyperparameters is fundamental both for the efficiency and the accuracy of neural networks for a given application.
There are two approaches for exploring and tuning hyperparameters:
{\em manual tuning} or {\em automated tuning} -- by algorithmic exploration of the multidimensional space of hyperparameters
and for each sample evaluating the generalization error.
The three main strategies for automated tuning are

\begin{itemize}

\item {\em random search} -- by defining a distribution for each hyperparameter, sampling configurations, and evaluating them;

\item {\em grid search} -- as opposed to random search, exploration is systematic on a small set of values for each hyperparameter; and

\item {\em model-based optimization} -- by building a model of the generalization error and running an optimization algorithm over it.

\end{itemize}

The challenge of automated tuning is its computational cost, although trials may be run in parallel.
We will not detail these approaches here;
however, further information can be found in \cite[Section~11.4]{goodfellow:deep:learning:book:2016}.


Note that this tuning activity is more objective for conventional tasks such as prediction and classification
because the evaluation measure is objective, being the error rate for the validation set\index{Validation!set}.
When the task is the generation of new musical content, tuning is more subjective because there is no preexisting evaluation measure.
It then turns out to be more {\em qualitative}, for instance through a manual evaluation of generated music by musicologists.
This evaluation issue will be addressed in Section~\ref{section:discussion:evaluation}.

\subsection{Platforms and Libraries}
\label{section:architecture:platforms}

Various platforms\footnote{See, for example,
	the survey in \cite{parvat:survey:deep:platforms:icisc:2017}.},
such as CNTK, MXNet, PyTorch and TensorFlow,
are available as a foundation for developing and running deep learning systems\footnote{There are also
	more general libraries
	for machine learning and data analysis, such as the SciPy library for the Python language,
	or the language R and its libraries.}.
They include libraries of

\begin{itemize}

\item basic architectures, such as the ones we are presenting in this chapter;

\item components, for example optimization algorithms;

\item runtime interfaces for running models on various hardware, including GPUs or distributed Web runtime facilities; and

\item visualization and debugging facilities.

\end{itemize}

Keras is an example of a higher-level framework to simplify development, with CNTK, TensorFlow and Theano as possible backends.
ONNX is an open format for representing deep learning models
and was designed to ease the transfer of models between different platforms and tools.

\section{Autoencoder}
\label{section:architecture:autoencoder}

\begin{figure}
\includegraphics[scale=0.4]{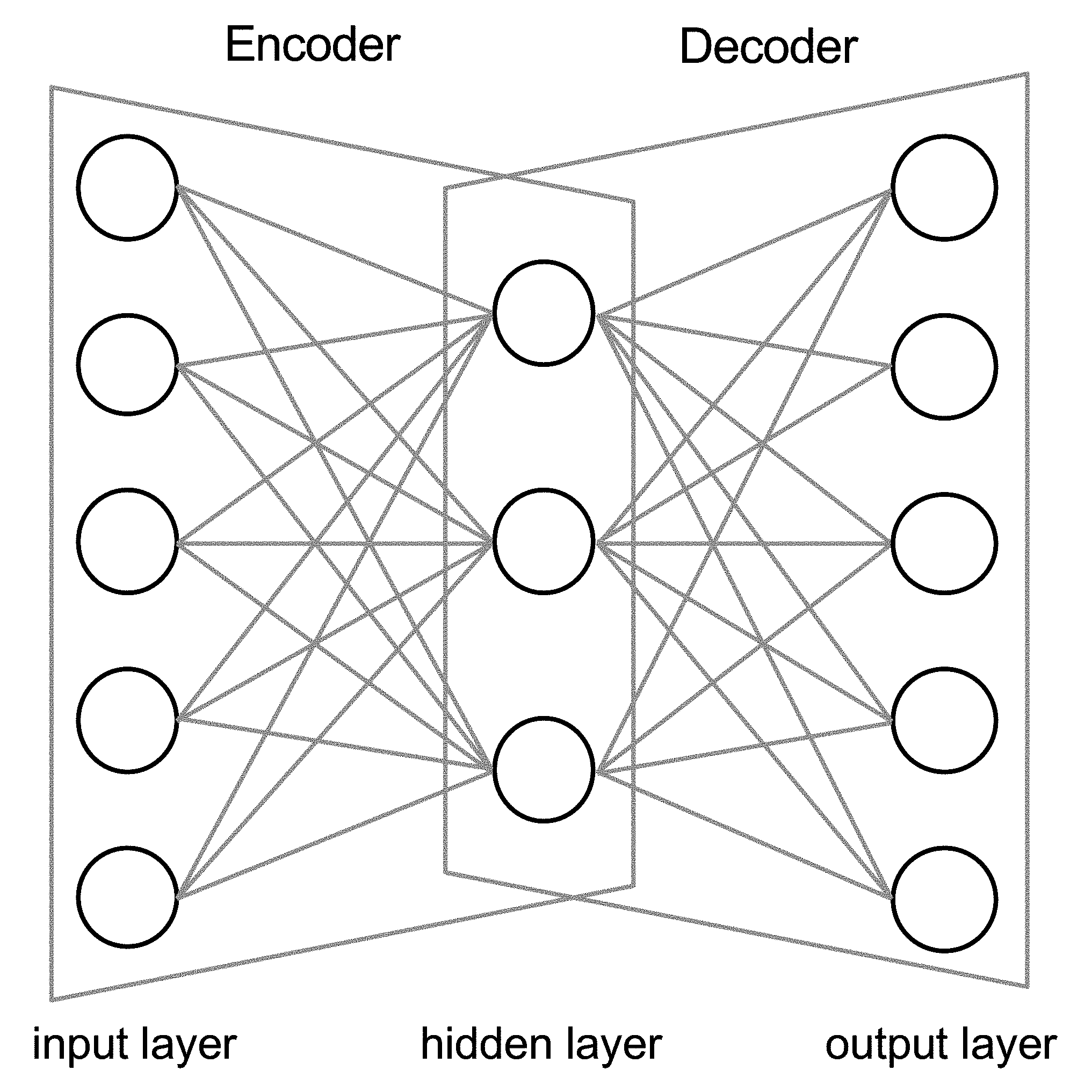}
\caption{Autoencoder architecture}
\label{figure:autoencoder}
\end{figure}

An {\em autoencoder\index{Autoencoder}} is a neural network with one hidden layer and with an additional {\em constraint}:
the number of output nodes\index{Output!node} is equal to
the number of input nodes\index{Input!node}\footnote{The bias is not counted/considered here
	as it is an implicit additional input node.}.
The output layer actually {\em mirrors} the input layer.
It is shown in Figure~\ref{figure:autoencoder}, with its peculiar symmetric\index{Symmetric} diabolo (or sand-timer) shape aspect.

Training an autoencoder represents a case of {\em unsupervised learning\index{Unsupervised learning}},
as the examples do not contain any additional label\index{Label} information
(the effective value or class to be predicted).
But the trick is that this is implemented using conventional supervised learning\index{Supervised learning} techniques,
by presenting output data equal to the input data\footnote{This is sometimes
	called
	{\em self-supervised\index{Self!-supervised learning}}
	learning \cite{ng:high:level:features:unsupervised:learning:2012}.}.
In practice, the autoencoder tries to learn the identity\index{Identity} function.
As the hidden layer usually has fewer nodes than the input layer,
the {\em encoder\index{Encoder}} component (shown in yellow in Figure~\ref{figure:autoencoder})
must {\em compress} information
while the {\em decoder\index{Decoder}} (shown in purple)
has to {\em reconstruct}, as accurately as possible, the initial information\footnote{Compared to traditional dimension reduction algorithms,
	such as principal component analysis\index{Principal component analysis} (PCA\index{PCA}),
	this approach has two advantages:
	1) feature extraction is nonlinear
	(the case of {\em manifold learning\index{Manifold!learning}},
	see \cite[Section~5.11.3]{goodfellow:deep:learning:book:2016} and Section~\ref{section:architecture:vae})
	and
	2) in the case of a sparse autoencoder (see next section),
	the number of features may be arbitrary (and not necessarily smaller than the number of input parameters).}.
This forces the autoencoder to {\em discover} significant (discriminating) {\em features\index{Feature!extraction}} to encode\index{Encoding} useful information
into the hidden layer nodes (also named the {\em latent variables\index{Latent!variable}}\footnote{In statistics,
	{\em latent variables} are variables that are not directly observed but are rather inferred
	(through a mathematical model)
	from other variables that are observed (directly measured).
	They can serve to reduce the dimensionality of data.}).
Therefore, autoencoders may be used to automatically extract high-level {\em features\index{Feature!extraction}}
\cite{ng:high:level:features:unsupervised:learning:2012}.
The set of features extracted are often named an {\em embedding\index{Embedding}}\footnote{See
	the definition of embedding in Section~\ref{section:representation:feature:extraction}.}.
Once trained, in order to extract features from an input,
one just needs to feedforward the input data and gather the activations of the hidden layer (the values of the latent variables).

Another interesting use of decoders is the high-level control of content generation.
The latent variables\index{Latent!variable} of an autoencoder constitute
a compact representation
of the common features of the learnt examples.
By instantiating these latent variables and decoding the embedding, we can generate a new musical content
corresponding to the values of the latent variables.
We will explore this strategy
in Section~\ref{section:strategy:decoder:feedforward}.

\subsection{Sparse Autoencoder}
\label{section:architecture:sparse:autoencoder}

A {\em sparse autoencoder\index{Sparse autoencoder}}
is an autoencoder with a {\em sparsity\index{Sparsity}} constraint,
such that its hidden layer units are inactive most of the time.
The objective is to
enforce the {\em specialization} of each unit in the hidden layer
as a specific {\em feature detector\index{Feature!detector}}.

\begin{figure}
\includegraphics[scale=0.8]{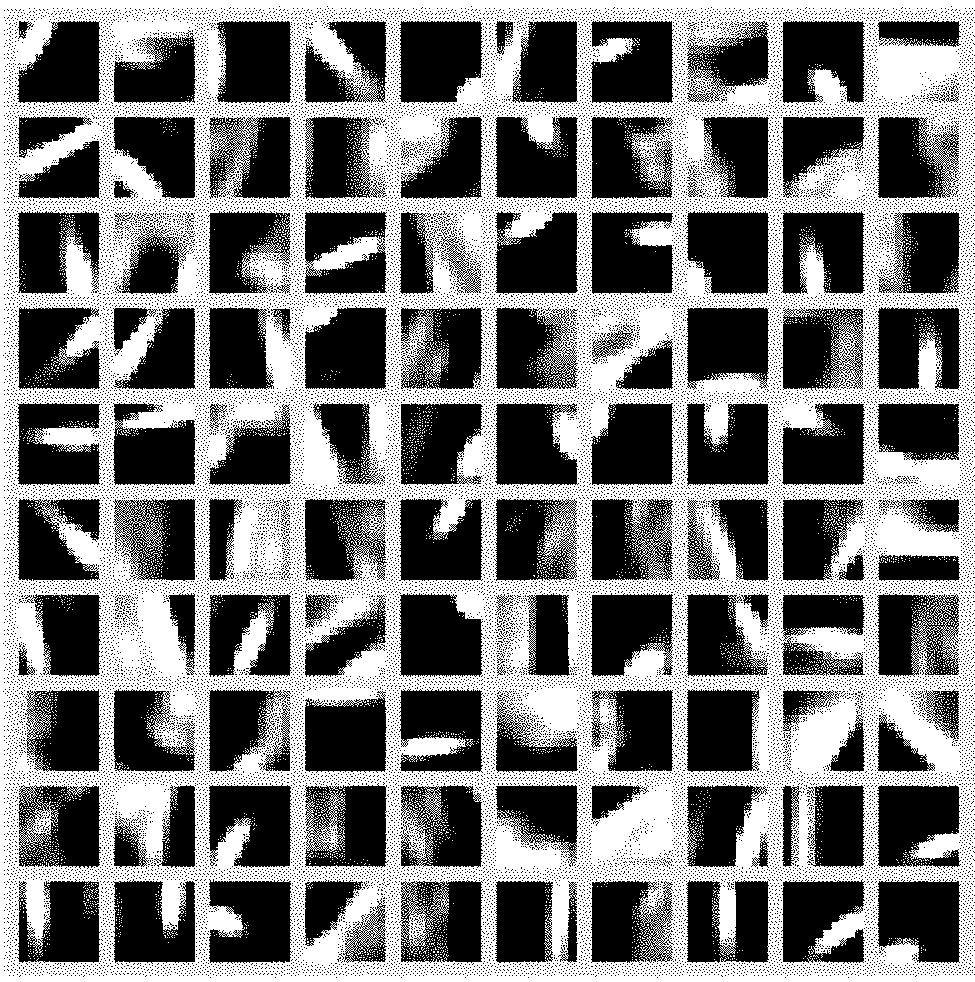}
\caption{Visualization of the input image motives that maximally activate each of the hidden units of a sparse autoencoder architecture.
Reproduced from \cite{ng:course:notes:sparse:autoencoder:2011} with permission of the author}
\label{figure:sparse:autoencoder:example}
\end{figure}

For instance, a sparse autoencoder with 100 units in its hidden layer
and trained on 10$\times$10 pixel images
will learn to detect edges at different positions and orientations in images, as shown in Figure~\ref{figure:sparse:autoencoder:example}.
When applied to other input domains, such as audio or symbolic music data, this algorithm will learn useful features for those domains too.

The sparsity constraint is implemented by adding an additional term to the cost\index{Cost} function to be minimized,
see more details in \cite{ng:course:notes:sparse:autoencoder:2011}
or~\cite[Section~14.2.1]{goodfellow:deep:learning:book:2016}.

\subsection{Variational Autoencoder}
\label{section:architecture:vae}

A {\em variational autoencoder\index{Variational!autoencoder}} (VAE\index{VAE}) \cite{kingma:vae:arxiv:2014}
has the added constraint that the encoded representation,
the latent variables\index{Latent!variable},
by convention\index{Notation convention} denoted by variable z,
follow some prior probability distribution $P(\text{z})$.
Usually, a {\em Gaussian distribution\index{Gaussian distribution}\index{Probability!distribution}}\footnote{Also named
	{\em normal distribution\index{Normal distribution}}.}
is chosen for its generality.

This constraint is implemented by adding a specific term to the cost\index{Cost} function,
by computing the cross-entropy\index{Cross-entropy} between the values of the latent variables
and the prior distribution\footnote{The actual implementation
	is more complex and has some tricks
	(e.g., the encoder actually generates a mean vector and a standard deviation vector)
	that we will not detail here.}.
For more details about VAEs, an example of tutorial could be found in \cite{doersch:tutorial:variational:autoencoder:2016}
and there is a nice introduction of its application to music in \cite{roberts:hierarchical:latent:icml:2018}.



As with an autoencoder\index{Autoencoder}, a VAE\index{VAE} will learn the identity\index{Identity} function,
but furthermore the decoder part will learn the relation between a Gaussian distribution of the latent variables\index{Latent!variable}
and the learnt examples.
As a result, sampling\index{Sampling} from the VAE is immediate,
one just needs to

\begin{itemize}

\item sample a value for the
latent variables\index{Latent!variable} $\text{z} \sim P(\text{z})$, i.e. z following distribution $P(\text{z})$;

\item input it into the decoder; and

\item feedforward the decoder to generate an output corresponding to the distribution of the examples,
following $P(\text{x} | \text{z})$ conditional probability\index{Conditional!probability} distribution learnt by the decoder.

\end{itemize}

This is in contrast to the need for indirect and computationally expensive strategies such as Gibbs sampling\index{Gibbs sampling}
for other architectures such as RBM,
to be introduced in Section~\ref{section:architecture:rbm}.


By construction, a variational autoencoder is representative of the dataset that it has learnt,
that is, for any example in the dataset,
there is at least one setting of the latent variables which causes the model to generate something very similar to that example \cite{doersch:tutorial:variational:autoencoder:2016}.

A very interesting characteristic
of the variational autoencoder architecture
for generation purposes
-- therefore often considered as one type of a class of models named {\em generative models\index{Generative!model}} --
is in the meaningful exploration of the latent space\index{Latent!space},
as a variational autoencoder is able to learn a ``smooth''\footnote{That is,
	a small change in the latent space will correspond to a small change in the generated examples,
	without any discontinuity or jump.
	For a more detailed discussion about which (and how) interesting effects
	(smoothness,
	parsimony
	and axis-alignment between data and latent variability)
	a VAE has on the latent representation
	(the vector of latent variables)
	learnt,
	see, e.g., \cite{wild:disentangled:vae:web:2018}.}
latent space mapping to realistic examples.
Note that this general objective is named {\em manifold learning\index{Manifold!learning}}
and more generally {\em representation learning\index{Representation!learning}} \cite{bengio:representation:learning:ieee:pami:2013},
that is the learning of a representation capturing the topology of a set of examples.
As defined in \cite[Section~5.11.3]{goodfellow:deep:learning:book:2016},
a {\em manifold\index{Manifold}} is a connected set of points (examples) that can be approximated by a smaller number of dimensions,
each one corresponding to a local direction of variation.
An intuitive example is a 2D map capturing the topology of cities dispersed on the 3D earth,
where a movement on the map corresponds to a movement on the earth.

\begin{figure}
\includegraphics[width=\textwidth]{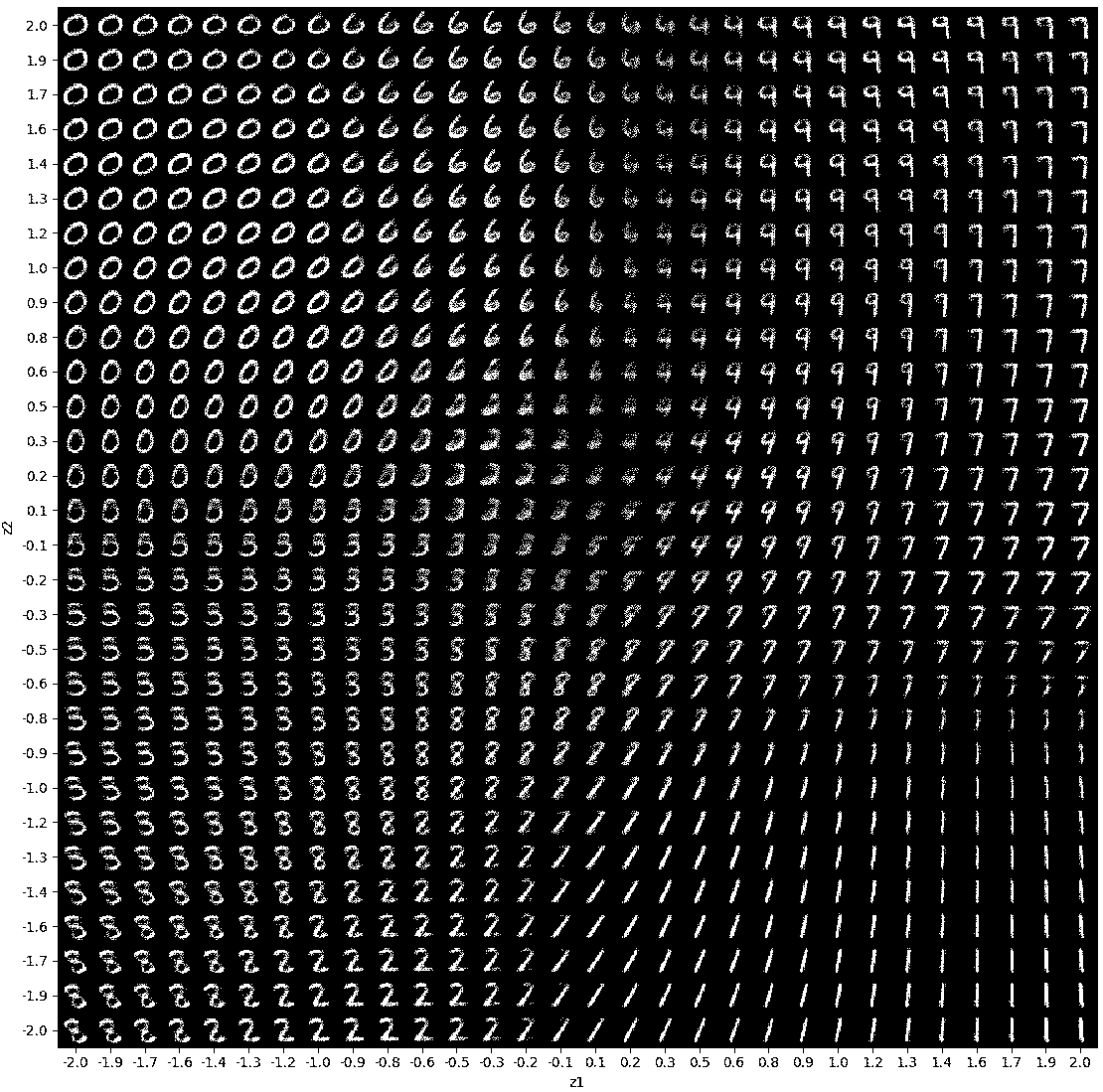}
\caption{Various digits generated by decoding sampled latent points at regular intervals on the MNIST handwritten digits database}
\label{figure:music:vae:example:mnist}
\end{figure}

To illustrate the possibilities, let us train a VAE with only two latent variables
on the MNIST\index{MNIST} handwritten digits database dataset \cite{lecun:mnist:web:1998} (with 60.000 examples,
each one being an image of 28$\times$28 pixels).
Then, we scan the latent two-dimension plane,
sampling latent values for the two latent variables (i.e. sampling points within the 2-dimension latent space)
at regular intervals
and generating the corresponding artificial digits by decoding the latent points\footnote{As
	proposed and implemented in \cite{chollet:keras:building:autoencoders:2016}.}.
Figure~\ref{figure:music:vae:example:mnist} shows examples of artificial digits generated.

Note that training the VAE has forced it to compress information about the actual examples by splitting
(though the encoder)
information in two subsets:

\begin{itemize}

\item the {\em specific} (discriminative) part, encoded within the latent variables\index{Latent!variable}; and

\item the {\em common} part, encoded within the weights of the decoder, in order to be able to reconstruct as close as possible each original data\footnote{Indeed,
	there is no magic here, the reversible compression from 28$\times$28 variables to 2 variables
	must have extracted
	and stored missing information somewhere.}.

\end{itemize}

The VAE actually has been forced to find out dimensions of variations for the dataset examples.
By looking at the figure, we can guess that the two dimensions {\em could} be:

\begin{itemize}

\item from angular to round elements for the 1st variable (z$_1$, horizontally represented),
and

\item the size of the compound element (circle or angle) for the second latent variable (z$_2$, vertically represented).

\end{itemize}

Note that we cannot expect/force the VAE towards the semantics (meaning) of specific dimensions,
as the VAE will automatically extract them (this depends on the dataset as well as on the training configuration),
and we can only try to interpret them {\em a posteriori}\footnote{However, we will see
	that
	we can construct arbitrary characteristic attributes from a subset of examples
	and impose them on other examples,
	by doing {\em attribute vector arithmetics},
	as defined in the immediately following list,
	and as will be illustrated in Figure~\ref{figure:music:vae:example:density}
	in Section~\ref{section:system:music:vae}.}.
Examples of possible dimensions for music generation could be:
the number of notes\footnote{This will be illustrated in Figure~\ref{figure:glsr:vae:space:examples}
	in Section~\ref{section:experiment:glsr:vae}.},
the range
(the distance from the lowest to the highest pitch),
etc.

Once learnt by a VAE, the latent representation (a vector of latent variables) can be used to explore the latent space\index{Latent!space}
with various operations
to control/vary the generation of content.
Some examples of operations on the latent space,
as proposed in \cite{roberts:hierarchical:latent:icml:2018} and \cite{roberts:hierarchical:latent:arxiv:2018}
for the MusicVAE system described in Section~\ref{section:system:music:vae}, are

\begin{itemize}

\item {\em translation};

\item {\em interpolation\index{Interpolation}};

\item {\em averaging\index{Average}} of some points;

\item {\em attribute vector arithmetics\index{Attribute!vector arithmetics}},
	by addition or subtraction of an attribute vector capturing a given characteristic\footnote{This attribute vector is computed as
	the average latent vector for a collection of examples sharing that attribute\index{Attribute} (characteristic).}.

\end{itemize}

\begin{figure}
\includegraphics[width=\textwidth]{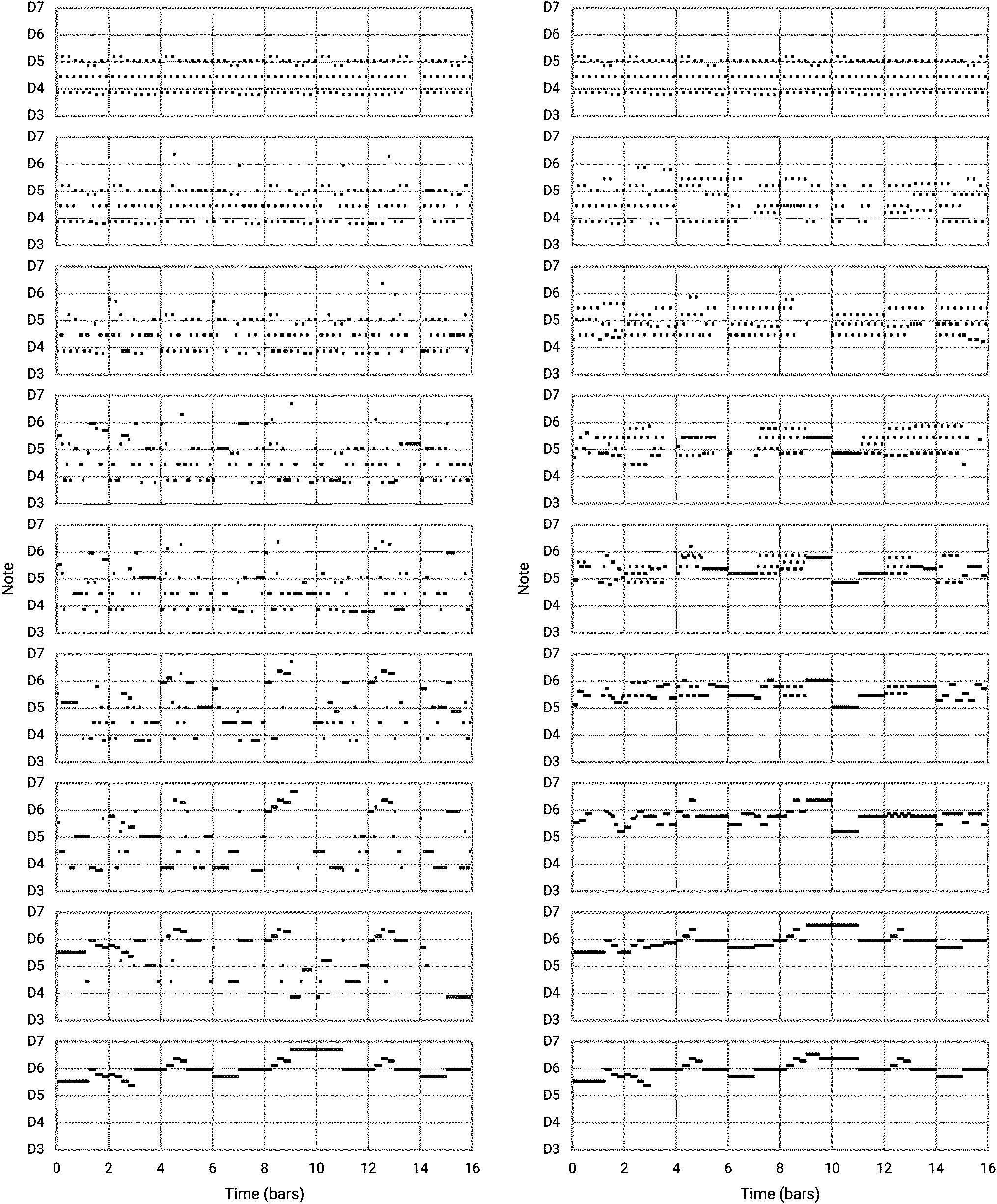}
\caption{Comparison of interpolations between the top and the bottom melodies by
(left) interpolating in the data (melody) space
and
(right) interpolating in the latent space and decoding it into melodies.
Reproduced from \cite{roberts:hierarchical:latent:arxiv:2018} with permission of the authors}
\label{figure:music:vae:example:interpolation}
\end{figure}

Figure~\ref{figure:music:vae:example:interpolation} shows an interesting comparison of melodies resulting from

\begin{itemize}

\item interpolation in the {\em data space}, that is the space of representation of melodies; and

\item interpolation in the {\em latent space}, which is then decoded into the corresponding melodies.

\end{itemize}

The interpolation in the latent space produces more meaningful and interesting melodies
than the interpolation in the data space (which basically just varies the ratio of notes from the two melodies),
as can be heard in \cite{roberts:music:vae:material:web} and \cite{roberts:music:vae:web}.
More details about these experiments will be provided in Section~\ref{section:system:music:vae}.


Variational autoencoders\index{Variational!autoencoder} are therefore elegant and promising models,
and as a result they are currently among the hot approaches explored for generating content with controlled variations.
Application to music generation will be illustrated
in Sections~\ref{section:systems:strategy:sampling:architecture:variational:recurrent}
and~\ref{section:system:music:vae}.

\subsection{Stacked Autoencoder}
\label{section:architecture:compound:stacked:autoencoders}

\begin{figure}
\includegraphics[scale=0.4]{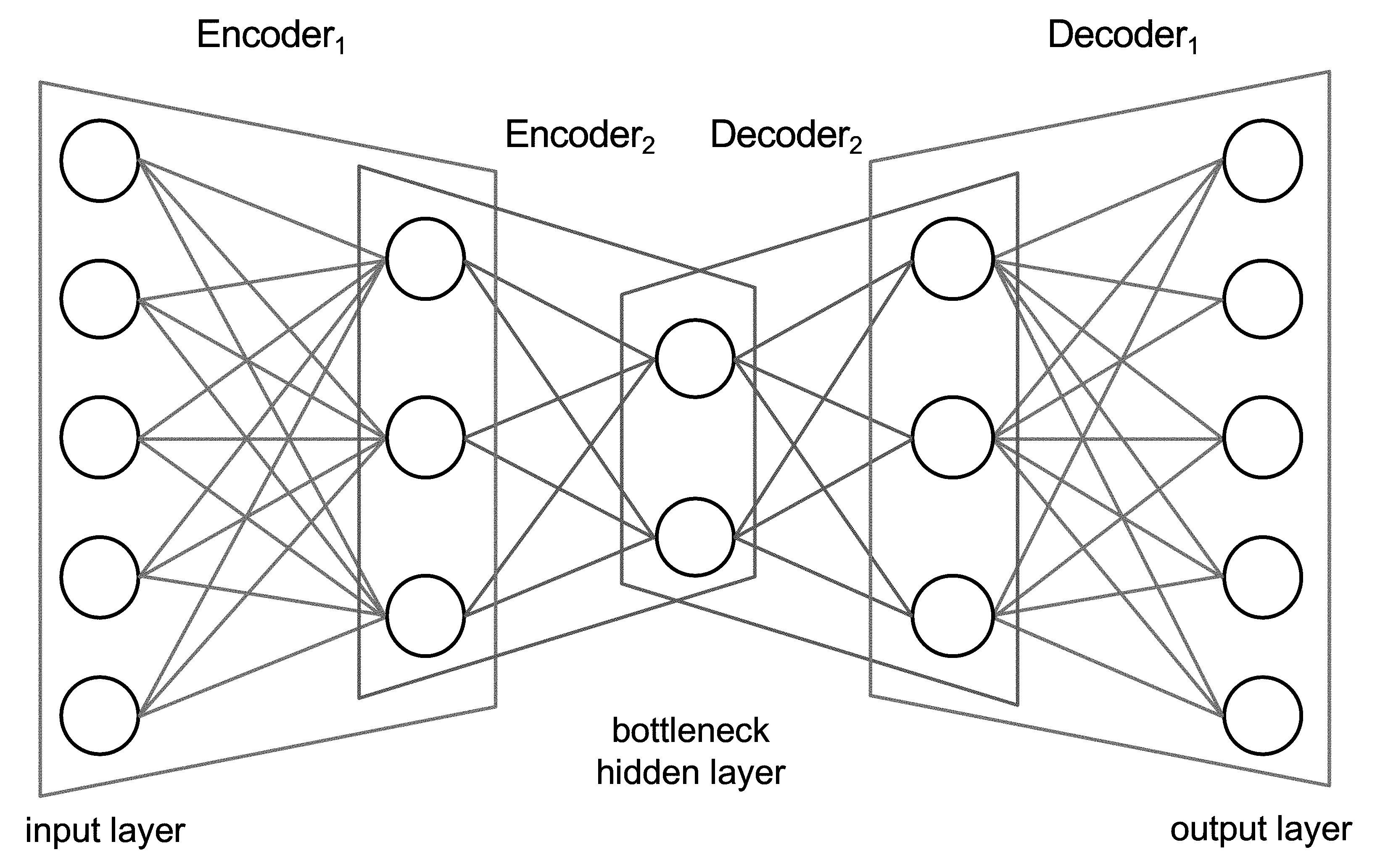}
\caption{A 2-layer stacked autoencoder architecture, resulting in a 4-layer full architecture}
\label{figure:stacked:autoencoders:architecture}
\end{figure}

The idea of a {\em stacked autoencoder\index{Stacked autoencoder}}
is to hierarchically\index{Hierarchical} nest successive autoencoders\index{Autoencoder}
with decreasing numbers of hidden layer units\index{Hidden!layer unit}.
An example of a 2-layer stacked autoencoder\footnote{Note that the convention in this case is to count
	and notate the number of nested autoencoders, i.e. the number of hidden layers.
	This is different from the depth of the {\em whole} architecture, which is double.
	For instance, a 2-layer stacked autoencoder results in a 4-layer whole architecture,
	as shown in Figure~\ref{figure:stacked:autoencoders:architecture}.},
i.e. two nested autoencoders
that we could notate as Autoencoder$^2$,
is illustrated in Figure~\ref{figure:stacked:autoencoders:architecture}.

The chain of encoders\index{Encoder} will increasingly compress data and extract higher-level\index{Higher!level} features\index{Feature}.
Stacked autoencoders, which are indeed deep networks\index{Deep!network},
are therefore used for feature extraction\index{Feature!extraction}
(an example will be introduced in Section~\ref{section:system:brentan:unit:selection}).
They are also useful for music generation, as we will see in Section~\ref{section:strategy:decoder:feedforward}.
This is because the {\em innermost hidden layer\index{Innermost hidden layer}}, sometimes named the {\em bottleneck hidden layer\index{Bottleneck hidden layer}},
provides a compact and high-level encoding\index{Encoding} (embedding\index{Embedding}) as a seed\index{Seed} for generation
(by the chain of decoders\index{Decoder}). 


\section{Restricted Boltzmann Machine (RBM)}
\label{section:architecture:rbm}

\begin{figure}
\includegraphics[scale=0.2]{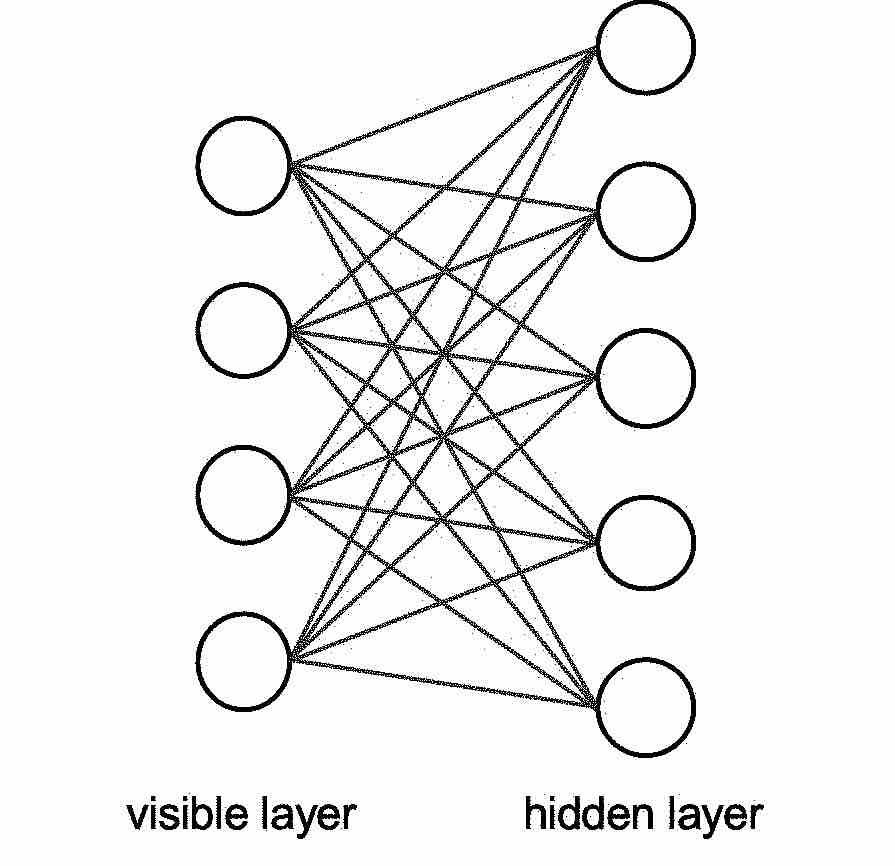}
\caption{Restricted Boltzmann machine (RBM) architecture}
\label{figure:rbm:architecture}
\end{figure}

A {\em restricted Boltzmann machine\index{Restricted Boltzmann machine}} (RBM\index{RBM}) \cite{hinton:rbm:science:2006}
is a {\em generative\index{Generative!model}} {\em stochastic}
artificial neural network that can learn a {\em probability distribution\index{Probability!distribution}} over its set of inputs.
Its name comes from the fact that it is a restricted (constrained) form\footnote{Which actually makes RBM practical,
	as opposed to the general form, which besides its interest suffers from a learning scalability limitation.}
of a (general) {\em Boltzmann machine\index{Boltzmann machine}} \cite{hinton:boltzmann:machines:pdp:1986},
named after the {\em Boltzmann distribution} in statistical mechanics, which is used in its sampling\index{Sampling} function.
The architectural restrictions of an RBM (see Figure~\ref{figure:rbm:architecture}) are that

\begin{itemize}

\item it is organized in {\em layers}, just as for a feedforward network or an autoencoder,
and more precisely two layers:

\begin{itemize}

\item the {\em visible} layer (analog to both the input layer and the output layer of an autoencoder); and

\item the {\em hidden} layer (analog to the hidden layer of an autoencoder);

\end{itemize}

\item as for a standard neural network, there cannot be connections between nodes within the same layer.

\end{itemize}

An RBM bears some similarity in spirit and objective to an autoencoder\index{Autoencoder}.
However, there are some important differences:

\begin{itemize}

\item an RBM has {\em no ouput} -- the input also acts as the output;

\item an RBM is {\em stochastic\index{Stochastic}}
	(and therefore {\em not deterministic\index{Deterministic}}, as opposed to a feedforward network or an autoencoder);

\item an RBM is trained in an {\em unsupervised learning\index{Unsupervised learning}} manner, with a specific algorithm
	(named {\em contrastive divergence\index{Contrastive divergence}}, see Section~\ref{section:architecture:rbm:training}),
	whereas an autoencoder is trained using a standard supervised learning\index{Supervised learning} method,
	with the same data as input and output; and

\item the values manipulated are {\em booleans\index{Boolean}}\footnote{Although
		there are extensions with multinoulli\index{Multinoulli} (categorical\index{Categorical}) or continuous values, see Section~\ref{section:architecture:rbm:variables:types}.}.
	
\end{itemize}
	
RBMs became popular after Hinton designed a specific fast learning algorithm for them,
named {\em contrastive divergence\index{Contrastive divergence}} \cite{hinton:contrastive:divergence:neural:computation:2002},
and used them for {\em pre-training\index{Pre-training}} deep neural networks \cite{erhan:pre:training:jmlr:2010}
(see Section~\ref{section:architectures:history}).

An RBM is an architecture dedicated to learning distributions.
Moreover, it can learn efficiently from only a few examples.
For musical applications, this is interesting for learning (and generating) chords,
as the combinatorial nature of possible notes forming a chord is large and the number of examples is usually small.
We will see an example of such an application in Section~\ref{section:experiment:rbm}.

\subsection{Training}
\label{section:architecture:rbm:training}

Training an RBM
has some similarity to training an autoencoder,
with the practical difference that,
because there is no decoder part,
the RBM will alternate between two steps\index{Step}:

\begin{itemize}

\item the {\em feedforward step} -- to encode the input (visible layer) into the hidden layer,
by making predictions about hidden layer node activations; and

\item the {\em backward step} -- to decode/reconstruct the input (visible layer),
by making predictions about visible layer node activations.

\end{itemize}



We will not detail here the learning technique behind RBMs,
see, for example, \cite[Section~20.2]{goodfellow:deep:learning:book:2016}.
Note that the reconstruction process is an example of {\em generative learning}
(and not {\em discriminative learning},
as for training autoencoders which is based on regression)\footnote{See, for example, a nice introduction
	to generative learning (and the difference with discriminative learning)
	in \cite{ng:course:notes:generative:learning:2016}.}.


%
%
%
%

\subsection{Sampling}
\label{section:architecture:rbm:sampling}

After the training phase has been completed, in the {\em generation} phase,
a {\em sample\index{Sample}} can be drawn from the model by randomly initializing visible layer vector $v$
(following a standard uniform distribution)
and running {\em sampling}\footnote{More precisely {\em Gibbs sampling\index{Gibbs sampling}} {\em (GS\index{GS})},
	see \cite{lam:mcmc:tutorial:web}.
	Sampling will be introduced in Section~\ref{section:sampling:basics}.}
until convergence.
To this end,
hidden nodes and visible nodes are alternately updated (as during the training phase).

In practice, convergence is reached when the energy stabilizes.
The {\em energy\index{Energy}} of a {\em configuration} (the pair of visible and hidden layers)
is expressed\footnote{For more details, see,
	for example, \cite[Section~16.2.4]{goodfellow:deep:learning:book:2016}.}
in the Equation~\ref{equation:rbm:energy},
where

\begin{equation}
E(\text{v}, \text{h}) = -a^\mathrm{T} \text{v} - b^\mathrm{T} \text{h} -\text{v}^\mathrm{T} W \text{h}
\label{equation:rbm:energy}
\end{equation}

\begin{itemize}

\item v and h, respectively, are column vectors representing the visible and the hidden layers;

\item $W$ is the matrix of weights associated with the connections between visible and hidden nodes;

\item $a$ and $b$, respectively, are column vectors representing the bias weights for visible and hidden nodes,
	with $a^\mathrm{T}$ and $b^\mathrm{T}$ being their respective transpositions into row vectors; and

\item $v^\mathrm{T}$ is the transposition of v into a row vector.

\end{itemize}


\subsection{Types of Variables}
\label{section:architecture:rbm:variables:types}

Note that there are actually three possibilities for the nature of RBM variables
(units, visible or hidden):

\begin{itemize}

\item {\em Boolean\index{Boolean}} or {\em Bernoulli\index{Bernoulli}} -- this is the case of standard RBMs\index{RBM},
	in which units\index{Unit} (visible\index{Visible unit} and hidden\index{Hidden!unit|see{Hidden layer unit}}) are Boolean,
	with a {\em Bernoulli distribution} (see \cite[Section~3.9.2]{goodfellow:deep:learning:book:2016});

\item {\em multinoulli\index{Multinoulli}} -- an extension with {\em multinoulli} units\footnote{As explained by
		Goodfellow {\em et al.} in \cite[Section~3.9.2]{goodfellow:deep:learning:book:2016}:
		````Multinoulli'' is a term that was recently coined by Gustavo Lacerdo and popularized by Murphy in \cite{murphy:ml:book:2012}.
		The multinoulli distribution is a special case of the multinomial distribution.
		A multinomial\index{Multinomial} distribution is the distribution over vectors in $\{0, . . . , n\}^k$
		representing how many times each of the $k$ categories is visited when $n$ samples are drawn from a multinoulli distribution.
		Many texts use the term ``multinomial'' to refer to multinoulli distributions without clarifying that they refer only to the $n = 1$ case.''},
	i.e. with more than two possible discrete values; and
%
%


\item {\em continuous\index{Continuous}} -- another extension with continuous \index{Continuous} units,
	taking arbitrary real\index{Real} values (usually within the $[0, 1]$ range\index{Range}).
	An example is the C-RBM\index{C-RBM} architecture analyzed in Section~\ref{section:experiment:c:rbm}.

\end{itemize}

\section{Recurrent Neural Network (RNN)}
\label{section:architecture:recurrent:network}

A {\em recurrent neural network\index{Recurrent!neural network}} (RNN\index{RNN}) is a feedforward neural network
extended with {\em recurrent connexions\index{Recurrent!connexion}} in order to learn series of items
(e.g., a melody as a sequence of notes).
The input of the RNN is an element x$_t$\footnote{This x$_t$ notation\index{Notation convention}
	-- or sometimes s$_t$ to stress the fact that it is a sequence --
	is very common but unfortunately introduces possible confusion with the notation of x$_i$ as the $i$th input variable.
	The context -- recurrent versus nonrecurrent network -- usually helps to discriminate,
	as well as the use of the letter $t$ (for time) as the index.
	An example of an exception is the RNN-RBM system analyzed in Section~\ref{section:experiment:rnn:rbm},
	which uses the x$^{(t)}$ notation.}
of the sequence,
where $t$ represents the {\em index\index{Index}} or the {\em time\index{Time}},
and the expected output is next element x$_{t+1}$.
In other words the RNN will be trained to predict the next element of a sequence.

In order to do so, the output of the hidden layer\index{Hidden!layer}
{\em reenters} itself
as an additional input (with a specific corresponding weight matrix).
This way, the RNN can learn, not only based on the {\em current} item but also on its {\em previous} own state,
and thus, recursively, on the whole of the previous sequence.
Therefore, an RNN can learn sequences, notably {\em temporal sequences\index{Temporal!sequence}},
as in the case of musical content.

%

\begin{figure}
\includegraphics[scale=0.88]{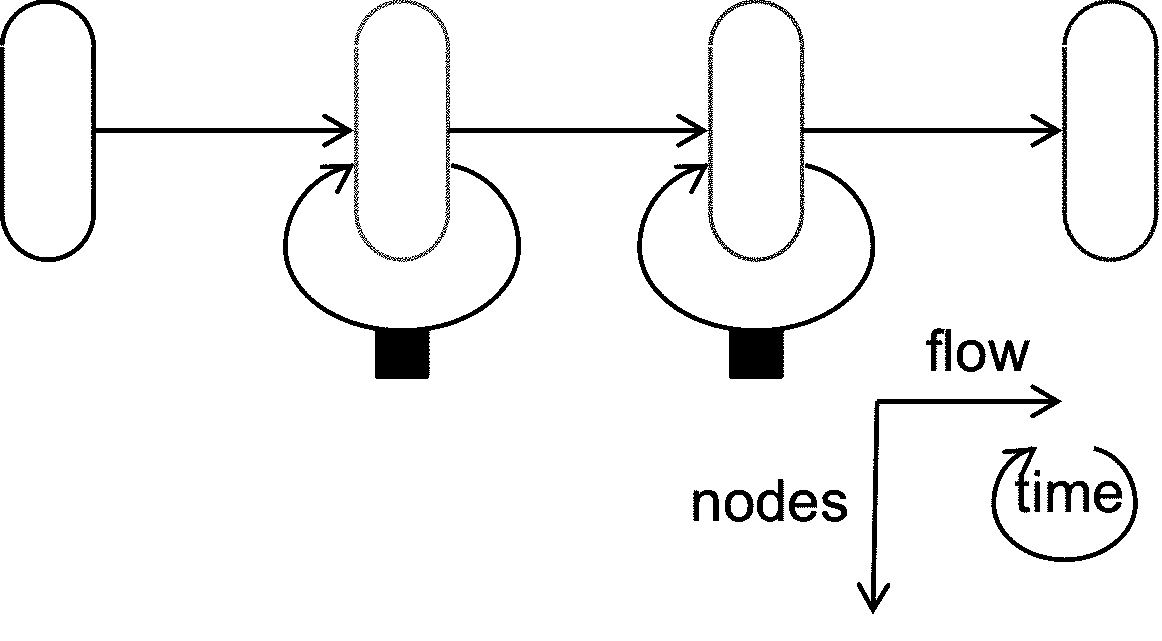}
\caption{Recurrent neural network (folded)}
\label{figure:recurrent:network:folded:abstract}
\end{figure}

\begin{figure}
\includegraphics[scale=0.9]{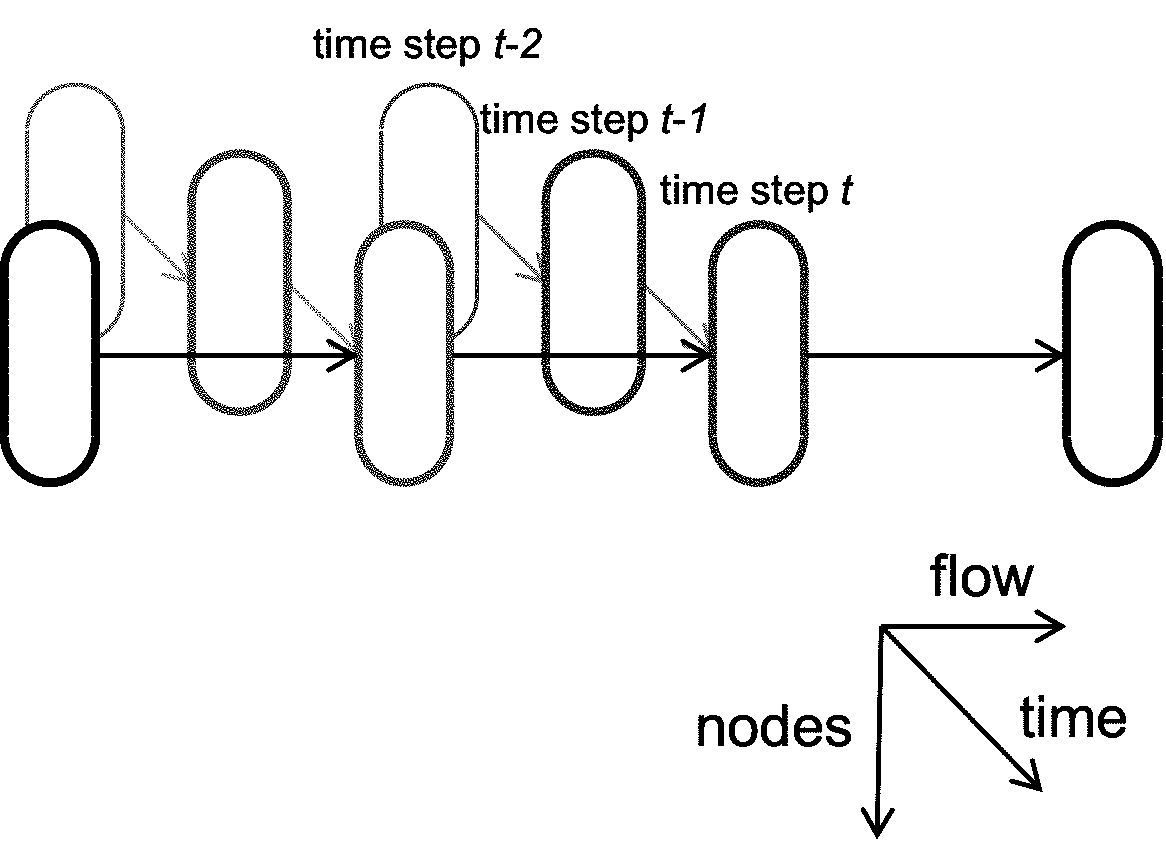}
\caption{Recurrent neural network (unfolded)}
\label{figure:recurrent:network:unfolded:abstract}
\end{figure}

\begin{figure}
\includegraphics[scale=1.1]{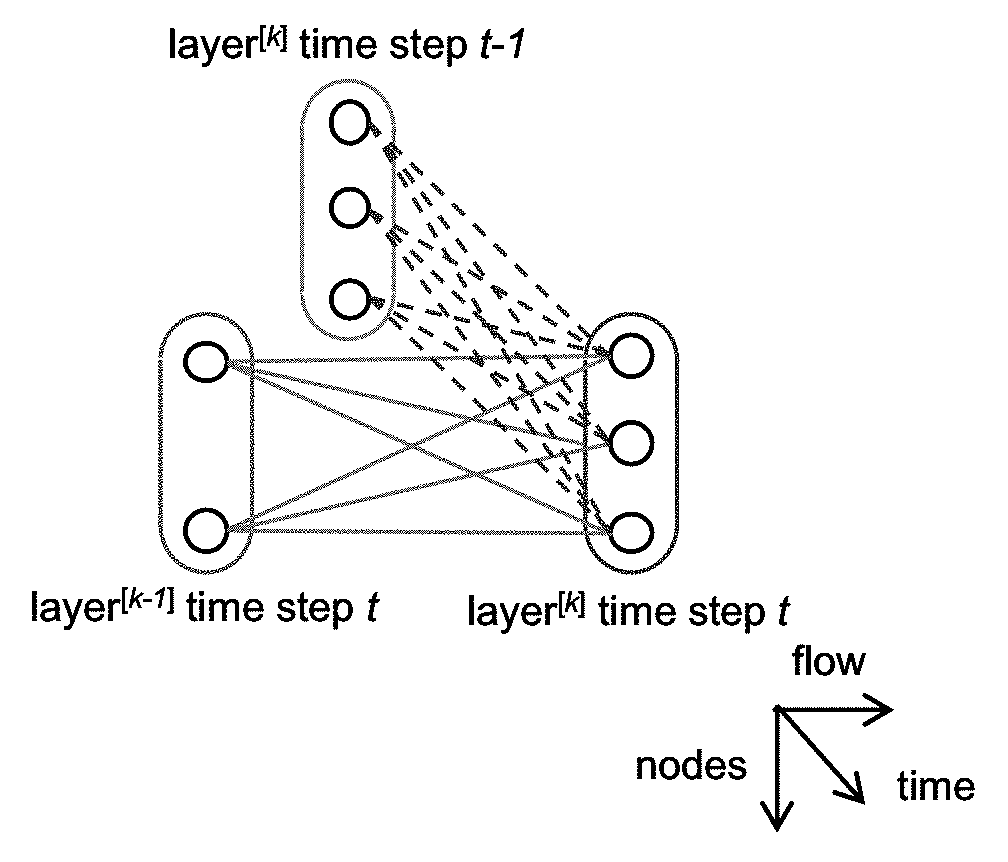}
\caption{Standard connexions versus recurrent connexions (unfolded)}
\label{figure:unfolded:recurrent:layer:architecture}
\end{figure}

An example of RNN (with two hidden layers) is shown in Figure~\ref{figure:recurrent:network:folded:abstract}.
Recurrent connexions\index{Recurrent!connexion} are signaled with a solid square,
in order to distinguish them from standard connexions\footnote{Actually, there are some variations of this basic architecture,
	depending on the exact nature and location of the recurrent connexions.
	The most standard case is a recurrent connexion for each hidden unit, as shown in Figure~\ref{figure:recurrent:network:folded:abstract}.
	But there are some other cases, see for example in \cite[Section~10.2]{goodfellow:deep:learning:book:2016}.
	An example of a music generation architecture with recurrent connexions from the output to a special context input will be introduced
	in Section~\ref{section:experiment:todd:sequential}.}.
The {unfolded\index{Unfolded}} version of the visual representation is in Figure~\ref{figure:recurrent:network:unfolded:abstract},
with a new diagonal axis representing the time dimension,
in order to illustrate the previous step value of each layer (in thinner and lighter color).
Note that,
as for standard connexions (shown in yellow solid lines),
recurrent connexions (shown in purple dashed lines)
fully connect (with a specific weight matrix)
all nodes corresponding to the previous step nodes to the nodes corresponding to the current step,
as illustrated in Figure~\ref{figure:unfolded:recurrent:layer:architecture}.

An RNN can learn a probability distribution\index{Probability!distribution} over a sequence by being trained
to predict the next element at time step $t$ in a sequence
as being the conditional probability\index{Conditional!probability} distribution $P(\text{s}_t | \text{s}_{t-1},... , \text{s}_1)$,
also notated as $P(\text{s}_t | \text{s}_{<t})$,
that is the probability distribution $P(\text{s}_t)$ given all previous elements generated s$_1$, s$_2$, \ldots~, s$_{t-1}$.
In summary, recurrent networks (RNNs) are good at learning sequences
and therefore are routinely used for natural text processing and for music generation.

\subsection{Visual Representation}
\label{section:architecture:recurrent:network:notation}

\begin{figure}
\includegraphics[scale=0.93]{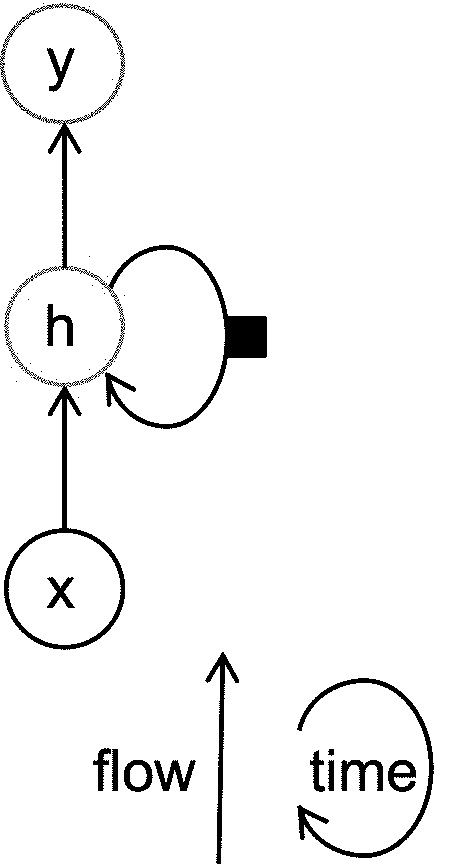}
\caption{Recurrent neural network (folded)}
\label{figure:recurrent:network:folded:abstract:alternative}
\end{figure}

\begin{figure}
\includegraphics[scale=0.22]{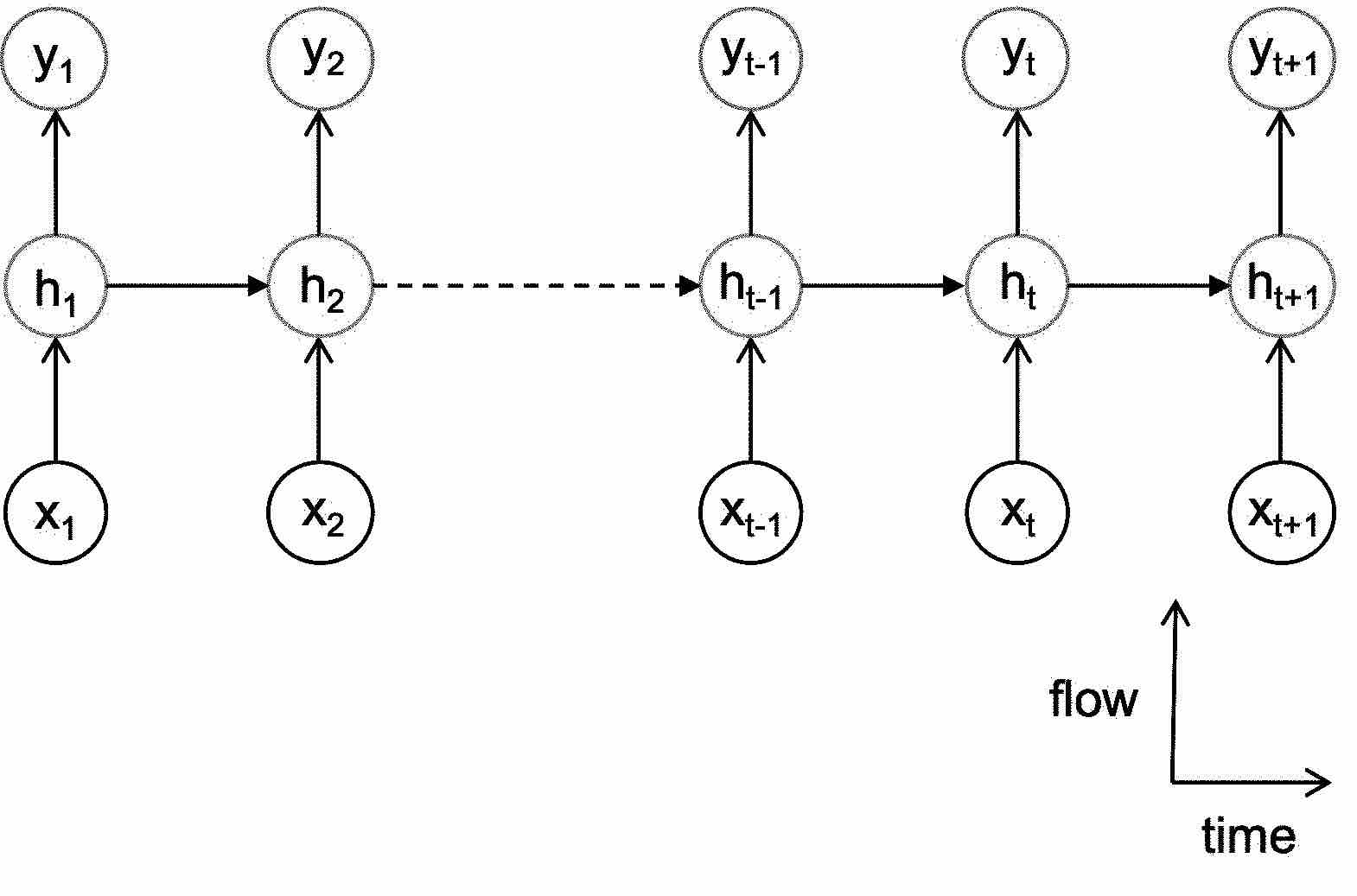}
\caption{Recurrent neural network (unfolded)}
\label{figure:recurrent:network:unfolded:abstract:alternative}
\end{figure}

A more frequent visual representation for an RNN is actually showing the flow upwards and time rightwards,
see the folded version (of an RNN with only one hidden layer) in Figure~\ref{figure:recurrent:network:folded:abstract:alternative}
and the unfolded version in Figure~\ref{figure:recurrent:network:unfolded:abstract:alternative},
with h$_t$ being the value of the hidden layer at step\index{Step} $t$,
and x$_t$ and y$_t$ being the values of the input and output at step $t$.

\subsection{Training}
\label{section:architecture:rnn:training}

\begin{figure}
\includegraphics[scale=0.22]{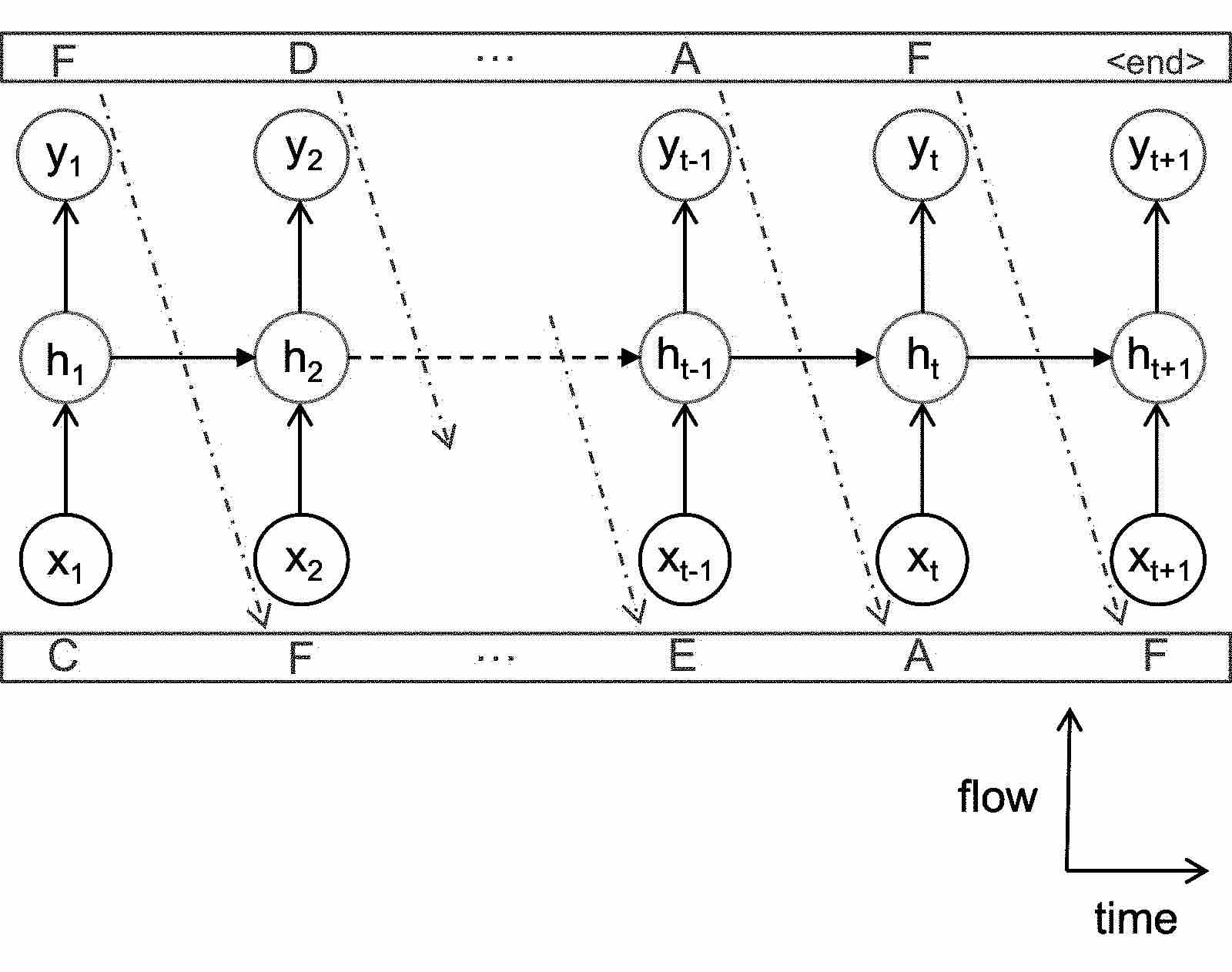}
\caption{Training a recurrent neural network}
\label{figure:recurrent:network:training}
\end{figure}

A recurrent network is not trained in exactly the same manner as a feedforward network.
The idea is to present an example element of a sequence (e.g., a note within a melody) as the input x$_t$
and the next element of the sequence (the next note) x$_{t+1}$ as the output y$_t$.
This will train the recurrent network to predict the next element of the sequence.
In practice, an RNN is rarely trained element by element
but with a sequence as an input and the same sequence shifted left by one step/item as the output.
See an example in Figure~\ref{figure:recurrent:network:training}\footnote{The end of the sequence
	is marked by a special symbol.}.
Therefore, the recurrent network will learn to predict\footnote{Pictured
	as dashed arrows.}
the next element for all successive elements of the sequence.

The backpropagation\index{Backpropagation} algorithm to compute gradients for feedforward networks, introduced in Section~\ref{architecture:network:training},
has been extended into a {\em backpropagation through time\index{Backpropagation!through time}} (BPTT\index{BPTT}) algorithm
for recurrent networks.
The intuition is in unfolding the RNN through time and considering an ordered sequence of input-output pairs,
but with every unfolded copy of the network sharing the same parameters,
and then applying the standard backpropagation algorithm.
More details may be found, for example, in \cite[Section~10.2.2]{goodfellow:deep:learning:book:2016}.

Note that,
a RNN has usually an output layer identical to its input layer\footnote{RNNs are actually more general
	and there are actually some rare cases of an RNN with an arbitrary output different from the input
	(as for a feedforward network).
	An example is a RNN-based architecture to generate a chord-based accompaniment,
	to be analyzed in Section~\ref{section:experiment:blstm:chord}.
	As Karpathy puts it in \cite{karpathy:unreasonable:effectiveness:rnn:blog2015}:
	``Depending on your background you might be wondering:
	What makes Recurrent Networks so special?
	A glaring limitation of Vanilla Neural Networks (and also Convolutional Networks) is that their API is too constrained:
	they accept a fixed-sized vector as input (e.g. an image) and produce a fixed-sized vector as output (e.g. probabilities of different classes).
	Not only that: These models perform this mapping using a fixed amount of computational steps (e.g. the number of layers in the model).
	The core reason that recurrent nets are more exciting is that they allow us to operate over sequences of vectors:
	Sequences in the input, the output, or in the most general case both.''},
as a recurrent network predicts the next item,
which will be used iteratively as the next input in a recursive\index{Recursion} way in order to produce a sequence.

Note also that training a recurrent network is usually considered as a case of supervised learning\index{Supervised learning}
as, for each item, the next item is presented as the expected prediction,
although it is not an additional label\index{Label} information
(effective value or class to be predicted)
but only the recurrent information about the next item ({\em intrinsically} present within a sequence).







\subsection{Long Short-Term Memory (LSTM)}
\label{section:architecture:lstm}

Recurrent networks suffered from a training problem
caused by the difficulty of estimating gradients
because in backpropagation through time\index{Backpropagation!through time}
recurrence brings repetitive multiplications,
and could thus lead to over {\em amplify} or {\em minimize} effects\footnote{This has been coined as
	the {\em vanishing\index{Vanishing gradient problem} or exploding gradient problem\index{Exploding gradient problem}}
	and also as the {\em challenge of long-term dependencies\index{Long!-term dependency}}
	(see, for example, \cite[Section~10.7]{goodfellow:deep:learning:book:2016}).}.
This problem has been addressed and resolved by the {\em long short-term memory\index{Long!short-term memory}} (LSTM\index{LSTM}) architecture,
proposed by Hochreiter and Schmidhuber in 1997
\cite{hochreiter:lstm:1997}.
As the solution has been quite effective, LSTM has become the de facto standard for recurrent networks\footnote{Although,
	there are a few subsequent but similar proposals, such as {\em gated recurrent units\index{Gated recurrent unit}} (GRUs\index{GRU}).
	See a comparative analysis of LSTM and GRU in \cite{chung:evaluation:gru:arxiv:2014}.}.

The idea behind LSTM is to secure information in memory {\em cells\index{Cell|see{LSTM cell}}},
within a {\em block}\footnote{Cells within the same block\index{Block|see{LSTM block}}
	{\em share} input, output and forget gates.
	Therefore, although each cell might hold a different value in its memory\index{Memory},
	all cell memories within a block are read, written or erased {\em all at once} \cite{hochreiter:lstm:1997}.},
protected from the standard data flow of the recurrent network.
Decisions about {\em writing} to, {\em reading} from and {\em forgetting} ({\em erasing}) the values of cells within a block 
are performed by the opening or closing of {\em gates}
and are expressed at a distinct control level ({\em meta-level\index{Meta-level}}), while being learnt during the training process.
Therefore, each gate is modulated by a {\em weight\index{Weight}} parameter, and thus is
suitable for backpropagation\index{Backpropagation} and standard training process.
In other words,
each LSTM block learns how to maintain its memory as a function of its input in order to minimize loss.

\begin{figure}
\includegraphics[scale=0.7]{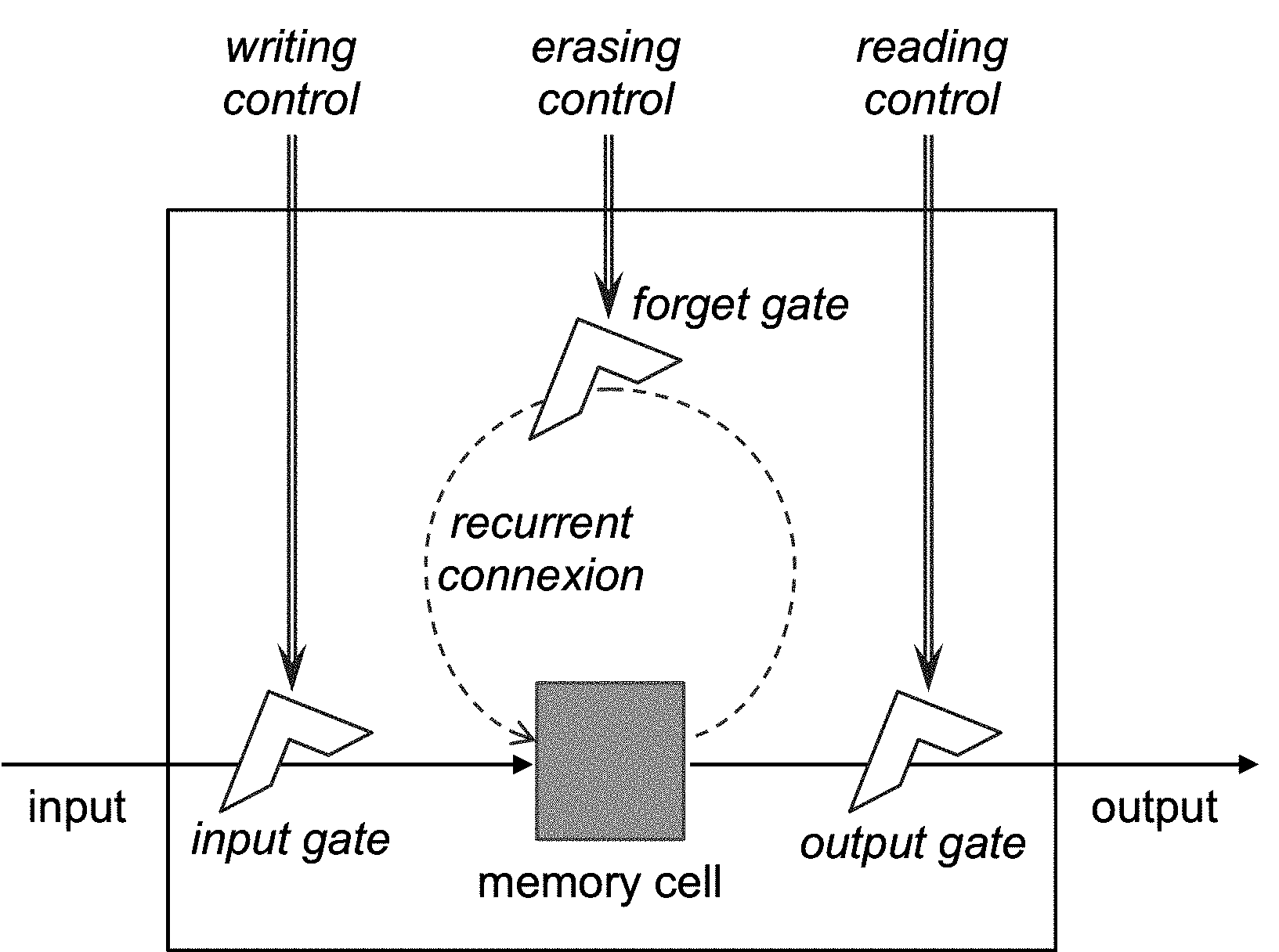}
\caption{LSTM architecture (conceptual)}
\label{figure:lstm:cell:conceptual}
\end{figure}

See a conceptual view of an LSTM cell in Figure~\ref{figure:lstm:cell:conceptual}.
We will not further detail here the inner mechanism of an LSTM cell (and block)
because we may consider it here as a {\em black box}
(please refer to, for example, the original article \cite{hochreiter:lstm:1997}).


Note that
a more general model of memory with access customized through training has recently been proposed:
neural Turing machines\index{Neural!Turing machine} (NTM\index{NTM}) \cite{graves:neural:turing:machines:2014}.
In this model, memory is global and  has {\em read} and {\em write operations} with differentiable controls,
and thus is subject to learning through backpropagation.
The memory to be accessed, specified by {\em location} or by {\em content},
is controlled via an {\em attention}\index{Attention mechanism}
mechanism (introduced in next section).

\subsection{Attention Mechanism}
\label{section:architecture:attention:mechanism}

The motivation for an {\em attention mechanism\index{Attention mechanism}}
has been inspired by the human visual system ability to efficiently track and recognize objects
by focusing its {\em attention}.
It has therefore been first introduced into neural network architectures for image recognition\index{Image!recognition},
as for instance for object tracking \cite{denil:attend:image:tracking:2011}.
It has then been adapted to recurrent architectures for natural language processing\index{Natural language processing}
(and more specifically for translation\index{Translation} tasks)
and has showed significant improvement for the management of long-term dependencies\index{Long!-term dependency}.

The idea of an attention mechanism is to focus at each time step\index{Time!step} on some specific elements of the input sequence.
This is modeled by weighted connexions onto the sequence elements (or onto the sequence of hidden units).
Therefore it is differentiable and subject to backpropagation-based\index{Backpropagation} learning at a meta-level,
as with LSTM\index{LSTM} gate control described in previous section.
For more details, see, for example, \cite[Section~12.4.5.1]{goodfellow:deep:learning:book:2016}.


Interestingly, a novel architecture for
translation\index{Translation} of sequences,
named {\em Transformer\index{Transformer}}
and which is {\em solely} based on an attention mechanism\index{Attention mechanism}\footnote{The architecture
	introduces {\em multi-head attention}
	which allows the model to jointly attend to information from different representation subspaces at different positions
	\cite{vaswani:attention:transformer:arxiv:2017}.},
has recently being proposed and shows promising results \cite{vaswani:attention:transformer:arxiv:2017}.
Its very recent application to music generation will be shortly discussed in Section~\ref{section:discussion:convolution:vs:recurrent}.

\section{Convolutional Architectural Pattern}
\label{section:architecture:convolution}

{\em Convolutional neural network\index{Convolutional!neural network}} (CNN\index{CNN} or ConvNet\index{ConvNet})
architectures for deep learning
have become common place for image\index{Image} applications.
The concept was originally inspired by both a model of human vision
and the {\em convolution\index{Convolution}} mathematical operator\footnote{In mathematics,
	a convolution is a mathematical operation on two functions sharing the same domain (usually noted $f*g$)
	that produces a third function which is the integral (or the sum in the discrete case -- the case of images made of pixels)
	of the pointwise multiplication of the two functions varying within the domain in an opposing way.
	In the case of a continuous domain $[low~~high]$: $$(f * g)(\text{x}) = \int_{low}^{high} f(\text{x} - \text{t})g(\text{t})d\text{t}$$
	In the discrete case: $$(f * g)(\text{n}) = \sum_{m=low}^{high} f(\text{n} - \text{m})g(\text{m})$$}.
It has been carefully adapted to neural networks and improved by LeCun,
at first for handwritten character and object recognition \cite{le:cun:convolutional:handbook:1998}.
This resulted in efficient and accurate architectures for pattern recognition,
exploiting the spatial local {\em correlation\index{Correlation}} present in natural images.

\subsection{Principles}
\label{section:architecture:convolution:principles}

The basic idea\footnote{Inspired
	by the nice intuitive explanation provided by Karn in \cite{karn:convolutional:web:2016}.
	For more technical details see, for example, \cite{li:course:notes:convolutional:2016} or \cite[Chapter~9]{goodfellow:deep:learning:book:2016}.}
is to

\begin{itemize}

\item
{\em slide} a matrix (named a {\em filter\index{Filter}}, a {\em kernel\index{Kernel}} or a {\em feature detector\index{Feature!detector}})
through the entire image (seen as the input matrix); and

\item
for each mapping position:

\begin{itemize}

\item 
compute the dot product of the filter with each mapped portion of the image; and

\item then sum up all elements of the resulting matrix;

\end{itemize}

\item
resulting in a new matrix (composed of the different sums for each sliding/mapping position),
named {\em convolved feature\index{Convolved feature}},
or also {\em feature map\index{Feature!map}}.

\end{itemize}

The size of the feature map is controlled by three hyperparameters\index{Hyperparameter}:

\begin{itemize}

\item {\em depth\index{Depth}} -- the number of filters used;

\item {\em stride\index{Stride}} -- the number of pixels by which we slide the filter matrix over the input matrix; and

\item {\em zero-padding\index{Zero-padding}} -- the padding of the input matrix with zeros around its border\footnote{Zero-padding allows
	mapping of the filter
	up to the borders of the image.
	It also avoids shrinking the representation, which otherwise would be problematic when using multiple consecutive convolutional layers
	\cite[Section~9.5]{goodfellow:deep:learning:book:2016}.}.

\end{itemize}

\begin{figure}
\includegraphics[scale=0.7]{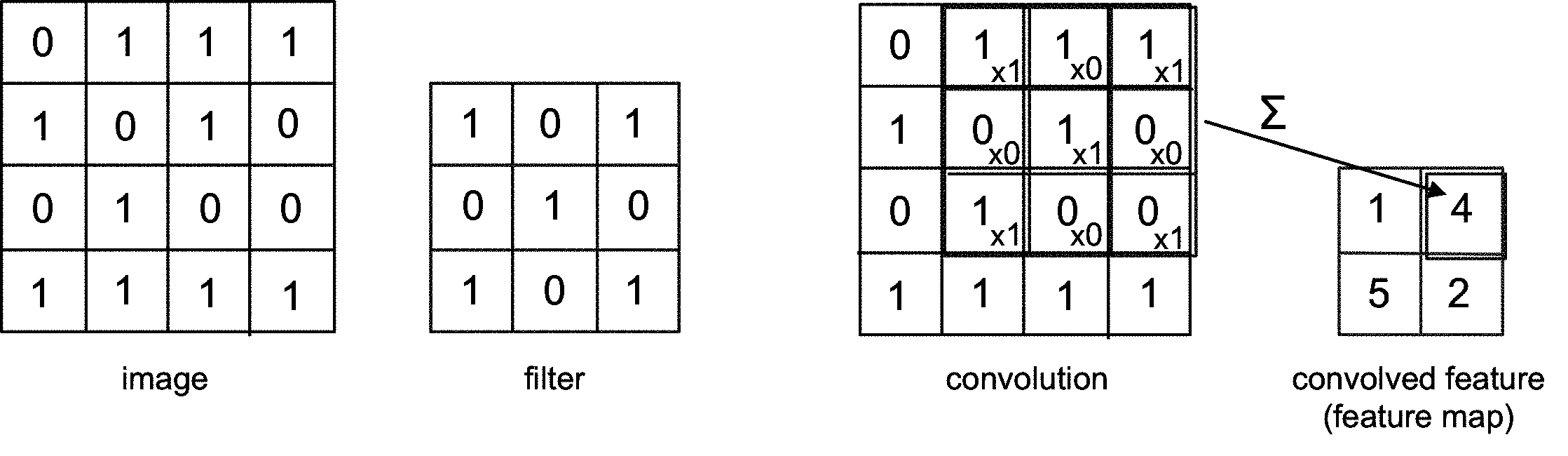}
\caption{Convolution, filter and feature map.
Inspired by Karn's data science blog post
\cite{karn:convolutional:web:2016}}
\label{figure:convolution:convolution}
\end{figure}

An example is illustrated in Figure~\ref{figure:convolution:convolution}
with some simple settings: depth
= 1, stride = 1 and no zero-padding.
Various filter matrixes can be used with different objectives, such as detection of different features (e.g., edges or curves)
or other operations such as sharpening or blurring.

The parameter sharing\index{Parameter!sharing} used by the convolution (because of the shared fixed filter) brings the important property of
{\em equivariance\index{Equivariance}} to translation,
i.e. a motif in an image can be detected independently of its location \cite[Chapter~9]{goodfellow:deep:learning:book:2016}.

\subsection{Stages}
\label{section:architecture:convolution:stages}

A convolution usually consists of three successive {\em stages}:

\begin{itemize}

\item a {\em convolution stage}, as described in Section~\ref{section:architecture:convolution:principles};

\item a {\em nonlinear rectification stage}, sometimes named detector stage,
which applies a nonlinear operation, usually ReLU\index{ReLU}; and

\item a {\em pooling stage}, also named {\em subsampling\index{Subsampling}}, to reduce the dimensionality.

\end{itemize}

\subsection{Pooling}
\label{section:architecture:convolution:pooling}

\begin{figure}
\includegraphics[scale=0.65]{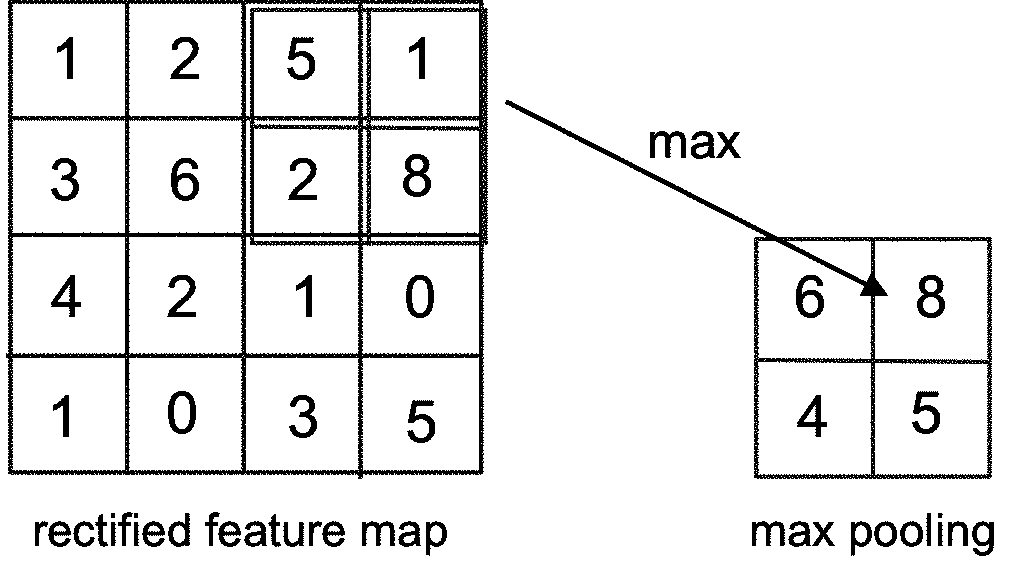}
\caption{Pooling.
Inspired by Karn's data science blog post
\cite{karn:convolutional:web:2016}}
\label{figure:convolution:pooling}
\end{figure}

The motivation for {\em pooling\index{Pooling}} is to reduce the dimensionality of each feature map
while retaining significant information.
Operations used for pooling are, for example, max, average and sum.
In addition to reducing the dimensionality of data, pooling brings the important property of
the {\em invariance\index{Invariance}} to small transformations, distortions and translations in the input image.
This provides an overall robustness\index{Robustness} to the processing \cite{karn:convolutional:web:2016}.
Like convolution, pooling has hyperparameters to control the process.
A simple example of max pooling with stride = 2 is illustrated in Figure~\ref{figure:convolution:pooling}.

\subsection{Multilayer Convolutional Architecture}
\label{section:architecture:convolution:multilayer}

\begin{figure}
\includegraphics[width=\textwidth]{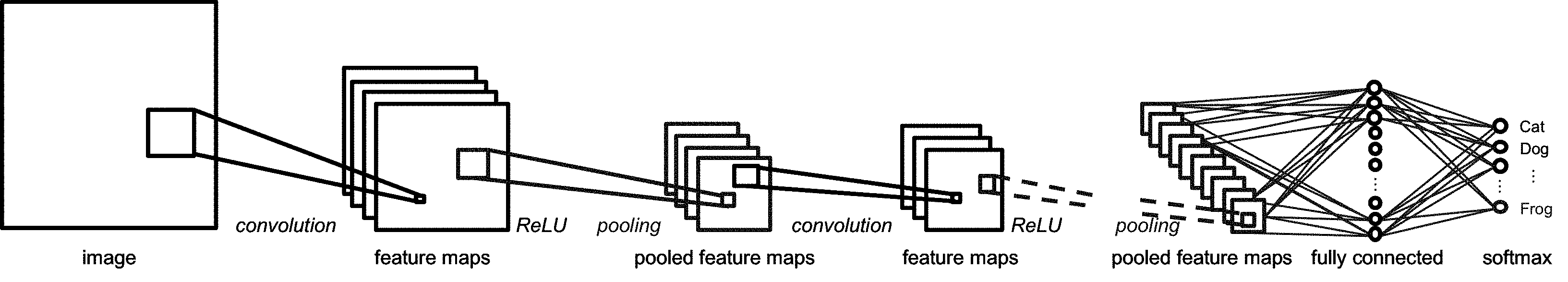}
\caption{Convolutional deep neural network architecture.
Inspired by Karn's data science blog post
\cite{karn:convolutional:web:2016}}
\label{figure:convolution:full}
\end{figure}

A typical example of a convolutional architecture
with successive layers -- each one including the three stages
of convolution, nonlinearity and pooling --
is illustrated in Figure~\ref{figure:convolution:full}.
The final layer is a fully connected layer, like in standard feedforward networks,
and typically ends up in a softmax
in order to classify image types.

Note that a convolution is an {\em architectural pattern\index{Architectural!pattern}},
as it may be applied internally to almost any architecture listed.

\subsection{Convolution over Time}
\label{section:architecture:convolution:time}

For musical applications, it could be interesting to apply convolutions to the {\em time dimension}\footnote{This approach
	is actually the basis for {\em time-delay neural networks\index{Time!-delay neural network}}
	\cite{lang:time:delay:network:nn:1990}.},
in order to model temporally invariant motives.
Therefore, the convolution operation will share parameters across time \cite[page~374]{goodfellow:deep:learning:book:2016},
like for RNNs\index{RNN}\footnote{Indeed, RNNs are invariant in time,
	as remarked in \cite{johnson:web:hexahedria:composing:music:recurrent:neural:2015}.}.
However, the sharing of parameters is {\em shallow}, as it applies only to a small number of temporal neighboring members of the input,
in contrast to RNNs that share parameters in a {\em deep} way, for {\em all} time steps.
RNNs are indeed much more frequent than convolutional networks\index{Convolutional!network} for musical applications.

That said, we have noticed the recent occurrence of some convolutional architectures as an alternative to RNN architectures,
following the pioneering WaveNet\index{WaveNet} architecture for audio\index{Audio} \cite{oord:wavenet:arxiv:2016},
described in Section~\ref{section:systems:wavenet}.
WaveNet presents a stack of causal convolutional\index{Convolutional} layers,
somewhat analogous to recurrent\index{Recurrent} layers.
Another example is the C-RBM\index{C-RBM} architecture,
described in Section~\ref{section:experiment:c:rbm}.

If we now consider the {\em pitch\index{Pitch} dimension}, in most cases pitch intervals are not considered
invariants,
and thus convolutions should not {\em a priori} apply to the pitch dimension\footnote{An exception is Johnson's architecture
	\cite{johnson:web:hexahedria:composing:music:recurrent:neural:2015},
	analyzed in Section~\ref{section:experiment:hexahedria},
	which explicitly looks for invariance in pitch (although this seems to be a rare choice)
	and accordingly uses an RNN\index{RNN} over the pitch dimension.}.

This issue of convolution\index{Convolution} versus recurrence (recurrent networks\index{Recurrent!network})
for musical applications will be further discussed in Section~\ref{section:discussion:convolution:vs:recurrent}.

\section{Conditioning Architectural Pattern}
\label{section:architecture:conditioning}

The idea of a {\em conditioning\index{Conditioning}\index{Conditioning!architecture}}
(sometimes also named {\em conditional\index{Conditional!architecture|see{Conditioning architecture}}}) architecture
is to parametrize the architecture based on some extra {\em conditioning} information,
which could be arbitrary, e.g., a class label\index{Label} or data from other modalities\index{Modality}.
The objective is to have some control\index{Control} over the data generation\index{Generation} process.
Examples of conditioning information are

%
%
%
%

\begin{itemize}

\item a {\em bass line\index{Bass!line}} or a {\em beat\index{Beat} structure}
in the rhythm generation system to be described in Section~\ref{section:systems:makris:rhythm};

\item a {\em chord progression\index{Chord!progression}}
in the MidiNet\index{MidiNet} system to be described in Section~\ref{section:systems:midinet};


\item some {\em positional constraints on notes}
in the Anticipation-RNN\index{Anticipation-RNN} system to be described in Section~\ref{section:systems:anticipation:rnn}; and

\item a {\em musical genre\index{Musical!genre}} or an {\em instrument\index{Instrument}}
in the WaveNet\index{WaveNet} system to be described in Section~\ref{section:systems:wavenet}.


\end{itemize}

\begin{figure}
\includegraphics[scale=0.1]{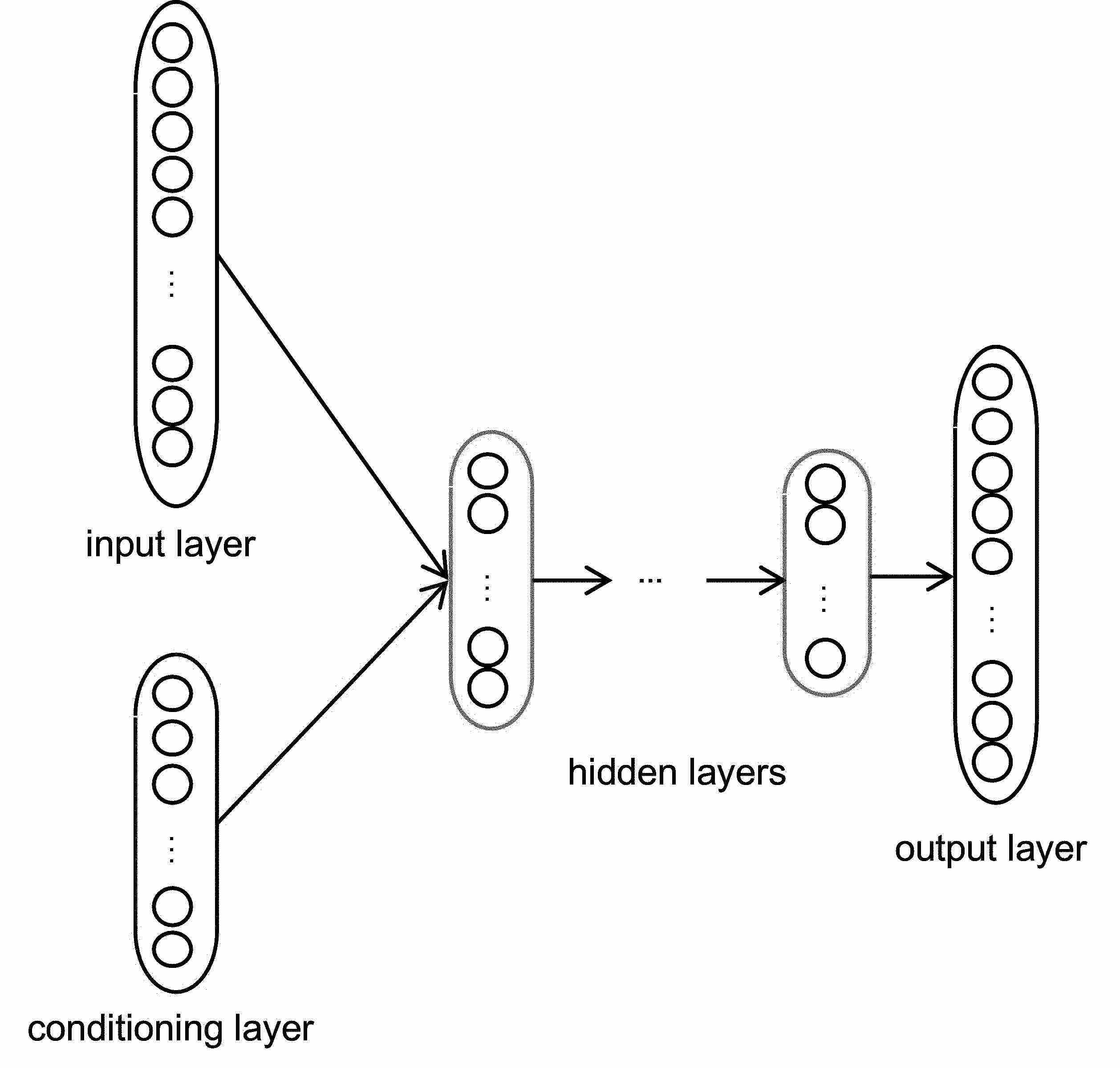}
\caption{Conditioning architecture}
\label{figure:conditioning:architecture}
\end{figure}

In practice, the conditioning information
is usually fed into the architecture as an additional and specific input layer\index{Input!layer},
shown in purple in Figure~\ref{figure:conditioning:architecture}.

The conditioning layer could be

\begin{itemize}

\item a simple input layer.
An example is a tag\index{Tag} specifying a musical genre or an instrument
in the WaveNet\index{WaveNet} system to be described in Section~\ref{section:systems:wavenet}; or

\item some output of some architecture, being

\begin{itemize}

\item the same architecture, as a way to condition the architecture on some history\footnote{This is close in spirit
	to a recurrent architecture (RNN).}.
An example is the MidiNet\index{MidiNet} system to be described in Section~\ref{section:systems:midinet},
in which history information from previous measure(s) is injected back into the architecture; or

\item another architecture.
An example is the DeepJ\index{DeepJ} system to be described in Section~\ref{section:systems:deepj},
in which two successive transformation layers of a style\index{Style} tag produce an embedding\index{Embedding}
used as the conditioning input\index{Conditioning!input}.

\end{itemize}

\end{itemize}



In the case of {\em conditioning} a time-invariant architecture
-- recurrent or convolutional over time --
there are two options

\begin{itemize}

\item {\em global conditioning} --
if the conditioning input\index{Conditioning!input} is shared for all time steps\index{Time!step}; and

\item {\em local conditioning} --
if the conditioning input is specific to each time step.

\end{itemize}

The WaveNet\index{WaveNet} architecture,
which is convolutional over time (see Section~\ref{section:architecture:convolution:time}),
offers the two options, as will be analyzed in Section~\ref{section:systems:wavenet}.

\section{Generative Adversarial Networks (GAN) Architectural Pattern}
\label{section:architecture:gan}

A significant conceptual and technical innovation was introduced in 2014 by Goodfellow {\em et al.} with
the concept of {\em generative adversarial networks\index{Generative!adversarial networks}} (GAN\index{GAN}) \cite{goodfellow:gan:arxiv:2014}.
The idea is to train simultaneously two neural networks\index{Neural!network}\footnote{In the original version,
	two feedforward networks\index{Feedforward!network|see{Feedforward neural network}} are used.
	But we will see that other networks may be used,
	e.g., recurrent networks\index{Recurrent!network} in the C-RNN-GAN\index{C-RNN-GAN} architecture
	(Section~\ref{section:systems:c:rnn:gan})
	and convolutional feedforward networks\index{Convolutional!network} in the MidiNet\index{MidiNet} architecture
	(Section~\ref{section:systems:midinet}).},
as illustrated in Figure~\ref{figure:gan:architecture}:

\begin{figure}
\includegraphics[scale=0.21]{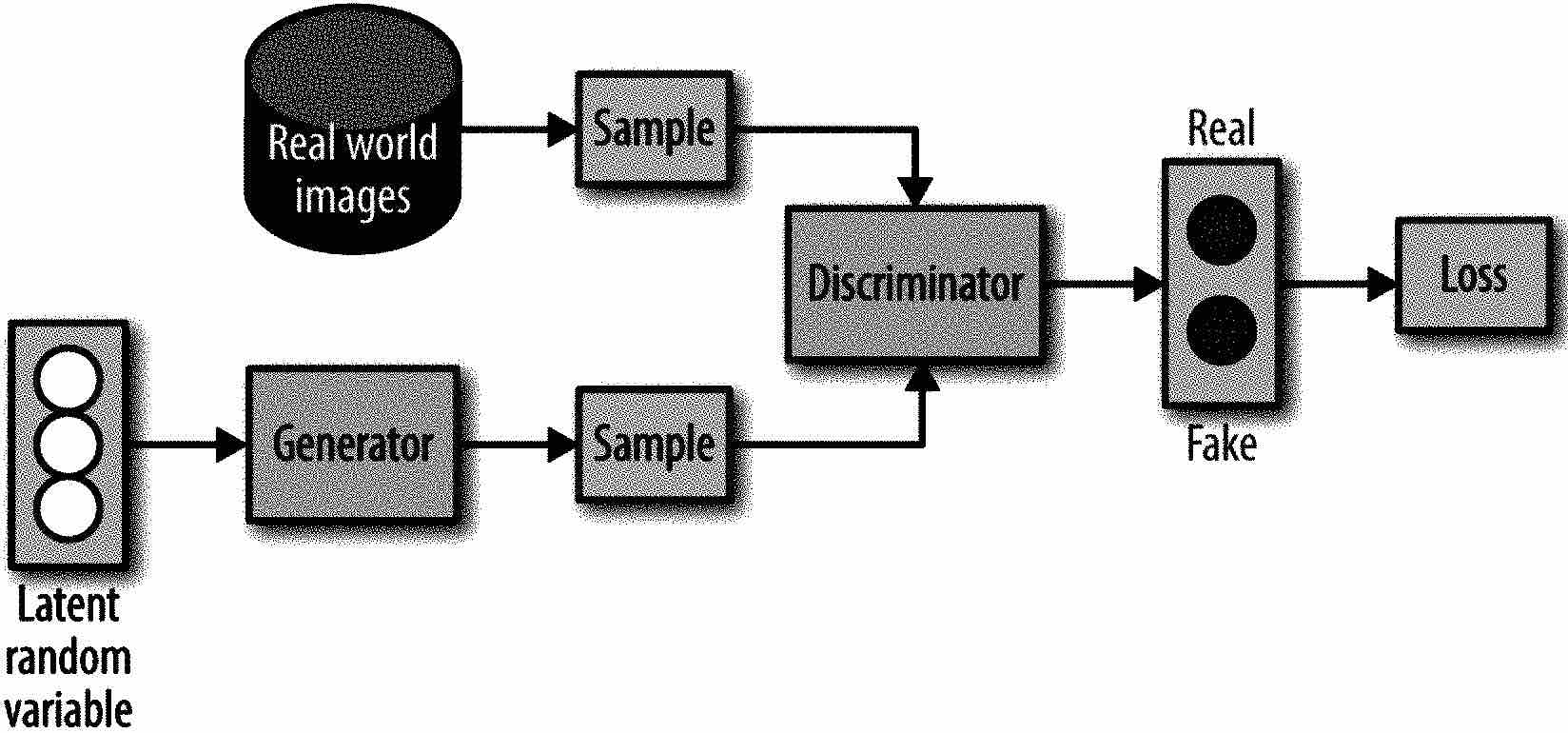}
\caption{Generative adversarial networks (GAN) architecture.
Reproduced from \cite{ramsundar:tensor:flow:deep:learning:2018} with permission of O'Reilly Media}
\label{figure:gan:architecture}
\end{figure}

\begin{itemize}

\item a {\em generative model} (or {\em generator\index{Generator}}) G,
whose objective is to transform a random noise vector into a synthetic (faked) {\em sample\index{Sample}},
which resembles real samples drawn from a distribution of real images; and

\item a {\em discriminative model} (or {\em discriminator\index{Discriminator}}) D,
which estimates\index{Estimation} the probability that a sample came from the real data rather than from the generator G\footnote{In some ways,
	a GAN represents an automated Turing test\index{Turing!test} setting,
	with the discriminator being the evaluator and the generator being the hidden actor.}.

\end{itemize}

This corresponds to a {\em minimax\index{Minimax}} two-player game,
with one unique (final) solution\footnote{It corresponds
	to the Nash equilibrium\index{Nash equilibrium} of the game.
	In game theory\index{Game theory}, the intuition of a Nash equilibrium
	is a solution where no player can benefit by changing strategies while the other players keep theirs unchanged,
	see, for example, \cite{osborne:course:game:theory:book:1994}.}:
G recovers the training data distribution and D outputs $1/2$ everywhere.
The generator is then able to produce user-appealing synthetic samples from noise vectors.
The discriminator may then be discarded.

The minimax relationship is defined in Equation~\ref{equation:minimax:gan}.

\begin{equation}
\underset{G}{min}~ \underset{D}{max}~ V(G, D) = \mathbb{E}_{\text{x} \sim P_{\text{Data}}}[\text{log}~D(\text{x})] + \mathbb{E}_{\text{z} \sim P_{\text{z}}(\text{z})}[\text{log}(1 - D(G(\text{z})))]
\label{equation:minimax:gan}
\end{equation}

%

\begin{itemize}

\item $D(\text{x})$ represents the probability
that x came from the real data
(i.e. the {\em correct} estimation by D); and

\item $\mathbb{E}_{\text{x} \sim p_{\text{Data}}}[\text{log}~D(\text{x})]$
is the expectation\footnote{The expectation has been introduced
	in Section~\ref{section:architecture:neural:network:entropy}.}
of $\text{log}~D(\text{x})$ with respect to x being drawn from the real data.

\end{itemize}

It is thus D's objective to estimate correctly {\em real data},
that is to maximize the $\mathbb{E}_{\text{x} \sim p_{\text{Data}}}[\text{log}~D(\text{x})]$ term.

\begin{itemize}

\item $D(G(\text{z}))$ represents the probability
that $G(\text{z})$ came from the real data
(i.e. the {\em uncorrect} estimation by D);

\item $1 - D(G(\text{z}))$ represents the probability
that $G(\text{z})$ did not come from the real data, i.e. that it was generated by G
(i.e. the {\em correct} estimation by D); and

\item $\mathbb{E}_{\text{z} \sim p_{\text{z}}(\text{z})}[\text{log}(1 - D(G(\text{z})))]$
is the expectation of $\text{log}(1 - D(G(\text{z})))$ with respect to $G(\text{z})$ being produced by G from z random noise.

\end{itemize}

It is thus also D's objective to estimate correctly {\em synthetic data},
that is to maximize the $\mathbb{E}_{\text{z} \sim p_{\text{z}}(\text{z})}[\text{log}(1 - D(G(\text{z})))]$ term.

In summary, it is D's objective to estimate correctly both {\em real data}
and {\em synthetic data}
and thus to maximize both
$\mathbb{E}_{\text{x} \sim p_{\text{Data}}}[\text{log}~D(\text{x})]$
and $\mathbb{E}_{\text{z} \sim p_{\text{z}}(\text{z})}[\text{log}(1 - D(G(\text{z})))]$ terms,
i.e. to maximize $V(G, D)$.
On the opposite side, G's objective is to minimize $V(G, D)$.
Actual training\index{Training} is organized with successive turns between the training of the generator and the training of the discriminator.

One of the initial motivations for GAN was for classification\index{Classification} tasks to prevent adversaries
from manipulating deep networks to force misclassification\index{Misclassification} of inputs
(this vulnerability is analyzed in detail in \cite{szegedy:intriguing:properties:arxiv:2014}).
However, from the perspective of content generation (which is our interest),
GAN improves the generation of samples\index{Sample}, which become hard to distinguish from the actual corpus examples.

To generate music, random noise is used as an input to the generator G,
whose goal is to transform random noises into the objective, e.g., melodies\footnote{In that respect, generation from a GAN
	has some similarity with generation by decoding hidden layer variables of a variational autoencoder
	(Section~\ref{section:architecture:vae}),
	as in both cases generation is done from latent variables.
	An important difference is that, by construction, a variational autoencoder is representative of the whole dataset that it has learnt,
	that is, for any example in the dataset,
	there is at least one setting of the latent variables which causes the model to generate something very similar to that example
	\cite{doersch:tutorial:variational:autoencoder:2016}.
	A GAN does not offer such guarantee
	and does not offer a smooth generation control interface over the latent space
	(by, e.g., interpolation or attribute arithmetics, see Section~\ref{section:architecture:vae}),
	but it can usually generate better quality (better resolution)
	images than a variational autoencoder \cite{mallat:gan:vae:personal:communication:2018}.
	Note that the resolution limitation for a VAE may be a problem too for audio generation of music,
	but it appears {\em a priori} less a direct concern in the case of symbolic generation of music.}.
An example of the use of GAN for generating music is the MidiNet\index{MidiNet} system,
to be described in Section~\ref{section:systems:midinet}.

\subsection{Challenges}
\label{section:architecture:gan:challenges}

Training based on a minimax objective is known to be challenging to optimize \cite{yan:attribute:image:arxiv:2016},
with a risk of nonconverging oscillations.
Thus, careful selection of the model and its hyperparameters\index{Hyperparameter} is important
\cite[page~701]{goodfellow:deep:learning:book:2016}.
There are also some newer techniques, such as {\em feature matching\index{Feature!matching}}\footnote{Feature matching
	changes the objective for the generator (and accordingly its cost function)
	to minimize the statistical difference between the features of the real data and the generated samples,
	see more details in \cite{salimans:improved:training:gan:arxiv:2016}.},
among others, to improve training \cite{salimans:improved:training:gan:arxiv:2016}.

A recent proposed alternative both to GANs and to autoencoders is
{\em generative latent optimization\index{Generative!latent optimization}} {(GLO\index{GLO})} \cite{bojanowski:glo:arxiv:2017}.
It is an approach to train a generator without the need to learn a discriminator,
by learning a mapping from noise vectors to images.
GLO can thus be viewed both as an encoder\index{Encoder}-less autoencoder\index{Autoencoder},
and as a discriminator\index{Discriminator}-less GAN\index{GAN}.
It can also be used, as for a VAE\index{VAE} (variational autoencoder) introduced in Section~\ref{section:architecture:vae},
to control generation by exploring the latent space.
GLO has been tested on images but not yet on music
and needs more evaluation.


\section{Reinforcement Learning}
\label{section:architecture:reinforcement:learning}

{\em Reinforcement learning\index{Reinforcement!learning}} (RL\index{RL})
may appear at first glance to be outside of our interest in deep learning architectures,
as it has distinct objectives and models.
However, the two approaches have recently been combined.
The first move, in 2013, was to use deep learning architectures to efficiently implement reinforcement learning techniques,
resulting in {\em deep reinforcement learning\index{Deep!reinforcement learning}} \cite{mnih:deep:reinforcement:learning:arxiv:2013}.
The second move, in 2016, is directly related to our concerns,
as it explored the use of reinforcement learning to control music generation,
resulting in the RL-Tuner\index{RL-Tuner} architecture \cite{jaques:rl:tuner:arxiv:2016}
to be described in Section~\ref{section:systems:rl-tuner}.

\begin{figure}
\includegraphics[scale=0.26]{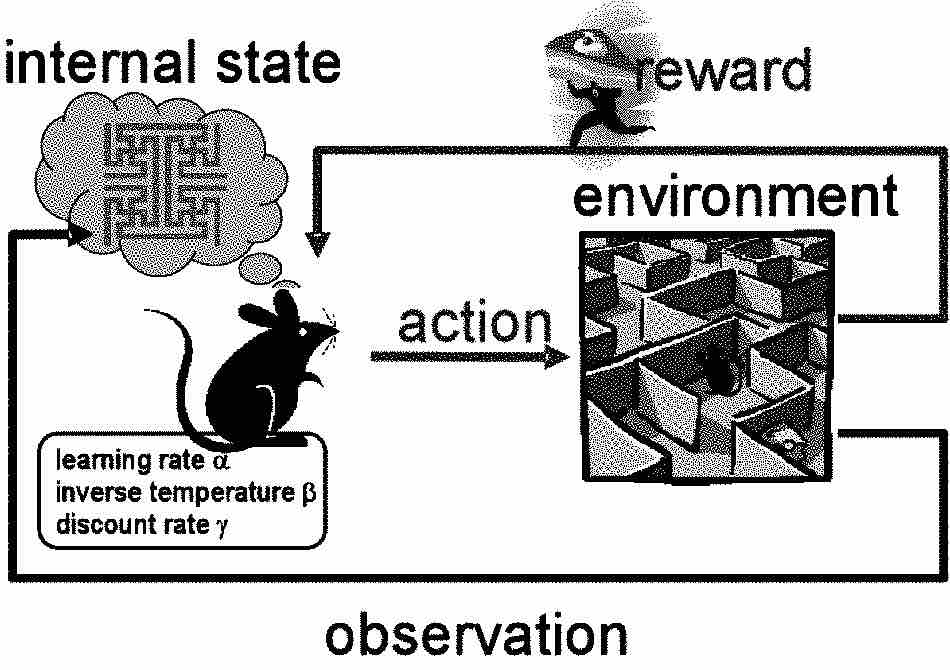}
\caption{Reinforcement learning -- conceptual model.
Reproduced from \cite{cyber:rodent:2005} with permission of SAGE Publications, Inc./Corwin}
\label{figure:reinforcement:learning:conceptual:architecture}
\end{figure}

Let us start with a reminder of the basic concepts of reinforcement learning,
illustrated in Figure~\ref{figure:reinforcement:learning:conceptual:architecture}


\begin{itemize}

\item an {\em agent\index{Agent}} within an {\em environment\index{Environment}} sequentially selects and performs
{\em actions\index{Action}} in an environment;

\item where each action performed brings it to a new {\em state\index{State}};

\item the agent receives a {\em reward\index{Reward}} ({\em reinforcement signal\index{Reinforcement!signal}}),
which represents the {\em fitness} of the action to the environment (current situation);

\item the objective
of the agent being to learn a near optimal {\em policy\index{Policy}} (sequence of actions) in order to maximize its {\em cumulated rewards\index{Cumulated rewards}} (named its {\em gain\index{Gain}}). 

\end{itemize}

Note that the agent does not know beforehand the model of the environment and the reward,
thus it needs to balance between {\em exploring} to learn more
and {\em exploiting} (what it has learned) in order to improve its gain\index{Gain} --
this is the {\em exploration exploitation dilemma\index{Exploration!exploitation dilemma}}.

There are many approaches and algorithms for reinforcement learning (for a more detailed presentation,
please refer, for example, to \cite{kaelbling:survey:rl:1996}).
Among them, {\em Q-learning\index{Q!-learning}} \cite{watkins:q:learning:ml:1992} turned out
to be a relatively simple and efficient method, and thus is widely used.
The name comes from the objective to learn (estimate\index{Estimation}) the {\em Q} function\index{Q!function} $Q^*(\text{s}, \text{a})$,
which represents the expected gain for a given pair $(\text{s}, \text{a})$, where s is a state and a an action,
for an agent choosing actions optimally (i.e. by following the optimal policy $\pi^*$).
The agent will manage a table, called the {\em Q-table\index{Q!-table}}, with values corresponding to all possible pairs.
As the agent\index{Agent} explores the environment\index{Environment},
the table is incrementally updated, with estimates becoming more accurate.

A recent combination of reinforcement learning (more specifically Q-learning) and deep learning,
named {\em deep reinforcement learning\index{Deep!reinforcement learning}},
has been proposed \cite{mnih:deep:reinforcement:learning:arxiv:2013}
in order to make learning\index{Learning} more efficient.
As the Q-table could be huge\footnote{Because of the high combinatorial nature
	when the number of possible states and possible actions is huge.},
the idea is to use a deep neural network\index{Deep!neural network} in order to approximate the expected values of the Q-table
through the learning of many replayed\index{Replay} experiences.

A further optimization, named {\em double Q-learning\index{Double Q-learning}} \cite{double:q:learning:arxiv:2015}
{\em decouples} the {\em action selection\index{Action!selection}} from the {\em evaluation\index{Action!evaluation}}, in order to avoid value overestimation\index{Overestimation}.
The task of the first network, named the Target Q-Network, is to estimate the gain (Q),
while the task of the Q-Network is to select the next action.

Reinforcement learning appears to be a promising approach for incremental\index{Incremental} adaptation\index{Adaptation} of the music to be generated,
e.g., based on the {\em feedback\index{Feedback}} from listeners
(this issue will be addressed in
Section~\ref{section:adaptability}).
Meanwhile, a significant move has been made in using reinforcement learning to inject some control\index{Control}
into the generation\index{Generation} of music by deep learning architectures,
through the reward\index{Reward} mechanism,
as described in Section~\ref{section:strategy:reinforcement}.

\section{Compound Architectures}
\label{section:architecture:compound}

Often {\em compound\index{Compound architecture}} architectures are used.
Some cases are {\em homogeneous\index{Homogeneous}} compound architectures,
combining various instances of the same architecture,
e.g., a stacked autoencoder\index{Stacked autoencoder} (see Section~\ref{section:architecture:compound:stacked:autoencoders}),
and most cases are {\em heterogeneous\index{Heterogeneous}} compound architectures, combining various types of architectures,
e.g., an RNN Encoder-Decoder\index{RNN Encoder-Decoder} which combines an RNN and an autoencoder,
see Section~\ref{section:architecture:compound:recurrent:autoencoder}.

\subsection{Composition Types}
\label{section:architecture:compound:composition:types}

We will see that, from an architectural\index{Architectural} point of view,
various types of composition\footnote{We are taking inspiration
	from concepts and terminology in programming languages and software architectures \cite{shaw:software:architecture:book:1996},
	such as {\em refinement}, {\em instantiation}, {\em nesting} and {\em pattern} \cite{gamma:design:patterns:book:1994}.}
may be used:

\begin{itemize}

\item {\em Composition\index{Composition}} --
at least two architectures, of the same type or of different types, are combined, such as

\begin{itemize}


\item a bidirectional\index{Bidirectional} RNN\index{Bidirectional!RNN|see{Bidirectional recurrent neural network}} (Section~\ref{section:architecture:compound:bidirectional:rnn})
combining two RNNs, forward and backward in time; and

\item the RNN-RBM\index{RNN-RBM} architecture (Section~\ref{section:architecture:compound:polyphonic:recurrent:network})
combining an RNN\index{RNN} architecture and an RBM\index{RBM} architecture.

\end{itemize}

\item {\em Refinement\index{Refinement}} --
one architecture is {\em refined} and {\em specialized\index{Specialization}}
through some additional constraint(s)\footnote{Both cases are refinements of the standard autoencoder\index{Autoencoder} architecture
	through additional constraints, in practice adding an extra term onto the cost function\index{Cost!function}.},
such as

\begin{itemize}

\item a sparse autoencoder\index{Sparse autoencoder} architecture (Section~\ref{section:architecture:sparse:autoencoder}); and

\item a variational autoencoder\index{Variational!autoencoder} (VAE\index{VAE}) architecture (Section~\ref{section:architecture:vae}).

\end{itemize}

\item {\em Nested} --
one architecture is nested\index{Nested} into the other one, for example

\begin{itemize}

\item a stacked autoencoder\index{Stacked autoencoder} architecture
(Section~\ref{section:architecture:compound:stacked:autoencoders}); and

\item the RNN Encoder-Decoder\index{RNN Encoder-Decoder} architecture
(Section~\ref{section:architecture:compound:recurrent:autoencoder}),
where two RNN\index{RNN} architectures
are nested within the encoder\index{Encoder} and decoder\index{Decoder} parts of an autoencoder\index{Autoencoder},
which we could therefore also notate as Autoencoder(RNN, RNN)\index{Autoencoder(RNN, RNN)}.

\end{itemize}

%
%
%


\item {\em Pattern instantiation\index{Pattern!instantiation}} --
an architectural pattern\index{Architectural!pattern} is instantiated\index{Instantiate} onto a given architecture(s), for example

\begin{itemize}

\item the C-RBM\index{C-RBM} architecture (Section~\ref{figure:crbmc:architecture})
that instantiates the convolutional architectural pattern onto an RBM\index{RBM} architecture,
which we could notate as Convolutional(RBM)\index{Convolutional(RBM)};

\item the C-RNN-GAN\index{C-RNN-GAN} architecture (Section~\ref{section:systems:c:rnn:gan}),
where the GAN\index{GAN}
architectural pattern is instantiated onto an RNN\index{RNN} architecture,
which we could
notate as GAN(RNN, RNN)\index{GAN(RNN, RNN)}; and


\item the Anticipation-RNN\index{Anticipation-RNN} architecture
(Section~\ref{section:systems:anticipation:rnn})
that instantiates the conditioning architectural pattern onto an RNN
with the output of another RNN as the conditioning input\index{Conditioning!input},
which we could notate as Conditioning(RNN, RNN)\index{Conditioning(RNN, RNN)}.


\end{itemize}

\end{itemize}


\subsection{Bidirectional RNN}
\label{section:architecture:compound:bidirectional:rnn}

Bidirectional recurrent neural networks\index{Bidirectional!recurrent neural network}
(bidirectional RNNs\index{Bidirectional!RNN})
were introduced by
Schuster and Paliwal
\cite{schuster:bidirectional:rnn:tsp:1997}
to handle the case when the prediction depends not only on the previous elements
but also on the {\em next} elements,
as for instance with speech recognition.
In practice, a bidirectional RNN\index{RNN} combines\footnote{See more details
	in \cite{schuster:bidirectional:rnn:tsp:1997}.}

\begin{itemize}

\item a first RNN that moves {\em forward} through time and begins from the {\em start} of the sequence; and

\item a second symmetric\index{Symmetric} RNN that moves {\em backward} through time
and begins from the {\em end} of the sequence.

\end{itemize}

The output y$_t$ of the bidirectional RNN at step $t$ combines

\begin{itemize}

\item the output h$^f_t$ at step $t$ of the hidden layer of the ``forward RNN'', and

\item the output h$^b_{N-t+1}$ at step $N-t+1$ of the hidden layer of the ``backward RNN''.

\end{itemize}

\begin{figure}
\includegraphics[width=\textwidth]{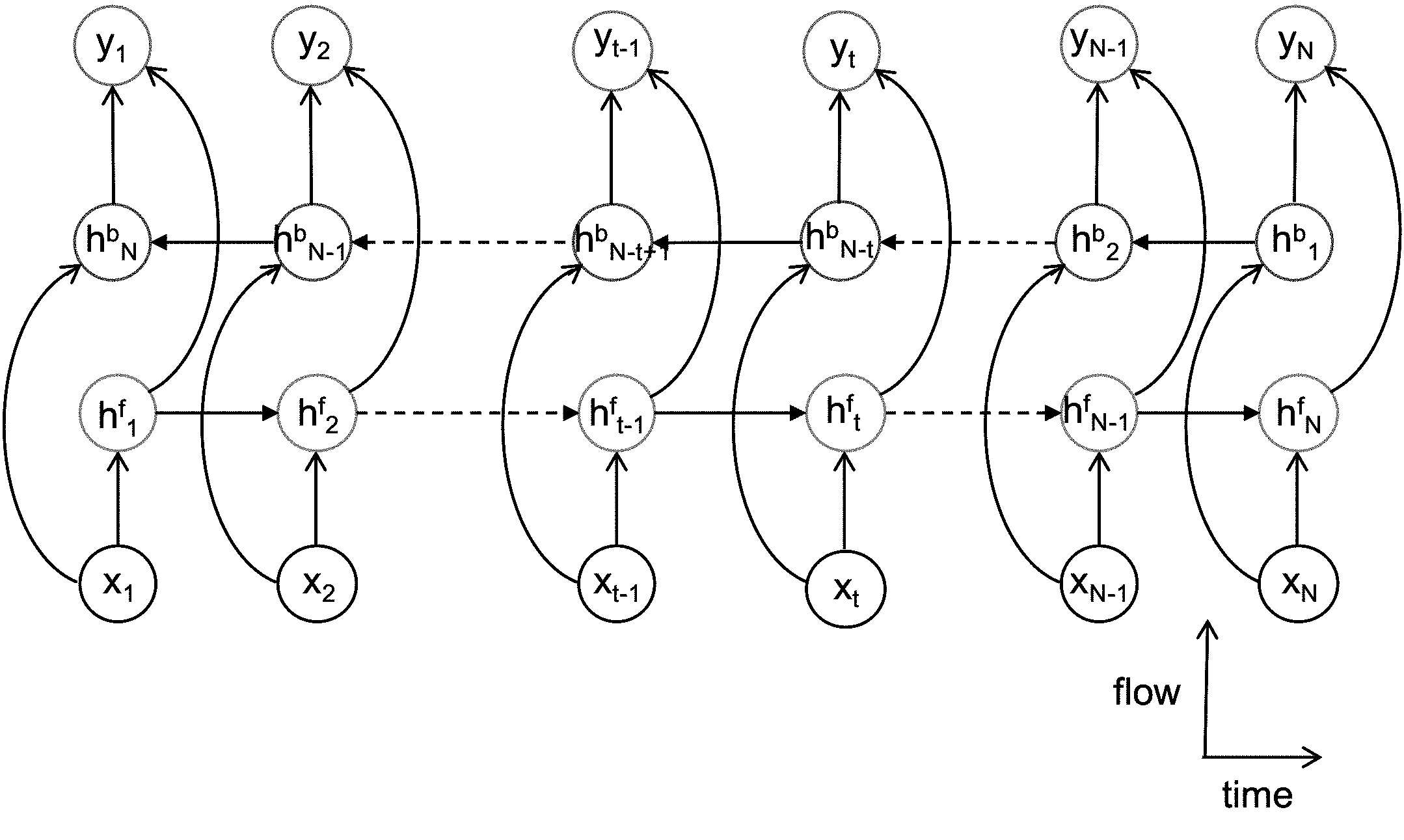}
\caption{Bidirectional RNN architecture}
\label{figure:rnn:bidirectional}
\end{figure}

An illustration is in Figure~\ref{figure:rnn:bidirectional}.
Examples of use are

\begin{itemize}

\item the BLSTM\index{BLSTM} architecture
(Section~\ref{section:experiment:blstm:chord});

\item the C-RNN-GAN\index{C-RNN-GAN} architecture
(Section~\ref{section:systems:c:rnn:gan})
that encapsulates a bidirectional RNN into the discriminator of a GAN\index{GAN}; and

\item the MusicVAE\index{MusicVAE} architecture
(Section~\ref{section:system:music:vae})
that encapsulates a bidirectional RNN into the encoder of a VAE\index{VAE}
(variational autoencoder\index{Variational!autoencoder}).

\end{itemize}

\subsection{RNN Encoder-Decoder}
\label{section:architecture:compound:recurrent:autoencoder}

The idea of encapsulating\index{Encapsulate} two identical recurrent networks\index{Recurrent!network} (RNNs\index{RNN})
into an autoencoder\index{Autoencoder},
named the {\em RNN Encoder-Decoder\index{RNN Encoder-Decoder}}\footnote{We could also notate it as
	Autoencoder(RNN, RNN)\index{Autoencoder(RNN, RNN)}.},
was initially proposed in \cite{cho:rnn:encoder:decoder:arxiv:2014}
as a technique to encode\index{Encoding} a variable length\index{Variable!length} sequence\index{Sequence}
learnt by a recurrent network into another variable length sequence produced by another recurrent network\footnote{This is named
	{\em sequence-to-sequence learning\index{Sequence!-to-sequence learning}}
	\cite{sutskever:sequence:2:sequence:nips:2014}.}.
The motivation and application target is the translation\index{Translation} from one language\index{Language} to another,
resulting in sentences of possibly different lengths\index{Length}.

\begin{figure}
\includegraphics[scale=0.22]{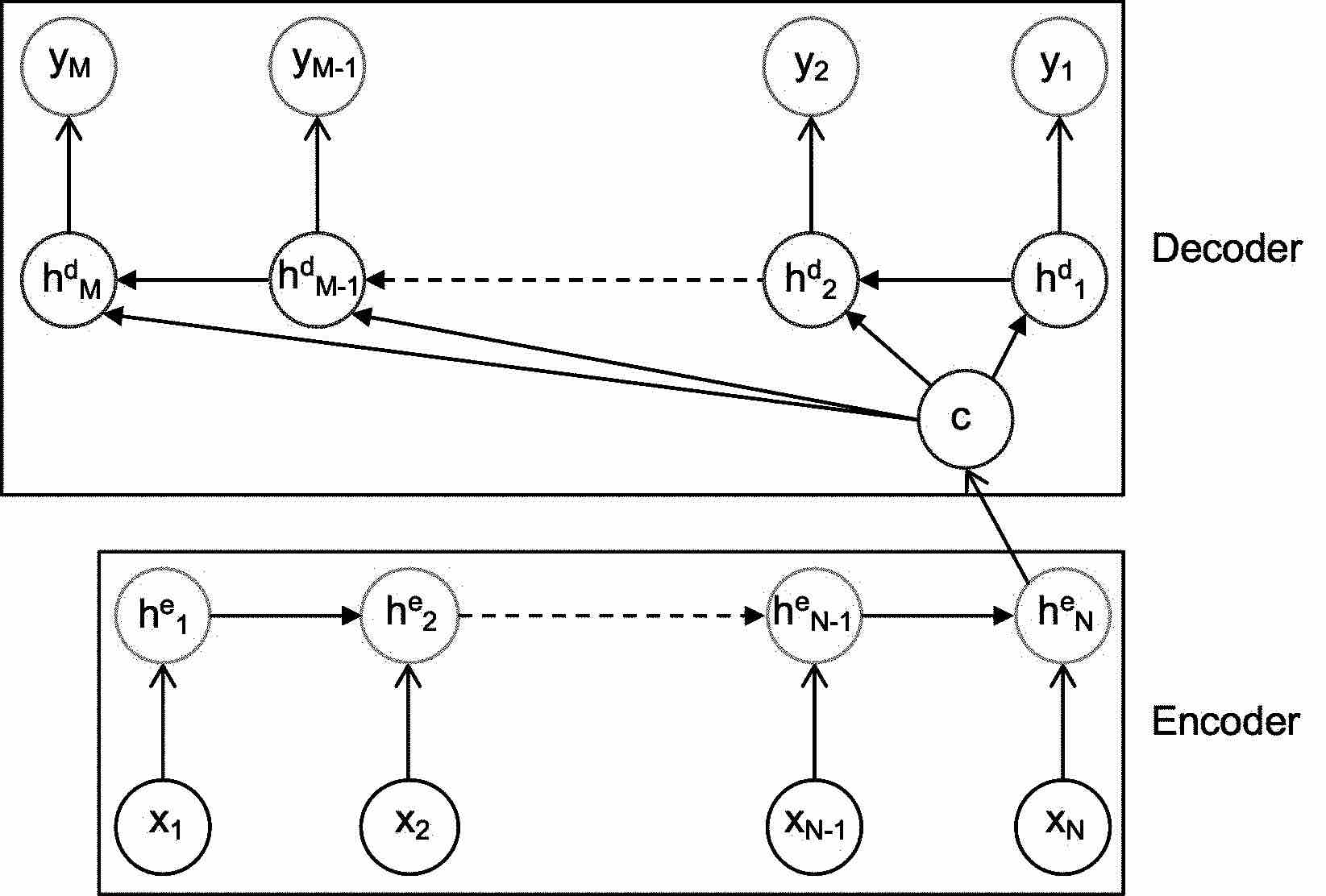}
\caption{RNN Encoder-Decoder architecture.
Inspired from \cite{cho:rnn:encoder:decoder:arxiv:2014}}
\label{figure:rnn:encoder:decoder:architecture}
\end{figure}

The idea is to use a fixed-length\index{Fixed-length} vector representation as a {\em pivot} representation\index{Pivot representation}
between an encoder\index{Encoder} and a decoder\index{Decoder} architecture,
see the illustration in Figure~\ref{figure:rnn:encoder:decoder:architecture}.
The hidden layer(s) $h^e_t$ of the encoder will act as a memory which

\begin{itemize}

\item iteratively accumulates information about some input sequence of length $N$,
while reading its successive x$_t$ elements\footnote{The end of the sequence
	is marked by a special symbol, as when training an RNN, see Section~\ref{section:architecture:rnn:training}.},
resulting in a final state h$^e_N$;

\item which is passed to the decoder as the summary c of the whole input sequence; and

\item the decoder then iteratively generates the output sequence of length $M$,
by predicting the next item y$_t$ given its hidden state h$^d_t$ and the summary
(as a conditioning\index{Conditioning!input} additional input) c\footnote{As noted
	by Goodfellow {\em et al.} in \cite[Section~10.4]{goodfellow:deep:learning:book:2016},
	an alternative is to use the summary c
	only to initialize the initial hidden state of the decoder h$^d_0$.
	This is, for instance, the strategy chosen in the GLSR-VAE\index{GLSR-VAE} architecture
	described in Section~\ref{section:experiment:glsr:vae}.}.

\end{itemize}

\begin{figure}
\includegraphics[scale=0.2]{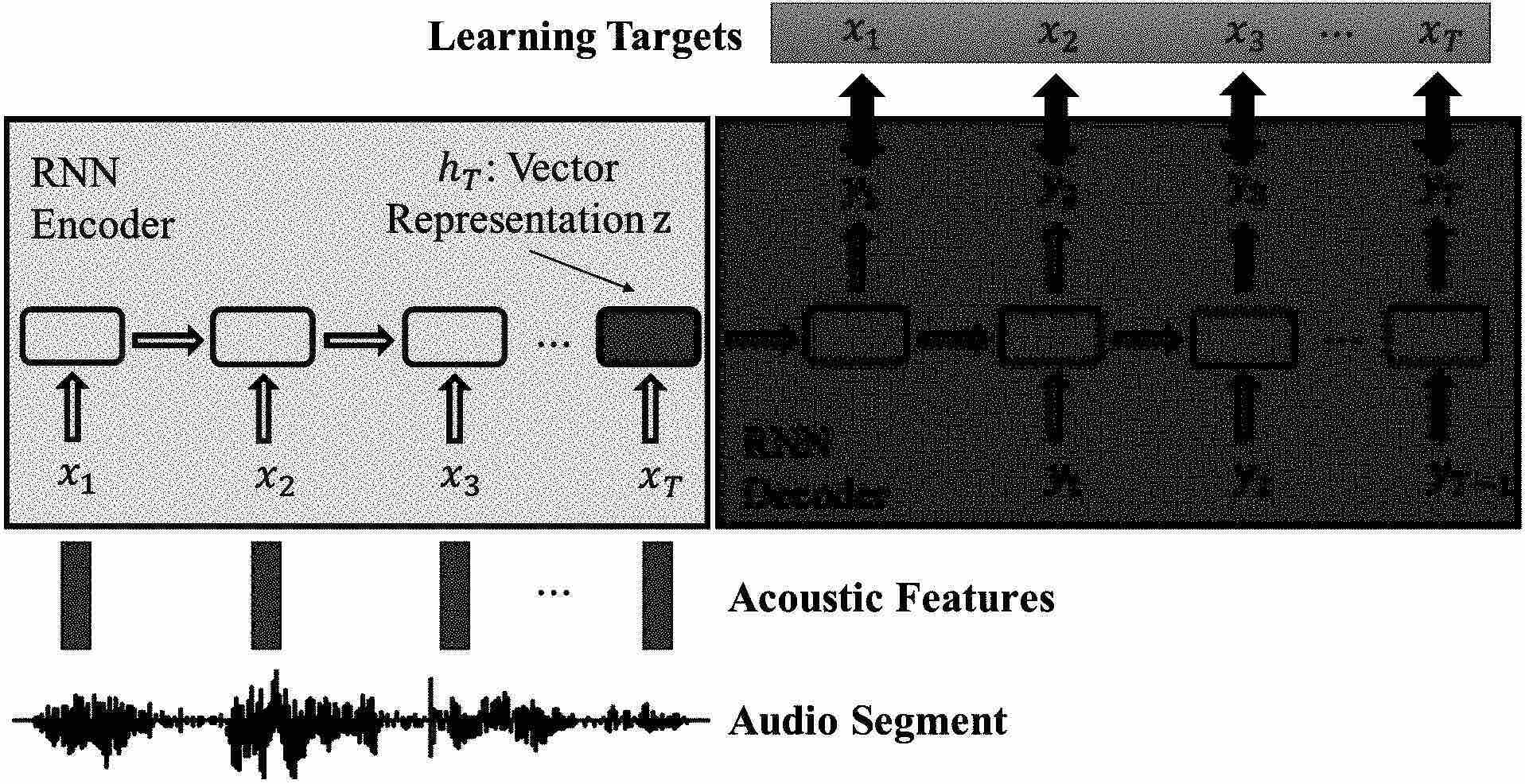}
\caption{RNN Encoder-Decoder audio Word2Vec architecture.
Reproduced from \cite{chung:audio:word2vec:arxiv:2016} with permission of the authors}
\label{figure:rnn:encoder:decoder:architecture:2}
\end{figure}

The two components of the RNN Encoder-Decoder are jointly trained
to minimize the cross-entropy\index{Cross-entropy!cost} between input and output.
See in Figure~\ref{figure:rnn:encoder:decoder:architecture:2}
the example of the Audio Word2Vec\index{AudioWord2Vec} architecture
for processing audio phonetic structures
\cite{chung:audio:word2vec:arxiv:2016}.

One limitation of the RNN Encoder-Decoder approach
is the difficulty for the summary to memorize very long sequences\footnote{In text translation applications,
	sentences have a limited size.}.
Two possible directions are

\begin{itemize}

\item using an {\em attention mechanism\index{Attention mechanism}}\footnote{Introduced
	in Section~\ref{section:architecture:attention:mechanism}.};
and

\item using a {\em hierarchical\index{Hierarchical} model}, as proposed in the MusicVAE\index{MusicVAE} architecture,
to be introduced in Section~\ref{section:system:music:vae}. 

\end{itemize}

\subsection{Variational RNN Encoder-Decoder}
\label{section:architecture:compound:recurrent:autoencoder:variational}

%
%
%
%

An interesting development is a {\em variational\index{Variational}} version of the RNN Encoder-Decoder,
in other words a variational autoencoder\index{Variational!autoencoder} (VAE\index{VAE})
encapsulating two RNNs.
We could notate it as
Variational(Autoencoder(RNN, RNN))\index{Variational(Autoencoder(RNN, RNN))}.
The objective is to combine

\begin{itemize}

\item the {\em variational} property of the VAE for controlling the generation\footnote{See
	Section~\ref{section:architecture:vae}.}; and

\item the {\em sequence} generation property of the RNN.

\end{itemize}

Examples of its application to music generation will be introduced in Section~\ref{section:systems:strategy:sampling:architecture:variational:recurrent}.

\subsection{Polyphonic Recurrent Networks}
\label{section:architecture:compound:polyphonic:recurrent:network}

The RNN-RBM\index{RNN-RBM} architecture,
to be introduced in Section~\ref{section:experiment:rnn:rbm},
combines an RBM\index{RBM} architecture and a recurrent\index{Recurrent!network} (LSTM\index{LSTM}) architecture
by {\em coupling} them to associate the vertical perspective (simultaneous notes\index{Note})
with the horizontal perspective (temporal sequences\index{Temporal!sequence} of notes) of a polyphony\index{Polyphony}
to be generated.


\subsection{Further Compound Architectures}
\label{section:architecture:compound:further}

It is possible to further combine architectures that are already compound, for example

\begin{itemize}



\item the WaveNet\index{WaveNet} architecture (Section~\ref{section:systems:wavenet}),
which is a conditioning convolutional feedforward architecture
with some tag\index{Tag} as the conditioning input\index{Conditioning!input},
which we could notate as Conditioning(Convolutional(Feedforward), Tag); and

\item the VRASH\index{VRASH} architecture (Section~\ref{section:experiment:vrash}),
which is a variational autoencoder encapsulating RNNs
with the decoder being conditioned on history,
which we could notate as Variational(Autoencoder(RNN, Conditioning(RNN, History))).

\end{itemize}

There are also some more specific (ad hoc\index{Ad hoc}) compound architectures, for example

\begin{itemize}

\item
Johnson's Hexahedria architecture (Section~\ref{section:experiment:hexahedria}),
which combines
two layers recurrent on the time dimension
with two other layers recurrent on the pitch dimension,
as an integrated alternative to the RNN-RBM\index{RNN-RBM} architecture; and

\item
The DeepBach\index{DeepBach} architecture (Section~\ref{section:experiment:deep:bach}),
which combines two feedforward\index{Feedforward!network} architectures
with two recurrent\index{Recurrent!network} architectures.

\end{itemize}

\subsection{The Limits of Composition}
\label{section:architecture:compound:limits}


There is a natural tendency to explore possible combinations of different architectures
with the hope of combining their respective features and merits.
An example of a sophisticated compound architecture is the VRASH\index{VRASH} architecture
(Section~\ref{section:experiment:vrash}),
which combines

\begin{itemize}

\item variational autoencoder;

\item recurrent networks; and

\item conditioning (on the decoder).

\end{itemize}

However, note that

\begin{itemize}

\item not all combinations make sense.
For instance, recurrence and convolution over the time dimension would compete, as discussed in Section~\ref{section:architecture:convolution}; and

\item there is no guarantee that combining a maximal variety of types will make a sound and accurate architecture\footnote{As in the case of a good cook,
	whose aim is not to simply mix {\em all} possible ingredients
	but to discover original successful combinations.}.

\end{itemize}

We will see in Chapter~\ref{section:chapter:challenges:strategies} that an important additional design dimension
is the {\em strategy\index{Strategy}},
which governs how an architecture will process representations in order to reach a given objective with some expected properties (the {\em challenges}).

%% file: challenge-strategy.tex
\chapter{Challenge and Strategy}
\label{section:chapter:challenges:strategies}
\label{section:chapter:challenges}
\label{section:chapter:strategies}
\label{section:challenges}
\label{section:strategies}

\abstract*{Chapter~\ref{section:chapter:challenges:strategies} Challenge and Strategy presents the fourth and the fifth dimensions of the conceptual framework proposed in this book to analyze,
classify and compare various deep learning-based music generation systems.
We analyze successive limitations and challenges occurring when applying deep learning techniques to music generation.
Examples of challenges are: content variability, control, structure, originality and interactivity.
For each challenge, we present alternative strategies for addressing it.
Examples of strategies are: single-step feedforward, iterative feedforward, decoder feedforward, input manipulation and sampling.
The analysis is illustrated by numerous examples of various deep learning-based music generation systems.
Each system is summarized along the conceptual framework presented.
This chapter is the core of the book.}

We are now reaching the core of this book.
This chapter will analyze in depth how to apply the architectures presented in Chapter~\ref{section:chapter:architecture}
to learn and generate music.
We will first start with a naive, straightforward strategy, 
using the basic prediction\index{Prediction} task of a neural network\index{Neural!network}
to generate an accompaniment for a melody.

We will see that, although this simple direct strategy does work, it suffers from some limitations.
We then will study these limitations, some relatively simple to solve, some more difficult and profound --
the
challenges\index{Challenge}.
We will analyze various strategies\index{Strategy}\footnote{Remember, and this will be important for the following sections,
	that, as stated in Chapter~\ref{section:chapter:method},
	we consider here the strategy related to the {\em generation phase}
	and not the training phase (which could be different).}
for each challenge, and illustrate them though different systems\footnote{As proposed
	in Chapter~\ref{section:method},
	we use the term {\em systems}\index{System} for various proposals -- architectures, models, prototypes,
	systems and related experiments --
	for deep learning-based music generation, collected from the related literature.}
taken from the relevant literature.
This also provides an opportunity to study the possible relationships\index{Relationship} between architectures\index{Architecture} and strategies.

\section{Notations for Architecture and Representation Dimensions}
\label{section:challenge:strategy:architecture:notation:depth}

At first, let us introduce some compact notations\index{Notation convention} for the dimension of an architecture and for the size of a representation:

\begin{itemize}

\item {\em Architecture-type}$^n$ for a $n$-layer architecture\footnote{This notation has actually already been introduced
	in Section~\ref{section:architecture:feedforward:depth}.},
	e.g., Feedforward$^2$ for the 2-layer feedforward architecture
	of the MiniBach system to be introduced in Section~\ref{section:experiment:mini:bach},

\item {\em Architecture-type}${\times}n$ for a $n$-instance compound architecture,
	e.g., RNN$\times$2 for the double RNN compound architecture of RL-Tuner to be introduced in Section~\ref{section:systems:rl-tuner},
	and

\item One-hot${\times}n$ for a multi-one-hot encoding representation, such as:

\begin{itemize}

%

\item a $n$-time steps one-hot encoding,
	e.g., One-hot$\times$64 for the 64-time steps representation of the DeepHear$_M$ system to be introduced in Section~\ref{section:experiment:deep:hear:melody},

\item a $n$-voice one-hot encoding,
	e.g., One-hot$\times$2 for the melody+chords representation of the Blues$_{MC}$ system
	to be introduced in Section~\ref{section:experiment:eck:blues:lstm:second:experiment},
	or

\item a combination of a multi-time steps encoding and a multivoice encoding,
	e.g., One-hot$\times$64$\times$(1+3) for the 64-time steps 1-voice input and 3-voices output representation of the MiniBach system
	to be introduced in Section~\ref{section:experiment:mini:bach}.


\end{itemize}

\end{itemize}

An example of a combination of the two notations is
LSTM$^2\times$2 for the double 2-layer RNN compound architecture of the Anticipation-RNN system to be introduced in Section~\ref{section:systems:anticipation:rnn}.

\section{An Introductory Example}
\label{section:challenges:strategies:introductory:example}
\label{section:introductory:example}


\subsection{Single-Step Feedforward Strategy}
\label{section:single:step:feedforward}
\label{section:strategy:single:step:feedforward}
\label{section:strategy:one:step:feedforward}

The most direct strategy is using the prediction\index{Prediction} or the classification\index{Classification} task
of a neural network\index{Neural!network} in order to generate\index{Generate} musical content\index{Content}.
Let us consider the following objective:
for a given melody
we want to generate an accompaniment\index{Accompaniment},
for example, a counterpoint\index{Counterpoint}.
We will consider a dataset\index{Dataset} of examples, each one being a pair $(melody, counterpoint\,melody(ies))$.
We then train a feedforward neural network\index{Feedforward!neural network} architecture
in a supervised learning\index{Supervised learning} manner on this dataset.
Once trained, we can choose an arbitrary melody and feedforward it into the architecture
in order to produce a corresponding counterpoint accompaniment, in the style of the dataset.
Generation is completed in a single-step\index{Single!-step} of feedforward\index{Feedforward} processing.
Therefore, we have named this strategy the {\em single-step feedforward strategy\index{Single!-step feedforward strategy}}.

\subsection{Example: MiniBach Chorale Counterpoint Accompaniment Symbolic Music Generation System}
\label{section:experiment:mini:bach}

Let us consider the following objective\index{Objective}:
generating a counterpoint\index{Counterpoint} accompaniment to a given melody\index{Melody}
for a soprano voice,
through three matching parts, corresponding to alto, tenor and bass voices.
We will use as a corpus the set of J. S. Bach's polyphonic chorales\index{Chorale} \cite{bach:chorales:book}.
As we want this first introductory system to be simple, we consider only 4 measures long excerpts from the corpus.
The dataset is constructed by extracting all possible 4 measures long excerpts from the original 352 chorales, also transposed in all possible keys.
%
Once trained on this dataset, the system may be used to generate three counterpoint voices corresponding to an arbitrary 4 measures long
melody provided as an input.
Somehow, it does capture the practice of J. S. Bach, who chose various melodies for a soprano and composed the three additional voices melodies
(for alto, tenor and bass) in a counterpoint\index{Counterpoint} manner.

%
First, we need to decide the input as well as the output representations\index{Representation}.
We represent
four measures
of 4/4 music.
Both the input and the output representations are symbolic, of the piano roll type, with one-hot encoding
for each voice, i.e. a multi-one-hot encoding for the output representation.
The three first voices (soprano, alto and tenor) have a scope of 20 possible notes
plus an additional token to encode a hold\footnote{See
	Section~\ref{section:representation:note:ending}.
	Note that, as a simplification, MiniBach does not consider rests.},
while the last voice (bass) has a scope of 27 possible notes plus the hold symbol.
%
%
Time quantization (the value of the time step) is set at the sixteenth note,
which is the minimal note duration\index{Note!duration} used in the corpus.
The input representation has a size of 21 possible notes $\times$ 16 time steps $\times$ 4 measures, i.e. $21\times16\times4 = 1,344$,
while the output representation has a size of $(21 + 21 + 28)\times16\times4 = 4,480$.

\begin{figure}
\includegraphics[width=\textwidth]{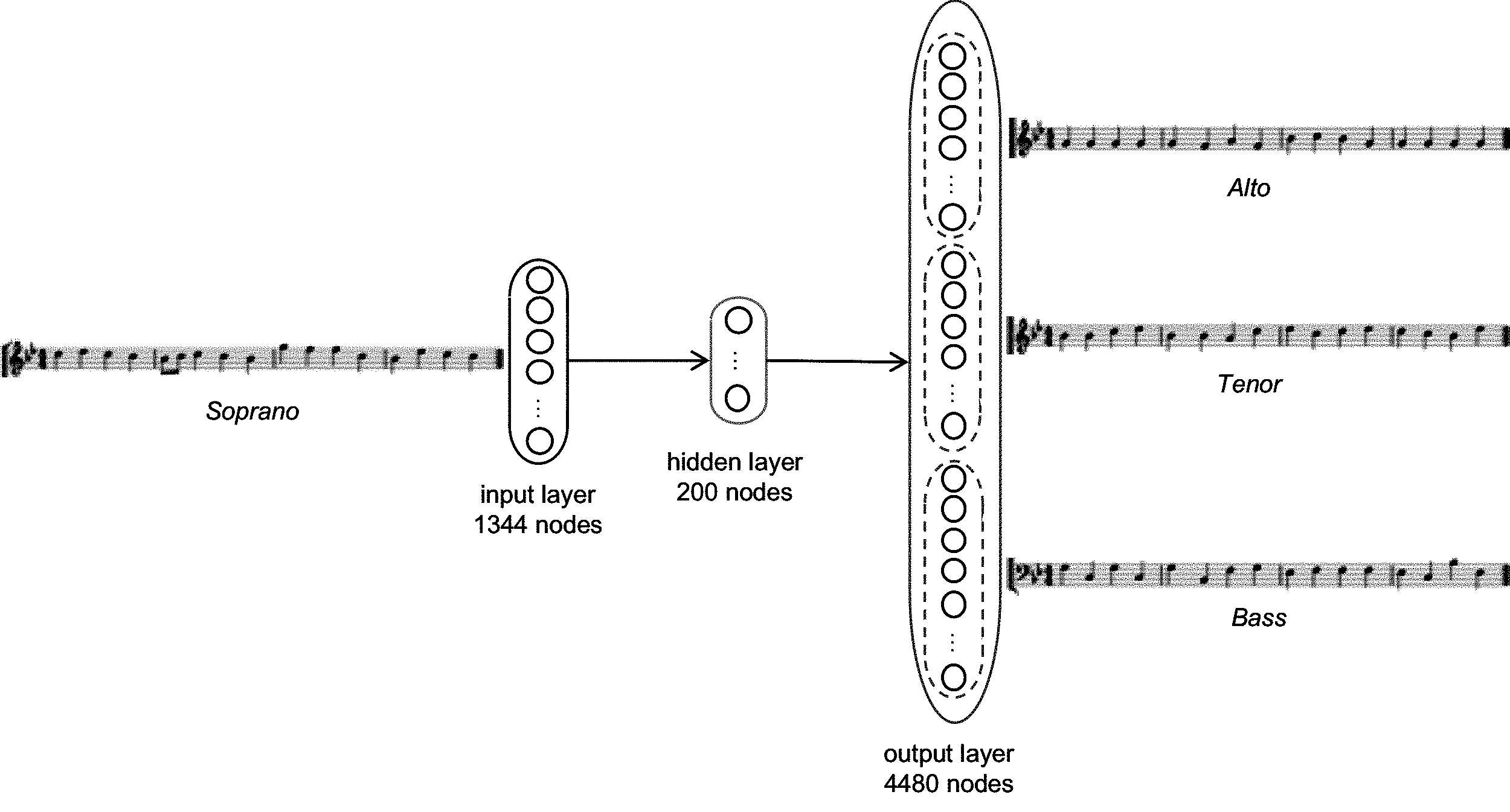}
\caption{MiniBach architecture}
\label{figure:architecture:mini:bach}
\end{figure}

The architecture, a feedforward network, is shown in Figure~\ref{figure:architecture:mini:bach}.
As explained previously and because of the mapping between the representation and the architecture,
the input layer has 1,344 nodes and the output layer 4,480.
There is a single hidden layer with 200 units\footnote{This is
	an arbitrary choice.}.
The nonlinear activation function used for the hidden layer is ReLU.
The output layer activation function is sigmoid
and the cost function used is binary cross-entropy
(this is a case of multi$^2$ multiclass single label,
see Section~\ref{section:architecture:neural:network:cost:function}).

\begin{figure}
\includegraphics[width=\textwidth]{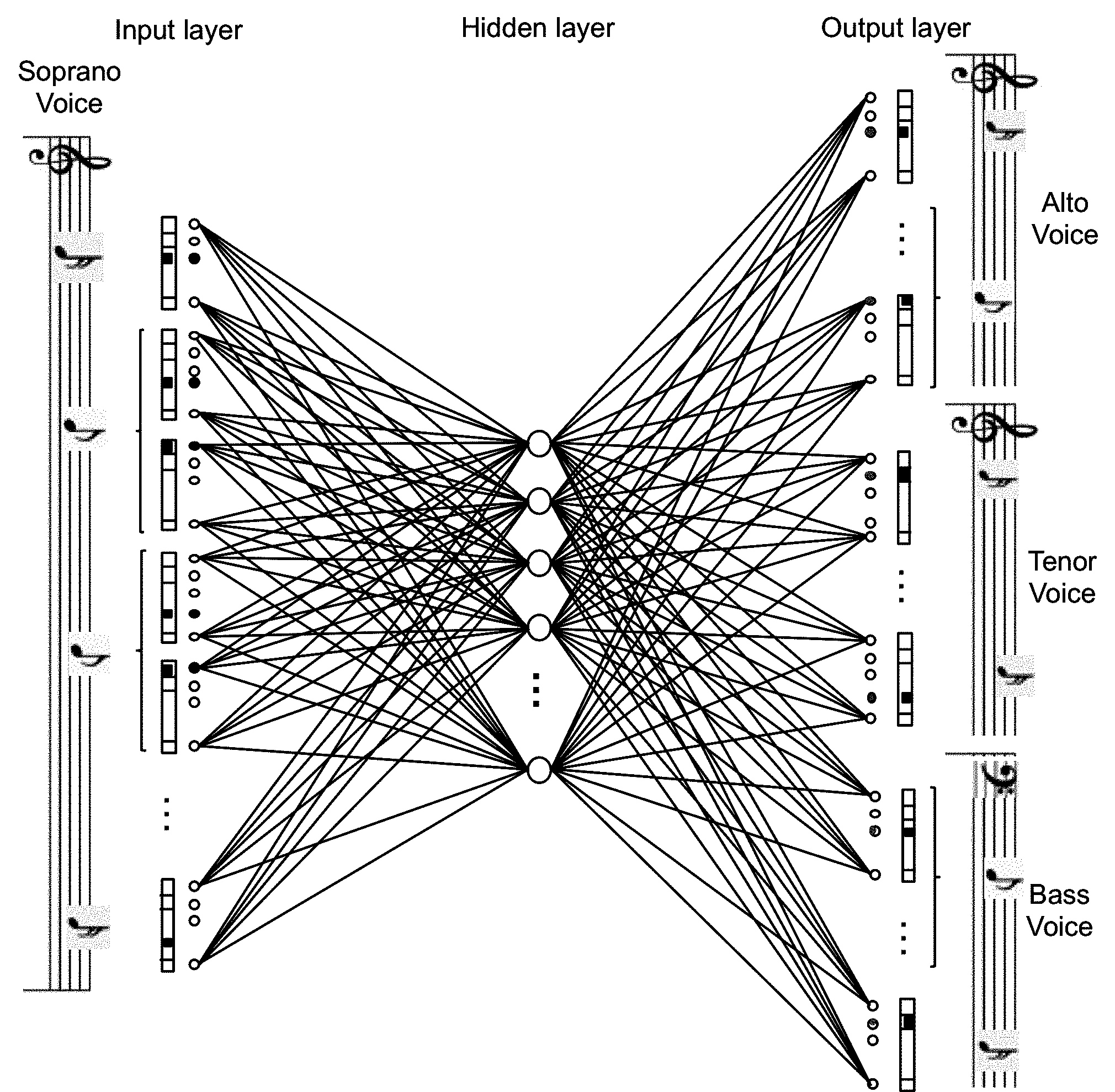}
\caption{MiniBach architecture and encoding}
\label{figure:architecture:mini:bach:detailed}
\end{figure}

The detail of the architecture and the encoding is shown in Figure~\ref{figure:architecture:mini:bach:detailed}.
It shows the encoding of successive music time slices into successive one-hot vectors directly mapped to the input nodes (variables).
In the figure, each blackened vector element as well as each corresponding blackened input node element
illustrate the specific encoding (one-hot vector index) of a specific note time slice, depending of its actual pitch
(or a hold in the case of a longer note, shown with a bracket). 
The dual process happens at the output.
Each grey output node element illustrates the chosen note (the one with the highest probability),
leading to a corresponding one-hot index, leading ultimately to a sequence of notes for each counterpoint voice.

\begin{figure}
\includegraphics[width=\textwidth]{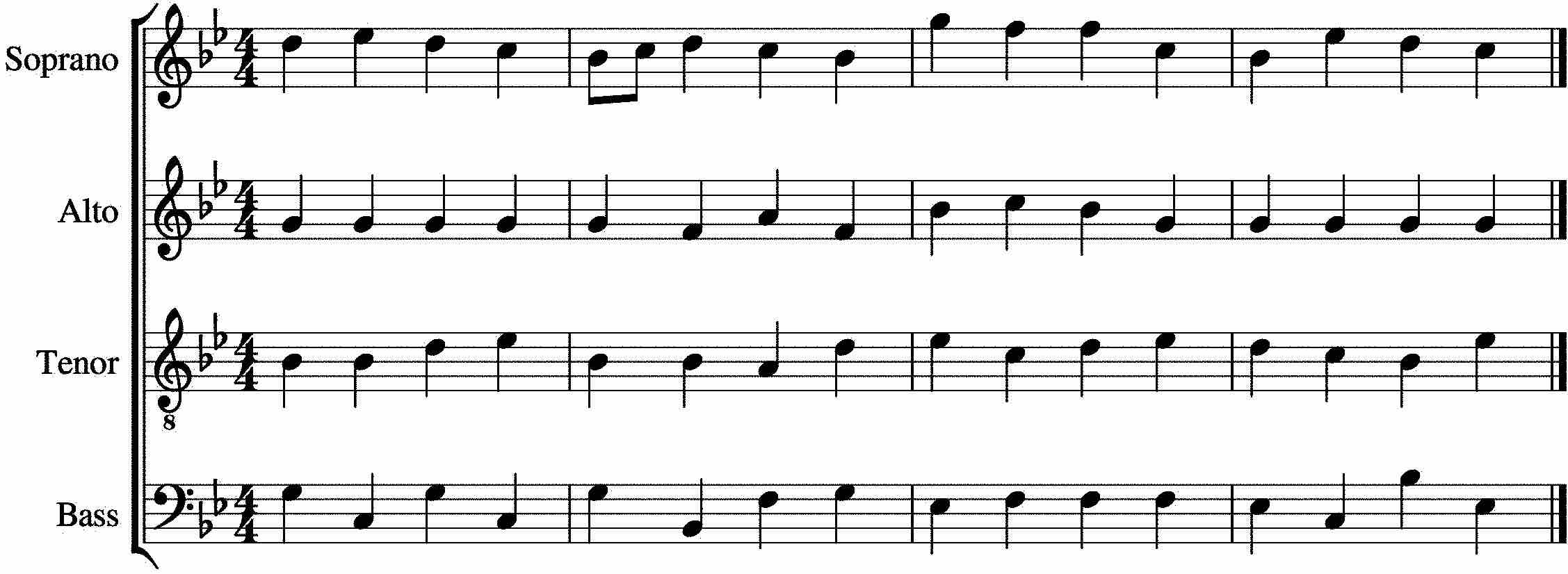}
\caption{Example of a chorale counterpoint generated by MiniBach from a soprano melody}
\label{figure:example:mini:bach}
\end{figure}

The characteristics of this system, named MiniBach\index{MiniBach}\footnote{MiniBach is actually a strong simplification
	-- but with the same objective, corpus and representation principles --
	of the DeepBach\index{DeepBach} system to be introduced in Section~\ref{section:experiment:deep:bach}.},
are summarized in our multidimensional
conceptual framework (as defined in Chapter~\ref{section:chapter:method} Method)
in Table~\ref{table:dimensions:mini:bach}.
The notation\footnote{These notations,
	introduced in Section~\ref{section:challenge:strategy:architecture:notation:depth},
	will be summarized in Section~\ref{section:analysis:notations}.}
One-hot$\times$64$\times$(1+3) means an encoding with 1 input + 3 output voices,
each with 64 (for 4 measures of 16 time steps each) one-hot encodings of notes.
The notation Feedforward$^2$ means a 2-layer feedforward architecture (with 1 hidden layer).
An example of a chorale counterpoint generated from a soprano melody is shown in Figure~\ref{figure:example:mini:bach}.

\begin{table}
\begin{tabular}{|l|l|}
\hline
{\em Objective}			&Accompaniment; Counterpoint; Chorale; Bach\\
\hline
{\em Representation}	&Symbolic; Piano roll; One-hot$\times$64$\times$(1+3); Hold\\
\hline
{\em Architecture}		&Feedforward$^2$\\
\hline
{\em Strategy}			&Single-step feedforward\\
\hline
\end{tabular}
\caption{MiniBach summary}
\label{table:dimensions:mini:bach}
\end{table}


\subsection{A First Analysis}
\label{section:experiment:mini:bach:analysis}

The chorales produced by MiniBach look convincing at first glance.
%
%
But, independently of a qualitative musical evaluation,
where an expert could detect some defects,
objective limitations of MiniBach appear:

\begin{itemize}

\item A structural limitation is that the music produced (as well as the input melody)
has a {\em fixed size} (one cannot produce a longer or shorter piece of music).

\item The same melody will always produce exactly the {\em same} accompaniment
because of the {\em deterministic} nature of a feedforward neural network architecture.

\item The generated accompaniment is produced in a {\em single atomic step},
without any possibility of human intervention (i.e. without {\em incrementality} and {\em interactivity}).

\end{itemize}



\section{A Tentative List of Limitations and Challenges}
\label{section:challenges:strategies:list:challenges}

Let us now introduce a tentative list of limitations (in most cases, properties not fulfilled) and challenges\footnote{Our shallow distinction
	between a limitation\index{Limitation} and a challenge\index{Challenge} is as follows:
	{\em limitations} have relatively well-understood solutions,
	whereas {\em challenges} are more profound and still the subject of open research.}:

\begin{itemize}

\item
{\em Ex nihilo} generation (vs accompaniment);

\item Length variability (vs fixed length);

\item Content variability (vs determinism);

\item Expressiveness (vs mechanization);

\item Melody-harmony consistency;

\item Control (e.g., tonality conformance, maximum number of repeated notes\ldots);

\item Style transfer;

\item Structure;

\item Originality (vs imitation);

\item Incrementality (vs one-shot generation);

\item Interactivity (vs automation);

\item Adaptability (vs no improvement through usage); and

\item Explainability (vs black box).

\end{itemize}

We will analyze them with possible matching solutions
and illustrate them through various examples systems.

\section{{\em Ex Nihilo} Generation}
\label{section:challenges:strategies:input:less}


The MiniBach system is good at generating an accompaniment (a counterpoint composed of three distinct melodies) matching an input melody.
This is an example of supervised learning\index{Supervised learning},
as training examples include both an input (a melody) and a corresponding output (accompaniment).

Now suppose that our objective is to generate a melody on its own
-- not as an accompaniment of some input melody --
while being based on a style learnt from a corpus of melodies.
A standard feedforward architecture\index{Feedforward!network}
and its companion single-step feedforward strategy\index{Single!-step feedforward strategy},
such as those used in MiniBach (described in Section~\ref{section:experiment:mini:bach}),
are not appropriate for such an objective.

Let us introduce some
strategies to generate new music content
{\em ex nihilo}
or from minimal {\em seed\index{Seed}}
information, such as a starting note or a high-level description.

%
%
%
%

%


\subsection{Decoder Feedforward}
\label{section:challenges:strategies:input:less:decoding}
\label{section:input:less:decoding}
\label{section:challenges:strategies:input:less:decoding:generating}
\label{section:strategy:decoder:feedforward}

The first strategy is based on an autoencoder\index{Autoencoder} architecture.
As explained in Section~\ref{section:architecture:autoencoder},
through the training phase an autoencoder will specialize its hidden layer\index{Hidden!layer} into
a detector of features\index{Feature} characterizing the type of music learnt
and its variations\footnote{To enforce this specialization,
	sparse autoencoders\index{Sparse autoencoder} are often used
	(see Section~\ref{section:architecture:sparse:autoencoder}).}.
One can then use these features as an {\em input interface\index{Interface}} to {\em parameterize\index{Parameterization}} the generation of musical content.
The idea is then to:

\begin{itemize}

\item {\em choose} a {\em seed\index{Seed}} as a vector of values corresponding to the hidden layer units;

\item {\em insert} it in the
hidden layer; and

\item {\em feedforward\index{Feedforward}} it through the decoder\index{Decoder}.

\end{itemize}

This strategy,
that we name {\em decoder feedforward\index{Decoder!feedforward strategy}},
will produce a {\em new} musical content corresponding to the features, in the same format as the training examples.


In order to have a minimal and high-level vector of features,
a stacked autoencoder (see Section~\ref{section:architecture:compound:stacked:autoencoders}) is often used.
The seed is then inserted at the {\em bottleneck hidden layer\index{Bottleneck hidden layer}} of the stacked autoencoder\footnote{In other words,
	at the exact middle of the encoder/decoder stack, as shown in Figure~\ref{figure:hierarchy:autoencoders:generation}.}
and feedforwarded through the chain of decoders.
Therefore, a simple seed information can generate an arbitrarily long, although fixed-length, musical content.

\subsubsection{\#1 Example: DeepHear Ragtime Melody Symbolic Music Generation System}
\label{section:experiment:deep:hear}
\label{section:experiment:deep:hear:melody}

\begin{figure}
\includegraphics[scale=0.14]{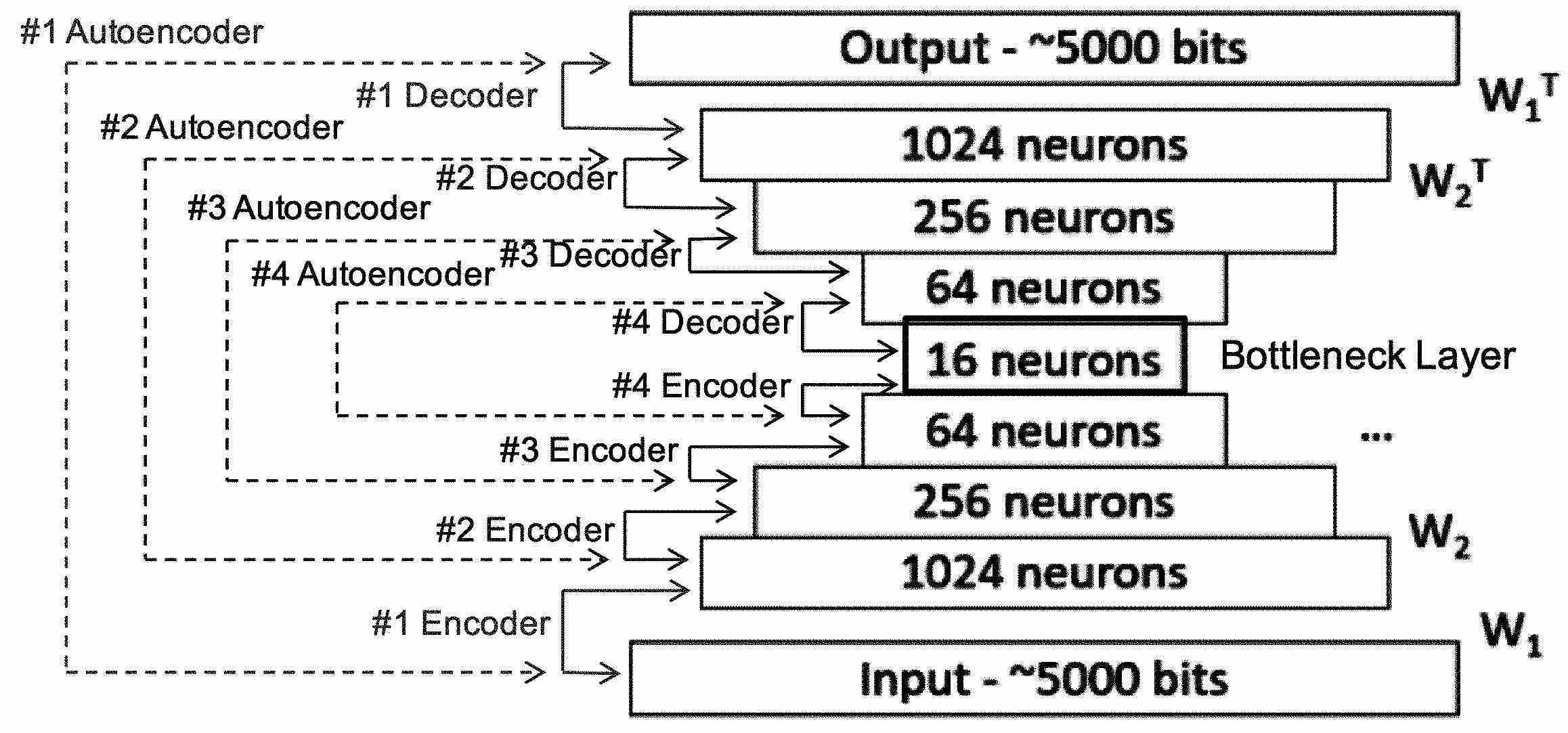}
\caption{DeepHear stacked autoencoder architecture.
Extension of a figure reproduced from \cite{sun:deep:hear} with permission of the author}
\label{figure:hierarchy:autoencoders}
\end{figure}

An example of this strategy is the
DeepHear\index{DeepHear} system by
Sun \cite{sun:deep:hear}.
The corpus used is 600 measures of Scott Joplin's ragtime\index{Ragtime} music, split into 4 measures long segments.
The representation used is piano roll with a multi-one-hot encoding.
The quantization (time step) is a sixteenth note, thus the representation includes $4\times16 = 64$ time steps (notated as One-hot$\times$64).
The number of input nodes is around 5,000,
which provides a vocabulary of about 80 possible note values.
The architecture is shown in Figure~\ref{figure:hierarchy:autoencoders}
and is a 4-layer stacked autoencoder (notated as Autoencoder$^4$)
with a decreasing number of hidden units, down to 16 units.

\begin{figure}
\includegraphics[scale=0.16]{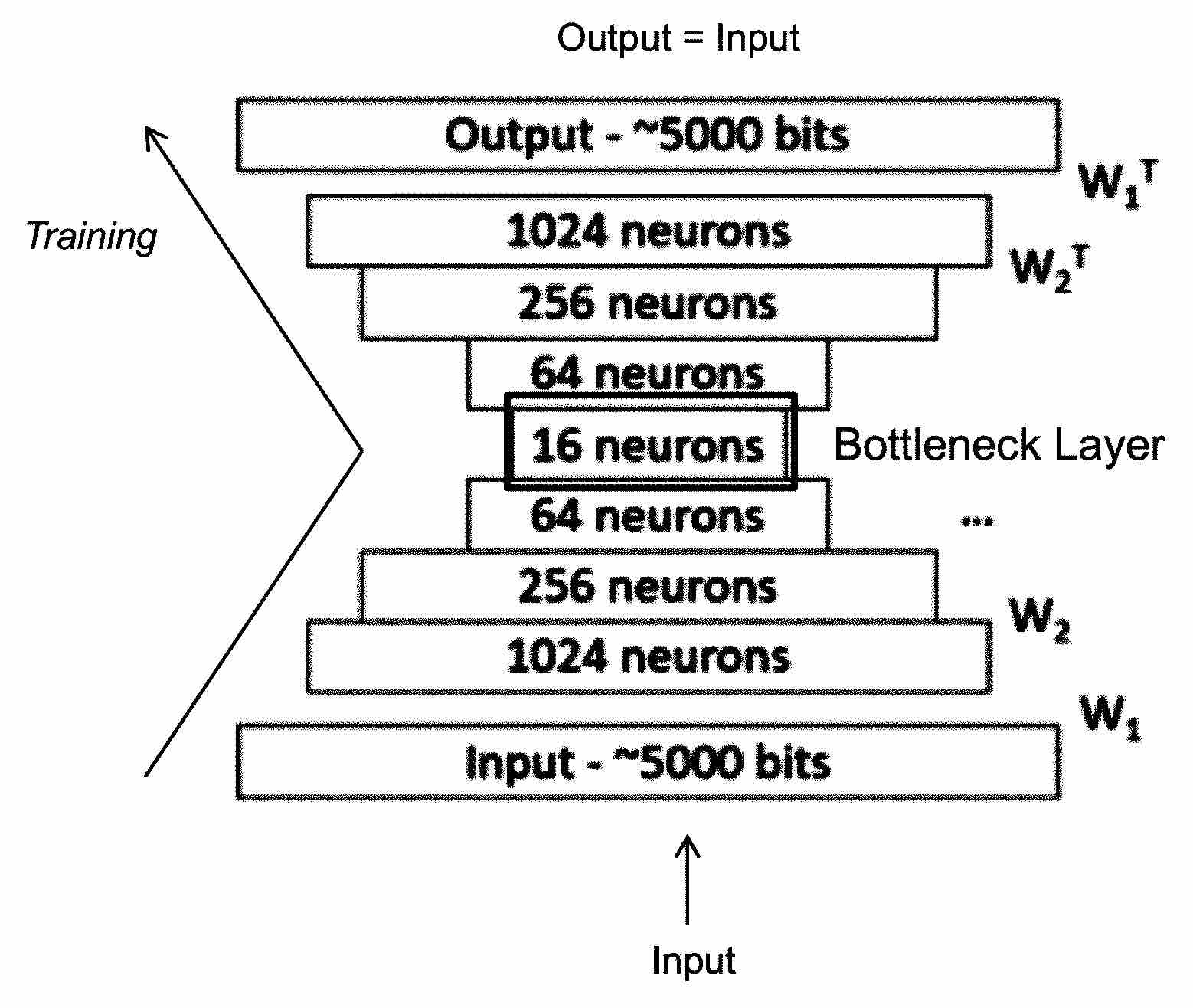}
\caption{Training DeepHear.
Extension of a figure reproduced from \cite{sun:deep:hear} with permission of the author}
\label{figure:hierarchy:autoencoders:training}
\end{figure}

After a pre-training phase\footnote{We do not detail pre-training here,
	please refer to, for example, \cite[page~528]{goodfellow:deep:learning:book:2016}.},
final training is performed, with each provided example used both as an input and as an output,
in the self-supervised learning\index{Self!-supervised learning} manner
(see Section~\ref{section:architecture:autoencoder})
shown in Figure~\ref{figure:hierarchy:autoencoders:training}.

\begin{figure}
\includegraphics[scale=0.16]{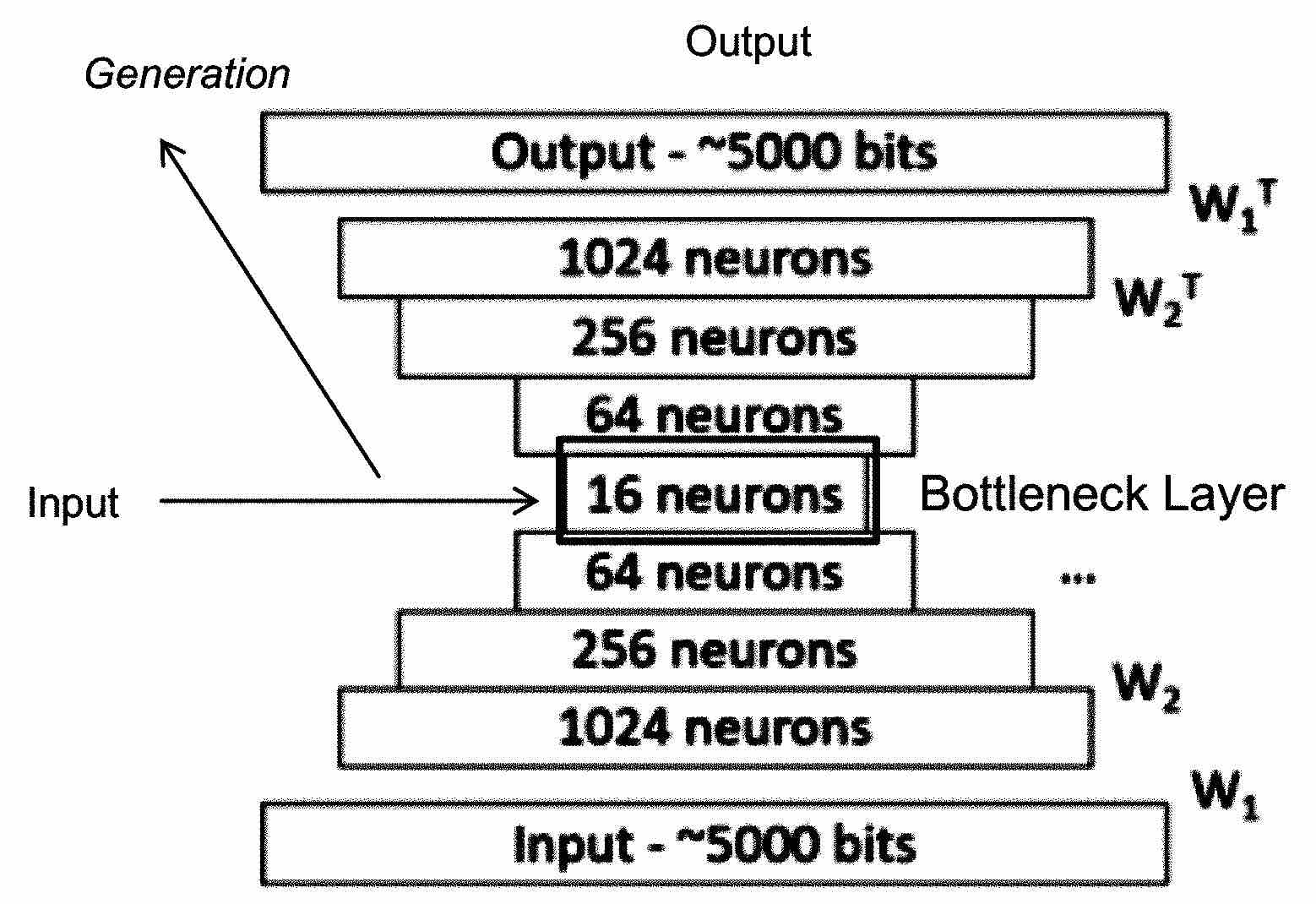}
\caption{Generation in DeepHear.
Extension of a figure reproduced from \cite{sun:deep:hear} with permission of the author}
\label{figure:hierarchy:autoencoders:generation}
\end{figure}

Generation is performed by inputing random\index{Random} data as the seed\index{Seed}
into the 16 bottleneck hidden layer units\footnote{The units of the hidden layer represent an embedding\index{Embedding}
	(see Section~\ref{section:representation:embedding}),
	of which an arbitrary instance is named by Sun a {\em label\index{Label}}.}
(shown within a red rectangle)
and then by feedforwarding it into the chain of decoders
to produce an output (in the same 4 measures long format as the training examples),
as shown in Figure~\ref{figure:hierarchy:autoencoders:generation}.
We summarize the characteristics of DeepHear$_M$\footnote{We notate DeepHear$_M$ this DeepHear melody generation system,
	where {\small $M$} stands for melody,
	because another experiment with the same DeepHear architecture but with a different objective
	will be presented later on in Section~\ref{section:experiment:deep:hear:harmonize}.}
in Table~\ref{table:dimensions:deep:hear:m}.

\begin{table}
\begin{tabular}{|l|l|}
\hline
{\em Objective}			&Melody; Ragtime\\
\hline
{\em Representation}	&Symbolic; Piano roll; One-hot$\times$64\\
\hline
{\em Architecture}		&Stacked autoencoder = Autoencoder$^4$\\
\hline
{\em Strategy}			&Decoder feedforward\\
\hline
\end{tabular}
\caption{DeepHear$_M$ summary}
\label{table:dimensions:deep:hear:m}
\end{table}

In \cite{sun:deep:hear}, Sun remarks that the system produces a certain amount of plagiarism\index{Plagiarism}.
Some generated music is almost recopied from the corpus.
He states that this is because of the small size of the bottleneck hidden layer (only 16 nodes) \cite{sun:deep:hear}.
He measured the similarity\index{Similarity} (defined as the percentage of notes in a generated piece that are also in one of the training pieces)
and found that, on average, it is 59.6\%,
which is indeed quite high, although it does not prevent most of generated pieces from sounding different.

\subsubsection{\#2 Example: deepAutoController Audio Music Generation System}
\label{section:experiment:deep:auto:controller}
\label{section:experiment:sarroff}

The deepAutoController\index{deepAutoController} system,
by
Sarroff and
Casey \cite{sarroff:audio:synthesis:2014},
is similar to DeepHear\index{DeepHear$_M$} (see Section~\ref{section:experiment:deep:hear:melody})
in that it also uses a stacked autoencoder.
But the representation is {\em audio},
more precisely a spectrum generated by Fourier transform,
see \cite{sarroff:audio:synthesis:2014} for more details.
The dataset is composed of 8,000 songs of 10 musical genres\index{Musical!genre},
leading to 70,000 frames of magnitude Fourier transforms\index{Fourier transform}\footnote{As the authors
	state in \cite{sarroff:audio:synthesis:2014}:
	``We chose to use frames of magnitude FFTs (Fast Fourier transforms\index{Fast!Fourier transform}) for our models
	because they may be reconstructed exactly into the original time domain signal
	when the phase information is preserved, the Fourier coefficients are not altered,
	and appropriate windowing and overlap-add is applied.
	It was thus easier to subjectively evaluate the quality of reconstructions that had been processed by the autoencoding models.''}.
The entire data is normalized to the $[0, 1]$ range.
The cost function used is mean squared error.
The architecture is a 2-layer stacked autoencoder, the bottleneck hidden layer having 256 units
and the input and output layers having 1,000 nodes. 
The authors report that increasing the number of hidden units does not appear to improve the model performance.

The system, summarized in Table~\ref{table:dimensions:deep:auto:controller},
also provides a user interface\index{User!interface},
analyzed in Section~\ref{section:interactivity},
to interactively control\index{Control} the generation,
e.g.,
selecting a given input (to be inserted at the bottleneck hidden layer),
generating a random input, and
controlling (by scaling or muting) the activation of a given unit.

\begin{table}
\begin{tabular}{|l|l|}
\hline
{\em Objective}			&Audio; User interface\\
\hline
{\em Representation}	&Audio; Spectrum\\
\hline
{\em Architecture}		&Stacked autoencoder = Autoencoder$^2$\\
\hline
{\em Strategy}			&Decoder feedforward\\
\hline
\end{tabular}
\caption{deepAutoController summary}
\label{table:dimensions:deep:auto:controller}
\end{table}

\subsection{Sampling}
\label{section:challenges:strategies:sampling}
\label{section:challenges:strategies:input:less:rbm}
\label{section:input:less:rbm}
\label{section:strategy:sampling}

Another strategy is based on sampling.
{\em Sampling\index{Sampling}} is the action of generating an element (a {\em sample\index{Sample}})
from a {\em stochastic\index{Stochastic}} model according to a {\em probability distribution\index{Probability!distribution}}.

\subsubsection{Sampling Basics}
\label{section:challenges:strategies:sampling:basics}
\label{section:sampling:basics}

The main issue for sampling is to ensure that the samples generated match a given distribution\index{Distribution}.
The basic idea is to generate a sequence of sample values in such a way that,
as more and more sample values are generated, the distribution of values more closely approximates the target distribution.
Sample values are thus produced {\em iteratively},
with the distribution of the next sample being dependent only on the current sample value.
Each successive sample is generated through a {\em generate-and-test\index{Generate!-and-test}} strategy,
i.e. by generating a prospective candidate, accepting or rejecting it (based on a defined {\em probability density})
and, if needed, regenerating it.
Various sampling strategies have been proposed:
Metropolis-Hastings\index{Metropolis-Hastings} algorithm, Gibbs sampling\index{Gibbs sampling} (GS\index{GS}),
block Gibbs sampling\index{Block!Gibbs sampling}, etc.
Please see, for example, \cite[Chapter~17]{goodfellow:deep:learning:book:2016} for more details about sampling algorithms.

%

\subsubsection{Sampling for Music Generation}
\label{section:challenges:strategies:sampling:musical:context}

For musical content, we may consider two different levels of probability distribution (and sampling):


\begin{itemize}

\item {\em item-level} or {\em vertical} dimension -- at the level of a compound musical item, e.g., a chord\index{Chord}.
In this case, the distribution is about the relations between the components of the chord, i.e. describing the probability of notes\index{Note} to occur together; and

\item {\em sequence\index{Sequence}}-level or {\em horizontal} dimension -- at the level of a sequence of items, e.g., a melody\index{Melody} composed of successive notes.
In this case, the distribution is about the sequence of notes,
i.e. it describes the probability of the occurrence of a specific note after a given note.

\end{itemize}

An RBM\index{RBM} (restricted Boltzmann machine\index{Restricted Boltzmann machine}) architecture
is generally\footnote{A counterexample is
	the C-RBM\index{C-RBM} convolutional RBM architecture,
	to be introduced in Section~\ref{section:systems:c-rbm},
	which models both the vertical dimension (simultaneous notes) and the horizontal dimension (sequence of notes)
	for single-voice polyphonies.}
used to model the vertical dimension, i.e. which notes should be played together.
As noted in Section~\ref{section:architecture:rbm},
an RBM architecture is dedicated to learning distributions and can learn efficiently from few examples.
This is particularly interesting for learning and generating chords,
as the combinatorial nature of possible notes forming a chord is large and the number of examples is usually small.
An example of a {\em sampling strategy\index{Sampling!strategy}} applied on an RBM for the horizontal dimension
will be presented in Section~\ref{section:experiment:rbm}.

An RNN\index{RNN} (recurrent neural network\index{Recurrent!neural network}) architecture is often used for the horizontal dimension,
i.e. which note is likely to be played after a given note,
as will be described in Section~\ref{section:challenges:strategies:iterative:forward}.
As we will see in Section~\ref{section:challenges:strategies:variability:sampling},
a sampling strategy may be also added to enforce variability.

We will see in Section~\ref{section:experiment:rnn:rbm} that
a compound architecture\index{Compound architecture} named RNN-RBM\index{RNN-RBM}
may combine and {\em articulate}\footnote{This issue of how to articulate vertical and horizontal dimensions, i.e. harmony\index{Harmony} with melody\index{Melody},
	will be further analyzed in Section~\ref{section:challenges:strategies:melody:harmony:consistency}.}
these two different approaches:

\begin{itemize}

\item an RBM architecture with a sampling strategy for the vertical dimension; and

\item an RNN architecture with an iterative feedforward strategy for the horizontal dimension.

\end{itemize}

An alternative approach is to use sampling as the {\em unique} strategy for both dimensions, as witnessed by the DeepBach\index{DeepBach} system
to be analyzed in Section~\ref{section:experiment:deep:bach}.

\subsubsection{Example: RBM-based Chord Music Generation System}
\label{section:harmonization}
\label{section:experiment:rbm}

In \cite{boulanger:temporal:dependencies:icml:2012},
Boulanger-Lewandowski {\em et al.}
propose to use a
restricted Boltzmann machine (RBM)
\cite{hinton:rbm:science:2006}
to model polyphonic music.
Their
objective is actually
to improve the transcription\index{Transcription}
of polyphonic music from audio.
But prior to that, the authors discuss the generation of samples from the model that has been learnt
as a qualitative evaluation and also for music generation
\cite{boulanger:rnn:rbm:generation:2015}.
In their first experiment, the RBM learns from the corpus the distribution of possible simultaneous notes, i.e. a repertoire of chords.

The corpus is the set of J. S. Bach\index{Bach}'s chorales (as for MiniBach, described in Section~\ref{section:experiment:mini:bach}).
The polyphony (number of simultaneous notes) varies from 0 to 15 and the average polyphony is 3.9.
The input representation has 88 binary visible units that span the whole range of piano from A$_0$ to C$_8$,
following a many-hot encoding.
The sequences are aligned (transposed) onto a single common tonality (e.g., C major/minor) to ease the learning process.
 
\begin{figure}
\includegraphics[scale=0.35]{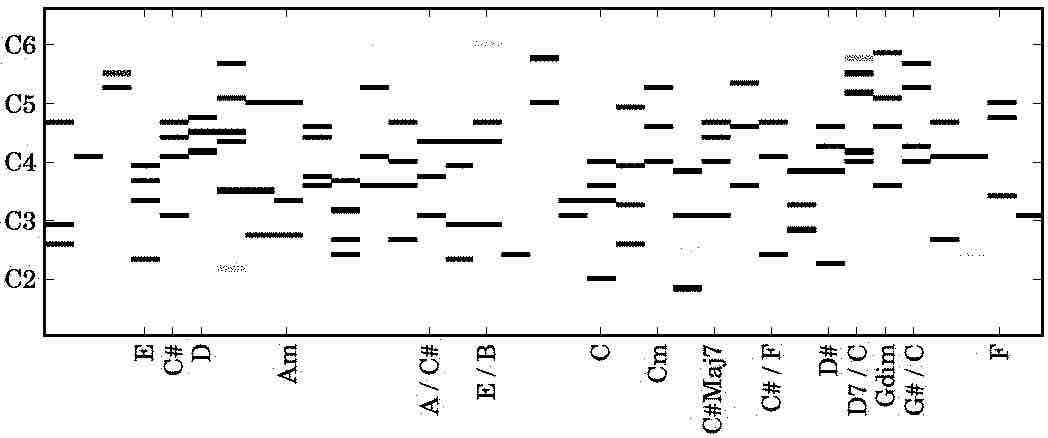}
\caption{Samples generated by the RBM trained on J. S. Bach chorales.
Reproduced from \cite{boulanger:temporal:dependencies:icml:2012} with permission of the authors}
\label{figure:rbm:polyphonic:example}
\end{figure}

One can sample from the RBM through block Gibbs sampling\index{Block!Gibbs sampling},
by performing alternative steps of sampling the hidden layer nodes (considered as variables) from the visible layer nodes
(see Section~\ref{section:architecture:rbm}).
Figure~\ref{figure:rbm:polyphonic:example} shows various examples of samples.
The vertical axis represents successive possible notes. 
Each column represents a specific sample composed of various simultaneous notes,
with the name of the chord written below when the analysis is unambiguous.
Table~\ref{table:dimensions:rbm:chords} summarizes this RBM-based chord generation system, which we notate RBM$_C$
(where {\small $C$} stands for chords).

\begin{table}
\begin{tabular}{|l|l|}
\hline
{\em Objective}			&Simultaneous notes (Chord)\\
\hline
{\em Representation}	&Symbolic; Many-hot\\
\hline
{\em Architecture}		&RBM\\
\hline
{\em Strategy}			&Sampling\\
\hline
\end{tabular}
\caption{RBM$_C$ summary}
\label{table:dimensions:rbm:chords}
\end{table}

\section{Length Variability}
\label{section:challenges:strategies:variable:length}
\label{section:variable:length}

An important limitation of the
single-step feedforward strategy (Section~\ref{section:single:step:feedforward})
and of the decoder feedforward strategy (Section~\ref{section:input:less:decoding})
is that the length\index{Length} of the music generated (more precisely the number of times steps or measures) is {\em fixed}.
It is actually fixed by the architecture, namely the number of nodes of the output layer\footnote{In the case of an RBM\index{RBM},
	the number of nodes of the input layer (which also has the role of an output layer).}.
To generate a longer (or shorter) piece of music,
one needs to reconfigure the architecture and its corresponding representation.

\subsection{Iterative Feedforward}
\label{section:challenges:strategies:iterative:forward}
\label{section:strategy:iterative:feedforward}
\label{section:challenges:strategies:input:less:rnn}
\label{section:input:less:rnn}
\label{section:challenges:strategies:input:less:rnn:iterative:generation}

The standard solution to this limitation is to use a recurrent neural network (RNN).
%
The typical usage, as initially described for text generation by Graves in \cite{graves:generating:sequences:rnn:arxiv:2014},
is to

\begin{itemize}

\item select some {\em seed\index{Seed}} information as the {\em first} item (e.g., the first note of a melody);

\item {\em feedforward} it into the recurrent network in order to produce the {\em next} item (e.g., next note);

\item use this next item as the next input to produce the {\em next next} item; and

\item repeat this process iteratively until a {\em sequence} (e.g., of notes, i.e. a melody) of the desired length is produced.

\end{itemize}

Note the {\em iterative}
aspect of the generation,
processed element by element.
Therefore, we name this approach the {\em iterative time step feedforward} strategy,
abbreviated as the {\em iterative feedforward} strategy\index{Iterative feedforward strategy}.
Actually, a {\em recursion\index{Recursion}}
-- current output reenters as the next input --
is also often present.
However, there are a few rare exceptions, as we will see, e.g.,
in Sequential\index{Sequential} (Section~\ref{section:experiment:todd:sequential})
and in BLSTM\index{BLSTM} (Section~\ref{section:experiment:blstm:chord}) architectures,
where there is an iteration but {\em no} recursion.


Note that the iterative feedforward strategy,
as the decoder feedforward strategy\index{Decoder!feedforward strategy} (Section~\ref{section:strategy:decoder:feedforward}),
is one kind of {\em seed-based generation\index{Seed!-based generation}}
(see Section~\ref{section:challenges:strategies:input:less}),
as the full sequence (e.g., a melody) is generated iteratively from an initial seed\index{Seed} item (e.g., a starting note).

\subsubsection{\#1 Example: Blues Chord Sequence Symbolic Music Generation System}
\label{section:experiment:eck:blues:lstm:first:experiment}
\label{section:experiment:eck:blues:lstm}

In \cite{eck:composition:lstm:2002},
Eck and
Schmidhuber describe a double experiment
undertaken with a recurrent network architecture using LSTMs\footnote{This was actually the first experiment
	in using LSTMs to generate music.}.
In their first experiment, the objective is to learn and generate chord sequences.
The format of representation is piano roll, with two types of sequences: melody and chords,
although chords are represented as notes.
The melodic range as well as the chord vocabulary is strongly constrained,
as the corpus consists of 12 measures long blues\index{Blues}
and is handcrafted (melodies and chords). 
The 13 possible notes extend from middle C (C$_4$) to tenor
C (C$_5$).
The 12 possible chords extend from C to B.



\begin{figure}
\includegraphics[width=\textwidth]{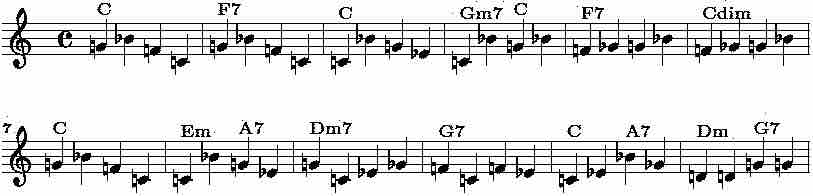}
\caption{A chord training example for blues generation.
Reproduced from \cite{eck:composition:lstm:2002} with permission of the authors}
\label{figure:eck:chord:training:example}
\end{figure}

A one-hot encoding is used.
Time quantization (time step) is set at the eighth note, half of the minimal note duration used in the corpus,
which is a quarter note.
With 12 measures long music this equates to 96 time steps.
An example of chord sequence training example is shown in Figure~\ref{figure:eck:chord:training:example}.

The architecture for this first experiment is:
an input layer with 12 nodes (corresponding to a one-hot encoding of the 12 chord vocabulary),
a hidden layer with four LSTM blocks containing two cells each\footnote{See
	in Section~\ref{section:architecture:lstm}
	for the difference between LSTM cells and blocks.}
and an output layer with 12 nodes (identical to the input layer).

\begin{figure}
\includegraphics[scale=0.7]{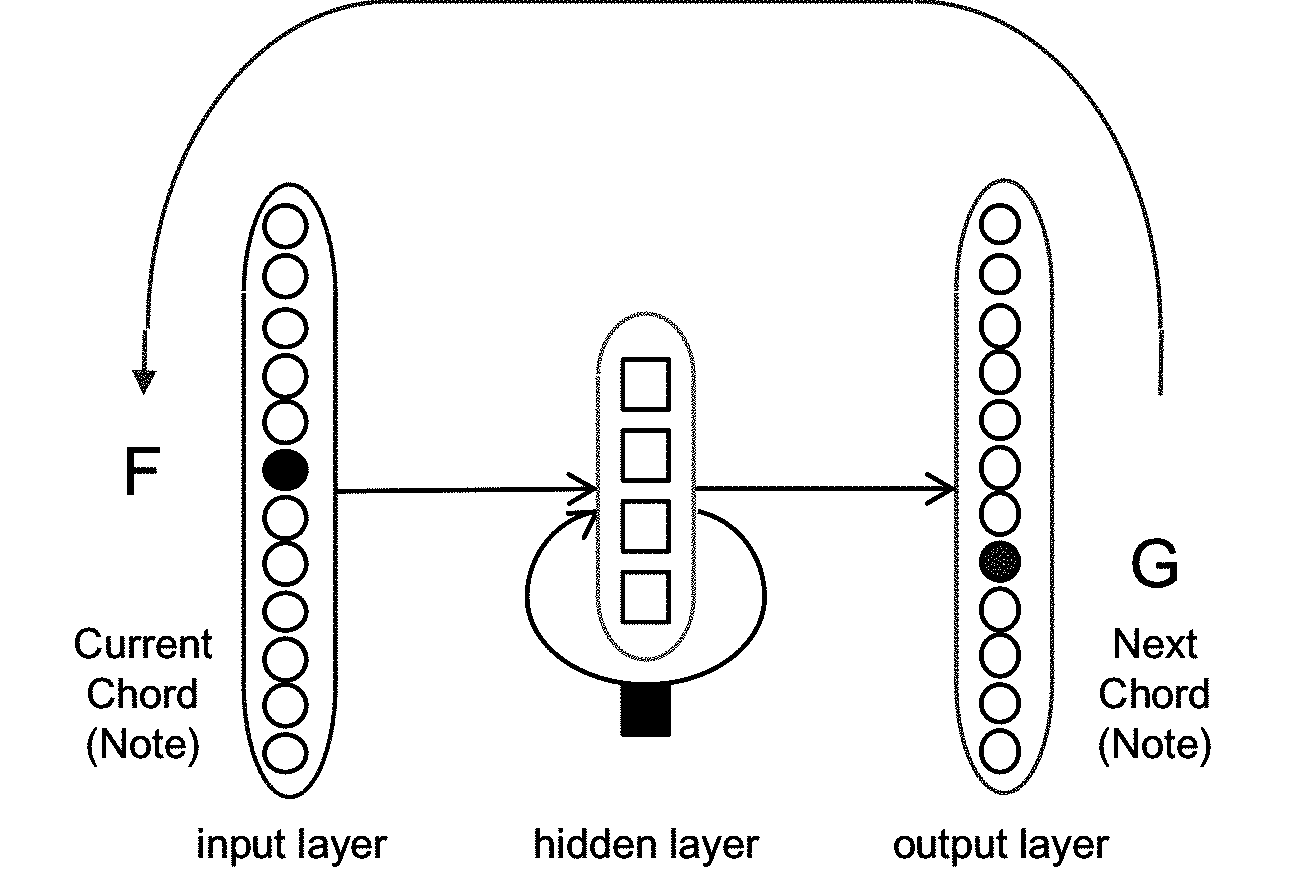}
\caption{Blues chord generation architecture}
\label{figure:eck:chord:architecture}
\end{figure}

Generation is performed by presenting a {\em seed\index{Seed}} chord (represented by a note) and by iteratively feedforwarding the network,
producing the prediction of the next time step chord, using it as the next input and so on, until a sequence of chords has been generated.
The architecture and the iterative generation is illustrated in Figure~\ref{figure:eck:chord:architecture}.
This system, which we notate Blues$_C$ (where {\small $C$} stands for chords), 
is summarized in Table~\ref{table:dimensions:blues:c}.

\begin{table}
\begin{tabular}{|l|l|}
\hline
{\em Objective}			&Chord sequence; Blues\\
\hline
{\em Representation}	&Symbolic; One-hot; Note end; Chord as note\\
\hline
{\em Architecture}		&LSTM\\
\hline
{\em Strategy}			&Iterative feedforward\\
\hline
\end{tabular}
\caption{Blues$_C$ summary}
\label{table:dimensions:blues:c}
\end{table}

\subsubsection{\#2 Example: Blues Melody and Chords Symbolic Music Generation System}
\label{section:experiment:eck:blues:lstm:second:experiment}

In
Eck and Schmidhuber's
second experiment \cite{eck:composition:lstm:2002},
the objective is to simultaneously generate melody and chord sequences.
The new architecture is an extension of the previous one:
it has an input layer with 25 nodes
(corresponding to a one-hot encoding of the 12 chord vocabulary and
to a one-hot encoding of the 13 melody note vocabulary),
a hidden layer with eight LSTM blocks
(four chord blocks and four melody blocks,
as we will see below),
containing two cells each,
and an output layer with 25 nodes (identical to the input layer).

The separation between chords and melody is ensured as follows:

\begin{itemize}

\item chord blocks are fully connected to the input nodes and to the output nodes corresponding to chords;

\item melody blocks are fully connected to the input nodes and to the output nodes corresponding to melody;

\item chord blocks have recurrent connections to themselves {\em and} to the melody blocks; and

\item melody blocks have recurrent connections {\em only} to themselves.

\end{itemize}

\begin{figure}
\includegraphics[width=\textwidth]{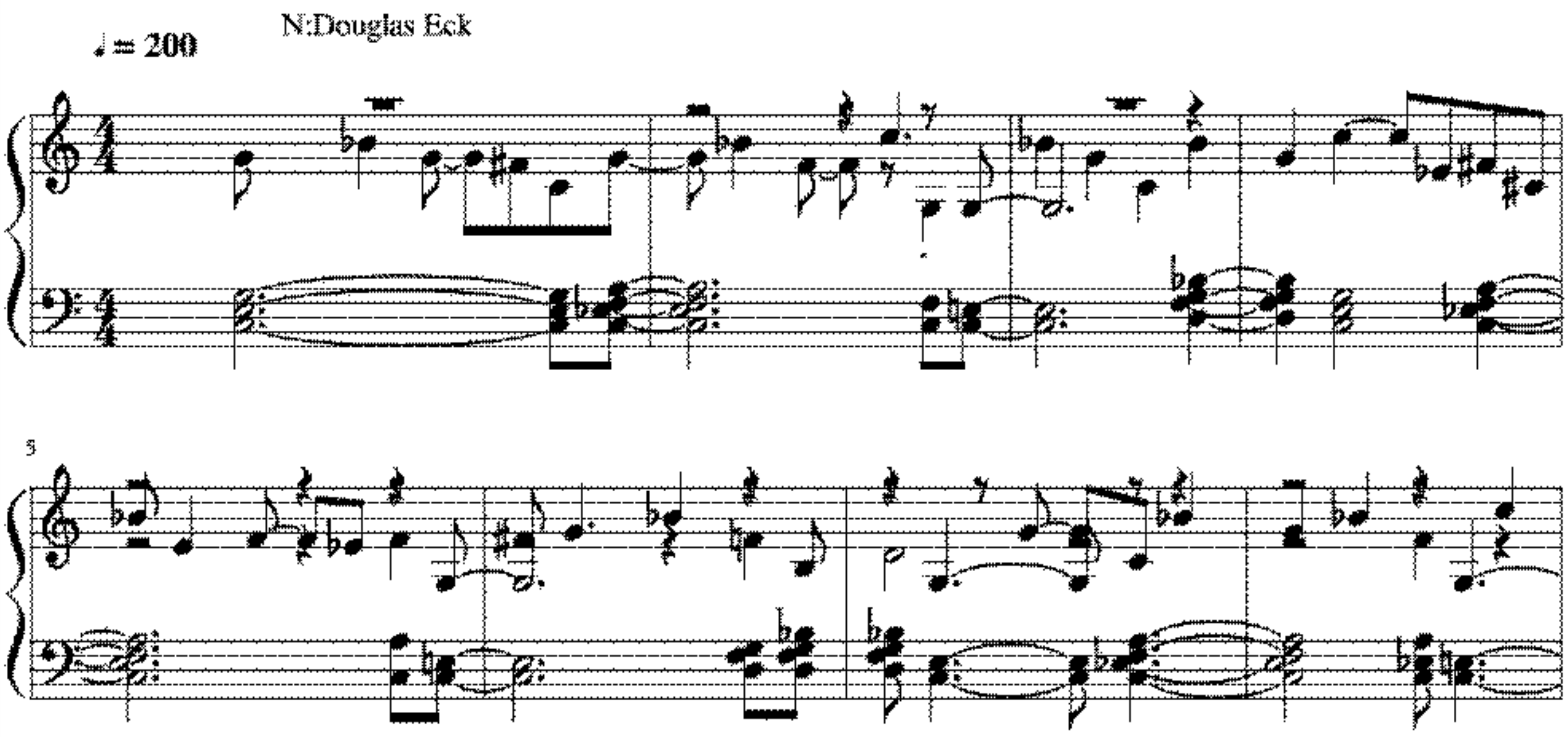}
\caption{Example of blues generated (excerpt).
Reproduced with permission of the authors}
\label{figure:eck:example}
\end{figure}

Generation is performed by presenting a seed (note and chord) and by recursively feedforwarding it into the network,
producing the prediction of the next time step note and chord, and so on, until a sequence of notes with chords is generated.
Figure~\ref{figure:eck:example} shows an example of the melody and chords generated.
Table~\ref{table:dimensions:blues:m:c} summarizes this second system,
which we notate Blues$_{MC}$ (where {\small $MC$} stands for melody and chords).

\begin{table}
\begin{tabular}{|l|l|}
\hline
{\em Objective}			&Melody + Chords; Blues\\
\hline
{\em Representation}	&Symbolic; One-hot$\times$2; Note end; Chord as note\\
\hline
{\em Architecture}		&LSTM\\
\hline
{\em Strategy}			&Iterative feedforward\\
\hline
\end{tabular}
\caption{Blues$_{MC}$ summary}
\label{table:dimensions:blues:m:c}
\end{table}

This second experiment is interesting in that it {\em simultaneously} generates melody {\em and} chords.
Note that in this second architecture, recurrent connexions are {\em asymmetric\index{Asymmetry}}
as the authors wanted to ensure the preponderant role of chords.
Chord blocks have recurrent connexions to themselves but also to melody blocks, whereas melody blocks do not have recurrent connexions to chord blocks.
This means that chord blocks will receive previous step information about chords {\em and} melody,
whereas melody blocks cannot use previous step information about chords.
This somewhat {\em ad hoc\index{Ad hoc}} configuration of the recurrent connexions in the architecture
is a way to control the interaction between harmony and melody in a master-slave manner.
The control of the interaction and consistency between melody and harmony is indeed an effective issue
and it will be further addressed in Section~\ref{section:challenges:strategies:melody:harmony:consistency}
where we will analyze alternative approaches.

%


\section{Content Variability}
\label{section:challenges:strategies:variability}

A limitation of the iterative feedforward strategy on an RNN,
as illustrated by the blues generation experiment described in Section~\ref{section:experiment:eck:blues:lstm:second:experiment},
is that generation is {\em deterministic}.
Indeed, a neural network is deterministic\footnote{There are stochastic versions of artificial neural networks -- an RBM is an example --
	but they are not mainstream.}.
As a consequence, feedforwarding the {\em same input} will always produce the {\em same output}.
As the generation of the next note, the next next note, etc., is deterministic,
the {\em same} seed note will lead to the {\em same} generated series of notes\footnote{The actual length of the melody generated
	depends on the number of iterations.}.
Moreover, as there are only 12 possible input values (the 12 pitch classes), there are only 12 possible melodies.

\subsection{Sampling}
\label{section:challenges:strategies:variability:sampling}
\label{section:variability:sampling}

\begin{figure}
\includegraphics[scale=0.8]{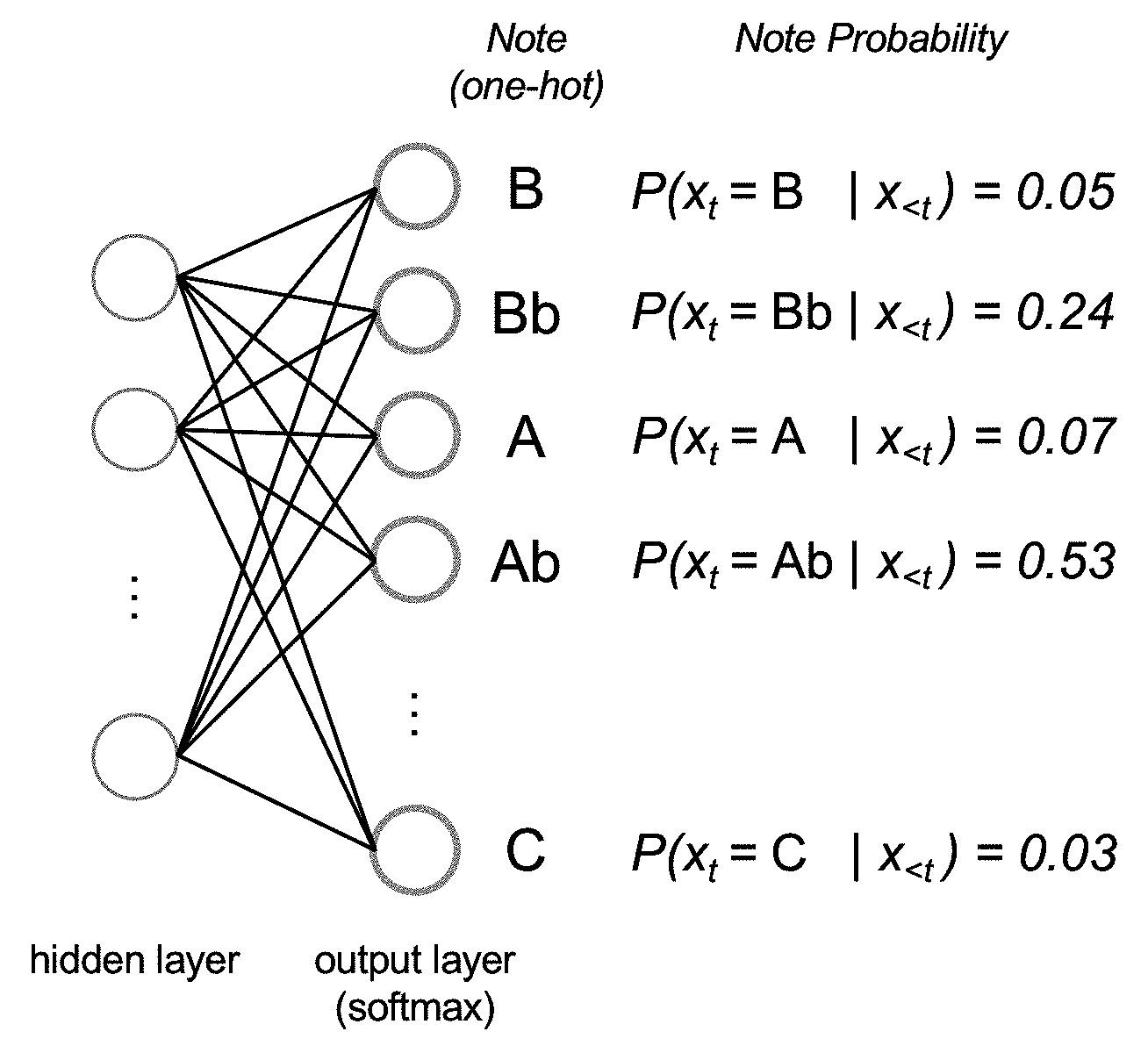}
\caption{Sampling the softmax output}
\label{figure:sampling:output}
\end{figure}

Fortunately, as we will see, the usual solution is quite simple.
The assumption is that the output representation of the melody is one-hot encoded.
In other words, the output representation is of a piano roll type,
the output activation layer is softmax
and generation is modeled as a classification\index{Classification!task} task.
See an example in Figure~\ref{figure:sampling:output},
where $P(\text{x}_t = \text{C} | \text{x}_{<t})$
represents the conditional probability for the element (note) x$_t$ at step $t$ to be a C
given the previous elements x$_{<t}$ (the melody generated so far).

The default {\em deterministic} strategy consists in choosing the class (the note) with the {\em highest probability},
i.e. $\text{argmax}\index{Argmax}_{\text{x}_t} P(\text{x}_t | \text{x}_{<t})$, that is A$\flat$ in
Figure~\ref{figure:sampling:output}.
We can then easily switch to a {\em nondeterministic} strategy,
by {\em sampling} the output which corresponds (through the softmax function)
to a probability distribution between possible notes.
By sampling a note following the distribution generated\footnote{The chance of sampling a given class/note
	is its corresponding probability. In the example shown in Figure~\ref{figure:sampling:output},
	A$\flat$ has around one chance in two of being selected and B$\flat$ one chance in four.},
we introduce {\em stochasticity} in the process and thus {\em variability} in the generation.

\subsubsection{\#1 Example: CONCERT Bach Melody Symbolic Music Generation System}
\label{section:experiment:concert}

CONCERT (an acronym for CONnectionist Composer of ERudite Tunes)
developed by Mozer \cite{mozer:composition:prediction:1994} in 1994,
was actually one of the first systems for generating music based on recurrent networks (and before LSTM).
It is aimed at generating melodies, possibly with some chord progression as an accompaniment.

\begin{figure}
\includegraphics[scale=0.6]{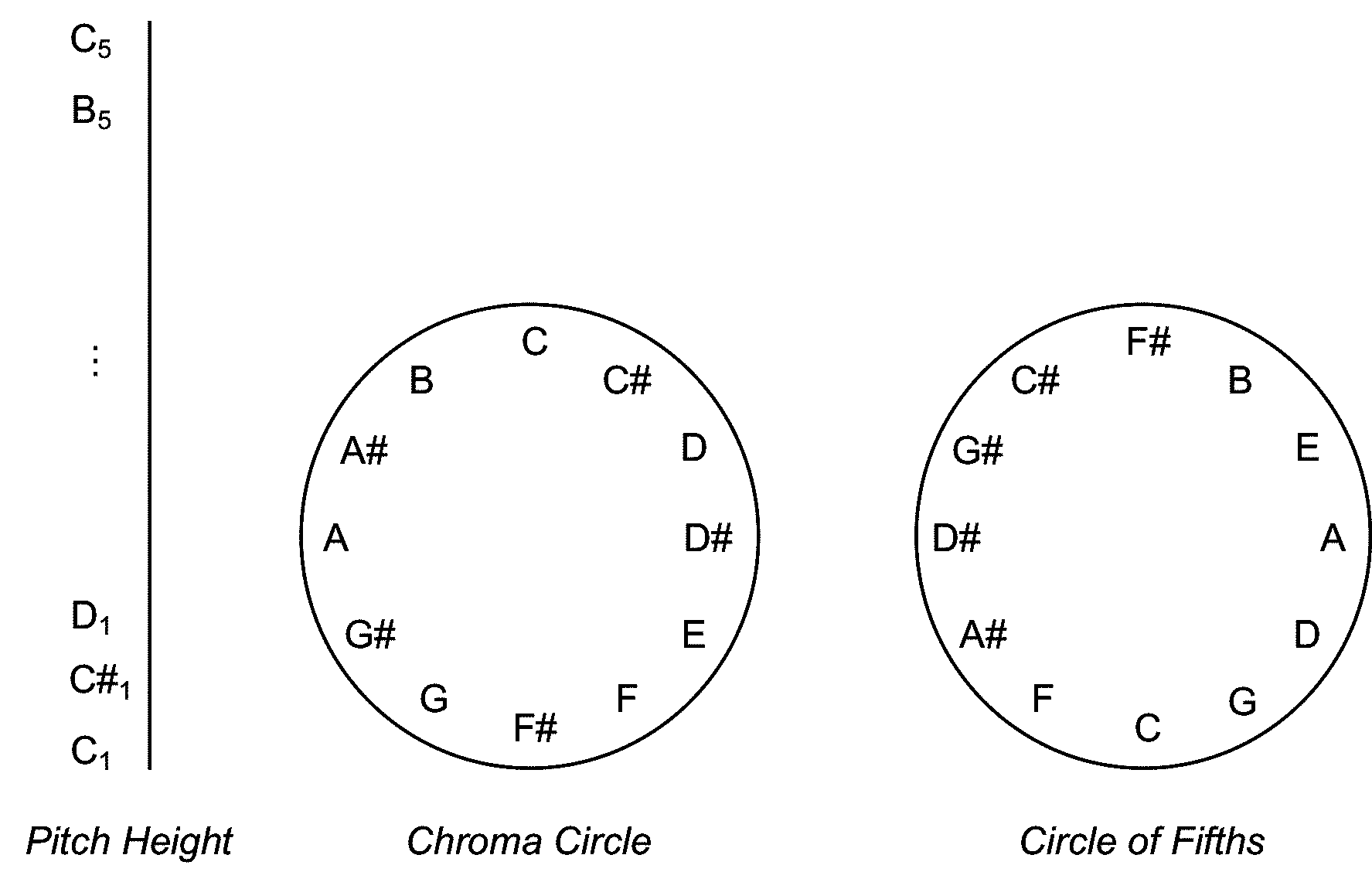}
\caption{CONCERT PHCCCH pitch representation.
Inspired by \cite{shepard:geometric:pitch:psychological:review:1982} and \cite{mozer:composition:prediction:1994}}
\label{figure:concert:pitch:representation}
\end{figure}

The input and output representation includes three aspects of a note:
pitch, duration and {\em harmonic chord accompaniment}.
The representation of a pitch, named PHCCCH,
is inspired by the psychological pitch representation space of Shepard
\cite{shepard:geometric:pitch:psychological:review:1982},
and is based on five dimensions, as illustrated in Figure~\ref{figure:concert:pitch:representation}.

The three main components are as follows:
\begin{itemize}

\item the pitch height (PH),

\item the (modulo) chroma circle (CC) cartesian coordinates, and

\item the (harmonic) circle of fifths (CH) cartesian coordinates.

\end{itemize}

The motivation is in having a more musically meaningful representation of the pitch
by capturing the similarity of octaves and also the harmonic similarity between a note and its fifth.
The proximity of two pitches is determined by computing the Euclidean distance between their PHCCCH representations, that distance being invariant under transposition.
The encoding of the pitch height is through a scalar variable scaled to range from -9.798 for C$_1$ to +9.798 for C$_5$.
The encoding of the chroma circle and of the circle of fifths is through a six binary value vector, for the reasons detailed in \cite{mozer:composition:prediction:1994}.
The resulting encoding includes 13 input variables, with some examples shown in Table~\ref{table:concert:pitch:representation}.
Note that a rest is encoded as a pitch with a unique code.

\begin{table}
\begin{tabular}{|l||l||c|c|c|c|c|c||c|c|c|c|c|c|}
\hline
\multicolumn{1}{|c||}{\em Pitch}
			&\multicolumn{1}{c||}{\em PH}
						&\multicolumn{6}{c||}{\em CC}		&\multicolumn{6}{c|}{\em CH}\\
\hline
C$_1$		&-9.798		&+1	&+1	&+1	&-1	&-1	&-1		&-1	&-1	&-1	&+1	&+1	&+1\\
\hline
F$\sharp_1$	&-7.349		&-1	&-1	&-1	&+1	&+1	&+1		&+1	&+1	&+1	&-1	&-1	&-1\\
\hline
G$_2$		&-2.041		&-1	&-1	&-1	&-1	&+1	&+1		&-1	&-1	&-1	&-1	&+1	&+1\\
\hline
C$_3$		&0			&+1	&+1	&+1	&-1	&-1	&-1		&-1	&-1	&-1	&+1	&+1	&+1\\
\hline
D$\sharp_3$	&1.225		&+1	&+1	&+1	&+1	&+1	&+1		&+1	&+1	&+1	&+1	&+1	&+1\\
\hline
E$_3$		&1.633		&-1	&+1	&+1	&+1	&+1	&+1		&+1	&-1	&-1	&-1	&-1	&-1\\
\hline
A$_4$		&8.573		&-1	&-1	&-1	&-1	&-1	&-1		&-1	&-1	&-1	&-1	&-1	&-1\\
\hline
C$_5$		&9.798		&+1	&+1	&+1	&-1	&-1	&-1		&-1	&-1	&-1	&+1	&+1	&+1\\
\hline
Rest			&0			&+1	&-1	&+1	&-1	&+1	&-1		&+1	&-1	&+1	&-1	&+1	&-1\\
\hline
\end{tabular}
\caption{Examples of PHCCCF pitch representation}
\label{table:concert:pitch:representation}
\end{table}

\begin{figure}
\includegraphics[scale=0.6]{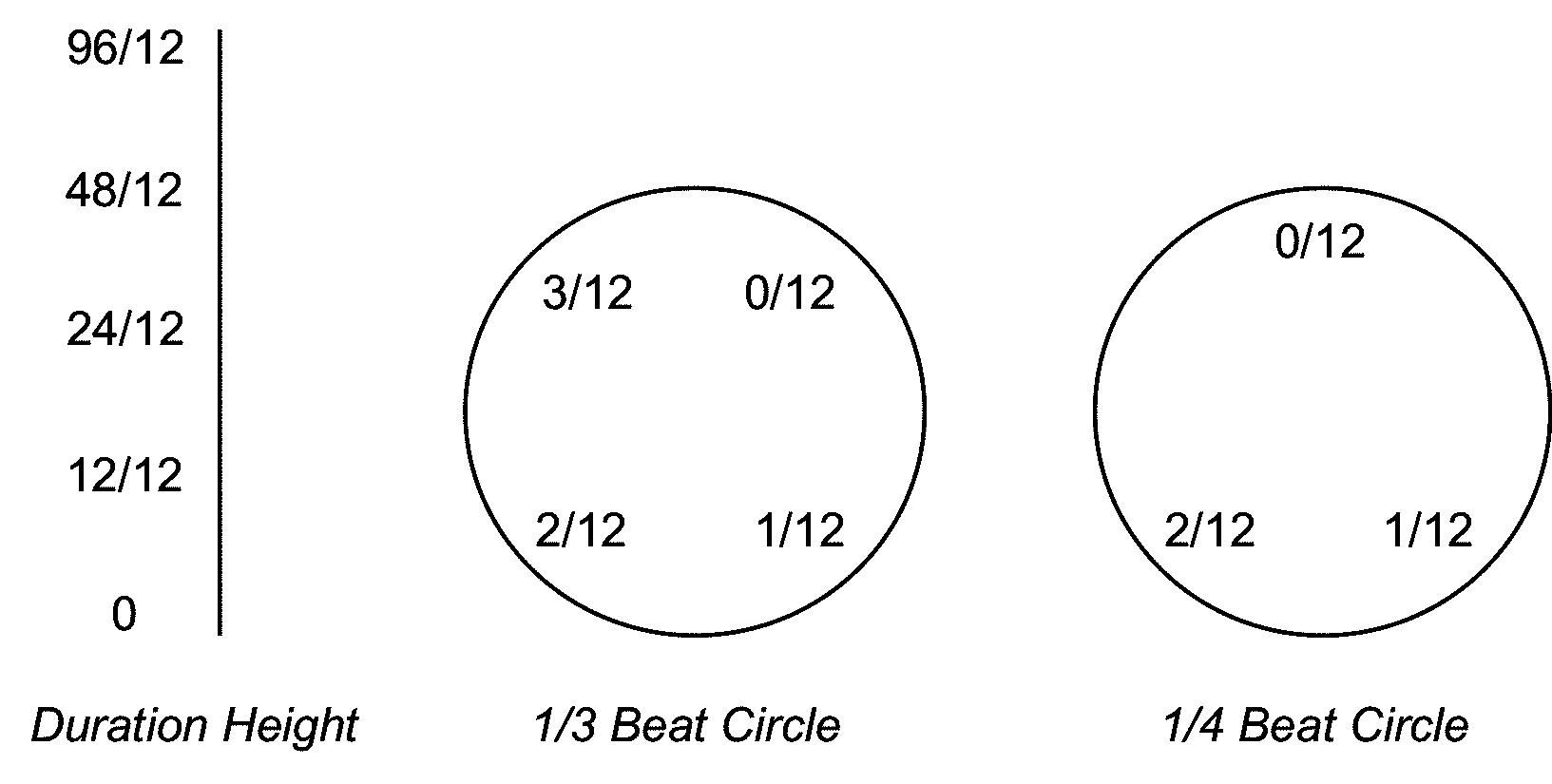}
\caption{CONCERT duration representation.
Inspired by \cite{mozer:composition:prediction:1994}}
\label{figure:concert:duration:representation}
\end{figure}

Durations are considered at a very fine-grain level, each beat (a quarter note) being divided into twelfths,
thus having a duration of 12/12.
This choice allows to represent binary (two or four divisions) as well as ternary (three divisions) rhythms.
In a similar way to the representation of pitch,
a duration is represented through a scalar and two circle coordinates, for 1/4 and 1/3 beat cycles,
as illustrated in Figure~\ref{figure:concert:duration:representation},
resulting in five dimensions directly encoded through a five binary value vector
(see more details in \cite{mozer:composition:prediction:1994}).
The temporal scope is a {\em note step\index{Note!step}},
that is the granularity of processing by the architecture is a {\em note}\footnote{And not a fixed time step
	as for most of recurrent architectures, e.g., in Section~\ref{section:experiment:eck:blues:lstm}.
	The various types of temporal scope have been introduced in Section~\ref{section:representation:temporal:scope}.}.

\begin{figure}
\includegraphics[scale=0.8]{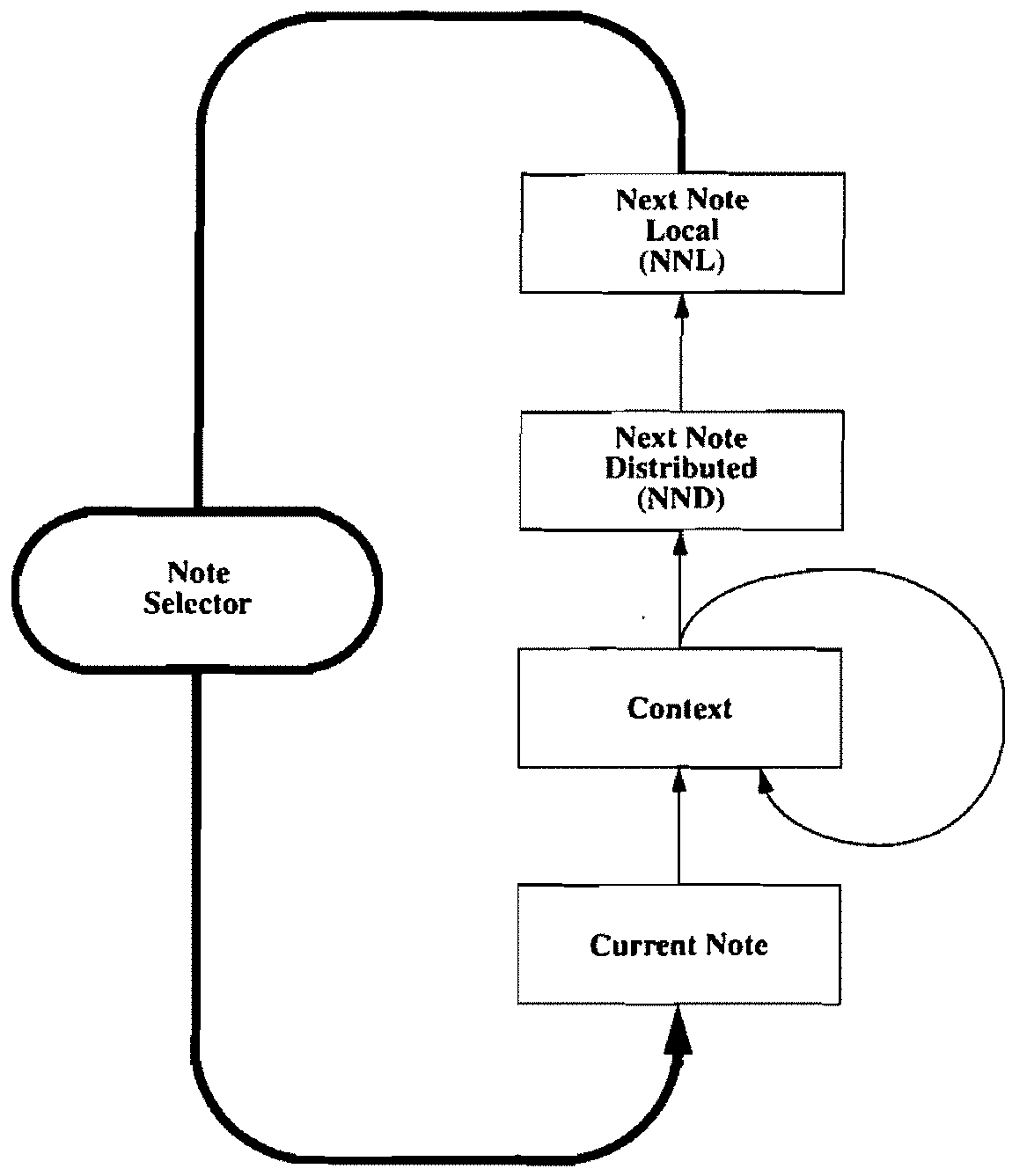}
\caption{CONCERT architecture.
Reproduced from \cite{mozer:composition:prediction:1994} with permission of Taylor \& Francis (www.tandfonline.com)}
\label{figure:concert:architecture}
\end{figure}

Chords are represented in an extensional\index{Extensional} way as a triad or a tetrachord,
through the root, the third (major or minor) and the fifth (perfect, augmented or diminished),
with the possible addition of a seventh component
(minor or major).
To represent the next note to be predicted, the CONCERT system actually uses both this rich and distributed representation
(named next-node-distributed, see Figure~\ref{figure:concert:architecture})
and a more concise and traditional representation (named next-node-local), in order to be more intelligible.
The activation function is the sigmoid function rescaled to the $[-1, +1]$ range and the cost function is mean squared error.

In the generation phase, the output is interpreted as a probability distribution\index{Probability!distribution} over a set of possible notes
as a basis for deciding the next note in a nondeterministic way,
following the {\em sampling\index{Sampling!strategy}} strategy.

\begin{figure}
\includegraphics[width=\textwidth]{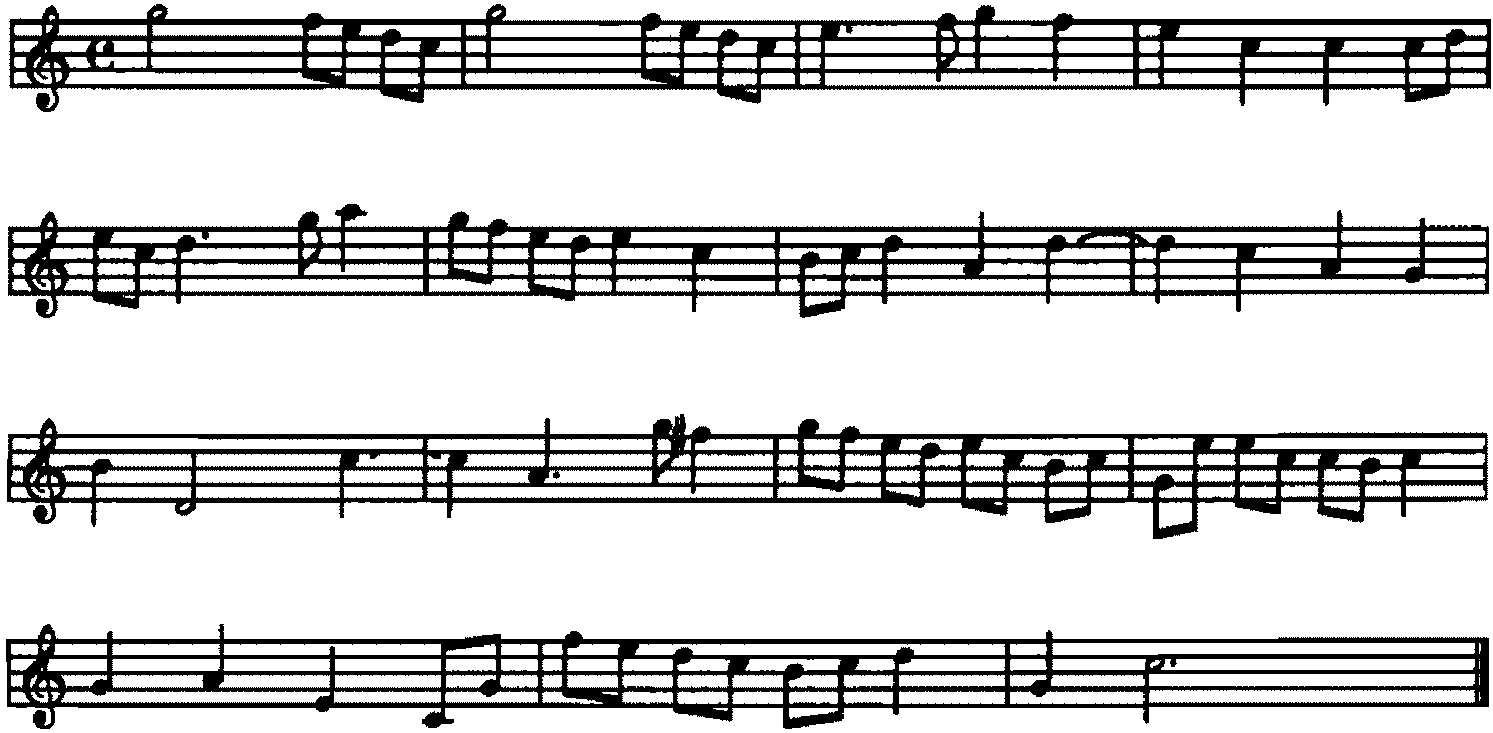}
\caption{Example of melody generation by CONCERT based on the J. S. Bach training set.
Reproduced from \cite{mozer:composition:prediction:1994} with permission of Taylor \& Francis (www.tandfonline.com)}
\label{figure:concert:example}
\end{figure}

CONCERT has been tested on different examples, notably after training with melodies of J. S. Bach\index{Bach}.
Figure~\ref{figure:concert:example} shows an example of a melody generated based on the Bach training set.
Although now a bit dated, CONCERT has been a pioneering model and the discussion in the article about representation issues
is still relevant.

Note also that CONCERT (which is summarized in Table~\ref{table:dimensions:concert})
is representative of the early generation systems, before the advent of deep learning architectures,
when representations\index{Representation} were designed with rich handcrafted features\index{Handcrafted!feature}.
One of the benefits of using deep learning architectures is that this kind of rich and deep representation
may be automatically extracted and managed by the architecture.

\begin{table}
\begin{tabular}{|l|l|}
\hline
{\em Objective}			&Melody + Chords\\
\hline
{\em Representation}	&Symbolic; Harmonics; Harmony; Beat\\
\hline
{\em Architecture}		&RNN\\
\hline
{\em Strategy}			&Iterative feedforward; Sampling\\
\hline
\end{tabular}
\caption{CONCERT summary}
\label{table:dimensions:concert}
\end{table}

\subsubsection{\#2 Example: Celtic Melody Symbolic Music Generation System}
\label{section:experiment:sturm:celtic:lstm}

Another representative example is the system by Sturm {\em et al.} to generate Celtic\index{Celtic} music melodies \cite{sturm:celtic:melody:csmc:2016}.
The architecture used is a recurrent network with three hidden layers,
which we could notate\footnote{Note that,
	as explained in Section~\ref{section:architecture:feedforward:depth},
	we notate the number of hidden layers
	without considering the input layer.}
as LSTM$^3$,
with 512 LSTM cells in each layer.

The corpus comprises folk\index{Folk} and Celtic monophonic melodies retrieved from a repository and discussion platform named The Session \cite{web:the:session}.
Pieces that were too short, too complex (with varying meters) or contained errors were filtered out,
leaving a dataset of 23,636 melodies.
All melodies are aligned (transposed) onto the single C key.
One of the specificities is that the representation chosen is {\em textual},
namely the token-based {\em folk-rnn\index{Folk!-rnn}} notation,
a transformation of the character-based ABC notation\index{ABC notation} (see Section~\ref{section:representation:text}).
The number of input and output nodes is equal to the number of tokens in the vocabulary
(i.e. with a one-hot encoding), in practice equal to 137.
The output of the network is a probability distribution over its vocabulary.

Training the recurrent network is done in an iterative way, as the network learns to predict the next item.
Once trained, the generation is done iteratively by inputing a random\index{Random} token or a specific token
(e.g., corresponding to a specific starting note),
feedforwarding it to generate the output,
sampling from this probability distribution,
and recursively using the selected vocabulary element as a subsequent input,
in order to produce a melody element by element.

\begin{figure}
\includegraphics[width=\textwidth]{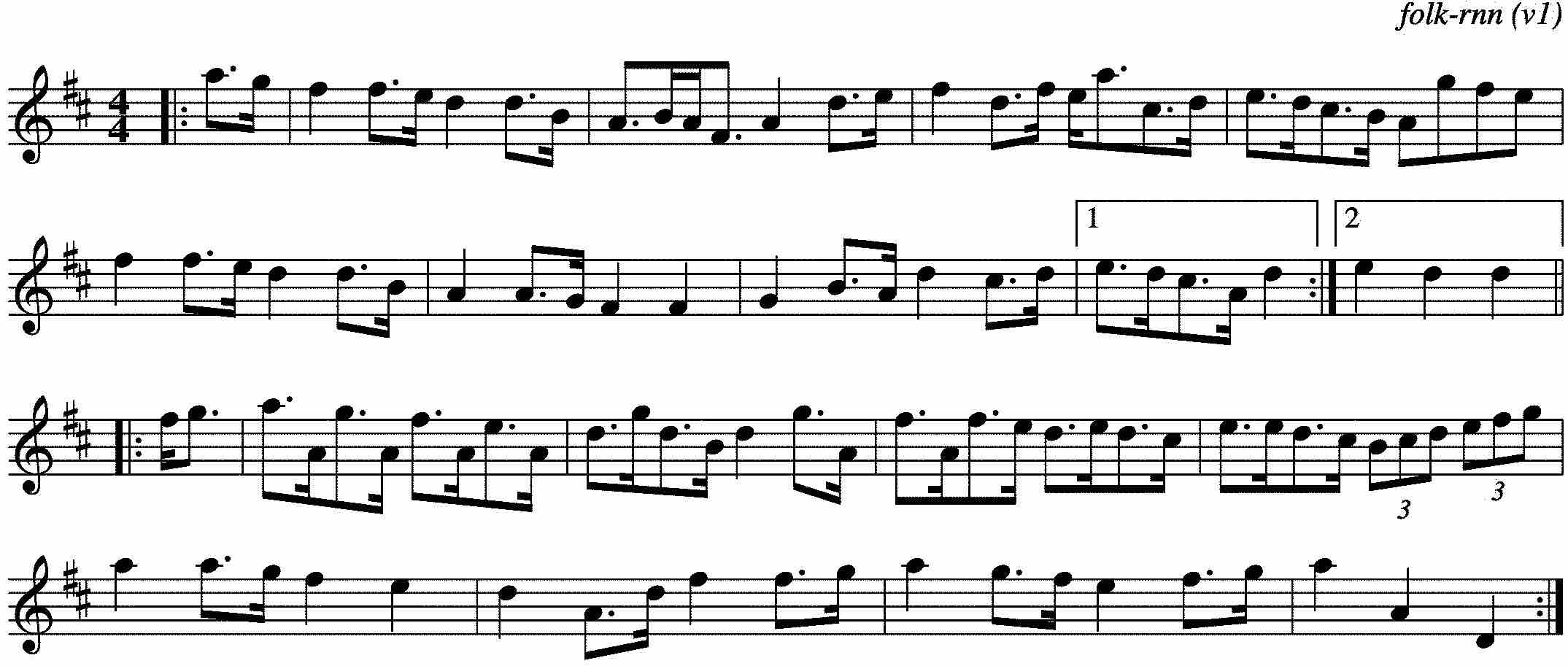}
\caption{Score of ``The Mal's Copporim'' automatically generated.
Reproduced from \cite{sturm:celtic:melody:csmc:2016} with permission of the authors}
\label{figure:score:mal:copporim}
\end{figure}

The final step is to decode the folk-rnn representation generated into a MIDI format melody to be played.
See in Figure~\ref{figure:score:mal:copporim} for an example of a melody generated.
One may also see and listen to results on \cite{web:endless:session}.
The results are very convincing, with melodies generated in a clear Celtic style.
The system is summarized in Table~\ref{table:dimensions:celtic}.

As observed in \cite{hadjeres:thesis:2018}:
``It is interesting to note that in this approach the bar lines and the repeat bar lines are given explicitly and are to be predicted as well.
This can cause some issues, since there is no guarantee that the output sequence of tokens would represent a valid song in ABC format.
There could be too many notes in one bar for example, but according to the authors, this rarely occurs.
This would tend to show that such an architecture is able to learn to count.''\footnote{On this issue,
	see also \cite{gers:rnn:time:count:ijcnn:2000}.}

\begin{table}
\begin{tabular}{|l|l|}
\hline
{\em Objective}			&Melody\\
\hline
{\em Representation}	&Symbolic; Text; Token-based; One-hot\\
\hline
{\em Architecture}		&LSTM$^3$\\
\hline
{\em Strategy}			&Iterative feedforward; Sampling\\
\hline
\end{tabular}
\caption{Celtic system summary}
\label{table:dimensions:celtic}
\end{table}

\section{Expressiveness}
\label{section:challenges:strategies:expressiveness}

One limitation of most existing systems is that they consider fixed dynamics (amplitude) for all notes
as well as an exact quantization\index{Quantization} (a fixed tempo\index{Tempo}), which makes the music generated
too mechanical, without {\em expressiveness\index{Expressiveness}} or {\em nuance\index{Nuance}}.

A natural approach resides in considering representations recorded from real performances and not simply scores,
and therefore with musically grounded (by skilled human musicians) variations of tempo\index{Tempo} and of dynamics\index{Dynamics},
as discussed in Section~\ref{section:representation:expressiveness}.

Note that an alternative approach is to automatically {\em augment} the generated music information (e.g., a standard MIDI piece)
with slight transformations on the amplitude and/or the tempo.
An example is the Cyber-Jo\~ao\index{Cyber-Jo\~ao} system \cite{dahia:cyber:joao:sbcm:2003},
which performs bossa nova\index{Bossa nova} guitar\index{Guitar} accompaniment with expressiveness,
through automatic retrieval\index{Retrieval}\footnote{By a mixed use of
	production rules\index{Rule} and case-based reasoning\index{Case-based reasoning} (CBR\index{CBR}).}
and application of rhythmic patterns\index{Rhythmic pattern}\footnote{These patterns have been manually extracted
	from a corpus of performances\index{Performance} by the guitarist and singer Jo\~ao Gilberto,
	one of the inventors of the Bossa nova\index{Bossa nova} style.
	One could also consider automatic extraction\index{Extraction}, as, for example, in \cite{lima:rhythmic:extraction:ismir:2008}.}.

As noted in Section~\ref{section:representation:expressiveness:audio},
in the case of an audio representation, expressiveness is implicit to the representation.
However, it is difficult\footnote{But not impossible to achieve,
	regarding recent achievements made on audio source separation through deep learning techniques,
	as has been pointed out in Section~\ref{section:representation:expressiveness:audio}.}
to separately control the expressiveness (dynamics or tempo) of a single instrument or voice
as the representation is global.

\subsection{Example: Performance RNN Piano Polyphony Symbolic Music Generation System}
\label{section:experiment:performance:rnn}

In \cite{simon:performance:rnn:web:2017},
Simon and Oore present their architecture and methodology named Performance RNN\index{Performance RNN}.
It is an LSTM-based recurrent neural network architecture.
One of the specificities is in the dataset characteristics, as the corpus is composed of recorded human performances\index{Performance},
with records of exact timing\index{Timing} as well as dynamics\index{Dynamics} for each note played.
The corpus used is the Yamaha e-Piano Competition dataset,
whose participants MIDI performance records are made available to the public
\cite{yamaha:e-piano:competition:web}.
It captures more than 1,400 performances\index{Performance} by skilled pianists.
To create additional training examples, some time stretching (up to 5\% faster or slower) as well as some transposition (up to a major third) is applied.

The representation is adapted to the objective.
At first look, it resembles a piano roll with MIDI note numbers but it is actually a bit different.
Each time slice is a multi-one-hot vector of the possible values for each of the following possible events:

\begin{itemize}

\item start of a new note -- with 128 possible values (MIDI pitches),

\item end of a note -- with 128 possible values (MIDI pitches),

\item time shift -- with 100 possible values (from 10 miliseconds to 1 second), and

\item dynamics -- with 32 possible values (128 MIDI velocities quantized into 32 bins\footnote{See the description of the
	binning transformation in Section~\ref{section:representation:input:encoding}.}).

\end{itemize}

\begin{figure}
\includegraphics[width=\textwidth]{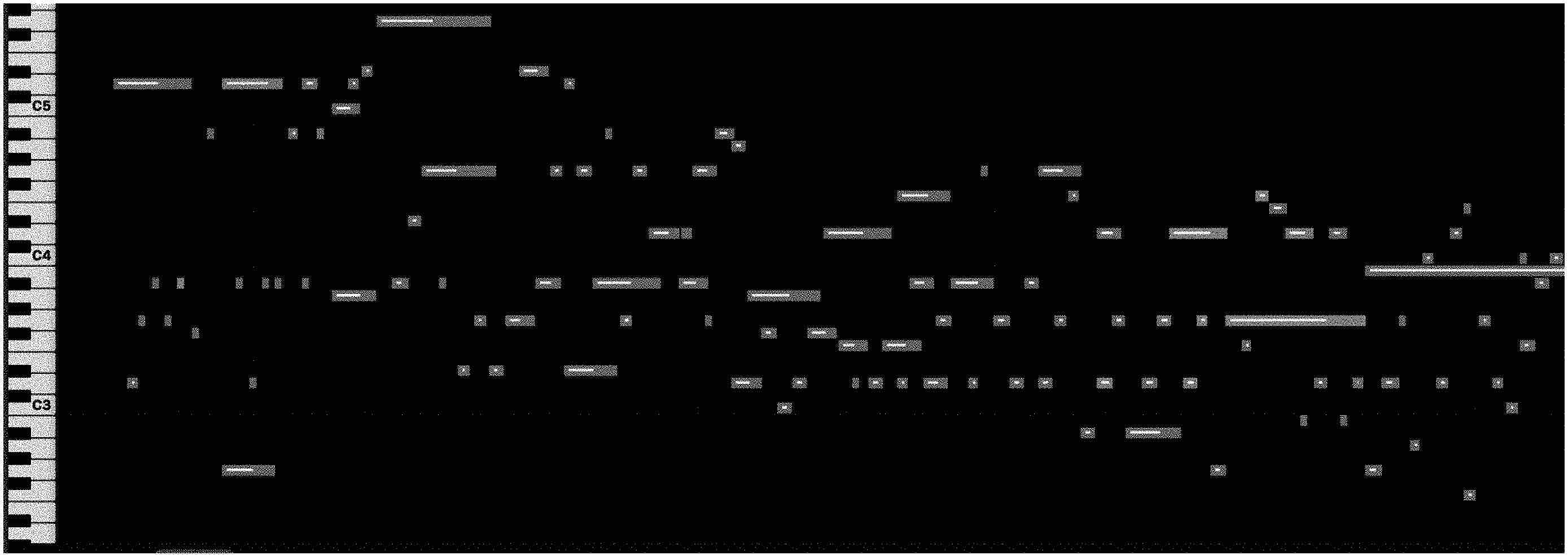}
\caption{Example of Performance RNN representation.
Reproduced from \cite{simon:performance:rnn:web:2017} with permission of the authors}
\label{figure:performance:rnn:representation:example}
\end{figure}

An example of a performance representation is shown in Figure~\ref{figure:performance:rnn:representation:example}.

Some control\index{Control} is made available to the user, referred to as the {\em temperature},
which controls the randomness\index{Randomness} of the generated events in the following way:

\begin{itemize}

\item a temperature of 1.0 uses the exact distribution\index{Distribution} predicted,

\item a value smaller than 1.0 reduces the randomness and thus increases the repetition\index{Repetition} of patterns, and

\item a larger value increases the randomness and decreases the repetition of patterns.

\end{itemize}

Examples are available on the web page \cite{simon:performance:rnn:web:2017}.
Performance RNN is summarized in Table~\ref{table:dimensions:performance:rnn}.

\begin{table}
\begin{tabular}{|l|l|}
\hline
{\em Objective}			&Polyphony; Performance control\\
\hline
{\em Representation}	&Symbolic; One-hot$\times$4; Time shift; Dynamics\\
\hline
{\em Architecture}		&LSTM\\
\hline
{\em Strategy}			&Iterative feedforward; Sampling\\
\hline
\end{tabular}
\caption{Performance RNN summary}
\label{table:dimensions:performance:rnn}
\end{table}

\section{RNN and Iterative Feedforward Revisited}
\label{section:challenges:strategies:first:discussion}
\label{section:strategies:rnn:iterative:revisited}

As we saw in previous examples,
the iterative feedforward strategy\index{Iterative feedforward strategy}
is based on the idea of the recurrent neural network (RNN) architecture to iteratively generate successive item of a sequence.
It looks like a recurrent neural network architecture and the iterative feedforward strategy are strongly coupled.
Indeed, almost all RNN-based systems use an iterative feedforward strategy and recursively\index{Recursion}
reenter the output produced (next time step generated)
into the input.
But we will introduce in this section some exceptions.


\subsection{\#1 Example: Time-Windowed Melody Symbolic Music Generation System}
\label{section:experiment:todd:time:windowed}

The experiments by Todd in \cite{todd:connectionist:composition:1989}
were one of the very first attempts (in 1989) at exploring how to use artificial neural networks to generate music.
Although the architectures he proposed are not directly used nowadays,
his experiments and discussion were pioneering and are still an important source of information. 

Todd's objective was to generate a monophonic melody in some iterative way.
He has experimented with different choices for representing
the notes (see Section~\ref{section:representation:interval}) and the durations,
but finally had decided to use a conventional pitch note representation with a one-hot encoding
and a time step temporal scope approach.
The time step is set at the duration of an eighth note.
In most of experiments, input melodies used for the training are 34 time steps long (that is, four measures and a half long),
padded at the end with rests.
A note begin is represented with a specific token and is encoded as an additional value encoding node
(see Sections~\ref{section:representation:note:ending} and~\ref{section:representation:input:encoding:hold:rest}).
Rests are not encoded explicitly but as the absence of a note, i.e. as the note one-hot encoding being all filled with null values
(see Section~\ref{section:representation:input:encoding:hold:rest}).

The first experiment is what the author named Time-Windowed\index{Time-Windowed} architecture,
where a sliding window of successive time-periods of fixed size is considered.
In practice this sliding window of a melody segment is one measure long, i.e. 8 time steps.
Its representation may be considered as a piano roll,
like in the MiniBach architecture (see Section~\ref{section:experiment:mini:bach}),
with the successive one-hot encodings of notes for the 8 successive time steps, notated as One-hot$\times$8.

\begin{figure}
\includegraphics[scale=1]{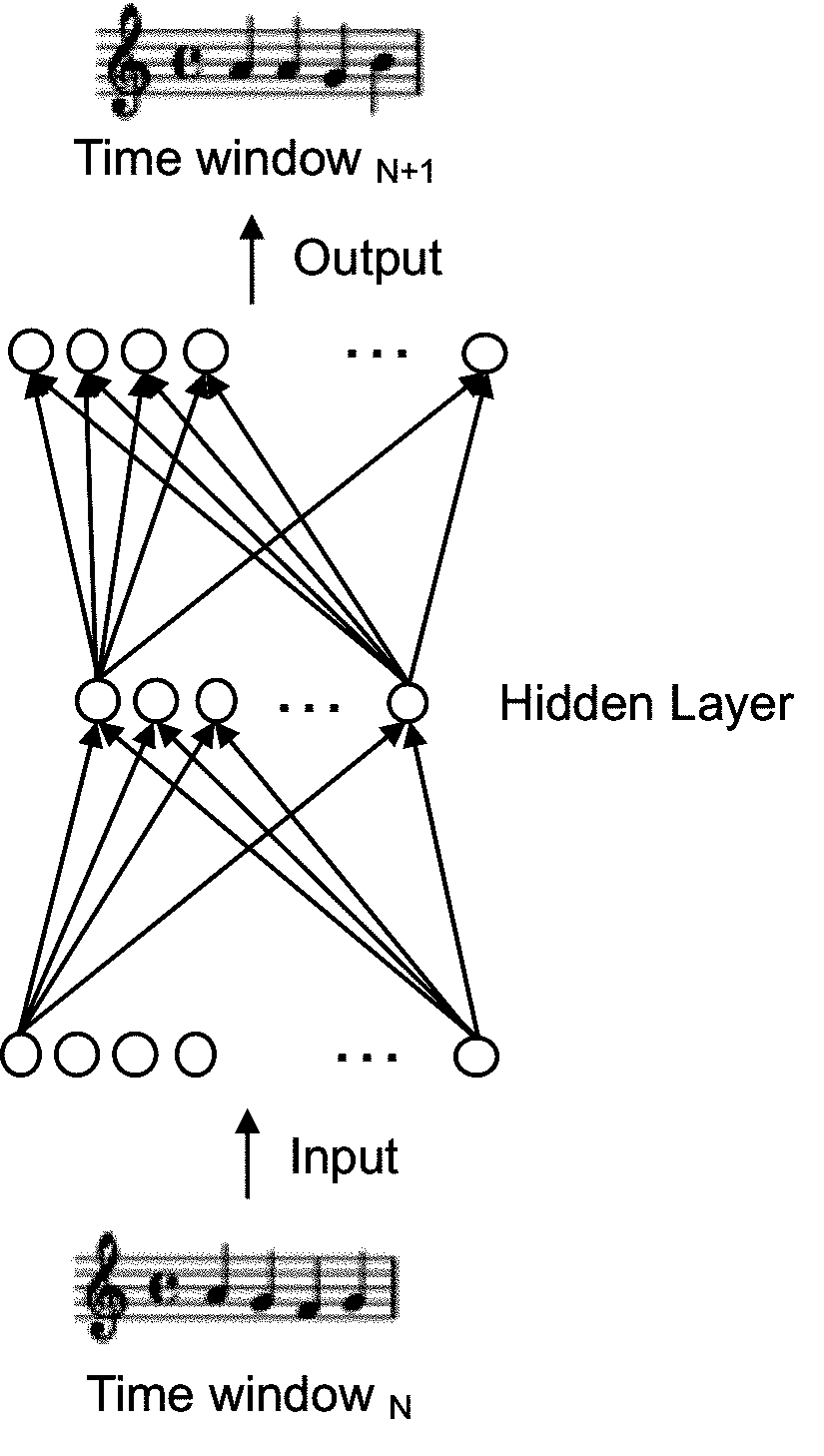}
\caption{Time-Windowed architecture.
Inspired from \cite{todd:connectionist:composition:1989}}
\label{figure:todd:time:windowed:architecture}
\end{figure}

The architecture is a feedforward network (and not an RNN),
with a melody segment as its input, the next melody segment as its output
and with a single hidden layer.
Generation is conducted iteratively (and recursively), melody segment by melody segment.
The architecture is illustrated in Figure~\ref{figure:todd:time:windowed:architecture}.

For each time step of the melody segment, the predicted note is the one with the highest probability.
Because of the zero-hot encoding of a rest (i.e. as all values being null), there is an ambiguity
between the case of every possible note has a low probability and the case of a rest (see Section~\ref{section:representation:input:encoding:hold:rest}).
For that reason, a probability threshold is introduced, namely 0.5.
Thus, the predicted note is the one with the highest probability if it is greater than 0.5
and is a rest otherwise.

The network is trained in a supervised way by presenting a melody segment as an input and its corresponding next segment
as the output, and repeating this for various segments.
Note that, as the architecture is not recurrent, although the network will learn the pairwise correlations between
two successive melody segments\footnote{In that respect, the Time-Windowed model is analog to an order 1 Markov model
	(considering only the previous state) at the level of a melody measure.},
there is no explicit memory for learning long term correlations such as in the case of a recurrent network architecture.
Thus, although the author does not show a comparison with its next experiment (see next section),
the architecture appears to have a low ability to learn long term correlations.
The Time-Windowed architecture
is summarized in Table~\ref{table:dimensions:todd:time:windowed}.

\begin{table}
\begin{tabular}{|l|l|}
\hline
{\em Objective}			&Melody\\
\hline
{\em Representation}	&Symbolic; Piano roll; One-hot$\times$8; Note begin; Implicit rest\\
\hline
{\em Architecture}		&Feedforward\\
\hline
{\em Strategy}			&Iterative feedforward\\
\hline
\end{tabular}
\caption{Time-Windowed summary}
\label{table:dimensions:todd:time:windowed}
\end{table}

\subsection{\#2 Example: Sequential Melody Symbolic Music Generation System}
\label{section:experiment:todd:sequential}

\begin{figure}
\includegraphics[scale=1]{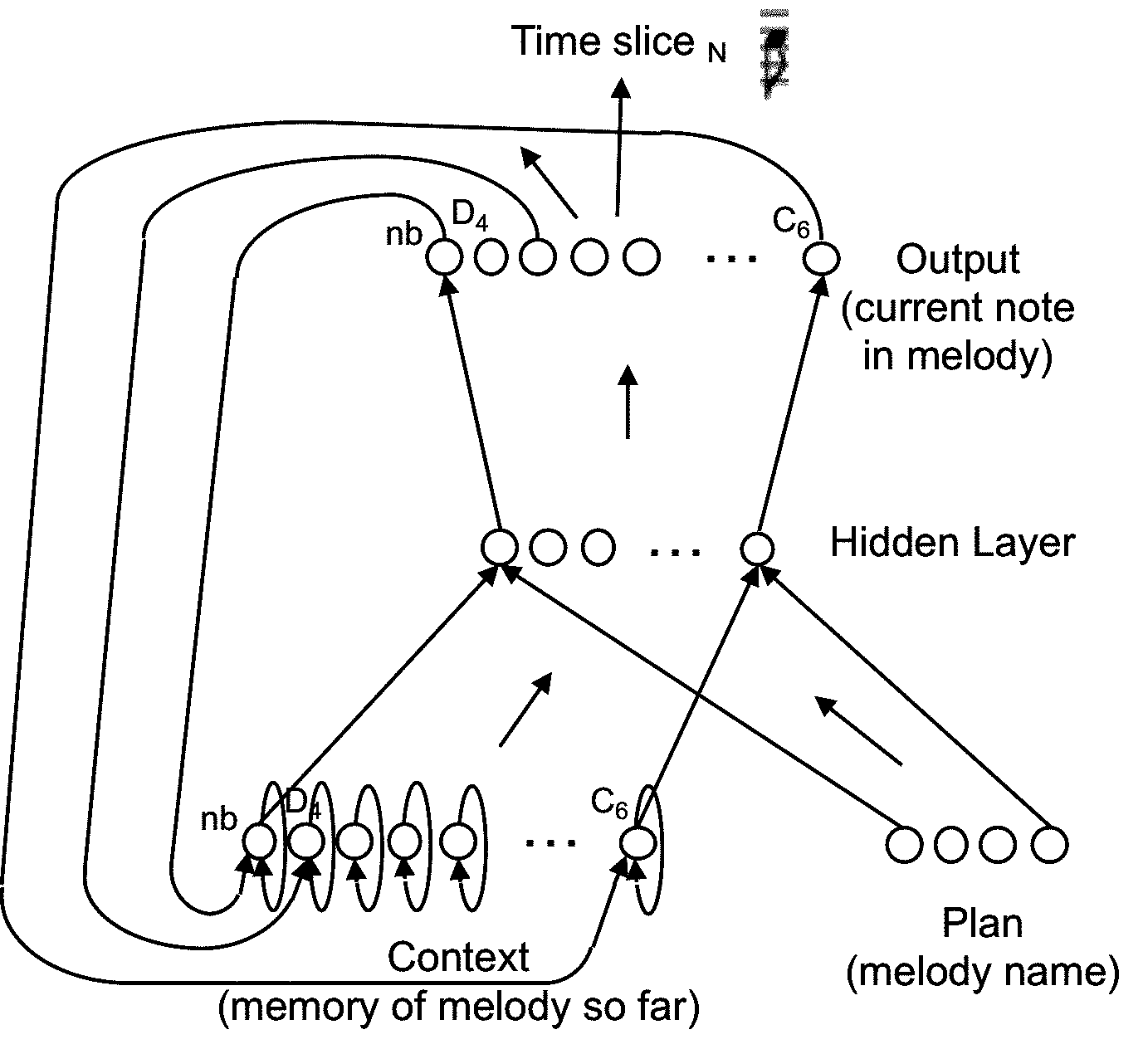}
\caption{Sequential architecture.
Inspired from \cite{todd:connectionist:composition:1989}}
\label{figure:todd:final:architecture}
\end{figure}

In \cite{todd:connectionist:composition:1989},
Todd proposed another architecture, that he named Sequential\index{Sequential}, as notes are generated in a sequence.
It is illustrated in Figure~\ref{figure:todd:final:architecture}.

The input layer is divided in two parts, named the {\em context} and the {\em plan}.
The context is the actual memory (of the melody generated so far)
and consists in units corresponding to each note (D$_4$ to C$_6$),
plus a unit about the note begin information (notated as ``nb'' in Figure~\ref{figure:todd:final:architecture}).
Therefore, it receives information from the output layer
which produces next note,
with a reentering connexion corresponding to each unit\footnote{Note that
	the output layer is isomorphic to the context layer.}.
In addition, as Todd explains it: ``A memory of more than just the single previous output (note)
is kept by having a self-feedback connection on each individual context unit.''\footnote{This is
	a peculiar characteristic of this architecture,
	as in a standard recurrent network architecture
	recurrent connexions are encapsulated within the hidden layer
	(see Figures~\ref{figure:unfolded:recurrent:layer:architecture} and~\ref{figure:lstm:cell:conceptual}).
	The argument by Todd in \cite{todd:connectionist:composition:1989}
	is that context units are more interpretable than hidden units:
	``Since the hidden units typically compute some complicated,
	often uninterpretable function of their inputs,
	the memory kept in the context units will likely also be uninterpretable.
	This is in contrast to [this] design,
	where, as described earlier,
	each context unit keeps a memory of its corresponding output unit, which is interpretable.''}
	
The plan represents a melody (among many) that the network has learnt.
Todd has experimented with various encodings, one-hot
or distributed (through a many-hot embedding).


Training is done by selecting a plan (melody) to be learnt.
The activations of the context units are initialized to 0
in order to begin with a clean empty context.
The network is then feedforwarded
and its output, corresponding to the first time step note, is compared to the first time step note of the melody to be learnt,
resulting in adjustment of the weights.
The output values\footnote{Actually, as an optimization,
	Todd proposes in the following of his description to pass back the target values and not the output values.}
are passed back to the current context.
And then, the network is feedforwarded again,
leading to the next time step note, again compared to the melody target,
and so on until the last time step of the melody.
This process is then repeated for various plans (melodies).

Generation of new melodies is conducted by feedforwarding the network with a new plan embedding, corresponding to a new melody
(not part of the training plans/melodies).
The activations of the context units are initialized to 0
in order to begin with a clean empty context.
The generation takes place iteratively, time step after time step.
Note that, as opposed to most cases of the iterative feedforward strategy (Section~\ref{section:strategy:iterative:feedforward}),
in which the output is explicitly reentered (recursively) into the input of the architecture,
in Todd's Sequential architecture the reentrance is implicit because of the specific nature of the recurrent connexions:
the output is reentered into the context units while the input -- the plan melody -- is constant.

\begin{figure}
{\large o)} \includegraphics[scale=0.71]{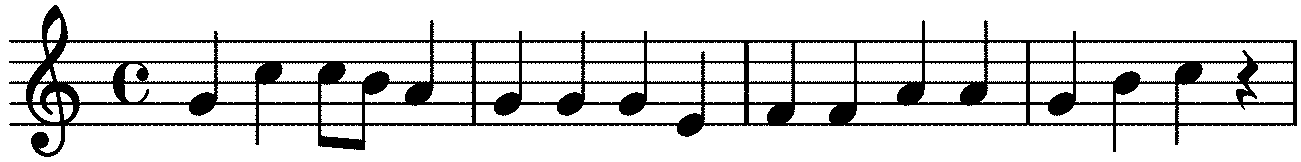}\\
{\large e$_1$)} \includegraphics[scale=0.71]{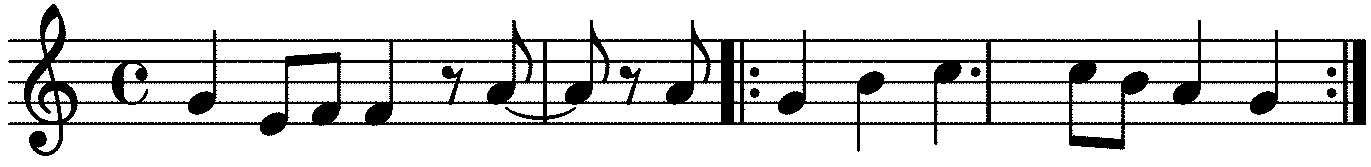}\\
{\large e$_2$)} \includegraphics[scale=0.71]{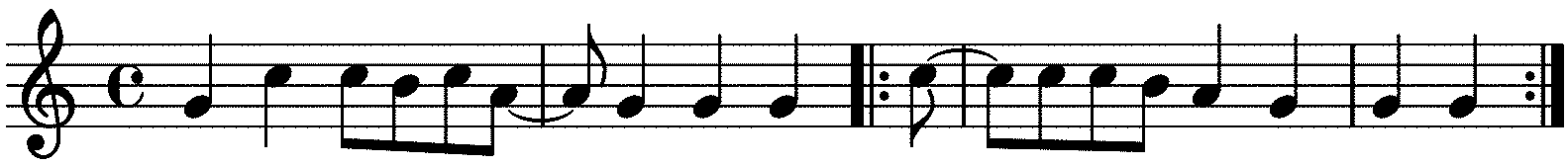}
\caption{Examples of melodies generated by the Sequential architecture.
(o) Original plan melody learnt.
(e$_1$ and e$_2$) Melodies generated by extrapolating from a new plan melody.
Inspired from \cite{todd:connectionist:composition:1989}}
\label{figure:todd:example:extrapolation}
\end{figure}

After having trained the network on a plan melody,
various melodies may be generated by extrapolation by inputing new plans,
as shown in Figure~\ref{figure:todd:example:extrapolation}.
A repeat sign {\bf :} indicates when the network output goes into a fixed loop.

\begin{figure}
{\large o$_A$)} \includegraphics[scale=0.71]{./score-todd-extrapolation-original.png}\\
{\large o$_B$)} \includegraphics[scale=0.71]{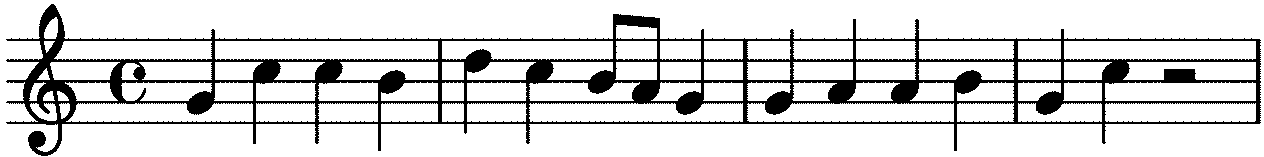}\\
{\large i$_1$)} \includegraphics[scale=0.71]{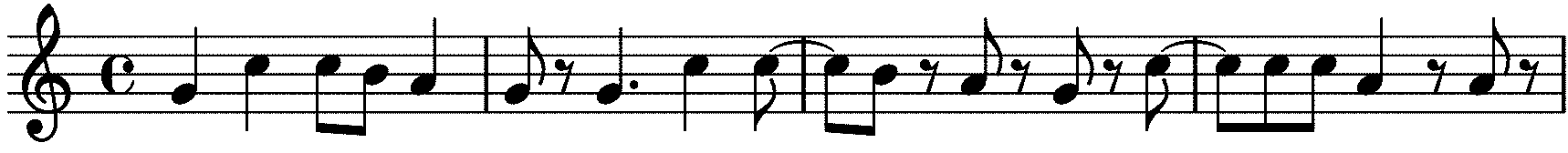}\\
{\large i$_2$)} \includegraphics[scale=0.71]{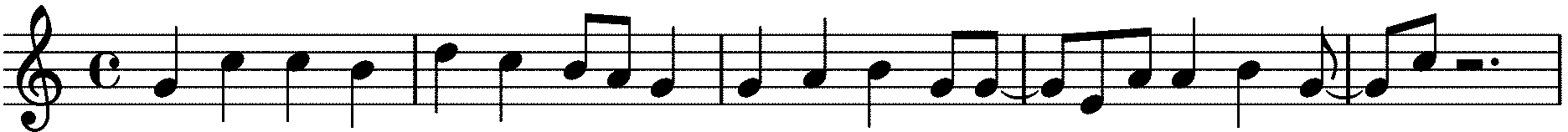}
\caption{Examples of melodies generated by the Sequential architecture.
(o$_A$ and o$_B$) Original plan melodies learnt.
(i$_1$ and i$_2$)  Melodies generated by interpolating between o$_A$ plan and o$_B$ plan melodies.
Inspired from \cite{todd:connectionist:composition:1989}}
\label{figure:todd:example:interpolation}
\end{figure}

One could also do interpolation between several (two or more) plans melodies that have been learnt\footnote{Note that this way of doing is actually some
	precursor of doing interpolation on embeddings of melodies
	to be generated by combining a decoder feedforward strategy and an iterative feedforward strategy,
	such as for example in the VRAE or the MusicVAE systems,
	to be described in Sections~\ref{section:experiment:vrae} and~\ref{section:system:music:vae}, respectively.}.
Examples are shown in Figure~\ref{figure:todd:example:interpolation}.
The Sequential architecture
is summarized in Table~\ref{table:dimensions:todd:sequential}.

\begin{table}
\begin{tabular}{|l|l|}
\hline
{\em Objective}			&Melody\\
\hline
{\em Representation}	&Symbolic; Piano roll; One-hot; Note begin; Implicit rest\\
\hline
{\em Architecture}		&RNN\\
\hline
{\em Strategy}			&Iterative feedforward\\
\hline
\end{tabular}
\caption{Sequential architecture summary}
\label{table:dimensions:todd:sequential}
\end{table}

\subsection{\#3 Example: BLSTM Chord Accompaniment Symbolic Music Generation System}
\label{section:experiment:blstm:chord}

The BLSTM\index{BLSTM} (Bidirectional LSTM\index{Bidirectional!LSTM})
chord accompaniment system by Lim et al. \cite{lim:chord:generation:from:melody:ismir:2017}
is a rare and interesting case\footnote{As noted
	in Sections~\ref{section:architecture:rnn:training} and~\ref{section:strategy:iterative:feedforward}.}
of an accompaniment system based on a recurrent architecture.
The objective is to generate a progression (sequence) of chords\index{Chord}\index{Chord!progression} as an accompaniment to a melody (specified symbolically).

The corpus is imported from a now defunct lead sheet public data base.
The authors selected 2,252 selected lead sheets of various western modern music
(rock\index{Rock}, pop\index{Pop}, country\index{Country}, jazz\index{Jazz}, folk\index{Folk}, R\&B\index{R\&B|see{Rhythm and blues}},
children's song, etc.),
all in major key and the majority with a single chord per measure (otherwise only the first chord is considered).
This results in a training set of 1,802 songs (making a total of 72,418 measures)
and a test set of 450 songs (17,768 measures).
All songs are transposed (aligned) to C major key.

\begin{figure}
\includegraphics[scale=1.5]{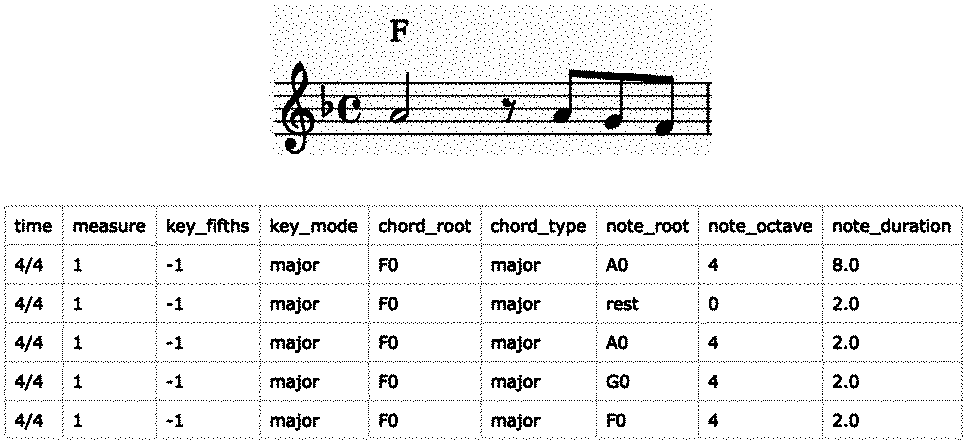}
\caption{Example of extracted data from a single measure.
Reproduced from \cite{lim:chord:generation:from:melody:ismir:2017}
under a CC BY 4.0 licence}
\label{figure:representation:blstm}
\end{figure}

Desired characteristics are extracted from the original XML\index{XML} files
and converted to a CSV\index{CSV|see{Comma-separated values}}\footnote{CSV
	stands for Comma-separated values.}
(spreadsheet) matrix format,
as shown in Figure~\ref{figure:representation:blstm}.
The specificities (simplifications) of the representation are as follows:

\begin{itemize}

\item for the melody\footnote{And obviously
	also for the chords.},
only pitch classes\index{Pitch!class} are considered (and octaves are not),
resulting in a 12 notes one-hot encoding (named 12-semitone-vector) plus the rest; and

\item for the chords, only their primary triads\index{Triad} are considered,
with only two types: major and minor, resulting in a 24 chords one-hot encoding.

\end{itemize}

\begin{figure}
\includegraphics[scale=1.3]{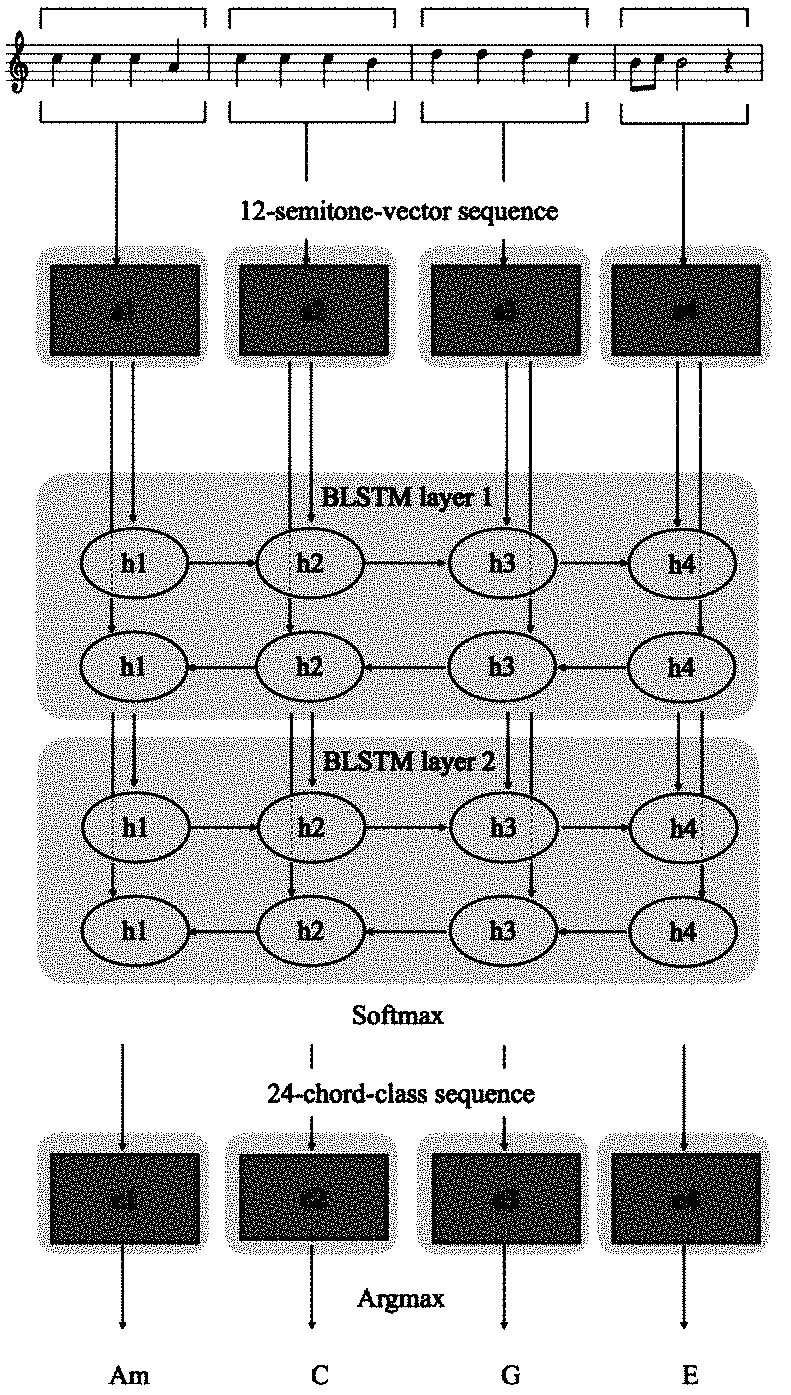}
\caption{BLSTM architecture.
Reproduced from \cite{lim:chord:generation:from:melody:ismir:2017}
under a CC BY 4.0 licence}
\label{figure:architecture:blstm}
\end{figure}

The architecture is a bidirectional\index{Bidirectional} LSTM with two LSTM layers, each one with 128 units.
The motivation is to provide the network with the musical context backward and also forward in time.
The time step considered by the architecture is four measures long, as shown in Figure~\ref{figure:architecture:blstm}.
The tanh function is used as the non linear activation function for the hidden layers
and softmax is used as the output layer activation function,
with categorical cross-entropy as its associated cost function.

Training is done with various four measures long samples as input and their associated four chords as output,
generated by sliding a four measures long window over each training song.
Generation is done by iteratively feedforwarding successive four measures long melody fragments (time slices)
of a song and concatenating the resulting four measures long chord progression fragments.

The architecture is peculiar in that, although recurrent, generation is not recursive
and the output data has a different nature and structure
(chords) than the input data (notes).
Furthermore, note that, although the strategy is iterative and the architecture is recurrent,
the granularity of each iterative step is quite coarse as it is 4 measures long,
as opposed to most of systems based on recurrent architectures and iterative feedforward strategy
which consider the time step at the level of the smallest notre duration (see, e.g.,
the system analyzed in Section~\ref{section:experiment:eck:blues:lstm:first:experiment}).
This kind of mixed architecture/strategy between forward/single step and recurrent/iterative
may have been motivated by the objective of capturing sufficiently the history of horizontal correlations
(between notes of the melody and between chords of the accompaniment)
as the LSTM cells focus on capturing the history of vertical correlations (between notes and chords).

\begin{figure}
\includegraphics[width=\textwidth]{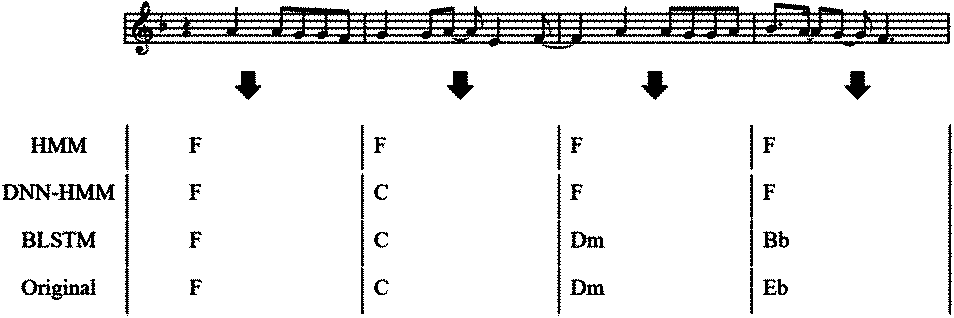}
\caption{Comparison of generated chord progressions (HMM, DNN-HMM, BLSTM and original).
Reproduced from \cite{lim:chord:generation:from:melody:ismir:2017}
under a CC BY 4.0 licence}
\label{figure:comparison:blstm}
\end{figure}

The system has been evaluated
by comparing to some hidden Markov model\index{Hidden!Markov model} (HMM\index{HMM}) model
and to some deep neural network--HMM hybrid model
(named DNN-HMM,
see details in \cite{vesely:sequence:discriminative:training:interspeech:2013}),
both quantitatively (by comparing the accuracies and through confusion matrixes),
and qualitatively (through a web-based survey of 25 musically untrained participants).
Results are showing a better accuracy and preference for the BLSTM model,
see a simple example in Figure~\ref{figure:comparison:blstm}. 
The authors note that the evaluation also shows that,
when songs are unknown, the preference for the BLSTM model is weaker.
They conjecture that this is because BLSTM often generates a more diverse chord sequence than the original.
The BLSTM system is summarized in Table~\ref{table:dimensions:blstm}.

\begin{table}
\begin{tabular}{|l|l|}
\hline
{\em Objective}			&Accompaniment; Chord sequence; Western modern music\\
\hline
{\em Representation}	&Symbolic; CSV; One-hot$\times$(12$\times$* + 24$\times$4); Rest\\
\hline
{\em Architecture}		&LSTM$^2$\\
\hline
{\em Strategy}			&Iterative feedforward\\
\hline
\end{tabular}
\caption{BLSTM summary}
\label{table:dimensions:blstm}
\end{table}

\subsection{Summary}
\label{section:challenges:strategies:first:reflexion}
\label{section:challenges:strategies:summary}
%

In summary, we have seen that an RNN architecture is usually coupled to an iterative feedforward strategy\index{Iterative feedforward strategy},
which allows a recursive\index{Recursion} seed-based\index{Seed!-based generation}
variable length\index{Variable!length} generation, as discussed in Section~\ref{section:variable:length}.
However, there are some exceptions:

\begin{itemize}

\item the Time-Windowed system by Todd (Section~\ref{section:experiment:todd:time:windowed})
uses an iterative feedforward strategy\index{Iterative feedforward strategy} on a feedforward architecture
in order to generate a melody,
and


\item the BLSTM\index{BLSTM} system (Section~\ref{section:experiment:blstm:chord})
uses an iterative feedforward strategy\index{Iterative feedforward strategy} on a recurrent architecture
in order to generate a chord accompaniment to a melody.

\end{itemize}


We will see further (with the VRAE\index{VRAE} system to be described in Section~\ref{section:experiment:vrae})
the use of an RNN Encoder-Decoder\index{RNN Encoder-Decoder} compound architecture (Section~\ref{section:architecture:compound:recurrent:autoencoder}),
as a way to decouple the length of the input sequence with the length of the output sequence,
by combining the decoder feedforward strategy with the iterative feedforward strategy.


Some other examples of couplings between architectures and strategies,
or between challenges,
will be discussed in Section~\ref{section:challenges:strategies:discussion}.
Before that, we will continue to analyze challenges and possible solutions or directions.


\section{Melody-Harmony Interaction}
\label{section:challenges:strategies:melody:harmony:consistency}

When the objective is to generate simultaneously a melody\index{Melody} with an accompaniment,
expressed through some harmony or counterpoint\footnote{Harmony and counterpoint
	are dual approaches of accompaniment with different focus and priorities.
	Harmony\index{Harmony} focuses on the {\em vertical} relations between simultaneous notes,
	as objects on their own ({\em chords\index{Chord}}),
	and then considers the horizontal relations between them (e.g., harmonic cadences\index{Cadence}).
	Conversely, counterpoint\index{Counterpoint} focuses on the {\em horizontal} relations between successive notes
	for each simultaneous melody (a {\em voice\index{Voice}}),
	and then considers the vertical relations between their progression (e.g., to avoid parallel fifths).
	Note that,
	although their perspectives are different,
	the analysis and control of relations between vertical and horizontal dimensions are their shared objectives.},
an issue is the musical consistency
between the melody and the harmony.
Although a general architecture such as MiniBach\index{MiniBach} (Section~\ref{section:experiment:mini:bach})
is supposed to have learnt correlations,
interactions between vertical and horizontal dimensions are not
explicitly considered.

We have analyzed in Section~\ref{section:experiment:eck:blues:lstm:second:experiment}
an example of a specific architecture
to generate simultaneously melody and chords,
with explicit relations between them
(i.e. chords can use previous step information about melody but not the opposite).
However, this architecture is a bit {\em ad hoc\index{Ad hoc}}.
In the following sections, we will analyze some more general architectures having in mind interactions between melody and harmony.


\subsection{\#1 Example: RNN-RBM Polyphony Symbolic Music Generation System}
\label{section:experiment:rnn:rbm}

In \cite{boulanger:temporal:dependencies:icml:2012},
Boulanger-Lewandowski {\em et al.} have associated
to the RBM-based architecture introduced in Section~\ref{section:experiment:rbm}
a recurrent network (RNN)
in order to represent the temporal sequence of notes.
The idea is to {\em couple\index{Couple}} the RBM
to a deterministic RNN with a single hidden layer, such that

\begin{itemize}

\item the RNN models the {\em temporal sequence} to produce successive outputs,
corresponding to successive time steps,

\item which are {\em parameters}, more precisely the {\em biases}, of an RBM
that models the {\em conditional probability\index{Conditional!probability} distribution}
of the {\em accompaniment notes},
i.e. which notes should be played together.

\end{itemize}

In other words, the objective is to combine a {\em horizontal view} (temporal sequence)
and a {\em vertical view} (combination of notes for a particular time step).
The resulting architecture named RNN-RBM\index{RNN-RBM} is shown in Figure~\ref{figure:rnn:rbm:architecture},
and can be interpreted as follows:

\begin{figure}
\includegraphics[scale=0.6]{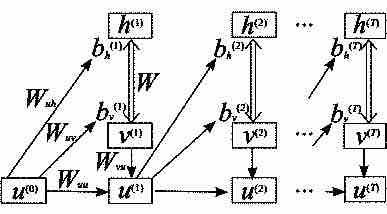}
\caption{RNN-RBM architecture.
Reproduced from \cite{boulanger:rnn:rbm:generation:2015} with permission of the authors}
\label{figure:rnn:rbm:architecture}
\end{figure}

\begin{itemize}

\item the bottom line represents the temporal sequence of RNN hidden units u$^{(0)}$, u$^{(1)}$, \ldots, u$^{(t)}$,
where u$^{(t)}$ notation means\footnote{Note that the usual notation would be u$_t$,
	as the u$^{(t)}$ notation is usually reserved to index dataset examples ($t$th example),
	see Section~\ref{section:architecture:recurrent:network}.}
the value of the RNN hidden layer u at time (index) $t$; and

\item the upper part represents the sequence of each RBM instance at time $t$,
which we could notate RBM$^{(t)}$,
with

\begin{itemize}

\item v$^{(t)}$ its visible layer with $b_\text{v}^{(t)}$ its bias,

\item h$^{(t)}$ its hidden layer with $b_\text{h}^{(t)}$ its bias, and

\item $W$ the weight matrix of connexions between the visible layer v$^{(t)}$ and the hidden layer h$^{(t)}$.

\end{itemize}

\end{itemize}

There is a specific training algorithm,
which we will not detail here, please refer to \cite{boulanger:temporal:dependencies:icml:2012}.
During generation, each $t$ time step of processing is as follows:

\begin{itemize}

\item compute the biases $b_\text{v}^{(t)}$ and $b_\text{h}^{(t)}$ of RBM$^{(t)}$,
via Equations~\ref{equation:rnn:rbm:bv:formula} and~\ref{equation:rnn:rbm:bh:formula} respectively,

\item sample from RBM$^{(t)}$ by using block Gibbs sampling\index{Block!Gibbs sampling}\index{Sampling}
to produce v$^{(t)}$, and

\item feedforward the RNN with v$^{(t)}$ as the input,
using the RNN hidden layer value u$^{(t-1)}$,
in order to produce the RNN new hidden layer value u$^{(t)}$
via Equation~\ref{equation:rnn:rbm:rnn:formula},
where

\begin{itemize}

\item $W_{\text{vu}}$ is the weight matrix and $b_\text{u}$ the bias for the connexions between the input layer of the RBM and the hidden layer of the RNN; and

\item $W_{\text{uu}}$ is the weight matrix for the recurrent connexions of the hidden layer of the RNN.

\end{itemize}

\end{itemize}

\begin{equation}
b_\text{v}^{(t)} = b_\text{v} + W_{\text{uv}} \text{u}^{(t-1)}
\label{equation:rnn:rbm:bv:formula}
\end{equation}

\begin{equation}
b_\text{h}^{(t)} = b_\text{h} + W_{\text{uh}} \text{u}^{(t-1)}
\label{equation:rnn:rbm:bh:formula}
\end{equation}

\begin{equation}
\text{u}^{(t)} = \text{tanh} (b_\text{u} + W_{\text{uu}} \text{u}^{(t-1)} + W_{\text{vu}} \text{v}^{(t)})
\label{equation:rnn:rbm:rnn:formula}
\end{equation}


Note that the biases $b_\text{v}^{(t)}$ and $b_\text{h}^{(t)}$ of RBM$^{(t)}$ are variable for each time step,
in other words they are {\em time dependent},
whereas the weight matrix $W$ for the connexions between the visible and the hidden layer of RBM$^{(t)}$
is {\em shared} for all time steps (for all RBMs),
i.e. it is {\em time independent}\footnote{$W_{\text{uv}}$, $W_{\text{uh}}$, $W_{\text{uu}}$ and $W_{\text{vu}}$ weight matrices
	are also shared and thus time independent.}.

\begin{figure}
\includegraphics[scale=0.4]{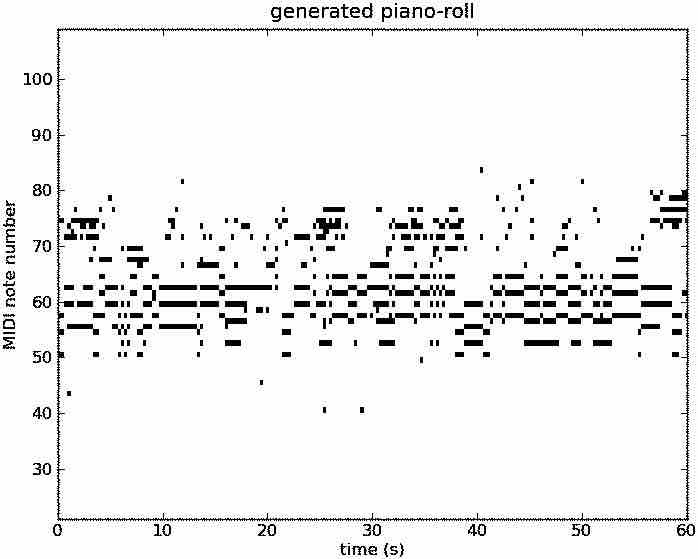}
\caption{Example of a sample generated by RNN-RBM trained on J. S. Bach chorales.
Reproduced from \cite{boulanger:rnn:rbm:generation:2015} with permission of the authors}
\label{figure:rnn:rbm:polyphonic:example}
\end{figure}

Four different corpora have been used in the experiments:
classical\index{Classical} piano\index{Piano}, folk\index{Folk} tunes, orchestral classical music
and J. S. Bach\index{Bach} chorales\index{Chorale}.
Polyphony varies from 0 to 15 simultaneous notes, with an average value of 3.9.
A piano roll representation is used with many-hot encoding of 88 units representing pitches\index{Pitch} from A$_0$ to C$_8$.
Discretization (time step) is a quarter note.
All examples are aligned onto a single common tonality: C major or minor.
An example of a sample generated in a piano roll representation is shown in Figure~\ref{figure:rnn:rbm:polyphonic:example}.
The quality of the model has made RNN-RBM,
summarized at Table~\ref{table:dimensions:rnn:rbm},
one of the reference architectures for polyphonic music generation.

\begin{table}
\begin{tabular}{|l|l|}
\hline
{\em Objective}			&Polyphony\\
\hline
{\em Representation}	&Symbolic; Many-hot\\
\hline
{\em Architecture}		&RBM-RNN\\
\hline
{\em Strategy}			&Iterative feedforward; Sampling\\
\hline
\end{tabular}
\caption{RNN-RBM summary}
\label{table:dimensions:rnn:rbm}
\end{table}

\subsubsection{Other RNN-RBM Systems}
\label{section:experiment:rnn:rbm:other}

There have been a few systems following on and extending the RNN-RBM architecture,
but they are not significantly different and furthermore they have not been thoroughly evaluated.
However, it is worth mentioning the following:

\begin{itemize}

\item the RNN-DBN\index{RNN-DBN} architecture\footnote{This is apparently
			the state of the art for the J. S. Bach Chorales dataset in terms of cross-entropy loss\index{Cross-entropy!cost}.},
	using multiple hidden layers \cite{goel2014polyphonic}; and

\item the LSTM-RTRBM architecture, using an LSTM instead of an RNN \cite{lyu2015}.
 

\end{itemize}


\subsection{\#2 Example: Hexahedria Polyphony Symbolic Music Generation Architecture}
\label{section:experiment:hexahedria}

The system for polyphonic music
proposed by Johnson in his Hexahedria blog \cite{johnson:web:hexahedria:composing:music:recurrent:neural:2015}
is hybrid and original in that it {\em integrates\index{Integration}}
into the same architecture

\begin{itemize}

\item a first part made of two RNNs (actually LSTM) layers, each with 300 hidden units,
recurrent over the {\em time dimension},
which are in charge of the {\em temporal} horizontal aspect, that is the relations between notes in a sequence.
Each layer has connections across time steps, while being independent across notes; and

\item a second part made of two other RNN (LSTM) layers, with 100 and 50 hidden units,
recurrent over the {\em note dimension},
which are in charge of the {\em harmony\index{Harmony}} vertical aspect,
that is the relations between simultaneous notes within the same time step.
Each layer is independent between time steps
but has transversal directed connexions between notes.

\end{itemize}

We can notate this architecture as LSTM$^{2+2}$
in order to highlight the two successive 2-level recurrent layers, recurrent in two different dimensions (time and note).
The architecture is actually a kind of integration within a single architecture\footnote{We will see in Section~\ref{section:experiment:biaxial}
	an alternative architecture, named Bi-Axial LSTM,
	where each of the 2-level time-recurrent layers is encapsulated into a different architectural module.}
of the RNN-RBM\index{RNN-RBM} architecture described in previous section.
The main originality is in using recurrent networks not only on the time dimension
but also on the note dimension, more precisely on the pitch class dimension.
This latter type of recurrence is used to model the occurence of a simultaneous note based on other simultaneous notes.
Like for the time relation, which is oriented towards the future,
the pitch class relation is oriented towards higher pitch, from C to B.

%
%
%

\begin{figure}
\includegraphics[scale=0.3]{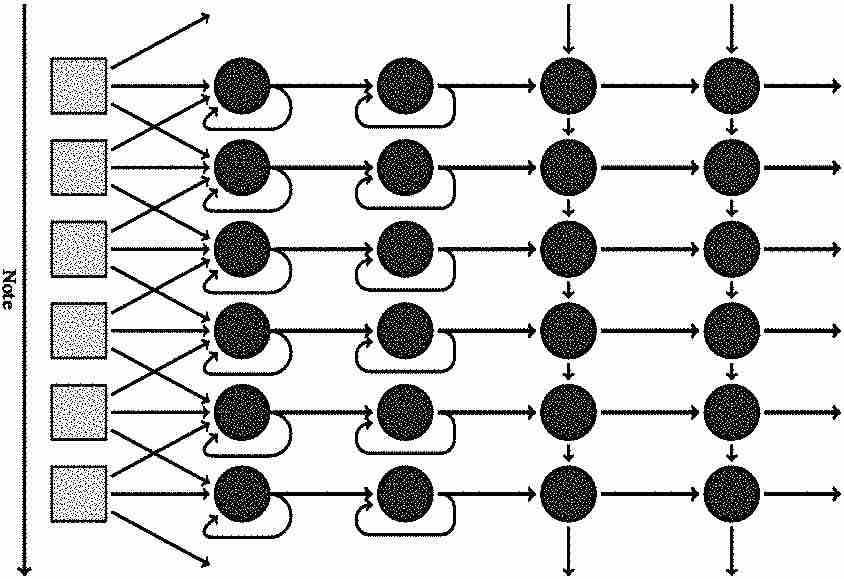}
\caption{Hexahedria architecture (folded).
Reproduced from \cite{johnson:web:hexahedria:composing:music:recurrent:neural:2015} with permission of the author}
\label{figure:hexahedria:final:architecture:folded}
\end{figure}

\begin{figure}
\includegraphics[scale=0.47]{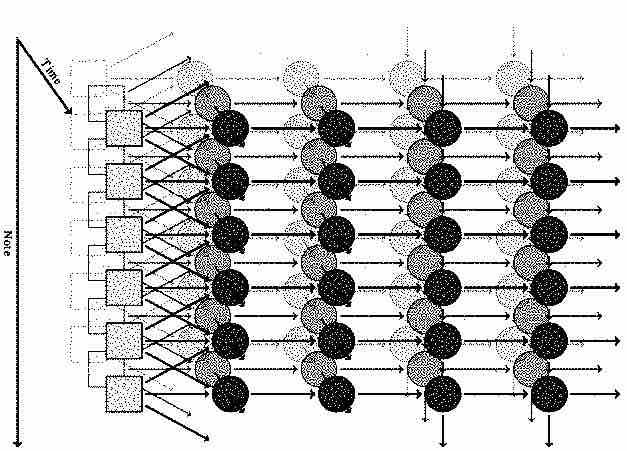}
\caption{Hexahedria architecture (unfolded).
Reproduced from \cite{johnson:web:hexahedria:composing:music:recurrent:neural:2015} with permission of the author}
\label{figure:hexahedria:final:architecture:unfolded}
\end{figure}

The resulting architecture is shown in its folded form in Figure~\ref{figure:hexahedria:final:architecture:folded}
and in its unfolded form\footnote{Our unfolded pictorial representation of an RNN
	shown in
	Figure~\ref{figure:recurrent:network:unfolded:abstract}
	was actually inspired by Johnson's Hexahedria pictorial representation.}
in Figure~\ref{figure:hexahedria:final:architecture:unfolded},
with three axes\index{Axis} represented:

\begin{itemize}

\item the {\em flow\index{Flow} axis}, shown horizontally and directed from left to right,
represents the flow of (feedforward) computation
through the architecture, from the input layer to the output layer;

\item the {\em note axis}, shown vertically and directed from top to bottom,
represents the connexions between units corresponding to successive notes
of each of the two last (note-oriented) recurrent hidden layers; and

\item the {\em time axis}, only in the unfolded Figure~\ref{figure:hexahedria:final:architecture:unfolded},
shown diagonally and directed from top left to bottom right,
represents the time steps and the propagation\index{Propagation} of the memory\index{Memory}
within a same unit of the two first (time-oriented) recurrent hidden layers.

\end{itemize}

The dataset is constructed by extracting 8 measures long parts from MIDI files from the
Classical\index{Classical} piano\index{Piano} MIDI database\index{Classical piano MIDI database} \cite{piano-midi.de:web}.
The input representation used is piano roll, with the pitch represented as the MIDI note number.
More specific information is added:
the pitch class,
the previous note played (as a way to represent a possible hold),
how many times a pitch class has been played in the previous time step
and the beat (the position within the measure, assuming a 4/4 time signature).
The output representation is also a piano roll, in order to represent the possibility of more than one note at the same time.
Generation is done in an iterative way (i.e. following the iterative feedforward strategy), as for most recurrent networks.
The system is summarized in Table~\ref{table:dimensions:hexahedria}.

\begin{table}
\begin{tabular}{|l|l|}
\hline
{\em Objective}			&Polyphony\\
\hline
{\em Representation}	&Symbolic; Piano roll; Hold; Beat\\
\hline
{\em Architecture}		&LSTM$^{2+2}$\\
\hline
{\em Strategy}			&Iterative feedforward; Sampling\\
\hline
\end{tabular}
\caption{Hexahedria summary}
\label{table:dimensions:hexahedria}
\end{table}

\subsection{\#3 Example: Bi-Axial LSTM Polyphony Symbolic Music Generation Architecture}
\label{section:experiment:biaxial}

Johnson recently proposed an evolution of his original Hexahedria architecture,
described in Section~\ref{section:experiment:hexahedria},
named Bi-Axial LSTM\index{Bi-Axial LSTM} (or BALSTM\index{BALSTM}) \cite{johnson:tied:evomusart:2017}.

The representation used is piano roll, with note hold and rest tokens added to the vocabulary.
Various corpora are used:
the JSB Chorales dataset\index{JSB Chorales dataset}, a corpus of 382 four-part chorales by J. S. Bach \cite{allan2005harmonising};
the MuseData library\index{MuseData library}, an electronic classical music library from CCARH in Stanford \cite{ccarh:musedata:web};
the Nottingham database\index{Nottingham database},
a collection of 1,200 folk tunes in ABC notation
\cite{foxley:nottingham:database:web};
and the Classical piano MIDI database\index{Classical piano MIDI database} \cite{piano-midi.de:web}. 
Each dataset is transposed (aligned) into the key of C major or C minor.

The probability of playing a note depends on two types of information:

\begin{itemize}

\item all notes at previous time steps -- this is modeled by the {\em time-axis module}; and

\item all notes within the current time step that have already been generated (the order being lowest to highest)
-- this is modeled by the {\em note-axis module}.

\end{itemize}

\begin{figure}
\includegraphics[width=\textwidth]{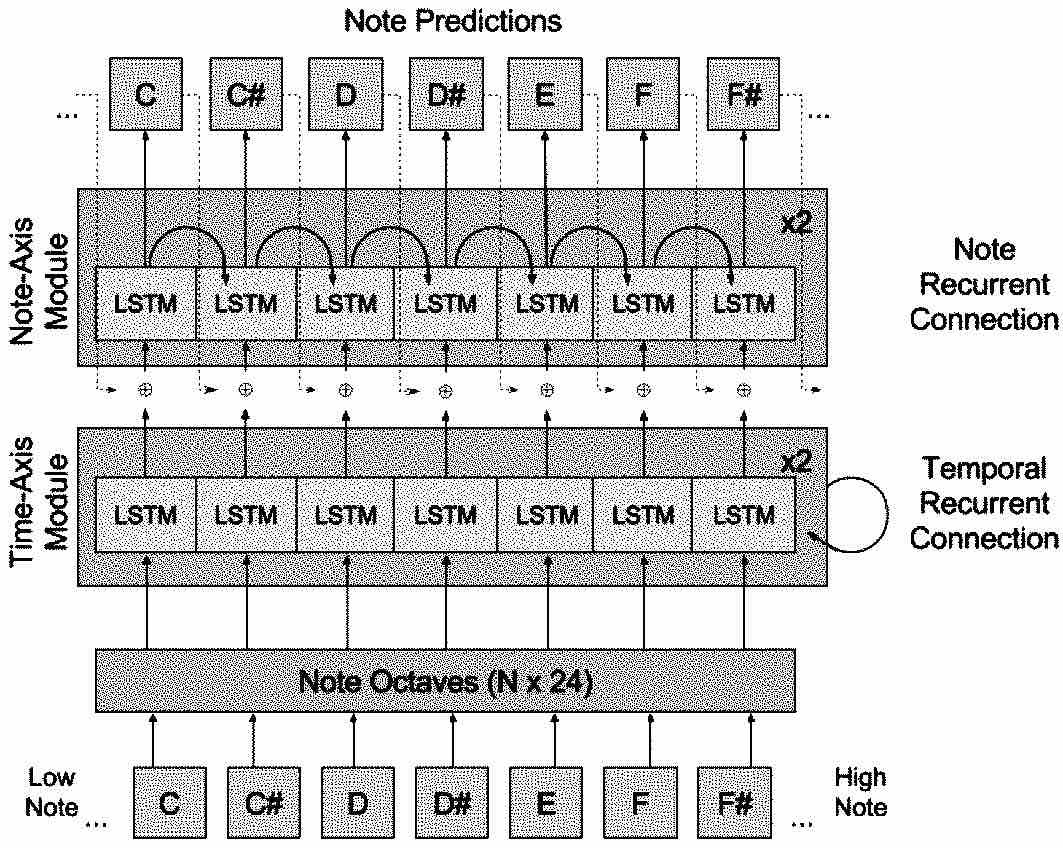}
\caption{Bi-Axial LSTM architecture.
Reproduced from \cite{mao:deepj:arxiv:2018} with permission of the authors}
\label{figure:biaxial:architecture:deepj}
\end{figure}

There is an additional front end layer, named ``Note Octaves'',
which transforms each note into a vector of all its possible corresponding octave\index{Octave} notes
(i.e. an extensional version of pitch classes).
The resulting architecture is illustrated in Figure~\ref{figure:biaxial:architecture:deepj}\footnote{This figure
	comes from the description of another system based on the Bi-Axial LSTM architecture,
	named DeepJ\index{DeepJ}, which will be described in Section~\ref{section:systems:deepj}.}.
The ``x2'' represents the fact that each module is stacked twice (i.e. has two layers).

The time-axis module is recurrent in time (as for a classical RNN), the LSTM weights being shared across notes in order to gain note transposition invariance\index{Transposition!invariance}.
The note-axis module\footnote{Note that, as opposed to Johnson's first architecture
	(that we refer to as Hexahedria, and which has been introduced in Section~\ref{section:experiment:hexahedria}),
	which integrates the 2-level time-recurrent layers with the 2-level note-recurrent layers within a single architecture
	and therefore notated as LSTM$^{2+2}$,
	the Bi-Axial LSTM architecture explicitly separates each 2-level time-recurrent layers into distinct architectural modules
	and is therefore notated as LSTM$^2\times$2.}
is recurrent in note.
For each note input of the note-axis module,
$\oplus$ represents the concatenation of the corresponding output from the time-axis module with the already predicted lower notes.
Sampling (into a binary value, by using a coin flip)
is applied to each note output probability in order to compute the final prediction
(whether that note is played or not).

As pointed out by Johnson \cite{johnson:tied:evomusart:2017},
during the training phase, as all the notes at all time steps are known,
the training process may be accelerated by processing each layer independently (e.g., on a GPU),
by running input through the two time-axis layers in parallel across all notes,
and using the two note-axis layers to compute probabilities in parallel across all time steps.

\begin{figure}
\includegraphics[width=\textwidth]{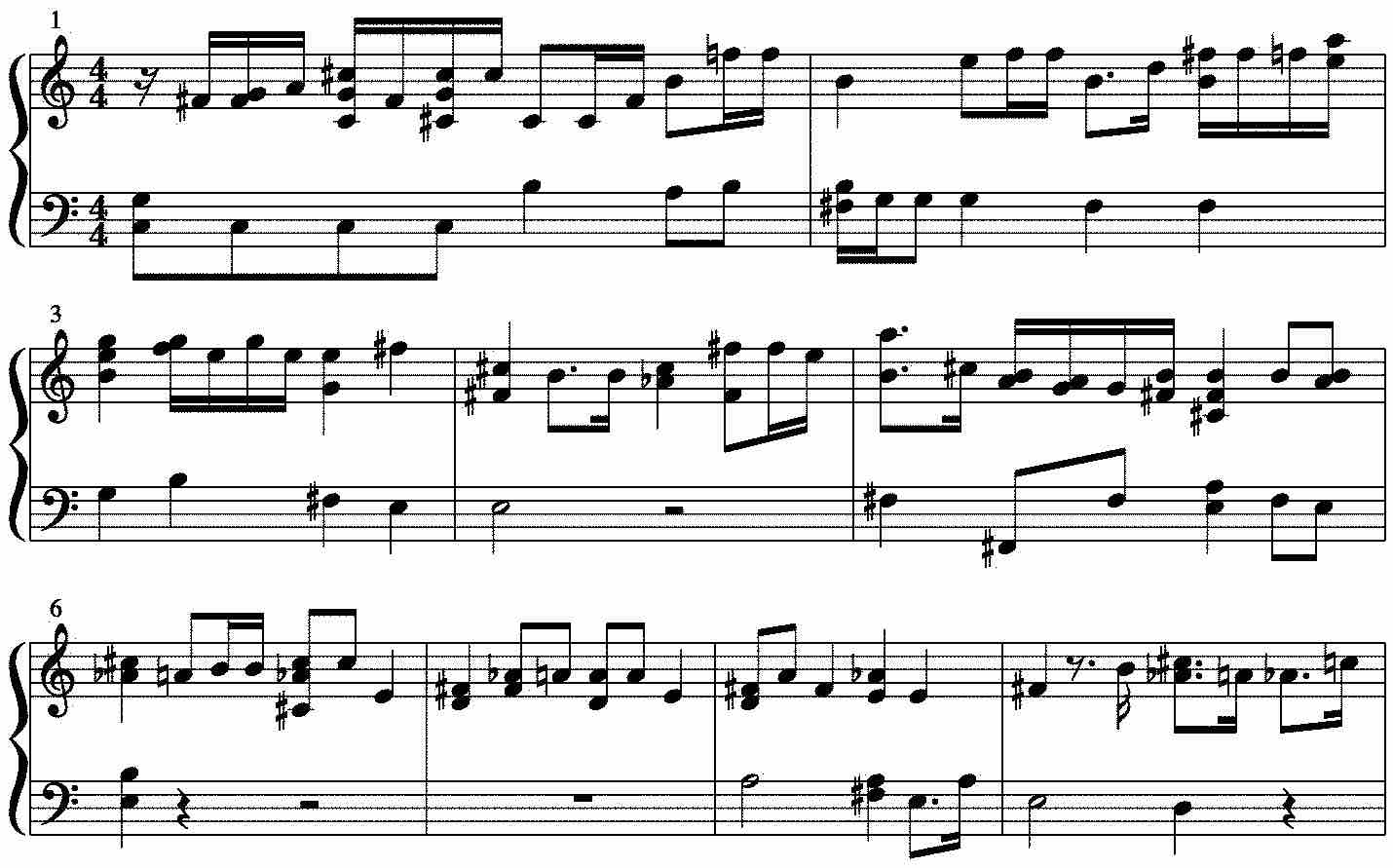}
\caption{Example of Bi-Axial LSTM generated music (excerpt).
Reproduced from \cite{johnson:tied:evomusart:2017}}
\label{figure:biaxial:example}
\end{figure}

The generation phase is sequential for each time step
(by following both the iterative feedforward strategy and the sampling strategy).
An excerpt of
music generated is shown in Figure~\ref{figure:biaxial:example}.

The Bi-Axial LSTM system, summarized in Table~\ref{table:dimensions:biaxial},
has been evaluated and compared to some other architectures.
The author reports noticeably better results with Bi-Axial LSTM,
the greatest improvements being on
the MuseData\index{MuseData library} \cite{ccarh:musedata:web}
and the Classical piano MIDI database\index{Classical piano MIDI database} \cite{piano-midi.de:web} datasets,
and states in \cite{johnson:tied:evomusart:2017} that:
``It is likely due to the fact that those datasets contain many more complex musical structures in different keys,
which are an ideal case for a translation-invariant architecture.''
Note that an extension of the Bi-Axial LSTM architecture with conditioning, named DeepJ\index{DeepJ},
will be introduced in Section~\ref{section:systems:deepj}.

\begin{table}
\begin{tabular}{|l|l|}
\hline
{\em Objective}			&Polyphony\\
\hline
{\em Representation}	&Symbolic; Piano roll; Hold; Rest\\
\hline
{\em Architecture}		&Bi-Axial LSTM = LSTM$^2\times$2\\
\hline
{\em Strategy}			&Iterative feedforward; Sampling\\
\hline
\end{tabular}
\caption{Bi-Axial LSTM summary}
\label{table:dimensions:biaxial}
\end{table}

\section{Control}
\label{section:challenges:strategies:control}

A deep architecture generates musical content matching the style learnt from the corpus.
This capacity of induction from a corpus without any explicit modeling or programming is an important ability,
as discussed in Chapter~\ref{section:introduction}
and also in \cite{fiebrink:ml:creative:tool:arxiv:2016}.
However, like a fast car that needs a good steering wheel,
control is also needed as
musicians usually want to {\em adapt} ideas and patterns {\em borrowed} from other contexts
to their own objective and context, e.g., transposition to another key, minimizing the number of notes,
finishing with a given note, etc.

\subsection{Dimensions of Control Strategies}
\label{section:challenges:strategies:control:dimensions:strategies}

Arbitrary control is a
difficult issue for deep learning architectures and techniques
because neural networks have not been designed to be controlled.
In the case of Markov chains,
they have an operational model
on which one can attach constraints to control the generation\footnote{Two examples are Markov constraints \cite{pachet:markov:constraints:constraints:2011}
	and factor graphs \cite{pachet:variations:structured:ismir:2017}.}.
However, neural networks
do not offer such an operational entry point
and the distributed nature of their representation does not provide a
clear relation
to the structure of the content generated.
Therefore, as we will see, most of strategies for controlling deep learning generation rely on {\em external} intervention at various
{\em entry points\index{Entry point}} (hooks\index{Hook}) and {\em levels}:

\begin{itemize}

\item input,

\item output,

\item input {\em and} output, and

\item encapsulation/reformulation.

\end{itemize}

Various control {\em strategies\index{Strategy}} can be employed:

\begin{itemize}

\item sampling,

\item conditioning,

\item input manipulation,

\item reinforcement, and

\item unit selection.

\end{itemize}

We will also see that some strategies
(such as
sampling,
see Section~\ref{section:challenges:strategies:control:sampling})
are more {\em bottom-up}
and others
(such as structure imposition, see Section~\ref{section:systems:c-rbm},
or unit selection, see Section~\ref{section:control:unit:selection})
are more {\em top-down}.
Lastly, there is also a continuum between {\em partial} solutions
(such as conditioning/parametrization, see Section~\ref{section:control:conditioning})
and more {\em general} approaches
(such as reinforcement, see Section~\ref{section:control:reinforcement}).

\subsection{Sampling}
\label{section:challenges:strategies:control:sampling}

\label{section:challenges:strategies:control:sampling:rbm}

Sampling\index{Sampling} from a stochastic architecture\index{Stochastic}
(such as a restricted Boltzmann machine\index{Restricted Boltzmann machine} (RBM\index{RBM}),
see Section~\ref{section:input:less:rbm}),
or from a deterministic architecture\index{Deterministic} (in order to introduce {\em variability},
see Section~\ref{section:variability:sampling}),
may be an entry point for control
if we introduce {\em constraints\index{Constraint}} into the sampling process.
This is called {\em constrained sampling\index{Constrained sampling}},
see for example the C-RBM\index{C-RBM} system in Section~\ref{section:systems:c-rbm}.

Constrained sampling is usually implemented by a {\em generate-and-test\index{Generate!-and-test}} approach,
where valid solutions are picked from a set of random\index{Random} samples\index{Sample} generated from the model.
But this could be a very costly process and, moreover, with no guarantee of success.
A key and difficult issue is therefore how to {\em guide} the sampling process in order to fulfill the constraints.

\subsubsection{Sampling for Iterative Feedforward Generation}
\label{section:challenges:strategies:control:sampling:iterative}

In the case of an iterative feedforward strategy\index{Iterative feedforward strategy} on a recurrent network\index{Recurrent!network},
some refinements in the sampling procedure can be made.

In Section~\ref{section:challenges:strategies:variability:sampling}, we introduced the technique of sampling the softmax output of a recurrent network
in order to introduce content variability.
However, this may sometimes lead to the generation of an unlikely note (with a low probability).
Moreover, as noted in \cite{hadjeres:thesis:2018},
generating such a ``wrong'' note can have a cascading effect on the remaining of the generated sequence.

A counter measure consist in adjusting a learnt RNN model
(conditional probability\index{Conditional!probability} distribution $P(\text{s}_t | \text{s}_{<t})$,
as defined in Section~\ref{section:architecture:recurrent:network})
by not considering notes with a probability under a certain threshold. 
The new model, with a probability distribution $P_{threshold}(\text{s}_t | \text{s}_{<t})$,
is defined in Equation~\ref{equation:rnn:model:threshold} following \cite{whorley:generation:statistical:harmony:jnmr:2016},
where:

\begin{equation}
P_{threshold}(\text{s}_t | \text{s}_{<t}) :=
\begin{cases}
0~\text{if}~P(\text{s}_t | \text{s}_{<t}) / \text{max}_{\text{s}_t} P(\text{s}_t | \text{s}_{<t}) < threshold,\\
P(\text{s}_t | \text{s}_{<t}) / z~\text{otherwise}.
\end{cases}
\label{equation:rnn:model:threshold}
\end{equation}
\begin{itemize}

\item $\text{max}_{\text{s}_t} P(\text{s}_t | \text{s}_{<t})$ is the note maximum probability,

\item $threshold$ is the threshold hyperparameter\index{Hyperparameter}, and

\item $z$ is a normalization constant.

\end{itemize}


A slightly more sophisticated version interpolates between the original distribution $P(\text{s}_t | \text{s}_{<t})$
and the $\text{argmax\index{Argmax}}_{\text{s}'_t} P(\text{s}'_t | \text{s}_{<t})$ deterministic variant\footnote{See
	Section~\ref{section:challenges:strategies:variability:sampling}.},
with some temperature user control hyperparameter (see more details in \cite[Section~4.1.1.3]{hadjeres:thesis:2018}).

This technique will be further generalized
and combined with the conditioning strategy
in order to control the generation of notes at specific positions via positional constraints\index{Positional constraint}.
This will be exemplified by the Anticipation-RNN\index{Anticipation-RNN} system to be introduced in Section~\ref{section:systems:anticipation:rnn}.

\subsubsection{Sampling for Incremental Generation}
\label{section:challenges:strategies:control:sampling:incremental}

In the case of an incremental generation (to be introduced in Section~\ref{section:challenges:strategies:incrementality}),
the user may select

\begin{itemize}

\item on which part (e.g., a given part of a melody and/or a given voice) sampling will occur (or reoccur), and

\item the interval of possible values on which sampling will occur.

\end{itemize}

In the case of the DeepBach\index{DeepBach} system (to be introduced in Section~\ref{section:interactivity:deep:bach}),
this will be the basis for introducing user control on the generation,
notably
to regenerate only some parts of a music,
to restrict note range,
and to impose some basic rhythm.

\subsubsection{Sampling for Variational Decoder Feedforward Generation}
\label{section:challenges:strategies:control:sampling:variational}

Another interesting case is the use of sampling for {\em generative models\index{Generative!model}},
such as variational autoencoders\index{Variational!autoencoder} (VAEs\index{VAE})
and generative adversarial networks\index{Generative!adversarial networks} (GANs\index{GAN}),
to be introduced in Section~\ref{section:challenges:strategies:control:sampling:and:adversarial}.
We will see that some nice control of the sampling, e.g., to produce an interpolation, averaging or attribute\index{Attribute} modification,
will produce meaningful variations in the content generated by the decoder feedforward strategy\index{Decoder!feedforward strategy}.
%
%
%
%
%
%
\label{section:challenges:strategies:control:sampling:and:variational}
%
%
%
%
%
%
Moreover, as has been discussed in Section~\ref{section:architecture:vae},
a variational autoencoder (VAE)
is interesting for its ability for controlling generation over significant dimensions that have been learnt.

\paragraph{\#1 Example: VRAE Video Game Melody Symbolic Music Generation System}
\label{section:experiment:vrae}
\label{section:experiment:rnn:encoder:decoder}
\label{section:systems:strategy:sampling:architecture:variational:recurrent}


In \cite{fabius:vrae:arxiv:2015}, Fabius and van Amersfoort propose the extension of the RNN Encoder-Decoder architecture
to the case of a variational autoencoder (VAE),
which is therefore named a variational recurrent autoencoder\index{Variational!recurrent autoencoder} (VRAE\index{VRAE}).
Both the encoder and the decoder encapsulate an RNN (actually an LSTM),
as has been explained in Section~\ref{section:architecture:compound:recurrent:autoencoder}.
In terms of strategy, the VRAE combines the iterative feedforward strategy
with the decoder feedforward strategy and the sampling strategy.


%

The corpus used in the experiment
is a set of MIDI files of eight video game\index{Video game} songs
from the 1980s and 1990s (Sponge Bob, Super Mario, Tetris\ldots),
which are divided into various shorter parts of 50 time steps.
A one-hot encoding of 49 possible pitches is used (pitches with too few occurrences of notes were not considered).
Experiments have been conducted with 2 or 20 hidden layer units (latent variables).
Training takes place as for training recurrent networks, i.e. for each input note presenting the next note as the output.

\begin{figure}
\includegraphics[scale=0.1]{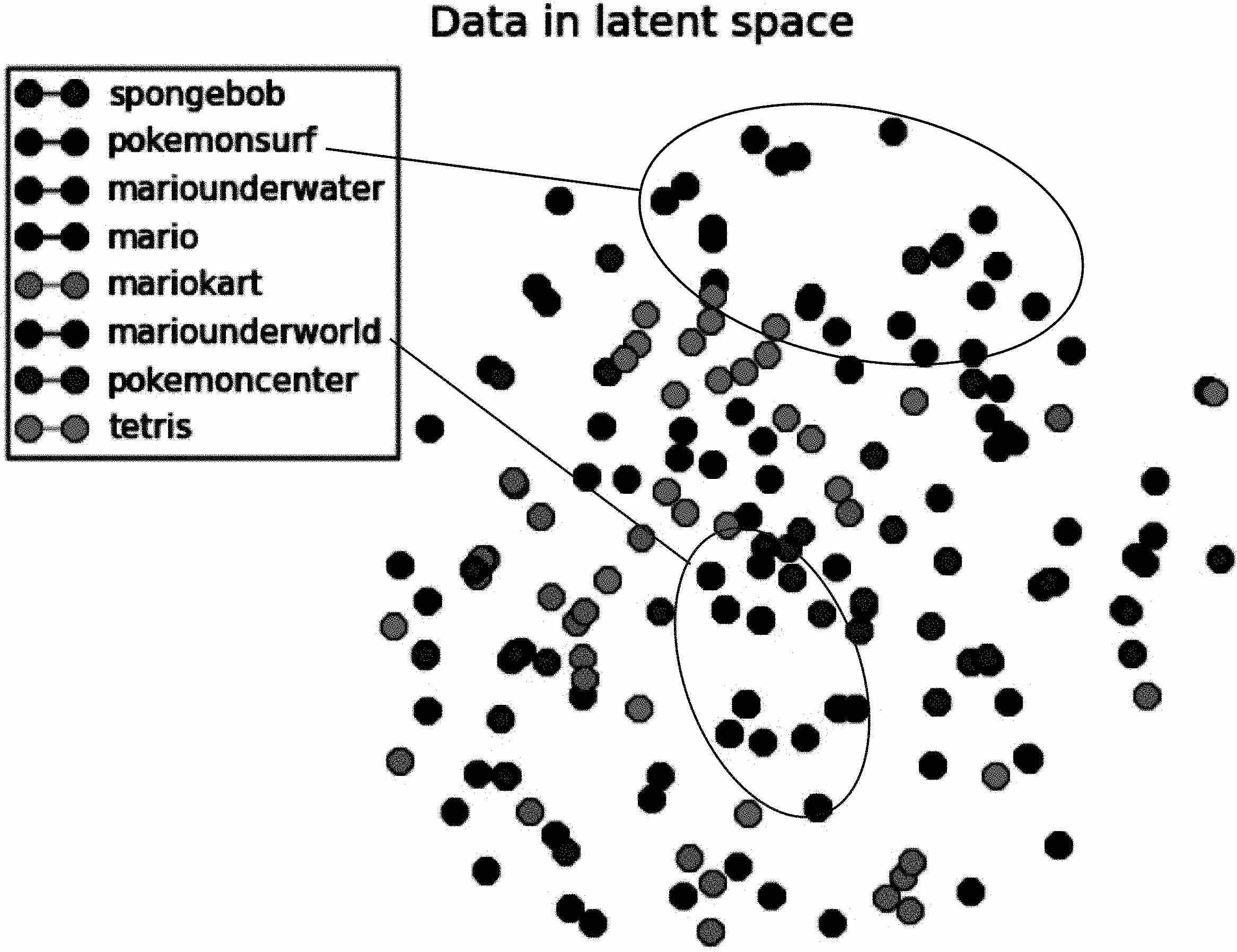}
\caption{Visualization of the VRAE latent space encoded data.
Extended from \cite{fabius:vrae:arxiv:2015} with permission of the authors}
\label{figure:vrae:space}
\end{figure}

After the training phase, the latent space vector can be sampled and used by the RNN encapsulated within the decoder
to generate iteratively a melody.
This could be done by random sampling or also by interpolating between the values of the latent variables corresponding to different songs that have been learnt,
creating a sort of ``medley'' of these songs.
Figure~\ref{figure:vrae:space} visualizes the organisation of the encoded data in the latent space,
each color representing the data points from one song.
The result is positive, but the low musical quality of the corpus hampers a careful evaluation.
The VRAE system is summarized in Table~\ref{table:dimensions:vrae}.

\begin{table}
\begin{tabular}{|l|l|}
\hline
{\em Objective}			&Melody; Video game songs\\
\hline
{\em Representation}	&Symbolic; MIDI; One-hot\\
\hline
{\em Architecture}		&Variational(Autoencoder(LSTM, LSTM))\\
\hline
{\em Strategy}			&Decoder feedforward; Iterative feedforward; Sampling\\
\hline
\end{tabular}
\caption{VRAE summary}
\label{table:dimensions:vrae}
\end{table}

\paragraph{\#2 Example: GLSR-VAE Melody Symbolic Music Generation System}
\label{section:experiment:glsr:vae}

The architecture proposed by Hadjeres and Nielsen in \cite{hadjeres:glsr:vae:arxiv:2017}
is based on a variational autoencoder (VAE) architecture
(Section~\ref{section:architecture:vae}),
but it
proposes an improvement in the control of the variation in the generation,
named {\em geodesic latent space regularization\index{Geodesic latent space regularization}} (GLSR\index{GLSR}),
with a system named GLSR-VAE\index{GLSR-VAE}.

%
%

%

The starting point is that a straight line between two points in the latent space
will not necessarily produce the {\em best} interpolation in the generated content domain space.
The idea is to introduce a
regularization
to relate variations in the latent space to variations in the attributes\index{Attribute} of the decoded elements.
The details of the definition of the added cost term may be found in \cite{hadjeres:glsr:vae:arxiv:2017}.

The experiment consists in generating chorale melodies in the style of J. S. Bach.
The dataset comprises monophonic soprano voices from the J. S. Bach chorales corpus \cite{bach:chorales:book}.

GLSR-VAE shares the principles of representation initiated by the DeepBach\index{DeepBach} system
(Section~\ref{section:experiment:deep:bach}),
that is

\begin{itemize}

\item one-hot encoding of a note,

\item with the addition to the vocabulary
of the hold\index{Hold} symbol ``\_\_'' and the rest\index{Rest} symbol to specify,
respectively, a note repetition and a rest
(see Section~\ref{section:representation:input:encoding:hold:rest}), and

\item using the names of the notes (with no enharmony,
e.g., F$\sharp$ and G$\flat$ are considered to be different, see Section~\ref{section:note:encoding}).

\end{itemize}

Quantization is at the level of a sixteenth note.
The latent variable space is set to 12 dimensions (12 latent variables).

\begin{figure}
\includegraphics[width=\textwidth]{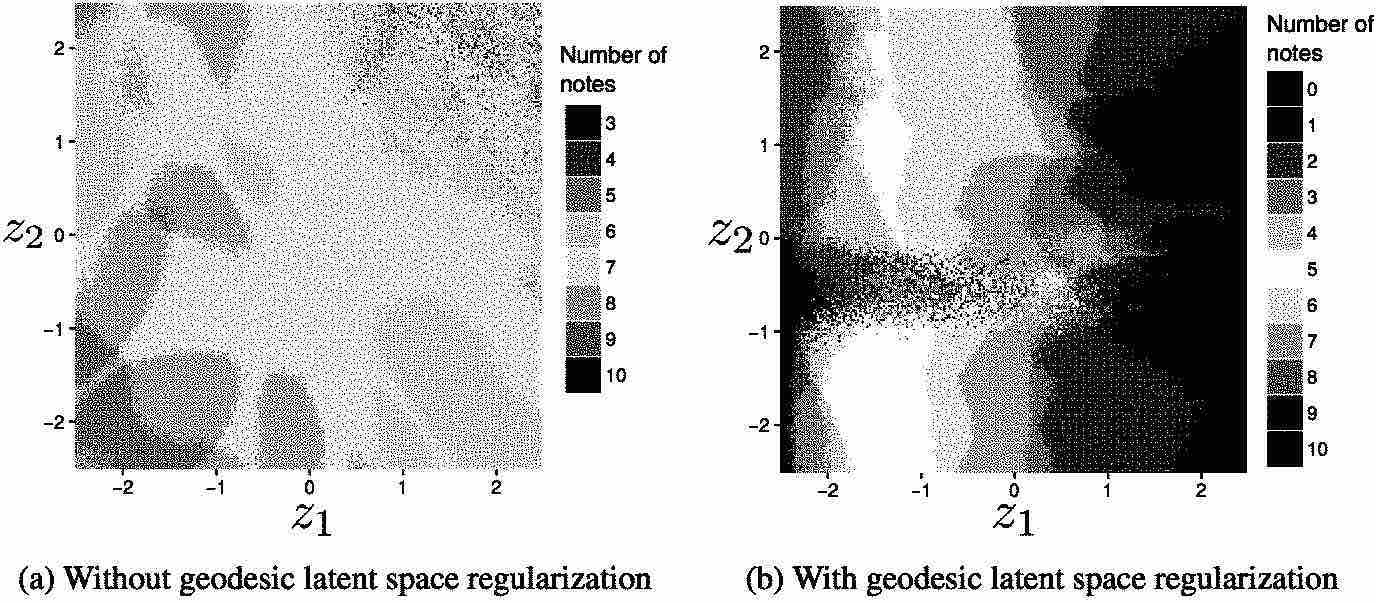}
\caption{Visualization of GLSR-VAE latent space encoded data.
Reproduced from \cite{hadjeres:glsr:vae:arxiv:2017} with permission of the authors}
\label{figure:glsr:vae:space:architecture}
\end{figure}
\begin{figure}
\includegraphics[width=\textwidth]{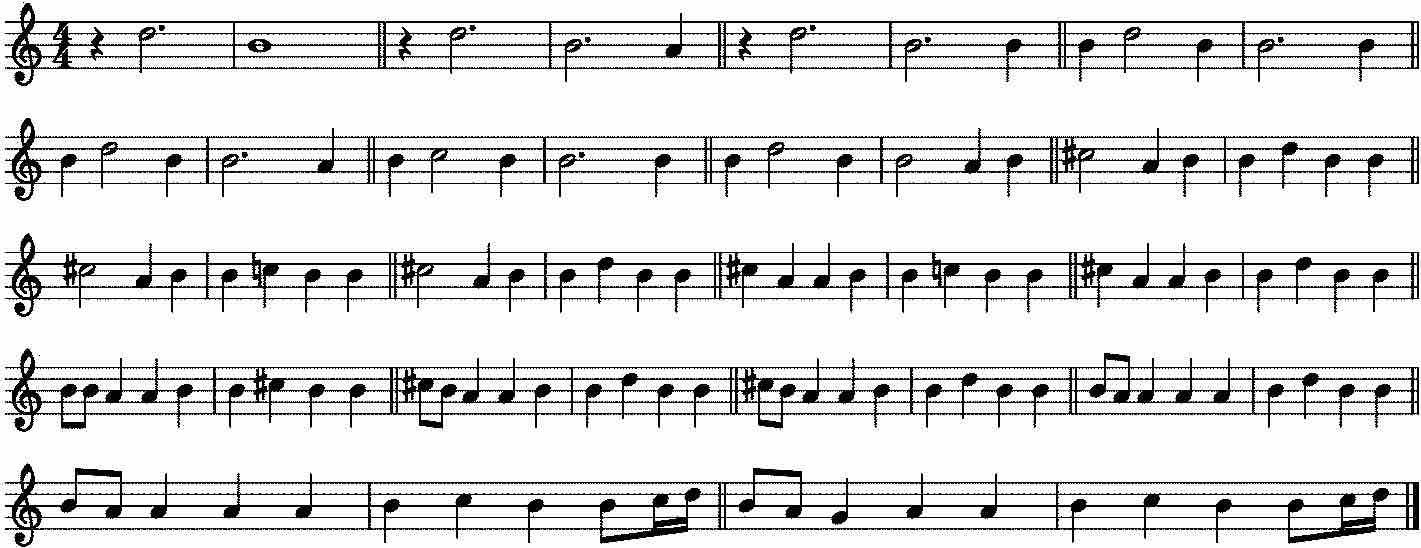}
\caption{Examples of 2 measures long melodies (separated by double bar lines) generated by GLSR-VAE.
Reproduced from \cite{hadjeres:glsr:vae:arxiv:2017} with permission of the authors}
\label{figure:glsr:vae:space:examples}
\end{figure}

In the experiments conducted, regularization is executed on a first dimension which has been found\footnote{See
	Section~\ref{section:architecture:vae}.}
to represent the number of notes (named z$_1$).
Figure~\ref{figure:glsr:vae:space:architecture} shows the organisation of the encoded data in the latent space,
with the number of notes z$_1$ being the abscissa axis,
with from left to right an effective progressive increase in the number of notes (shown with scales of colors).
Figure~\ref{figure:glsr:vae:space:examples} shows examples of the melodies generated
(each 2 measures long, separated by double bar lines)
while increasing z$_1$, showing a progressive correlated densification of the melodies generated.

GLSR-VAE is summarized in Table~\ref{table:dimensions:glsr:vae}.
More examples of sampling from variational autoencoders will be described in Section~\ref{section:system:music:vae}.

\begin{table}
\begin{tabular}{|l|l|}
\hline
{\em Objective}			&Melody; Bach\\
\hline
{\em Representation}	&Symbolic; Piano roll; One-hot; Hold; Rest; Fermata; No enharmony\\
\hline
{\em Architecture}		&Variational(Autoencoder(LSTM, LSTM)); Geodesic regularization\\
\hline
{\em Strategy}			&Decoder feedforward; Sampling\\
\hline
\end{tabular}
\caption{GLSR-VAE summary}
\label{table:dimensions:glsr:vae}
\end{table}




\subsubsection{Sampling for Adversarial Generation}
\label{section:challenges:strategies:control:sampling:and:adversarial}

\label{section:system:strategy:adversarial:feedforward}

\label{section:system:strategy:adversarial:feedforward:architecture:gan:rnn}



Another example of a generative model\index{Generative!model}
is a generative adversarial networks\index{Generative!adversarial networks} (GAN\index{GAN}) architecture.
In such an architecture, after having trained the generator in an adversarial way, generation of content is done by sampling latent random variables.


\paragraph{Example: Mogren's C-RNN-GAN Classical Polyphony Symbolic Music Generation System}
\label{section:systems:c:rnn:gan}

The objective of Mogren's C-RNN-GAN\index{C-RNN-GAN} \cite{mogren:c-rnn-gan:arxiv:2016} system
is the generation of single voice polyphonic music.
The representation chosen is inspired by MIDI and models each musical event (note)
via four attributes\index{Attribute}: duration, pitch, intensity and time elapsed since the previous event,
each attribute being encoded as a real value scalar.
This allows the representation of simultaneous\index{Simultaneous} notes (in practice up to three).
The musical genre of the corpus is classical music,
retrieved in MIDI format from the Web and contains 3,697 pieces from 160 composers.

\begin{figure}
\includegraphics[scale=0.4]{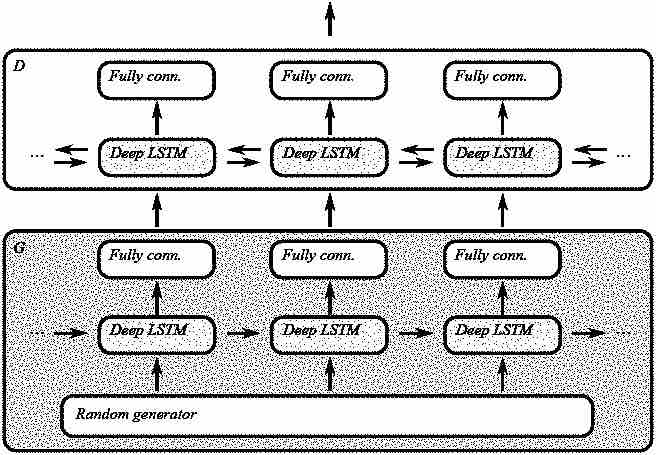}
\caption{C-RNN-GAN architecture.
Reproduced from \cite{mogren:c-rnn-gan:arxiv:2016} with permission of the authors}
\label{figure:c:rnn:gan:architecture}
\end{figure}

C-RNN-GAN is based on a generative adversarial networks (GAN) architecture,
with both the generator and the discriminator being recurrent networks\footnote{This generative GAN
	architecture encapsulates two recurrent networks,
	in the same spirit that the generative VRAE\index{VRAE} variational autoencoder architecture encapsulates two recurrent networks
	as explained in Section~\ref{section:experiment:vrae}.},
more precisely each having two LSTM layers with 350 units each.
A specificity is that the discriminator (but not the generator) has a bidirectional recurrent architecture\index{Bidirectional!recurrent neural network},
in order to take context\index{Context} from both the past and the future for its decisions.
The architecture is shown in Figure~\ref{figure:c:rnn:gan:architecture}
and summarized in Table~\ref{table:dimensions:c:rnn:gan}.

The discriminator is trained, in parallel to the generator, to classify if a sequence input is coming from the real data.
Similar to the case of the encoder part of the RNN Encoder-Decoder,
which summarizes a musical sequence into the values of the hidden layer
(see Section~\ref{section:architecture:compound:recurrent:autoencoder}),
the bidirectional RNN decoder part of the C-RNN-GAN summarizes the sequence input into the values of the two hidden layers
(forward sequence and backward sequence)
and then classifies them.

\begin{figure}
\includegraphics[width=\textwidth]{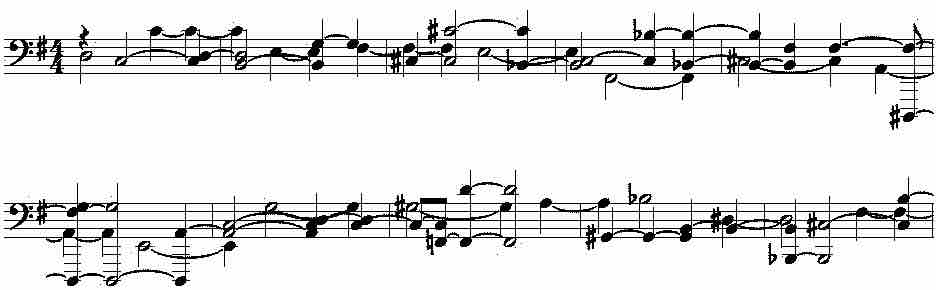}
\caption{C-RNN-GAN generated example (excerpt).
Reproduced from \cite{mogren:c-rnn-gan:arxiv:2016} with permission of the authors}
\label{figure:c:rnn:gan:example}
\end{figure}

An example of generated music is shown in Figure~\ref{figure:c:rnn:gan:example}.
The author conducted a number of measurements
on the generated music.
He states that the model trained with feature matching\index{Feature!matching}\footnote{A regularization technique for improving GANs,
	see Section~\ref{section:architecture:gan:challenges}.}
achieves a better trade-off\index{Trade-off}
between structure and surprise than the other variants.
Note that this is consonant with the use of the feature matching regularization technique
to control creativity\index{Creativity} in MidiNet\index{MidiNet}
(to be introduced in Section~\ref{section:systems:midinet}).
C-RNN-GAN is summarized in Table~\ref{table:dimensions:c:rnn:gan}.

\begin{table}
\begin{tabular}{|l|l|}
\hline
{\em Objective}			&Polyphony\\
\hline
{\em Representation}	&Symbolic; MIDI; Value encoding$\times$4\\
\hline
{\em Architecture}		&GAN(Bidirectional-LSTM$^2$, LSTM$^2$)\\
\hline
{\em Strategy}			&Iterative feedforward; Sampling\footnote{Note that we do not consider
	GAN (adversarial) as a strategy, because we consider that the essence of GAN is
	about the {\em architecture} and the strategy about the {\em training phase},
	whereas we are here concerned with the strategy about the {\em generation phase}.
	After having trained the generator, generation uses a conventional iterative feedforward strategy for an RNN
	with the addition of some sampling for the variability,
	as described in Section~\ref{section:challenges:strategies:variability:sampling}.}\\
\hline
\end{tabular}
\caption{C-RNN-GAN summary}
\label{table:dimensions:c:rnn:gan}
\end{table}

\subsubsection{Sampling for Other Generation Strategies}
\label{section:challenges:strategies:control:sampling:other}


Sampling may also be combined with other strategies for content generation,
as for instance

\begin{itemize}

\item {\em Conditioning}, as a way to {\em parametrize} generation with constraints, in Section~\ref{section:systems:anticipation:rnn}, or

\item {\em Input manipulation}, as a way to {\em correct} the manipulation performed
in order to {\em realign} the samples with the learnt distribution,
in Section~\ref{section:control:input:manipulation:sampling}.

\end{itemize}

\subsection{Conditioning}
\label{section:challenges:strategies:control:conditioning}
\label{section:control:conditioning}


The idea of {\em conditioning\index{Conditioning}\index{Conditioning!architecture}}
(sometimes also named {\em conditional architecture\index{Conditional!architecture}})
is to condition the architecture on some extra
information,
which could be arbitrary, e.g., a class label\index{Label} or data from other modalities\index{Modality}.
Examples are

\begin{itemize}

\item a {\em bass line\index{Bass!line}} or a {\em beat\index{Beat} structure},
in the rhythm generation architecture (Section~\ref{section:systems:makris:rhythm}),

\item a {\em chord progression\index{Chord!progression}},
in the MidiNet\index{MidiNet} architecture (Section~\ref{section:systems:midinet}),

\item the {\em previously generated note},
in the VRASH\index{VRASH} architecture (Section~\ref{section:experiment:vrash}),

\item some {\em positional constraints\index{Positional constraint} on notes},
in the Anticipation-RNN\index{Anticipation-RNN} architecture (Section~\ref{section:systems:anticipation:rnn}),

\item a {\em musical genre\index{Musical!genre}} or an {\em instrument\index{Instrument}},
in the WaveNet\index{WaveNet} architecture (Section~\ref{section:systems:wavenet}), and

\item a {\em musical style},
in the DeepJ\index{DeepJ} architecture (Section~\ref{section:systems:deepj}).

\end{itemize}


In practice, the conditioning information
is usually fed into the architecture as an additional input layer\index{Input!layer}
(for example, see 
Figure~\ref{figure:conditioning:architecture}).
This distinction between {\em standard input} and {\em conditioning input\index{Conditioning!input}}
follows a good architectural\index{Architectural} modularity\index{Modularity} principle\footnote{Note that we do not consider
	conditioning as a strategy because we consider that the essence of conditioning
	relates to the {\em conditioning architecture\index{Conditioning!architecture}}.
	Generation uses a conventional strategy (e.g., single-step feedforward, iterative feedforward\ldots)
	depending on the type of the architecture (e.g., feedforward, recurrent\ldots).}.
Conditioning is a way to have some degree of parametrized control over the generation\index{Generation} process.



 

The conditioning layer could be

\begin{itemize}

\item a simple input layer.
An example is a tag\index{Tag} specifying a musical genre or an instrument
in the WaveNet\index{WaveNet} system (Section~\ref{section:systems:wavenet}),

\item some output of some architecture, being

\begin{itemize}

\item the same architecture, as a way to condition the architecture on some history\footnote{This is close in spirit
	to a recurrent architecture (RNN).}
	-- an example is the MidiNet\index{MidiNet} system (Section~\ref{section:systems:midinet})
	in which history information from previous measure(s) is injected back into the architecture, or

\item another architecture
	-- examples are the rhythm generation system (Section~\ref{section:systems:makris:rhythm})
	in which a feedforward network in charge of the bass line and the metrical structure information
	produces the conditioning 	input\index{Conditioning!input},
	and the DeepJ\index{DeepJ} system (Section~\ref{section:systems:deepj})
	in which two successive transformation layers of a style\index{Style} tag\index{Tag}
	produce an embedding\index{Embedding} used as the conditioning input.

\end{itemize}

\end{itemize}



If the architecture is time-invariant
-- i.e. recurrent or convolutional over time --,
there are two options

\begin{itemize}

\item {\em global conditioning} --
if the conditioning input\index{Conditioning!input} is shared for all time steps\index{Time!step}, or

\item {\em local conditioning} --
if the conditioning input is specific to each time step.

\end{itemize}

The WaveNet\index{WaveNet} architecture,
which is convolutional over time (see Section~\ref{section:architecture:convolution:time}),
offers the two options, as will be analyzed in Section~\ref{section:systems:wavenet}.

\subsubsection{\#1 Example: Rhythm Symbolic Music Generation System}
\label{section:systems:makris:rhythm}
\label{section:experiment:makris:rhythm}

The system proposed by 
Makris {\em et al.} \cite{makris:rhythm:composition:2017} is specific in that it is dedicated
to the generation of sequences of rhythm\index{Rhythm}.
Another specificity is the possibility to condition\index{Conditioning} the generation relative to some particular information,
such as a given beat\index{Beat} or bass line\index{Bass!line}.

The corpus includes 45 drum and bass patterns, each 16 measures long in 4/4 time signature,
from three different rock bands and converted to MIDI.
The representation of drums is described in
Section~\ref{section:representation:rhythm:drums}
and summarized as follows.
Different drum components (kick, snare, toms, hi-hat, cymbals)
are considered as distinct simultaneous voices, following a many-hot approach,
and encoded in text as a binary word of length 5,
e.g., $10010$ represents the simultaneous playing of kick and high-hat.

The representation also includes a condensed representation of the bass line\index{Bass!line} part.
It captures the voice leading perspective of the bass\footnote{The voice leading of the bass has proven a valuable aspect
	in harmonization systems, see, e.g., \cite{harnel:chornet:jnmr:2004}.},
by specifying the pitch difference direction for the bass between two successive time steps.
This is represented in a binary word of length 4,
the first digit specifying the existence of a bass event ($1$) or a rest event ($0$),
while the three remaining digits specify the 3 possible directions for voice leading:
steady ($000$), upward ($010$) and downward ($001$).
Last, the representation includes some additional information
representing the metrical structure (the beat\index{Beat} structure), also through binary words.
See further details in \cite{makris:rhythm:composition:2017}.

\begin{figure}
\includegraphics[width=\textwidth]{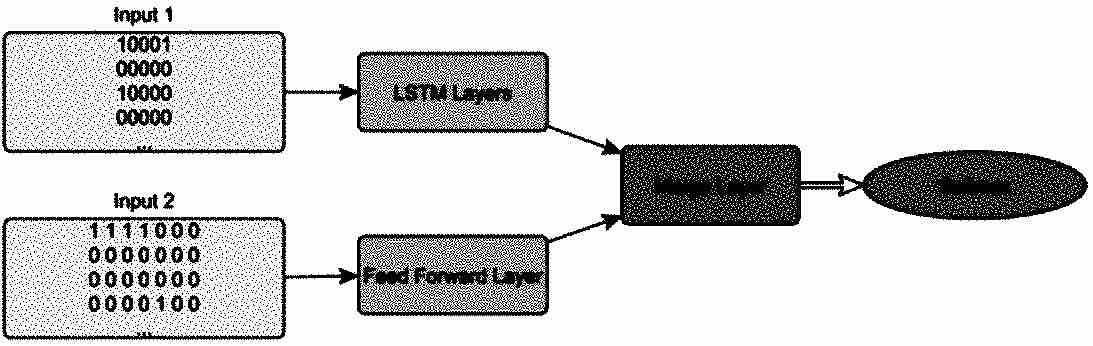}
\caption{Rhythm generation architecture.
Reproduced from \cite{makris:rhythm:composition:2017} with permission of the authors}
\label{figure:rhythm:architecture}
\end{figure}

The architecture is a combination of a recurrent network (more precisely, an LSTM)
and a feedforward network\index{Feedforward!network}, representing the conditioning layer.
The LSTM
(two stacked LSTM layers with 128 or 512 units)
is in charge of the drums\index{Drums} part,
while the feedforward network is in charge of the bass line\index{Bass!line} and the metrical structure
information.
The outputs of these two networks are then merged\footnote{Note that in this system,
	the conditioning layer
	is added to the main architecture at its output level and not at its input level.
	Therefore an additional feedforward merge layer is introduced.
	We could notate the resulting architecture as Conditioning(Feedforward(LSTM$^2$), Feedforward).},
resulting in the architecture illustrated in Figure~\ref{figure:rhythm:architecture}.
The authors report that the conditioning layer (bass line and beat information) improves the quality of the learning and of the generation.
It may also be used in order to mildly influence the generation.
More details may be found in the article \cite{makris:rhythm:composition:2017}.
The architecture is summarized in Table~\ref{table:dimensions:rhythm}.

\begin{figure}
\includegraphics[scale=1.5]{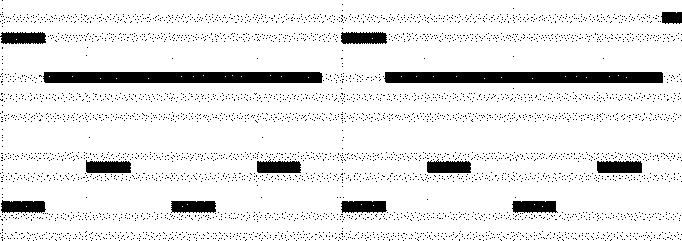}
\caption{Example of a rhythm pattern generated.
The five lines of the piano roll correspond (downwards) to:
kick, snare, toms, hi-hat and cymbals.
Reproduced from \cite{makris:rhythm:composition:2017} with permission of the authors}
\label{figure:rhythm:example}
\end{figure}

\begin{figure}
\includegraphics[scale=1.5]{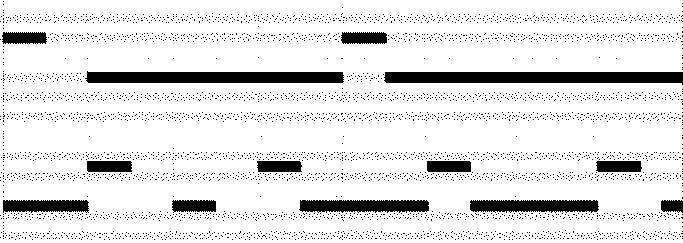}
\caption{Example of a rhythm pattern generated with a specific bass line as the conditioning input.
Reproduced from \cite{makris:rhythm:composition:2017} with permission of the authors}
\label{figure:rhythm:example:bass}
\end{figure}

An example of a rhythm pattern generated is shown in Figure~\ref{figure:rhythm:example}
with in Figure~\ref{figure:rhythm:example:bass} the use of a specific and more complex bass line as a conditioning input
which produces a rhythm more elaborate.
The piano roll like visual representation shows in its five successive lines (downwards)
the kick, snare, toms, hi-hat and cymbals components events.

\begin{table}
\begin{tabular}{|l|l|}
\hline
{\em Objective}			&Multivoice; Rhythm; Drums\\
\hline
{\em Representation}	&Symbolic; Beat; Drums; Many-hot; Bass line; Note; Rest; Hold\\
\hline
{\em Architecture}		&Conditioning(Feedforward(LSTM$^2$), Feedforward)\\
\hline
{\em Strategy}			&Iterative feedforward; Sampling\\
\hline
\end{tabular}
\caption{Rhythm system summary}
\label{table:dimensions:rhythm}
\end{table}

\subsubsection{\#2 Example: WaveNet Speech and Music Audio Generation System}
\label{section:systems:wavenet}

WaveNet\index{WaveNet}, by
van der Oord {\em et al.} \cite{oord:wavenet:arxiv:2016},
is a system for generating raw audio\index{Audio} waveforms\index{Waveform},
quite innovative in that respect.
It has been tested in three audio domains:
multi-speaker\index{Speaker}, text-to-speech\index{Text!-to-speech} (TTS\index{TTS})
and music.

\begin{figure}
\includegraphics[width=\textwidth]{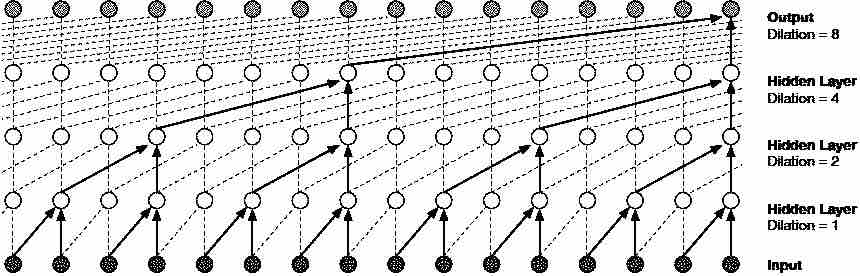}
\caption{WaveNet architecture.
Reproduced from \cite{oord:wavenet:arxiv:2016} with permission of the authors}
\label{figure:wavenet:architecture}
\end{figure}

The architecture
is based on a convolutional\index{Convolutional!network}
feedforward network with no pooling\index{Pooling} layer.
Convolutions\index{Convolution} are constrained in order to ensure that the prediction only depends on previous time steps,
and are therefore named {\em causal convolutions\index{Causal convolution}}.
The actual implementation is optimized through the use of
{\em dilated convolution\index{Dilated convolution}} (also called ``\`a trous''),
where the convolution filter is applied over an area larger than its length by skipping input values with a certain step.
Incrementally dilated successive convolution layers\footnote{The dilation is doubled for every layer up to a limit and then repeated,
	e.g., 1, 2, 4, \ldots, 512, 1, 2, 4, \ldots, 512, \ldots}
enable networks to have very large receptive fields with just a few layers
while preserving the input resolution throughout the network as well as computational efficiency
(see \cite{oord:wavenet:arxiv:2016} for more details).
The architecture is illustrated in Figure~\ref{figure:wavenet:architecture}.

Another specificity of WaveNet is in the training/generation asymmetry\index{Asymmetry}: 
during the training\index{Training} phase, predictions for all time steps can be made in parallel\index{Parallel},
whereas during the generation\index{Generation} phase, predictions are sequential
(following the iterative feedforward strategy).

The WaveNet architecture is made conditioning\index{Conditioning}, as a way to guide the generation, by adding an additional tag\index{Tag}
as a conditioning input\index{Conditioning!input}.
We could thus notate the architecture as Conditioning(Convolutional(Feedforward), Tag).

There are actually two options:

\begin{itemize}

\item {\em global conditioning\index{Global!conditioning}},
if the conditioning input\index{Conditioning!input} is shared for {\em all} time steps\index{Time!step}; and

\item {\em local conditioning\index{Local!conditioning}},
if the conditioning input is specific to {\em each} time step.

\end{itemize}


An example of conditioning for a text-to-speech\index{Text!-to-speech} application domain
is to feed in linguistic features from different speakers, e.g., North American or Mandarin Chinese English speakers,
in order to generate speech with a specific prosody.

The authors also conducted preliminary work on conditioning\index{Conditioning} models
to generate music given a set of tags specifying,
for example, genre\index{Genre|see{Musical genre}} or instruments\index{Instrument}.
They state (without further details) that their preliminary attempt is promising \cite{oord:wavenet:arxiv:2016}.
WaveNet is summarized in Table~\ref{table:dimensions:wavenet}.

Last, let us mention, a recent proposal as an offspring from WaveNet,
which uses a symbolic representation (associated to the audio input) as the conditioning model/input,
in order to better guide and structure the generation of (audio) music
(see details in \cite{manzelli:combining:raw:symbolic:audio:networks:ismir:2018}).

\begin{table}
\begin{tabular}{|l|l|}
\hline
{\em Objective}			&Audio\\
\hline
{\em Representation}	&Audio; Waveform\\
\hline
{\em Architecture}		&Conditioning(Convolutional(Feedforward), Tag); Dilated convolutions\\
\hline
{\em Strategy}			&Iterative feedforward; Sampling\\
\hline
\end{tabular}
\caption{WaveNet summary}
\label{table:dimensions:wavenet}
\end{table}

\subsubsection{\#3 Example: MidiNet Pop Music Melody Symbolic Music Generation System}
\label{section:systems:midinet}

In \cite{yang:midinet:ismir:2017},
Yang {\em et al.} propose the MidiNet\index{MidiNet} architecture,
which is both
adversarial
and convolutional,
for the generation of single or multitrack pop\index{Pop} music monophonic melodies.

The corpus used is
a collection of 1,022 pop music songs from the TheoryTab\footnote{Tabs are piano roll-like leadsheets,
	including melody, lyrics and notation of chords.}
online database \cite{theorytab:web:2017}	
that provides two channels per tab, one for the melody and the other for the underlying chord progression.
This allows two versions of the system:
one with only the melody channel
and another that additionally uses chords to condition melody generation.
After all the preprocessing steps, the dataset is composed of 526 MIDI tabs (representing 4,208 measures).
Data augmentation is then performed by circularly shifting all melodies and chords to any of the 12 keys,
leading to a final dataset of 50,496 measures of melody and chord pairs for training.

The representation is obtained by transforming each channel of MIDI files
into a one-hot encoding of 8 measures long piano roll representations,
using one of the encodings to represent silence (rest)
and neglecting the velocity of the note events. 
The time step is set at the smallest note, a sixteenth note.
All melodies have been transposed in order to fit within the two-octave interval $[$C$_4,$~B$_5]$\footnote{However,
	the authors considered all the 128 MIDI note numbers
	(corresponding to the $[$C$_0,$~G$_{10}]$ interval) in a one-hot encoding and state in \cite{yang:midinet:ismir:2017} that:
	``In doing so, we can detect model collapsing more easily, by checking whether the model generates notes outside these octaves.''}.
Note that the current representation does not distinguish between a long note and two short repeating notes,
and the authors mention considering future extensions
in order to emphasize the note onsets\index{Note!onset}.

For chords,
instead of using a many-hot\index{Many-hot encoding} vector extensional\index{Extensional} representation
of dimension 24 (for the two octaves),
the authors state that they found it more efficient to use an intensional\index{Intensional} representation of dimension 13:
12 for the pitch-class (key) and 1 for the chord type (major or minor).

\begin{figure}
\includegraphics[width=\textwidth]{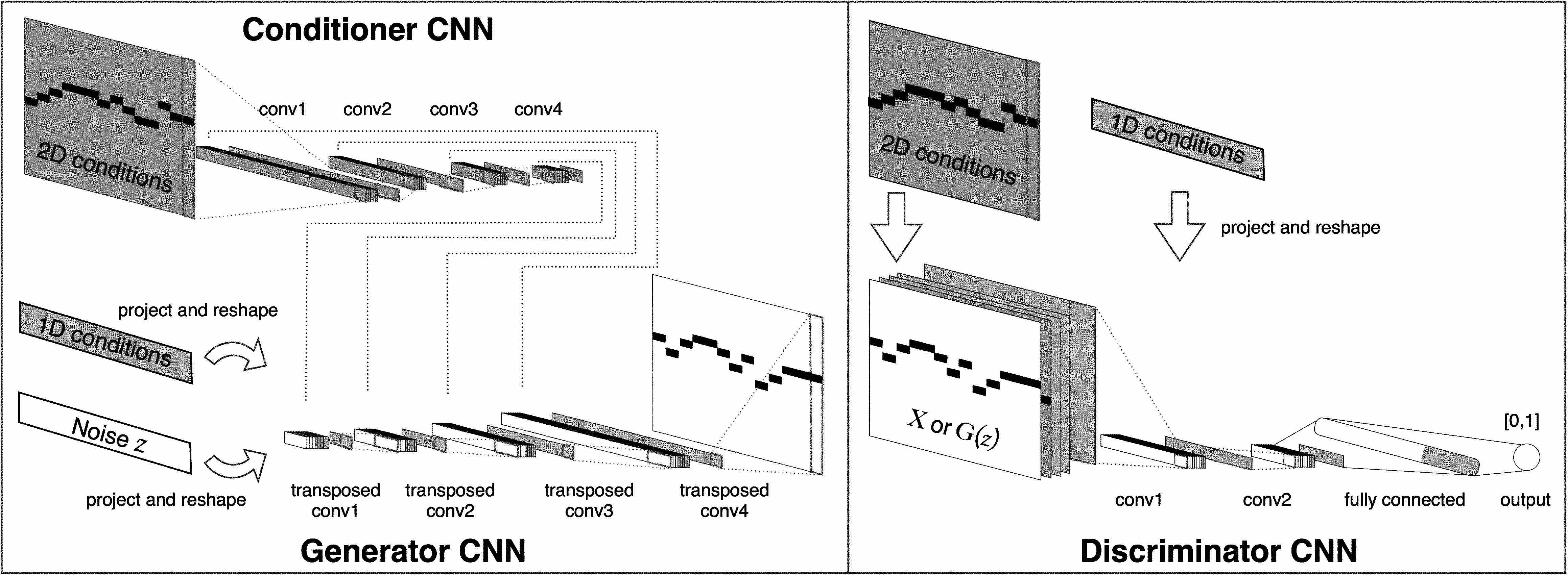}
\caption{MidiNet architecture.
Reproduced from \cite{yang:midinet:ismir:2017} with permission of the authors}
\label{figure:midinet:architecture}
\end{figure}

\begin{figure}
\includegraphics[width=\textwidth]{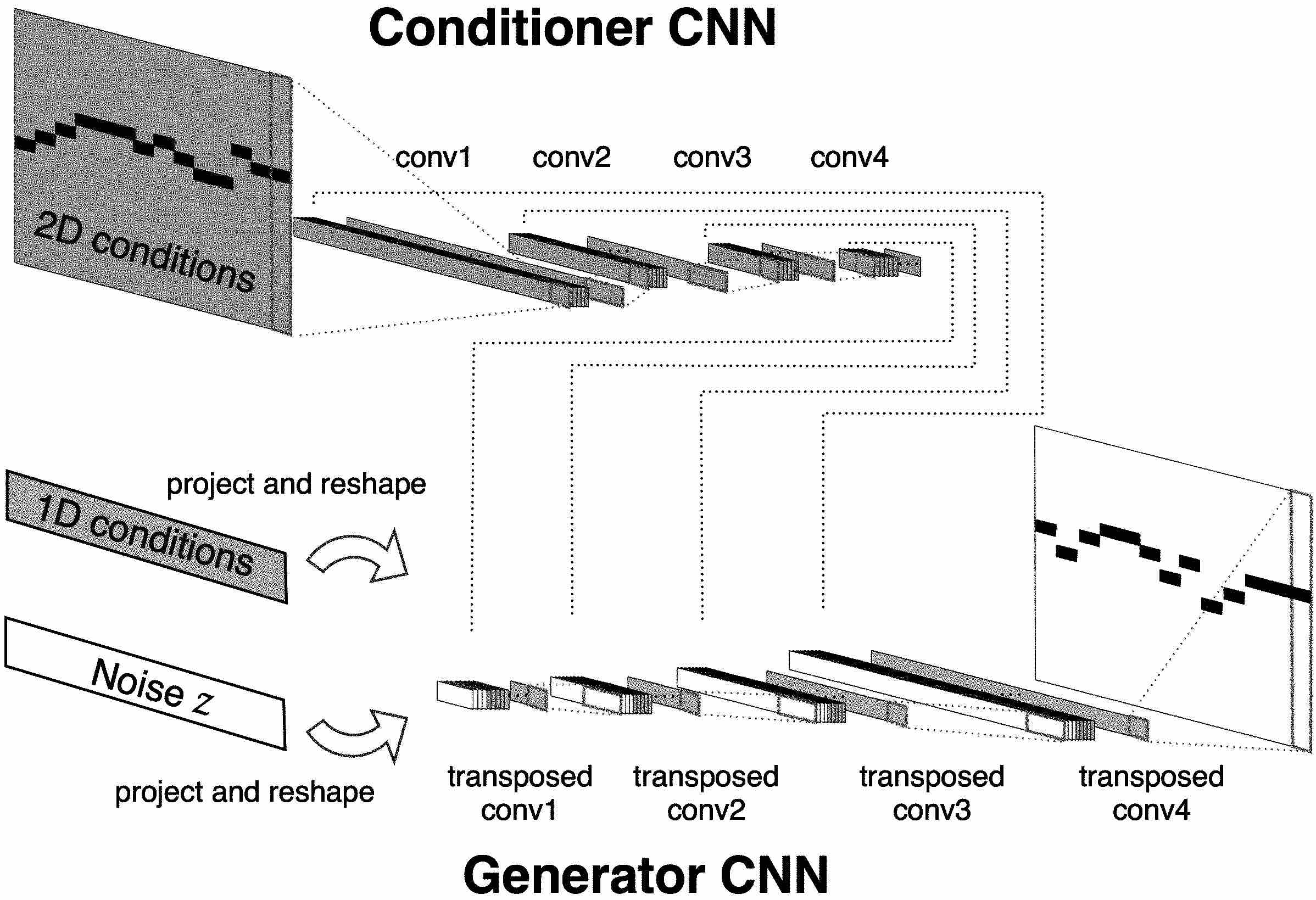}
\caption{Architecture of the MidiNet generator.
Reproduced from \cite{yang:midinet:ismir:2017} with permission of the authors}
\label{figure:midinet:architecture:generator}
\end{figure}

The architecture\footnote{The architecture is complex,
	please see further details in \cite{yang:midinet:ismir:2017}.}
is illustrated in Figures~\ref{figure:midinet:architecture}
and~\ref{figure:midinet:architecture:generator}.
It is composed of a generator
and a discriminator,
which are both convolutional networks.
The generator includes two fully-connected layers (with 1,024 and 512 units respectively) followed by four convolutional layers.
Generation takes place iteratively, by sampling one measure after one measure until reaching 8 measures.
The generator is conditioned by a module
(named Conditioner CNN in Figure~\ref{figure:midinet:architecture})
which includes four convolutional layers with a reverse\index{Reverse} architecture.
The conditioning\index{Conditioning} mechanism incorporates

\begin{itemize}

\item history information from previous measures
(as a memory\index{Memory} mechanism, analog to a RNN),
and

\item the chord sequence (only for the generator).
The discriminator includes two convolutional layers followed by some fully connected layers and the final output activation function
is cross-entropy.

\end{itemize}

The discriminator is also conditioned, but without specific conditioner layers.
We could thus notate the architecture as

\begin{tabular}{l}
GAN(Conditioning(Convolutional(Feedforward),\\
\hspace{1.4cm}Convolutional(Feedforward(History, Chord sequence))),\\
\hspace{0.9cm}Conditioning(Convolutional(Feedforward), History)).
\end{tabular}

The conditioning information could be

\begin{itemize}

\item only about the previous measure -- named ``1D conditions''
(shown in yellow in Figures~\ref{figure:midinet:architecture} and~\ref{figure:midinet:architecture:generator}); or

\item about various previous measures -- named ``2D conditions''
(shown in blue).

\end{itemize}

Both cases are illustrated in Figure~\ref{figure:midinet:architecture}.
The authors report experiments performed with different variants:



\begin{itemize}

\item melody generation with conditioning on the previous measure
(with previous measure as 2D conditions for the generator and as 1D conditions for the discriminator\footnote{To ensure that the discriminator
	distinguishes between real and generated melodies only from the present measure.});

\item melody generation with conditioning on the previous measure and on the chord sequence
(with chord sequence as 1D conditions for the generator,
or alternatively also as 2D conditions only for its last convolutional layer
in order to highlight the chord condition); and

\item melody generation with conditioning on the previous measure and on the chord sequence in a creative mode
(with chord sequence as 2D conditions for all convolutional layers of the generator).

\end{itemize}

For the second variant, which they name {\em stable mode}, the authors report that the generation is more chord-dominant and stable,
in other words it closely follows the chord progression and seldom generates notes violating chord constraints.
For the third variant, named {\em creative mode}\footnote{On the challenge of creativity\index{Creativity},
	see Section~\ref{section:originality}.},
the generator sometimes violates the constraint imposed by the chords,
to better adhere to the melody of the previous measure.
In other words, the creative mode allows a better balance between melody following over chord following.
The authors state in \cite{yang:midinet:ismir:2017} that:
``Such violations sometimes sound unpleasant, but can be sometimes creative.
Unlike the previous two variants, we need to listen to several melodies generated by this model to handpick good ones.
However, we believe such a model can still be useful for assisting and inspiring human composers.''

MidiNet is summarized in Table~\ref{table:dimensions:midinet}.



%
%
%

\begin{table}
\begin{tabular}{|l|l|}
\hline
{\em Objective}			&Melody + Chords; Pop music; Melody vs chords following balance\\
\hline
{\em Representation}	&Symbolic; Chords; Piano roll; One-hot; Rest\\
\hline
{\em Architecture}
					&GAN(Conditioning(Convolutional(Feedforward),\\
					&\hspace{1.2cm}Convolutional(Feedforward(History, Chord sequence))),\\
					&\hspace{0.75cm}Conditioning(Convolutional(Feedforward), History))\\
\hline
{\em Strategy}			&Iterative feedforward; Sampling\\
\hline
\end{tabular}
\caption{MidiNet summary}
\label{table:dimensions:midinet}
\end{table}

\subsubsection{\#4 Example: DeepJ Style-Specific Polyphony Symbolic Music Generation System}
\label{section:systems:deepj}

In \cite{mao:deepj:arxiv:2018},
Mao {\em et al.} propose 
a system named DeepJ\index{DeepJ},
with the objective of being able to control the style\index{Style}
of music generated.
In their experiment, they consider 23 styles, each corresponding to a different composer\index{Composer}
(from Johann Sebastian Bach\index{Bach} to Pyotr Ilyich Tchaikovsky\index{Tchaikovsky})
with his/her specific style\footnote{In other words,
	they identify a style to a composer.}.
They encode the style -- or a combination of styles\footnote{In the case of a combination of several styles,
	the vector must be normalized in order for its sum to be equal to 1.}
-- as a
many-hot
representation over all possible styles (i.e. composers).
Composers are grouped into musical genres\index{Genre}.
Thus a genre is specified (extensionally\index{Extensional}) as an equal combination of the styles (composers) of that genre. 
For example, if the Baroque\index{Baroque} genre is defined by composers 1 to 4,
the Baroque style would be equal to $[0.25, 0.25, 0.25, 0.25, 0, 0,~...]$.
We will see below, when detailing the architecture,
that this somewhat simplistic user-defined\index{User!-defined} style encoding
will be automatically transformed through the learning phase into an
adaptive distributed representation.

The foundation of the architecture is the Bi-Axial LSTM\index{Bi-Axial LSTM} architecture proposed by Johnson in \cite{johnson:tied:evomusart:2017}
(see Section~\ref{section:experiment:biaxial}).
Music representation is based on piano roll, modeling a note through its MIDI note number,
within a truncated range
(originally within the $\{0, 1,\ldots~127\}$ discrete set, truncated to $\{36, 37,\ldots~84\}$, i.e. four octaves)
in order to reduce note input dimensionality.
Quantization is 16 time steps per measure, i.e. a time step with the value of a sixteenth note.
The representation is similar to that for Bi-Axial LSTM\index{Bi-Axial LSTM}.
DeepJ representation uses a replay matrix,
dual to the piano roll matrix of notes,
in order to distinguish between a held note and a replayed note.
DeepJ representation also includes information about dynamics
through a scalar variable\footnote{The authors comment that
	they have also tried an alternate representation of dynamics as a categorical value (one-hot encoding) with 128 bins\index{Bin}
	(as in WaveNet\index{WaveNet}, see Section~\ref{section:systems:wavenet}),
	which is actually the original MIDI\index{MIDI} discretization\index{Discretization}.
	But: ``Contrary to WaveNet's results, our experiments concluded that the scalar representation yielded results that were more harmonious.''
	\cite{mao:deepj:arxiv:2018}}
within the $[0, 1]$ interval.
But the main addition is the use of {\em style\index{Style} conditioning\index{Conditioning}},
via global conditioning\index{Global!conditioning}\footnote{This means that the conditioning input\index{Conditioning!input}
	is shared for {\em all} time steps\index{Time!step},
	see Section~\ref{section:systems:wavenet}.},
as in WaveNet\index{WaveNet}.

\begin{figure}
\includegraphics[scale=0.4]{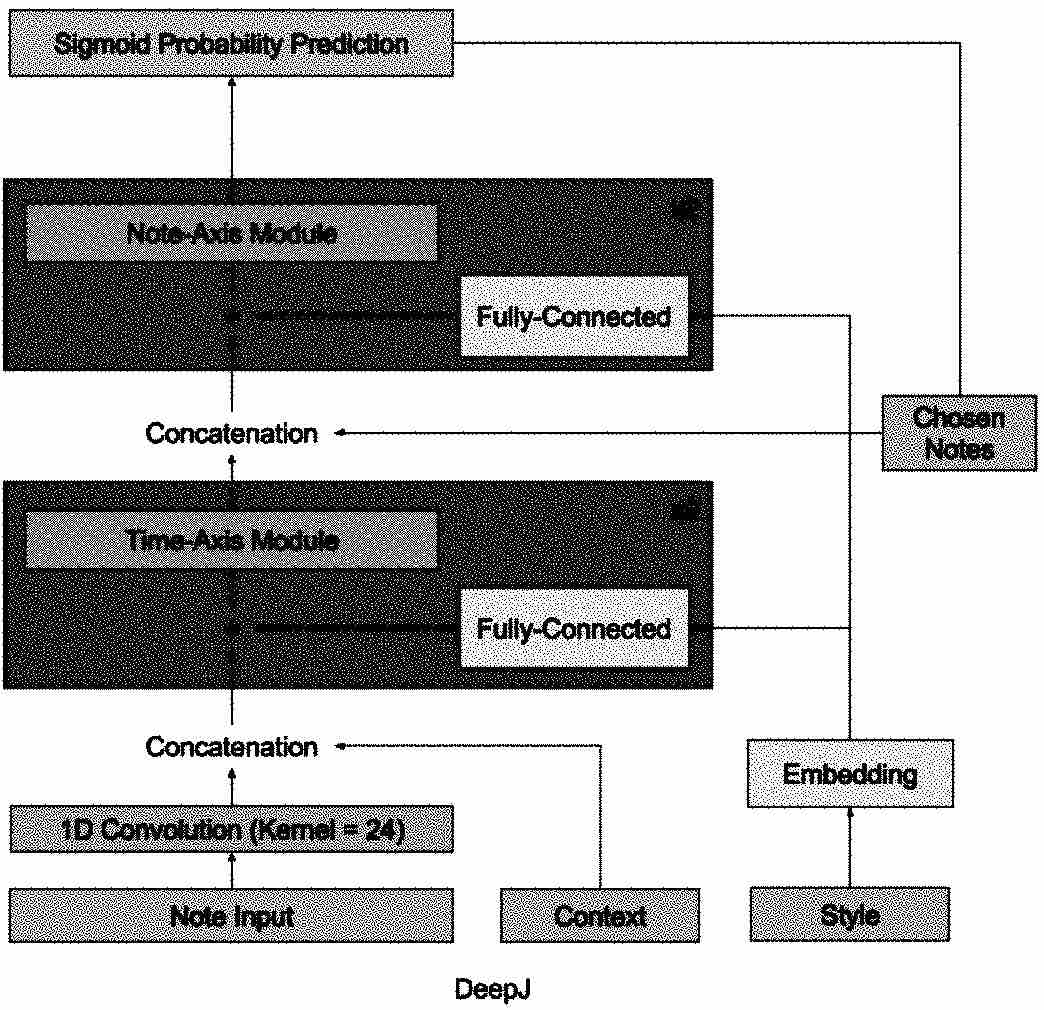}
\caption{DeepJ architecture.
Reproduced from \cite{mao:deepj:arxiv:2018} with permission of the authors}
\label{figure:deepj:architecture}
\end{figure}

As has been noted, the user-defined style encoding is too simplistic to be used as it is.
Musical styles are not necessarily orthogonal to each other and may share many characteristics.
The first transformation layer
linearly transforms the user-defined many-hot encoding of the style into a first embedding\index{Embedding}
(a set of hidden/latent variables,
pictured as the yellow Embedding box in Figure~\ref{figure:deepj:architecture}).
The second transformation layer
transforms this first embedding in a nonlinear way (through a tanh activation\footnote{Hyperbolic
	tangent function.})
into a second embedding of the style
(pictured as the lower yellow Fully-Connected box)
to be added as a conditioning input\index{Conditioning!input} to the time-axis module.
A similar transformation and conditioning is performed for the note-axis module.
Further details and discussion may be found in \cite{mao:deepj:arxiv:2018}.
DeepJ is summarized in Table~\ref{table:dimensions:deepj}.


\begin{figure}
\includegraphics[width=\textwidth]{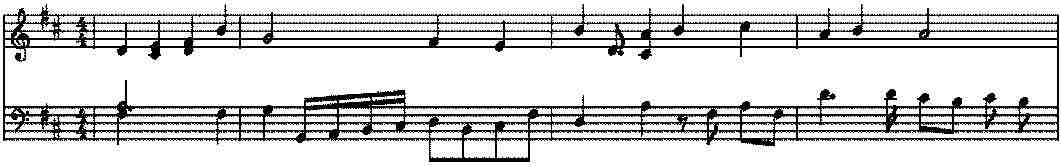}
\caption{Example of baroque music generated by DeepJ.
Reproduced from \cite{mao:deepj:arxiv:2018} with permission of the authors}
\label{figure:deepj:example}
\end{figure}

\begin{figure}
\includegraphics[scale=0.15]{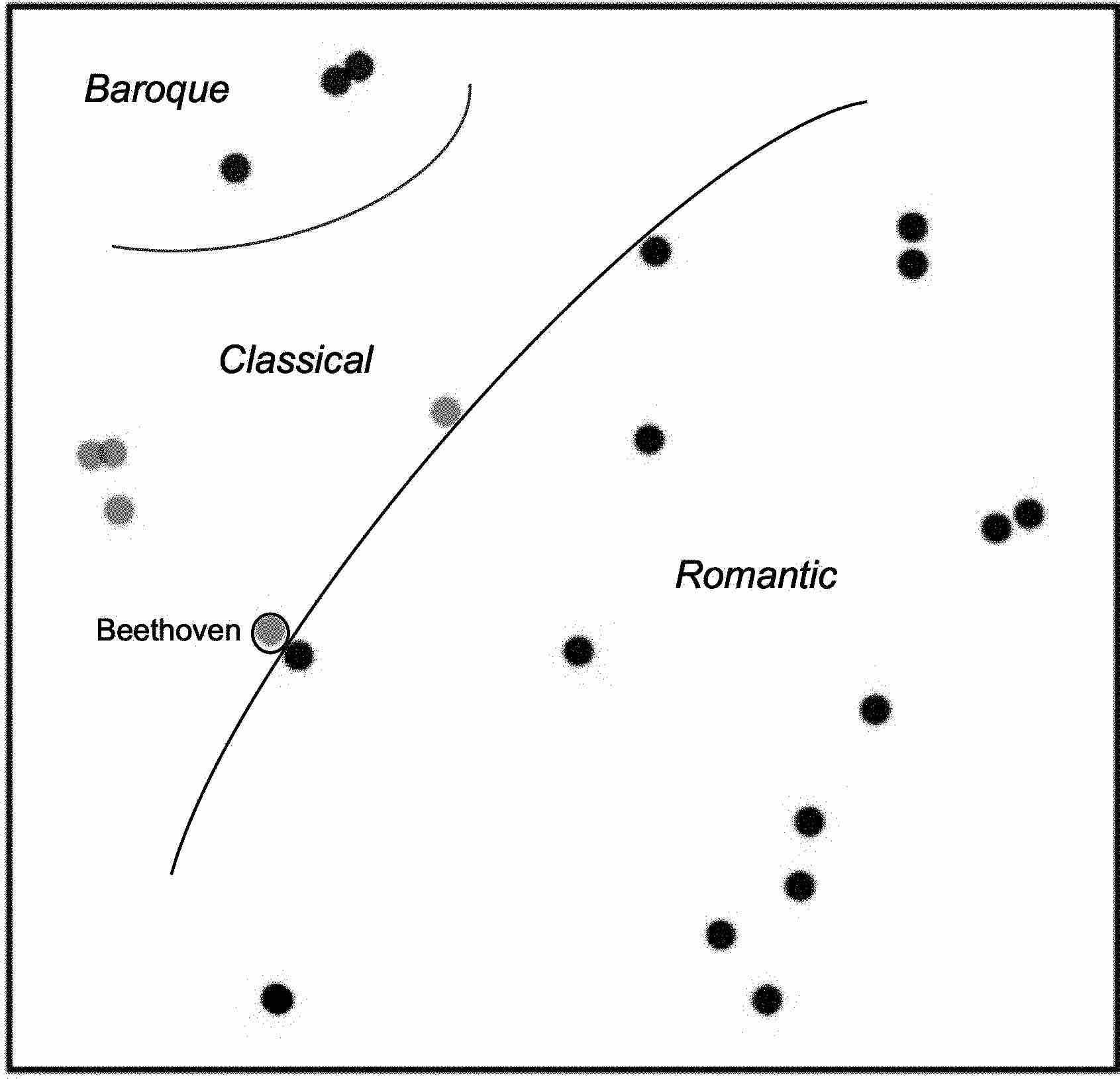}
\caption{Visualization of DeepJ embedding space.
Extended from \cite{mao:deepj:arxiv:2018} with permission of the authors}
\label{figure:deepj:space}
\end{figure}

The authors have conducted an initial subjective evaluation with human listeners comparing music generated by DeepJ
(an example is shown in Figure~\ref{figure:deepj:example})
and by Bi-Axial LSTM.
They report that DeepJ compositions were usually preferred and they comment that the style conditioning
makes generated music more stylistically consistent.
They also conducted a second subjective evaluation in order to verify
whether DeepJ can generate stylistically distinct music (correctly identified by human listeners).
The authors report no statistically significant differences
between the classification accuracy for DeepJ music and real composers music.
A more objective analysis has also been undertaken by visualizing the style\index{Style} embedding\index{Embedding} space,
shown in Figure~\ref{figure:deepj:space},
with each composer pictured as a dot and each cluster as a color
(blue, yellow and red are for baroque, classical and romantic clusters, respectively).
The authors found that composers from similar periods do cluster together (same color)
and point out the interesting result that Ludwig van Beethoven\index{Beethoven} appears at the limit
between the classical and romantic clusters.

\begin{table}
\begin{tabular}{|l|l|}
\hline
{\em Objective}			&Polyphony; Classical; Style\\
\hline
{\em Representation}	&Symbolic; Piano roll; Replay matrix; Rest; Style; Dynamics\\
\hline
{\em Architecture}		&Conditioning(Bi-Axial LSTM, Embedding)\\
					&= Conditioning(LSTM$^2\times$2, Embedding)\\
\hline
{\em Strategy}			&Iterative feedforward; Sampling\\
\hline
\end{tabular}
\caption{DeepJ summary}
\label{table:dimensions:deepj}
\end{table}

\subsubsection{\#5 Example: Anticipation-RNN Bach Melody Symbolic Music Generation System}
\label{section:systems:anticipation:rnn}

In \cite{hadjeres:anticipation:rnn:arxiv:2017},
Hadjeres and Nielsen propose a system named Anticipation-RNN\index{Anticipation-RNN} for generating melodies
with
unary constraints on notes (to enforce a given note at a given time position to have a given value).
The limitation when using a standard iterative feedforward strategy for generation
is that enforcing the constraint\index{Constraint} at time $i$ may retrospectively invalidate
the distribution of the previously generated items\footnote{As the authors
	put it, imposing a constraint on time index $i$
	``twists'' the conditional probability\index{Conditional!probability} distribution $P(\text{s}_t | \text{s}_{<t})$ for $t < i$.},
as shown in \cite{pachet:markov:constraints:ijcai:2011}.
The idea is then to condition the RNN on information summarizing the set of further (in time) constraints,
as a way to anticipate oncoming constraints, in order to generate notes with a correct distribution.

Therefore, a second RNN architecture, named Constraint-RNN, is used that functions backward in time.
Its outputs are used as additional inputs for the main RNN, which the authors name Token-RNN.
The complete architecture is illustrated in Figure~\ref{figure:anticipation:rnn:architecture},
with the following notation and meaning:

\begin{figure}
\includegraphics[width=\textwidth]{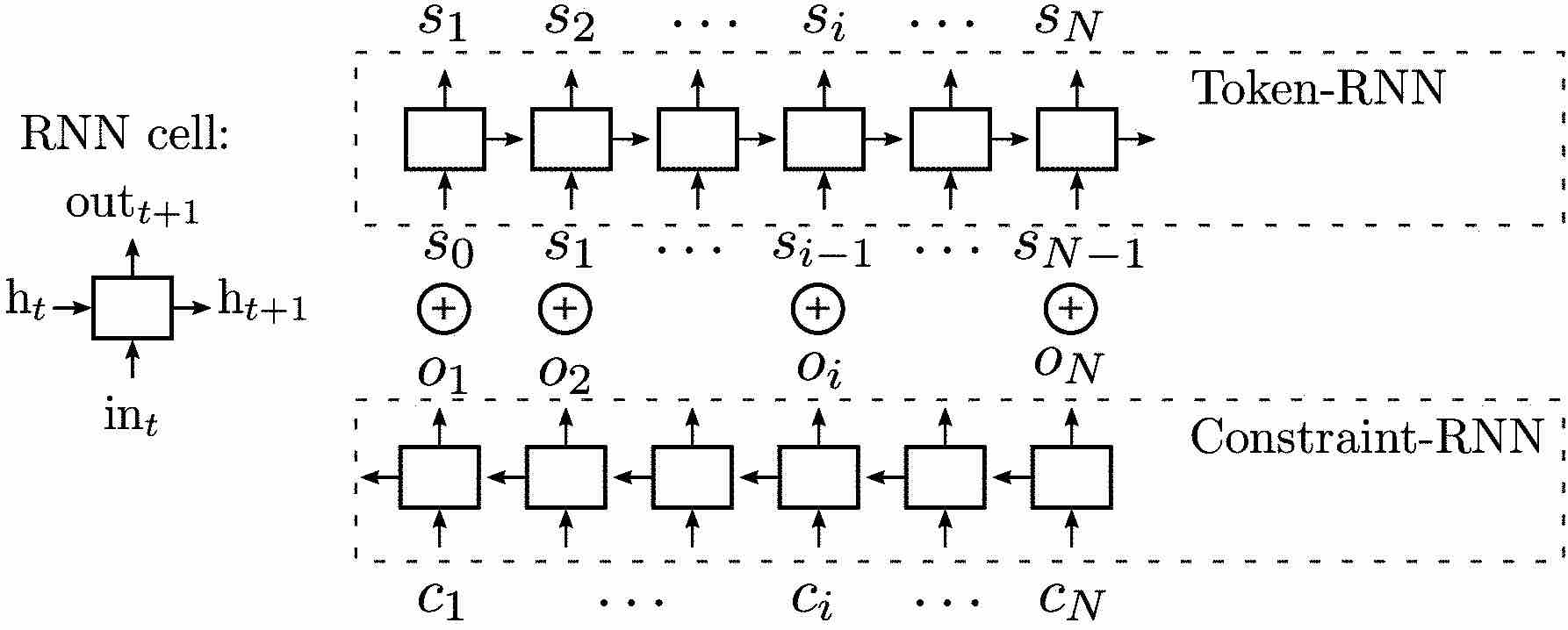}
\caption{Anticipation-RNN architecture.
Reproduced from \cite{hadjeres:anticipation:rnn:arxiv:2017} with permission of the authors}
\label{figure:anticipation:rnn:architecture}
\end{figure}
\begin{itemize}

\item c$_i$ is a {\em positional constraint\index{Positional constraint}}; and

\item o$_i$ is the output at index $i$ (after $i$ iterations) of Constraint-RNN
-- it summarizes constraints information from step $i$ to the final step $N$ (the end of the sequence).
It will be concatenated (the $\oplus$ circled plus sign) to input s$_{i-1}$ of Token-RNN in order to predict the next item s$_i$.

\end{itemize}

Note that Anticipation-RNN is not a symmetric\index{Symmetric} bidirectional recurrent architecture\index{Bidirectional!recurrent neural network},
as in the case of the BLSTM\index{BLSTM} (Section~\ref{section:experiment:blstm:chord})
or the C-RNN-GAN\index{C-RNN-GAN} (Section~\ref{section:systems:c:rnn:gan}) architectures
because what is processed backwards is another sequence (of the constraints associated to the first sequence). 
Both RNNs (Constraint-RNN and Token-RNN) are implemented as a 2-layer LSTM.

The corpus used is the set of soprano voice melodies extracted from the four-voice Chorales of J. S. Bach.
Data synthesis is performed by transposing in all keys
within the original voice range
and by pairing them with some sorted set of constraints\footnote{This is done to reduce the combinatorial explosion,
	as one does not need to construct all possible pairs $(melody, constraint)$
	as long as the coverage is sufficient for good learning.}.

Anticipation-RNN shares the principles of representation initiated by the DeepBach\index{DeepBach} system
to be presented in Section~\ref{section:experiment:deep:bach},
that is
one-hot encoding
with the addition of the hold\index{Hold} symbol ``\_\_''
and the rest symbol to specify, respectively, a note repetition and a rest,
and using the names of the notes with no enharmony.
Quantization is at the level of a sixteenth note.

\begin{figure}
\includegraphics[width=\textwidth]{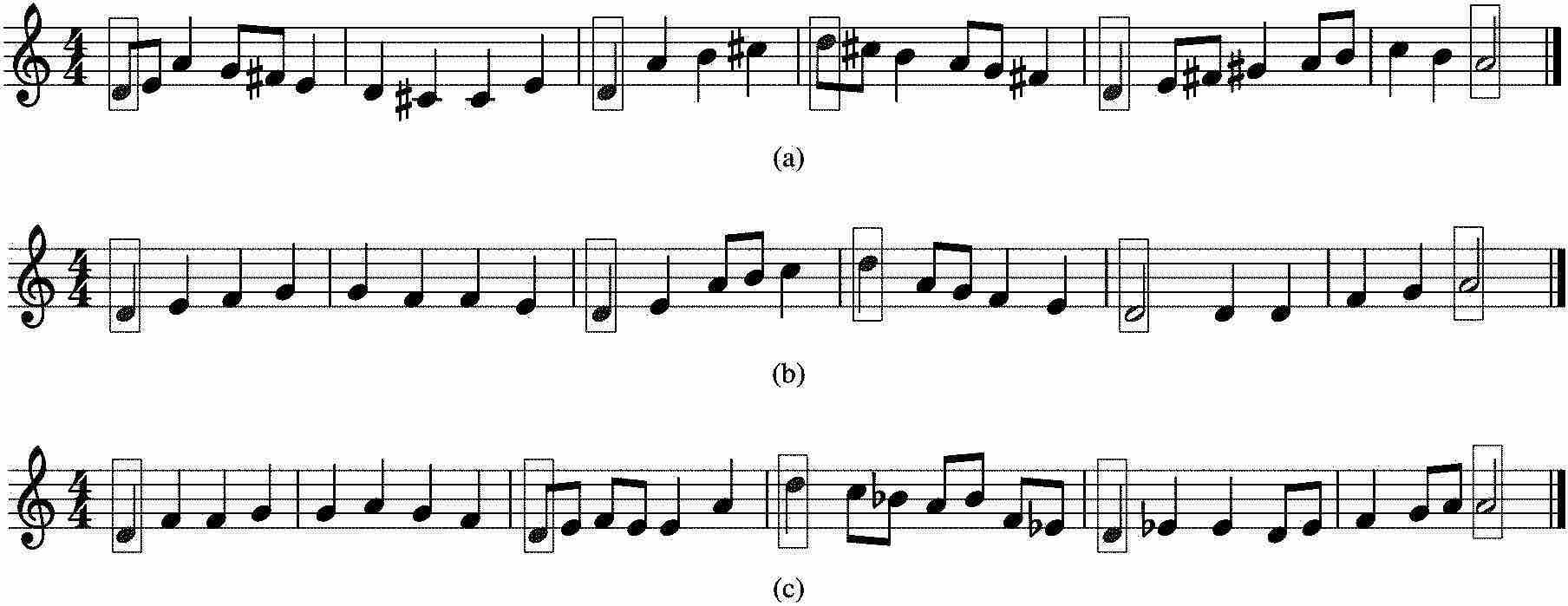}
\caption{Examples of melodies generated by Anticipation-RNN.
Reproduced from \cite{hadjeres:anticipation:rnn:arxiv:2017} with permission of the authors}
\label{figure:anticipation:rnn:example}
\end{figure}

Three examples of melodies generated with the same set of positional constraints
(each one indicated with a green note within a green rectangle)
are shown in Figure~\ref{figure:anticipation:rnn:example}.
The model is indeed able to anticipate each positional constraint by adjusting its direction towards the target (lower-pitched or higher-pitched note).
Further details and analysis of the results are provided in \cite{hadjeres:anticipation:rnn:arxiv:2017}.
Anticipation-RNN is summarized in Table~\ref{table:dimensions:anticipation:rnn}.

\begin{table}
\begin{tabular}{|l|l|}
\hline
{\em Objective}			&Melody; Bach\\
\hline
{\em Representation}	&Symbolic; One-hot; Hold; Rest; No enharmony\\
\hline
{\em Architecture}		&Conditioning(LSTM$^2$, LSTM$^2$)\\
\hline
{\em Strategy}			&Iterative feedforward; Sampling\\
\hline
\end{tabular}
\caption{Anticipation-RNN summary}
\label{table:dimensions:anticipation:rnn}
\end{table}

\subsubsection{\#6 Example: VRASH Melody Symbolic Music Generation System}
\label{section:experiment:vrash}

The system described by Tikhonov and Yamshchikov in \cite{tikhonov:generation:vae:history:arxiv:2017},
although similar to VRAE (see Section~\ref{section:experiment:vrae}),
uses a different representation, separately encoding in a multi-one-hot manner the pitch,
the octave and the duration.
The training set is composed of various songs (different epochs and genres), derived from MIDI files
following filtering and normalization (see the details in \cite{tikhonov:generation:vae:history:arxiv:2017}).
The architecture has four LSTM\footnote{To be more precise, a recent evolution
	named recurrent highway networks\index{Recurrent!highway network} (RHNs\index{RHN}) \cite{zilly:rhn:arxiv:2017}.}
layers for the encoder and for the decoder.

\begin{figure}
\includegraphics[width=\textwidth]{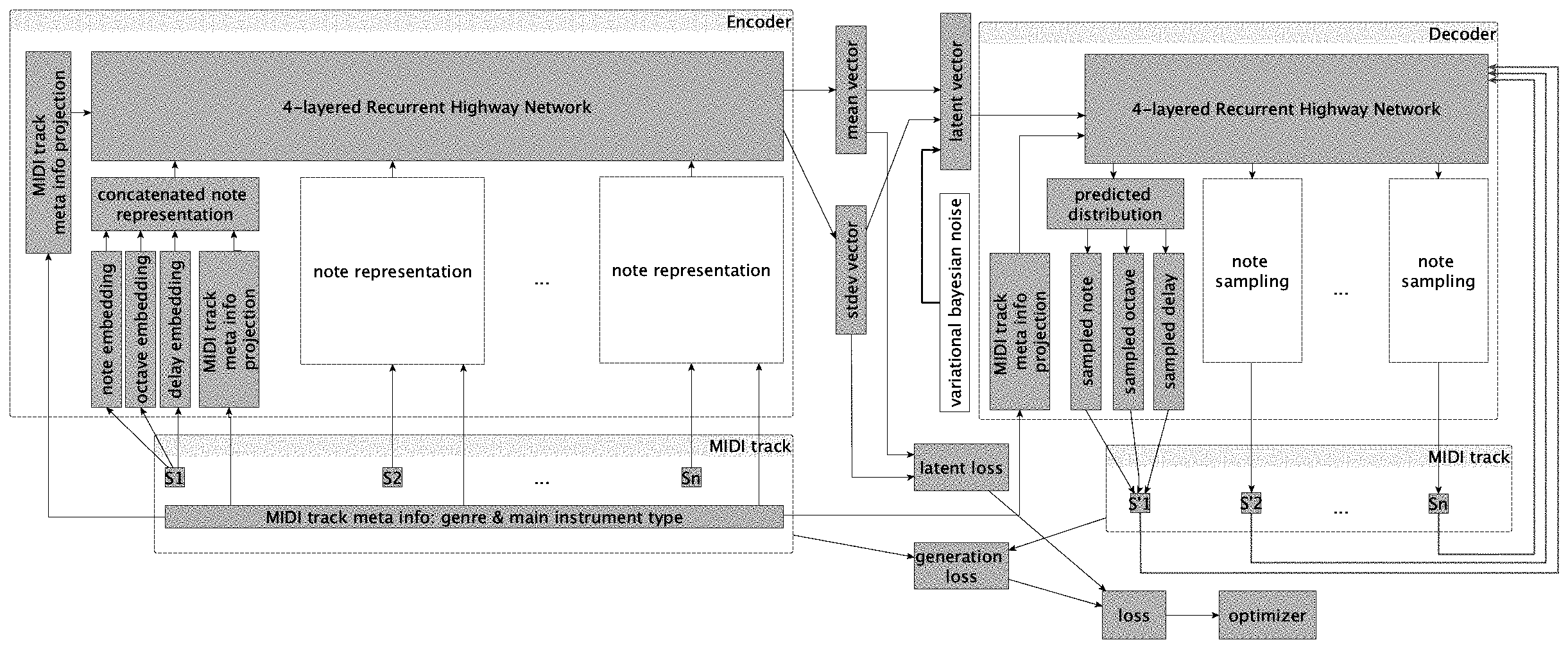}
\caption{VRASH architecture.
Reproduced from \cite{tikhonov:generation:vae:history:arxiv:2017} with permission of the authors}
\label{figure:vrash:architecture}
\end{figure}

\begin{figure}
\includegraphics[width=\textwidth]{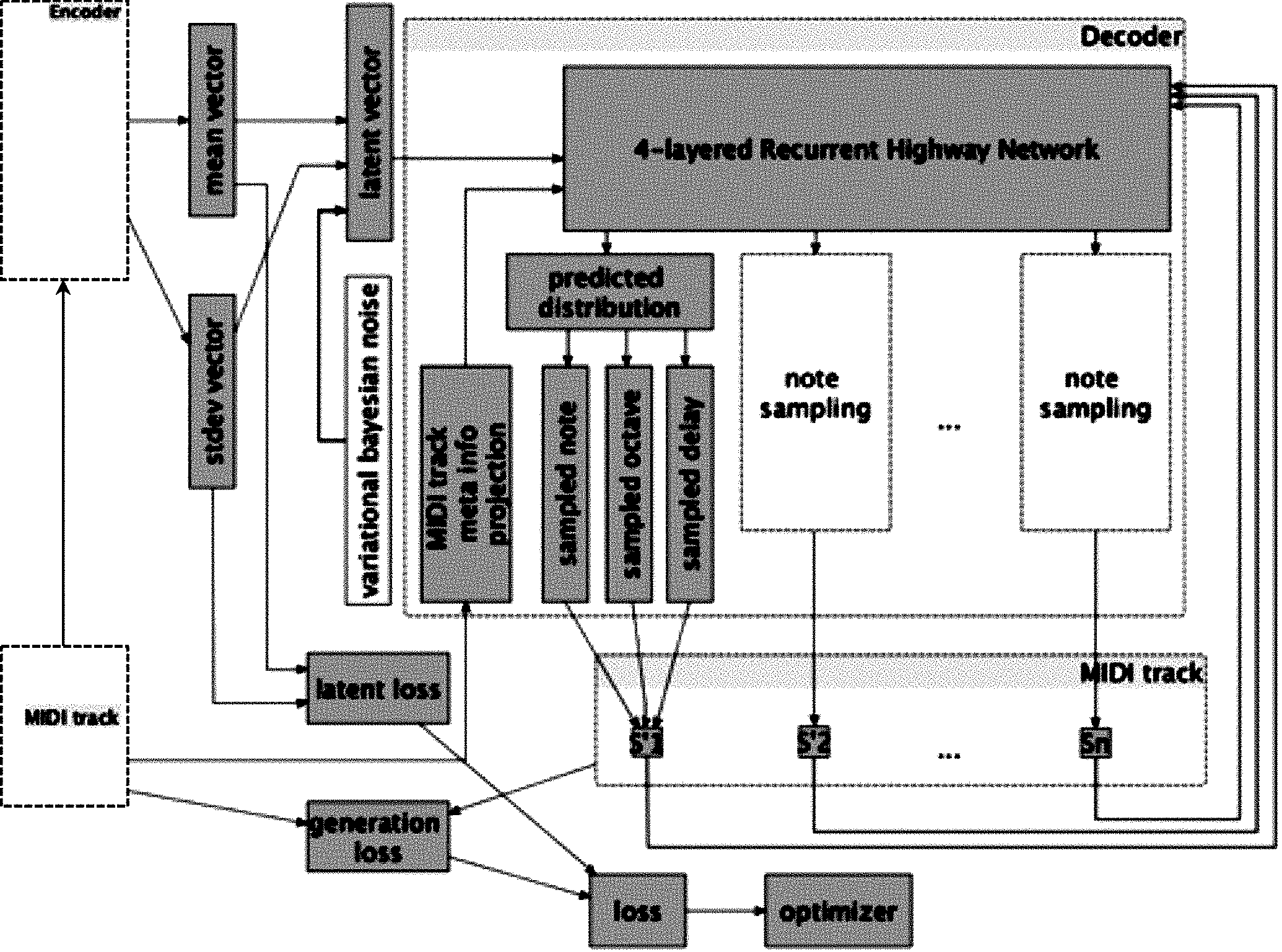}
\caption{VRASH architecture with a focus on the decoder.
Extended from \cite{tikhonov:generation:vae:history:arxiv:2017} with permission of the authors}
\label{figure:vrash:architecture:decoder}
\end{figure}

The authors have experimented with feeding the output of the decoder back into the decoder
as a way of including the previously generated note
as an additional information
(therefore, they have named their final architecture VRASH\index{VRASH},
for variational recurrent autoencoder supported by history\index{Variational!recurrent autoencoder supported by history}).
It is illustrated in Figures~\ref{figure:vrash:architecture} and~\ref{figure:vrash:architecture:decoder}
and summarized in Table~\ref{table:dimensions:vrash}.
In their evaluation,
the authors state that the melodies generated are only slightly closer to the corpus
(using a cross-entropy\index{Cross-entropy!cost} measure)
than when not adding history information,
but that qualitatively the results are better.

\begin{table}
\begin{tabular}{|l|l|}
\hline
{\em Objective}			&Melody\\
\hline
{\em Representation}	&Symbolic; MIDI; Multi-one-hot\\
\hline
{\em Architecture}		&Variational(Autoencoder(LSTM$^4$, Conditioning(LSTM$^4$, History)))\\
\hline
{\em Strategy}			&Decoder feedforward; Iterative feedforward; Sampling\\
\hline
\end{tabular}
\caption{VRASH summary}
\label{table:dimensions:vrash}
\end{table}

\subsection{Input Manipulation}
\label{section:challenges:strategies:control:input:manipulation}

\label{section:strategy:input:manipulation}

The {\em input manipulation\index{Input!manipulation strategy}} strategy
was pioneered for images\index{Image} by Deep Dream\index{Deep Dream}.
The idea is that the initial input content, or a brand new (randomly\index{Random} generated) input content,
is incrementally {\em manipulated} in order to match a target {\em property}.
Note that control of the generation is {\em indirect},
as it is not applied to the output but to the input, {\em before} generation.
Examples of target properties are

\begin{itemize}

\item maximizing the {\em similarity\index{Similarity}} to a given {\em target\index{Target}},
in order to create a consonant\index{Consonant} melody,
as in DeepHear$_C$\index{DeepHear$_C$}
(Section~\ref{section:experiment:deep:hear:harmonize});

\item maximizing\index{Maximize} the {\em activation} of a specific {\em unit},
to amplify some visual element associated to this unit,
as in Deep Dream\index{Deep Dream}
(Section~\ref{section:systems:deep:dream});

\item maximizing the {\em content similarity} to some initial image {\em and} the {\em style similarity} to a reference style image,
to perform {\em style transfer\index{Style!transfer}}
(Section~\ref{section:system:gatys:style:transfer}); and

\item maximizing the {\em similarity} of the {\em structure} to some reference music,
to perform {\em style imposition\index{Style!imposition}}
(Section~\ref{section:systems:c-rbm}).

\end{itemize}

Interestingly, this is done by reusing
standard training mechanisms,
namely backpropagation\index{Backpropagation} to compute the gradients,
as well as gradient descent\index{Gradient!descent} (or ascent) to minimize the cost
(or to maximize the objective).

\subsubsection{\#1 Example: DeepHear Ragtime Counterpoint Symbolic Music Generation System}
\label{section:experiment:deep:hear:harmonize}

In \cite{sun:deep:hear},
in addition to the generation of melodies (described in Section~\ref{section:experiment:deep:hear:melody}),
Sun proposed to use DeepHear\index{DeepHear} for a different objective:
to harmonize a melody, while using the {\em same} architecture as well as what has already been learnt\footnote{It is a simple example
	of {\em transfer learning\index{Transfer learning}} (see \cite[Section~15.2]{goodfellow:deep:learning:book:2016}),
	using the same domain and the same training but for a different task.}.
We notate this second experiment DeepHear$_C$, where {\small $C$} stands for counterpoint,
in order to distinguish it from DeepHear$_M$ for melody generation (Section~\ref{section:experiment:deep:hear:melody}).
%
%

The idea is to find a label\index{Label} instance of the embedding\index{Embedding},
i.e. a set of values for the 16 units of the bottleneck hidden layer\index{Bottleneck hidden layer} of the stacked autoencoder,
which will result in a decoded output resembling a given melody.
Therefore, a simple distance\index{Distance} (error) function is defined to represent the distance (similarity\index{Similarity})
between two melodies (in practice, the number of unmatched notes).
Then a gradient descent\index{Gradient!descent} is conducted on the variables of the embedding,
guided by the gradients corresponding to the error function,
until a sufficiently similar decoded melody is found.

Although this is not a real counterpoint\index{Counterpoint}\footnote{As, for example, in the case
	of MiniBach\index{MiniBach} (Section~\ref{section:experiment:mini:bach})
	or DeepBach\index{DeepBach} (Section~\ref{section:experiment:deep:bach})
	for real counterpoint generation.},
but rather the generation of a similar\index{Similarity} (consonant\index{Consonant}) melody,
the results (tested on ragtime melodies)
do produce a naive counterpoint with a ragtime\index{Ragtime} flavor.

Note that in DeepHear$_C$ (summarized in Table~\ref{table:dimensions:deep:hear:c}),
the input manipulated is the input of the innermost decoder (the starting point of the chain of decoders)
and not the main input of the full architecture.
Whereas, in the case of the Deep Dream\index{Deep Dream} system to be introduced in Section~\ref{section:systems:deep:dream},
this is the main input of the full (feedforward) architecture which is manipulated.

\begin{table}
\begin{tabular}{|l|l|}
\hline
{\em Objective}			&Accompaniment; Ragtime\\
\hline
{\em Representation}	&Symbolic; Piano roll; One-hot$\times$64\\
\hline
{\em Architecture}		&Autoencoder$^4$\\
\hline
{\em Strategy}			&Input manipulation; Decoder feedforward\\
\hline
\end{tabular}
\caption{DeepHear$_C$ summary}
\label{table:dimensions:deep:hear:c}
\end{table}

\subsubsection{Relation to Variational Autoencoders}

Note that in the case of the manipulation of the hidden layer units of an autoencoder (or a stacked autoencoder,
the case of DeepHear$_C$),
the input manipulation strategy
does have some analogy with variational autoencoders\index{Variational!autoencoder}, such as for instance
the VRAE\index{VRAE} system (Section~\ref{section:experiment:vrae}) or
the GLSR-VAE\index{GLSR-VAE} system (Section~\ref{section:experiment:glsr:vae}).
Indeed in both cases, there is some exploration of possible values for the hidden units in order to generate variations of musical content by the decoder
(or the chain of decoders).
The important difference is that

\begin{itemize}

\item in the case of a variational autoencoder,
the exploration of values is {\em user-directed},
although it could be guided by some principle, e.g., geodesic in GLSR-VAE,
interpolation or attribute vector arithmetics in MusicVAE\index{MusicVAE} (Section~\ref{section:system:music:vae}), whereas

\item in the case of input manipulation, the exploration of values is {\em automatically guided} by the gradient descent (or ascent) mechanism,
the user having previously specified a cost function to be minimized (or an objective to be maximized).

\end{itemize}

\subsubsection{\#2 Example: Deep Dream Psychedelic Images Generation System}
\label{section:control:input:manipulation:deep:dream}
\label{section:systems:deep:dream}

\begin{figure}
\includegraphics[width=\textwidth]{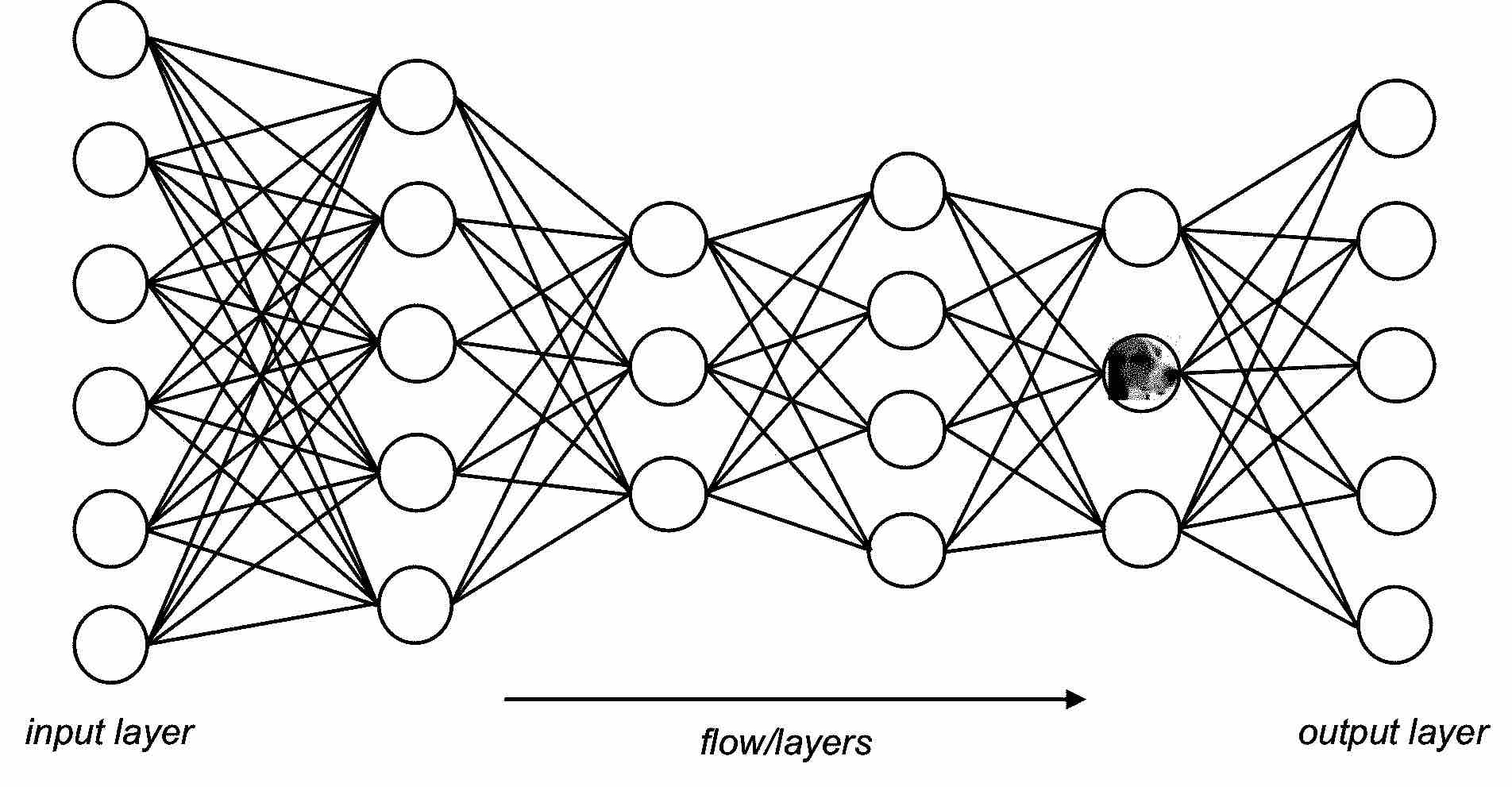}
\caption{Deep Dream architecture (conceptual)}
\label{figure:deep:dream:architecture:higher:level}
\end{figure}

Deep Dream\index{Deep Dream},
by
Mordvintsev {\em et al.} \cite{google:dream:web:2015},
has become famous for generating psychedelic\index{Psychedelic} versions of standard images\index{Image}.
The idea is to use a deep convolutional feedforward neural network\index{Feedforward!network} architecture
(see Figure~\ref{figure:deep:dream:architecture:higher:level})
and to use it to {\em guide} the incremental alteration of an initial input image\index{Image},
in order to maximize the potential occurrence of a specific visual\index{Visual} motif\index{Motif}\footnote{To create a
	{\em pareidolia} effect,
	where a pareidolia\index{Pareidolia} is a psychological phenomenon in which the mind responds to a stimulus, like an image or a sound,
	by perceiving a familiar pattern where none exists.}
correlated to the activation of a given unit.

\begin{figure}
\includegraphics[width=\textwidth]{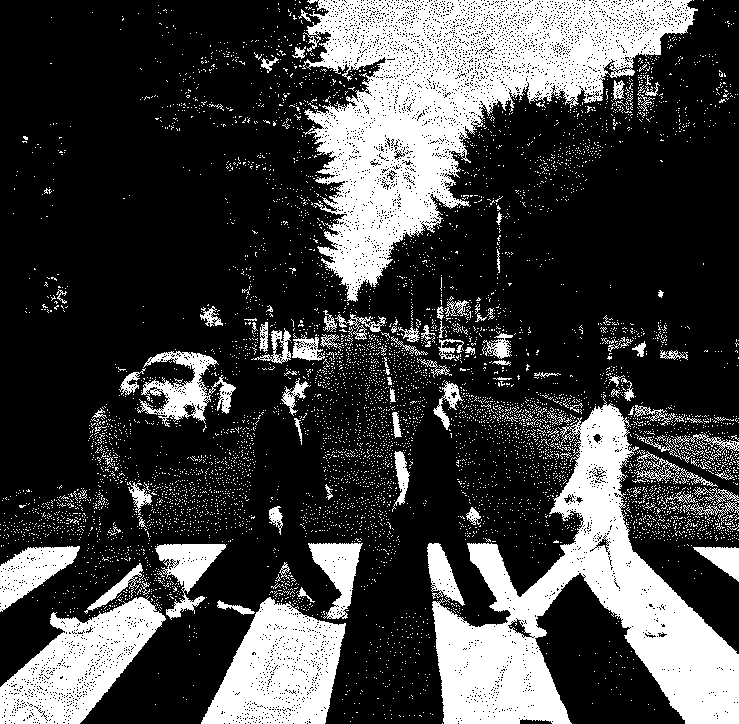}
\caption{Deep Dream. Example of a higher-layer unit maximization transformation.
Created by Google's Deep Dream.
Original picture: Abbey Road album cover, Beatles, Apple Records (1969). Original photograph by Iain Macmillan}
\label{figure:deep:dream:higher:level:example}
\end{figure}

The method is as follows:

\begin{itemize}

\item the network is first trained on a large dataset\index{Dataset} of images;

\item instead of minimizing the cost function\index{Cost!function},
the objective is to {\em maximize} the {\em activation\index{Activation}} of some specific {\em unit}(s)
which has (have) been identified to activate for some specific visual feature(s), e.g., a dog's face,
see Figure~\ref{figure:deep:dream:architecture:higher:level}\footnote{Instead of exactly prescribing which feature(s) we want the network to amplify,
	an alternative is to let the network make that decision,
	by picking a layer and asking the network to enhance whatever it has detected \cite{google:dream:web:2015}.};

\item an initial image
is {\em iteratively} slightly altered
(e.g., by jitter\index{Jitter}\footnote{Adding
	a small random noise\index{Random!noise} displacement of pixels\index{Pixel}.}),
under {\em gradient ascent\index{Gradient!ascent}} control,
in order to maximize the activation of the specific unit(s).
This will favor the emergence
of the correlated visual motif\index{Motif} (motives),
see Figure~\ref{figure:deep:dream:higher:level:example}.

\end{itemize}


Note that

\begin{figure}
\includegraphics[width=\textwidth]{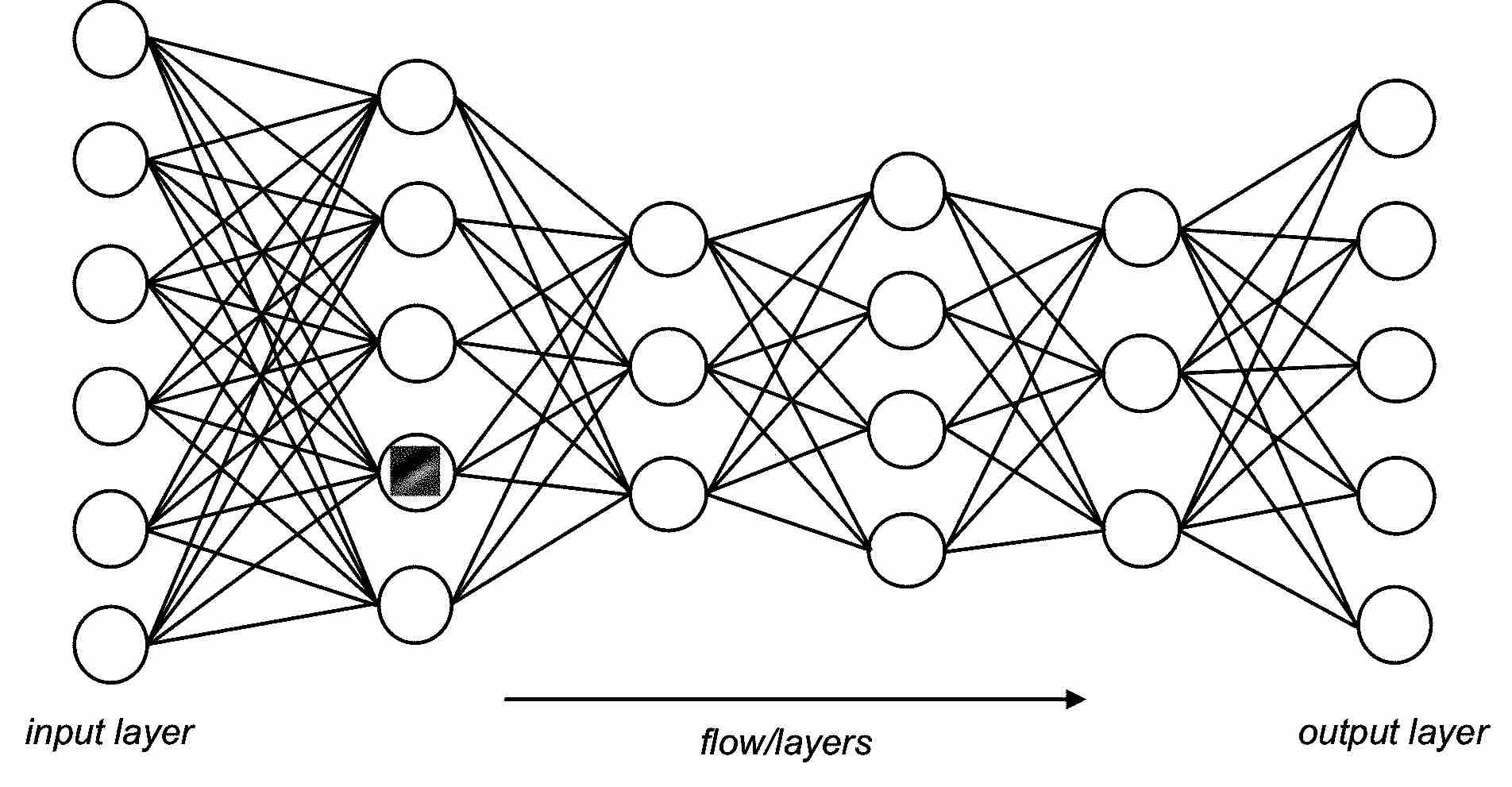}
\caption{Deep Dream architecture focusing on a lower-level unit}
\label{figure:deep:dream:architecture:lower:level}
\end{figure}

\begin{figure}
\includegraphics[scale=0.56]{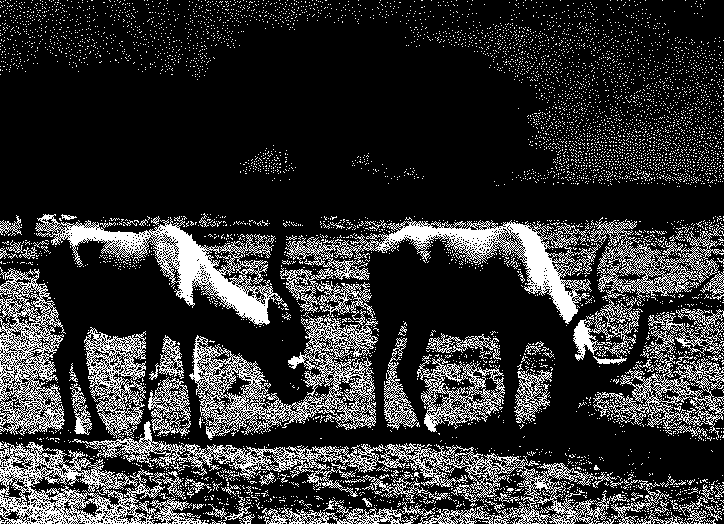}\\
\includegraphics[scale=0.564]{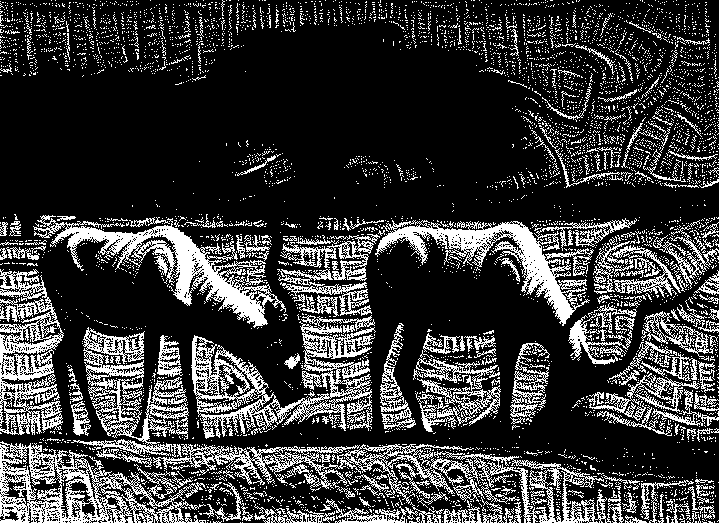}
\caption{Deep Dream. Example of a lower-layer unit maximization transformation.
Reproduced from \cite{google:dream:web:2015} under a CC BY 4.0 licence. Original photograph by Zachi Evenor}
\label{figure:deep:dream:lower:level:example}
\end{figure}

\begin{itemize}

\item the activation maximization of a {\em higher-layer} unit(s),
as in Figure~\ref{figure:deep:dream:architecture:higher:level},
will favor the emergence\index{Emergence}
in the image of a correlated {\em high-level\index{High-level}} motif (motives),
like a dog's face
(see Figure~\ref{figure:deep:dream:higher:level:example}\footnote{As the authors put it in \cite{google:dream:web:2015}:
	``The results are intriguing --
	even a relatively simple neural network can be used to over-interpret an image, just like as children we enjoyed watching clouds and interpreting the random shapes.
	This network was trained mostly on images of animals, so naturally it tends to interpret shapes as animals.''}); whereas

\item the activation maximization of a {\em lower-layer} unit(s),
as in Figure~\ref{figure:deep:dream:architecture:lower:level},
will result in {\em texture\index{Texture} insertion}
(see Figure~\ref{figure:deep:dream:lower:level:example}).

\end{itemize}


One may
imagine a direct transposition of the Deep Dream approach to music,
by maximizing the activation of a specific node\footnote{Particularly if the role of a node has been identified,
	through a correlation analysis
	between node/layer activations and musical motives,
	as, for example, in
	Section~\ref{section:explainability:example:bachbot}.}.



\subsubsection{\#3 Example: Style Transfer Painting Generation System}
\label{section:system:gatys:style:transfer}

\label{section:experiment:style:transfer}
\label{section:challenges:style:transfer}

The idea in this approach, named {\em style transfer\index{Style!transfer}}, pioneered by
Gatys {\em et al.} \cite{gatys:neural:style:2015}
and designed for images\index{Image},
is to use a deep convolutional feedforward architecture to independently capture

\begin{itemize}

\item the features of a first image (named the {\em content}), and

\item the style\index{Style} (as a correlation\index{Correlation} between features) of a second image (named the {\em style}).

\end{itemize}

Gradient-based learning is then used to guide the incremental\index{Incremental} modification of an initially random\index{Random} third image,
with the double objective of matching both the content {\em and} the style descriptions.
More precisely, the method is as follows:

\begin{itemize}

\item capture the {\em content} information of the first image (the content reference)
by feed-forwarding\index{Feedforward} it into the network
and by storing {\em units activations\index{Activation}} for each layer;

\item capture the {\em syle} information of the second image (the style reference)
by feed-forwarding it into the network
and by storing {\em feature spaces}, which are {\em correlations\index{Correlation}} between units activations for each layer; and

\item synthesize a hybrid image.

\end{itemize}

The hybrid image is created by generating a random image,
defining it as current image,
and then iterating the following loop
until the {\em two targets} ({\em content similarity} and {\em style similarity}) are reached:

\begin{itemize}

\item capture the {\em contents} and the {\em style} information of the current image,

\item compute the {\em content cost} (distance between reference and current content)
and the {\em style cost} (distance between reference and current style),

\item compute the corresponding {\em gradients\index{Gradient}} through standard backpropagation\index{Backpropagation}, and

\item {\em update} the current image guided by the gradients.

\end{itemize}

\begin{figure}
\includegraphics[width=\textwidth]{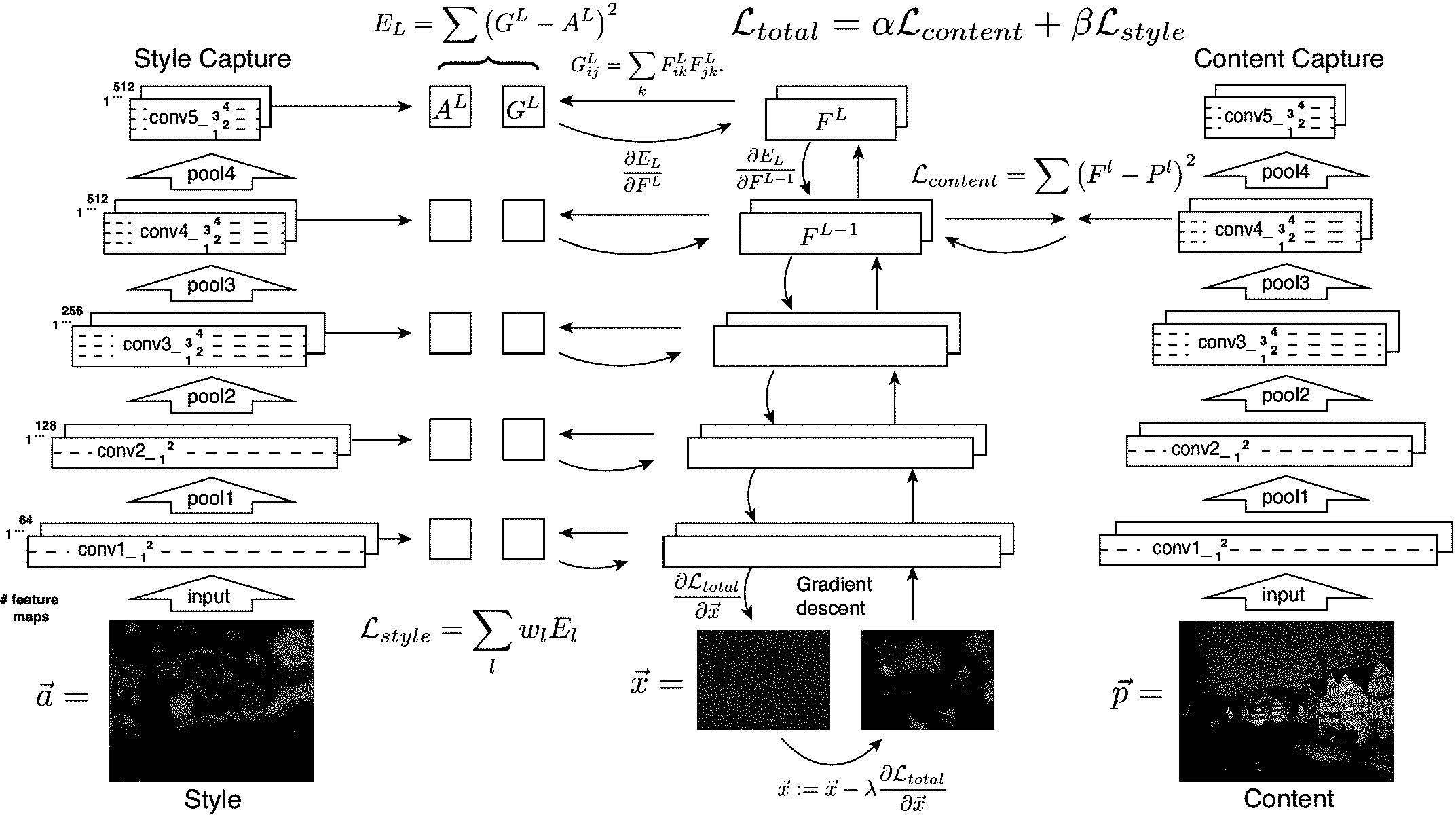}
\caption{Style transfer full architecture/process.
Extension of a figure reproduced from \cite{gatys:neural:style:2015} with permission of the authors}
\label{figure:style:transfer:architecture}
\end{figure}

\begin{figure}
\includegraphics[width=\textwidth]{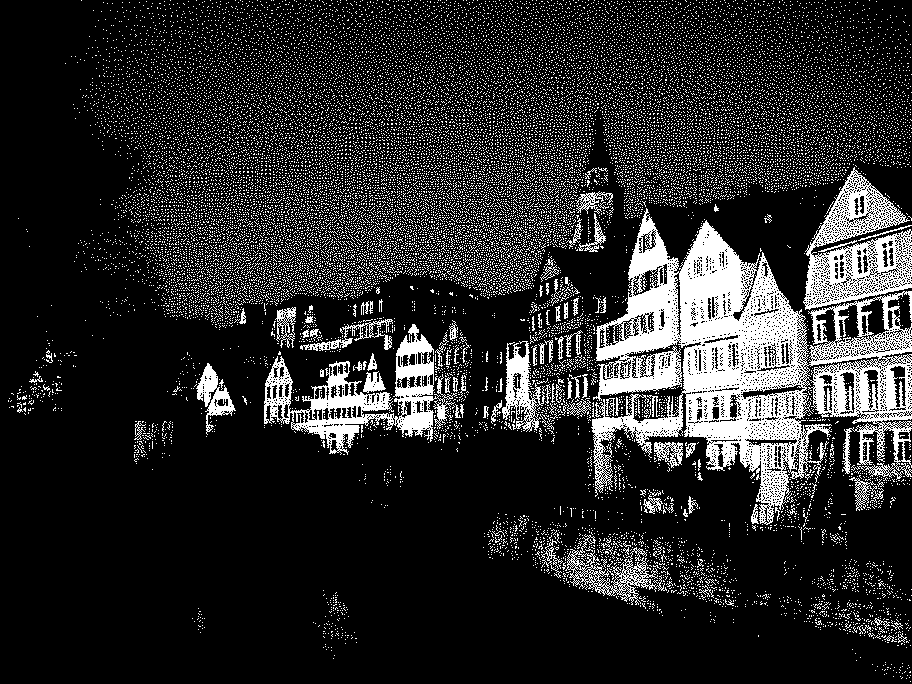}
\caption{T\"ubingen's Neckarfront. Photograph by Andreas Praefcke.
Reproduced from \cite{gatys:neural:style:2015} with permission of the authors}
\label{figure:style:transfer:neckarfront}
\end{figure}

The architecture and process are summarized in Figure~\ref{figure:style:transfer:architecture}
(more details may be found in \cite{gatys:style:transfer:cvf:2016}).
The content image (on the right) is a photograph of T\"ubingen's Neckarfront in Germany\footnote{The location
	of the researchers.}
(shown in Figure~\ref{figure:style:transfer:neckarfront})
and the style image (on the left) is the painting ``The Starry Night'' by Vincent van Gogh (1889).

\begin{figure}
\includegraphics[width=\textwidth]{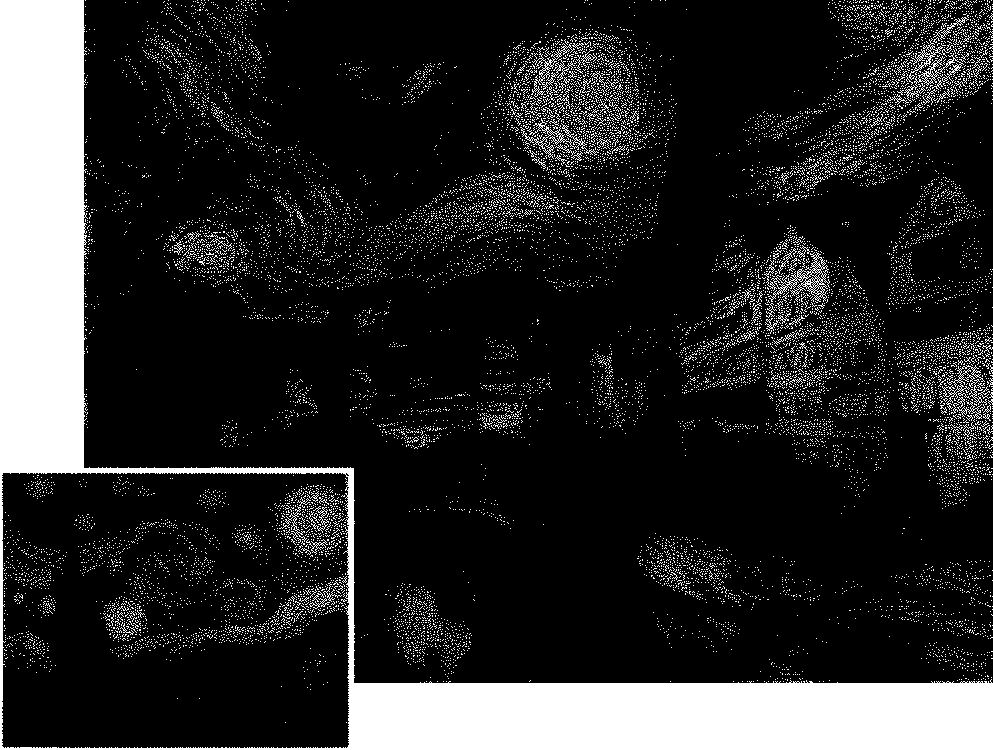}
\caption{Style transfer of ``The Starry Night'' by Vincent van Gogh (1889) on T\"ubingen's Neckarfront photograph. 
Reproduced from \cite{gatys:neural:style:2015} with permission of the authors}
\label{figure:style:transfer:van:gogh}
\end{figure}

\begin{figure}
\includegraphics[width=\textwidth]{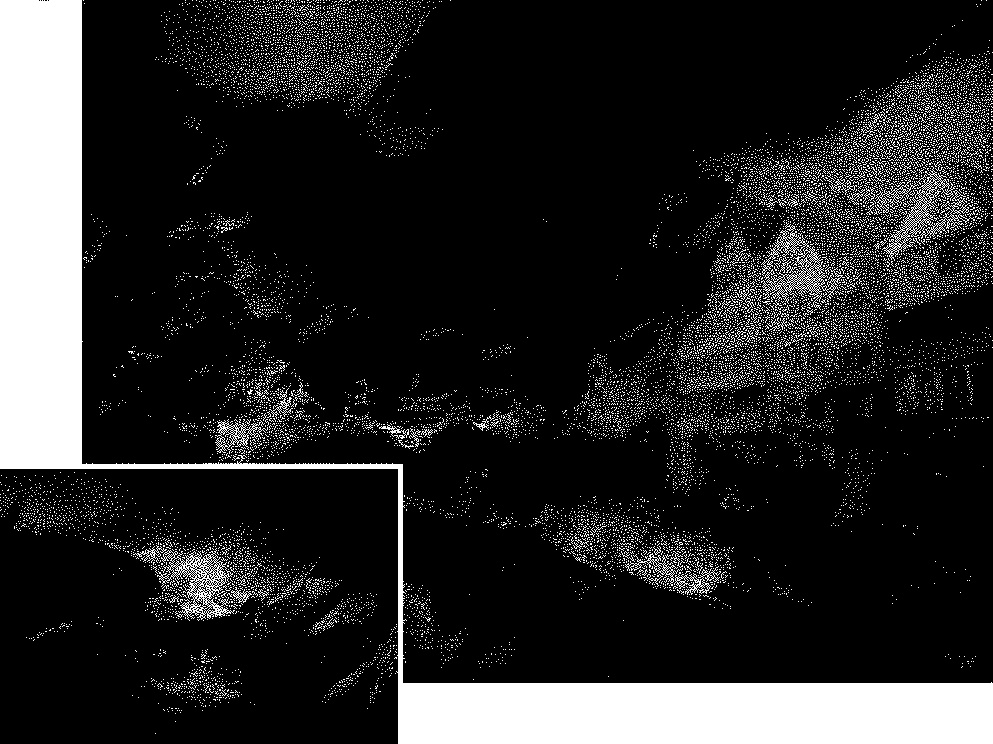}
\caption{Style transfer of ``The Shipwreck of the Minotaur'' by J. M. W. Turner (1805) on T\"ubingen's Neckarfront photograph.
Reproduced from \cite{gatys:neural:style:2015} with permission of the authors}
\label{figure:style:transfer:turner}
\end{figure}

Examples of transfer for the same content (T\"ubingen's Neckarfront) and the styles
``The Starry Night'' by Vincent van Gogh (1889) and
``The Shipwreck of the Minotaur'' by J. M. W. Turner (1805)
are shown in Figures~\ref{figure:style:transfer:van:gogh} and~\ref{figure:style:transfer:turner},
respectively.

\begin{figure}
\includegraphics[width=\textwidth]{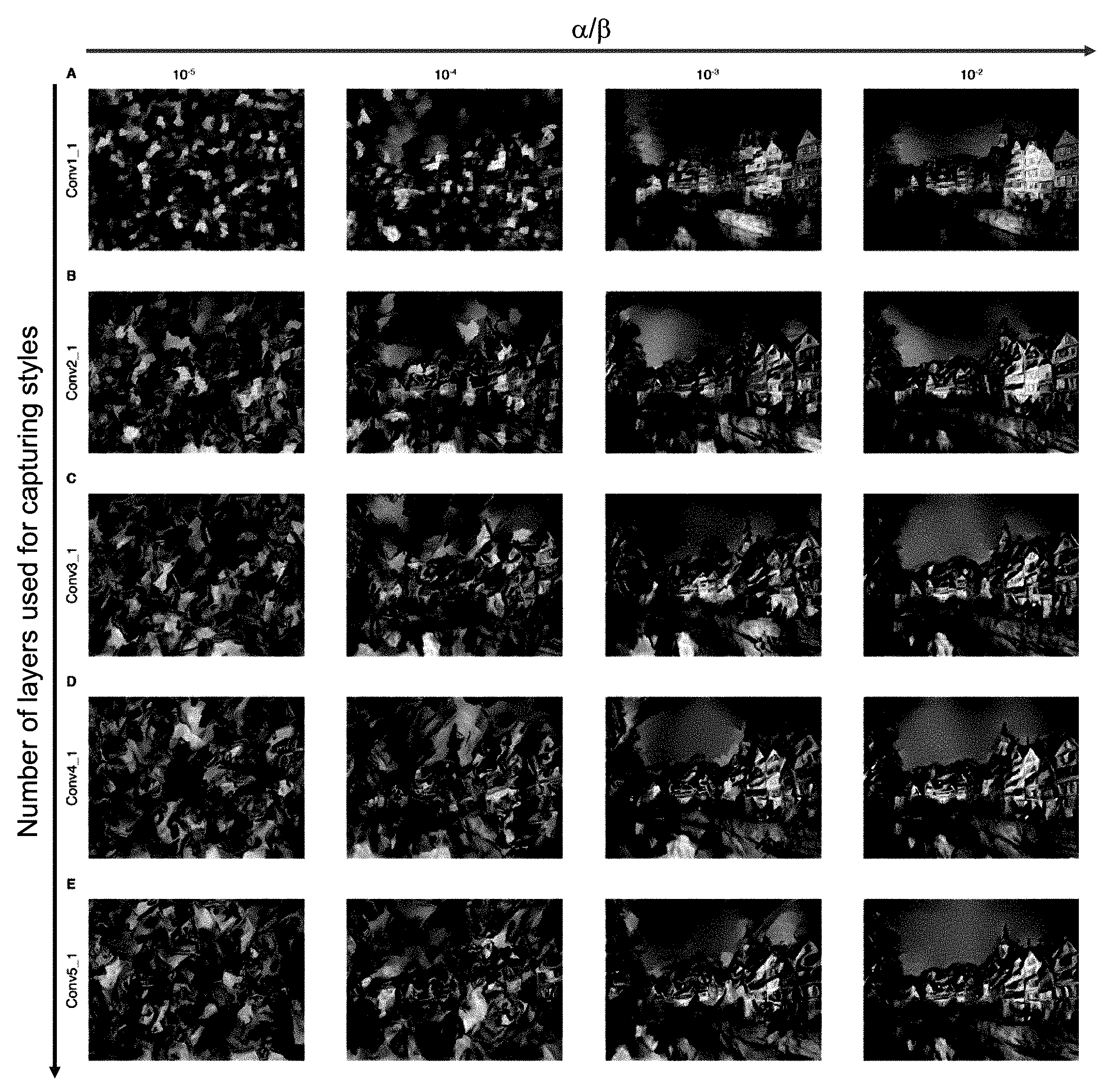}
\caption{Variations on the style transfer of ``Composition VII'' by Wassily Kandinsky (1913) on T\"ubingen's Neckarfront photograph.
Reproduced from \cite{gatys:neural:style:2015} with permission of the authors}
\label{figure:style:transfer:variations}
\end{figure}

Note that one may balance content and style targets\footnote{Through the $\alpha$ and $\beta$ parameters,
	see at the top of Figure~\ref{figure:style:transfer:architecture}
	the total loss defined as $\mathcal{L}_{total} = \alpha \mathcal{L}_{content} + \beta \mathcal{L}_{style}$.}
($\alpha/\beta$ ratio)
in order to favor content or style.
In addition, the complexity of the capture may also be adjusted
via the number of hidden layers used.
These variations are shown in Figure~\ref{figure:style:transfer:variations}:
rightwards
an increasing
$\alpha/\beta$ content/style objectives ratio
and
downwards
an increasing
number of hidden layers used (from 1 to 5) for capturing the style.
The style image is the painting ``Composition VII'' by Wassily Kandinsky (1913).

\subsubsection{Style Transfer vs Transfer Learning}
\label{section:style:transfer:vs:transfer:learning}

Note that although style transfer\index{Style!transfer} shares some of the general objectives
of {\em transfer learning\index{Transfer learning}},
it is actually different (in terms of objective and techniques).
Transfer learning is about reusing what has been learnt by a neural network architecture for a specific task
and applying it to {\em another} task and/or domain (e.g., another type of classification).
Transfer learning issues will be touched upon in Section~\ref{section:discussion:transfer}.


\subsubsection{\#4 Example: Music Style Transfer}
\label{section:control:input:manipulation:musical:style:transfer}

Transposing this style transfer technique to music ({\em music style transfer\index{Music!style transfer}})
is a tempting direction.
However, as we will see, the style of a piece of music is more multidimensional
and could be related to different types of music representation (composition, performance, sound, etc.),
and is thus more difficult to capture via such a simple correlation of activations.
Therefore, we will analyze this issue as a specific challenge in Section~\ref{section:challenges:strategies:style:transfer}.

\subsection{Input Manipulation and Sampling}
\label{index}
\label{section:control:input:manipulation:sampling}
\label{section:strategy:input:manipulation:sampling}

An example of the combination of the input manipulation strategy\index{Input!manipulation strategy}
with the sampling strategy\index{Sampling!strategy},
thus acting both on the input and the output\footnote{Interestingly,
	the input is actually equal to the output
	because the architecture used is an RBM
	(see Section~\ref{section:architecture:rbm}),
	where the visible layer acts both as input {\em and} output.},
is exemplified in the following section.

\subsubsection{Example: C-RBM Polyphony Symbolic Music Generation System}
\label{section:systems:c-rbm}
\label{section:experiment:c:rbm}

In the system presented by
Lattner {\em et al.} in \cite{lattner:structure:polyphonic:generation:jcms:2018},
the starting point is to use a restricted Boltzmann machine (RBM) to learn the local structure,
seen as the {\em musical texture},
of a corpus of musical pieces.
The additional idea is to impose, through constraints\index{Constraint},
a more {\em global\index{Global} structure\index{Structure}}
(form\index{Form}, e.g., AABA, as well as tonality\index{Tonality}),
seen as a {\em structural template\index{Template}} inspired by an existing musical piece,
on the new piece to be generated.
This is called {\em structure imposition}\footnote{This is an example
	of score-level {\em composition style transfer\index{Composition!style transfer}}
	(see Section~\ref{section:control:input:manipulation:musical:style:transfer}).},
also earlier coined as {\em templagiarism\index{Templagiarism}}
(short for template plagiarism)
by Hofstadter \cite{hofstadter:staring:emi:eye:virtual:music:2001}.
These constraints, concerning structure, tonality and meter\index{Meter}, will guide an iterative generation
through a search\index{Search} process, manipulating the input, based on gradient descent.

The actual objective is the generation of polyphonic music.
The representation used is piano roll, with 512 time steps and a range of 64 notes
(corresponding to MIDI note numbers 28 to 92).
The corpus is Wolfgang Amadeus Mozart\index{Mozart}'s sonatas\index{Sonata}.
Each piece is transposed into all possible keys in order to have sufficient training data for all possible keys
(this also helps reduce sparsity\index{Sparsity} in the training data).
The architecture is a convolutional restricted Boltzmann machine\index{Convolutional!restricted Boltzmann machine} {(C-RBM\index{C-RBM})}
\cite{lee:c:rbm:icml:2009},
i.e. an RBM with convolution,
with 512$\times$64 = 32,768 input nodes and 2,048 hidden units.
Units have continuous and not Boolean values, as for standard RBMs (see Section~\ref{section:architecture:rbm:sampling}).
Convolution is only performed on the {\em time dimension},
in order to model temporally invariant\index{Invariance} motives\index{Motif}
but not pitch invariant motives
(there are correlations between notes over the whole pitch range),
which would break the notion of tonality\footnote{As the authors state in \cite{lattner:structure:polyphonic:generation:jcms:2018}:
	``Tonality is another very important higher-order\index{Higher!-order} property in music.
	It describes perceived tonal relations between notes and chords.
	This information can be used to, for example, determine the key of a piece or a musical section.
	A key is characterized by the distribution of pitch classes in the musical texture within a (temporal) window\index{Window} of interest.
	Different window lengths may lead to different key estimations\index{Estimation}, constituting a hierarchical\index{Hierarchical} tonal structure
	(on a very local level, key estimation is strongly related to chord estimation).''}.

Training of the C-RBM\index{C-RBM} is undertaken using the RBM-specific
contrastive divergence\index{Contrastive divergence} algorithm
(see Section~\ref{section:architecture:rbm},
more precisely a more advanced version
named persistent contrastive divergence).
Generation is performed by sampling with some constraints.
Three types of constraints are considered:

\begin{itemize}

\item {\em Self-similarity} -- the purpose is to specify a {\em global structure\index{Structure}} (e.g. AABA) in the generated music piece.
This is modeled by minimizing the distance\index{Distance}
(measured through a mean squared error)
between the self-similarity\index{Similarity}\index{Self!-similarity}
matrixes of the reference target and of the intermediate solution.

\item {\em Tonality constraint} -- the purpose is to specify a {\em key} (tonality).
To estimate the key in a given temporal window,
the distribution\index{Distribution} of pitch classes in the window is compared
with the so-called key profiles\index{Profile}
of the reference
(i.e. paradigmatic relative pitch-class strengths for specific scales\index{Scale} and modes\index{Mode}
\cite{temperley:cognition:musical:structures:2011},
in practice the major\index{Major} and minor\index{Minor} modes).
They are repeated in the time and pitch dimensions of the piano roll matrix,
with a modulo octave shift in the pitch dimension.
The resulting key estimation\index{Estimation} vectors are combined (see the article for more details) to obtain an overall key estimation vector.
In the same way as for self-similarity, the distance between the target and the intermediate solution key estimations is minimized.

\item {\em Meter constraint} -- the purpose is to impose a specific {\em meter\index{Meter}} (also named a {\em time signature\index{Time!signature}},
e.g., 3/4, 4/4, see Section~\ref{section:representation:rhythm}) and its related rhythmic pattern\index{Pattern}
(e.g., relatively strong accents\index{Accent} on the first and the third beat of a measure in a 4/4 meter).
As note intensities are not encoded in the data, only note onsets\index{Note!onset} are considered.
The relative occurrence of note onsets
within a measure is constrained to follow that of the reference.

\end{itemize}

\begin{figure}
\includegraphics[width=\textwidth]{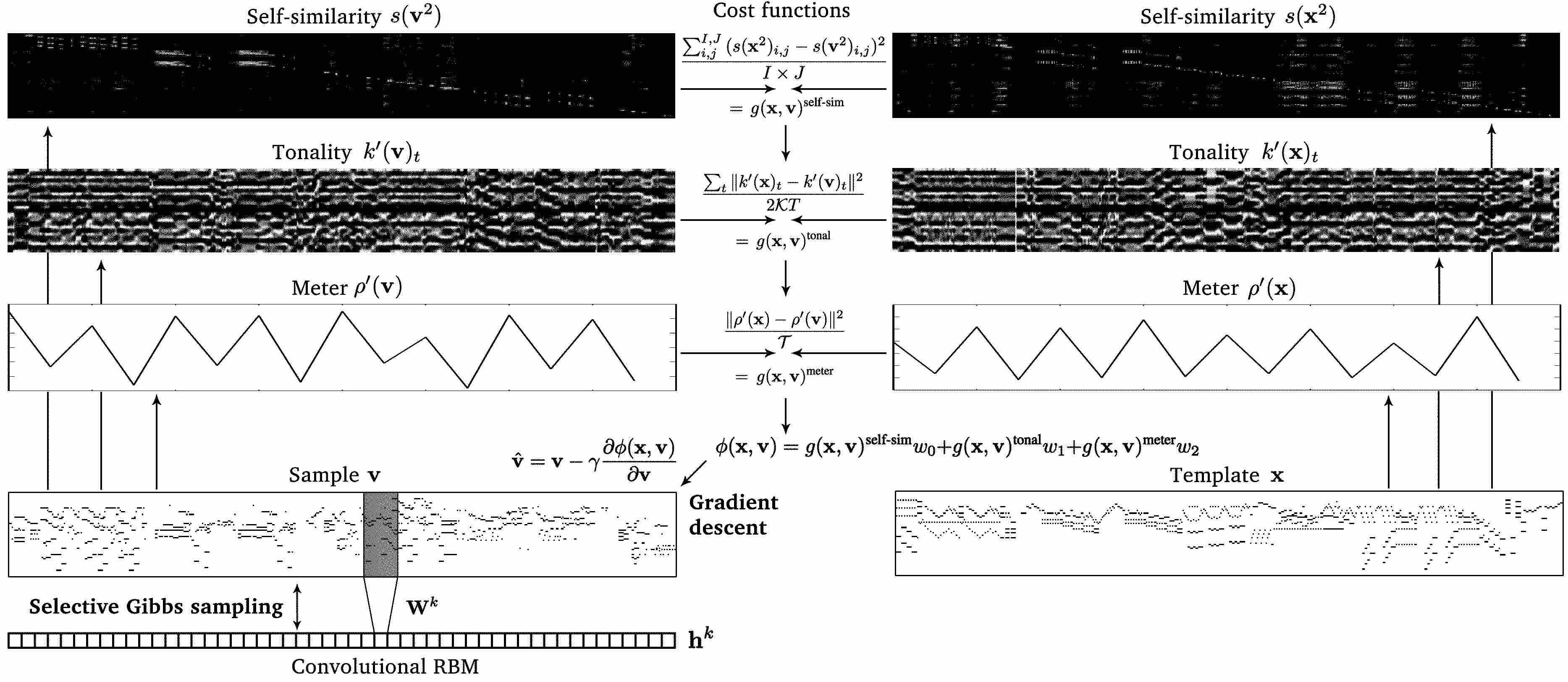}
\caption{C-RBM architecture.
Reproduced from \cite{lattner:structure:polyphonic:generation:jcms:2018} with permission of the authors}
\label{figure:crbmc:architecture}
\end{figure}

Generation\index{Generation} is performed via {\em constrained sampling\index{Constrained sampling}} (CS\index{CS}),
a mechanism used to restrict the set of possible solutions in the sampling process
according to some pre-defined constraints.
%
The principles of the process, illustrated in Figure~\ref{figure:crbmc:architecture}, are as follows:

\begin{itemize}

\item A sample is randomly initialized following the standard uniform distribution.

\item A step of constrained sampling (CS) is performed comprising

\begin{itemize}

\item $n$ runs of gradient descent (GD) optimization
to impose the high-level structure, and

\item $p$ runs of selective Gibbs sampling\index{Selective Gibbs sampling} (SGS\index{SGS})\footnote{Selective Gibbs sampling (SGS)
	is the authors' variant of Gibbs sampling\index{Gibbs sampling} (GS\index{GS}).}
to selectively realign the sample onto the learnt distribution.

\end{itemize}

\item A simulated annealing\index{Simulated annealing} algorithm is applied in order to decrease exploration\index{Exploration}
in relation to a decrease of variance\index{Variance} over solutions.

\end{itemize}

\begin{figure}
\includegraphics[width=\textwidth]{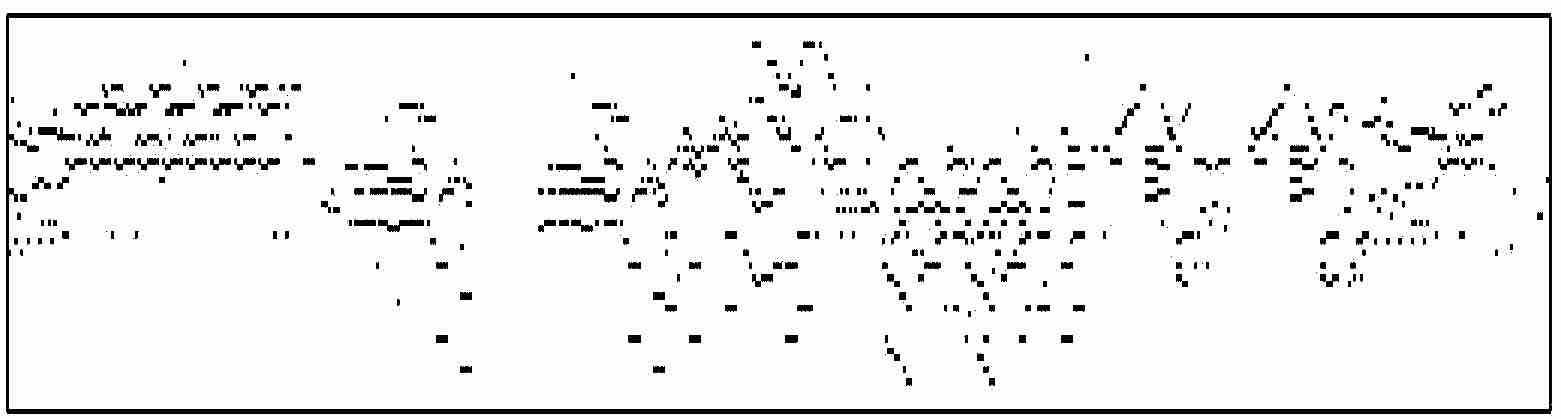}
\caption{Piano roll sample generated by C-RBM.
Reproduced with permission of the authors}
\label{figure:constrained:sampling:example:simplified}
\end{figure}

The different steps of constrained sampling\index{Constrained sampling}
are further detailed in \cite{lattner:structure:polyphonic:generation:jcms:2018}.
Figure~\ref{figure:constrained:sampling:example:simplified} shows an example of a generated sample in piano roll format.
%

The results for the C-RBM (summarized in Table~\ref{table:dimensions:c:rbm})
are interesting and promising.
One current limitation, stated by the authors, is that constraints only apply to the high-level\index{High-level} structure.
Initial attempts at imposing low-level\index{Low-level} structure constraints are challenging because,
as constraints are never purely content-invariant, when trying to transfer low-level structure,
the template piece can be exactly reconstructed in the GD phase.
Therefore, creating constraints for low-level structure would have to be accompanied by an increase in their content invariance.
Another issue is convergence\index{Convergence} and satisfaction\index{Constraint!satisfaction} of the constraints.
As discussed by the authors,
their approach is not exact, as opposed to the Markov constraints\index{Markov!constraint} approach
(for Markov chains\index{Markov!chain})
proposed in \cite{pachet:markov:constraints:constraints:2011}.

\begin{table}
\begin{tabular}{|l|l|}
\hline
{\em Objective}			&Polyphony; Style imposition\\
\hline
{\em Representation}	&Symbolic; Piano-roll; Rest; Many-hot; Meter\\
\hline
{\em Architecture}		&Convolutional(RBM)\\
\hline
{\em Strategy}			&Input manipulation; Sampling\\
\hline
\end{tabular}
\caption{C-RBM summary}
\label{table:dimensions:c:rbm}
\end{table}

\subsection{Reinforcement}
\label{section:challenges:strategies:control:reinforcement}
\label{section:control:reinforcement}
\label{section:strategy:reinforcement}

The idea of the {\em reinforcement strategy\index{Reinforcement!strategy}} is to reformulate the generation of musical content
as a {\em reinforcement learning problem}:
using the similarity to the output of a recurrent network trained on the dataset as a reward
and adding user defined constraints, e.g., some tonality rules according to music theory, as an additional reward.
\label{section:control:reinforcement:control}


Let us consider the case of a monophonic
melody formulated as a reinforcement learning problem:

\begin{itemize}

\item the {\em state} represents the musical content (a partial melody) generated so far, and

\item the {\em action} represents the selection of the next note to be generated.

\end{itemize}

Let us now consider a recurrent neural network (RNN) trained on the chosen corpus of melodies.
Once trained, the RNN will be used as a {\em reference} for the reinforcement learning architecture.
The reward of the reinforcement learning architecture is defined as a combination of two objectives:

\begin{itemize}

\item adherence to {\em what has been learnt}, by measuring the similarity of the action selected,
i.e. the next note to be generated,
to the note predicted by the recurrent network in a similar state (i.e. the partial melody generated so far); and

\item adherence to {\em user-defined constraints}
(e.g., consistency with current tonality, avoidance of excessive repetitions, etc.),
by measuring how well they are fulfilled.

\end{itemize}

In summary, the reinforcement learning architecture is rewarded to {\em mimic} the RNN,
while also being rewarded to enforce some user-defined constraints.

\subsubsection{Example: RL-Tuner Melody Symbolic Music Generation System}
\label{section:systems:rl-tuner}
\label{section:experiment:rl-tuner}

\begin{figure}
\includegraphics[width=\textwidth]{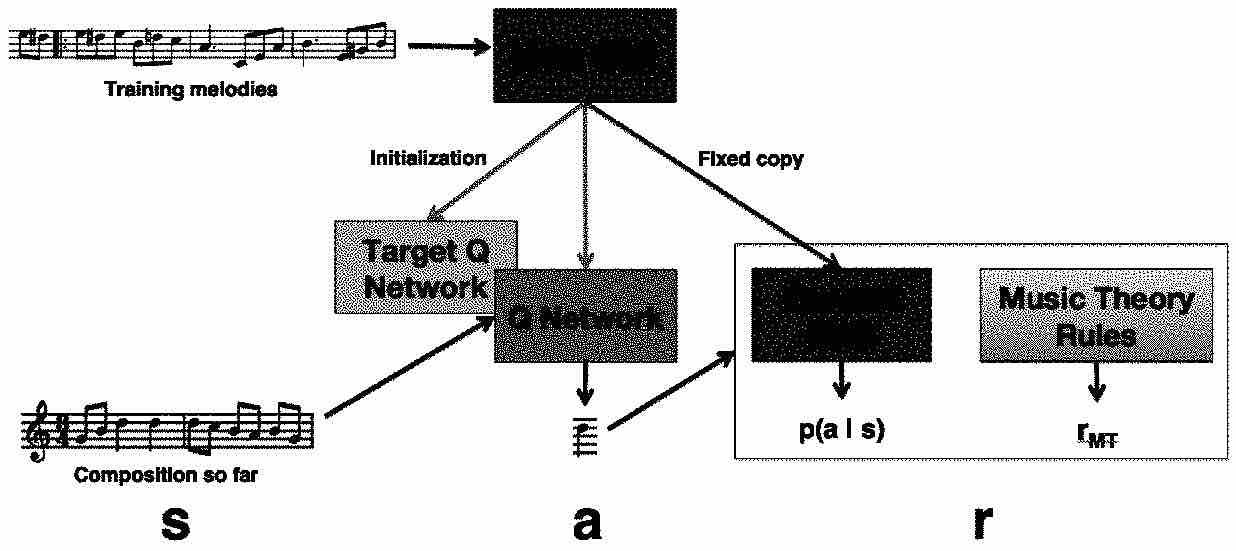}
\caption{RL-Tuner architecture.
Reproduced from \cite{jaques:rl:tuner:arxiv:2016} with permission of the authors}
\label{figure:rl-tuner:architecture}
\end{figure}

The reinforcement strategy was pioneered in the RL-Tuner\index{RL-Tuner} architecture
by
Jaques {\em et al.} \cite{jaques:rl:tuner:arxiv:2016}.
This architecture, illustrated in Figure~\ref{figure:rl-tuner:architecture},
consists in two deep Q network\index{Q!network} reinforcement learning architectures\footnote{An implementation of the Q-learning
	reinforcement learning strategy
	through a deep learning architecture \cite{double:q:learning:arxiv:2015}.}
and two recurrent neural network
(RNN) architectures.

\begin{itemize}

\item The initial RNN, named Note RNN, is trained on the dataset of melodies
for the task of predicting and generating the next note,
following the iterative feedforward strategy.

\item A fixed copy of Note RNN is made, named Reward RNN, which will be used by the reinforcement learning architecture
as a reference.

\item The Q Network architecture task is to learn to select the next note (next action a)
from the generated (partial) melody so far (current state s).

\item The Q Network is trained in parallel to the other Q Network, named Target Q Network,
which estimates\index{Estimation} the value of the gain\index{Gain} (accumulated rewards\index{Reward})
and which has been initialized from what Note RNN has learnt.

\item Q Network's reward r combines two rewards, as defined in
previous section:

\begin{itemize}

\item adherence to {\em what has been learnt}, measured by the probability of Reward RNN to play that note,
in practice $\text{log}\,P(\text{a} | \text{s})$, the log probability for the next note being a given a melody s; and

\item adherence to {\em music theory constraints},
in practice\footnote{This list of musical theory constraints
	has been selected from \cite{gauldin:counterpoint:book:1988},
	see more details in \cite{jaques:rl:tuner:arxiv:2016}.}:

\begin{itemize}

\item staying in key,

\item beginning and ending with the tonic note,

\item avoiding excessively repeated notes,

\item preferring harmonious intervals,

\item resolving large leaps,

\item avoiding continuously repeating extrema notes,

\item avoiding high auto-correlation,

\item playing motifs, and

\item playing repeated motifs.

\end{itemize}

\end{itemize}

\end{itemize}

The total reward $\text{r}(\text{s}, \text{a})$\footnote{Which means the reward to be received when
	from state s action a is chosen.}
is defined by Equation~\ref{equation:rl:reward},
where r$_{MT}$ is the reward concerning music theory
and $c$ is a parameter controlling the balance between the two competing constraints.

\begin{equation}
\text{r}(\text{s}, \text{a}) = \text{log}\,P(\text{a} | \text{s}) + \text{r}_{MT}(\text{a}, \text{s})/c
\label{equation:rl:reward}
\end{equation}

%
%
%

\begin{figure}
\includegraphics[width=\textwidth]{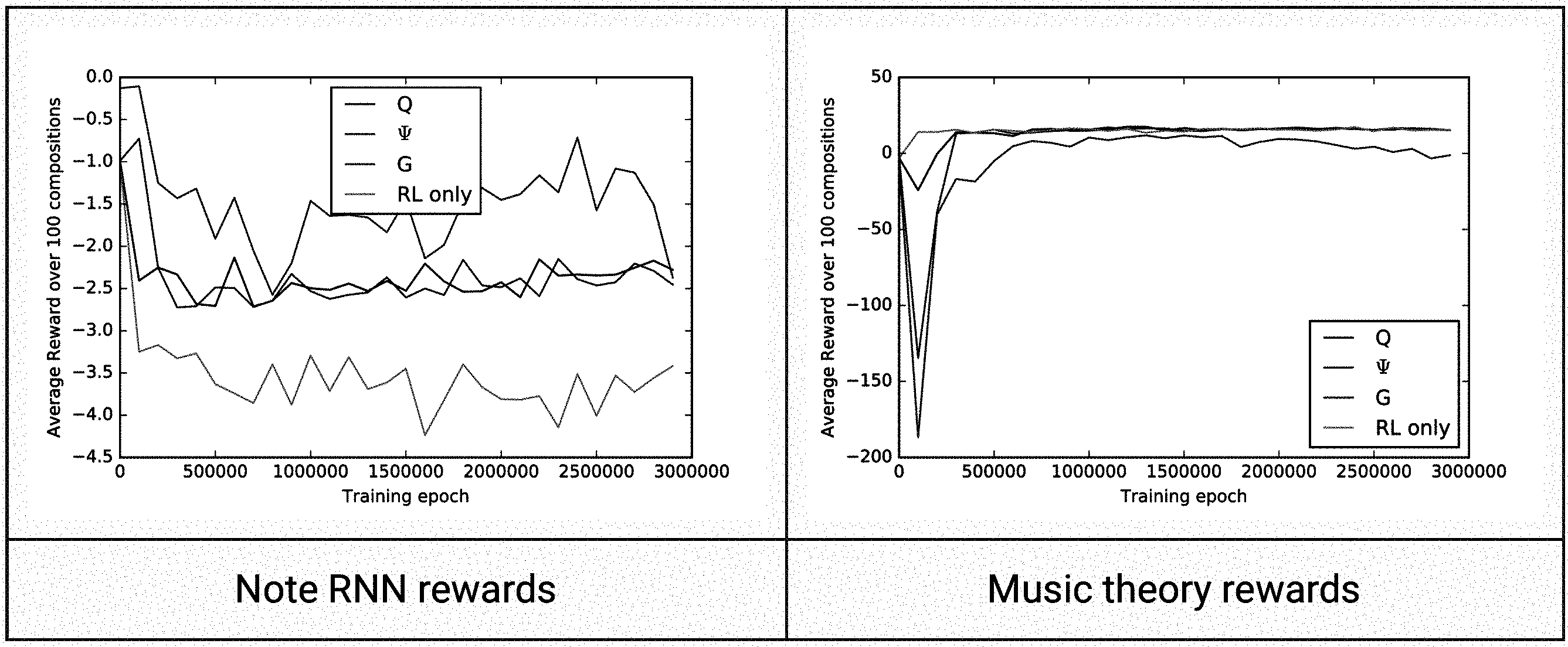}
\caption{Evolution during training of the two types of rewards for the RL-Tuner architecture.
Reproduced from \cite{jaques:rl:tuner:arxiv:2016} with permission of the authors}
\label{figure:rl-tuner:rewards}
\end{figure}

Figure~\ref{figure:rl-tuner:rewards} shows the evolution during the training phase of the two types of rewards (adherence to Note RNN and to music theory), with three different reinforcement learning algorithms:
Q-learning, $\Psi$-learning and G-learning (see details in \cite{jaques:rl:tuner:arxiv:2016}).

The corpus used for the experiments is a set of monophonic melodies extracted from a corpus of 30,000 MIDI songs.
The time step is set at a sixteenth note.
The one-hot encoding (of dimension 38) considers three octaves of notes plus two special events: note off (encoded as 0)
and no note (a rest, encoded as 1).
The MIDI note number is translated in order to start the lowest note (C$_3$) as a 2 (B$_5$ is encoded as 37)
and have special events smoothly integrated within the integer encoding.
Note that, as melodies are monophonic, playing a different note implicitly ends the last played note without requiring an explicit note off event,
which results in a more compact representation.
Note RNN (and its copy Reward RNN) have one LSTM layer with 100 cells.

In summary, the reinforcement strategy\index{Reinforcement!strategy}
allows arbitrary user given constraints (control\index{Control})
to be combined with a style\index{Style} learnt by the recurrent network.

Note that in the case of RL-Tuner, the reward\index{Reward} is known beforehand and dual purpose:
{\em handcrafted} for the music theory rules and {\em learnt} from the dataset by an RNN for the musical style.
Therefore, there is an opportunity to add another type of reward,
an  {\em interactive} feedback by the {\em user\index{User}}
(see Section~\ref{section:interactivity}).
However, a feedback\index{Feedback} at the granularity\index{Granularity} of each note generated may be too demanding
and, moreover, not that accurate\index{Accurate}\footnote{As Miles Davis\index{Miles Davis} coined it:
	``If you hit a wrong note, it's the next note that you play that determines if it's good or bad.''}.
We will discuss in Section~\ref{section:adaptability} the issue of learning from user feedback.
RL-Tuner is summarized in Table~\ref{table:dimensions:rl:tuner}.

\begin{table}
\begin{tabular}{|l|l|}
\hline
{\em Objective}			&Melody\\
\hline
{\em Representation}	&Symbolic; One-hot; Note-off; Rest\\
\hline
{\em Architecture}		&LSTM$\times$2 + RL\\
\hline
{\em Strategy}			&Iterative feedforward; Reinforcement\\
\hline
\end{tabular}
\caption{RL-Tuner summary}
\label{table:dimensions:rl:tuner}
\end{table}

%
%

\subsection{Unit Selection}
\label{section:challenges:strategies:control:unit:selection}
\label{section:control:unit:selection}
\label{section:strategy:unit:selection}

The {\em unit selection strategy\index{Unit!selection strategy}} is about querying successive musical units\index{Musical!unit}
(e.g., one measure long melodies)
from a database and concatenating\index{Concatenation} them in order to generate a sequence according to some user characteristics.
Querying\index{Query} is using features\index{Feature} which have been automatically extracted\index{Extraction} by an autoencoder.
Concatenation\index{Concatenation}, i.e. ``what unit next?'', is controlled by two LSTMs,
each one for a different criterium\index{Criterium},
in order to achieve a balance between {\em direction\index{Sense of direction}} and {\em transition\index{Transition}}.

This strategy, as opposed to most of the other ones,
which are bottom-up\index{Bottom-up},
is {\em top-down\index{Top-down}},
as it starts with a structure\index{Structure} and fills\index{Fill} it.

\subsubsection{Example: Unit Selection and Concatenation Symbolic Melody Generation System}
\label{section:system:brentan:unit:selection}
\label{section:experiment:brentan:unit:selection}

This strategy was pioneered by
Bretan {\em et al.} \cite{bretan:unit:selection:iccc:2017}.
The idea is to generate music from a concatenation\index{Concatenation} of musical units\index{Music!unit}, queried\index{Query} from a database\index{Database}.
The key process here is unit selection\index{Selection}, which is based on two criteria:
{\em semantic relevance\index{Semantic relevance}} and {\em concatenation cost\index{Concatenation!cost}}.
The idea of unit selection to generate sequences was actually inspired by a technique commonly used in text-to-speech\index{Text!-to-speech} (TTS\index{TTS}) systems.

The objective is to generate melodies.
The corpus\index{Corpus} considered is a dataset\index{Dataset} of
4,235
lead sheets in various musical styles\index{Style}
(jazz\index{Jazz}, folk\index{Folk}, rock\index{Rock}\ldots)
and
120
jazz solo\index{Solo} transcriptions\index{Transcription}.
The granularity\index{Granularity} of a musical unit is a measure.
This means there are roughly 170,000 units in the dataset.
The dataset is restricted to a five octaves range (MIDI note numbers 36 to 99)
and augmented by transposing each unit in all keys so that all possible pitches are covered.

The architecture includes one autoencoder and two LSTM recurrent networks.
The first step is feature extraction\index{Feature!extraction}:
10 features\index{Feature}, manually handcrafted\index{Handcrafted}, are considered,
following a {\em bag-of-words\index{Bag-of-words}} (BOW\index{BOW}) approach
(see Section~\ref{section:representation:bag:of:words}),
e.g., counts of a certain pitch class,
counts of a certain pitch class rhythm tuple\index{Tuple},
whether the first note is tied\index{Tied!note} to the previous measure, etc.
This results in 9,675 actual features.
Most of features have integer values,
with the exception of rests being represented using a negative pitch value
and of some Boolean features.
Therefore, each unit is described (indexed\index{Index}) as a feature vector of size 9,675.

\begin{figure}
\includegraphics[scale=0.29]{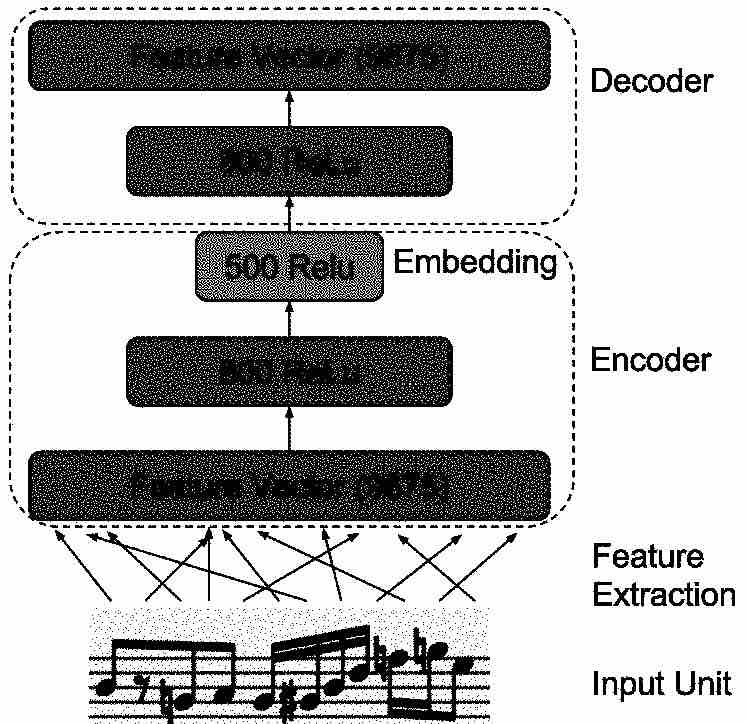}
\caption{Unit selection indexing architecture.
Reproduced from \cite{bretan:unit:selection:iccc:2017} with permission of the authors}
\label{figure:unit:selection:autoencoder:architecture}
\end{figure}

The autoencoder used has a 2-layer stacked autoencoder architecture,
as illustrated in Figure~\ref{figure:unit:selection:autoencoder:architecture}.
Once trained on the set of feature vectors,
in the usual self-supervised\index{Self!-supervised learning} way for autoencoders (see Section~\ref{section:architecture:autoencoder}),
the autoencoder becomes a features extractor\index{Feature!extractor}
encoding a feature vector of size 9,675
into an {\em embedding\index{Embedding}} vector of size 500. 

There is one remaining issue for generating a melody:
how to select the best (or at least, a very good) candidate
from a given (current, named seed by the authors) musical unit
as a successor musical unit?
Two criteria\index{Criterium} are considered:

\begin{itemize}

\item {\em Successor semantic relevance\index{Semantic relevance}} -- based on a model of transition\index{Transition} between units,
as learnt by an LSTM recurrent network.
In other words, relevance\index{Relevance} is based on the distance\index{Distance} to the (ideal) next unit as predicted by the model.
This first LSTM architecture has two hidden layers,
each with 128 units.
The input and output layers have 512 units (corresponding to the format of the embedding\index{Embedding}).

\item {\em Concatenation\index{Concatenation} cost\index{Cost}} -- based on another model of transition\footnote{At a more fine-grained\index{Fine-grained}
	note-level transition
	than the previous model.}
between the last note of current unit and the first note of the next unit, as learnt by another LSTM recurrent network.
This second LSTM architecture is multilayer and its input and output layers have about 3,000 units,
corresponding to a multi-one-hot encoding of the characterization of an individual note
(as defined by its pitch and its duration).

\end{itemize}

\begin{figure}
\includegraphics[scale=0.32]{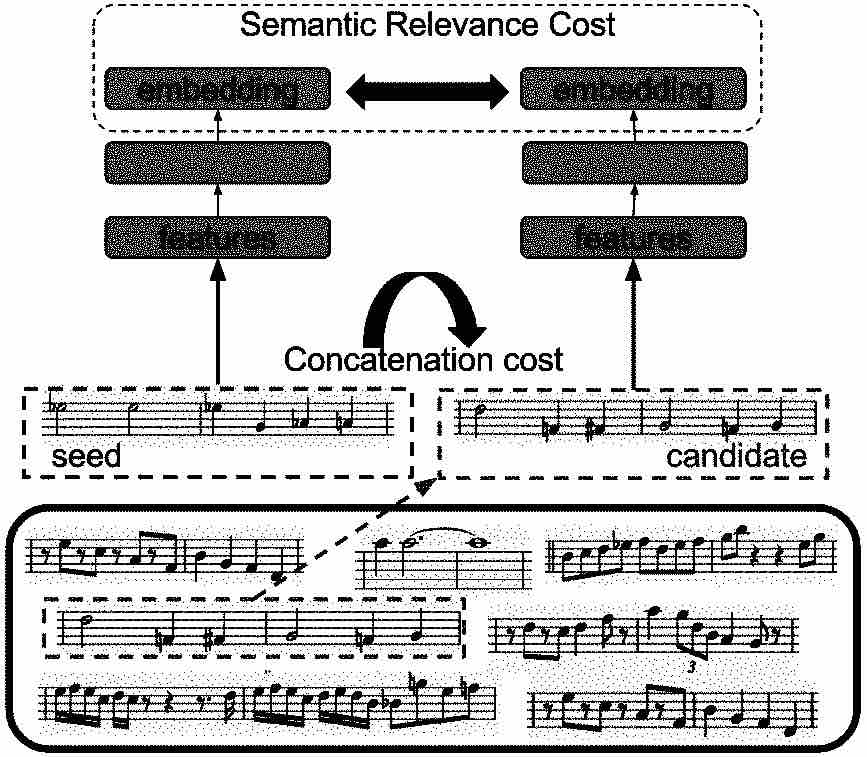}
\caption{Unit selection based on semantic cost.
Reproduced from \cite{bretan:unit:selection:iccc:2017} with permission of the authors}
\label{figure:unit:selection:semantic-cost}
\end{figure}

The combination of the two criteria
(illustrated in Figure~\ref{figure:unit:selection:semantic-cost},
with current (seed) unit in blue and next (candidate) unit in red)
is handled by a heuristic-based\index{Heuristic} dynamic ranking\index{Rank} process\index{Process}:


\begin{enumerate}

\item rank\index{Rank} all musical units according to their successor semantic relevance\index{Successor semantic relevance} with current musical unit\footnote{The initial musical unit of a melody to be generated
	may be chosen by the user or sorted.};

\item take the top 5\% and re-rank them according to the combination of their successor semantic relevance
and their concatenation\index{Concatenation} cost\index{Concatenation!cost}\index{Cost};
and


\item select the musical unit with the highest combined rank.

\end{enumerate}

The process is iterated in order to generate successive musical units and thus a melody of arbitrary length.
This may at first look like a standard iterative feedforward\index{Iterative feedforward strategy} generation
from a recurrent network
(see Section~\ref{section:experiment:eck:blues:lstm:second:experiment}),
but there are two important differences:

\begin{itemize}

\item the label\index{Label} (instance of the embedding\index{Embedding})
of the next musical unit is computed through a multicriteria\index{Multicriteria} ranking\index{Rank} algorithm\index{Algorithm}; and

\item the actual unit is queried\index{Query} from a database\index{Database} with the label as the index\index{Index}.

\end{itemize}


Initial external human evaluation\index{Evaluation} has been conducted by the authors.
They found that music generated using one or two measures long units
tend to be ranked higher according to naturalness and likeability
than four measures long units or note-level
generation,
with an ideal unit length appearing to be one measure.

Note that the unit selection strategy does not directly provide control,
but it does provide {\em entry points} for control
as one may extend the selection framework
(currently based on two criteria: successor semantic relevance and concatenation cost)
with user defined constraints/criteria.
The system is summarized in Table~\ref{table:dimensions:unit:selection}.

\begin{table}
\begin{tabular}{|l|l|}
\hline
{\em Objective}			&Melody\\
\hline
{\em Representation}	&Symbolic; Rest; BOW Features\\
\hline
{\em Architecture}		&Autoencoder$^2$ + LSTM$\times$2\\
\hline
{\em Strategy}			&Unit selection; Iterative feedforward\\
\hline
\end{tabular}
\caption{Unit selection summary}
\label{table:dimensions:unit:selection}
\end{table}

\section{Style Transfer}
\label{section:challenges:strategies:style:transfer}

In Section~\ref{section:control:input:manipulation:musical:style:transfer} we introduced
style transfer\index{Style!transfer}
as one example of using the input manipulation\index{Input!manipulation strategy} strategy to control content generation.
The style transfer technique for images
(proposed by Gatys {\em et al.} \cite{gatys:neural:style:2015} and described in Section~\ref{section:system:gatys:style:transfer})
is effective and relatively straightforward to apply.
However, as opposed to paintings, where the common representation is two-dimensional and uniformly digitalized in terms of pixels,
music is a much more complex object with various levels and models of representation
(see Chapter~\ref{section:chapter:representation} and also Section~\ref{section:musical:style:transfer:timbre:challenges}).

In their recent analysis, Dai {\em et al.} \cite{dai:music:style:transfer:arxiv:2018}
consider three main levels (or dimensions) of representation and associated types of music style transfer:

\begin{itemize}

\item {\em score\index{Score}}-level, which they name {\em composition style transfer\index{Composition!style transfer}};

\item {\em sound\index{Sound}}-level, which they name {\em timbre style transfer\index{Timbre!style transfer}}; and

\item {\em performance\index{Performance} control\index{Control}}-level, which they name {\em performance style transfer\index{Performance!style transfer}}.

\end{itemize}

They state that music style transfer for each level
(namely, composition, timbre and performance style transfer)
are very different in nature.
They also point out the issue of the interrelation and the {\em entanglement\index{Entanglement}}
of these different levels (and nature) of representation.
Therefore, they point out the need for automated learning of the disentanglement\index{Disentanglement}\footnote{Disentanglement
	is the objective of separating the different factors governing variability in the data
	(e.g., in the case of human images,
	identity of the individual and facial expression,
	see, for example, \cite{desjardins:disentangling:arxiv:2012}).
	Recent work on {\em disentanglement learning\index{Disentanglement!learning}}
	can be found, for example, in \cite{ldr:disentangled:workshop:nips:2017}.
	Also note that variational autoencoders\index{Variational!autoencoder} (VAEs\index{VAE}, see Section~\ref{section:architecture:vae})
	are currently among the promising approaches for disentanglement learning
	because, as Goodfellow {\em et al.} put it in \cite[Section~20.10.3]{goodfellow:deep:learning:book:2016}:
	``Training a parametric encoder in combination with the generator network forces the
	model to learn a predictable coordinate system that the encoder can capture.''}
of different levels of music representation,
in order to ease music style transfer.

\subsection{Composition Style Transfer}
\label{section:musical:style:transfer:composition}

Style transfer at the composition level means working on symbolic\index{Symbolic} representations.
An example is {\em structure imposition\index{Structure!imposition}},
i.e. transferring some existing structure (e.g., an AABA global structure)
from an initial composition into another newly generated composition.
The C-RBM\index{C-RBM} system, presented in Section~\ref{section:experiment:c:rbm},
implements this kind of structure imposition by considering separately three kinds of structures and associated constraints:
global structure (e.g., AABA), tonality and meter (rhythm).

One may think that such structure descriptors are too low level to define a style.
But they are an interesting first step, as one may consider higher-level style descriptors by aggregating such structure descriptors.
Let us imagine, for instance, describing (and later on transferring) the style of a composer like Michel Legrand\index{Michel Legrand}
with his own way of repeating transpositions of motives.

Note that in the DeepJ\index{DeepJ} system for controlling the style of the generation (Section~\ref{section:systems:deepj})
the objective is different, as the style is explicitly specified
by the user via a set of musical examples,
learnt and applied during generation time through conditioning\index{Conditioning}.

\subsection{Timbre Style Transfer}
\label{section:musical:style:transfer:timbre}

For timbre style transfer\index{Timbre!style transfer}, based on audio\index{Audio} representations, some researchers have straightforwardly
applied Gatys {\em et al.}'s technique (Section~\ref{section:system:gatys:style:transfer})
to sound (audio), using various kinds of sources (various styles of music as well as speech),
as explained in next section.

\subsubsection{Examples: Audio Timbre Style Transfer Systems}
\label{section:musical:style:transfer:timbre:examples}

Examples of style transfer systems for audio (timbre\index{Timbre}) are:

\begin{itemize}

\item Ulyanov and Lebedev's system in \cite{ulyanov:audio:style:transfer:web:2016}, and

\item Foote {\em et al.}'s system in \cite{foote:audio:style:transfer:2016}.

\end{itemize}

These two systems both use a spectrogram (and not a direct wave signal) as their input representation\footnote{For a comparison
	of various audio representations for audio style transfer,
	see the recent analysis by Wyse
	\cite{wyse:spectrogram:convolutional:dlm:2017}.}.
In \cite{wyse:spectrogram:convolutional:dlm:2017}, Wyse points out two specificities (which he calls ``two remarkable aspects'')
in the architecture of Ulyanov and Lebedev's system
that differentiate it from the
image style transfer technique:

\begin{itemize}

\item the network uses only a single layer.
Therefore,
the only difference between content and style comes from the difference
between first-order and second-order (correlations) measures of activation; and

\item the network was not pre-trained and uses random weights.

\end{itemize}

Wyse further adds in \cite{wyse:spectrogram:convolutional:dlm:2017}:
``The blog post claims this unintuitive approach generated results as good as any other, and the sound examples posted are indeed compelling.''
We also found convincing the examples of audio style transfer,
although not as interesting as painting style transfer,
as it sounds similar to some sound modulation/merging of both style and content signals.
In their own analysis in \cite{foote:audio:style:transfer:2016},
Foote {\em et al.} summarize the difficulty as follows:
``On this level we draw one main conclusion:
audio is dissimilar enough from images that we shouldn't expect work in this domain to be as simple as changing 2D convolutions to 1D.''
We will try to analyze some possible reasons for this in the
next section.

Table~\ref{table:dimensions:musical:style:transfer} summarizes the main features of these audio (timbre) style transfer systems
(which we will reference to as AST in Chapter~\ref{section:chapter:analysis}).

\begin{table}
\begin{tabular}{|l|l|}
\hline
{\em Objective}			&Audio style transfer (AST)\\
\hline
{\em Representation}	&Audio; Spectrum\\
\hline
{\em Architecture}		&Convolutional(Feedforward)\\
\hline
{\em Strategy}			&Input manipulation; Single-step feedforward\\
\hline
\end{tabular}
\caption{Audio (timbre) style transfer (AST) summary}
\label{table:dimensions:musical:style:transfer}
\end{table}

\subsubsection{Limits and Challenges}
\label{section:musical:style:transfer:timbre:challenges}

\begin{figure}
\includegraphics[width=\textwidth]{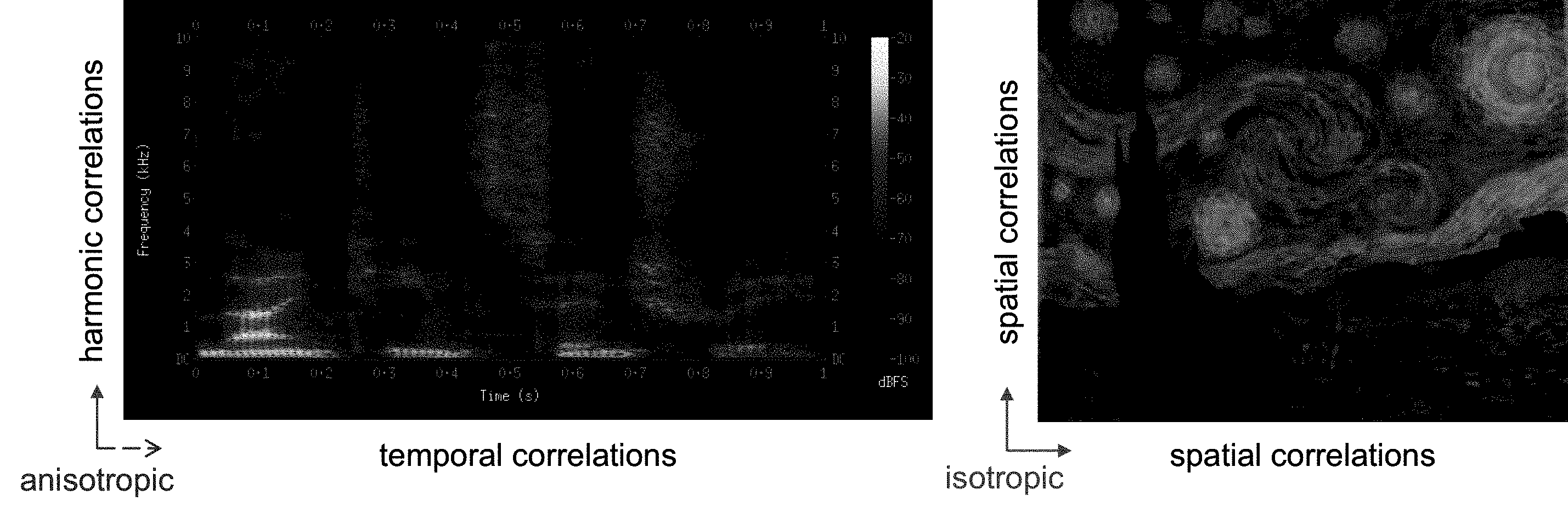}
\caption{Anisotropic music vs an isotropic image.
Incorporating Aquegg's original image from ``https://en.wikipedia.org/wiki/Spectrogram''
and the painting ``The Starry Night'' by Vincent van Gogh (1889)}
\label{figure:anisotropic}
\end{figure}

We believe that, in part, the difficulty of directly transposing image style transfer to music
comes from the {\em anisotropy}\footnote{Isotropy means invariance of properties regardless of the direction,
	whereas anisotropy
	means direction dependence.}
of global audio music content representation.
In the case of a natural image,
the correlations between visual elements (pixels) are equivalent whatever the direction
(horizontal axis, vertical axis, diagonal axis or any arbitrary direction), i.e. correlations are {\em isotropic}.
In the case of a global representation of audio data
(where the horizontal dimension represents time
and the vertical dimension represents the notes),
this uniformity no longer holds
as horizontal correlations represent {\em temporal} correlations and vertical correlations represent {\em harmonic} correlations,
which are very different in nature
(see an illustration in Figure~\ref{figure:anisotropic}).

One direction could be to reformulate the capture of the style information,
and therefore the nature of the correlations,
in order to take into account the time dimension. 
Another (also hypothetical) direction could be to use a ``time-compressed'' representation,
by considering the summary learnt by an RNN Encoder-Decoder\index{RNN Encoder-Decoder}
(see Section~\ref{section:experiment:vrae}).



\subsection{Performance Style Transfer}
\label{section:musical:style:transfer:performance}



Although it does not directly address performance style transfer,
the Performance RNN system described in Section~\ref{section:experiment:performance:rnn}
provides a background representation for modeling performance (note onsets\index{Note!onset} as well as dynamics).
What remains to be undertaken to develop a {\em performance imposition} system could be along the following lines:

\begin{itemize}

\item model mappings between performance and other(s) features (e.g., mean duration of notes, modulation, etc.);

\item learn mappings for a given corpus (musician, context, etc.) through correlation analysis,
revealing the performance style of a given musician (and corpus); and

\item transpose a mapping to an existing piece in order to transfer the performance style.

\end{itemize}

As noted by Dai {\em et al.} in \cite{dai:music:style:transfer:arxiv:2018}, performance style transfer is closely related to expressive performance rendering,
see, for instance, the example of the Cyber-Jo\~ao\index{Cyber-Jo\~ao} system \cite{dahia:cyber:joao:sbcm:2003}
introduced in Section~\ref{section:challenges:strategies:expressiveness}
and a recent system based on deep learning in \cite{malik:translation:musical:style:arxiv:2017}.
But it also requires the disentanglement of control (style) and score information (content)
as well as the learning of the mappings discussed above.
Thus, this is still a direction to be explored. 

\subsection{Example: FlowComposer Composition Support Environment}
\label{section:systems:flow:composer}

A good example of an interactive music composition environment addressing style transfer in different dimensions
is the FlowComposer\index{FlowComposer} system \cite{pachet:flow:composer:ecai:2014,papadopoulos:flow:composer:cp:2016},
developed by Pachet {\em et al.}
during the Flow Machines\index{Flow Machines} project \cite{csl:flow:machines:web:2012}.
Note that it is based on Markov chain\index{Markov!chain} models and not (yet) deep learning models.

\begin{figure}
\includegraphics[width=\textwidth]{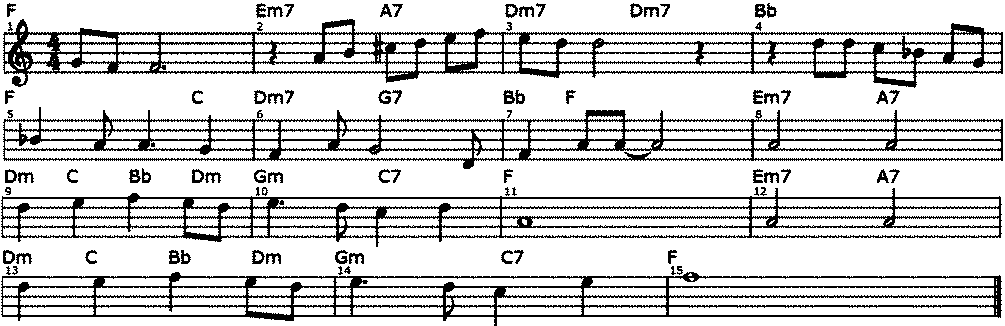}
\caption{Yesterday (Lennon/McCartney) (first 15 measures) -- original harmonization.
Reproduced from \cite{papadopoulos:flow:composer:cp:2016} with permission of the authors}
\label{figure:flow:composer:yesterday}
\end{figure}

\begin{figure}
\includegraphics[width=\textwidth]{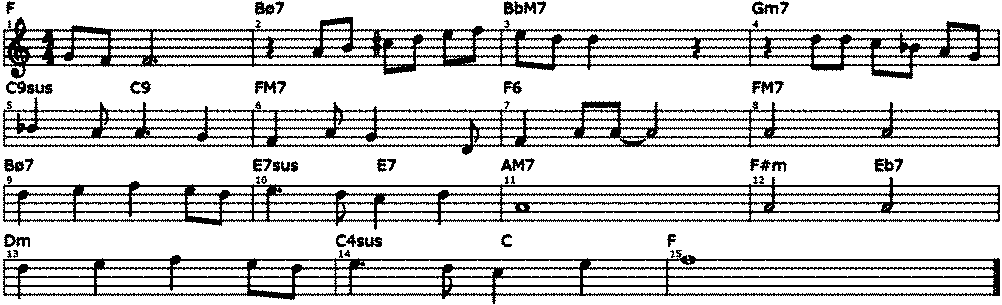}
\caption{Yesterday (Lennon/McCartney) (first 15 measures) -- reharmonization by FlowComposer in the style of Michel Legrand.
Reproduced from \cite{papadopoulos:flow:composer:cp:2016} with permission of the authors}
\label{figure:flow:composer:yesterday:legrand}
\end{figure}

\begin{figure}
\includegraphics[width=\textwidth]{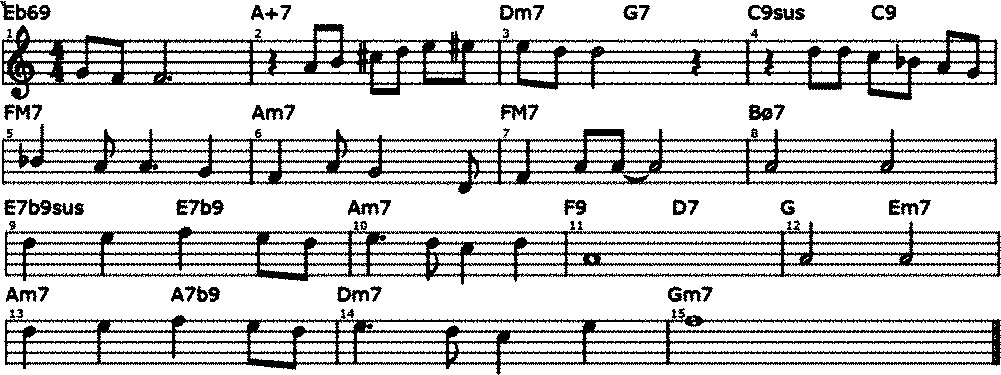}
\caption{Yesterday (Lennon/McCartney) (first 15 measures) -- reharmonization by FlowComposer in the style of Bill Evans.
Reproduced from \cite{papadopoulos:flow:composer:cp:2016} with permission of the authors}
\label{figure:flow:composer:yesterday:evans}
\end{figure}

FlowComposer provides possibilities for music style transfer at the following levels:

\begin{itemize}

\item Composition style level --
some style transfer may be performed,
e.g., automated reharmonization\index{Harmonization} based on a style (corpus) of selected music.
See the examples of the automatic reharmonization of Yesterday by John Lennon and Paul McCartney (Figure~\ref{figure:flow:composer:yesterday})
in the style of Michel Legrand\index{Michel Legrand} (Figure~\ref{figure:flow:composer:yesterday:legrand})
and Bill Evans\index{Bill Evans} (Figure~\ref{figure:flow:composer:yesterday:evans}).

\item Timbre and performance style levels\footnote{Jointly,
	as there is no possibility yet for separating/disentangling these two concerns.} --
rendering\index{Rendering} may be done via style transfer, by automated mapping and extrapolation from a library of various
instrumental audio performances (through the ReChord\index{ReChord} component \cite{ramona:re:chord:ijcai:2015}).

\end{itemize}

\begin{figure}
\includegraphics[width=\textwidth]{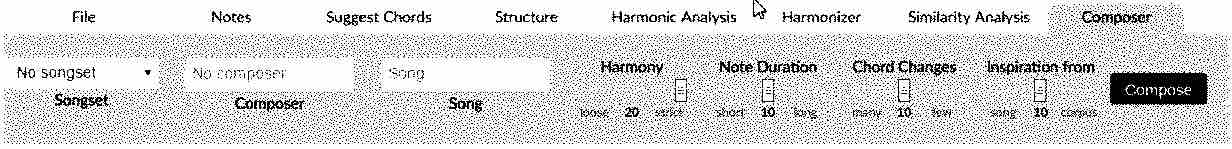}
\caption{Flow Composer control panel.
Reproduced from \cite{papadopoulos:flow:composer:cp:2016} with permission of the authors}
\label{figure:flow:composer:control:panel}
\end{figure}

\begin{figure}
\includegraphics[width=\textwidth]{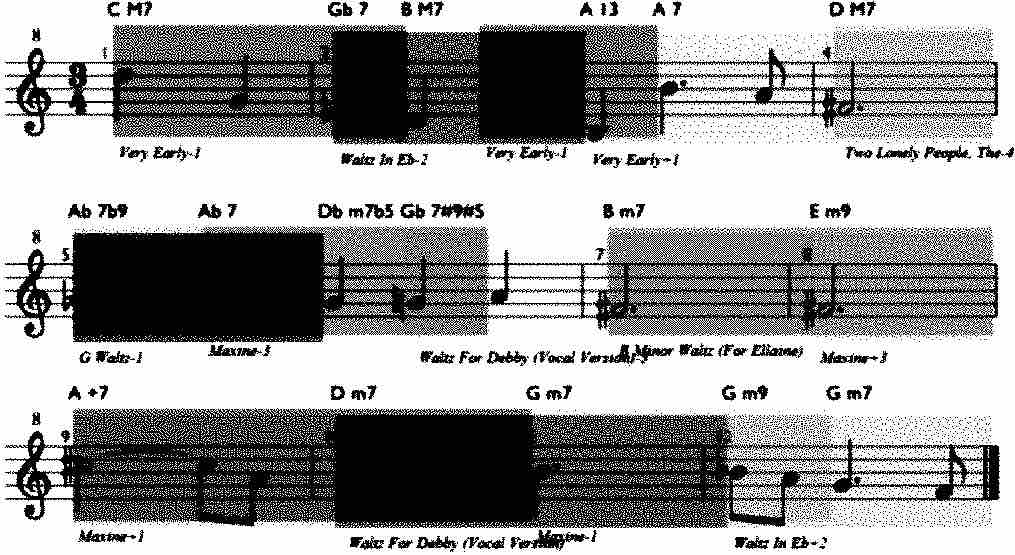}
\caption{Example of a Flow Composer interactively generated lead sheet.
Reproduced from \cite{pachet:flow:composer:ecai:2014} with permission of the authors}
\label{figure:flow:composer:example}
\end{figure}

The FlowComposer control panel includes various fields to select composition style and sliders to set harmonisation conformance, inspiration,
average note duration and chord changes, as shown in Figure~\ref{figure:flow:composer:control:panel}.
An example of a lead sheet generated
in the style of Bill Evans\index{Bill Evans} is shown in Figure~\ref{figure:flow:composer:example},
with the following user defined characteristics:
a 3/4 time signature,
a constraint on the first (C7) and last (G7) chords,
and a ``max order'' of four beats\footnote{This very interesting feature controls the maximum amount of successive notes (actually beats) copied from the corpus.
	It relies on the integration of a new constraint named MaxOrder
	\cite{papadopoulos:maxorder:universality:book:2016}
	in the Markov constraints framework \cite{pachet:markov:constraints:constraints:2011} underlying FlowComposer.
	This is one possible way to control originality (see Section~\ref{section:originality}).}.
Color backgrounds indicate sequences of notes extracted as a whole from a given song in the chosen corpus
(here all of Bill Evans' compositions in 3/4).

\section{Structure}
\label{section:challenge:structure}

One challenge is that most existing systems have a tendency to generate music with no clear structure\index{Structure}
or ``sense of direction\index{Sense of direction}''\footnote{Beside the technical improvements
	brought by LSTMs\index{LSTM} on the learning of long-term dependencies (see Section~\ref{section:architecture:lstm}).}.
In other words, although the style of the generated music corresponds to the corpus learnt,
the music appears to wander without any higher organization\index{Organization},
as opposed to human composed music which has some global organization
(usually named a {\em form\index{Form}}) and identified components,
such as

\begin{itemize}

\item an overture, an allegro, an adagio or a finale in classical\index{Classical} music;

\item an AABA or an AAB form in jazz\index{Jazz};

\item a refrain, a verse or a bridge in song music.

\end{itemize}

Note that there are various possible levels of structure.
For instance, an example of a finer-grain structure is at the level of a melodic motif\index{Motif} that can be repeated,
often being transposed in order to adapt to a new harmonic structure.

The reinforcement strategy (used by RL-Tuner\index{RL-Tuner} in Section~\ref{section:experiment:rl-tuner})
and the structure imposition approach (used by C-RBM\index{C-RBM} in Section~\ref{section:systems:c-rbm})
can both enforce (and/or transfer, see Section~\ref{section:challenges:strategies:style:transfer})
some constraints, possibly high-level, on the generation.
About structure imposition, see also
a recent proposal combining two graphical models, one for chords and one for melody,
for the generation of lead sheets with an imposed structure
\cite{pachet:variations:structured:ismir:2017}.
An alternative top-down approach is followed by the unit selection strategy\index{Unit!selection strategy}
(see Section~\ref{section:control:unit:selection}),
by generating an abstract sequence structure and filling it with musical units,
although the structure is not yet very high-level as it effectively stays at the level of a measure.

A related challenge is not about the {\em imposition} of {\em preexisting} high-level structures,
but about the capacity for {\em learning} high-level structures
and, moreover, the capacity for {\em invention} (emergence) of high-level structures.
Therefore, a natural direction is to explicitly consider and process different levels (hierarchies) of temporality and structure.

\subsection{Example: MusicVAE Multivoice Hierarchical Symbolic Music Generation System}
\label{section:system:music:vae}

In
\cite{roberts:hierarchical:latent:icml:2018},
Roberts {\em et al.} propose an architecture named MusicVAE\index{MusicVAE},
based on a variational recurrent autoencoder\index{Variational!recurrent autoencoder} (VRAE\index{VRAE})
with a 2-level hierarchical RNN within the decoder

The corpus comprises MIDI files collected from the web,
from which three types of musical examples are extracted:

\begin{itemize}

\item monophonic melodies\footnote{MusicVAE has since
	been extended to an arbitrary number of polyphonic tracks,
	see details in \cite{simon:latent:space:arxiv:2018}.},
	2 or 16 measures long;

\item drum patterns, 2 or 16 measures long; and

\item trio sequences with three different voices (melody, bass line and drum pattern), 16 measures long.

\end{itemize}

Encoding of monophonic melodies and bass lines is through tokens representing MIDI events:
the 128 ``Note on'' events corresponding to the 128 possible MIDI note numbers (pitches) of the defined interval,
the single\footnote{Only one ``Note off'' event is needed for all possible pitches as the melody is monophonic.
	It is used to differentiate a note held from two successive identical notes,
	see Section~\ref{section:representation:note:ending}.}
``Note off'' event and the rest (silence) token.
Encoding of drum patterns is done by mapping MIDI standard drum classes through a binning\index{Binning}
into 9 canonical classes, leading to $2^9 = 512$ categorical tokens representing all possible combinations.
Quantization is at the sixteenth note.

The architecture follows the principles of a variational autoencoder encapsulating recurrent networks such as VRAE\index{VRAE},
with two differences:

\begin{itemize}

\item the encoder is a bidirectional recurrent network\index{Bidirectional!recurrent neural network}
(see Section~\ref{section:architecture:compound:bidirectional:rnn})
-- an LSTM with the input and output layers having 2,048 nodes and a single hidden layer of 512 cells; and

\item the decoder is a hierarchical\index{Hierarchical} 2-level recurrent network, composed of

\begin{itemize}

\item a high-level RNN named the conductor
 -- an LSTM with the input and output layers having 512 nodes and a single hidden layer of 1,024 cells --
that produces a sequence of embeddings\index{Embedding}; and

\item a bottom-layer RNN
-- an LSTM with two hidden layers of 1,024 cells --
that uses each embedding as an initial state and also as an additional input concatenated to its previously generated token\footnote{Along
	the iterative feedforward strategy.}
to produce each subsequence.
In order to prioritize the conductor RNN over the bottom-layer RNN,
its initial state is reinitialized with the decoder generated embedding for each new subsequence.
In the case of a multivoice trio (melody, bass and drums), there are three LSTMs, one for each voice.

\end{itemize}

\end{itemize}

\begin{figure}
\includegraphics[width=\textwidth]{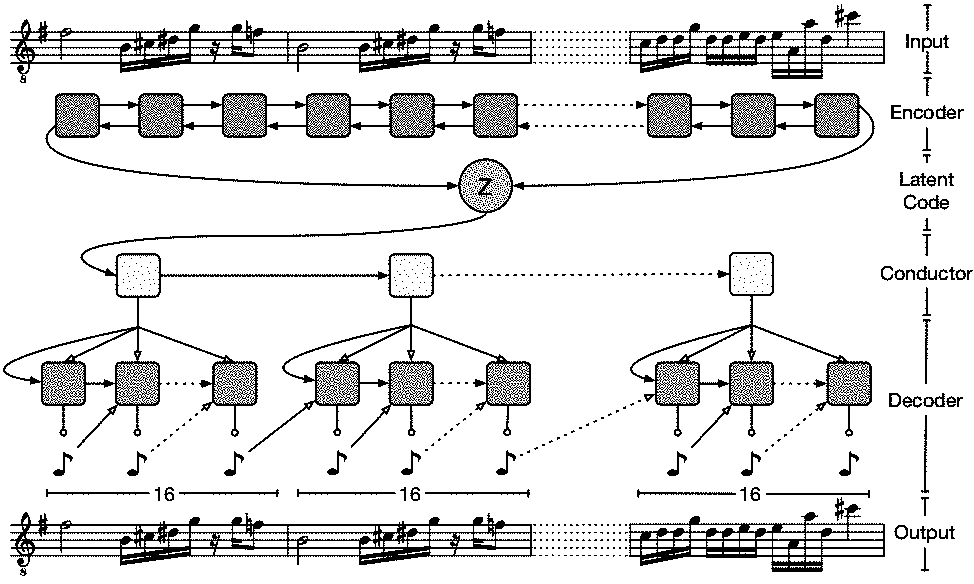}
\caption{MusicVAE architecture.
Reproduced from \cite{roberts:hierarchical:latent:icml:2018} with permission of the authors}
\label{figure:music:vae:architecture}
\end{figure}

The MusicVAE architecture is illustrated in Figure~\ref{figure:music:vae:architecture}.
The authors report that an equivalent ``flat'' MusicVAE architecture (without hierarchy), although accurate in modeling the style
in the case of 2 measures long examples, was inaccurate in the case of 16 measures long examples, with a 27\% error increase
for the autoencoder reconstruction (0.883 accuracy for the flat architecture and 0.919 accuracy for the hierarchical architecture).

\begin{figure}
\includegraphics[width=\textwidth]{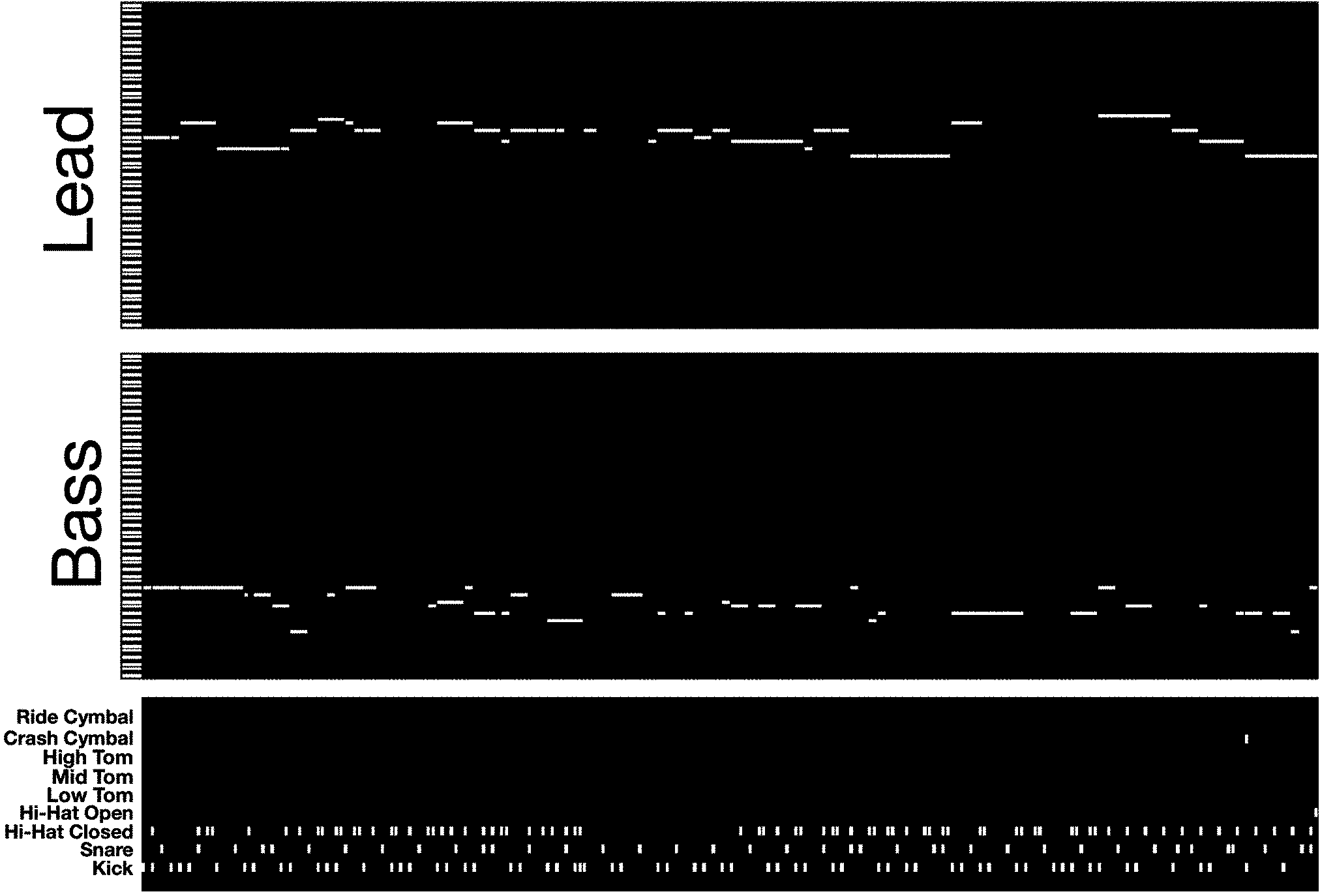}
\caption{Example of a trio music generated by MusicVAE.
Reproduced from \cite{roberts:hierarchical:latent:arxiv:2018} with permission of the authors}
\label{figure:music:vae:example:trio}
\end{figure}

An example of trio music generated is shown in Figure~\ref{figure:music:vae:example:trio}.
A preliminary evaluation has been conducted with listeners comparing three versions
(flat architecture, hierarchical architecture and real music)
for three types of music: melody, trio and drums.
The results show a very significant gain with the hierarchical architecture,
see more details in \cite{roberts:hierarchical:latent:icml:2018}.

An interesting feature of the variational autoencoder architecture is in the capacity for exploring the latent space via various operations
such as

\begin{figure}
\includegraphics[scale=1]{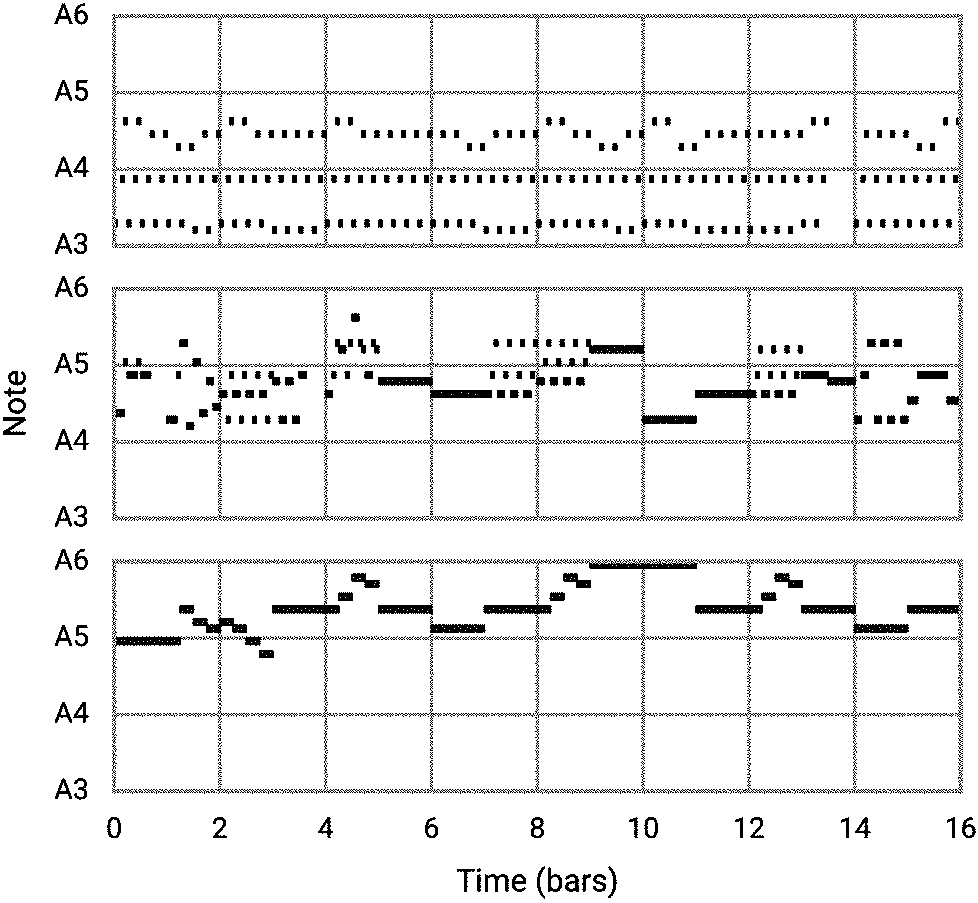}
\caption{Example of a melody generated (middle) by MusicVAE by averaging the latent spaces of two melodies (top and bottom).
Reproduced from \cite{roberts:hierarchical:latent:icml:2018} with permission of the authors}
\label{figure:music:vae:example:averaging}
\end{figure}

\begin{figure}
\includegraphics[scale=1]{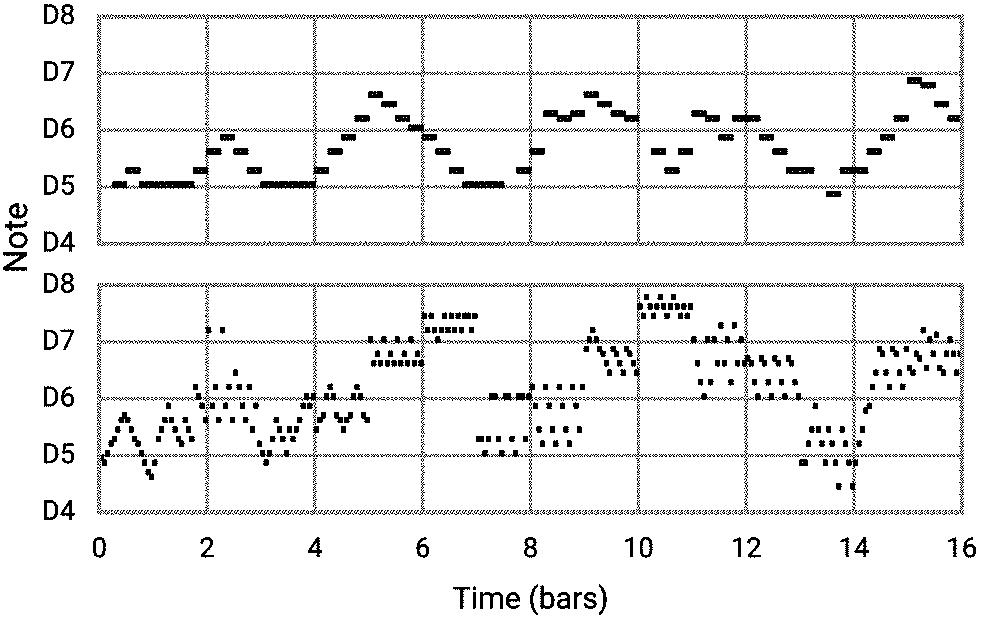}
\caption{Example of a melody generated (bottom) by MusicVAE
by adding a ``high note density'' attribute vector to the latent space of an existing melody (top).
Reproduced from \cite{roberts:hierarchical:latent:arxiv:2018} with permission of the authors}
\label{figure:music:vae:example:density}
\end{figure}

\begin{itemize}

\item {\em translation};

\item  {\em interpolation\index{Interpolation}} --
Figure~\ref{figure:music:vae:example:interpolation} in Section~\ref{section:architecture:vae}
shows an interesting comparison of melodies resulting from
interpolation in the data space (that is the space of representation of melodies)
and interpolation in the latent space which is then decoded into the corresponding melodies.
One can see (and hear) that the interpolation in the latent space produces much more meaningful and interesting melodies;

\item  {\em averaging\index{Average}} --
Figure~\ref{figure:music:vae:example:averaging} shows an example of a melody (in the middle of the figure)
generated from the combination (averaging) of the latent spaces of two melodies (at the top and bottom of the figure);

\item {\em attribute vector arithmetics\index{Attribute!vector arithmetics}},
by {\em addition or subtraction of an attribute vector capturing a given characteristic} --
Figure~\ref{figure:music:vae:example:density} shows an example of a melody (at the bottom of the figure) generated
when a ``high note density'' attribute vector is added to the latent space of an existing melody (at the top of the figure).
An attribute vector is computed as the average of the latent vectors for a collection of examples sharing that characteristic (attribute)
(e.g., high density of notes, rapid change, high register, etc.).

\end{itemize}


\begin{figure}
\includegraphics[width=\textwidth]{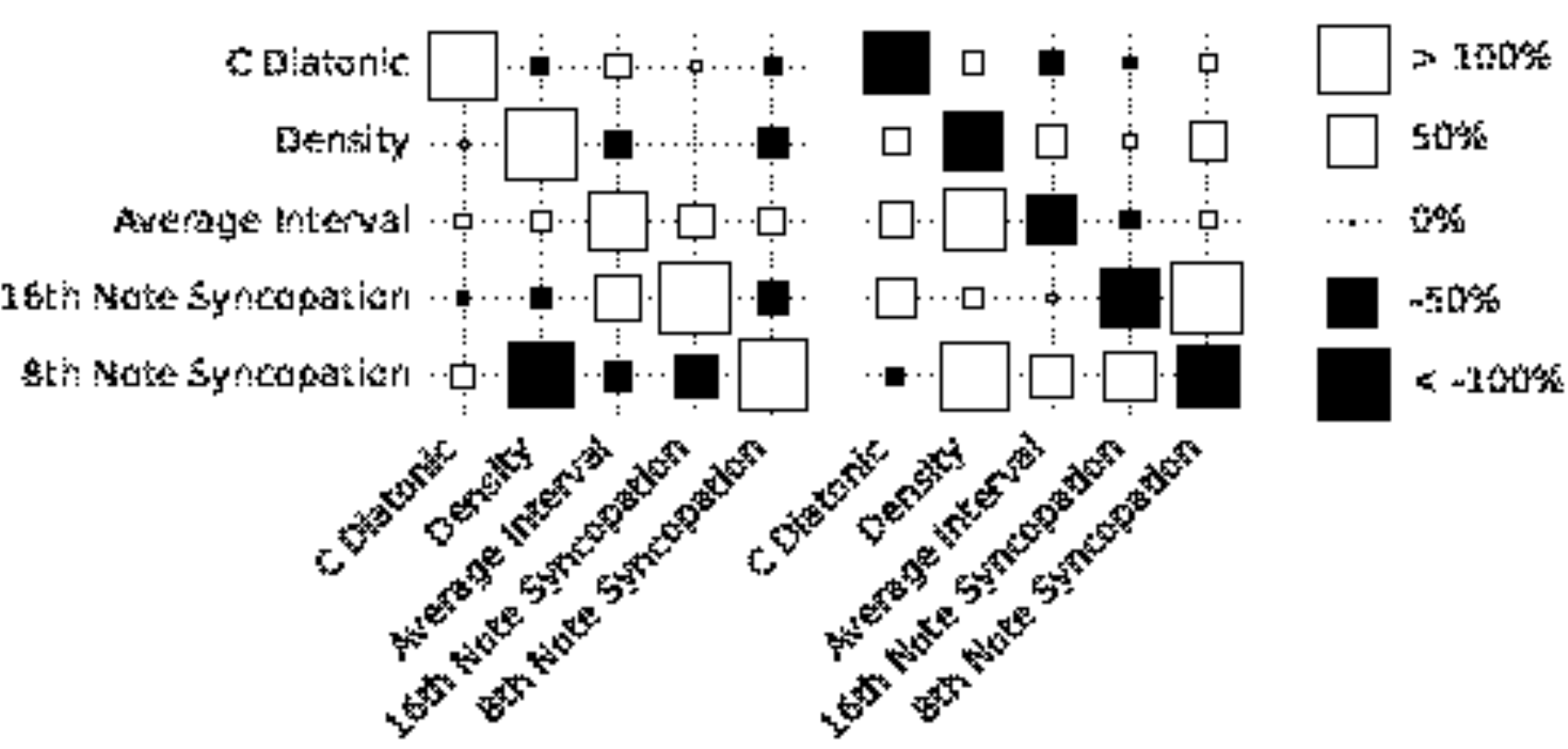}
\caption{Correlation matrices of the effect of adding (left) of subtracting (right) an attribute to other attributes in MusicVAE.
Reproduced from \cite{roberts:hierarchical:latent:icml:2018} with permission of the authors}
\label{figure:music:vae:example:adding:subtracting:attribute}
\end{figure}

Furthermore,
Figure~\ref{figure:music:vae:example:adding:subtracting:attribute}
shows the effect, as a percent change,
of modifying individual attributes\index{Attribute} of 16 measures long melodies by adding (left matrix), or respectively subtracting (right matrix),
attribute vectors in the latent space.
The vertical axis of each correlation matrix denotes the attribute vector applied
and the horizontal axis denotes the attribute measured. 
These correlation matrixes show that individual attributes can be modified without effecting others,
except for the cases when correlations are expected, as for instance between the eighth and sixteenth note syncopations. 


%

Audio examples are available in \cite{roberts:music:vae:web} and \cite{roberts:music:vae:material:web}.
MusicVAE is summarized in Table~\ref{table:dimensions:music:vae}.

\begin{table}
\begin{tabular}{|l|l|}
\hline
{\em Objective}			&Melody; Trio (Melody, Bass, Drums)\\
\hline
{\em Representation}	&Symbolic; Drums; Note end; Rest\\
\hline
{\em Architecture}		&Variational Autoencoder(Bidirectional-LSTM, Hierarchical$^2$-LSTM)\\
\hline
{\em Strategy}			&Iterative feedforward; Sampling; Latent variables manipulation\\
\hline
\end{tabular}
\caption{MusicVAE summary}
\label{table:dimensions:music:vae}
\end{table}

\subsection{Other Temporal Architectural Hierarchies}
\label{section:challenge:structure:other:hierarchies}

There are some alternative solutions to organize temporal hierarchies\index{Hierarchical} within a deep learning architecture.
A first example is ClockworkRNN\index{ClockworkRNN}, by Koutn\'{\i}k {\em et al.} \cite{koutnik:clockworkrnn:arxiv:2014}.
The idea is to partition the hidden recurrent layer into various modules,
all fully connected (in a parallel way) to input layer nodes and to output layer nodes,
but each module with a different clock rate
with interconnections between modules according to their clock rates
(neurons of a faster module are fully connected to neurons of a slower module).

Another example is SampleRNN\index{SampleRNN}, by Mehri {\em et al.} \cite{mehri:samplernn:arxiv:2017}.
It is an extension of WaveNet\index{WaveNet} architecture (Section~\ref{section:systems:wavenet})
inspired from the idea of different clock rates from ClockworkRNN, but with some external modules (full networks)
organized via some conditioning\footnote{Conditioning
	has been introduced in Section~\ref{section:architecture:conditioning}.},
as opposed to ClockworkRNN's internal modules.
Each module is a deep RNN which summarizes the history of its inputs (successive waveform frames)
into a conditioning vector for the next module downward
which operates on frames of shorter duration.
More details of these two architectures may be found in their respective articles,
\cite{koutnik:clockworkrnn:arxiv:2014} and \cite{mehri:samplernn:arxiv:2017}.

These examples show that there is an active ongoing research activity
to explore various ways to organize temporal hierarchies in deep learning architectures
(for audio or symbolic contents),
in order to try to better capture longer term structure of music.

\section{Originality}
\label{section:challenges:strategies:originality}
\label{section:originality}

The issue of the {\em originality\index{Originality}} of the music generated is not only an artistic issue ({\em creativity})
but also an economic one,
because it raises the issue of
the copyright\footnote{On this issue,
	see the recent paper by Deltorn \cite{deltorn:deep:creations:digital:humanities:2017}.}.

One approach is {\em a posteriori}, by ensuring that the generated music is not too similar\index{Similarity}
(e.g., in not having recopied a significant number of notes of a melody)
to an existing piece of music.
Therefore, existing algorithms to detect similarities in texts may be used.

Another approach, more systematic but even more challenging, is {\em a priori},
by ensuring that the music generated will not recopy\index{Recopy} a given portion of music
from the training corpus\index{Training!corpus}\footnote{Note that this addresses the issue
	of
	significant recopying from the training corpus,
	but it does not prevent a system from {\em reinventing} existing music outside of the training corpus.}.
A solution for music generation from Markov chains\index{Markov!chain} has been proposed
\cite{papadopoulos:maxorder:universality:book:2016}.
It is based on a variable order Markov model
and constraints over the order of the generation through some min order and max order constraints,
in order to attain some sweet spot between junk and plagiarism.
However, there is not yet a solution for deep learning architectures.

Let us now analyze some recent directions for favoring originality in the generated musical content.
 
\subsection{Conditioning}
\label{section:challenges:strategies:originality:conditioning}

\subsubsection{Example: MidiNet Melody Generation System}
\label{section:originality:conditioning:midinet}

In their description of MidiNet \cite{yang:midinet:ismir:2017}
(see Section~\ref{section:systems:midinet}),
the authors discuss two methods to control\index{Control} creativity\index{Creativity}:

\begin{itemize}

\item restricting the conditioning\index{Conditioning} by inserting the conditioning data only
in the intermediate convolution\index{Convolution} layers of
the generator architecture; and

\item decreasing the values of the two control parameters\index{Control!parameter}
of feature\index{Feature} matching\index{Feature!matching} regularization\index{Regularization},
in order to reduce the requirement for
the closeness of the distributions
of real data and generated distributions\index{Distribution} of real and generated data.

\end{itemize}

These experiments are interesting but they remain at the level of {\em ad hoc} tuning
of the hyper-parameters of the architecture.


\subsection{Creative Adversarial Networks}
\label{section:challenges:strategies:originality:can}

Another more systematic and conceptual direction is the concept of
{\em creative adversarial networks\index{Creative!adversarial networks}} {\em (CAN\index{CAN})}
proposed by Elgammal {\em et al.} \cite{elgammal:can:arxiv:2017},
as an extension of the generative adversarial networks\index{Generative!adversarial networks} (GAN\index{GAN}) architecture
(introduced in Section~\ref{section:architecture:gan}).


\subsubsection{Creative Adversarial Networks Painting Generation System}
\label{section:experiment:can}

Elgammal {\em et al.} propose in \cite{elgammal:can:arxiv:2017} to address the issue of {\em creativity}
by extending a generative adversarial networks\index{Generative!adversarial networks} (GAN\index{GAN}) architecture
into a creative adversarial networks\index{Creative!adversarial networks} (CAN\index{CAN}) architecture
to ``generate art\index{Art} by learning about styles\index{Style} and deviating from style norms\index{Norm}.'' \cite{elgammal:can:arxiv:2017}.

Their assumption is that in a standard GAN architecture,
the generator\index{Objective} objective is to generate images that fool the discriminator\index{Discriminator}
and, as a consequence, the generator is trained to be {\em emulative\index{Emulative}} but not {\em creative\index{Creative}}.
In the proposed creative adversarial networks (CAN) (illustrated in Figure~\ref{figure:can:architecture}),
the generator\index{Generator} receives from the discriminator\index{Discriminator} not just one but {\em two} signals:

\begin{figure}
\includegraphics[width=\textwidth]{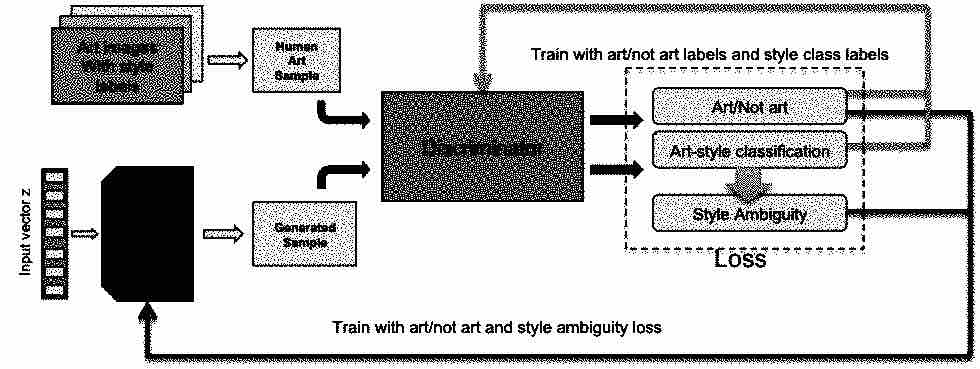}
\caption{Creative adversarial networks (CAN) architecture.
Reproduced from \cite{elgammal:can:arxiv:2017} with permission of the authors}
\label{figure:can:architecture}
\end{figure}
\begin{itemize}

\item the first signal is analog to the case of the standard GAN
(see Equation~\ref{equation:minimax:gan})
and is the discriminator's estimation whether the generated sample is real or faked art; and

\item the second signal is about how easily the discriminator can {\em classify} the generated sample into
predefined {\em established styles}.
If the generated sample is {\em style-ambiguous} (i.e. the various classes are {\em equiprobable\index{Equiprobable}}),
this means that the sample is difficult to fit within the existing art styles,
which may be interpreted as the creation of a {\em new style}.

\end{itemize}

These two signals are contradictory forces which push the generator to explore the space for generating items
that are
close to the distribution of existing art pieces {\em and} style-ambiguous.

\begin{figure}
\includegraphics[width=\textwidth]{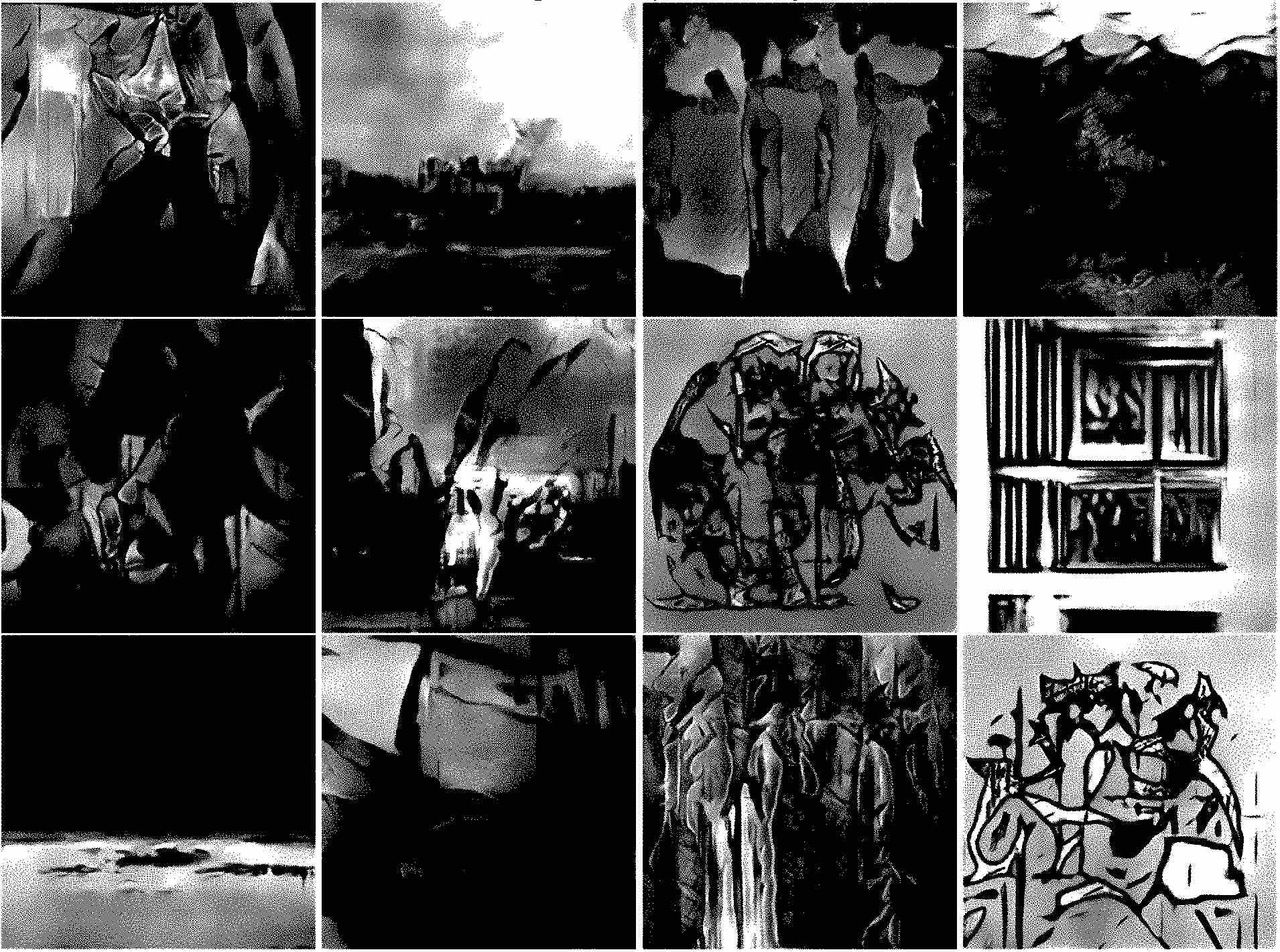}
\caption{Examples of images generated by CAN.
Reproduced from \cite{elgammal:can:arxiv:2017} with permission of the authors}
\label{figure:can:examples}
\end{figure}

Experiments have been done with paintings from a WikiArt dataset \cite{wikiart:web:2017}.
This collection has images of 81,449 paintings from 1,119 artists
ranging from the fifteenth century to the twentieth century.
It has been tagged\index{Tag} with 25 possible painting styles
(e.g., cubism, fauvism, high-renaissance, impressionism, pop-art, realism, etc.).
Some examples of images generated by CAN are shown in Figure~\ref{figure:can:examples}.

As the authors discuss, the generated images are not recognized like traditional art, in terms of standard
genres (portraits, landscapes, religious paintings, still lifes, etc.),
as shown by a preliminary external human evaluation
and also a preliminary analysis of their approach.
Note that the CAN approach assumes the existence of a prior style classification
and reduces the idea of creativity to exploring new styles
(which indeed has some grounding in the art history).
The necessary prior classification between different styles
does have an important role and it will be interesting to experiment
with other types of classification, including styles which are automatically constructed.
Experimenting with the transposition of the CAN approach to music generation appears as a tempting direction.

Note that, as opposed to most of techniques applying control constraints during the {\em generation} phase while leaving the training phase
untouched (see, e.g., style imposition in C-RBM system in Section~\ref{section:experiment:c:rbm}),
in the CAN approach, the incentive for creativity is applied during the {\em training} phase.

The important issue of originality (and creativity) will be further discussed in Section~\ref{section:discussion:evaluation}.


\section{Incrementality}
\label{section:challenges:strategies:incrementality}
\label{section:incrementality}

A straightforward use of deep architectures for generation leads to a one-shot generation of a musical content as a whole
in the case of a feedforward or autoencoder network architecture,
or to an iterative generation of time slices of a musical content in the case of a recurrent network architecture.
This is a strong limitation if we compare this to the way a human composer creates and generates music, in most cases
very incrementally, though successive refinements of arbitrary parts.

\subsection{Note Instantiation Strategies}
\label{section:challenges:strategies:incrementality:strategies}

Let us review how notes are instantiated during generation.
There are three main strategies:

\begin{itemize}

\item
{\em Single-step feedforward\index{Single!-step feedforward strategy}} --
a feedforward architecture processes in a single processing step
a global representation which includes all time steps.
An example is MiniBach\index{MiniBach}
(Section~\ref{section:experiment:mini:bach}).


\item
{\em Iterative feedforward\index{Iterative feedforward strategy}} --
a recurrent architecture iteratively processes
a local representation corresponding to a single time step.
An example is CONCERT\index{CONCERT}
(Section~\ref{section:experiment:concert}).


\item
{\em Incremental sampling\index{Sampling!strategy}} --
a feedforward architecture incrementally processes
a global representation which includes all time steps,
by incrementally instantiating its variables (each variable corresponding to the possibility of a note at a specific time step).
An example is DeepBach\index{DeepBach}
(Section~\ref{section:experiment:deep:bach}).

\end{itemize}

\begin{figure}
\includegraphics[width=\textwidth]{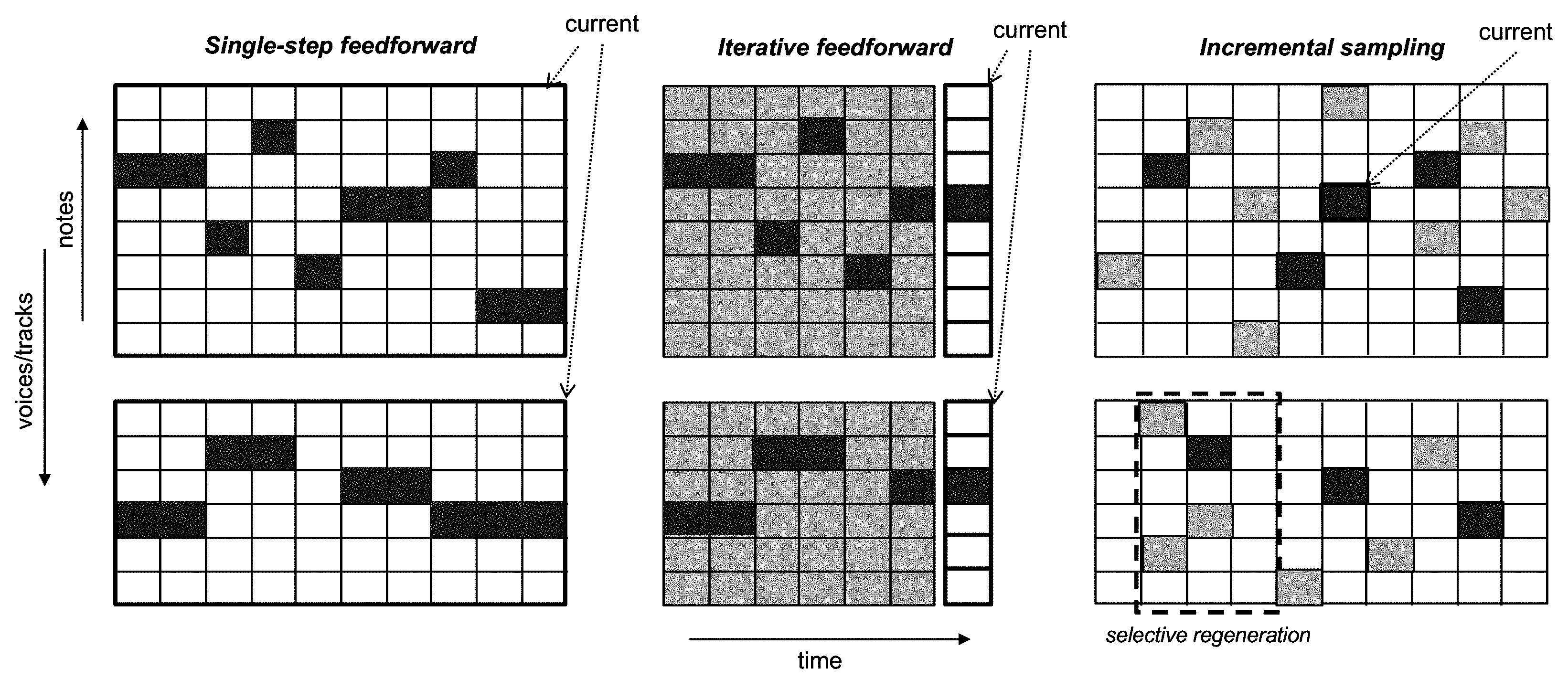}
\caption{Note generation/instantiation -- three main strategies}
\label{figure:generation:strategies}
\end{figure}

These three strategies are compared and illustrated in Figure~\ref{figure:generation:strategies}.
The representation is piano roll type with two simultaneous voices (or tracks).
The cells in blue are the notes to be played.
The rectangles with a thick line labeled as ``current'' indicate the parts being processed,
whereas the parts in light grey indicate the parts already processed.

In the case of the incremental sampling strategy (right part of Figure~\ref{figure:generation:strategies}),
at each processing step
a new cell representing a triplet $(voice, note, time\,step)$,
labeled as ``current'',
is randomly chosen and instantiated.
Triplets already instantiated are blue-filled if a note is to be played and light grey-filled otherwise.

Note that, with this incremental sampling strategy,
it is possible to only generate or to {\em regenerate} an arbitrary part (slice) of the musical content,
for a specific time interval between two time steps
and/or for a specific subset of voices/tracks,
without the need for regenerating the whole content.
In Figure~\ref{figure:generation:strategies}, the dashed rectangle indicates a zone selected by the user
to perform a selective regeneration\footnote{With the single-step feedforward strategy,
	one could imagine selecting only the desired slice from the regenerated content
	and ``copy/pasting'' it into the previously generated content,
	but with the obvious absence of a guarantee that the old and the new parts will be consistent.}.



\subsection{Example: DeepBach Chorale Multivoice Symbolic Music Generation System}
\label{section:experiment:deep:bach}

\begin{figure}
\includegraphics[scale=0.28]{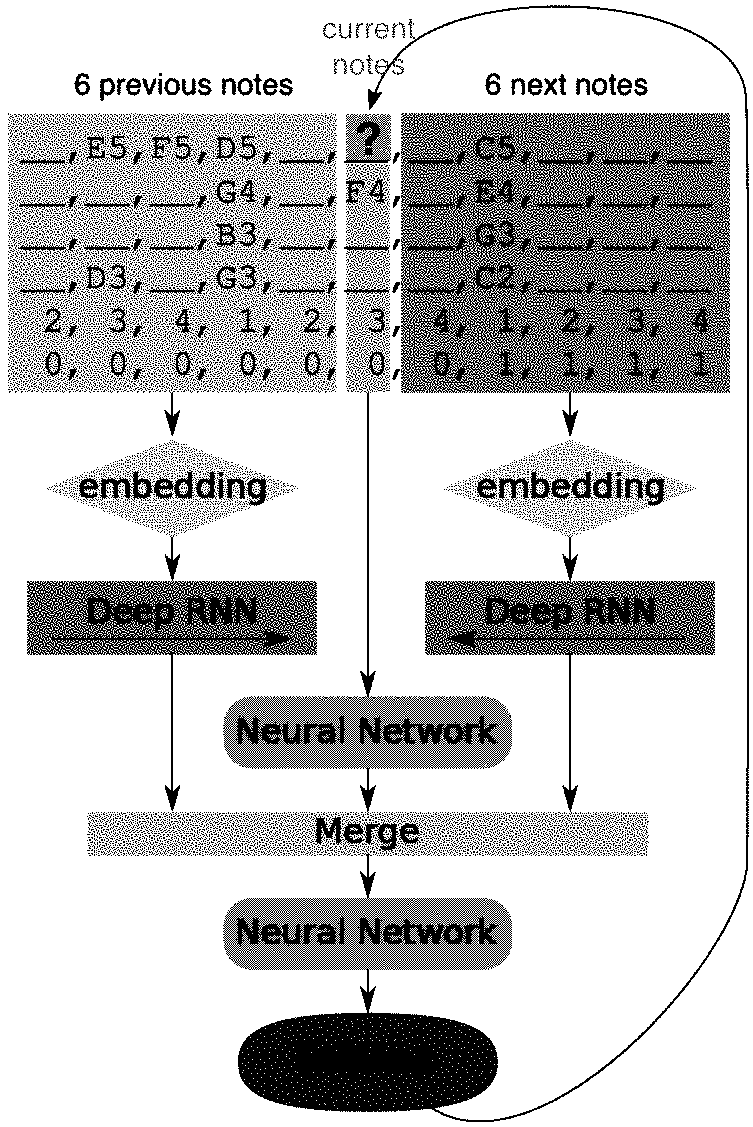}
\caption{DeepBach architecture.
Reproduced from \cite{hadjeres:deep:bach:arxiv:2017} with permission of the authors}
\label{figure:deep:bach:architecture:sampling}
\end{figure}

Hadjeres {\em et al.} have proposed the DeepBach\index{DeepBach}
architecture\footnote{The MiniBach\index{MiniBach} architecture
	described in Section~\ref{section:experiment:mini:bach}
	is actually a deterministic single-step feedforward (major) simplification of the DeepBach architecture.}
for the generation of J. S. Bach\index{Bach} chorales\index{Chorale} \cite{hadjeres:deep:bach:arxiv:2017}.
The architecture, shown in
Figure~\ref{figure:deep:bach:architecture:sampling},
combines two recurrent networks
(LSTMs)
and two feedforward networks.
As opposed to the standard use of recurrent networks where a single time direction is considered\footnote{An exception
	is, for example, some bidirectional recurrent architecture\index{Bidirectional!recurrent neural network}
	used in the BLSTM\index{BLSTM} and in the C-RNN-GAN\index{C-RNN-GAN} systems
	analyzed, respectively,
	in Sections~\ref{section:experiment:blstm:chord} and~\ref{section:systems:c:rnn:gan}.},
DeepBach architecture considers two directions:
{\em forward\index{Forward}} in time and {\em backward\index{Backward}} in time\footnote{The authors state that
	this architectural choice somewhat matches the real compositional practice of Bach chorales.
	Indeed, when reharmonizing\index{Harmonization} a given melody, it is often simpler to start from the cadence\index{Cadence}
	and write music {\em backward\index{Backward}} \cite{hadjeres:deep:bach:arxiv:2017}.}.
Therefore, two recurrent networks (more precisely, LSTMs with 200 cells)
are used,
one summing up past information and another summing up information coming from the future,
together with a nonrecurrent network in charge of notes occurring at the same time.
Their three outputs are merged and passed to the input of a final feedforward neural network,
with one hidden layer with 200 units.
The final output activation function is softmax.

The initial corpus is the set of J. S. Bach's\index{Bach} polyphonic (multivoice) chorales\index{Chorale}
\cite{bach:chorales:book},
where the composer chose various given melodies for a soprano and composed the three additional ones (for alto, tenor and bass\index{Bass})
in a {\em counterpoint\index{Counterpoint}} manner.
The initial dataset (352 chorales)
is augmented by adding all chorale transpositions which fit within the vocal ranges defined by the initial corpus.
This leads to a total corpus of 2,503 chorales.
The vocal ranges contain up to 28 different pitches for each voice\footnote{21 for the soprano, alto and tenor parts
	and 28 for the bass part.}.

The choice of the representation in DeepBach has some specificities.
A hold symbol ``\_\_'' is used to indicate whether a note is being held (see Section~\ref{section:representation:note:ending}).
The authors emphasize in \cite{hadjeres:deep:bach:arxiv:2017} that this representation is well-suited to the sampling method used,
more precisely that the fact that they obtain good results using Gibbs sampling relies exclusively on their choice to integrate
the hold symbol into the list of notes.
Another specificity is that the representation consists in encoding notes using their real names
and not their MIDI note numbers (e.g., F$\sharp$ is considered separately from G$\flat$,
see Section~\ref{section:note:encoding}).
Last, the fermata\index{Fermata} symbol for Bach chorales is explicitly considered as it helps to produce structure and coherent phrases.

The first four lines of the example data at top of Figure~\ref{figure:deep:bach:architecture:sampling} correspond to the four voices.
The two bottom lines correspond to metadata\index{Metadata} (fermata and beat information).
Actually this architecture is replicated four times, one for each voice\index{Voice} (four in a chorale).



Training, as well as generation, is not done in the conventional way for neural networks.
The objective is to predict the value of the current note for a given voice
(shown
in light green
with a
red
``?'', at top center of
Figure~\ref{figure:deep:bach:architecture:sampling}),
using as input information the surrounding contextual notes and their associated metadata,
more precisely

\begin{itemize}

\item the three current notes for the three other voices (the thin rectangle
in light blue
in top center);

\item the six previous notes (the rectangle
in light turquoise blue
in top left) for all voices; and

\item the six next notes (the rectangle
in light grey blue
in top right) for all voices.

\end{itemize}

The training set is formed on-line by repeatedly randomly selecting a note in a voice from an example of the corpus
and its surrounding context (as previously defined).


Generation is performed
by incremental sampling\index{Sampling!strategy},
using a pseudo-Gibbs sampling\index{Pseudo-Gibbs sampling} algorithm
analog to but computationally simpler than Gibbs sampling\index{Gibbs sampling} algorithm\footnote{The difference
	with Gibbs sampling 
	(based on the non-assumption of
	compatibility of conditional probability\index{Conditional!probability} distributions)
	and the algorithm are detailed and discussed in \cite{hadjeres:deep:bach:arxiv:2017}.}
(see Section~\ref{section:sampling:basics}),
to produce a set of values (each note) of a polyphony, following the distribution that the network has learnt.
The algorithm for generation by incremental sampling is shown in Figure~\ref{algorithm:sampling:deep:bach}
and has been illustrated in Figure~\ref{figure:generation:strategies}.


\begin{figure}
\begin{framed}
Create four lists V = (V$_1$; V$_2$; V$_3$; V$_4$) of length $L$;\\
Initialize them with random notes drawn from
the ranges of the corresponding voices\\
(sampled uniformly or from the marginal distributions of the notes);\\
{\bf for} $m$ from $1$ to $max\,number\,of\,iterations$ {\bf do}\\
Choose voice $i$ uniformly between $1$ and $4$;\\
Choose time $t$ uniformly between $1$ and $L$;\\
Re-sample V$_i^t$ from $P_i(\text{V}_i^t | \text{V}_{\symbol{92} i, t}, \theta_i)$\\
{\bf end for}
\end{framed}
\caption{DeepBach incremental generation/sampling algorithm}
\label{algorithm:sampling:deep:bach}
\end{figure}

\begin{figure}
\includegraphics[width=\textwidth]{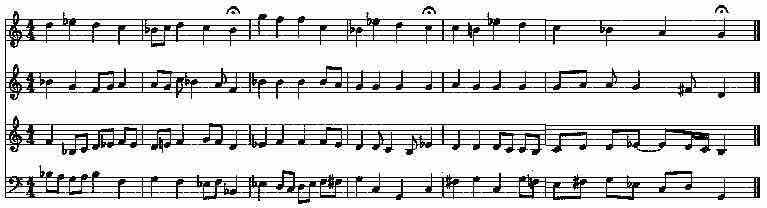}
\caption{Example of a chorale generated by DeepBach.
Reproduced from \cite{hadjeres:deep:bach:arxiv:2017} with permission of the authors}
\label{figure:deep:bach:example}
\end{figure}

An example of a chorale generated\footnote{We will see in Section~\ref{section:interactivity:deep:bach}
	that DeepBach may also be used for a different objective: counterpoint accompaniment.}
is shown in Figure~\ref{figure:deep:bach:example}.
As opposed to many experiments, a systematic evaluation\index{Evaluation} in a Turing-type test\index{Turing!test} has been conducted
(with more than 1,200 human subjects, from experts to novices, via a questionnaire on the Web\index{Web}\footnote{An evaluation was also
	conducted during a live program on a Dutch TV channel.})
and the results are very positive, showing a significant difficulty to discriminate between chorales composed by Bach
and chorales generated by DeepBach.
DeepBach is summarized in Table~\ref{table:dimensions:deep:bach}.


\begin{table}
\begin{tabular}{|l|l|}
\hline
{\em Objective}			&Multivoice; Counterpoint; Chorale; Bach\\
\hline
{\em Representation}	&Symbolic; Piano roll; Multi$^2$-one-hot; Hold; Rest; Fermata\\
\hline
{\em Architecture}		&Feedforward$\times$2 + LSTM$\times$2\\
\hline
{\em Strategy}			&Sampling\\
\hline
\end{tabular}
\caption{DeepBach summary}
\label{table:dimensions:deep:bach}
\end{table}

\section{Interactivity}
\label{section:challenges:strategies:interactivity}
\label{section:interactivity}


An important issue is that, for most current systems, generation of musical content is an automated and autonomous process.
Some {\em interactivity} with a human user(s) is fundamental to obtaining a companion system
to help humans in their musical tasks (composition, counterpoint, harmonization, analysis, arranging, etc.) in an incremental and interactive manner.
An example, already introduced in Section~\ref{section:systems:flow:composer},
is the FlowComposer prototype \cite{papadopoulos:flow:composer:cp:2016}.

A couple of examples of partially interactive incremental systems based on deep network architectures are
deepAutoController\index{deepAutoController}
(Section~\ref{section:experiment:deep:auto:controller})
and DeepBach\index{DeepBach} (Section~\ref{section:experiment:deep:bach}).

\subsection{\#1 Example: deepAutoController Audio Music Generation System}
\label{section:interactivity:deep:auto:controller}

\begin{figure}
\includegraphics[scale=0.25]{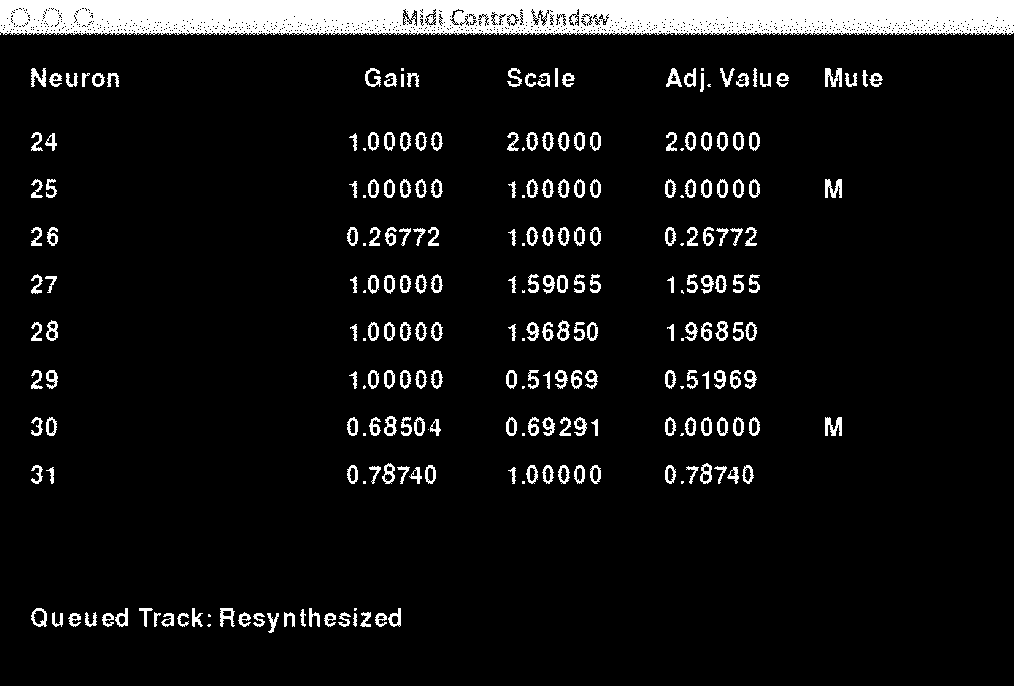}
\caption{Snapshot of a deepAutoController information window showing hidden units.
Reproduced from \cite{sarroff:audio:synthesis:2014} with permission of the authors}
\label{figure:deep:auto:controller:snapshot}
\end{figure}

The deepAutoController system
\cite{sarroff:audio:synthesis:2014} introduced in Section~\ref{section:experiment:deep:auto:controller}
provides a user interface\index{User!interface},
shown in Figure~\ref{figure:deep:auto:controller:snapshot},
to interactively\index{Interactive} control\index{Control} the generation,
for instance by

\begin{itemize}

\item selecting a given input,

\item generating a random input to be feedforwarded into the decoder stack, or

\item controlling (by scaling or muting) the activation of a given unit.

\end{itemize}

\subsection{\#2 Example: DeepBach Chorale Symbolic Music Generation System}
\label{section:interactivity:deep:bach}

\begin{figure}

\includegraphics[width=\textwidth]{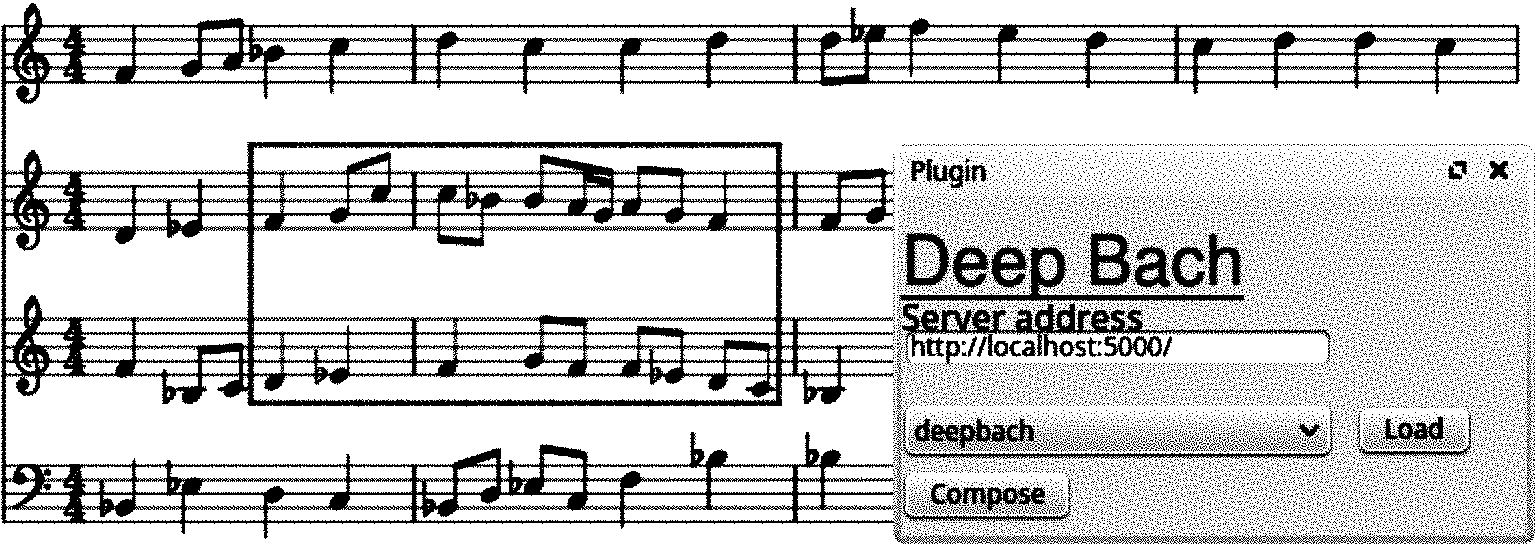}
\caption{DeepBach user interface.
Reproduced from \cite{hadjeres:deep:bach:arxiv:2017} with permission of the authors}
\label{figure:deep:bach:user:interface}
\end{figure}

The user interface of DeepBach \cite{hadjeres:deep:bach:arxiv:2017} (see Section~\ref{section:experiment:deep:bach})
is implemented as a plugin for the MuseScore music editor (see Figure~\ref{figure:deep:bach:user:interface}).
It helps the human user to interactively select and control partial regeneration of chorales.
This is made possible by the incremental nature of the generation (see Section~\ref{section:incrementality}).
Moreover, the user can enforce some user-defined constraints, such as

\begin{itemize}

\item freezing a voice (e.g., the soprano) and resampling the other voices
	in order to reharmonize the fixed melody\footnote{In practice,
		this means changing from the original objective of generating a {\em 4-voice polyphony} from scratch
		as discussed in Section~\ref{section:experiment:deep:bach},
		to generating a {\em 3-voice counterpoint accompaniment} for a given melody.};

\item modifying the fermata list in order to impose an end to musical phrases at specific places;

\item restricting the note range for a given voice and a given temporal interval; and

\item imposing a rhythm by restricting the note range to the hold symbol (as it is considered as a note) in specific parts.

\end{itemize}

\subsection{Interface Definition}
\label{section:challenges:strategies:interactivity:interface}

Let us finally mention,
at the junction between {\em control} and {\em interactivity},
the interesting discussion by Morris {\em et al.} in \cite{morris:exposing:parameters:aaai:08}
on the issue of what control parameters
(for music generation by a Markov chain trained model) should be {\em exposed} at the human user level.
Some examples
of user-level control parameters
they have experimented with are as follows

\begin{itemize}

\item major vs minor;

\item following melody vs following chords; and

\item locking a feature (e.g., a chord).

\end{itemize}

\section{Adaptability}
\label{section:challenges:strategies:adaptability}
\label{section:adaptability}

One fundamental limitation of current deep learning architectures for the generation of musical content is that
they paradoxically do {\em not} learn or adapt.
Learning\index{Learning} is applied during the {\em training} phase of the network,
but no learning or adaptation\index{Adaptation} occurs during the {\em generation\index{Generation}} phase.
However, one can imagine some feedback from a user,
e.g., the composer, producer, listener,
about the quality and the adequacy of the generated music.
This feedback may be explicit, which puts a task on the user,
but it could also be, at least partly, implicit and automated.
For instance, the fact that the user quickly stops listening to the music just generated
could be interpreted as negative feedback\index{Feedback}.
On the contrary, the fact that the user selects a better rendering after a first quick listen to some initial reproduction
could be interpreted as positive feedback.

Several approaches are possible.
The most straightforward approach, considering the nature of neural networks and supervised learning,
would be to add the newly generated musical piece to the training set\index{Training!set}
and eventually\footnote{Immediately,
	after some time,
	or after some amount of new feedback,
	as with a minibatch\index{Minibatch} (see Section~\ref{section:training:algorithm}).}
retrain the network\footnote{This could be done
	in the background.}.
This would reinforce the number of positive examples and gradually update the learnt model and,
as a consequence,
future generations.
However, there is no guarantee that the overall generation quality would improve.
This could also lead the model to overfit\index{Overfitting} and loose some generalization\index{Generalization}.
Moreover, there is no direct possibility of negative feedback,
as one cannot remove a badly generated example from the dataset\index{Dataset}
because there is almost no chance that it was already present in the dataset.

At the junction between {\em adaptability} and {\em interactivity},
an interesting approach is that of interactive machine learning for music generation, as discussed by Fiebrink and Caramiaux \cite{fiebrink:ml:creative:tool:arxiv:2016}.
They report on experience with a toolkit they designed, named Wekinator, to allow users to interactively modify the training examples.
For instance, they argue in \cite{fiebrink:ml:creative:tool:arxiv:2016} that:
``Interactive machine learning can also allow people to build accurate models from very few training examples:
by iteratively placing new training examples in areas of the input space that are most needed to improve model accuracy
(e.g., near the desired decision boundaries between classes), users can allow complicated concepts to be learned more efficiently
than if all training data were representative of future data.''

Another approach is to work not on the training\index{Training} dataset but on the generation\index{Generation} phase.
This leads us back to the issue of control (see Section~\ref{section:challenges:strategies:control}),
via, for example, a constrained sampling strategy, an input manipulation strategy
or, obviously, a reinforcement strategy.
%
The RL-Tuner\index{RL-Tuner} framework (Section~\ref{section:systems:rl-tuner})
is an interesting step in this direction.
Although the initial motivation for RL-Tuner was to introduce musical constraints on the generation,
by encapsulating them into an additional reward\index{Reward},
this approach could also be used to introduce user feedback as an additional reward.

\section{Explainability}
\label{section:challenges:strategies:explainability}


A common critique of sub-symbolic\index{Sub-symbolic} approaches
of Artificial Intelligence\index{Artificial!intelligence} (AI\index{AI})\footnote{As opposed to
	symbolic approaches, see Section~\ref{section:introduction:motivation:symbolic:versus}.},
such as neural networks\index{Neural!network} and deep learning\index{Deep!learning},
is their {\em black box\index{Black box}} nature,
which makes it difficult to explain and justify their decisions
\cite{castelvecchi:black:box:ai:nature:2016}.
Explainability\index{Explainability} is indeed a real issue,
as we would like to be able to understand and explain what (and how) a deep learning system has learned from a corpus
as well as why it ends up generating a given musical content.

\subsection{\#1 Example: BachBot Chorale Polyphonic Symbolic Music Generation System}
\label{section:explainability:example:bachbot}

\label{section:experiment:bach:bot}

Although preliminary, an interesting study conducted with the BachBot system
concerns the analysis of the specialization of some of the units (neurons) of the network,
through a correlation analysis with some specific motives and progressions.

BachBot\index{BachBot},
by Liang \cite{liang:bach:bot:masters:2016,liang2016bachbot},
is a system designed to generate chorales in the style of J. S. Bach,
an objective shared by DeepBach (Section~\ref{section:experiment:deep:bach}).
All examples from the dataset are aligned onto the same key.
The initial representation is piano roll but it is encoded in text, in a similar way to the Celtic melody generation system
described in Section~\ref{section:experiment:sturm:celtic:lstm}.

One of the specificities of the encoding is the way simultaneous notes are encoded as a sequence of tokens,
with a special delimiter symbol ``{\tt |||}'' indicating the next time frame\index{Time!frame},
with a constant time step of an eighth note.
Actually, a chorale
is considered in BachBot as a single-voice polyphony\index{Single!-voice polyphony}
and not as a multivoice polyphony\index{Multivoice!polyphony},
as for instance in the cases of DeepBach\index{DeepBach}
(Section~\ref{section:experiment:deep:bach})
and MiniBach\index{MiniBach} (Section~\ref{section:experiment:mini:bach}).
Rests are encoded as empty frames.
Notes are ordered in a descending pitch
and are represented by their MIDI note number,
with a boolean indicating if it is tied\index{Tied!note} to a note at the same pitch from previous time step\footnote{This is equivalent
	to a hold ``\_\_'' indication.}.
An example is shown in Figure~\ref{figure:example:score:encoding:bach:bot},
encoding two successive chords:

\begin{itemize}

\item notes B$_3$, G$\sharp_3$, E$_3$ and B$_2$,
corresponding to a E major with a B as the bass (often notated as E/B in jazz)
with the duration of a quarter note, and repeated with a tied note;

\item a fermata\index{Fermata},
notated as ``(.)''; and

\item notes A$_3$, E$_3$, C$_3$ and A$_2$,
corresponding to a A minor with the duration of an eighth note.

\end{itemize}

\begin{figure}
\begin{verbatim}
(59, False)
(56, False)
(52, False)
(47, False)
|||
(59, True)
(56, True)
(52, True)
(47, True)
|||
(.)
(57, False)
(52, False)
(48, False)
(45, False)
|||
\end{verbatim}
\caption{Example of score encoding in BachBot.
Reproduced from \cite{liang:bach:bot:masters:2016}}
\label{figure:example:score:encoding:bach:bot}
\end{figure}

The architecture is a recurrent network (LSTM).
The author used a grid search\index{Grid search} in order to select the optimal setting for hyperparameters\index{Hyperparameter}
of the architecture
(number of layers, number of units, etc.).
The selected architecture has three layers and as the author notes in \cite{liang:bach:bot:masters:2016}:
``Depth matters! Increasing num\_layers can yield up to 9\% lower validation loss.
The best model is 3 layers deep, any further and overfitting occurs.''
Generation is done time step by time step, following the iterative feedforward strategy.

As for DeepBach\index{DeepBach} (Sections~\ref{section:experiment:deep:bach} and~\ref{section:interactivity:deep:bach}),
BachBot may be readapted
from the initial 4-voice multivoice\index{Multivoice} chorale generation objective
to a melody 3-voice counterpoint\index{Counterpoint} accompaniment\index{Accompaniment} objective,
see details in \cite[Section~6.1]{liang:bach:bot:masters:2016}.
However, as opposed to DeepBach architecture and representation which stay unchanged,
in the case of BachBot both the architecture, the representation temporal scope\index{Temporal!scope}
and the strategy have to be {\em structurally changed} from a time step/iterative feedforward generation approach
to a global/single-step feedforward generation approach
(similar to MiniBach\index{MiniBach}).

\begin{figure}
\includegraphics[width=\textwidth]{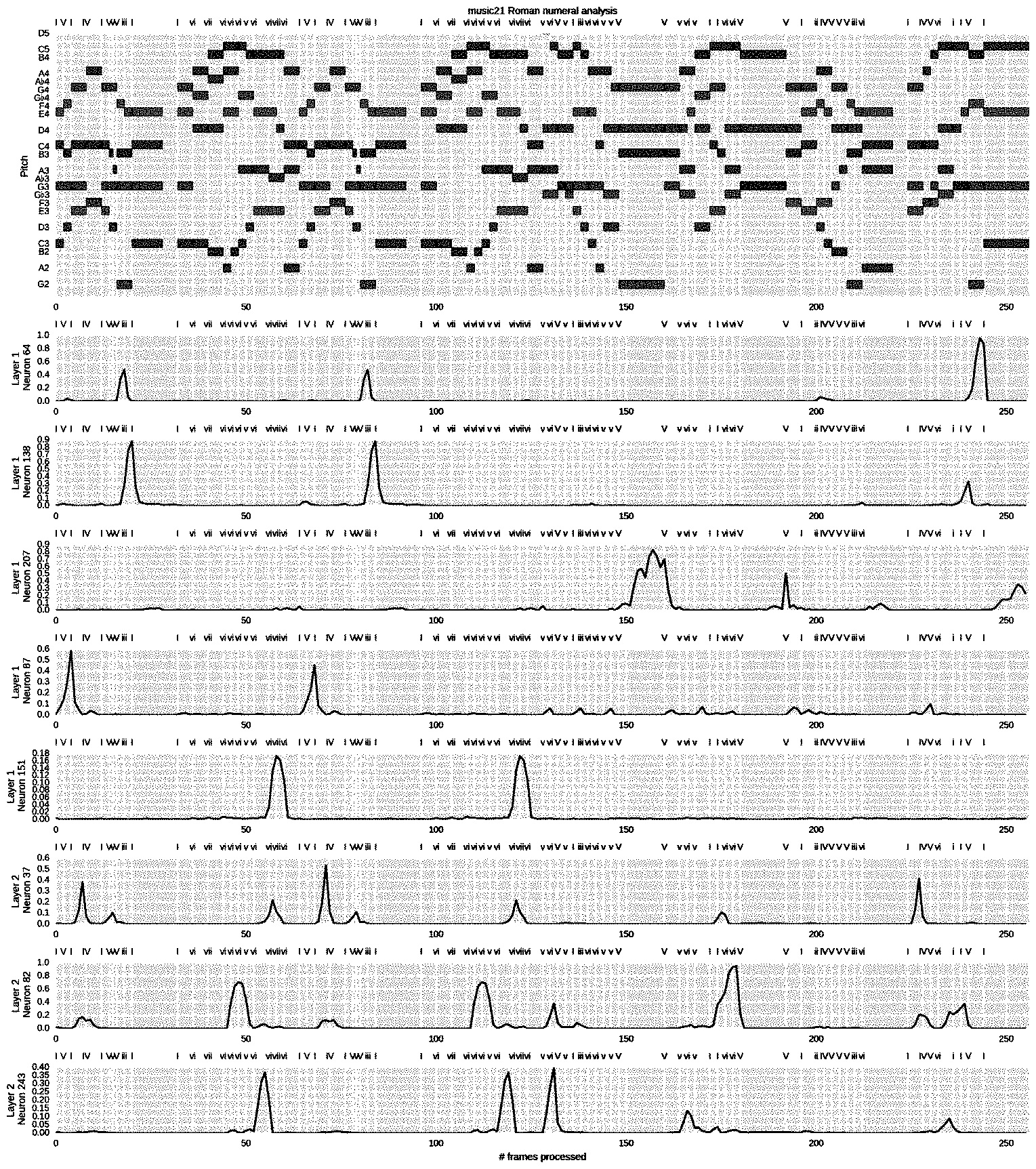}
\caption{Correlation analysis of BachBot layer/unit activation.
Reproduced from \cite{liang:bach:bot:masters:2016}}
\label{figure:analysis:activation:bachbot}
\end{figure}

An interesting preliminary study by the author
was the invitation of a musicologist
to manually search for possible correlations\index{Correlation} between unit activation
and specific motives and progressions,
as shown in Figure~\ref{figure:analysis:activation:bachbot}.
Some examples of the correlations found\footnote{More details
	may be found in \cite[chapter 5]{liang:bach:bot:masters:2016}.}
are as follows:

\begin{itemize}

\item
Neurons 64 and 138 of Layer 1 seem to detect (specifically) perfect cadences (V--I)
with root position chords in the tonic key.


\item
Neuron 87 of Layer 1 seems to detect an I (first degree) chord on the first downbeat and its reprise four measures later.

\item
Neuron 151 of Layer 1 seems to detect A minor
cadences that end phrases 2 and~4.

\item Neuron 37 of Layer 2 seems to be looking for I chords: strong peak for a full I and weaker for other similar chords (same bass).

\end{itemize}

BachBot is summarized in Table~\ref{table:dimensions:bach:bot}.

\begin{table}
\begin{tabular}{|l|l|}
\hline
{\em Objective}			&Multivoice; Chorale; Bach\\
\hline
{\em Representation}	&Symbolic; Text; One-hot; Hold; Fermata\\
\hline
{\em Architecture}		&LSTM$^3$\\
\hline
{\em Strategy}			&Iterative feedforward; Sampling\\
\hline
\end{tabular}
\caption{BachBot summary}
\label{table:dimensions:bach:bot}
\end{table}

\subsection{\#2 Example: deepAutoController Audio Music Generation System}
\label{section:explainability:example:deep:audio:controller}

In \cite{sarroff:audio:synthesis:2014}, the authors of deepAutoController
(Section~\ref{section:experiment:deep:auto:controller})
discuss the musical effects of different controls over the units of the architecture:
``The optimal parameters of the models were mostly inhibitory.
Therefore the deactivation of a unit in a hidden layer yields a denser mixture of sounds at the output.
Learning to play such an interface may prove difficult for new users,
as one typically expects the opposite behavior from a musical synthesizer.
$<$\ldots$>$ We explored models having non-negative weights
by using an asymmetric weight decay as shown in \cite{lemme:non-negative:sparse:autoencoders:neural:networks:2012}.
The results are not presented here as they are preliminary.
Reconstruction error in such models is worse than without non-negativity constraints.
But we find informally that the models are somewhat more intuitive to play as synthesizers.''

\subsection{Towards Automated Analysis}
\label{section:explainability:automated}

The two previous examples in Sections~\ref{section:explainability:example:bachbot}
and~\ref{section:explainability:example:deep:audio:controller}
are examples of a preliminary manual correlation analysis.
Meanwhile, an active area of research relates to the understanding of the way deep learning architectures work
and the explanation of their predictions or decisions via automated analyses.
An example of such an approach is using saliency maps\index{Saliency map} with the three following categories\footnote{Following
	Kindermans {\em et al.}'s study in \cite{kindermans:unreliability:saliency:arxiv:2017},
	actually a critique of the reliability of saliency methods.}:
\begin{itemize}

\item gradient sensitivity,
to estimate how a small change to the input can affect the classification task
(see, for example, \cite{baehrens:explain:jmlr:2010});

\item signal methods,
to isolate input patterns that stimulate neuron activation in higher layers
(see, for example, \cite{kindermans:explain:arxiv:2017}); and

\item attribution methods,
to decompose the value at a specific output neuron into contributions from the individual input dimensions\footnote{With an approach
	analog to reverse correlation,
	which is used in neurophysiology for studying how sensory neurons add up signals from different
	sources and
	sum up stimuli
	at different times to generate a response
	(see, for example, \cite{ringach:reverse:correlation:cognitive:science:2004}).}
(see, for example, \cite{montavon:deep:taylor:pattern:2017}).

\end{itemize}

Note that this type of analysis could also be used with a different objective:
to optimize the configuration of the architecture
by removing components that are considered to make no contribution and are therefore unnecessary,
see, for example, \cite{lecun:brain:damage:nips:1990} (with its provocative title).

Last, let us mention some recent work
targeted for image recognition which are showing interesting direction and prospects,
like for instance
to interactively explore
activation atlases of the features the network has learned\footnote{Using a first processing stage
	inspired by Deep Dream\index{Deep Dream}
	feature visualization by optimization, see Section~\ref{section:systems:deep:dream}.}
\cite{carter:activation:atlas:2019}
or
to automatically explore incorrect behaviors by generating test counterexamples \cite{pei:deepxplore:arxiv:2017}.

\section{Discussion}
\label{section:challenges:strategies:discussion}


We can observe that the various limitations and challenges that we have analyzed
may be partially dependent on one another and, furthermore, conflicting.
Thus, resolving one may damper another.
For instance, the sampling strategy\index{Sampling!strategy}
used by DeepBach\index{DeepBach} (Section~\ref{section:experiment:deep:bach})
provides incrementality but 
the length of the generated music is fixed,
whereas the iterative feedforward strategy\index{Iterative feedforward strategy}
allows variable and unbounded length but incrementality is only forward in time.



There is probably no general solution
and, as for multicriteria decisions, the selection of architectures and strategies depends on preferences and priorities.
Also, as already noted in Section~\ref{section:architecture:compound:limits},
there is no guarantee that combining a variety of different architectures and/or strategies will make a sound and accurate system.
As for a good cook, the best outcome is not achieved by simply mixing together all the possible ingredients.
Therefore, it is important to continue to deepen our understanding and to explore solutions
as well as their possible articulations and combinations.
We hope that the survey and analysis conducted in this chapter and in the two next chapters provide a contribution to this understanding.


%% file: analysis.tex
\chapter{Analysis}
\label{section:chapter:analysis}

\abstract*{Chapter~\ref{section:chapter:analysis} Analysis summarizes the analysis of the various deep learning-based music generation systems
considered in this book.
For that purpose, various tables are proposed.
Correlation tables are also introduced, in order to highlight the relations between the dimensions.}

\label{section:analysis}

We now present a preliminary analysis\index{Analysis} and summary\index{Summary} of the various systems surveyed,
following our proposed five dimensions referential\index{Referential}, through various tables.
This provides material for an analysis of the relations\index{Relation} between the different dimensions and the corresponding design decisions.



\section{Referencing and Abbreviations}
\label{section:analysis:referencing}


At first, we reference in Table~\ref{table:analysis:references:systems} the various systems that we have analyzed.
Then, because of space limitations, we introduce abbreviations for the various possible types for each dimension: 

\begin{itemize}

\item {\em objectives} in Table~\ref{table:analysis:abbreviation:objective};

\item {\em representations} in Table~\ref{table:analysis:abbreviation:representation};

\item {\em architectures} in Table~\ref{table:analysis:abbreviation:architecture};

\item {\em challenges} in Table~\ref{table:analysis:abbreviation:challenge}; and

\item {\em strategies} in Table~\ref{table:analysis:abbreviation:strategy}.

\end{itemize}

\begin{table}[htbp]
\begin{tabular}{|l||l|l|l|l|}
\hline
\multicolumn{5}{|c|}{\bf System}\\
\hline
{\em Reference name}	& {\em Original name}	& {\em Authors}		& {\em Reference}		& {\em Section}\\
\hline
\hline
Anticipation-RNN\index{Anticipation-RNN}		&Anticipation-RNN	&G. Hadjeres \& F. Nielsen	&\cite{hadjeres:anticipation:rnn:arxiv:2017}	&\ref{section:systems:anticipation:rnn}\\
\hline
AST\index{AST} (Audio Style &		&D. Ulyanov \& V. Lebedev	&\cite{ulyanov:audio:style:transfer:web:2016}	&\ref{section:musical:style:transfer:timbre:examples}\\
\cline{3-5}
Transfer)	&					&D. Foote {\em et al.}		&\cite{foote:audio:style:transfer:2016}	&\ref{section:musical:style:transfer:timbre:examples}\\
\hline
BachBot\index{BachBot}	&BachBot		&F. Liang		&\cite{liang:bach:bot:masters:2016}		&\ref{section:experiment:bach:bot}\\
\hline
Bi-Axial LSTM\index{Bi-Axial LSTM}		&Bi-Axial LSTM		&D. Johnson		&\cite{johnson:tied:evomusart:2017}	&\ref{section:experiment:biaxial}\\
\hline
BLSTM\index{BLSTM}	&BLSTM		&H. Lim {\em et al.}	&\cite{lim:chord:generation:from:melody:ismir:2017}		&\ref{section:experiment:blstm:chord}\\
\hline
Blues$_C$\index{Blues$_M$}		&		&D. Eck \& J. Schmidh\"uber	&\cite{eck:composition:lstm:2002}	&\ref{section:experiment:eck:blues:lstm:first:experiment}\\
\hline
Blues$_{MC}$\index{Blues$_{MC}$}	&		&D. Eck \& J. Schmidh\"uber	&\cite{eck:composition:lstm:2002}	&\ref{section:experiment:eck:blues:lstm:second:experiment}\\
\hline
Celtic\index{Celtic}		&				&B. Sturm {\em et al.}	&\cite{sturm:celtic:melody:csmc:2016}	&\ref{section:experiment:sturm:celtic:lstm}\\
\hline
CONCERT\index{CONCERT}	&CONCERT	&M. Mozer	&\cite{mozer:composition:prediction:1994}		&\ref{section:experiment:concert}\\
\hline
C-RBM\index{C-RBM}	& C-RBM		&S. Lattner {\em et al.}	&\cite{lattner:structure:polyphonic:generation:jcms:2018}	&\ref{section:experiment:c:rbm}	\\
\hline
C-RNN-GAN\index{C-RNN-GAN}	&C-RNN-GAN		&O. Mogren	&\cite{mogren:c-rnn-gan:arxiv:2016}		&\ref{section:systems:c:rnn:gan}\\
\hline
deepAutoController\index{deepAutoController}	&deepAutoController		&A. Sarroff \& M. Casey	&\cite{sarroff:audio:synthesis:2014}	&\ref{section:experiment:sarroff}\\
\hline
DeepBach	\index{DeepBach}	&DeepBach	&G. Hadjeres {\em et al.}		&\cite{hadjeres:deep:bach:arxiv:2017}	&\ref{section:experiment:deep:bach}\\
\hline
DeepHear$_C$\index{DeepHear$_C$}	&DeepHear	&F. Sun	&\cite{sun:deep:hear}	&\ref{section:experiment:deep:hear:harmonize}	\\
\hline
DeepHear$_M$\index{DeepHear$_M$}	&DeepHear\index{DeepHear}	&F. Sun	&\cite{sun:deep:hear}	&\ref{section:experiment:deep:hear:melody}\\
\hline
DeepJ\index{DeepJ}				&DeepJ		&L.-C. Mao {\em et al.}	&\cite{mao:deepj:arxiv:2018}	&\ref{section:systems:deepj}\\
\hline
GLSR-VAE\index{GLSR-VAE}		&GLSR-VAE	&G. Hadjeres \& F. Nielsen	&\cite{hadjeres:glsr:vae:arxiv:2017}			&\ref{section:experiment:glsr:vae}\\
\hline
Hexahedria\index{Hexahedria}	&	&D. Johnson	&\cite{johnson:web:hexahedria:composing:music:recurrent:neural:2015}	&\ref{section:experiment:hexahedria}\\
\hline
MidiNet\index{MidiNet}	&MidiNet		&L.-C. Yang {\em et al.}	&\cite{yang:midinet:ismir:2017}		&\ref{section:systems:midinet}\\
\hline
MiniBach\index{MiniBach}		&MiniBach	&G. Hadjeres \& al. 	&		&\ref{section:experiment:mini:bach}\\
\hline
MusicVAE\index{MusicVAE}	&MusicVAE	&A. Roberts {\em et al.}	&\cite{roberts:hierarchical:latent:icml:2018}	&\ref{section:system:music:vae}\\
\hline
Performance RNN\index{Performance RNN}		&Performance RNN		&I. Simon \& S. Oore		&\cite{simon:performance:rnn:web:2017}	&\ref{section:experiment:performance:rnn}\\
\hline
RBM$_C$	\index{RBM$_C$}		&	&N. Boulanger-Lewandowski {\em et al.}		&\cite{boulanger:temporal:dependencies:icml:2012}		&\ref{section:experiment:rnn:rbm}\\
\hline
Rhythm\index{Rhythm}	&	&D. Makris {\em et al.}	&\cite{makris:rhythm:composition:2017}		&\ref{section:experiment:makris:rhythm}\\
\hline
RL-Tuner\index{RL-Tuner}	&RL-Tuner		&N. Jaques {\em et al.}	&\cite{jaques:rl:tuner:arxiv:2016}	&\ref{section:experiment:rl-tuner}\\	
\hline
RNN-RBM\index{RNN-RBM}	&RNN-RBM	&N. Boulanger-Lewandowski {\em et al.}		&\cite{boulanger:temporal:dependencies:icml:2012}		&\ref{section:experiment:rnn:rbm}\\
\hline
Sequential\index{Sequential}		&Sequential	&P. Todd	&\cite{todd:connectionist:composition:1989}	&\ref{section:experiment:todd:sequential}\\
\hline
Time-Windowed\index{Time-Windowed}	&Time-Windowed	&P. Todd	&\cite{todd:connectionist:composition:1989}	&\ref{section:experiment:todd:time:windowed}\\
\hline
UnitSelection\index{UnitSelection}	&		&M. Bretan {\em et al.}	&\cite{bretan:unit:selection:iccc:2017}	&\ref{section:experiment:brentan:unit:selection}	\\
\hline
VRAE\index{VRAE}	& VRAE		&O. Fabius \& J. van Amersfoort	&\cite{fabius:vrae:arxiv:2015}		&\ref{section:experiment:vrae}\\
\hline
VRASH\index{VRASH}	&VRASH		&A. Tikhonov  \& I. Yamshchikov		&\cite{tikhonov:generation:vae:history:arxiv:2017}	&\ref{section:experiment:vrash}	\\
\hline
WaveNet\index{WaveNet}		&WaveNet	&A. van der Oord {\em et al.}	&\cite{oord:wavenet:arxiv:2016}	&\ref{section:systems:wavenet}	\\
\hline
\end{tabular}
\caption{Systems referencing}
\label{table:analysis:references:systems}
\end{table}



%
%
%
%
%
%
%


\begin{table}[htbp]
\begin{tabular}{|l|l|}
\hline
{\em Abbreviation}	&\multicolumn{1}{c|}{\bf Objective}\\
\hline
\multicolumn{2}{|l|}{\em Type}\\
\hline
Au		&Audio\\
\hline
Me		&Melody (Single-voice monophonic melody)\\
\hline
Po		&Polyphony (Single-voice polyphony)\\
\hline
CP		&Chord progression (sequence)\\
\hline
MV		&Multivoice (Multivoice/multitrack polyphony)\\
\hline
Dr		&Drums\\
\hline
Co		&Counterpoint accompaniment\\
\hline
CA		&Chord (progression) accompaniment\\
\hline
ST		&Style transfer\\
\hline
\multicolumn{2}{|l|}{\em Destination \& Use}\\
\hline
AR		&Audio reproduction\\
\hline
SP		&Software processing\\
\hline
HI		&Human interpretation\\
\hline
\multicolumn{2}{|l|}{\em Mode}\\
\hline
AG		&Autonomous generation\\
\hline
IG		&Interactive generation\\
\hline
\end{tabular}
\caption{Abbreviations for the types of objective}
\label{table:analysis:abbreviation:objective}
\end{table}

\begin{table}[htbp]
\begin{tabular}{|l|l|}
\hline
{\em Abbreviation}	&\multicolumn{1}{c|}{\bf Representation}\\
\hline
\multicolumn{2}{|l|}{\em Audio}\\
\hline
Wa		&Waveform\\
\hline
Sp		&Spectrum\\
\hline
\multicolumn{2}{|l|}{\em Symbolic}\\
\hline
\multicolumn{2}{|l|}{\em Concept}\\
\hline
No		&Note\\
\hline
Re		&Rest\\
\hline
Ch		&Chord\\
\hline
Rh		&Rhythm (Meter \& Beats)\\
\hline
Dr		&Drums\\
\hline
\multicolumn{2}{|l|}{\em Format}\\
\hline
MI		&MIDI\\
\hline
Pi		&Piano roll\\
\hline
Te		&Text\\
\hline
\multicolumn{2}{|l|}{\em Temporal Scope}\\
\hline
Gl		&Global\\
\hline
TS		&Time step\\
\hline
\multicolumn{2}{|l|}{\em Meta-Data}\\
\hline
NH		&Note hold or ending\\
\hline
NE		&No enharmony (Note denotation)\\
\hline
Fe		&Fermata\\
\hline
FE		&Feature extraction\\
\hline
\multicolumn{2}{|l|}{\em Expressiveness}\\
\hline
To		&Tempo\\
\hline
Dy		&Dynamics\\
\hline
\multicolumn{2}{|l|}{\em Encoding}\\
\hline
VE		&Value encoding\\
\hline
OH		&One-hot encoding\\
\hline
MH		&Many-hot encoding\\
\hline
\multicolumn{2}{|l|}{\em Dataset}\\
\hline
Tr		&Transposition\\
\hline
Al		&Alignment\\
\hline
\end{tabular}
\caption{Abbreviations for the types of representation}
\label{table:analysis:abbreviation:representation}
\end{table}

\begin{table}[htbp]
\begin{tabular}{|l|l|}
\hline
{\em Abbreviation}	&\multicolumn{1}{c|}{\bf Architecture}\\
\hline
Fd		&Feedforward\\
\hline
Ae		&Autoencoder\\
\hline
Va		&Variational\\
\hline
RB		&Restricted Boltzmann machine (RBM)\\
\hline
RN		&Recurrent (RNN)\\
\hline
Cv		&Convolutional\\
\hline
Cn		&Conditioning\\
\hline
GA		&Generative adversarial networks (GAN)\\
\hline
RL		&Reinforcement learning (RL)\\
\hline
Cp		&Compound\\
\hline
\end{tabular}
\caption{Abbreviations for the types of architecture}
\label{table:analysis:abbreviation:architecture}
\end{table}

\begin{table}[htbp]
\begin{tabular}{|l|l|}
\hline
{\em Abbreviation}	&\multicolumn{1}{c|}{\bf Challenge}\\
\hline
EN		&{\em Ex-nihilo} generation\\
\hline
LV		&Length variability\\
\hline
CV		&Content variability\\
\hline
Es		&Expressiveness\\
\hline
MH		&Melody-harmony consistency\\
\hline
Co		&Control\\
\hline
St		&Structure\\
\hline
Or		&Originality\\
\hline
Ic		&Incrementality\\
\hline
It		&Interactivity\\
\hline
Ad		&Adaptability\\
\hline
Ey		&Explainability\\
\hline
\end{tabular}
\caption{Abbreviations for the types of challenge}
\label{table:analysis:abbreviation:challenge}
\end{table}

\begin{table}[htbp]
\begin{tabular}{|l|l|}
\hline
{\em Abbreviation}	&\multicolumn{1}{c|}{\bf Strategy}\\
\hline
SF		&Single-step feedforward\\
\hline
DF		&Decoder feedforward\\
\hline
Sa		&Sampling\\
\hline
IF		&Iterative feedforward\\
\hline
IM		&Input manipulation\\
\hline
Re		&Reinforcement\\
\hline
US		&Unit selection\\
\hline
Cp		&Compound\\
\hline
\end{tabular}
\caption{Abbreviations for the types of strategy}
\label{table:analysis:abbreviation:strategy}
\end{table}

\clearpage

\section{System Analysis}
\label{section:analysis:tables:system}


We summarize in Tables\footnote{This table is split in two
	because of vertical space limitations.}~\ref{table:analysis:summary:1}
and~\ref{table:analysis:summary:2}
how each system is positioned in respect to each of the following four dimensions:
{\em objective}, {\em representation}, {\em architecture} and {\em strategy}.
We then analyze each system in a more detailed manner, dimension by dimension:

\begin{itemize}

\item {\em objective} in Table~\ref{table:analysis:system:objective};

\item {\em representation} in Tables\footnote{This table is split in two
	because of horizontal space limitations.}~\ref{table:analysis:system:representation:1}
and~\ref{table:analysis:system:representation:2};

\item {\em architecture} and {\em strategy} in Table~\ref{table:analysis:system:architecture:strategy}; and

\item {\em challenge} in Table~\ref{table:analysis:system:challenge}.

\end{itemize}

For each table,
which analyzes each system (line) in respect to the possible types (columns) for a given dimension,
the occurrence of an ``X'' at the crossing of a given line (system) and a given column (type)
means that this system does match that given type for that dimension (e.g., follows some representation facet, is based on some type of architecture,
fulfills some challenge\ldots).
Note that we base this analysis on how each system {\em is} presented in the literature referenced, and not as it could be further extended.

\label{section:analysis:notations}

Furthermore, we use notations\index{Notation convention}
such as X$^n$ and X${\times}n$ (introduced in Section~\ref{section:challenge:strategy:architecture:notation:depth}) to convey additional information
about the number of occurrences of a type.

\begin{table}[htbp]
\begin{tabular}{|l||l|l|l|l|}
\hline
\multicolumn{5}{|c|}{\bf System}\\
\hline
{\em Name}		&{\em Objective}			&{\em Representation}		&{\em Architecture}			&{\em Strategy}	\\
\hline
\hline
Anticipat-			&Melody;					&Symbolic; One-hot;			&Condition-				&Iterative feedforward;\\
ion-RNN			&Bach					&Hold; Rest;				&ing(LSTM$^2$,			&Sampling\\
				&						&No enharmony			&LSTM$^2$)				&\\
\hline
AST				&Audio style				&Audio; Spectrum			&Convolution-				&Input manipulation;\\
				&transfer					&						&al(Feedforward)			&Single-step\\
				&						&						&						&feedforward\\
\hline
BachBot			&Polyphony;		 		&Symbolic; Text;			&LSTM$^3$				&Iterative feedforward;\\
				&Chorale;					&One-hot; Hold;			&						&Sampling\\
				&Bach					&Fermata					&						&\\
\hline
Bi-Axial 			&Polyphony				&Symbolic; Piano roll;		&LSTM$\times$2			&Iterative feedforward;\\
LSTM			&						&Hold; Rest				&						&Sampling\\
\hline			
BLSTM			&Accompaniment;			&Symbolic; CSV;			&LSTM$^2$				&Iterative feedforward\\
				&Chord sequence;			&One-hot					&						&\\
				&Western music			&$\times$(12$\times$* + 24$\times$4); Rest	&			&\\
\hline
Blues$_C$		&Chord sequence; 			&Symbolic; One-hot;			&LSTM					&Iterative feedforward\\
				&Blues					&Note end; Chord as note		&						&\\
\hline
Blues$_{MC}$		&Melody 					&Symbolic; One-hot$\times$2;	&LSTM					&Iterative feedforward\\
				&+ Chords; Blues 			&Note end; Chord as note		&						&\\
\hline
Celtic			&Melody					&Symbolic; Text;	 		&LSTM$^3$				&Iterative feedforward;\\
				&						&Token-based; 	One-hot		&						&Sampling\\
\hline
CONCERT		&Melody	 				&Symbolic; Harmonics;		&RNN					&Iterative feedforward;\\
				&+ Chords				&Harmony; Beat			&						&Sampling\\
\hline
C-RBM			&Polyphony;				&Symbolic;				&Convolution-				&Input manipulation;\\
				&Style impos-				&Piano-roll; Rest;			&al(RBM)					&Sampling\\
				&ition					&Many-hot; Meter			&						&\\
\hline
C-RNN-			&Polyphony				&Symbolic; MIDI			&GAN(Birectio-				&Iterative feedforward;\\
GAN				&						&						&nal(LSTM$^2$),			&Sampling\\
				&						&						&LSTM$^2$)				&\\
\hline
deepAuto-			&Audio;					&Audio; Spectrum			&Autoencoder$^2$			&Decoder feedforward\\
Controller			&User interface				&						&						&\\
\hline
DeepBach			&Multivoice;				&Symbolic; Piano roll;		&Feedforward$\times$2		&Sampling\\		
				&Counterpoint; 				&Multi$^2$-one-hot; Hold; 	&+ LSTM$\times$2			&\\
				&Chorale; Bach				&Rest; Fermata;			&						&\\
				&						&No enharmony			&						&\\
\hline
DeepHear$_C$		&Melody		 			&Symbolic; Piano roll		&Autoencoder$^4$			&Input manipulation;\\
				&accompaniment			&One-hot$\times$64			&						&Decoder feedforward\\
\hline
DeepHear$_M$	&Melody; 					&Symbolic; Piano roll;		&Autoencoder$^4$			&Decoder feedforward\\
				&Ragtime					&One-hot$\times$64			&						&\\
\hline
DeepJ			&Polyphony;	 			&Symbolic; Piano roll; 		&Condition-				&Iterative feedforward;\\
				&Classical; Style			&Replay matrix; Rest;		&ing(LSTM$^2\times$2,		&Sampling\\
				&						&Style; Dynamics			&Embedding)				&\\
\hline
GLSR-VAE		&Melody;	 				&Symbolic; Piano roll; 		&Variational(Auto-			&Decoder feedforward;\\			
				&Bach					&One-hot; Hold; Rest		&encoder(LSTM,			&Sampling\\
				&						&Fermata;					&LSTM); Geodesic			&\\
				&						&No enharmony			&regularization				&\\
\hline
Hexahedria		&Polyphony				&Symbolic; Piano roll; 		&LSTM$^{2+2}$			&Iterative feedforward;\\
				&						&Hold; Beat				&						&Sampling\\
\hline
\end{tabular}
\caption{Systems summary (1/2)}
\label{table:analysis:summary:1}
\end{table}

\begin{table}[htbp]
\begin{tabular}{|l||l|l|l|l|}
\hline
\multicolumn{5}{|c|}{\bf System}\\
\hline
{\em Name}		&{\em Objective}			&{\em Representation}		&{\em Architecture}			&{\em Strategy}	\\
\hline
\hline
MidiNet			&Melody					&Symbolic; Chords			&GAN(Condition-			&Iterative\\
				&+ Chords; Pop;			&Piano roll; 				&ing(Convolution-			&feedforward;\\
				&Melody\,vs Chor-			&One-hot;					&al(Feedforward, 			&Sampling\\
				&ds following				&Rest					&Convolutional(Feed-		&\\
				&						&						&forward(History,			&\\
				&						&						&Chord sequence))),			&\\
				&						&						&Conditioning(Convol-		&\\
				&						&						&utional(Feedforward),		&\\
				&						&						&History))					&\\
\hline
MiniBach			&Accompaniment; 			&Symbolic; 				&Feedforward$^2$			&Single-step\\
				&Counterpoint;				&Piano roll;				&						&feedforward\\
				&Chorale; Bach				&One-hot$\times$64$\times$(1+3);	&					&\\
				&						&Hold					&						&\\
\hline
MusicVAE			&Melody;					&Symbolic; 				&Variational Auto-			&Iterative feedforward;\\
				&Trio (Melody,				&Drums; 					&encoder(Bidirectional-		&Sampling;\\
				&Bass, Drums)				&Note end; Rest			&LSTM,					&Latent variables\\
				&						&						&Hierarchical$^2$-LSTM)		&manipulation\\
\hline
Perform- 			&Polyphony;				&Symbolic; One-hot; 		&LSTM					&Iterative feedforward;\\
ance-RNN			&Performance				&Time shift;				&						&Sampling\\
				&control					&Dynamics				&						&\\
\hline
RBM$_C$			&Simultaneous 				&Symbolic;				&RBM					&Sampling\\
				&notes (Chord)				&Many-hot				&						&\\
\hline
Rhythm			&Multivoice;				&Symbolic; Beat;			&Conditioning(Feed-			&Iterative feedforward;\\
				&Rhythm; Drums			&Drums; Bass line;			&forward(LSTM$^2$),		&Sampling\\
				&						&Note; Rest; Hold			&Feedforward)				&\\
\hline
RL-Tuner			&Melody					&Symbolic; One-hot;			&LSTM$\times$2 + RL		&Iterative feedforward;\\
				&						&Note off; Rest				&						&Reinforcement\\
\hline
RNN-RBM		&Polyphony				&Symbolic; 				&RBM-RNN				&Iterative feedforward;\\
				&						&Many-hot				&						&Sampling\\
\hline
Sequential		&Melody					&Symbolic; Piano			&RNN					&Iterative feedforward\\
				&						&roll; One-hot; Note			&						&\\
				&						&begin; Implicit rest			&						&\\
\hline
Time-			&Melody					&Symbolic; Piano			&Feedforward				&Iterative feedforward\\
windowed			&						&roll; One-hot$\times$8;		&						&\\
				&						&Note begin;				&						&\\
				&						&Implicit rest				&						&\\
\hline
Unit-				&Melody					&Symbolic; Rest;			&Autoencoder$^2$			&Unit selection;\\
Selection			&						&BOW Features			&+ LSTM$\times$2			&Iterative feedforward\\
\hline
VRAE			&Melody; 					&Symbolic; 				&Variational(Autoen-			&Decoder feedforward;\\
				&Video game				&						&coder(LSTM, LSTM)		&Iterative feedforward;\\
				&songs					&						&						&Sampling\\
\hline
VRASH			&Melody					&Symbolic; MIDI;			&Variational(Autoen-			&Decoder feedforward;\\
				&						&Multi-one-hot				&coder(LSTM$^4$,			&Iterative feedforward;\\
				&						&						&Conditioning(LSTM$^4$,		&Sampling\\
				&						&						&History)))				&\\
\hline
WaveNet			&Audio					&Audio;					&Conditioning(Convol-		&Iterative feedforward;\\
				&						&Waveform				&utional(Feedforward),		&Sampling\\
				&						&						&Tag);					&\\
				&						&						&Dilated convolutions		&\\
\hline
\end{tabular}
\caption{Systems summary (2/2)}
\label{table:analysis:summary:2}
\end{table}

\begin{table}[htbp]
\begin{tabular}{|l||c|c|c|c|c|c|c|c|c||c|c|c||c|c|}
\hline
				&\multicolumn{14}{c|}{\bf Objective}\\
\hline
				&\multicolumn{9}{c||}{Type}
					&\multicolumn{3}{c||}{Dest./Use}			&\multicolumn{2}{c|}{Mode}\\
\hline
				&Au			&Me			&Po			&CP			&MV			&Dr			&Co			&CA			&ST
					&AR			&SP			&HI			&AG			&IG\\
\hline
\hline
\multicolumn{14}{|l|}{\bf System}\\
\hline
\hline
Anticipation-RNN	&			&X			&			&			&			&			&			&			&
					&			&X			&X			&X			&\\
\hline
AST				&X			&			&			&			&			&			&			&			&X
					&X			&			&			&X			&\\
\hline
BachBot			&			&			&X			&			&			&			&			&			&
					&			&X			&X			&X			&\\
\hline
Bi-Axial LSTM		&			&			&X			&			&			&			&			&			&
					&			&X			&X			&X			&\\
\hline
BLSTM			&			&			&			&X			&			&			&			&X			&
					&			&X			&X			&X			&\\
\hline
Blues$_C$		&			&			&			&X			&			&			&			&			&
					&			&X			&X			&X			&\\
\hline
Blues$_{MC}$		&			&X			&			&X			&X			&			&			&			&
					&			&X			&X			&X			&\\
\hline
Celtic			&			&X			&			&			&			&			&			&			&
					&			&X			&X			&X			&\\
\hline
CONCERT		&			&X			&			&X			&X			&			&			&			&
					&			&X			&X			&X			&\\
\hline
C-RBM			&			&			&X			&			&			&			&			&			&X
					&			&X			&X			&X			&\\
\hline
C-RNN-GAN		&			&			&X			&			&			&			&			&			&
					&			&X			&X			&X			&\\
\hline
deepAutoController	&X			&			&			&			&			&			&			&			&
					&X			&			&			&X			&X\\
\hline
DeepBach			&			&X$\times$3	&			&			&X			&			&X			&			&
					&			&X			&X			&X			&X\\
\hline
DeepHear$_C$		&			&X			&			&			&			&			&X			&			&
					&			&X			&X			&X			&\\
\hline
DeepHear$_M$	&			&X			&			&			&			&			&			&			&
					&			&X			&X			&X			&\\
\hline
DeepJ			&			&			&X			&			&			&			&			&			&
					&			&X			&X			&X			&\\
\hline
GLSR-VAE		&			&X			&			&			&			&			&			&			&
					&			&X			&X			&X			&\\
\hline
Hexahedria		&			&			&X			&X			&X			&			&			&			&
					&			&X			&X			&X			&\\
\hline
MidiNet			&			&X			&			&X			&X			&			&			&			&
					&			&X			&X			&X			&\\
\hline
MiniBach			&			&X$\times$3	&			&			&X			&			&X			&			&
					&			&X			&X			&X			&\\
\hline
MusicVAE			&			&X			&			&			&X 			&X			&			&			&
					&			&X			&X			&X			&\\
\hline
Performance RNN	&			&			&X			&			&			&			&			&			&
					&			&X			&X			&X			&\\
\hline
RBM$_C$			&			&			&			&X			&			&			&			&			&
					&			&X			&X			&X			&\\
\hline
Rhythm			&			&			&			&			&X			&X			&			&			&
					&			&X			&X			&X			&\\
\hline
RL-Tuner			&			&X			&			&			&			&			&			&			&
					&			&X			&X			&X			&\\
\hline
RNN-RBM		&			&			&X			&			&			&			&			&			&
					&			&X			&X			&X			&\\
\hline
Sequential		&			&X			&			&			&			&			&			&			&
					&			&X			&X			&X			&\\
\hline
Time-Windowed	&			&X			&			&			&			&			&			&			&
					&			&X			&X			&X			&\\
\hline
UnitSelection		&			&X			&			&			&			&			&			&			&
					&			&X			&X			&X			&\\
\hline
VRAE			&			&X			&			&			&			&			&			&			&
					&			&X			&X			&X			&\\
\hline
VRASH			&			&X			&			&			&			&			&			&			&
					&			&X			&X			&X			&\\
\hline
WaveNet			&X			&			&			&			&			&			&			&			&
					&X			&			&			&X			&\\
\hline
\end{tabular}
\caption{System $\times$ Objective}
\label{table:analysis:system:objective}
\end{table}


\begin{table}[htbp]
\begin{tabular}{|l||c|c||c|c|c|c|c||c|c|c||c|c|}
\hline
				&\multicolumn{12}{c|}{\bf Representation}\\
\hline
				&\multicolumn{2}{c||}{Audio}	&\multicolumn{5}{c||}{Concept}
					&\multicolumn{3}{c||}{Format}				&\multicolumn{2}{c|}{TmpS}\\
\hline
				&Wa			&Sp			&No			&Re			&Ch			&Rh			&Dr
					&MI			&Pi			&Te			&Gl			&TS\\
\hline
\hline
\multicolumn{13}{|l|}{\bf System}\\
\hline
\hline
Anticipation-RNN	&			&			&X			&X			&			&			&
					&			&X			&			&			&X\\
\hline
Audio Style Transfer (AST)
				&			&X			&			&			&			&			&
					&			&			&			&X			&\\
\hline
BachBot			&			&			&X			&X			&			&			&
					&			&			&X			&			&X\\
\hline
Bi-Axial LSTM		&			&			&X			&X			&X			&X			&
					&			&X			&			&			&X\\
\hline
BLSTM			&			&			&X			&X			&X			&			&
					&			&			&			&			&X\\
\hline
Blues$_C$		&			&			&X			&			&X			&			&
					&			&X			&			&			&X\\
\hline
Blues$_{MC}$		&			&			&X			&			&X			&			&
					&			&X			&			&			&X\\
\hline
Celtic			&			&			&X			&			&			&X			&
					&			&			&X			&			&X\\
\hline
CONCERT		&			&			&X			&X			&X			&X			&
					&			&X			&			&			&X\\
\hline
C-RBM			&			&			&X			&X			&			&X			&
					&			&X			&			&X			&\\
\hline
C-RNN-GAN		&			&			&X			&X			&			&			&
					&X			&			&			&			&X\\
\hline
deepAutoController	&			&X			&			&			&			&			&
					&			&			&			&X			&\\
\hline
DeepBach			&			&			&X			&X			&			&			&
					&			&X			&			&X			&\\
\hline
DeepHear$_C$		&			&			&X			&			&			&			&
					&			&X			&			&X			&\\
\hline
DeepHear$_M$	&			&			&X			&			&			&			&
					&			&X			&			&X			&\\
\hline
DeepJ			&			&			&X			&X			&X			&X			&
					&			&X			&			&			&X\\
\hline
GLSR-VAE		&			&			&X			&X			&			&			&
					&			&			&			&			&X\\
\hline
Hexahedria		&			&			&X			&			&X			&X			&
					&			&X			&			&			&X\\
\hline
MidiNet			&			&			&X			&X			&X			&			&
					&			&X			&			&X			&\\
\hline
MiniBach			&			&			&X			&			&			&			&
					&			&X			&			&X			&\\
\hline
MusicVAE			&			&			&X			&X			&			&X			&X
					&X			&			&X			&			&X\\
\hline
Performance RNN	&			&			&X			&X			&			&			&
					&X			&			&			&			&X\\
\hline
RBM$_C$			&			&			&X			&			&X			&			&
					&			&X			&			&X			&\\
\hline
Rhythm			&			&			&X			&X			&			&X			&X
					&			&X			&			&			&X\\
\hline
RL-Tuner			&			&			&X			&X			&			&			&
					&			&X			&			&			&X\\
\hline
RNN-RBM		&			&			&X			&			&X			&			&
					&			&X			&			&			&X\\
\hline
Sequential		&			&			&X			&			&			&			&
					&			&X			&			&			&X\\
\hline
Time-Windowed		&			&			&X			&			&			&			&
					&			&X			&			&			&\\
\hline
UnitSelection		&			&			&X			&X			&			&			&
					&			&X			&			&			&\\
\hline
VRAE			&			&			&X			&			&			&			&
					&			&			&			&X			&X\\
\hline
VRASH			&			&			&X			&			&			&			&
					&			&X			&			&X			&X\\
\hline
WaveNet			&X			&			&			&			&			&			&
					&			&			&			&			&X\\
\hline
\end{tabular}
\caption{System $\times$ Representation (1/2)}
\label{table:analysis:system:representation:1}
\end{table}

\begin{table}[htbp]
\begin{tabular}{|l||c|c|c|c||c|c||c|c|c||c|c|}
\hline
				&\multicolumn{11}{c|}{\bf Representation}\\
\hline
				&\multicolumn{4}{c||}{Meta-data}
						&\multicolumn{2}{c||}{Expr.}	&\multicolumn{3}{c||}{Encoding}			&\multicolumn{2}{c|}{DSet}\\
\hline
				&NH			&NE			&Fe			&FE
						&To			&Dy			&VE			&OH			&MH			&Tr			&Al\\
\hline
\hline
\multicolumn{12}{|l|}{\bf System}\\
\hline
\hline
Anticipation-RNN	&X			&X			&X			&
						&			&			&			&X			&			&X			&\\
\hline
Audio Style Transfer (AST)
				&			&			&			&
						&X			&X			&X			&			&			&			&\\
\hline
BachBot			&X			&			&X			&
						&			&			&			&X			&			&			&X\\
\hline
Bi-Axial LSTM		&X			&			&			&X
						&			&			&			&X			&X			&			&X\\
\hline
BLSTM			&X			&			&			&
						&			&			&	&X$\times$(12$\times$* + 24$\times$4)	&	&		&X\\
\hline
Blues$_C$		&X			&			&			&
						&			&			&			&X			&			&			&\\
\hline
Blues$_{MC}$		&X			&			&			&
						&			&			&			&X$\times$2	&			&			&\\
\hline
Celtic			&			&			&			&
						&			&			&			&X			&			&			&X\\
\hline
CONCERT		&			&			&			&
						&			&			&X			&X			&X			&			&\\
\hline
C-RBM			&X			&			&			&
						&			&			&			&			&X			&X			&\\
\hline
C-RNN-GAN		&X			&			&			&
						&			&			&X$\times$4	&			&			&			&\\
\hline
deepAutoController	&			&			&			&
						&X			&X			&X			&			&			&			&\\
\hline
DeepBach			&X			&X			&X			&
						&			&			&			&X			&			&X			&\\
\hline
DeepHear$_C$		&			&			&			&
						&			&			&			&X			&			&			&\\
\hline
DeepHear$_M$	&			&			&			&
						&			&			&			&X			&			&			&\\
\hline
DeepJ			&X			&			&			&X
						&			&X			&X			&X			&X			&			&\\
\hline
GLSR-VAE		&X			&X			&X			&
						&			&			&			&X			&			&X			&\\
\hline
Hexahedria		&X			&			&			&
						&			&			&X			&X			&X			&			&\\
\hline
MidiNet			&			&			&			&
						&			&			&			&X			&			&X			&\\
\hline
MiniBach			&X			&X			&			&
						&			&			&		&X$\times$64$\times$(1+3)	&	&X			&\\
\hline
MusicVAE			&X			&			&			&
						&			&			&			&X			&			&			&\\
\hline
Performance RNN	&X			&			&			&
						&X			&X			&X			&X$\times$4	&			&X			&\\
\hline
RBM$_C$			&			&			&			&
						&			&			&			&			&X			&			&X\\
\hline
Rhythm			&X			&			&			&
						&			&			&			&			&X			&			&\\
\hline
RL-Tuner			&X			&			&			&
						&			&			&			&X			&			&			&\\
\hline
RNN-RBM		&			&			&			&
						&			&			&			&			&X			&			&X\\
\hline
Sequential		&X			&			&			&
						&			&			&			&X			&			&			&X\\
\hline
Time-Windowed		&X			&			&			&
						&			&			&			&X$\times$8	&			&			&X\\
\hline
UnitSelection		&			&			&			&X
						&			&			&X			&X			&			&X			&\\
\hline
VRAE			&			&			&			&
						&			&			&			&			&			&			&\\
\hline
VRASH			&			&			&			&
						&			&			&			&X$\times$3	&			&			&\\
\hline
WaveNet			&			&			&			&
						&X			&X			&X			&X			&			&			&\\
\hline
\end{tabular}
\caption{System $\times$ Representation (2/2)}
\label{table:analysis:system:representation:2}
\end{table}


\begin{table}[htbp]
\begin{tabular}{|l||c|c|c|c|c|c|c|c|c|c||c|c|c|c|c|c|c|c|}
\hline
	&\multicolumn{10}{c||}{\bf Architecture}
		&\multicolumn{8}{c|}{\bf Strategy}\\
\hline
	&Fd			&Ae			&Va			&RB			&RN			&Cv			&Cn			&GA			&RL			&Cp
		&SF			&DF			&Sa			&IF			&IM			&Re			&US			&Cp\\
\hline
\hline
\multicolumn{19}{|l|}{\bf System}\\
\hline
\hline
Anticipation-RNN
	&			&			&			&			&X$^2\times$2	&			&X			&			&			&X
		&			&			&X			&X			&			&			&			&X\\
\hline
AST
	&X			&			&			&			&			&X			&			&			&			&X
		&X			&			&			&			&X			&			&			&X\\
\hline
BachBot
	&			&			&			&			&X$^3$		&			&			&			&			&
		&			&			&X			&X			&			&			&			&X\\
\hline
Bi-Axial LSTM
	&			&			&			&			&X$^2\times$2	&			&			&			&			&X	
		&			&			&X			&X			&			&			&			&X\\
\hline
BLSTM
	&			&			&			&			&X			&			&			&			&			&				&			&			&			&X			&			&			&			&\\
\hline
Blues$_C$
	&			&			&			&			&X			&			&			&			&			&
		&			&			&			&X			&			&			&			&\\
\hline
Blues$_{MC}$
	&			&			&			&			&X			&			&			&			&			&
		&			&			&			&X			&			&			&			&\\
\hline
Celtic
	&			&			&			&			&X$^3$			&			&		&			&			&
		&			&			&X			&X			&			&			&			&X\\
\hline
CONCERT
	&			&			&			&			&X			&			&			&			&			&
		&			&			&X			&X			&			&			&			&X\\
\hline
C-RBM
	&			&			&			&X			&			&X			&			&			&			&X
		&			&			&X			&			&X			&			&			&X\\
\hline
C-RNN-GAN
	&			&			&			&			&X$^2\times$2	&			&			&X			&			&X
		&			&			&X			&X			&			&			&			&X\\
\hline
deepAutoController
	&			&X$^2$		&			&			&			&			&			&			&			&X
		&			&X			&			&			&			&			&			&\\
\hline
DeepBach	
	&X$\times$2	&			&			&			&X$\times$2	&			&			&			&			&X
		&			&			&X			&			&			&			&			&\\
\hline
DeepHear$_C$	
	&			&X$^4$		&			&			&			&			&			&			&			&X
		&			&X			&			&			&X			&			&			&X\\
\hline
DeepHear$_M$
	&			&X$^4$		&			&			&			&			&			&			&			&X
		&			&X			&			&			&			&			&			&\\
\hline
DeepJ
	&			&			&			&			&X$^2\times$2	&			&X			&			&			&X
		&			&			&X			&X			&			&			&			&X\\
\hline
GLSR-VAE
	&			&X			&X			&			&X$^2$		&			&			&			&			&X
		&			&X			&X			&			&			&			&			&X\\
\hline
Hexahedria
	&			&			&			&			&X$^{2+2}$	&			&			&			&			&X
		&			&			&X			&X			&			&			&			&X\\
\hline
MidiNet
	&X			&			&			&			&			&X			&X			&X			&			&X
		&X			&			&X			&X			&			&			&			&X\\
\hline
MiniBach
	&X$^2$		&			&			&			&			&			&			&			&			&
		&X			&			&			&			&			&			&			&\\
\hline
MusicVAE	
	&			&X			&X			&			&X$\times$2+2 	&		&			&			&			&X
		&			&X			&X			&X			&			&			&			&X\\
\hline
Performance RNN
	&			&			&			&			&X			&			&			&			&			&
		&			&			&X			&X			&			&			&			&X\\
\hline
RBM$_C$	
	&			&			&			&X			&			&			&			&			&			&
		&			&			&X			&			&			&			&			&\\
\hline
Rhythm
	&X			&			&			&			&X$^2$		&			&X			&			&			&X
		&X			&			&X			&X			&			&			&			&X\\
\hline
RL-Tuner
	&			&			&			&			&X$\times$2	&			&			&			&X			&X
		&			&			&X			&X			&			&X			&			&X\\
\hline
RNN-RBM
	&			&			&			&X			&X			&			&			&			&			&X
		&			&			&X			&X			&			&			&			&X\\
\hline
Sequential
	&			&			&			&			&X			&			&			&			&			&
		&			&			&			&X			&			&			&			&\\
\hline
Time-Windower
	&X			&			&			&			&			&			&			&			&			&
		&			&			&			&X			&			&			&			&\\
\hline
UnitSelection
	&			&X$^2$		&			&			&X$\times$2	&			&			&			&			&X
		&			&			&			&X			&			&			&X			&X\\
\hline
VRAE
	&			&X			&X			&			&X$\times$2	&			&			&			&			&X
		&			&X			&X			&X			&			&			&			&X\\
\hline
VRASH
	&			&X			&X			&			&X$^4\times$2	&			&X			&			&			&X
		&			&X			&X			&X			&			&			&			&X\\
\hline
WaveNet
	&X			&			&			&			&			&X			&X			&			&			&X
		&			&			&X			&X			&			&			&			&X\\
\hline
\end{tabular}
\caption{System $\times$ Architecture \& Strategy}
\label{table:analysis:system:architecture:strategy}
\end{table}


\clearpage

\begin{table}[htbp]
\begin{tabular}{|l||c|c|c|c|c|c|c|c|c|c|c|c|}
\hline
				&\multicolumn{12}{c|}{\bf Challenge}\\
\hline
				&EN			&LV			&CV			&Es			&MH			&Co
					&St			&Or			&Ic			&It			&Ad			&Ey\\
\hline
\hline
\multicolumn{13}{|l|}{\bf System}\\
\hline
\hline
Anticipation-RNN	&X			&X			&X			&			&			&X
					&			&			&X			&			&			&\\
\hline
AST				&			&			&			&			&			&X
					&			&			&			&			&			&\\
\hline
BachBot			&X			&X			&X			&			&X			&
					&			&			&X			&			&			&X\\
\hline
Bi-Axial LSTM		&X			&X			&X			&			&X			&
					&			&			&X			&			&			&\\
\hline
BLSTM			&			&X			&			&			&X			&
					&			&			&X			&			&			&\\
\hline
Blues$_C$		&X			&X			&			&			&			&
					&			&			&X			&			&			&\\
\hline
Blues$_{MC}$		&X			&X			&			&			&X			&
					&			&			&X			&			&			&\\
\hline
Celtic			&X			&X			&X			&			&			&
					&			&			&X			&			&			&\\
\hline
CONCERT		&X			&X			&X			&			&			&
					&			&			&X			&			&			&\\
\hline
C-RBM			&X			&			&X			&			&X			&X
					&X			&			&X			&			&			&\\
\hline
C-RNN-GAN		&X			&X			&X			&			&			&
					&			&			&X			&			&			&\\
\hline
deepAutoController	&X			&			&			&			&			&X
					&			&			&			&X			&			&\\
\hline
DeepBach			&			&			&X			&			&X			&
					&			&			&X			&X			&			&\\
\hline
DeepHear$_C$		&X			&			&			&			&X			&
					&			&			&			&			&			&\\
\hline
DeepHear$_M$	&X			&			&			&			&			&
					&			&			&			&			&			&\\
\hline
DeepJ			&X			&X			&X			&X			&			&X
					&			&			&X			&			&			&\\
\hline
GLSR-VAE		&X			&X			&X			&			&			&X
					&			&			&X			&			&			&\\
\hline
Hexahedria		&X			&X			&X			&			&X			&
					&			&			&X			&			&			&\\
\hline
MidiNet			&X			&			&X			&			&X			&X
					&			&X			&			&			&			&\\
\hline
MiniBach			&			&			&			&			&X			&
					&			&			&			&			&			&\\
\hline
MusicVAE			&X			&X			&X			&			&			&X
					&X			&			&X			&			&			&\\
\hline
Performance RNN	&X			&X			&X			&X			&			&X
					&			&			&X			&			&			&\\
\hline
RBM$_C$			&X			&			&X			&			&			&
					&			&			&			&			&			&\\
\hline
Rhythm			&X			&X			&X			&X			&			&X
					&			&			&X			&			&			&\\
\hline
RL-Tuner			&X			&X			&X			&			&			&X
					&X			&			&X			&			&			&\\
\hline
RNN-RBM		&X			&X			&X			&			&X			&
					&			&			&X			&			&			&\\
\hline
Sequential		&X			&X			&			&			&			&
					&			&			&X			&			&			&\\
\hline
Time-Windowed		&X			&X			&			&			&			&
					&			&			&X			&			&			&\\
\hline
UnitSelection		&X			&X			&X			&			&			&X
					&X			&			&X			&			&			&\\
\hline
VRAE			&X			&X			&X			&			&			&
					&			&			&X			&			&			&\\
\hline
VRASH			&X			&X			&X			&			&			&
					&			&			&X			&			&			&\\
\hline
WaveNet			&X			&			&X			&X			&			&X
					&			&			&X			&			&			&\\
\hline
\end{tabular}
\caption{System $\times$ Challenge}
\label{table:analysis:system:challenge}
\end{table}

Note that, when considering the analysis regarding the challenges in Table~\ref{table:analysis:system:challenge},
we have to keep in mind that

\begin{itemize}

\item the limitations and challenges are not of equal importance and difficulty; and

\item the majority of the systems may be further extended in order to better address some of the challenges\footnote{For instance,
	the RL-Tuner system has the potential for addressing interactivity and adaptability challenges,
	although, to our knowledge, not yet experimented.}.

\end{itemize}

That said, we can see the emergence of some divide
between
systems using a global
versus a time step\index{Time!step} temporal scope representation
(see Section~\ref{section:representation:temporal:scope}),
depending on the following requirements:
{\em ex nihilo} generation and length variability.
This will be further discussed in Section~\ref{section:discussion:global:vs:sequence}.


\section{Correlation Analysis}
\label{section:analysis:correlation}

The last series of tables analyse some correlations between the dimensions:

\begin{itemize}

\item {\em representation} with respect to the {\em objective} in Table~\ref{table:analysis:representation:objective};

\item {\em architecture} and {\em strategy} with respect to the {\em objective} in Table~\ref{table:analysis:architecture:strategy:objective};

\item {\em objective}, {\em architecture} and {\em strategy} with respect to the {\em representation} in Table~\ref{table:analysis:architecture:strategy:representation};

\item {\em strategy} with respect to the {\em architecture} in Table~\ref{table:analysis:strategy:architecture}; and

\item {\em architecture} and {\em strategy} with respect to the {\em challenge} in Table~\ref{table:analysis:architecture:strategy:challenge}. 

\end{itemize}

The occurrence of an ``X'' at the crossing
of a given line (a given type for the first dimension)
and a given column (a given type for the second dimension)
means that there is (at least) a system\footnote{Within
	the set of systems analyzed in the book.}
which matches both types
(for instance, is based on a particular architecture and follows a particular strategy).
However, the absence of an ``X''
does not mean that the two types are incompatible or that such a system does not exist or may not be constructed.
Therefore, we add an additional ``x'' notation\index{Notation convention}
to represent an {\em a priori} potential compatibility between types
and furthermore a possible direction to be explored.


\begin{table}[htbp]
\begin{tabular}{|l||c|c|c|c|c|c|c|c|c||c|c|c||c|c|}
\hline
				&\multicolumn{14}{c|}{\bf Objective}\\
\hline
				&\multicolumn{9}{c||}{Type}
					&\multicolumn{3}{c||}{Dest./Use}			&\multicolumn{2}{c|}{Mode}\\
\hline
				&Au			&Me			&Po			&CP			&MV			&Dr			&Co			&CA			&ST
					&AR			&SP			&HI			&AG			&IG\\
\hline
\hline
\multicolumn{15}{|l|}{\bf Representation}\\
\hline
\hline
\multicolumn{15}{|l|}{\em Audio}\\
\hline
Waveform			&X			&			&			&			&			&			&			&			&x
					&X			&			&			&X			&x\\
\hline
Spectrum			&X			&			&			&			&			&			&			&			&X
					&X			&			&			&X			&x\\
\hline
\multicolumn{15}{|l|}{\em Symbolic}\\
\hline
\multicolumn{15}{|l|}{\em Concept}\\
\hline
Note				&			&X			&X			&X			&X			&			&X			&X			&X
					&			&X			&X			&X			&X\\
\hline
Rest				&			&X			&X			&X			&X			&X			&X			&X			&X
					&			&X			&X			&X			&X\\
\hline
Chord			&			&			&X			&X			&X			&			&			&X			&X
					&			&X			&X			&X			&x\\
\hline
Rhythm (Meter \& Beats)	&		&X			&X			&X			&X			&X			&X			&x			&X
					&			&X			&X			&X			&X\\
\hline
Drums			&			&			&X			&			&X			&X			&			&x			&x
					&			&X			&X			&X			&x\\
\hline
\multicolumn{15}{|l|}{\em Format}\\
\hline
MIDI				&			&X			&X			&X			&X			&X			&X			&x			&x
					&			&X			&X			&X			&x\\
\hline
Piano roll			&			&X			&X			&X			&X			&X			&X			&x			&X
					&			&X			&X			&X			&X\\
\hline
Text				&			&X			&X			&X			&			&X			&X			&			&x
					&			&X			&X			&X			&x\\
\hline
\multicolumn{15}{|l|}{\em Temporal Scope}\\
\hline
Global			&X			&X			&X			&X			&X			&X			&X			&x			&X
					&X			&X			&X			&X			&X\\
\hline
Time step			&X			&X			&X			&X			&X			&X			&X			&X			&x
					&X			&X			&X			&X			&x\\
\hline
\multicolumn{14}{|l|}{\em Meta-Data}\\
\hline
Note hold or ending	&			&X			&X			&X			&X			&X			&X			&x			&x
					&			&X			&X			&X			&X\\
\hline
No enharmony (Note denotation)	&	&X		&x			&x			&X			&			&X			&x			&x
					&			&X			&X			&X			&X\\
\hline
Fermata			&			&X			&X			&x			&X			&			&X			&x			&x
					&			&X			&X			&X			&X\\
\hline
Feature extraction	&x			&X			&x			&x			&x			&x			&x			&x			&x
					&X			&X			&X			&X			&x\\
\hline
\multicolumn{15}{|l|}{\em Expressiveness}\\
\hline
Tempo			&X			&x			&X			&x			&x			&x			&x			&x			&X
					&X			&X			&X			&X			&x\\
\hline
Dynamics			&X			&x			&X			&x			&x			&x			&x			&x			&X
					&X			&X			&X			&X			&x\\
\hline
\multicolumn{15}{|l|}{\em Encoding}\\
\hline
Value encoding		&X			&x			&x			&x			&x			&x			&x			&			&X
					&X			&X			&X			&X			&X\\
\hline
One-hot encoding	&X			&X			&			&X			&X$\times$n	&x$\times$n	&X$\times$n	&X$\times$n	&x
					&X			&X			&X			&X			&X\\
\hline
Many-hot encoding	&			&			&X			&x			&			&X			&x			&x			&X
					&			&X			&X			&X			&x\\
\hline
\multicolumn{15}{|l|}{\em Dataset}\\
\hline
Transposition		&			&X			&X			&X			&X			&			&X			&x			&x
					&			&X			&X			&X			&X\\
\hline
Alignment			&			&X			&X			&x			&x			&			&x			&X			&x
					&			&X			&X			&X			&x\\
\hline
\end{tabular}
\caption{Representation $\times$ Objective}
\label{table:analysis:representation:objective}
\end{table}



\begin{table}[htbp]
\begin{tabular}{|l||c|c|c|c|c|c|c|c|c||c|c|c||c|c|}
\hline
					&\multicolumn{14}{c|}{\bf Objective}\\
\hline
					&\multicolumn{9}{c||}{Type}
						&\multicolumn{3}{c||}{Dest./Use}			&\multicolumn{2}{c|}{Mode}\\
\hline
					&Au			&Me			&Po			&CP			&MV			&Dr			&Co			&CA			&ST
						&AR			&SP			&HI			&AG			&IG\\
\hline
\hline
\multicolumn{15}{|l|}{\bf Architecture}\\
\hline
\hline
Feedforward			&X			&X			&x			&x			&X			&x			&X			&x			&X
						&X			&X			&X			&X			&X\\
\hline
Autoencoder			&X			&X			&x			&x			&X			&X			&X			&			&
						&X			&X			&X			&X			&X\\
\hline
Variational				&x			&X			&X			&X			&X			&X			&			&			&
						&x			&X			&X			&X			&x\\
\hline
Restricted Boltzmann machine
					&x			&x			&X			&X			&x			&x			&			&			&
						&x			&X			&X			&X			&x\\
\hline
Recurrent (RNN)		&x			&X			&X			&X			&X			&X			&X			&X			&X
						&x			&X			&X			&X			&x\\
\hline
Convolutional			&X			&X			&X			&X			&X			&X			&X			&			&
						&X			&X			&X			&X			&x\\
\hline
Conditioning			&X			&X			&X			&X			&X			&X			&X			&x			&
						&X			&X			&X			&X			&x\\
\hline
Generative adversarial networks
					&x			&X			&X			&X			&X			&X			&			&			&
						&x			&X			&X			&X			&x\\
\hline
Reinforcement learning (RL)
					&x			&X			&X			&X			&X			&X			&			&			&
						&			&X			&X			&X			&x\\
\hline
\hline
\multicolumn{15}{|l|}{\bf Strategy}\\
\hline
\hline
Single-step feedforward	&X			&X			&x			&x			&X			&x			&X			&x			&X
						&X			&X			&X			&X			&x\\
\hline
Decoder feedforward		&X			&X			&x			&x			&x			&x			&x			&			&
						&X			&X			&X			&X			&X\\
\hline
Sampling				&X			&x			&x			&x			&X			&x			&X			&x			&
						&x			&X			&X			&X			&X\\
\hline
Iterative feedforward		&X			&X			&X			&X			&X			&X			&X			&X			&
						&			&X			&X			&X			&x\\
\hline
Input manipulation		&x			&x			&X			&x			&x			&x			&X			&			&X
						&x			&X			&X			&X			&x\\
\hline
Reinforcement			&			&X			&x			&x			&x			&x			&x			&			&
						&			&X			&X			&X			&x\\
\hline
Unit selection			&x			&X			&x			&x			&x			&x			&x			&			&
						&x			&X			&X			&X			&x\\
\hline
\end{tabular}
\caption{Architecture \& Strategy $\times$ Objective}
\label{table:analysis:architecture:strategy:objective}
\end{table}



\begin{table}[htbp]
\begin{tabular}{|l||c|c||c|c|c|c|c||c|c|c||c|c||c|c|c|c||c|c||c|c|c||c|c|}
\hline
	&\multicolumn{23}{c|}{\bf Representation}\\
\hline
	&\multicolumn{2}{c||}{Audio}
					&\multicolumn{5}{c||}{Concept}					&\multicolumn{3}{c||}{Format}	&\multicolumn{2}{c||}{TmpS}
		&\multicolumn{4}{c||}{Meta-Data}		&\multicolumn{2}{c||}{Expr.}
														&\multicolumn{3}{c||}{Encoding}
																				&\multicolumn{2}{c|}{DSet}\\
\hline
	&Wa		&Sp		&No		&Re		&Ch		&Rh		&Dr		&MI		&Pi		&Te		&Gl		&TS
		&NH		&NE		&Fe		&FE		&To		&Dy		&VE		&OH		&MH		&Tr		&Al\\
\hline
\hline
\multicolumn{24}{|l|}{\bf Objective}\\
\hline
\hline
\multicolumn{24}{|l|}{\em Type}\\
\hline
Au	&X		&X		&x		&x		&x		&x		&x		&x		&x		&x		&X		&x
		&		&		&		&x		&x		&x		&X		&X		&		&		&\\
\hline
Me	&		&		&X		&X		&		&X		&		&X		&X		&X		&X		&X
		&X		&X		&X		&X		&x		&x		&X		&X		&		&X		&X\\
\hline
Po	&		&		&X		&X		&X		&X		&		&X		&X		&X		&X		&X
		&X		&X		&X		&X		&X		&X		&X		&		&X		&X		&X\\
\hline
CP	&		&		&X		&X		&X		&X		&		&x		&X		&x		&X		&X
		&X		&X		&X		&x		&x		&x		&X		&X		&X		&X		&x\\
\hline
MV	&		&		&X		&X		&X		&X		&X		&X		&X		&X		&X		&X
		&X		&X		&X		&x		&x		&x		&X		&X$\times$n	&X	&X		&x\\
\hline
Dr	&		&		&		&X		&		&X		&X		&X		&X		&X		&X		&X
		&X		&		&X		&x		&x		&x		&x		&X		&X		&		&\\
\hline
Co	&		&		&X		&X		&x		&X		&		&x		&X		&X		&X		&X
		&X		&X		&X		&x		&x		&x		&x		&X		&x		&X		&x\\
\hline
CA	&		&		&X		&X		&X		&x		&		&x		&x		&x		&x		&X
		&X		&x		&x		&x		&x		&x		&x		&X$\times$n	&x	&x		&X\\
\hline
ST	&		&		&X		&X		&x		&X		&x		&x		&X		&x		&X		&x
		&X		&x		&x		&x		&X		&X		&X		&X		&x		&X		&x\\
\hline
\multicolumn{24}{|l|}{\em Destination \& Use}\\
\hline
AR	&X		&X		&x		&x		&x		&x		&x		&x		&x		&x		&X		&
		&		&		&		&x		&x		&x		&X		&X		&x		&		&\\
\hline
SP	&		&		&X		&X		&X		&X		&X		&X		&X		&X		&X		&X
		&X		&X		&X		&X		&X		&X		&X		&X		&X		&X		&X\\
\hline
HI	&		&		&X		&X		&X		&X		&X		&X		&X		&X		&X		&X
		&X		&X		&X		&X		&X		&X		&X		&X		&X		&X		&X\\
\hline
\multicolumn{24}{|l|}{\em Mode}\\
\hline
AG	&X		&X		&X		&X		&X		&X		&X		&X		&X		&X		&X		&X
		&X		&X		&X		&X		&X		&X		&X		&X		&X		&X		&X\\
\hline
IG	&x		&X		&X		&X		&x		&X		&x		&x		&X		&x		&X		&x
		&X		&X		&X		&x		&X		&X		&X		&X		&x		&X		&x\\
\hline
\hline
\multicolumn{24}{|l|}{\bf Architecture}\\
\hline
\hline
Fd	&X		&X		&X		&X		&X		&X		&X		&x		&X		&x		&X		&X
		&X		&X		&X		&x		&X		&X		&X		&X		&X		&X		&x\\
\hline
Ae	&x		&X		&X		&X		&x		&X		&X		&X		&X		&X		&X		&
		&X		&X		&X		&X		&X		&X		&X		&X		&x		&X		&x\\
\hline
Va	&x		&x		&X		&X		&x		&X		&X		&X		&X		&X		&X		&
		&X		&X		&X		&x		&X		&X		&x		&X		&x		&X		&x\\
\hline
RB	&x		&x		&X		&X		&X		&X		&x		&x		&X		&x		&X		&
		&X		&x		&X		&x		&X		&X		&x		&		&X		&X		&X\\
\hline
RN	&		&		&X		&X		&X		&X		&X		&X		&X		&X		&		&X
		&X		&X		&X		&x		&X		&X		&X		&X		&X		&X		&X\\
\hline
Cv	&X		&X		&X		&X		&X		&X		&x		&x		&X		&x		&X		&x
		&X		&x		&X		&x		&X		&X		&X		&X		&x		&X		&x\\
\hline
Cn	&X		&x		&X		&X		&X		&X		&X		&x		&X		&x		&X		&X
		&X		&X		&X		&X		&X		&X		&X		&X		&X		&X		&x\\
\hline
GA	&x		&x		&X		&X		&X		&x		&x		&X		&X		&x		&X		&
		&X		&x		&X		&x		&X		&X		&X		&X		&x		&X		&x\\
\hline
RL	&		&		&X		&X		&x		&x		&x		&x		&X		&x		&		&X
		&X		&x		&X		&x		&X		&X		&x		&X		&x		&X		&x\\
\hline
\hline
\multicolumn{24}{|l|}{\bf Strategy}\\
\hline
\hline
SF	&X		&x		&X		&X		&X		&X		&X		&X		&X		&X		&X		&
		&X		&X		&x		&x		&X		&X		&x		&X		&X		&X		&X\\
\hline
DF	&x		&X		&X		&X		&x		&X		&X		&X		&X		&X		&X		&
		&X		&X		&X		&x		&X		&X		&X		&X		&x		&X		&x\\
\hline
Sa	&X		&x		&X		&X		&X		&X		&X		&X		&X		&X		&X		&X
		&X		&X		&X		&x		&X		&X		&X		&X		&X		&X		&X\\
\hline
RF	&X		&		&X		&X		&X		&X		&X		&X		&X		&X		&		&X
		&X		&X		&X		&X		&X		&X		&X		&X		&X		&X		&X\\
\hline
IF	&X		&		&X		&X		&X		&X		&X		&X		&X		&X		&		&X
		&X		&X		&X		&X		&X		&X		&X		&X		&X		&X		&X\\
\hline
IM	&x		&X		&X		&X		&x		&x		&X		&X		&X		&X		&X		&
		&X		&x		&x		&x		&X		&X		&X		&X		&x		&X		&x\\
\hline
Re	&		&		&X		&X		&x		&x		&X		&X		&X		&X		&		&X
		&X		&x		&X		&x		&X		&X		&x		&X		&x		&X		&x\\
\hline
US	&x		&x		&X		&X		&x		&x		&X		&X		&X		&X		&		&X
		&X		&x		&x		&X		&X		&X		&X		&X		&x		&X		&x\\
\hline
\end{tabular}
\caption{Objective \& Architecture \& Strategy $\times$ Representation}
\label{table:analysis:architecture:strategy:representation}
\end{table}





\begin{table}[htbp]
\begin{tabular}{|l||c|c|c|c|c|c|c|c|c|}
\hline
					&\multicolumn{9}{c|}{\bf Architecture}\\
\hline
					&Fd			&Ae			&Va			&RB			&RN			&Cv			&Cn			&GA			&RL\\
\hline
\hline
\multicolumn{10}{|l|}{\bf Strategy}\\
\hline
\hline
Single-step feedforward	&X			&			&			&			&			&X			&X			&X			&\\
\hline
Decoder feedforward		&			&X			&X			&			&			&x			&X			&x			&\\
\hline
Sampling				&X			&X			&X			&X			&X			&X			&X			&X			&X\\
\hline
Iterative feedforward		&X			&			&X			&			&X			&X			&X			&X			&\\
\hline
Input manipulation		&X			&X			&x			&X			&x			&X			&x			&x			&\\
\hline
Reinforcement			&			&			&			&			&X			&			&x			&			&X\\
\hline
Unit selection			&			&X			&			&			&X			&			&x			&			&\\
\hline
\end{tabular}
\caption{Strategy $\times$ Architecture}
\label{table:analysis:strategy:architecture}
\end{table}







\begin{table}[htbp]
\begin{tabular}{|l||c|c|c|c|c|c|c|c|c|c|c|c|}
\hline
	&\multicolumn{12}{c|}{\bf Challenge}\\
\hline
						&EN		&LV		&CV		&Es		&MH		&Co		&St		&Or		&Ic		&It		&Ad		&Ey\\
\hline
\hline
\multicolumn{13}{|l|}{\bf Architecture}\\
\hline
\hline
Feedforward				&X		&		&		&		&		&X		&		&		&		&X		&		&\\
\hline
Autoencoder				&X		&		&X		&		&		&X		&		&		&		&X		&		&\\
\hline
Variational					&X		&		&X		&		&		&X		&		&		&		&X		&		&\\
\hline
Restricted Boltzmann machine (RBM)
						&X		&		&X		&		&		&X		&		&		&		&		&		&\\
\hline
Recurrent (RNN)			&X		&X		&X		&		&X		&X		&		&		&X		&X		&		&\\
\hline
Convolutional				&		&		&		&		&		&		&		&		&		&		&		&\\
\hline
Conditioning				&		&		&		&		&X		&X		&		&X		&		&		&		&\\
\hline
Generative adversarial networks (GAN)
						&X		&		&X		&		&		&x		&		&X		&		&		&		&\\
\hline
Reinforcement learning (RL)	&X		&X		&X		&		&		&X		&		&x		&X		&X		&X		&\\
\hline
\hline
\multicolumn{13}{|l|}{\bf Strategy}\\
\hline
\hline
Single-step feedforward		&		&		&		&		&		&		&		&		&		&		&		&\\
\hline
Decoder feedforward			&X		&		&		&		&		&X		&		&		&		&		&		&\\
\hline
Sampling					&X		&		&X		&		&		&X		&		&x		&X		&X		&		&\\
\hline
Iterative feedforward			&X		&X		&		&		&		&X		&X		&		&X		&x		&		&\\
\hline
Input manipulation			&X		&		&X		&		&X		&X		&X		&		&x		&x		&		&\\
\hline
Reinforcement				&X		&X		&X		&		&		&X		&X		&		&X		&x		&X		&\\
\hline
Unit selection				&X		&X		&X		&		&		&X		&X		&		&X		&x		&		&\\
\hline
\end{tabular}
\caption{Architecture \& Strategy $\times$ Challenge}
\label{table:analysis:architecture:strategy:challenge}
\end{table}

\clearpage

The analysis of the correlation tables allows us to draw a few first observations
about some design decisions and their consequences:

\begin{itemize}

\item audio versus symbolic representation
	(Table~\ref{table:analysis:architecture:strategy:representation});

\item global versus time step temporal scope representation
	(Tables~\ref{table:analysis:architecture:strategy:representation} and~\ref{table:analysis:strategy:architecture}).
	A global temporal scope representation is usually coupled with:
	a) a feedforward architecture and a single-step feedforward strategy,
	or b) an autoencoder architecture with a decoder feedforward strategy.
	A time step temporal representation is usually coupled with a recurrent architecture and
	an iterative feedforward strategy.
	However, there are some exceptions
	(e.g., Time-Windowed\index{Time-Windowed} in Section~\ref{section:experiment:todd:time:windowed}
	and BLSTM\index{BLSTM} in Section~\ref{section:experiment:blstm:chord});
	and

\item accompaniment objective versus seed-based generation
	(Table~\ref{table:analysis:architecture:strategy:objective}).
	DeepBach\index{DeepBach} is a notable exception, because thanks to its {\em sampling only strategy},
	it can generate an accompaniment as well as a complete chorale
	(see the discussion in Section~\ref{section:explainability:example:bachbot}).



\end{itemize}


Note that in Table~\ref{table:analysis:architecture:strategy:challenge},
the columns corresponding to the expressiveness
and explainability challenges are left empty.
This is because these challenges are not to be solved with the lone choice of an architecture or a strategy.

We do not comment further these tables in this book.
We consider them as a first version of analysis tools related to our proposed conceptual framework
which could be tried out and improved, for investigating current as well as future systems.




%% file: discussion-conclusion.tex
\chapter{Discussion and Conclusion}
\label{section:chapter:discussion:conclusion}

\abstract*{Chapter~\ref{section:chapter:discussion:conclusion} Discussion and Conclusion is the last chapter of the book.
It revisits some design decision issues introduced during the analysis of challenges and strategies
and it discusses some related prospects,
before concluding by wrapping up the contributions presented in this book.}

\label{section:discussion}

%
%

\label{section:discussion:good:practices}

We now revisit some design decision issues raised through our analysis and discuss related prospects.

\section{Global versus Time Step}
\label{section:discussion:global:vs:sequence}

As we have seen
in Sections~\ref{section:challenges:strategies:first:discussion} and~\ref{section:challenges:strategies:incrementality:strategies},
one important decision is to choose between the two main types of temporal scope representation:

\begin{itemize}

\item global, including all time steps --
typically coupled with a feedforward\index{Feedforward!neural network} or an autoencoder\index{Autoencoder} architecture; and

\item time step\index{Time!step}, representing a single time step --
typically coupled with a recurrent (neural network)\index{Recurrent!neural network} (RNN\index{RNN}) architecture\footnote{In general,
	the granularity of the time step\index{Time!step} is set at the level of the smallest notre duration,
	as discussed in Section~\ref{section:representation:temporal:granularity}.
	Time-Windowed and BLSTM architectures
	(respectively, Sections~\ref{section:experiment:todd:time:windowed} and~\ref{section:experiment:blstm:chord})
	are both peculiar cases of a coarse grained time step (respectively, one and four measures long).
	Time-Windowed architecture has the additional specificity that it is a feedforward and not a recurrent architecture.}.

\end{itemize}

The pros and cons are as follows:

\begin{itemize}

\item global

\begin{description}[+/--]

\item[+] allows arbitrary output, e.g., for the objective of generating some accompaniment\index{Accompaniment}
through the single-step feedforward strategy\index{Single!-step feedforward strategy} on a feedforward architecture,
as, for example, in the MiniBach\index{MiniBach} system (Section~\ref{section:experiment:mini:bach});

\item[+] supports incremental\index{Incremental} instantiation (via sampling\index{Sampling}),
as, for example, by the DeepBach\index{DeepBach} system (Section~\ref{section:experiment:deep:bach});

\item[--] does not allow variable length\index{Variable!length} generation;

\item[--] does not allow seed-based generation\index{Seed!-based generation} for a feedforward architecture;

\item[+] allows seed-based generation
through the decoder feedforward strategy\index{Decoder!feedforward strategy} on an autoencoder architecture,
as, for example, in the DeepHear\index{DeepHear} system (Section~\ref{section:experiment:deep:hear});

\end{description}

\item time step

\begin{description}[+/--]


\item[+/--] supports incremental instantiation, but only forward in time,
through the iterative feedforward strategy\index{Iterative feedforward strategy} on a recurrent network architecture;

\item[+] allows variable length generation, as, for example, in the CONCERT\index{CONCERT} system (Section~\ref{section:experiment:concert}).

\end{description}

\end{itemize}

Actually some attempt at combining ``the best of both worlds'' seems to lie in using an RNN Encoder-Decoder\index{RNN Encoder-Decoder} architecture, as

\begin{itemize}

\item generation is iterative,
which allows variable length content generation,

\item while allowing arbitrary output generation,
as the output sequence may have
an arbitrary length and content\footnote{As is the case
	for translation\index{Translation} tasks.},
and

\item allowing the manipulation of a global temporal scope representation
(the latent variables\index{Latent!variable}).

\end{itemize}

Moreover, in a variational\index{Variational!autoencoder} version, such as, for example, VRAE\index{VRAE},
the latent space\index{Latent!space} could be explored in a disciplined and meaningful manner,
as, for example, in the GLSR-VAE\index{GLSR-VAE} and MusicVAE\index{MusicVAE} systems
(Sections~\ref{section:experiment:glsr:vae} and~\ref{section:system:music:vae}). 

Note also that an alternative to a recurrent architecture is a convolutional architecture applied over the time dimension,
as discussed in the (next) Section~\ref{section:discussion:convolution:vs:recurrent}.

\section{Convolution versus Recurrent}
\label{section:discussion:convolution:vs:recurrent}

As noted in Section~\ref{section:architecture:convolution},
convolutional architectures\index{Convolutional!architecture}, while prevalent for image\index{Image} applications,
are more seldom used than recurrent neural network (RNN\index{RNN}) architectures\index{Recurrent!neural network} in music applications.
The few examples of nonrecurrent architectures using convolution\index{Convolution} on the time\index{Time} dimension
that we have encountered and analyzed are

\begin{itemize}

\item WaveNet\index{WaveNet}, a convolutional feedforward architecture for audio (Section~\ref{section:systems:wavenet});

\item MidiNet\index{MidiNet}, a GAN architecture encapsulating conditional convolutional architectures
	(Section~\ref{section:systems:midinet})\footnote{A comparison between
	MidiNet and C-RNN-GAN\index{C-RNN-GAN}, both using GANs
	but encapsulating a convolutional network versus a recurrent network, is also interesting.}; and

\item C-RBM\index{C-RBM}, a convolutional RBM (Section~\ref{section:experiment:c:rbm}).

\end{itemize}

Let us try to list and analyze the relative pros of cons of using recurrent architectures or convolutional architectures to model time
correlations:

\begin{itemize}

\item recurrent networks are popular and accurate, especially since the arrival of LSTM architectures;

\item convolution should be used {\em a priori} only on the time dimension because,
as opposed to images where motives\index{Motif} are invariant\index{Invariance} in all dimensions,
in music the pitch\index{Pitch} dimension is {\em a priori} not metrically invariant\footnote{Otherwise
	this could break the notion of tonality, see the rationale for the C-RBM system in Section~\ref{section:systems:c-rbm}.
	However, the system analyzed in Section~\ref{section:experiment:hexahedria}
	considers recurrence on the pitch class dimension in order to model simultaneous notes (chords).};

\item convolutional networks\index{Convolutional!network} are typically faster\index{Fast} to train\index{Training} and easier to parallelize
than recurrent networks\index{Recurrent!network} \cite{oord:pixel:cnn:decoders:arxiv:2016};

\item using convolution on the time dimension in place of using a recurrent network implies the multiplication of the number of input variables
by the number of time steps considered
and thus leads to a significant augmentation
of the volume of data to process and of the number of parameters to adjust\footnote{For that reason,
	recurrent networks are still the norm for learning time series of multi-dimensional data like, for example, 2-D images for weather prediction.};
	
\item sharing weights by convolutions only applies to a small number of temporal neighboring members of the input,
in contrast to a recurrent network that shares parameters in a deep way, for all time steps (see Section~\ref{section:architecture:convolution});

\item the authors of WaveNet\index{WaveNet}
argue that the layers of dilated convolutions allow the receptive field to grow longer in a much cheaper way than using LSTM\index{LSTM} units;

\item the authors of MidiNet\index{MidiNet}
argue that using a conditioning strategy for a convolutional architecture allows the incorporation of information
from previous measures\index{Measure} into intermediate layers\index{Layer}
and therefore considers history\index{History} as a recurrent network\index{Recurrent!network} would do.

\end{itemize}


A potential output of this initial comparative analysis is that,
as there are many systems using nonrecurrent architectures
(like feedforward networks\index{Feedforward!network} or autoencoders\index{Autoencoder}),
it may be interesting to study whether their extension into a convolutional architecture
on the time dimension could bring some effective gain, in terms of efficiency and/or accuracy.

Actually, this issue of using convolutional versus recurrent architecture is recently being challenged by the introduction of a novel architecture,
named Transformer\index{Transformer} \cite{vaswani:attention:transformer:arxiv:2017},
with an objective similar to that of a RNN Encoder-Decoder\index{RNN Encoder-Decoder} architecture
(introduced in Section~\ref{section:architecture:compound:recurrent:autoencoder}).
This architecture does not use convolutions or recurrence and is only based on an {\em attention mechanism\index{Attention mechanism}} (Section~\ref{section:architecture:attention:mechanism}).
A very recent\footnote{Too late to
	analyze it thoroughly in this book.}
application to music generation, named MusicTransformer\index{MusicTransformer}
has been presented in \cite{huang:music:transformer:arxiv:2018},
with apparent very good results about its capacity to
model long-term structure.

\section{Style Transfer and Transfer Learning}
\label{section:discussion:transfer}

Transfer learning\index{Transfer learning} is an important issue for deep learning and machine learning in general.
As training can be a tedious process,
the issue is to be able to {\em reuse\index{Reuse}}, at least partially,
what has been learnt in one context\index{Context} and use it in other contexts.
Various cases may be considered, e.g., similar source and target domains, similar task\index{Task}, etc.
This new research subdomain, named {\em transfer learning},
is about methodologies and techniques for the transfer of what has been learnt
\cite[Section~15.2]{goodfellow:deep:learning:book:2016}.

We have not addressed this important issue in our analysis
because it has not yet been specifically addressed for music generation,
although we think that it will become an area of investigation.
Meanwhile, an example,
although still simplistic,
is the way the DeepHear\index{DeepHear} architecture and what it has learnt is transfered from the objective of generating a melody to the objective of generating a counterpoint (see Section~\ref{section:experiment:deep:hear:harmonize}).
Another example is the way the DeepBach\index{DeepBach} architecture allows to adapt the objective
from {\em ex nihilo} chorale generation to multi-voice accompaniment (see Section~\ref{section:experiment:deep:bach}).

Last, let us remember
that style transfer\index{Style!transfer}
is a very specific case of transfer learning in terms of objective and techniques
(see Section~\ref{section:style:transfer:vs:transfer:learning})

%


\section{Cooperation}
\label{section:discussion:cooperation}

All the systems surveyed are basically lone systems (although the architecture may be compound).
A more cooperative approach is natural for handling complexity, heterogeneity, scalability and openness,
as, for example, pioneered by multi-agent systems \cite{wooldridge:intro:mas:book:2009}.

An example
is the system proposed by Hutchings and McCormack \cite{hutchings:autonomous:agents:evomusart:2017}.
It is composed of two agents:

\begin{itemize}

\item a {\em harmony} agent, based on an RNN (LSTM) architecture, in charge of the progression of chords; and

\item a {\em melody} agent, based on a rule-based system, in charge of the melody.

\end{itemize}

The two agents work in a cooperative way and alternate between leading and accompanying roles
(inspired by, for example, the way musicians function in a jazz band).
The authors relate the interesting dynamics between the two agents and also an intereresting balance between harmonic creativity and harmonic consistency\footnote{On this issue,
	see Section~\ref{section:originality}.}.
This approach appears to be an interesting direction to pursue and extend with more agents and roles.

\section{Specialization}
\label{section:discussion:specialization}

A general issue is the hyper-specialization\index{Specialization} of systems designed for a specific objective and/or a specific type of corpus.
This is witnessed by the diversity of the architectures and approaches surveyed.
Note that this is a known issue for Artificial Intelligence\index{Artificial!intelligence} (AI\index{AI}) research in general.
There is some tendency towards hyper-specialized systems solving specific problems,
especially in the case of competitions organized by conferences or other institutions,
with the risk of loosing the initial objective
of a general problem solving framework\footnote{An interesting counterexample
	is the ongoing research and competition about general game playing
	\cite{genesereth:ggp:ai:magazine:2013}.}.

Meanwhile, the general objective of generating interesting musical content is complex and still an opened issue.
Thus, we need to work both on general approaches for general problems and specific approaches for specific subproblems,
as well as top-down and bottom-up approaches,
while not losing interest in how to interpret, generalize and reuse advances and lessons learnt\footnote{We hope
	that the survey and analysis conducted in this book 
	will contribute to this research agenda.}.

\section{Evaluation and Creativity}
\label{section:discussion:evaluation}

Evaluation\index{Evaluation} of a system generating music mostly
consists in a qualitative evaluation of examples of generated music\footnote{About the possibility
	for more systematic objective criteria for evaluation,
	we can for example look at the analysis by Theis {\em et al.} for the case of image generation \cite{theis2015note}.
	The authors state that an evaluation of image generative models is {\em multicriteria\index{Multicriteria}}
	via different possible metrics, such as log-likelihood, Parzen window estimates, or qualitative visual fidelity,
	and that a good result with respect to one criterion does not necessarily imply a good result with respect to another criterion.}.
For many experiments, evaluation\index{Evaluation} is only preliminary, and in many cases, only conducted by the designers themselves.
There are of course some exceptions, with more systematic and external evaluations (by a more or less expert public).

When the corpus is very precise, e.g., in the case of J. S. Bach's\index{Bach} chorales
for the BachBot\index{BachBot} or the DeepBach\index{DeepBach} systems
(Sections~\ref{section:explainability:example:bachbot}
and~\ref{section:experiment:deep:bach}),
a Turing test\index{Turing!test} may be conducted:
a piece of music being presented to the public who has to guess if it is one of the original pieces
or music generated by a computer.
But this methodology is more limited when the objective is not to generate music highly conformant to a relatively narrow style (and corpus),
as in the case of J. S. Bach chorales\footnote{Bach chorales, and more generally speaking Bach music,
	are often used for experiments and evaluation,
	because the corpus
	is quite homogeneous regarding a given style (e.g., preludes, chorales\ldots)
	as well as quality.
	It also fits particularly well
	with algorithmic composition, of which Bach was somehow a precursor.},
but to generate more creative music.


Moreover, if we consider as a general objective for a system
the capacity to assist composers and musicians\footnote{As, for instance,
	pioneered by the FlowComposer prototype,
	introduced in Section~\ref{section:systems:flow:composer}.},
rather than to autonomously generate music
(see Section~\ref{section:introduction:motivation:assistance:versus}),
we should maybe consider as an evaluation criteria the satisfaction of the {\em composer}
(notably, if the assistance of the computer allowed him to compose and create music that he may consider not having been possible otherwise),
rather than the satisfaction of the {\em auditors} (who remain too often guided by some conformance to a current musical trend).

Some fundamental limitation is that there is no clear objective function associated to creativity\index{Creativity} and to art quality.
Therefore, the selection of a musical corpus used as a set of training examples
is a first fundamental step and decision\footnote{As noted
	in Section~\ref{section:representation:dataset}.}.
But in order to be able to learn something interesting, a relatively coherent/homogenous
corpus needs to be selected\footnote{Constructing a corpus with the ``best considered'' musical pieces,
	independently of the style (classical, jazz, pop, etc.)
	-- as could do a museum or an exhibition presenting in a single room
	its best artefacts of different nature and origin --,
	is not likely to produce interesting results
	because such a corpus is too much sparse and heterogeneous.}.
This will unfortunately favor the quality (actually, the conformance)
of the generated musical content regarding the learnt style,
rather than its intrinsic quality (interest).

Current experiments and directions to promote creativity rely mostly on constraints to avoid plagiarism and/or heuristics
to incentive a generation outside the
``comfort zone'' that the deep architecture has learned from the corpus,
while balancing elements of surprise with predictability/understandability.
Such creativity control may be applied during the training phase
(the case of the CAN architecture, see Section~\ref{section:challenges:strategies:originality:can}),
or (for most types of control, see Sections~\ref{section:challenges:strategies:control}
and~\ref{section:challenges:strategies:originality})
during the generation phase.

Some alternative (and complementary) direction to better model such an element of surprise
could be to include a model of an artificial listener with some model of expectation,
e.g., which could evaluate how well a new generated content could be ``explained'' in terms of references to
an existing memory,
as proposed, e.g., in \cite{dubnov:delegating:creativity:digital:da:vinci:2014}.

Last, some additional fundamental limitation is that current deep learning techniques
for learning and generating music are based on artefacts,
actual musical data, independently of the processes and the culture
that have led to them.
If we want to envision more profound systems,
it is likely that we will have to incorporate some modeling of the context and the process leading
to musical artefacts and not so the artefacts themselves.
Indeed, when considering art\index{Art} history, creation takes place within a historico-cultural context with refinements\footnote{E.g.,
	the extension of classical harmony based on triads (only root, third and fifth) to extended chords.}
as well as possible ruptures\footnote{E.g.,
	movements like dodecaphonism or free jazz.}.
One possible direction would then be to not just model content generation from a frozen artistic corpus outside of its history,
but to try to model a more dynamical process of creation including the historico-cultural context\footnote{The modeling of the context\index{Context}
	is one of the limitations of current deep learning architectures and is a topic of ongoing research.
	An illustrating real counterexample
	is the case of a Chinese woman (chairwoman of China's biggest air conditioners maker)
	who had found her face displayed in 2018 in the port city of Ningbo on a huge screen
	that displays images of people caught jaywalking by surveillance cameras.
	It was then found that the artificial intelligence-backed monitoring system
	had captured her face from an advertisement on the side of a moving bus
	\cite{tao:scmp:recognition:jaywalking:2018}.},
with the history and dynamics maybe addressed by recurrent architectures and/or reinforcement learning,
although this appears yet a long way to go.

\section{Conclusion}
\label{section:discussion:conclusion}

The use of deep learning techniques for the creation of musical content, and more generally speaking creative artistic content,
is nowadays getting increased attention.
This book presented a survey and an analysis of various strategies and techniques for using deep learning to generate musical content.
We have proposed a multi-criteria conceptual framework based on five dimensions: objective, representation, architecture, challenge and strategy.
We have analyzed and compared various systems and experiments proposed by various researchers in the literature.
We hope that the conceptual framework provided in this book will help in understanding the issues
and in comparing various approaches for using deep learning for music generation,
and therefore contribute to this research agenda.